\documentclass[12pt]{book}

\usepackage{axodraw}
\usepackage{pstricks}
\usepackage{color}
\usepackage{epsfig}
\usepackage{graphicx}
\usepackage{amssymb}
\usepackage{amsmath}
\usepackage{amsfonts}
\usepackage{sw20nmsu}
\usepackage[center,small,md]{caption2}
\usepackage{draftstamp}
\input{tcilatex}

\begin{document}
\begin{thesistitle}
{\Large Production of Charged and Neutral Higgs Bosons with Charginos and Neutralinos through different Propagators in the \it {MSSM}}
\end{thesistitle}

\begin{thesisauthor}
Hatim Hegab Ali Hegab
\end{thesisauthor}

\begin{dissertation}
Master of Science, Physics.
\end{dissertation}

\begin{MajorSubject}
Physics
\end{MajorSubject}

\begin{MinorSubject}
Theoretical High Energy Physics
\end{MinorSubject}

\begin{auniversity}
Cairo University

Giza

Egypt
\end{auniversity}

\begin{thesisdate}
March, 2006
\end{thesisdate}

\begin{tcopyright}
Copyright 2006 by Hatim Hegab

All rights reserved
\end{tcopyright}

\begin{committeesigna}
Prof. M. M. El-Khishen

Prof. of Theoretical Physics,

Faculty of Science,

Cairo University.
\end{committeesigna}

\begin{committeesignb}
Dr. Tareq Abdul Aziem

Lecturer,

Faculty of Science,

Cairo University.
\end{committeesignb}

\begin{approvaldate}
March, 2006
\end{approvaldate}

\begin{committeeincharge}
Prof. Dr. M. M. El-Khishen, Chair

\ \ \ \ \ \ \ Prof. of Theoretical Physics,

\ \ \ \ \ \ \ Faculty of Science,

\ \ \ \ \ \ \ Cairo University.

Dr. Tareq Abdul Aziem

\ \ \ \ \ \ \ lecturer,

\ \ \ \ \ \ \ Faculty of Science,

\ \ \ \ \ \ \ Cairo University.
\end{committeeincharge}

\begin{dedication}
This work is dedicated to my first teacher of mathematics, \textbf{My Mother}.\\%
 And to my father, the late \textbf{Prof. Hegab Ali Hegab, ex.
Prof. of Microbiology at The Technical Agricultural Institute in
Kaseem, Bureidah, Saudi Arabia} whose guidance, instructions and kind heart helped to shape and build my personality.\\
And finally, to my wife, and my daughhter, \textbf{Basma (A ``Smile''  in the Arabic Language)}, your existence in my life gave me the power to do much more than I could do alone.
\end{dedication}

\begin{acknowledgments}
The author would like to thank all those people who helped make
this work possible, I here want to express my deepest gratitude to
the following people for their help,

\begin{enumerate}
\item Prof. M. M. El-Khishen, for his kind supervision and his
helpful manners.

\item Dr. Tareq Abdul Aziem.

\item Prof. Alexander Pukhov, for his fruitful discussions, cooperative manners, without his help, I wouldn't be able to do this work.

\end{enumerate}
\end{acknowledgments}

\begin{chair}
Dr. Tarek Abdul Aziem
\end{chair}

\begin{generalabstract}
In this work, the following reactions have been studied;

\begin{enumerate}
\item $e^- e^+ \rightarrow
H^- \widetilde{\chi }_{1}^+ \widetilde{\chi }_{1}^o$

\item $e^{-}e^{+}\rightarrow h\widetilde{\chi }%
_{1}^{+}\widetilde{\chi }_{1}^{-}$

\item $e^{-}e^{+}\rightarrow
h\widetilde{\chi }_{1}^{o}\widetilde{\chi }_{1}^{o}$

\item $e^{-}e^{+}\rightarrow
hH^{+}H^{-}$.

where $\widetilde{\chi }_{1}^{\pm }$ is the chargino and $\widetilde{\chi }_{1}^{0}$ is the neutralino.
$h$ represent the lightest neutral Higgs boson, and $H_3$ (sometimes refered to as $A$) is the pseudoscalar Higgs boson. 
\end{enumerate}

The work went to calculate the differential cross section due to each
Feynman diagram, then, the \emph{total cross section} for each
reaction is calculated according to a carefully used set of
parameters. Results are graphed and tabulated.
\end{generalabstract}

\begin{toc}
\end{toc}

\part{Theoretical Framework}

\begin{center}
{\huge Introduction}
\end{center}

Progress in theoretical physics, as in all sciences, has almost
always been based on an interplay of two quite different
approaches of nature.

One starts by collecting and ordering observational, or
experimental, data \textquotedblleft\ Tycho\textquotedblright ,
then describes these data by a small number of empirical laws
\textquotedblleft\ Kepler\textquotedblright , and finally
\textit{explains} these laws by a \textbf{theory, }based on a
few principles \textbf{\textquotedblleft\ Newton\textquotedblright . }%
Theoretical predictions\textbf{, }or the outcome, of further, more
refined, observations and experiments can then be made (
\textquotedblleft\ discovery of Neptune\textquotedblright\ ).

the other approach starts from an idea, formulates it in terms of
a theory, and proceeds to make predictions which then acts as a
test of the theory and of its original idea. The later approach --
\ in its pure form-- has been most dramatically and singularly
successful in Einstein's development of the Theory of General
Relativity (TGR). \textbf{Su}per\textbf{sy}mmetry (\ \textbf{SUSY\
}) has started from an idea, and at the moment a huge work is
going on to confirm this idea.

Modern particle physics, in seeking a single unified\ theory of
all elementary particles and their fundamental interactions,
appear to be reaching the limits of this process and finds itself
forced, in part and often reluctantly, to revert to guidelines to
the medieval principles of symmetry and beauty.

Supersymmetric theories are highly symmetric and very beautiful.

They are remarkable in that they unify fermions (matter) with
bosons ( force carriers), either in flat space (SUSY) or in curved
space, supergravity, (SUGRA). \textbf{Su}per\textbf{Gra}vity,
\textbf{( SUGRA\ )} naturally unifies gravity with other
interactions. None of the present model theories is in any sense
complete; the hurdles on the way to experimental predictions-- and
thus to acceptance or rejection-- have not yet been cleared. What
naive immediate predictions can be made seem to be in disagreement
with nature. Yet, this particular field of research appears to
promise solutions of so many outstanding problems that it has
excited enthusiasm in large parts of the physics community ( and
equally large doubts in others). In a truly philosophical spirits
has been even said about the theory \textbf{that it is so
beautiful so it is hard to be incorrect. }

In high energy physics, or as it is sometimes called- elementary
particle physics, the hope is that we will eventually\ achieve a
unified scheme which combines all particles and all interactions
into one consisting theory. We wish to make further progress on
the way which started with Maxwell's unification of the magnetism
and electrostatics, and which has more recently led to unified
gauge theories (UGT) of electromagnetic and the weak, and perhaps
also of the strong interactions.

\textsl{Supersymmetry }is, by definition, a symmetry between
fermions and bosons. A supersymmetric field theoretical model
consists of a set of quantum fields and of a Lagrangian for them
which exhibit such a symmetry. The Lagrangian determines, through
the Action Principle, the equations of motion and hence the
dynamic behavior of the particles. A supersymmetric model which is
covariant under general coordinate transformations, or equally, a
model which posses a local ( \textquotedblleft\
gauged\textquotedblright\ ) supersymmetry is called a supergravity
model. Supersymmetric theories describe model worlds of particles,
created from the vacuum by the fields, and the interactions
between these particles. The supersymmetry manifests itself in the
particle spectrum and in the stringent relationship between
different interaction processes even if these processes involve
particles of different spin and statistics.

Both supersymmetry and supergravity aim at unified description of \textit{%
fermions} and \textit{bosons,} and hence of matter and
interactions. Supergravity is particularly promising at its
attempt to \textit{unify gravity with other interactions. }All
supersymmetric models succeed to some extent in these aims, but
they fail in describing the real world as we live in and
experience it and thus are models not theories. We are still to
struggle to find some contact between one of the models and the
physical world, reality, so that the model could become an
underlying theory for nature ant its \textquotedblleft\ most
fundamental level\textquotedblright .

By \textquotedblleft\ most fundamental level\textquotedblright\ we
mean at present the decomposition of matter into quarks and
leptons (fermions) and the understanding of all forces between
them as arising out of four types of basic interactions, namely,
\textbf{gravitational, weak, electromagnetic, and strong}. These
are described in terms of exchanged particles (bosons ). The
framework within which these building blocks make up a physical
theory is \textit{Relativistic Quantum Field Theory. }Seen at this
level, \textquotedblleft\ unification\textquotedblright\ ought to
include all four interactions. There is , however, a quantitative
and a qualitative difference between the gravitational interaction
and the others which has had profound consequences on both the
structure of the universe and on our understanding of it.

Unification of gravity with other forces is an illusive goal.
Since gravity is always attractive and a long range force, it
would make all complex and large physical objects collapse under
their own weight, however, it is not stronger enough at short
distances to have any remarkable effect with respect to other
forces, [\ interactions].

This difference in strength, necessary to the universe as we see,
has in turn set gravity so far apart from the rest of physics that
it is difficult to think of any experiment which could actually
test predictions of a unified field theory of all interactions,
and even less of one that could provide experimental input into
the construction of such a theory. The natural domain of Newtonian
gravity and of its modern replacement , Einstein's TGR, is the
world of large distances and massive accumulations of matter, that
of the other forces is the world of atoms, nuclei, and elementary
particles. A very large Order of magnitude separate them.

The \textit{strong, electromagnetic and weak interactions }are
fairly understood today. It has been found that their exchange
particles arise naturally in a quantum field theory, if that field
theory is required to be locally gauge invariant.

The theory for both gravitation and elementary particles'
interactions are well established within their respective domains.
On the submicroscopic level, for the masses and distances
involved, the deviation introduced by gravity from the flat
Minkowiskian metric are so minute that elementary particles, or
atoms, can safely be treated as if gravitation does not exist. Any
\textquotedblleft \textbf{\ true}\textquotedblright , that is,
generally covariant, theory should thus be closely approximated by
Lorentz--covariant, non-gravitational theories. We must, however
demand of the \textit{true} \ theory to be mathematically
consistent and that it predicts the correct flat limit. Any
quantum theory of gravitation so far fails to do so.

The energy at which gravity and quantum effects become of
comparable strength can be estimated from only an expression with
a dimension of an energy that can be formed from the constants of
nature $\hbar ,c,G,$\bigskip

\begin{equation*}
E_{Planck}=c^{2}\sqrt{\frac{\hbar c}{G}}\simeq 10^{19}GeV
\end{equation*}

It is in the region of this energy, \textit{Planck's energy},
where our present theories for the gravitational and other
interactions become incompatible with each other and where we
expect a possible unification of the interactions to become
manifest. A point particle with Plank mass would have a
Schwarzschild radius equal to twice its Compton wavelength. The
very remoteness of such an energy region eliminates all hope for a
direct experimental proof. Perhaps, if we are lucky enough, some
isolated prediction of such a unified theory could be testable on
a system that integrates a minute effect over a vast range (
proton decay experiments are of this type, where large numbers of
available protons can make the very small probabilities for decay
available and measurable). We can, however, not expect
experimental physics to give us much reliable guidance into the
\textit{Plank region.}

The $SU(3)\times SU(2)\times U(1)$ picture of the
non-gravitational forces is not yet a \textit{unified} one. The
only property which unites them is that they are each described by
a gauge theory. The fact that the direct product structure of
$SU(2)\times U(1)$ is \textquotedblleft skew\textquotedblright\
(by the weak mixing angle) against the natural distinction between
the weak and electromagnetic force may suggest some underlying
unified scheme which can predict that angle. A lot of work has.
over many years, gone into finding a larger gauge group which
would describe all three interactions at some high energy. If such
a grand unification occurred it is known that it must happen at
energies of about $10^{15}GeV,$ only four orders of magnitude less
than $E_{Planck}.$\textit{Grand Unified Theories} (GUTs) have had
some success (such as the prediction of the mixing angle) and some
failures (such as proton life time experiments). In any case, even
a GUT would at most unify different kinds of interactions (strong
and electroweak) with each other and different kinds of matter
(quarks and leptons) with each other. the unification of matter
with interactions is not one of the aims of GUTs.

What is it then that points in the direction of supersymmetric
theories for a \textit{solution to the unification problem?}
Already the most obvious difference between gravity and, say,
electrodynamics, namely the absence of negative gravitational
charges, can be shown to imply that only a supersymmetric theory
can unify them. As long as we do not dramatically deviate from
standard quantum field theory, and we hope that, that will not be
necessary, a purely attractive force must be carried by a field
with even integer spin. The electromagnetic force, on the other
hand, which is- of course- not always attractive, is carried be a
field of spin one. A number off no-go theorems about which we will
have to know more later, forbids any direct symmetry
transformations between fields of different integer spin and
actually leave supersymmetric theories as the only fields
theoretical models which achieve unification of all forces of
nature. Supersymmetry transformations do not directly relate
fields of different integer spin, rather they relate a graviton (a
quantum of gravity with spin 2) to a photon (spin 1) via a spin
$\frac{3}{2}$ intermediary, the \textit{gravitino}.

A partial unification of matter (fermions) with interactions
(bosons) thus arises naturally out of the attempt to unite gravity
with the other interactions.

The no-go theorems imply that \textit{supersymmetry} and \textit{supergravity%
} are only possibilities for unification within the framework of
quantum field theory. Failure of these theories to ultimately give
results which are compatible with the real world would force us to
give up either unification or quantum field theory.

A part from the no-go theorems there is a further, more technical
point that singles out supersymmetric theories; they may resolve
the nonrenormalizability problem of quantized gravity. In
perturbation quantum field theory fermions and bosons tend to
contribute with opposite signs to higher order corrections.
Supersymmetry almost always achieve a fine-tuning
between these contributions which makes some a aprori present and \textit{%
divergent terms vanish}. For a long time now there has been great
optimism that of the supergravity models may be entirely free of
infinities and thus be a consistent model of quantized gravitation
with no need for renormalization. This hope is slowly fading now,
but only to the extent that no obvious proof can be found for the
conjectured finiteness. What remains is a remarkably friendly
behavior of all supersymmetric theories when it comes to quantum
divergences, and the conviction of many theorists that it would be
surprising if nature did not make use of it. We will have to say
later more about \textit{cancellations of divergences} and the
enormous interest which they have aroused, not only for gravity
but also for the \textit{hierarchy problem} of GUTs, the thirteen
orders of magnitude between the GUT mass of $10^{15}GeV/c^{2}$and
the $W$ -boson mass. Normally, a gap of this size is not stable in
perturbation theory because of the considerable admixture of the
large mass to the small one through vacuum polarization. the gap
can only be maintained by repeated fine-tuning up to high orders
in the perturbation expansion. In supersymmetric versions of GUTs
new particles are exchanged and pair-created, and these new
processes cancel some of the effects of the old ones. Mass mixing
and consequently fine-tuning can usually be avoided, and the
hierarchy, once established, is stabilized.

\begin{center}
\textbf{\Large Outline of Part One:}
\end{center}

This part consists of four chapters of which the fourth chapter is
the main one.

In the \textbf{first chapter, }Introduction, an introduction to
the subjected presented in this part is given (this one).

In the \textbf{second chapter, }The \textit{Standard Model of
Electroweak Interactions }is presented with a brief background
examples of gauge theories and Higgs mechanism.

In the \textbf{third chapter, }The \textit{Supersymmetry concepts,
Supersymmetry Algebra, and Supersymmetric Models }are introduced.

And in the \textbf{fourth chapter, }The \textsl{Minimal
Supersymmetric
Standard Model was studied, }including\textsl{\ the extended Higgs model }and%
\textsl{\ the particle spectrum of the model.}

\chapter{Standard Model $of$ Electroweak Interactions}

\section{Introduction}

THE invention of unified renormalization theories of electroweak
interactions is actually one of the outstanding successes of
elementary particle physics. the first of this theories was the
theory of Glashow, Weinberg and Salaam ($\mathbf{GWS}$) and is
known as the standard electroweak theory.

In 1961 Glashow constructed a model for the weak and
electromagnetic interactions of leptons which was based on the
assumption that, together with the photon, there exist also a
charged $W$ and a neutral $Z$ intermediate bosons. The mass of the
$W$ and $Z$ bosons were introduced \textquotedblleft\ by
hand\textquotedblright , \textit{ad hoc.} As a result, the model
was unrenormalizable. In 1967-1968 Weinberg and Salam constructed
the $SU(2)\times U(1)$ model of electroweak interactions of
leptons including a spontaneous breakdown of the gauge symmetry.
In 1971-1972 it was proved by t'Hooft and others that the model of
this type were renormalizable. The model was generalized to quarks
using the mechanism proposed by Glashow, Iliopoulos, and Maiani.

The $GWS$ theory is based on the assumption of the existence of
charged and neutral intermediate vector bosons and it is
constructed so that, for massless fundamental fermions (leptons
and quarks), a local $SU(2)\times SU(1)$ gauge invariance takes
place. then the interaction (again locally gauge invariance) of
Higgs scalar fields with both gauge vector bosons and fermions, is
introduced. As a consequence of the spontaneous breakdown of the
underlying symmetry, leptons, quarks, and intermediate bosons all
acquire masses.

The only free parameter, which enters in the definition of the
neutral current in the $GWS$ theory, is $\sin ^{2}\theta _{W}$ (
where $\theta _{W}$ is the Weinberg angel)

Neutral currents were discovered at CERN in 1973 in an experiment
using the large bubble chamber \textquotedblleft
Gargamelle\textquotedblright . In
this experiment the process $\overline{\nu }_{\mu }+e\rightarrow \overline{%
\nu }_{\mu }+e$ was observed. After the pioneering work of the
\textquotedblleft Gargamelle\textquotedblright\ collaboration, a
large number of experiments were done investigating various
neutral current induced processes. After this work it became
possible to perform a complete phenomenological analysis of all
the neutral current data. As a result one could uniquely determine
all the coefficients appearing in the most general
phenomenological $V,$ $A$ expressions written for hadron and
lepton neutral
currents. It was shown that this unique solution is in agreement with the $%
GWS$ theory.

In 1980-1981, in experiments on the $e^{-}e+$ colliding beams,
information has been obtained on the contribution of neutral
currents to the cross sections of the process \
$e^{+}+e^{-}\rightarrow l^{+}+l^{-}(l=e,\mu ,\tau ).$ These data
also agreed with the \textsl{standard electroweak model.}

The $GWS$ theory predicts the values of the charged $(W)$ and
neutral $(Z)$ intermediate boson masses, namely, $\ m_{W}\sim
80GeV$ and $m_{Z}\sim 91GeV.$

the discovery in 1983 of the $W$ and $Z$ bosons at the $CERN$
$p\overline{p}$ collider, with exactly the predicted masses, was a
dramatic confirmation of the $GWStheory.$

In the $GWStheory$ $\sin ^{2}\theta _{W}$ is a free parameter. It
is related
to the value of $W$ and $Z$ masses as $\sin ^{2}\theta _{W}=1-(\frac{%
m_{W}^{2}}{m_{Z}^{2}})\sim 0.23$

\section{Gauge Invariance}

The concept of gauge invariance [1] grew out of the observation that of a
\textquotedblleft\ charge\textquotedblright\ (e.g. electric
charge, total energy, isospin, etc.) is conserved in a dynamically
system, then the Lagrangian for the system is invariant under
\textquotedblleft\ global gauge transformation\textquotedblright\
of the fields. For example, the electric charge is related to
invariance under
phase transformations $\psi \rightarrow e^{iq\theta }\psi $ for all fields $%
\psi $ which describe particles of charge $q$. Similarly, the
energy is related to time translations $\psi (t,x)\rightarrow \psi
(t+\Delta t,x).$ The converse is also true (Nether's theorem); if
the Lagrangian is invariant under some infinitesimal
transformation $\psi \rightarrow \psi +\delta \psi , $ then there
is a conserved current and a conserved charge associated with this
gauge invariance, (\textquotedblleft\ gauge\textquotedblright\ is
an unfortunate naming, originating in an attempt by H. Wyle in
1918 to relate the electric charge to re-scaling transformation
$\psi \rightarrow e^{\lambda }\psi ).$We call the transformation
global if their parameters do not depend on the space-time
coordinates, i.e. if $\theta =cons.$ This relationship between
conserved quantum numbers and global symmetries of the Lagrangian
led, in the 1960's, to a search for globally-invariant field
theories capable of describing and classifying all elementary
particles. The \textbf{\textquotedblleft\ 8-fold
way\textquotedblright } very much in this vein and it was in this
context that quarks were first postulated as building blocks of
strongly interacting particles.

The requirement of the \textsl{local gauge invariance} (also known
as \textquotedblleft\ gauge invariance of the 2$^{nd}$
kind\textquotedblright ) goes beyond that which can be inferred
from charge conservation. We now demand invariance of the
Lagrangian under transformations with a space-time dependent
parameter $\theta =\theta (x).$This interaction, which results in
the exchange of the field quanta, will generate forces between the
particles. \textquotedblleft\ Gauging\textquotedblright\ the phase
transformation associated with electric charge ( i.e. making them $%
x-dependent$ ) forces us to introduce the electromagnetic
four-vector potential and, as its quanta, the photons. The result
is quantum electrodynamics. Requiring other gauge invariances
requires additional gauge potentials which give rise to more
exchange particles and the other
interactions. These exchanged particles are the discovered $W^{\pm }$ and $%
Z^{0}$ for the weak force and the gluons for the strong
interactions. The latter have only been indirectly seen in their
effects on the distribution of the debris in high energy particle
collisions (\textsl{jets}). To sum up, \textquotedblleft\
gauging\textquotedblright\ an invariance of the Lagrangian will
always give rise to interaction and to forces.

Nowadays, the name gauge theory is used exclusively for theories
with local gauge invariance.

The gauge transformation under which the Lagrangian is invariant
forms a group as it fulfills the axioms of a group (in the
mathematical sense):

\ \ \ 1- Two subsequent invariances will again be an invariance,

\ \ \ 2- \textquotedblleft\ No transformation\textquotedblright\
is the identity element,

\ \ \ 3- There is exactly one inverse transformation for each
invariance transformation, and

\ \ \ 4- Three transformations are associative.

Using the standard terminology of groups, the respective gauge groups are $%
SU(3)$ for the strong interactions and $SU(2)\times U(1)$ for the
electroweak interactions. The $SU(3)$ transformations act on
triplets of quarks whose properties are very similar. They are
said to differ only in \textquotedblleft\
\textbf{color\textquotedblright }, hence the name
quantum-chromodynamics $(QCD)$ \ for the $SU(3)$ gauge theory of
the strong interactions. The success of gauge theories in
describing a variety of elementary particles phenomena eclipsed
the rule played by global invariance, and nowadays such global
symmetries are thought of as more or less accidental -if indeed
they are present. In this context it is already important to
mention that local (gauged) supersymmetry will imply supergravity.

\section{Renormalization}

\textsl{Renormalizatoin} is required in all quantum field theories
in order to make sense of divergent integrals which appear in the
perturbation expansions for physical processes. Such expansions
are unfortunately the only calculations tools currently available
for solving the equations of motion of the theory; they are
usually conceptualized in terms of vacuum polarizations and
virtual particle interactions and are illustrated by Feynman
Diagrams. In renormalizable theories, the divergences which appear
can be tested by redefining, in each order of the perturbation
expansion, a finite number of theoretical parameters in such a way
that the results of \textquotedblleft\ test
experiments\textquotedblright\ are reproduced. Other processes can
be calculated uniquely to the same order. In the lowest order, the
parameters which must be so renormalized typically present vacuum
energies, masses, coupling constants and factors which are
multiplied to the wave functions. Correspondingly, one speaks of
\textquotedblleft\ vacuum, mass, coupling constant, and wave
function renormalizatoin\textquotedblright . One of the strongest
motivations for gauge theories is their renormalizability.

A theory is called non-renormalizable if infinitely many
parameters must be redefined. Such a \textquotedblleft\
theory\textquotedblright\ can make no predictions and therefore
not a theory in the sense of exact science. I general, coupling
constants with negative mass dimensions (for $\hbar =c=1$) lead to
non- renormalizable theories. No matter how we attempt to quantize
gravity,we end up with a field theory whose coupling constant, \textbf{%
Newton's} \textsl{universal gravitational constant} $G$, has dimensions $%
\frac{1}{mass^{2}}$ in these units and quantum gravity is
therefore non-renormalizable.

\section{Quantum Electrodynamics}

Quantum Electrodynamics ($QED$) is the gauge invariance theory
which describes all relevant experimental data. As an example,
consider the electron field $\psi (x).$ The free Lagrangian of
this field has the
standard form,[2].

\begin{equation}
\mathfrak{L}=-\overline{\psi }(\gamma _{\alpha }\partial _{\alpha
}+m)\psi
\end{equation}

where $m$ is the mass of the electron, $\partial _{\alpha }\equiv \frac{%
\partial }{\partial x_{\alpha }}.$ The Lagrangian (1.1) is invariant with
respect to the global gauge transformation%
\begin{equation}
\psi (x)\rightarrow \psi ^{^{\prime }}(x)=e^{i\lambda }\psi (x),
\end{equation}

where $\lambda $ is an arbitrary real constant. It is obvious that
the Lagrangian $(1.1)$ is not invariant with respect to the local
gauge transformation

\begin{equation}
\psi (x)\rightarrow \psi ^{^{\prime }}(x)=U(x)\psi (x)
\end{equation}

where

\begin{equation*}
U(x)=\exp \{i\lambda (x)\}
\end{equation*}

and where $\lambda (x)$ is an arbitrary real function of $x$. The
derivative $\partial _{\alpha }\psi (x)$ is indeed not transformed
under (1.3) as the field $\psi (x)$ itself. Really, we have

\begin{equation*}
\partial _{\alpha }\psi ^{^{\prime }}(x)=U(x)\left( \partial _{\alpha
}+i\partial _{\alpha }\lambda (x)\right) \psi (x)
\end{equation*}

As is well known, the local gauge invariance (1.3) can be
maintained provided that the interaction of the field $\psi $ with
electromagnetic field $A_{\alpha }$ is introduced. Consider the
quantity $(\partial _{\alpha }-ieA_{\alpha })\psi $ ($e$ is the
electron charge), we will have

\begin{equation}
(\partial _{\alpha }-ieA_{\alpha }(x))\psi (x)=U^{-1}(x)(\partial
_{\alpha }-ieA_{\alpha }^{^{\prime }}(x))\psi ^{\prime
}(x)(\partial _{\alpha }-ieA_{\alpha })\psi
\end{equation}

where

\begin{equation}
A^{\prime }(x)=A_{\alpha }(x)+\frac{1}{e}\partial _{\alpha
}\lambda (x)
\end{equation}

From (1.4) it is obvious that the Lagrangian, which follows from
(1.1) by the substitution

\begin{equation}
\partial _{\alpha }\psi \rightarrow (\partial _{\alpha }-ieA_{\alpha })\psi
\end{equation}

is now invariant with respect to the gauge transformation (1.3)
and (1.5).

To construct the complete Lagrangian of the system under
consideration, we have to add also the gauge invariant Lagrangian
of the electromagnetic field. the tensor of the electromagnetic
field is given as

\begin{equation}
F_{\alpha \beta }=\partial _{\alpha }A_{\beta }-\partial _{\beta
}A_{\alpha }
\end{equation}

Clearly, $F_{\alpha \beta }^{\prime }=F_{\alpha \beta }\,$,
consequently, the gauge invariant Lagrangian of the fields of
electrons and photons takes the form

\begin{equation}
\mathfrak{L}=-\overline{\psi }[\gamma _{\alpha }(\partial _{\alpha
}-ieA_{\alpha })+m]\psi -\frac{1}{4}F_{\alpha \beta }F_{\alpha
\beta }
\end{equation}

The substitution of the derivative $\partial _{\alpha }\psi $ by
the covariant derivative $(\partial _{\alpha }-ieA_{\alpha })\psi
$ in the free Lagrangian of the field $\psi $ leads to the
following interaction Lagrangian for electrons and photons;

\begin{equation}
\mathfrak{L}_{i}=iej_{\alpha }A_{\alpha }
\end{equation}

where $j_{\alpha }=\overline{\psi }\gamma _{\alpha }\psi $ is the
electromagnetic current. Thus the substitution (1.6) fixes
uniquely the form of the interaction Lagrangian. Such an
interaction is called minimal electromagnetic interaction. Let us
note however that the principle of gauge invariance alone does not
fix the interaction Lagrangian uniquely. For example, the addition
of the Pauli term $\mu \overline{\psi }\sigma _{\alpha \beta }\psi
F_{\alpha \beta }$ to the Lagrangian (1.8) does not spoil the
gauge invariance of the theory, $(\mu $ is the anomalous magnetic
moment$)$.

All available experimental data confirm that the Lagrangian (1.9)
is the true Lagrangian which governs the interactions of electrons
and photons. It is also well known that electrodynamics, with the
minimal interaction (1.9), is a renormalizable theory.

\section{Yang-Mills Theory}

The modern theory of weak interactions is constructed in analogy
with quantum electrodynamics. We know from the experiment that the
Hamiltonian of weak interactions contains charged currents.
Therefore, to construct a theory of weak interactions we have to
start with a gauge theory containing fields of charged vector
particles. Such a theory does exit. It is the \textbf{Yang-Mills}
theory which we will now briefly present.

consider the doublet

\begin{equation*}
\psi =\binom{\psi ^{(1)}}{\psi ^{(-1)}}
\end{equation*}

of the group $SU(2)$ $(\psi ^{(1)},\psi
^{(-1)}arespinorfields).$The Lagrangian of the field $\psi $ is
written as

\begin{equation}
\mathfrak{L}_{0}=-\overline{\psi }(\gamma _{\alpha }\partial
_{\alpha }+m)\psi
\end{equation}

where $m$ is the common mass of particles, which corresponds to the fields $%
\psi ^{(1)},\psi ^{(-1)}.$ Obviously, The Lagrangian (1.10) is
left invariant with respect to the global $SU(2)$ transformation

\begin{equation}
\psi (x)\rightarrow \psi ^{\prime }(x)=\exp \{i\frac{1}{2}\tau
\lambda \}\psi (x)
\end{equation}

Here $\tau _{i}$ are Pauli matrices and $\lambda _{i}$ are real
constants.

We are now interested in the conditions under which the Lagrangian
of the system is invariant with respect to the local $SU(2)$
transformations

\begin{equation}
\psi (x)\rightarrow \psi ^{\prime }(x)=U(x)\psi (x)
\end{equation}

where

\begin{equation*}
U(x)=\exp \{i\frac{1}{2}\tau \lambda (x)\}
\end{equation*}

nd where $\lambda (x)$ are arbitrary real functions of $x$. It is
sufficient
to consider only the infinitesimal transformations (1.12). The parameters $%
\lambda _{i}$ will be taken as infinitesimal and in all expansions
in powers of $\lambda _{i}$ we shall keep only the linear terms.
Thus we have

\begin{equation}
U(x)\cong 1+i\frac{1}{2}\tau \lambda (x)
\end{equation}

Next, we get

\begin{equation}
\partial _{\alpha }\psi (x)=U^{-1}(x)\left( \partial _{\alpha }-i\frac{1}{2}%
\tau \lambda (x)\right) \psi ^{\prime }(x)
\end{equation}

It is clear from equation (1.14) that the Lagrangian (1.10) is not
invariant under the transformation of (1.12). To construct a gauge
invariant theory in
analogy with electrodynamics, we thus introduce, besides the field $\psi $%
,the vector field $A_{\alpha }$. consider the quantity

\begin{equation}
\left( \partial _{\alpha }-ig\frac{1}{2}\tau A_{\alpha }\right)
\psi (x)
\end{equation}

where $g$ is a dimensionless constant. Using equation (1.13) and
the
commutation relations $\left[ \frac{1}{2}\tau _{i},\frac{1}{2}\tau _{j}%
\right] =i\epsilon _{ijk\frac{1}{2}\tau _{k}},\,$\ we find

\begin{eqnarray}
\left( \partial _{\alpha }-ig\frac{1}{2}\tau A_{\alpha }\right)
\psi (x) &=&U^{-1}(x)U(x)\left( \partial _{\alpha
}-ig\frac{1}{2}\tau A_{\alpha
}\right) U^{-1}(x)\psi ^{\prime }(x)  \notag \\
&=&U^{-1}(x)\left( \partial _{\alpha }-ig\frac{1}{2}\tau A_{\alpha
}^{\prime }(x)\right) \psi ^{\prime }(x),
\end{eqnarray}

with

\begin{equation}
A_{\alpha }^{\prime }(x)=A_{\alpha }(x)+\frac{1}{g}\partial
_{\alpha }\lambda (x)-\lambda (x)\times A_{\alpha }(x)
\end{equation}

The field $A_{\alpha }$ is called a $Yang-Mills$ $field.$ It is
seen from
equation (1.17), that under the global $SU(2)$ transformations the fields $%
A_{\alpha }$ transforms as a triplet.

Thus, as it is seen from the expression (1.16), the covariant derivative $%
\partial _{\alpha }-ig\frac{1}{2}\tau A_{\alpha }$ applied to the filed $%
\psi ,$transforms under the gauge transformations (1.12) and
(1.17) as the field $\psi $ itself ( a primed quantity is obtained
from unprimed one through its multiplication by the matrix$U$).
This means that the substitution for the derivative $\partial
_{\alpha }\psi $ in the free Lagrangian by the covariant
derivative (1.15) leads to a Lagrangian, which is invariant with
respect to the gauge transformations (1.12) and (1.17).

To construct the gauge invariant Lagrangian of the field $A$,
consider the quantity

\begin{equation}
F_{\alpha \beta }=\partial _{\alpha }A_{\beta }-\partial _{\beta
}A_{\alpha }+gA_{\alpha }\times A_{\beta }
\end{equation}

With the help of equation (1.17) it is easy to check that

\begin{equation}
F_{\alpha \beta }^{\prime }=F_{\alpha \beta }-\lambda \times
F_{\alpha \beta }
\end{equation}

It is immediately seen that the quantity $F_{\alpha \beta
}F_{\alpha \beta }$ is a group scalar. In analogy with
electrodynamics we take the Lagrangian of the field $A_{\alpha
}$in the form

\begin{equation}
\mathfrak{L}_{0}^{\prime }=-\frac{1}{4}F_{\alpha \beta }F_{\alpha
\beta }
\end{equation}

Thus, if the interaction of the fields $\psi $ and $A_{\alpha }$
is introduced through the \textquotedblleft\
minimal\textquotedblright\ substitution $\ \partial _{\alpha }\psi
\rightarrow \left( \partial _{\alpha }-ig\frac{1}{2}\tau A_{\alpha
}\right) \psi ,$ the total Lagrangian of the system under
consideration has the form

\begin{equation}
\mathfrak{L}=-\overline{\psi }\left[ \gamma _{\alpha }\left(
\partial
_{\alpha }-ig\frac{1}{2}\tau A_{\alpha }\right) +m\right] \psi -\frac{1}{4}%
F_{\alpha \beta }F_{\alpha \beta }
\end{equation}

Consequently, the interaction Lagrangian of the fields $\psi
andA_{\alpha }$ is as follows:

\begin{equation}
\mathfrak{L}_{i}=ig\overline{\psi }\gamma _{\alpha
}\frac{1}{2}g\tau \psi A_{\alpha }
\end{equation}

The constant $g$ introduced before becomes the interaction
constant. Therefore the \textquotedblleft\
minimal\textquotedblright\ substitution $\
\partial _{\alpha }\psi \rightarrow \left( \partial _{\alpha }-ig\frac{1}{2}%
\tau A_{\alpha }\right) \psi $ fixes uniquely the interaction
Lagrangian of the fields $\psi $ and $A_{\alpha }$. We have
arrived at the \ \ \textquotedblleft\ minimal\textquotedblright\
interaction Lagrangian for the fields $\psi $ and $A_{\alpha }$,
which is compatible with gauge invariance. One should notice also
that owing to the non-linear term $gA_{\alpha }\times A_{\beta }$
which appears in the expression (1.18) written for the field
tensor $F_{\alpha \beta },$ the Lagrangian (1.21) contains terms
which are responsible for the self-interaction of the field
$A_{\alpha }$.

Notice that a mass term $-\frac{1}{2}m_{\gamma }^{2}A_{\alpha
}A_{\alpha }$ for the gauge field can not be added to the
Lagrangian of the fields of electrons and photons because its
presence would destroy the gauge invariance of the theory. this
means that the mass of the photon is equal to zero. In this case
of the Yang-Mills theory, the imposed gauge invariance also does
not allow a mass term of the $-\frac{1}{2}m_{\gamma }^{2}A_{\alpha
}A_{\alpha }$. Consequently, the particles of the fields
$A_{\alpha }$ are all massless.

We conclude this section with the following remark. Consider several fields $%
\psi _{i}(i=1,2,3,...,n)$ interacting with the gauge field
$A_{\alpha }$. We can write

\begin{equation}
\psi _{i}^{\prime }(x)=\exp \{i\lambda _{i}(x)\}\psi _{i}(x)
\end{equation}

and

\begin{equation}
\left( \partial _{\alpha }-ieA_{\alpha }(x)\right) \psi (x)=\exp
\{i\lambda _{i}(x)\}\left( \partial _{\alpha }-ieA_{\alpha
}(x)\right) \psi _{i}^{\prime }(x)
\end{equation}

where

\begin{equation}
A_{\alpha }^{\prime }(x)=A_{\alpha }(x)+\frac{1}{e_{i}}\partial
_{\alpha }\lambda _{i}(x)
\end{equation}

$e_{i}$ are constants of interaction between the fields $\psi
_{i}$ and the gauge fields $A_{\alpha }$. It is clear from
equation (1.25) that the local gauge invariance is guaranteed
provided that

\begin{equation}
\lambda _{i}(x)=e_{i}\lambda (x)
\end{equation}

($\lambda (x)$ is an arbitrary real function of $x$). Gauge
invariance does not impose any restriction on the coupling
constants $e_{i}.$

In a non-Abelian $Yang-Mills$ theory the situation is completely
different. If there are several field multiplets interacting with
one $\ Yang-Mills$ gauge field, the coupling constants of all the
fields with the gauge field are unique. It follows immediately
from the fact that the coupling constants enters into the
expression for the field tensor $F_{\alpha \beta }$ ( eq. (1.18))
because of the non-Abelian character of the Yang-Mills group.

\section{The Higgs Mechanism}

The Lagrangian mass terms are introduced into the $GWS$ theory via
the so called $Higgs$ mechanism for the spontaneous breakdown of
the gauge symmetry. To illustrate how this mechanism works, we
shall consider in this section some classical examples of
spontaneous symmetry breakdown in relativistic field theory.

Consider for instance the complex \ scalar field $\phi (x)$ with
the Lagrangian density[3]

\begin{equation}
\mathfrak{L}=-\partial _{\alpha }\phi ^{\ast }\partial _{\alpha
}\phi -V\left( \phi ^{\ast }\phi \right)
\end{equation}

where

\begin{equation}
V(\phi ^{\ast }\phi )=-\mu ^{2}\phi ^{\ast }\phi +\lambda \left(
\phi ^{\ast }\phi \right) ^{2}
\end{equation}

and where $\mu ^{2}and$ $\lambda $ are positive constants. The
Hamiltonian density obtained from equation (1.27) reads:

\begin{equation}
\mathfrak{H}=\partial _{0}\phi ^{\ast }\phi +\bigtriangledown \phi
^{\ast }\bigtriangledown \phi +V(\phi ^{\ast }\phi )
\end{equation}

We now look for the minimum of the energy of the system.
Obviously, The Hamiltonian (1.29) is minimal at $\phi =cont.,$ a
value obtained from the condition

\begin{equation*}
\frac{\partial V}{\partial \phi }=\phi ^{\ast }\left( -\mu
^{2}+2\lambda \phi ^{\ast }\phi \right) =0
\end{equation*}

Then we find that the energy of the field is minimal at

\begin{equation}
\left\vert \phi _{0}\right\vert ^{2}=\frac{\mu ^{2}}{\lambda }=\frac{%
\upsilon ^{2}}{2}
\end{equation}

i.e

\begin{equation}
\phi _{0}=\frac{\upsilon }{\sqrt{2}}e^{i\alpha }
\end{equation}

where $\alpha $ is an arbitrary real parameter. Thus the minimum
of the Hamiltonian (1.29) is infinitely degenerate. The degeneracy
is obviously connected with the fact the Lagrangian (1.27) is
invariant with respect to the global $U(1)$ transformations

\begin{equation}
\phi (x)\rightarrow \phi ^{\prime }(x)=e^{i\lambda }\phi (x)
\end{equation}

The energy minimum of the system under consideration corresponds
to an arbitrary value of $\alpha $ in equation (1.31). Due to the
gauge invariance of equation (1.32) it is always possible to take

\begin{equation}
\phi _{0}=\frac{\upsilon }{\sqrt{2}}
\end{equation}

This is the typical example of the spontaneouly broken symmetry;
the
Lagrangian of the field $\psi $ is invariant with respect to the global $%
U(1) $ transformations, while the value of the field $\phi $ is
invariant with respect to the global $U(1)$ transformations, while
the value of the field $\phi $, corresponding to the minimal
energy, is just one of many possible choices.

We further introduce two real fields $\chi _{1}$ and $\chi _{2}$
as

\begin{equation}
\phi =\frac{\upsilon }{\sqrt{2}}+\frac{1}{\sqrt{2}}(\chi
_{1}+i\chi _{2})
\end{equation}

It follows from equation (1.33) that the energy of the system
reaches its minimum value when the fields $\chi _{1},$ $\chi _{2}$
have vanishing values. Substituting (1.34) into equation (1.27),
and omitting the unimportant constant $\frac{\lambda \upsilon
^{4}}{4},$ we get the Lagrangian of the system in the following
form:

\begin{equation}
\mathfrak{L}=-\frac{1}{2}\partial _{\alpha }\chi _{1}\partial
_{\alpha }\chi
_{1}-\frac{1}{2}\partial _{\alpha }\chi _{2}\partial _{\alpha }\chi _{2}-%
\frac{1}{4}\lambda (4\upsilon ^{2}\chi _{1}^{2}+4\upsilon \chi
_{1}^{3}+\chi _{1}^{4}+4\upsilon \chi _{1}\chi _{2}^{2}+4\chi
_{1}^{2}\chi _{2}^{2}+\chi _{2}^{2})
\end{equation}

It now describes the interactions of two neutral scalar fields.
The mass term of the field $\chi _{1}$ is

\begin{equation}
2\lambda \upsilon ^{2}\chi _{1}^{2}=m_{\chi _{1}}^{2}\chi _{1}^{2}
\end{equation}

Consequently, in the case of quantized fields, the mass of the field quantum$%
\chi _{1}$ equals $m_{\chi _{1}}=\sqrt{2\lambda \mu }=\sqrt{2}\mu
.$ There is no term quadratic in the field $\chi _{2}.$ This means
that the particle corresponding to the quantum of the field $\chi
_{2}$ is massless.

We have assumed that the value of the constants $\lambda $ and
$\mu ^{2}$ in the Lagrangian (1.27) are positive. Consequently,
the quadratic in the field $\phi $ appears in equation (1.27) are
positive. , i.e. \textquotedblleft\ wrong\textquotedblright\ sign.
This leads to the spontaneous breaking of the symmetry. The
degeneracy of the ground state is a characteristic of this
phenomenon. We however introduced new real fields $\chi _{1}$ and
$\chi _{2}$ for which the ground state is not degenerate. This
leads to the spontaneous breakdown of the original $U(1)$ global
symmetry of the Lagrangian. As a result, the quanta of one field
are massive, while the mass of the second field is zero.

With spontaneous breakdown of continuous symmetry massless
spinless ( spin zero) particles always appear. This statement is
quite general, and it comprises the content of the goldstone
theorem. the corresponding massless particles are not observed.
This might imply that the ideas of spontaneous symmetry breakdown
are useless in constructing realistic physical theories in
elementary particle physics. However, it will be shown in the
following, how the spontaneous breakdown of local $gauge$ \
symmetry results in massive gauge quanta due to the disappearance
of Goldstone bosons.

Let us assume that the complex field $\phi $ with the Lagrangian
(1.27) interacts minimally with the gauge field $A_{\alpha }.$
This interaction is introduced by the substitution $\partial
_{\alpha }\phi \rightarrow \left(
\partial _{\alpha }-igA_{\alpha }\right) \phi $ in equation (1.27)

The complete Lagrangian of the system is

\begin{equation}
\mathfrak{L}=\left( \partial _{\alpha }+igA_{\alpha }\right) \phi
^{\ast
}\left( \partial _{\alpha }-igA_{\alpha }\right) \phi -V(\phi ^{\ast }\phi )-%
\frac{1}{4}F_{\alpha \beta }F_{\alpha \beta }
\end{equation}

where

\begin{equation}
F_{\alpha \beta }=\partial _{\alpha }A_{\beta }-\partial _{\beta
}A_{\alpha }
\end{equation}

the Lagrangian (1.37) is invariant with respect to the local gauge
transformations

\begin{eqnarray}
\phi (x) &\rightarrow &\phi ^{\ast }(x)=e^{i\lambda (x)}\phi ,  \notag \\
A_{\alpha }(x) &\rightarrow &A_{\alpha }^{\prime }=A_{\alpha }(x)+\frac{1}{g}%
\partial _{\alpha }\lambda (x),
\end{eqnarray}

where $\lambda (x)$ is an arbitrary function of $x$ .

As in the previous example, the minimum of the energy corresponds
to a value of the field $\phi $ equal to $\left( \frac{\upsilon
}{\sqrt{2}}\right)
e^{i\alpha }$ (where $\alpha $ is an arbitrary parameter, $\left( \frac{%
\upsilon }{\sqrt{2}}=\sqrt{\frac{\mu ^{2}}{2\lambda }}\right) .$
Due to the gauge invariance of the Lagrangian (1.37) the
\textquotedblleft\ vacuum\textquotedblright\ value of the field
$\phi $ can always be taken as

\begin{equation}
\phi _{0}=\frac{\upsilon }{\sqrt{2}}
\end{equation}

We shall write the field $\phi $ in the form

\begin{equation}
\phi \left( x\right) =\frac{1}{\sqrt{2}}\left( \upsilon +\chi (x)\right) e^{i%
\frac{\theta (x)}{\upsilon }}
\end{equation}

where $\chi (x)$ and $\theta (x)$ are real functions of $x$
defined so that zero values correspond to the minimum of $V.$

It is clear that due to the local gauge invariance of the theory
the function $\theta (x)$ appearing in equation (1.41) has no
physical meaning. It can always be eliminated by an appropriate
gauge transformation. Thus, we have

\begin{equation}
\phi (x)=\frac{\left( \upsilon +\chi (x)\right) }{\sqrt{2}}
\end{equation}

Substituting (1.42) and equation (1.37) and omitting the
unimportant constant, we get the Lagrangian of the system under
consideration in the following form

\begin{equation}
\mathfrak{L}=-\frac{1}{2}\partial _{\alpha }\chi \partial _{\alpha }\chi -%
\frac{1}{2}g^{2}\left( \upsilon +\chi \right) ^{2}A_{\alpha }A_{\alpha }-%
\frac{1}{4}\lambda \left( \chi +2\upsilon \right) ^{2}\chi ^{2}-\frac{1}{4}%
F_{\alpha \beta }F_{\alpha \beta }
\end{equation}

The Lagrangian (1.43) contains the term of the vector field
$A_{\alpha }\left( -\frac{1}{2}g^{2}\upsilon ^{2}A_{\alpha
}A_{\alpha }\right) $ and the mass term of the scalar field $\chi
\left( -\frac{1}{2}2\lambda \upsilon ^{2}\chi ^{2}\right) .$
Consequently, the masses of the fields $A_{\alpha }$ and $\chi $
are equal to $m_{A}=g\upsilon $ and $m_{\chi }=\sqrt{2\lambda
\upsilon ^{2}}=\sqrt{2}\mu ,$ respectively.

Before spontaneous symmetry breakdown the Lagrangian of the system
contained a complex scalar field ( two real functions) and a
massless gauge field (two independent real functions). After
spontaneous breakdown of the local symmetry we arrived at the
Lagrangian of an interacting real massive scalar field (one real
function) and a massive vector field (three real functions). The
degree of freedom, which would correspond to the massless
Goldstone boson (in the absence of the gauge field $A_{\alpha }$),
has been transformed through the spontaneous breakdown of the
local gauge symmetry of the Lagrangian (1.37), into the additional
degree of freedom ( masses of the vector field).

The mechanism thus discussed is called Higgs mechanism. The scalar
particle, corresponding to the quantum of the field $\chi ,$ is
called the Higgs particle.

We have explained the basic principles which are used in
constructing models of electroweak interactions. Now we turn to
the detailed discussion of the standard $SU(2)\times U(1)$ theory
of $\ Glashow,$ $Weinberg,$ $andSalam.$

\section{Glashow, Weinberg $and$ Salam Theory}

The phenomenological V-A ($current\times current$ )theory was
capable of describing the vast amount of existing experimental
data. Consequently, any new theory of weak interactions has to be
built up as to reproduce the results of the results of this
theory.

The GWS
theory[4],[5], [6] is based on the assumption that there exists intermediate vector
bosons. To reproduce the results of the V-A theory at low energies
it is therefore necessary to assume that at least part of the
\textquotedblleft\
true\textquotedblright\ weak interaction Lagrangian is of the form[7][8]

\begin{equation}
\mathfrak{L}=\frac{ig}{2\sqrt{2}}j_{\alpha }^{(+)}W_{\alpha }+h.c
\end{equation}

where $W_{\alpha }$ is the field of the vector bosons and
$j_{\alpha }^{(+)}$ is the charged weak currents. The
dimensionless coupling constant $g$ is related to the fermion
constant by

\begin{equation}
\frac{g^{2}}{8m_{W}^{2}}=\frac{G_{F}}{\sqrt{2}}
\end{equation}

where $m_{W}$ is the mass of the charged intermediate boson. The
charged current is the sum of lepton and hadron (quark) current.
In this study we
shall consider the $GWS$ theory of leptons\footnote{%
The study for the case of quarks can be found in reference [2] and
the references therein}. Consequently, we will be interested only
in the lepton current. It follows from all available data that the
charged lepton current is

\begin{equation}
j_{\alpha }^{(+)}=\overline{v}_{e}\gamma _{\alpha }\left( 1+\gamma
_{5}\right) e+\overline{v}_{\mu }\gamma _{\alpha }\left( 1+\gamma
_{5}\right) \mu +\overline{v}_{\tau }\gamma _{\alpha }\left(
1+\gamma _{5}\right) \tau
\end{equation}

where $e,\mu ,\tau $ are the field operators of the electron, muon
and $\tau -lepton.,$ respectively; $v_{e},v_{\mu },v_{\tau }are$
the field operators of the electron-, muon-, and tau-neutrinos,
respectively.

At the beginning we shall consider the case of massless fields. In
order to get the term (1.44) in the interaction Lagrangian of the
leptons and vector bosons we assume that

\begin{equation}
\psi _{lL}=\binom{v_{lL}^{^{\prime }}}{l_{L}^{^{\prime }}},\text{
\ \ \ \ \ \ \ \ \ \ \ \ \ \ \ \ \ \ }(l=e,\mu ,\tau )
\end{equation}

forms a doublet of the $SU(2)$ group and

\begin{equation}
l_{R}^{^{\prime }},v_{lR}^{^{\prime }}
\end{equation}

are singlets of this group\footnote{%
the primes put on lepton fields indicates that these fields do not
necessarily correspond \ to lepton fields with well defined masses
which will be generated later through spontaneous breakdown of the
underlying symmetry.}. Here

\begin{eqnarray}
\psi _{lL} &=&\frac{1}{2}\left( 1+\gamma _{5}\right) \binom{v_{i}^{\prime }}{%
l^{\prime }}  \notag \\
l_{R}^{\prime } &=&\frac{1}{2}(1-\gamma _{5})l^{\prime },  \notag \\
v_{lR}^{\prime } &=&\frac{1}{2}(1-\gamma _{5})v_{l}^{\prime }
\end{eqnarray}

are the left-handed (L) and the right-handed (R) components of the
corresponding fields.

The free field lepton Lagrangian

\begin{equation}
\mathfrak{L}=\dsum\limits_{l=e,\mu ,\tau }\left( \overline{\psi
}_{lL}\gamma _{\alpha }\partial _{\alpha }\psi
_{lL}+\overline{l}_{R}^{\prime }\gamma _{\alpha }\partial _{\alpha
}l_{R}^{\prime }+\overline{v}_{lR}^{\prime }\gamma _{\alpha
}\partial _{\alpha }v_{lR}^{\prime }\right)
\end{equation}

is clearly invariant with respect to the global $SU(2)group.$ We
demand now for massless fields the local $Yang-Mills$ invariance
with respect to

\begin{subequations}
\begin{eqnarray}
\psi _{lL}(x) &\rightarrow &\psi _{lL}^{\prime }(x)=\exp
\{i\frac{1}{2}\tau
\lambda (x)\}\psi _{L}(x)\}, \\
l_{R}^{\prime }(x) &\rightarrow &(l_{R}^{\prime }(x))^{\prime
}=l_{R}^{\prime }(x) \\
v_{R}^{\prime }(x) &\rightarrow &(l_{R}^{\prime }(x))^{\prime
}=l_{R}^{\prime }(x) \\
\QTR{sl}{A}_{\alpha }(x) &\rightarrow &\QTR{sl}{A}_{\alpha }^{\prime }(x)=%
\QTR{sl}{A}_{\alpha }(x)+\frac{1}{g}\partial _{\alpha }\lambda
(x)-\lambda (x)\times A_{\alpha }(x)
\end{eqnarray}

(where the $\lambda (x)$ are arbitrary real functions of $x$
($i=1,2,3$ ), and where $A_{\alpha }^{i}$ is a triplet of vector
fields). We assume the interaction of vector bosons to be minimal.
Such an interaction is introduced via the substitution (see
sec.(1.5))

\end{subequations}
\begin{equation}
\partial _{\alpha }\psi _{lL}\rightarrow \left( \partial _{\alpha }-ig\frac{1%
}{2}\tau A_{\alpha }\right) \psi _{lL}
\end{equation}

($g$ is the dimensionless constant). from equations (1.50) and
(1.52) we get the interaction Lagrangian of leptons and vector
bosons as

\begin{equation}
\mathfrak{L}_{i}=ig\mathbf{j}_{\alpha }A_{\alpha }
\end{equation}

where

\begin{equation}
\mathbf{j}_{\alpha }=\dsum\limits_{l}\overline{\psi }_{lL}\gamma _{\alpha }%
\frac{1}{2}\tau \psi _{lL}
\end{equation}

From (1.53) we can single out the interaction of leptons with
charged vector bosons:

\begin{equation}
\mathfrak{L}_{i}=\left( \frac{ig}{2\sqrt{2}}j_{\alpha
}^{(+)}W_{\alpha }+h.c.\right) +igj_{\alpha }^{3}A_{\alpha }^{3}
\end{equation}

where

\begin{equation}
W_{\alpha }=\frac{1}{\sqrt{2}}\left( A_{\alpha }^{1}-iA_{\alpha
}^{2}\right) =\frac{1}{\sqrt{2}}A_{\alpha }^{1-i2}
\end{equation}

is the field of charged vector bosons and

\begin{equation}
j_{\alpha }^{(+)}=2j_{\alpha }^{1+i2}=2\dsum\limits_{i}\overline{\psi }%
_{iL}\gamma _{\alpha }\tau _{+}\psi _{lL}=\dsum\limits_{l}\overline{%
v_{l}^{\prime }}\gamma _{\alpha }\left( 1+\gamma _{5}\right)
l^{\prime }
\end{equation}

is the charged current. therefore, the interaction Lagrangian
(1.55) which follows from the local gauge invariance does contain
the term (1.44) describing the interaction of leptons with a
charged intermediate boson.

The second term in the Lagrangian (1.55) describes the interaction
of neutrinos and charged leptons with the neutral vector boson:

\begin{equation}
\mathfrak{L}_{i}^{\prime }=ig\frac{1}{4}\dsum\limits_{l}\left( \overline{%
v_{l}^{\prime }}\gamma _{\alpha }\left( 1+\gamma _{5}\right)
v_{l}^{\prime }-l\overline{^{\prime }}\gamma _{\alpha }\left(
1+\gamma _{5}\right) l^{\prime }\right) A_{\alpha }^{3}
\end{equation}

The $GWS$ theory is a unified theory of weak and electromagnetic
interaction. Obviously, the interaction (1.58) is not
electromagnetic interaction. For a unification of weak and
electromagnetic interactions it is necessary, therefore, to
require the invariance of the Lagrangian of the system with
respect to a larger group than the local $SU(2).$ The simplest
possibility is the group $SU(2)\times U(1)$ which makes the basis of the $%
GWS $ theory.

To construct the locally $SU(2)\times U(1)$ invariant Lagrangian
we perform in equation (1.50) the minimal substitution ( section
(1.5))

\begin{subequations}
\begin{eqnarray}
\partial _{\alpha }\psi _{lL} &\rightarrow &\left( \partial _{\alpha }-ig%
\frac{1}{2}\tau A_{\alpha }-ig^{\prime }\frac{1}{2}y_{L}B_{\alpha
}\right)
\psi _{lL}, \\
\partial _{\alpha }l_{R}^{\prime } &\rightarrow &\left( \partial _{\alpha
}-ig^{\prime }\frac{1}{2}y_{R}^{(-1)}B_{\alpha }\right) l_{R}^{\prime }, \\
\partial _{\alpha }v_{lR}^{\prime } &\rightarrow &\left( \partial _{\alpha
}-ig^{\prime }\frac{1}{2}y_{R}^{(0)}B_{\alpha }\right)
v_{lR}^{\prime },
\end{eqnarray}

where $A_{\alpha }is$ a triplet of gauge fields with respect to the group $%
SU(2),B_{\alpha }is$ the gauge field associated with the symmetry group $%
U(1) $ and the $y$ constants are the corresponding hypercharge.
The complete gauge invariant Lagrangian of leptons and vector
bosons consequently becomes

\end{subequations}
\begin{eqnarray}
\mathfrak{L} &=&-\sum\limits_{l}\overline{\psi }_{lL}\gamma
_{\alpha }\left(
\partial _{\alpha }-ig\frac{1}{2}\tau A_{\alpha }-ig^{\prime }y_{L}B_{\alpha
}\right) \overline{\psi
}_{lL}-\sum\limits_{l}\overline{l}_{R}^{\prime
}\gamma _{\alpha }\left( \partial _{\alpha }-ig^{\prime }\frac{1}{2}%
y_{R}^{(-1)}B_{\alpha }\right) l_{R}^{\prime }  \notag \\
&&-\sum\limits_{l}\overline{v}_{lR}^{\prime }\gamma _{\alpha
}\left(
\partial _{\alpha }-ig^{\prime }\frac{1}{2}y_{R}^{(-1)}B_{\alpha }\right)
l_{R}^{\prime }-\frac{1}{4}\mathbf{F}_{\alpha \beta
}\mathbf{F}_{\alpha \beta }-\frac{1}{4}F_{\alpha \beta }F_{\alpha
\beta }
\end{eqnarray}

where

\begin{subequations}
\begin{eqnarray}
\mathbf{F}_{\alpha \beta } &=&\partial _{\alpha }A_{\beta
}-\partial _{\beta
}A_{\alpha }+gA_{\alpha }\times A_{\beta } \\
F_{\alpha \beta } &=&\partial _{\alpha }B_{\beta }-\partial
_{\beta }B_{\alpha }
\end{eqnarray}

The interaction Lagrangian of leptons and vector bosons, which
follows from equation (1.60), can be written as

\end{subequations}
\begin{equation}
\mathfrak{L}_{i}=ig\mathbf{j}_{\alpha }\mathbf{A}_{\alpha }+ig^{\prime }%
\frac{1}{2}j_{\alpha }^{y}B_{\alpha }
\end{equation}

The current $j_{\alpha }$ is given by (1.54), and

\begin{equation}
j_{\alpha }^{y}=\sum\limits_{l}y_{L}\overline{\psi }_{lL}\gamma
_{\alpha }\psi
_{lL}+\sum\limits_{l}y_{R}^{(-1)}\overline{l}_{R}^{\prime }\gamma
_{\alpha }l_{R}^{\prime
}+\sum\limits_{l}y_{R}^{(0)}\overline{v}_{R}^{\prime }\gamma
_{\alpha }l_{R}^{\prime }
\end{equation}

The $U(1)$ invariance does not impose any constraints on the
coupling constants between the leptons and the field $B_{\alpha }$
(see the discussion at the end of sec. (1.5)). This freedom in the
choice of the coupling constants for the $U(1)$ gauge group can
then be used to unify weak and electromagnetic interactions.

We will choose $y_{L},$ $y_{R}^{(-1)}$ and $y_{R}^{(0)}so$ as to
satisfy the Gell-Mann--Nishijima relation

\begin{equation}
Q=I_{3}+\frac{1}{2}y
\end{equation}

Here $Q$ is the electric charge in units of the proton charge,
$I_{3}$ is the third component of the weak isospin. It follows
that $y_{L}$ equals the sum of the charges of the
\textquotedblleft\ upper\textquotedblright\ and \textquotedblleft\
lower\textquotedblright components of the doublet $\psi _{lL}$

\begin{equation}
y_{L}=-1
\end{equation}

correspondingly, the weak hypercharges of the right-handed singlets $%
l_{R}^{\prime }$ and $v_{R}^{\prime }$ are equal to:

\begin{equation}
y_{R}^{(-1)}=-2,\text{ \ \ \ \ \ }y_{R}^{(0)}=0
\end{equation}

respectively. With the help of equations (1.64)-(1.66) it is easy
to check that

\begin{equation}
\frac{1}{2}j_{\alpha }^{(y)}=j_{\alpha }^{em}-j_{\alpha }^{3}
\end{equation}

where

\begin{equation}
j_{\alpha }^{em}=\sum\limits_{l}\left( -1\right)
\overline{l}^{\prime }\gamma _{\alpha }l^{\prime }
\end{equation}

is the electromagnetic current of leptons and where $j_{\alpha
}^{3}is$ the third component of the isovector \thinspace
$\mathbf{j}_{\alpha }.$

Using equation (1.67) the interaction Lagrangian (1.62) can be
written as

\begin{equation}
\mathfrak{L=}\left( \frac{ig}{2\sqrt{2}}j_{\alpha }^{(+)}W_{\alpha
}+h.c.\right) +\mathfrak{L}_{i}^{0}
\end{equation}

where

\begin{equation}
\mathfrak{L}_{i}^{0}=igj_{\alpha }^{3}A_{\alpha }^{3}+ig^{\prime
}\left( j_{\alpha }^{em}-j_{\alpha }^{3}\right) B_{\alpha }
\end{equation}

is the interaction Lagrangian of the leptons and the neutral
vector bosons.

To single out the electromagnetic interaction from equation
(1.70), we rewrite this expression as

\begin{equation}
\mathfrak{L}_{i}^{0}\mathfrak{=}i\sqrt{g^{2}+g^{\prime
2}}j_{\alpha
}^{3}\left( \frac{g}{\sqrt{g^{2}+g^{\prime 2}}}A_{\alpha }^{3}-\frac{%
g^{\prime }}{\sqrt{g^{2}+g^{\prime 2}}}B_{\alpha }\right)
+ig^{\prime }j_{\alpha }^{em}B_{\alpha }
\end{equation}

Instead of the fields $A_{\alpha }^{3}$ and $B_{\alpha }$ we
introduce the field

\begin{equation}
Z_{0}=\frac{g}{\sqrt{g^{2}+g^{\prime 2}}}A_{\alpha }^{3}-\frac{g^{\prime }}{%
\sqrt{g^{2}+g^{\prime 2}}}B_{\alpha }
\end{equation}

and the field

\begin{equation}
A_{\alpha }=\frac{g^{\prime }}{\sqrt{g^{2}+g^{\prime 2}}}A_{\alpha }^{3}-%
\frac{g}{\sqrt{g^{2}+g^{\prime 2}}}B_{\alpha }
\end{equation}

orthogonal to $Z_{\alpha }.$ Elementary algebra implies that the field $%
A_{\alpha }$ is coupled only to $j_{\alpha }^{em},$ while the field $%
Z_{\alpha }$ is coupled both with the current $j_{\alpha }^{3}$ and $%
j_{\alpha }^{em}.$ This means that the expression (1.71) contains
the Lagrangian of the electromagnetic interactions and that
$A_{\alpha }$ is the electromagnetic field.

Indeed, we have

\begin{equation}
\mathfrak{L}_{i}^{0}=i\frac{1}{2}\sqrt{g^{2}+g^{\prime
2}}j_{\alpha }^{0}Z_{0}+i\frac{gg^{\prime }}{\sqrt{g^{2}+g^{\prime
2}}}j_{\alpha }^{em}A_{\alpha }
\end{equation}

where

\begin{equation}
j_{\alpha }^{0}=2\left( j_{\alpha }^{3}-\frac{g^{\prime }}{g^{2}+g^{\prime 2}%
}j_{\alpha }^{em}\right)
\end{equation}

If the coupling constants $g$ and $g^{\prime }$ are related to the
charge of the proton as

\begin{equation}
\frac{gg^{\prime }}{\sqrt{g^{2}+g^{\prime 2}}}=e
\end{equation}

the second term $iej_{\alpha }^{em}A_{\alpha }$ in expression
(1.74) becomes the interaction Lagrangian between leptons and
photons.

Thus there are four vector bosons fields associated with the gauge $%
SU(2)\times U(1)$ group. Two fields correspond to charged vector bosons ($%
W^{+}andW^{-}$) and two fields correspond to neutral ones. One
neutral field is identified with the electromagnetic field, the
other is the field of the neutral intermediate boson.

Consequently, the unification of weak and electromagnetic
interactions based on the group $SU(2)\times U(1)$ is possible
provided that not only charged vector bosons and charged currents
but also neutral vector bosons and neutral currents, do exist.

Now we will continue our constructing of the unified electroweak theory of $%
GWS$. The Weinberg angle $\theta _{W}$ is introduced as follows:

\begin{equation}
\tan \theta _{W}=\frac{g^{\prime }}{g}
\end{equation}

For the neutral current $j_{\alpha }^{0}$ we get

\begin{equation}
j_{\alpha }^{0}=2j_{\alpha }^{3}-2\sin ^{2}\theta _{W}j_{\alpha
}^{em}
\end{equation}

and the relation (1.76) turns into

\begin{equation}
g=\frac{e}{\sin \theta _{W}}
\end{equation}

the complete interaction Lagrangian of leptons and gauge bosons
can be rewritten with the help of equations (1.69), (1.74) and
(1.77) as

\begin{equation}
\mathfrak{L}_{i}\mathfrak{=}\left( \frac{ig}{2\sqrt{2}}j_{\alpha
}^{(+)}W_{\alpha }+h.c\right) +i\frac{g}{2\cos \theta
_{W}}j_{\alpha }^{0}Z_{\alpha }+iej_{\alpha }^{em}A_{\alpha }
\end{equation}

The structure of the neutral current in the $GWS$ theory is
determined by the unifying weak and electromagnetic interactions.
the first term in equation (1.78) is the third component of the
isovector, whose \textquotedblleft\
plus-component\textquotedblright\ is identified with the charged
weak current. the parameter $\sin ^{2}\theta _{W}$ is thus the
only parameter which enters the expression for the neutral
current. Its value can be determined from the data on the neutral
current induced processes.

The theory we have considered so far satisfies the requirements of a local $%
SU(2)\times U(1)$ gauge invariance. Mass terms of the vector boson
fields can not be introduced into the Lagrangian of such a theory.
It is also obvious that the $SU(2)\times U(1)$ invariance with
left-handed fields in doublets $\psi _{lL}$ and right-handed
fields \thinspace $l_{R}$ in singlets also forbids the
introduction of lepton mass terms into the Lagrangian.

In the standard electroweak theory the Lagrangian mass terms of
the both the vector boson and fermion fields are introduced by the
Higgs mechanism of spontaneous breakdown of the gauge symmetry (
see sec. (1.6) ). The theory is built up so that, at the
beginning, the complete Lagrangian, including the Higgs sector, is
locally $SU(2)\times U(1)$ invariant. It is then necessary to
assume that the Higgs fields transformed according to a definite
(non-trivial) representation of the gauge group.\ Further, due to
the spontaneous breakdown of the gauge invariance charged ($W^{+}and$ $W^{-}$%
) as well as neutral $Z^{0}$ intermediate bosons have to acquire
masses. That is, three Goldstone degrees of freedom of Higgs field
can transform at the spontaneous breakdown of the gauge invariance
into the additional degrees of freedom of vector fields (three
masses). Thus, we are forced to assume that the Higgs fields form
at least doublet. It is this \textquotedblleft\
minimal\textquotedblright\ assumption which is at the bases of the
$GWS$ theory.

hence we assume that the Higgs fields forma doublet of the $SU(2)$
group

\begin{equation}
\phi =\binom{\phi _{+}}{\phi _{0}}
\end{equation}

where the complex function $\phi _{+},$ $\phi _{0}$ are the fields
of the charged and neutral bosons, respectively. Weak hypercharge
of the doublet (1.81) is defined so as to fulfill the
Gell-Mann--Nishijima \ relation (1.64).\ We have

\begin{equation}
y_{\phi }=1
\end{equation}

The Lagrangian of the Higgs field $\phi (x)$ is given as (see
sec.(1.6))

\begin{equation}
\mathfrak{L}^{0}=-\partial _{\alpha }\phi ^{+}\partial _{\alpha
}\phi -V(\phi ^{+}\phi ).
\end{equation}

Here

\begin{equation}
V(\phi ^{+}\phi )=-\mu ^{2}\phi ^{+}\phi +\lambda (\phi ^{+}\phi
)^{2}=\lambda \left( \phi ^{+}\phi -\frac{\mu ^{2}}{2\lambda }\right) ^{2}-%
\frac{\mu ^{4}}{4\lambda }
\end{equation}

where $\mu ^{2}$ and $\lambda $ are positive constants.

Taking into account (1.82), we get from (1.83) by the standard
substitution

\begin{equation*}
\partial _{\alpha }\phi \rightarrow \left( \partial _{\alpha }-ig\frac{1}{2}%
\tau A_{\alpha }-ig^{\prime }\frac{1}{2}B_{\alpha }\right) \phi
\end{equation*}

the Lagrangian

\begin{equation}
\mathfrak{L}=-\left[ \partial _{\alpha }\phi ^{+}+\phi ^{+}\left( ig\frac{1}{%
2}\tau A_{\alpha }+ig^{\prime }\frac{1}{2}B_{\alpha }\right)
\right] \left[
\partial _{\alpha }\phi +\left( ig\frac{1}{2}\tau A_{\alpha }+ig^{\prime }%
\frac{1}{2}B_{\alpha }\right) \phi \right] -V(\phi ^{+}\phi )
\end{equation}

is invariant with respect to the gauge group $SU(2)\times U(1).$
It is obvious from (1.84) that the potential $V(\phi ^{+}\phi )$
is minimal for

\begin{equation}
(\phi ^{+}\phi )_{0}=\frac{\mu ^{2}}{2\lambda }=\frac{\upsilon
^{2}}{2}
\end{equation}

For the minimal (vacuum) value of $\phi $ we choose

\begin{equation}
\phi _{vac.}=\binom{0}{\frac{\upsilon }{\sqrt{2}}}
\end{equation}

Further, the doublet $\phi $ can always be written in the form

\begin{equation}
\phi (x)=\exp \left\{ i\frac{1}{2}\tau \frac{\theta (x)}{\upsilon
}\right\} \binom{0}{\frac{\left\{ \upsilon +\chi (x)\right\}
}{\sqrt{2}}}
\end{equation}

where $\theta (x)$ and $\chi (x)$ are real functions. Finally, the
functions $\theta (x),$ which, correspond to the \textquotedblleft
would be\textquotedblright\ Goldstone bosons can always be
eliminated owing to the gauge invariance of the Lagrangian (1.85)
by appropriately fixing the gauge (the so called unitary gauge).
Thus we have

\begin{equation}
\phi (x)=\binom{0}{\frac{\left\{ \upsilon +\chi (x)\right\}
}{\sqrt{2}}}
\end{equation}

Let us substitute (1.89) into (1.85). Taking into account that

\begin{equation*}
\left( \mathbf{\tau A}_{\alpha }\right) \left( \mathbf{\tau
A}_{\alpha }\right) =2W_{\alpha }\overline{W}_{\alpha }+A_{\alpha
}^{3}A_{\alpha }^{3},
\end{equation*}

\begin{equation*}
\phi ^{+}\left( \mathbf{\tau A}_{\alpha }\right) B_{\alpha }\phi
=-A_{\alpha }^{3}B_{\alpha }\frac{1}{2}\left( \upsilon +\chi
\right) ^{2}
\end{equation*}

we get

\begin{equation}
\mathfrak{L}=\frac{1}{2}\partial _{\alpha }\chi \partial _{\alpha }\chi -%
\frac{1}{2}\left( \upsilon +\chi \right) ^{2}\left[ \frac{1}{4}%
g^{2}2W_{\alpha }\overline{W}_{\alpha }+\frac{1}{4}\left(
g^{2}+g^{\prime 2}\right) Z_{\alpha }Z_{\alpha }\right]
-\frac{1}{4}\lambda \chi ^{2}\left( \chi +2\upsilon \right) ^{2}
\end{equation}

Here

\begin{equation*}
W_{\alpha }=\frac{A_{\alpha }^{1-i2}}{\sqrt{2}}\text{ \ \ \ \ \ \
\ \ \ \ \ \ and }\overline{W}_{\alpha }=\frac{A_{\alpha
}^{1+i2}}{\sqrt{2}}
\end{equation*}

are the fields of the charged vector bosons and $Z_{\alpha }is$
the field of the neutral vector bosons.

As a results of the spontaneous breakdown of the symmetry, mass
term for the intermediate bosons have emerged from in the
Lagrangian

\begin{equation}
\mathfrak{L}_{m}=-m_{W}^{2}W_{\alpha }\overline{W}_{\alpha }-\frac{1}{2}%
m_{Z}^{2}Z_{\alpha }Z_{\alpha },
\end{equation}

where

\begin{equation}
m_{W}^{2}=\frac{1}{4}g^{2}\upsilon ^{2},\text{ \ \ \ \ \ \ \ \ }m_{Z}^{2}=%
\frac{1}{4}(g^{2}+g^{\prime 2})\upsilon ^{2}.
\end{equation}

Symmetry was broken in such a way that the photon remained a
massless particle.

The function $\chi \left( x\right) $ is a field of neutral scalar
particles (the so called Higgs particles). It follows from (1.90)
that their mass is equal to

\begin{equation}
m_{\chi }=\sqrt{2\lambda }\upsilon =\sqrt{2}\mu
\end{equation}

note that the Lagrangian (1.90) contains also a term describing
the interaction of the Higgs particles with the intermediate
bosons.

We find from (1.77) and (1.92) that the mass squared of the $Z$
boson is related to that of the $W$ boson and the parameter $\cos
^{2}\theta _{W}$ by

\begin{equation}
m_{Z}^{2}=\frac{m_{W}^{2}}{\cos ^{2}\theta _{W}}.
\end{equation}

It should be stressed that this relation is satisfied only if the
Higgs fields form doublets. In the case of higher Higgs multiplets
no relation between masses of neutral and charged intermediate
bosons does exist.

It follows from (1.45) and (1.92) that

\begin{equation}
\upsilon =\frac{1}{\left( \sqrt{2}G_{F}\right) ^{\frac{1}{2}}}
\end{equation}

Therefore, the theory enables us to calculate the parameter
$\upsilon .$ Substituting the numerical value of $G_{F:}$

\begin{equation*}
G_{F}=1.1664\times 10^{-5}GeV,
\end{equation*}

we find

\begin{equation}
\upsilon =246.2GeV.
\end{equation}

Also, it follows from (1.45) and (1.79) that

\begin{equation}
m_{W}=\left( \frac{\pi \alpha }{\sqrt{2}G_{F}}\right) ^{\frac{1}{2}}\frac{1}{%
\sin \theta _{W}}.
\end{equation}

the value of the parameter $\sin \theta _{W}$ is determined from
experimental data on neutral currents induced processes.
therefore, the theory enables us to predict the value of the $\ W$
boson mass.

From the analysis of the world data on deep inelastic processes,
one could deduce the value

\begin{equation}
\sin ^{2}\theta _{W}=0.2315.
\end{equation}

With the above value of $\sin ^{2}\theta _{W},$ the masses of
charged and neutral intermediate bosons using (1.97) and (1.94)
turn out to be

\begin{eqnarray}
m_{W} &=&80.330GeV \\
m_{Z} &=&91.187GeV
\end{eqnarray}

\chapter{Supersymmetry}

\section{Motivation $for$ Supersymmetry}

Ever since its discovery in the early seventies, supersymmetry has
been the focus of considerable attention. Although no compelling
supersymmetric model has yet emerged, and in spite of the fact
that there is no experimental evidence for supersymmetry, its
remarkable theoretical properties have provided sufficient
motivation for its study.

Supersymmetry, is a novel symmetry that interrelates bosons and
fermions, thereby providing a new level of synthesis. It is the
most general (known) symmetry of the $S-matrix$ consistent with
$Poincare^{\prime }$ invariance. Supersymmetry leads to an
improvement ( and sometimes even to elimination of divergencies
that occur in Quantum Field Theory (QFT), in particular, quadratic
divergencies are absent); these feature will play an important
role in our subsequent discussions. Since two successive
supersymmetry transformation involve a space-time translation,
local supersymmetry theories ( and past experience shows that
nature prefers local supersymmetry!) necessarily include
gravitation with the gauge fermion (the
\textsl{gravitino}) being the supersymmetric partner of the \textsl{%
graviton. }Further, because supersymmetric QFT's exhibit a better
ultraviolet behavior, they may provide a hope of eventually
obtaining a consistent quantum theory that include gravitation.
Finally, supersymmetry is an essential ingredient in the
construction of the most recent candidate for a \textsl{theory of
everything,TOE, the SUPERSTRING.}

At this point, one may ask why any extension of the Standard Model
(SM) needs to be considered [7]. After all, the $GWS$ theory seems
for all known electromagnetic and weak phenomena, and quantum
chromodynamics $(QCD)$ \ is generally accepted to be the theory of
strong interactions. Indeed, it
appears that all experiments are consistent \ with a gauge theory based on $%
SU(3)_{c}\times SU(2)_{L}\times U(1)_{Y},$ with the
$SU(2)_{L}\times U(1)_{Y} $ being spontaneously broken to
$U(1)_{em}.$

In the $GWS$ model, the spontaneous breakdown is brought about the
introduction of an elementary scalar field. This leads to the
prediction of (at least) an additional spin-zero particle, the
Higgs boson. With the discovery of the $W^{\pm }$ and $Z^{0}$
bosons at the$CERN$ $p\overline{p}$ collider, and the top quark at
the $FERMILAB$, only the Higgs boson remains to be discovered to
complete the particle content of the $GWS$ electroweak theory. It
should be pointed out that no elementary spin-zero particles have
ever been found. In fact, the $ad$ $hoc$ introduction of these
considered by many theorists to be an unpleasant feature of the
standard model.

The problem is the instability of the scalar particles masses
under radiative corrections. For example, one-loop radiative
corrections to these diverge quadratically, leading to corrections
of the form:

\begin{equation}
\delta m^{2}=O(\alpha /\pi )\Lambda ^{2}
\end{equation}

where $\Lambda $ is a cut-off parameter representing the scale of
the theory and $\alpha =e^{2}/hc$ $\approx \frac{1}{137}$ is the
fine structure
constant. The parameter $\Lambda $ may be the \textsl{Grand Unified Theory} (%
$GUT$) scale $O(10^{15}GeV/c^{2}),$ or the \textsl{Plank scale }$%
O(10^{19}GeV/c^{2})$ if we believe there is no new physics all the
way up to the scale associated with quantum gravity. On the other
hand, we know that for the scalar self-couplings to be sensibly
treated within perturbation theory, the scalar mass

\begin{equation}
m^{2}\leqslant O(m_{W}^{2}/\alpha )\sim 1TeV/c^{2}
\end{equation}

In other words, either the Higgs sector is strongly interacting,
or $\delta m^{2}\gg ,m^{2.}$ Such theories (where $\delta m^{2}\gg
,m^{2}$) have been technically referred to as \textquotedblleft\
unnatural\textquotedblright , because the parameters have to be
tuned with unusual precision in order to preserve the lightness of
the Higgs mass compared to the GUT scale $\Lambda
\symbol{126}O(1TeV/c^{2})$

One possible solution to the problem of naturalness is to imagine
that the
quarks, leptons and gauge bosons are all composites with associated scale $%
\Lambda \symbol{126}O(1TeV/c^{2}).$ While these solve the problem
at present energies, it does not really represent a solution as we
could ask the same questions of any underlying theory. Moreover,
it would be difficult to understand why the gauge principle seems
to work so well at least up to this point.

A different approach would be to eliminate the fundamental scalars
and
imagine that the Higgs is a composite of new fermions bound by a new force -%
\textsl{the technicolor force-} that becomes strong at a scale of $%
O(1TeV/c^{2})$. While this is an appealing idea, and while the
composite Higgs boson can indeed lead to masses for the $W^{\pm
}and$ $Z^{0},$ it does not account for quark and lepton masses.
This led to the introduction of the yet another interaction, the
extended technicolor. This appears to have phenomenological
problems, particularly with flavor-changing neutral currents
(FCNC). Although that it appears there is no reason for the
technicolor idea not to work, it seems fair to say that no viable
model has yet emerged.

Supersymmetry provides yet another to the naturalness question.
this may be simply understood by recalling that fermion and masses
are protected from large radiative corrections by chiral symmetry.
The analogue for equation (2.1) is

\begin{equation}
\delta m_{f}=O\left( \alpha /\pi \right) m_{t}\ln \left( \Lambda
/m_{t}\right)
\end{equation}

Thus, massless fermions do not acquire masses via radiative
corrections. This is a manifestation of the chiral symmetry The
naturalness problem arises because, unlike the case of fermions,
there is no symmetry to keep massless scalars from acquiring large
masses via radiative corrections.

In practice, at the one-loop level, this works because boson and
fermion loops both enter the scalar correction, but with a
relative minus sign. For supersymmetric theories, equation (2.1)
takes the form

\begin{equation}
\delta m^{2}\approx O(\alpha /\pi )\Lambda ^{2}-O(\alpha /\pi
)\Lambda ^{2}=0
\end{equation}

Exact cancellation requires that the bosons and fermions enter
with the same quantum numbers (this is ensured by supersymmetry)
so that their couplings are the same except for supersymmetry
Clebsch-Gordon factors, and also that they have the same masses.
Spontaneous breaking of supersymmetry as for any other symmetry,
maintains a relation between the couplings but breaks the mass
relations, and equation (2.4) takes the form

\begin{equation}
\delta m^{2}\approx O(\alpha /\pi )\left\vert
m_{B}^{2}-m_{f}^{2}\right\vert
\end{equation}

We see from expression (1.2) and (1.5) that for supersymmetry to
solve the naturalness problem,

\begin{equation}
\left\vert m_{B}^{2}-m_{f}^{2}\right\vert \leq O(1TeV/c^{2})^{2}
\end{equation}

where $m_{B}^{2}$ ($m_{f}^{2}$) is the boson (fermion) to have
masses $\leq $ $O(1TeV/c^{2})$ and hope that these may show up at
future LEP energies.

We emphasize that not one of all these reasons for considering
supersymmetry, no matter how compelling it may appear, requires
any particular mass scale for supersymmetric particles
(sparticles). It is only if we require supersymmetry to address
the naturalness question that the mass scale is fixed. We note
that the mass scale does address the question "
why is the scalar mass 12 (16) order of magnitude smaller than the $GUT$ ($%
Planck$ ) scale in the first place?". But in supersymmetric
theories, once this value has been set, either \textquotedblleft
by hand\textquotedblright\ or by any other mechanism, radiative
corrections preserve this hierarchy of scales.

\section{Rules $of$ Supersymmetry}

We are here to establish some of the rules of supersymmetry
treatment [8].

First and foremost we postulate the $existence$ of supersymmetry
between fermions and bosons which should underlay the laws of
physics. No experimental observation has yet revealed particles or
forces which manifestly show such a symmetry except for some rare
events [9]. therefore the development of a theory based on
supersymmetry requires an understanding not only of how the
various symmetry transformations affect each other ( the algebra )
but also of all possible systems ( \textit{multiplets }of
particles or quantum fields ) on which the supersymmetry
transformations can act. The symmetry operations will transform
different members of a multiplet into each other. More precisely,
the transformations are to be represented
by linear operators acting on the vector space. ( the \textquotedblleft $%
representation$ $space$\textquotedblright ) spanned by the
multiplet. Finally the theory must predict the time development of
interacting physical
systems. This is usually achieved by finding appropriate \textsl{%
Hamiltonians or Lagrangians. }The supersymmetry present in the
physical system will manifest itself in the invariance of this
\textsl{Lagrangian} - or rather its integral over all time,
\textbf{the action- }\ if all the fields undergo their respective
supersymmetry transformation. Because of the lack of experimental
input, a large fraction of the research effort of supersymmetry
theorists has, in fact, been devoted to the finding of, and
exploration of, possible supersymmetry respecting interactions.

The theoretical framework in which to construct supersymmetric
models in flat space-time is quantum field theory, and it must be
pointed out that \textsl{the standard concept of quantum field
theory allow for supersymmetry without any further assumptions.
}This introduction of supersymmetry is not a revolution in the way
one views physics. It is an additional symmetry that in otherwise
\textquotedblleft normal\textquotedblright\ field theoretical
model can have. As we shall see, all that is required for a field
theory to be supersymmetric is that it contains specified types
and numbers of fields in interactions with each other and that the
various interaction strengths and particle masses have properly
related events. As an example, consider the $SU(3)$ gauge theory
of gluons, which can be made supersymmetric by including a
massless neutral color octet of spin $\frac{1}{2}$ particles which
are their own anti-particles. Such spin $\frac{1}{2}$ partners of
the gluons are called \textquotedblleft
\textsl{gluinos}\textquotedblright . If our models contains not
only gluons but also quarks, we must also add corresponding
partners for them. These have spin $0$ and are commonly called
\textquotedblleft \textsl{squarks }\textquotedblright .
(Procedures like these are employed particularly in the
construction of supersymmetric Grand Unified Theories$susyGUTs$ or
$\sup erGUTs.$ )

Before we proceed to discuss the ingredients of supersymmetric
models, we must address the question of the $Fermi-Bose$
matter-force duality. After all, the wave particle duality of
quantum mechanics and the subsequent question of the
\textquotedblleft exchange particle\textquotedblright\ in
perturbative quantum field theory seemed to have abolished that
distinction for good. The recent triumph progress of gauge
theories has, however,
reintroduced it. forces are mediated by gauge potentials, i.e., from spin $%
\frac{1}{2}fermions$. The Higgs particles, necessary to mediate
the needed spontaneous breakdown of the gauge invariances (more
about this later), play a somewhat intermediate role. They must
have zero spin and are thus bosons, but they are not directly
related to any of the forces. Purists hope to see the arise as
bound states of the fermions (condensates). Supersymmetric
theories, and particularly supergravity theories,
\textquotedblleft unite\textquotedblright\ fermions and bosons
into multiplets and left the basic distinction between matter and
interaction. The gluinos, for example, are thought of as carriers
of the strong force as much as the gluons, except that as fermions
they must obey an exclusion principle an thus will never conspire
to form a coherent, measurable potential. the distinction between
forces and matter becomes phenomenological: bosons - and
particularly massless ones- manifest themselves as forces because
they can build up coherent classical fields; fermions are seen as
matter because no tow identical ones can occupy the same point in
space.

For some time it was thought that supersymmetry which would
naturally relate forces and fermionic matter would be in conflict
with field theory. The progress in understanding elementary
particles through the $SU(3)$ classification of the
\textquotedblleft eight-fold way\textquotedblright\ (a global
symmetry) had led to attempts to find a unifying symmetry which
would directly relate to each other several of the $SU(3)$
multiplets (baryon octet, decuplet,etc.) even if these had
different spins. The failure of attempts to make those
\textquotedblleft\ spin symmetry\textquotedblright\
relativistically covariant led to the formulation of a series of
no-go theorems, culminating in 1967 in a paper by $Coleman$ and
$Mandula$ which was widely understood to show that it is
impossible, within the theoretical framework of relativistic field
theory, to unify space-time symmetry with internal symmetries.
More precisely, say that the charge operators whose eigenvalues
represent the \textquotedblleft internal\textquotedblright\
quantum numbers such as electric charge, isospin, hypercharge,
etc., must be transitionally invariant. This means that these
operators commute with the energy, the momentum operators. indeed
the only symmetry generators which
transfer at all under both translations and rotations are those of the $%
Lorentz$ transformations themselves ( rotations and
transformations to coordinate systems which move with constant
velocity). The generators of
internal symmetries can not relate eigenstates with different eigenvalues $%
m^{2}$ and $l(l+1)\hbar ^{2}$ of the mass and spin operators. This
means that irreducible multiplets of symmetry groups can not
contain particles of different mass or of different spin. This
no-go theorem, seemed to rule out exactly the sort of unity which
was sought. One of the assumptions made in Coleman and Mandula's
proof did, however, turn out to be unnecessary: they had admitted
only those symmetry transformations which form $Lie$ group with
real parameters. Examples of such symmetries are space rotations with the $%
Euler$ angles as parameters and the phase transformations of
electrodynamics with a real phase angel $\theta $ about which we
talked earlier. The charge operators associated with such Lie
groups of symmetry transformations (their generators) obey
well-defined $commutation$ $relations$ with each other.
Perhaps the best-known example is the set of commutators $%
L_{x}L_{y}-L_{y}L_{x}\equiv \left[ L_{x},L_{y}\right] =i\hbar
L_{z},$ for the angular momentum operators which generate spatial
rotations.

Different spins in the same multiplet are allowed if one includes
symmetry
operations whose generators obey $anticommutation$ $relations$ of the form $%
AB+BA\equiv \{A,B\}=C.$ This was first proposed in 1971 by
Gol'fand and Likhtman, and followed up by Volkov and Akulov who
arrived at what we now call a non-linear realization of
supersymmetry. Their model was not renormalizable. In 1973, Wess
and Zumino presented a renormalizable theoretical model of a spin
$\frac{1}{2}$ particle in interaction with two spin $0$ particles
where the particles are related by symmetry transformations, and
therefore \textquotedblleft sit\textquotedblright\ in the same
multiplet. The limitation of the Coleman-Mandula no-go theorem had
been avoided by the introduction of a fermionic symmetry operator
which carried a spin $\frac{1}{2},$ and thus when acting on a
state of spin $j$
resulted in a linear combination of states with spin $j+\frac{1}{2}$ and $j-%
\frac{1}{2}.$ Such operators must and do observe anticommutator
relations with each other. They do not generate Lie groups and are
therefore not rules out by the Coleman-Mandula no-go theorem. In
the light of this discovery, Haag, Lopuszanski, and Sohnius
extended the results of Coleman-Mandula to include symmetry
operations which obey Fermi-statistics. they proved that in the
context of relativistic field theory the only model which can lead
to a solution of the unification problems are supersymmetric
theories, and space-time and internal symmetries can only be
related to each other by
fermionic symmetry operators $Q$ of spin $\frac{1}{2}(not$ $\frac{3}{2}or$ $%
higher$) whose properties are either exactly those of the
Wess-Zumino model or are at least closely related to them. Only in
the presence of supersymmetry can multiplets contain particles of
different spin, such as the graviton and the photon.

\section{Essentials $of$ Supersymmetry Algebra}

Supersymmetry transformations are generated by quantum operators
$Q$ which change fermionic states to bosonic ones and vice versa,

\begin{equation}
Q\left\vert fermion\right\rangle =\left\vert boson\right\rangle
;\text{ \ \ \ \ \ \ \ \ \ }Q\left\vert boson\right\rangle
=\left\vert fermion\right\rangle .
\end{equation}

Which particular bosons and fermions are related to each other by
the operation of some such $Q?$, how many $Q^{\prime }s$ there
are? and which properties other than statistics of the states are
changed by that operation depends on the supersymmetric model
under study. There are, however, a number of properties which are
common to the $Q^{\prime }s$ in all supersymmetric models.

By definition, the $Q^{\prime }s$ change the statistics and hence
the spin of the state. Spin is related to behavior under spatial
rotations, and thus supersymmetry is- in some since- a space-time
symmetry. Normally, and particularly so in models of
\textquotedblleft extended
supersymmetry\textquotedblright\ (supergravity is being one example), the $%
Q^{\prime }s$ also affect the internal quantum numbers of the
states. It is this property of combining internal with space-time
behavior that makes supersymmetric field theories interesting in
the attempts to unify all fundamental interactions.

As a simple illustration of the non-trivial space-time properties of the $%
Q^{\prime }s,$ consider the following. Because fermions and bosons
behave differently under rotations, the $Q$ can not be invariant
under such rotations. We can, for example, apply the unitary
operator $U(U^{-1}U=1)$ which, in Hilbert space, represents \ a
rotation in configuration space by 360$^{0}$ around some axis.
Since fermionic states pick up a minus sign when rotated through
360$^{0}$ and bosonic states do not, we have

\begin{equation}
U\left\vert fermion\right\rangle =-\left\vert fermion\right\rangle
,\text{ \ \ \ \ \ \ \ \ \ \ }U\left\vert boson\right\rangle
=\left\vert boson\right\rangle ,
\end{equation}

then from equation (2.1) we get

\begin{eqnarray}
Q\left\vert boson\right\rangle &=&\left\vert fermion\right\rangle
=-U\left\vert fermion\right\rangle =-UQ\left\vert
boson\right\rangle
=-UQU^{-1}U\left\vert boson\right\rangle  \notag \\
&=&-UQU^{-1}\left\vert boson\right\rangle ,
\end{eqnarray}

\begin{eqnarray}
Q\left\vert fermion\right\rangle &=&\left\vert boson\right\rangle
=U\left\vert boson\right\rangle =UQ\left\vert fermion\right\rangle
=UQU^{-1}U\left\vert fermion\right\rangle  \notag \\
&=&-UQU^{-1}\left\vert fermion\right\rangle ,
\end{eqnarray}

and since all fermionic and bosonic states, taken together, from a
basis in the Hilbert space, we easily see that we $must$ have

\begin{equation}
UQU^{-1}=-Q
\end{equation}

The rotated supersymmetry generator picks up a minus sign, just as
a fermionic state does. One can extend this analysis and show that
the behavior of the $Q^{\prime }s$ under any Lorentz
transformation -not only under rotations by 360$^{0}$- is
precisely that of a \textsl{spinor operator. \ }More technically
speaking, the $Q^{\prime }s$ transform like tensor operators of
spin $\frac{1}{2}$ and, in particular, they do not commute with
Lorentz transformations followed by a supersymmetry transformation
is different from that when the order of the transformation is
reserved.

It is not easy to illustrate, but it is nevertheless true that, on
the other hand, the $Q^{\prime }s$ are $invariant$ $under$
$transfomation.$ It does not matter whether we translate the
coordinate system before or after we perform a supersymmetry
transformation. In technical terms, this means that
we have a vanishing commutator of $Q$ with the energy of momentum operator $%
E $ $and$ $\mathbf{P}$, which generates space-time translations,

\begin{equation}
\left[ Q,E\right] =\left[ Q,\mathbf{P}\right] =0
\end{equation}

The structure of a set of symmetry operations is determined by the
result of two subsequent operations. For continuous symmetries
like space rotations or supersymmetry, this structure is best
described by the commutators of the generators, such as the ones
given above for the angular momentum operators. The commutator
structure of the $Q^{\prime }s$ with themselves can best be
examined by if they are viewed as products of operators which
annihilate fermions and create bosons instead, or vice versa. It
can be shown that the canonical quantization rules for creation
and annihilation operators of particles (and in particular the
$anti$commutator rules for fermions, which reflect Pauli's
exclusion principle) lead to the results that it is the anti
commutator of two $Q^{\prime }s$, not the commutator, which is
again a supersymmetry generator, albeit one of bosonic nature.

Let us consider the anticommutator of some \thinspace $Q$ with its
Hermitian adjoint $Q^{+}.$ As spinor components, the $Q^{\prime
}s$ are in general not Hermitian, but $\{Q,Q^{+}\}\equiv
QQ^{+}+Q^{+}Q$ is a Hermitian operator with positive definite
eigenvalues;

\begin{equation}
\left\langle ...\left\vert QQ^{+}\right\vert ...\right\rangle
+\left\langle ...\left\vert Q^{+}Q\right\vert ...\right\rangle
=\left\vert Q\left\vert ...\right\rangle \right\vert
^{2}+\left\vert Q^{2}\left\vert ...\right\rangle \right\vert
^{2}\geq 0.
\end{equation}

This can only be zero for all states $\left\vert ...\right\rangle
$ if $Q=0.$ A more detailed investigation will show that
$\{Q,Q^{+}\}$ must be a linear combination of the energy and
momentum operators;

\begin{equation}
\left\{ Q,Q^{+}\right\} =\alpha E+\beta \mathbf{P}
\end{equation}

This relation between the anticommutator of two generators of
supersymmetry
transformations on the one hand and the generators of $space-time$ $%
translations$ (namely energy and momentum) on the other, is
central to the entire field of supersymmetry and supergravity. It
means that the subsequent operations of two finite supersymmetry
transformations will include translations in space and time of the
states on which they operate.

There is a further important consequence of the form of equation
(2.14).
When summing this equation over all supersymmetry generators, we find that $%
\beta \mathbf{P}$ terms cancel while the $\alpha E$ terms add up,
so that

\begin{equation}
\sum\limits_{allQ}\left\{ Q,Q^{+}\right\} \propto E.
\end{equation}

Depending on the sign of the proportionality factor, the spectrum
for the energy would have to be either $\geq 0$ or $\leq 0$
because of the inequality (2.13). For a physical sensible theory
with energies bounded from below but not from above, the
proportionality factor will therefore be positive.

The equation (2.12) to (2.15) are crucial properties of the
supersymmetry generators, and many of the most important features
of supersymmetric theories, whether in flat or curved space-time,
can be derived from them.

One such feature is the \textit{positivity of energy, }\ as can be
seen from equation (2.15)in conjunction with (2.13), \emph{the
spectrum of the energy operator E (the Hamiltonian) in a theory
with supersymmetry contains no negative eigenvalues. }\ We denote
the state (or the family of states) with
the lowest energy by $\left\vert 0\right\rangle $ and call it the \textit{%
vacuum. }The vacuum will have zero energy

\begin{subequations}
\begin{equation}
E\left\vert 0\right\rangle =0
\end{equation}

if and only if

\begin{equation}
Q\left\vert 0\right\rangle =0\text{ \ \ \ \ \ \ \ \ \ and \ \ \ \ \ \ }%
Q^{+}\left\vert 0\right\rangle =0\text{ \ \ \ \ \ \ \ for all }Q
\end{equation}

Any state whose energy is not zero, e.g. any one-particle state,
can not be invariant under supersymmetry transformations. This
means that there must be
one ( or more) superpartner state $Q\left\vert 1\right\rangle $ or $%
Q^{+}\left\vert 1\right\rangle $ for every one-particle state
$\left\vert 1\right\rangle .$ thus: \emph{each supermultiplet must
contain at least one boson and one fermion whose spins differ by
}$\frac{1}{2}.$ A supermultiplet is a set of quantum states (or,
in different context, of quantum fields) which can be transformed
into one another by one or more supersymmetry transformations.
This is exactly analogous to the concept of \textquotedblleft
multiplet\textquotedblright\ known from atomic, nuclear and
elementary particle physics where e.g., the proton and the neutron
from an isospin doublet and can be transformed into each other by
an isospin rotation.

The translational invariance of $Q,$ expressed by equation (2.12)
implies that $Q$ does not change energy and momentum and that
therefore:\emph{\ all states in a multiplet of unbroken symmetry
have the same mass. }Experiments do not show elementary particles
to be accompanied by superpartners. with different spins but
identical mass. Thus, if supersymmetry is fundamental to nature,
it can only be realized as \emph{a spontaneously broken symmetry}.

The term \textquotedblleft spontaneously broken
symmetry\textquotedblright\ is used when the interaction
potentials in a theory, and therefore the basic dynamics, are
symmetric but the states with lowest energy, the ground state or
vacuum, is not. If a generator of such symmetry acts on the vacuum
the result will not be zero. Perhaps the most familiar example of
a spontaneously broken symmetry is the occurrence of
ferromagnetism in spite of the spherical symmetry of the laws of
electrodynamics. because the dynamics retain essential symmetry of
the theory, states with very high energy tend to lose the memory
of the asymmetry of the ground state and the \textquotedblleft
spontaneously broken symmetry\textquotedblright\ gets
re-established. high energy may mean high temperatures,
essentially because the statistics of the occupation of states is
different from fermions and bosons.

If supersymmetry is spontaneously broken, the ground state will
not be invariant under all supersymmetric operations: $Q\left\vert
0\right\rangle \neq 0$ or $Q^{+}\left\vert 0\right\rangle \neq 0$
for some $Q$. From what we said above in equation (2.16), we
conclude that:\emph{\ supersymmetry is spontaneously broken if and
only if the energy of the lowest lying state (the vacuum) is not
exactly zero. }Whereas spontaneous supersymmetry breaking may lift
the mass degeneracy of the supermultiblets by giving different
masses to different members of the multiplet spectrum itself will
remain intact. In particular, we still need \textquotedblleft
superpartners\textquotedblright\ for all known elementary
particles, although these may not be superheavy or otherwise
experimentally unobtainable. The superpartners carry a new quantum
number (called R-charge). It has been shown that the highly
desirable property of super GUTs model, mentioned in the
introduction, namely that they stabilize the GUT hierarchy, is
closely associated with a strict conservation law for this quantum
number. If nature works that way the lightest particle with a
non-zero R-charge must be stable. Whereas this particle may be so
weakly interacting that it has not yet been observed, its presence
in the universe could crucially and measurably influence
cosmology.

As a matter of convention, fermionic superpartners of known bosons
are denoted by the suffix -ino ( hence \textquotedblleft\
gravitino, photino, gluino\textquotedblright ); the bosonic
superpartners of fermions are denoted by the prefixes s-(
\textquotedblleft squark, slepton\textquotedblright ). The
discovery of any such bosinos or sfermions would confirm the
important prediction of superpartners which is common to all
supersymmetric models. it would be a major breakthrough and would
establish supersymmetry as an important property of the physics of
nature rather than just an attractive hypothesis.

We have not yet specified \textquotedblleft how
much\textquotedblright\ supersymmetry there should be. Do we
propose one spin $\frac{1}{2}$ photino as a partner of a physical
photon, or two, or how many? Different supersymmetric models give
different answers, depending on how many supersymmetric generators
$Q$ are present, as conserved charges, in the model. As already
said, the $Qs$ are spinor operators, and a spinor in four
space-time dimensions must have at least four real components. The
total number of $Q^{\prime }s$ must therefore be a multiple of
four. A theory with minimal supersymmetry would be invariant under
the transformations generated
by just the four independent components of a single spinor operator $%
Q_{\alpha }$ with $\alpha =1,...,4$. We call this \emph{a theory
with N=1 supersymmetry, }\ and it would give rise to, e.g., a
single uncharged massless spin $\frac{1}{2}$ photino which is its
own antiparticle ( a \textquotedblleft Majorana
neutrino\textquotedblright ). If there is more supersymmetry, then
there will be several spinor generators with four
components each, $Q_{\alpha i}$ with $i=1,...N,$ and we speak of a \emph{%
theory with N-extended supersymmetry, }\ which will give rise to
N-photinos. The fundamental relationship (2.14) between the
generators of supersymmetry is now replaced by

\end{subequations}
\begin{equation}
\{Q_{i},Q_{j}^{+}\}=\delta _{ij}\left( \alpha E+\beta
\mathbf{P}\right)
\end{equation}

\section{The Algebra of N=1 supersymmetry}

Coleman and Mandula (1967) showed that under very general
assumptions, a Lie group that contains both the Poincare$^{\prime
}$ group \textit{P }and an
internal symmetry group \textit{G }must be just a direct product of $P$ and $%
G$ [10]. the generators of the Poincare$^{\prime }$ group are the
four-momentum $P_{\mu }(P_{0}=E=H,$ $P_{1}=P_{x},$ $P_{2}=P_{y},$ $%
P_{3}=P_{z}),$ which produces space-time translations, and the
antisymmetric tensors $M_{\mu \upsilon },$ which generates
space-time rotations, that is

\begin{equation}
\mathbf{J\equiv }\left( M_{23},M_{31},M_{12}\right) ;\text{ \ \ \ \ \ \ \ \ }%
\mathbf{K\equiv (}M_{01},M_{02},M_{03}),
\end{equation}

Where the regular momentum operator $J_{i}generates$ space
rotations about the $i-axis$ and $K_{i}$ generates Lorentz boosts
along the $i-axis.$ So, if
the generators of the internal supersymmetry group $G$ are denoted by $%
T_{a}, $ the Coleman-Mandula theorem requires that

\begin{equation}
\left[ P_{\mu },T_{a}\right] =\left[ M_{\mu \upsilon
},T_{a}\right] =0
\end{equation}

This no-go theorem shows that it is impossible to mix internal and
Lorentz space-time symmetries in a non-trivial way. Supersymmetry
escapes this \textquotedblleft no-go\textquotedblright\ theorem
because, in addition to the generators $P_{\mu },$ $M_{\mu
\upsilon },$ $T_{a}$ which satisfy commutation relations, it
involves fermionic generators $Q$ that satisfy anti-commutation
relations. If we call the generators $P_{\mu },$ $M_{\mu \upsilon
},$ $T_{a}$ \textquotedblleft even\textquotedblright , and $Q$
\textquotedblleft odd\textquotedblright , then the supersymmetry
algebra has the general structure

\begin{eqnarray}
\left[ even,even\right] &=&even,  \notag \\
\left[ odd,odd\right] &=&even,  \notag \\
\left[ even,odd\right] &=&odd,
\end{eqnarray}

The above is called \emph{graded Lie Algebra.}

We now present the simplest form of supersymmetry algebra (N=1
supersymmetry). We introduce four generators $Q_{\alpha }(\alpha
=1,...,4),$ which form a four-component Majorana spinor. Majorana
spinors are the simplest possible type of spinor. They are
self-conjugate, i.e.

\begin{equation}
Q=Q^{c}=C\overline{Q^{\dagger }},
\end{equation}

and hence have only half as many degrees of freedom as a Dirac
spinor. Indeed, any Dirac spinor $\psi $ may be written

\begin{equation}
\psi =\frac{1}{\sqrt{2}}\left( \psi _{1}+i\psi _{2}\right) ,
\end{equation}

where

\begin{equation}
\psi _{1}=\frac{1}{\sqrt{2}}(\psi +\psi ^{c}),\text{ \ \ \ \
}and\text{ \ \ \ }\psi _{2}=-\frac{i}{\sqrt{2}}(\psi -\psi ^{c})
\end{equation}

are two independent Majorana spinors that satisfy $\psi _{i}=\psi
_{i}^{c}.$

Since $Q_{\alpha }$ is a spinor, it must satisfy

\begin{equation}
\left[ Q_{\alpha },M_{\mu \upsilon }\right] =\frac{1}{2}\left(
\sigma _{\mu \upsilon }\right) _{\alpha \beta }Q_{\beta }
\end{equation}

This relation expresses the fact that the $Q_{\alpha }$transform
as a spinor under the rotation generated by $M_{\mu \upsilon }$
(recall that $\sigma _{\mu \upsilon }$ when sandwiched between
spinors, transform as an antisymmetric tensor). The Jacobi
identity of commutators

\begin{equation}
\left[ \left[ Q_{\alpha },P_{\mu }\right] ,P_{\nu }\right] +\left[
\left[
P_{\nu },Q_{\alpha }\right] ,P_{\mu }\right] +\left[ \left[ P_{\mu },P_{\nu }%
\right] ,Q_{\alpha }\right] =0
\end{equation}

requires that $Q_{\alpha }$must be transnational invariant, that
is

\begin{equation}
\left[ Q_{\alpha },P_{\mu }\right] =0.
\end{equation}

It is the remaining anticommutation relation

\begin{equation}
\left\{ Q_{\alpha },\overline{Q_{\beta }}\right\} =2(\gamma ^{\mu
})_{\alpha \beta }P_{\mu }
\end{equation}

which we shall derive later, and which closes the algebra, that
has the most interesting consequences. Clearly the anticommutator
has to yield an even generator, which might be either $P_{\mu }$
or $M_{\mu \nu }.$ A term of the form $\sigma ^{\mu \nu }M_{\mu
\nu }on$ the right-hand side would violate a
generalized Jacobi identity involving $Q_{\alpha },$ $Q_{\beta }$ and $%
P_{\mu }$ and the algebra would not close. Indeed, if go back to
the \textquotedblleft\ no-go\textquotedblright\ theorem and allow
for anticommutators as well as commutators, we find that the only
allowed supersymmetries (apart from trivial extension) are those
based on the graded Lie algebra defined by eqs. (2.24)-(2.27).

we choose $Q_{\alpha }$ to be Majorana spinor with four
independent (real) parameters, but we could have used a Weyl
spinor with two complex components equally well. In fact, we shall
find it more convenient to work with a left-handed Weyl spinor
$\psi _{\alpha }$ with $\alpha =1,2,$ and the chiral
representation of the Dirac matrices in which

\begin{equation}
\gamma =%
\begin{pmatrix}
0 & \sigma \\
-\sigma & 0%
\end{pmatrix}%
,\gamma _{0}=%
\begin{pmatrix}
0 & I \\
I & 0%
\end{pmatrix}%
,\gamma _{5}=%
\begin{pmatrix}
-I & 0 \\
0 & I%
\end{pmatrix}%
,C\gamma _{0}^{T}=%
\begin{pmatrix}
0 & i\sigma _{2} \\
-i\sigma _{2} & 0%
\end{pmatrix}%
.
\end{equation}

Using the two component Weyl spinor $\psi _{\alpha },$ we can
construct a Majorana spinor in this chiral representation we find

\begin{equation}
Q=Q^{c}=\QATOPD\{ \} {\psi }{0}+C\gamma _{0}^{T}\QATOPD\{ \} {\psi
^{\ast }}{0}-\QATOPD\{ \} {\psi }{-i\sigma \psi ^{\ast }}.
\end{equation}

We then look for possible supersymmetric representations that
contain massless particles. these should be the relevant
multiplets, the particles we observe are thought to acquire their
masses only as a result of spontaneous symmetry breaking. The
procedure we employ is to evaluate \ the
anticommutator (2.27) for a massless particle moving along the z-axis with $%
P_{\mu }=(E,0,0,E).$ On substituting (2.29) into equation (2.27),
we find

\begin{equation}
\{\psi _{a},\psi _{b}^{+}\}=2E(1-\sigma _{3})_{ab}
\end{equation}

with $a,b=1,2,$ giving

\begin{equation}
\left\{ \psi _{1},\psi _{2}^{+}\right\} =0,\left\{ \psi _{1},\psi
_{1}^{+}\right\} =0,\left\{ \psi _{2},\psi _{2}^{+}\right\} =4E.
\end{equation}

We see that $\psi _{2}^{+}$ and $\psi _{2}$ act as creation and
annihilation operators.

Now, a massless particle of spin $s$ can only have helicities
$\lambda =\pm s,$ so, starting from the left-handed state
$\left\vert s,\lambda =-s\right\rangle ,$ which is annihilated by
$\psi _{2},$ only one new state can be formed, i.e., $\psi
_{2}^{+}\left\vert s,-s\right\rangle .$ This describes a particle
of spin $s+\frac{1}{2}$ and helicity $-(s+\frac{1}{2}),$ and by
virtue of (2.26) it is also massless. Then, acting again with
$\psi _{1}^{+}$ or with $\psi _{2}^{+}$ gives states with zero
norm by virtue of (2.31) and $\psi _{2}^{+}\psi _{2}^{+}=0$ (which
follows from the fermionic nature of $\psi _{2}$). So the
resulting massless irreducible representation consists of just two
states. Hence, the possible supersymmetric multiplets of interest
to us are

\begin{center}
\begin{tabular}{|c|c|}
\hline \textbf{Chairal multiplet} & \textbf{Vector (or gauge)
multiplet} \\ \hline
fermion $\left\vert \frac{1}{2},\frac{1}{2}\right\rangle $ & gauge boson $%
\left\vert 1,1\right\rangle $ \\ \hline
sfermion $\left\vert 0,0\right\rangle $ & gaugino $\left\vert \frac{1}{2},%
\frac{1}{2}\right\rangle $ \\ \hline
\end{tabular}
\end{center}

To maintain \emph{CPT }invariance we must add the antiparticle
states that have opposite helicity, thus giving a total of four
states, $\left\vert s\pm \frac{1}{2},\pm
(s+\frac{1}{2})\right\rangle ,$ $\left\vert s,\pm s\right\rangle
,$in each multiplet.

All the particles in such multiplets carry the same gauge quantum
numbers. For this reason, the know fermions ($i.e$ the quarks and
leptons) must be partnered by spin 0 particles (called
"sfermions"), not spin 1 bosons. This is because the only spin 1
bosons allowed in a renormalizable theory are the gauge bosons and
they have to belong to the adjoin representation of the gauge
group, but the quarks and leptons do not. Instead, the gauge
bosons are partnered by new spin-$\frac{1}{2}$ "gauginos"
(spin-$\frac{3}{2}$ being rules out by the requirement of
renormalizability). The need to introduce new supersymmetry
partners, rather than interrelate the known bosons and fermions,
is undoubtedly a major setback for supersymmetry.

For completeness, we briefly consider also supermultiplets of
particles with nonzero mass $\mathit{M}$. in the particle's rest
frame, $P_{\mu }=(M,0,0,0)$ so the anticommutator (2.27) becomes

\begin{equation}
\{\psi _{a},\psi _{b}^{+}\}=2M\delta _{ab}.
\end{equation}

We see that $\frac{\psi _{a}^{+}}{\sqrt{2M}}$and $\frac{\psi
_{a}}{\sqrt{2M}} $ act as creation and annihilation operators,
repetitively, for both $a=1$ and $2.$ Starting from a spin state
$\left\vert s,s_{3}\right\rangle ,$which is annihilated by the
$\psi _{a},$we reach three other states by the action of $\psi
_{1}^{+},\psi _{2}^{+},$ and $\psi _{1}^{+}\psi _{2}^{+}=-\psi
_{2}^{+}\psi _{1}^{+}.$

\section{The Wess-Zumino Model}

We are now going to consider the construction of supersymmetric
field theories. We shall begin with the model of Wess and Zumino
(1974) of the massless spin 0, spin-$\frac{1}{2}$ multiplet.
Indeed, probably the most intuitive way of introducing
supersymmetry is to explore, through this simple model, possible
\textit{Fermi-Bose }symmetries of the Lagrangian. It cold
therefore equally well have been the starting point for our
discussion of supersymmetry.

The simplest multiplet in which to search for supersymmetry
consists of a two-component \textit{Weyl spinor} (or equivalently
a four-component of \textit{Majorana spinor}) together with two
real scalar fields. To be specific, we take a massless Majorana
spinor field $\psi ,$and massless scalar and pseudoscalar fields
$A$ and $B$, respectively. The Kinetic energy is

\begin{equation}
\mathfrak{L}=\frac{1}{2}(\partial ^{\mu }A)(\partial _{\mu }A)+\frac{1}{2}%
(\partial ^{\mu }B)(\partial _{\mu }B)+\frac{1}{2}i\overline{\psi
}\gamma _{\mu }\partial ^{\mu }\psi .
\end{equation}

The unfamiliar factor of $\frac{1}{2}$ in the fermion term arises because $%
\psi $ is a Majorana spinor; a Dirac spinor is a linear
combination of two Majorana spinors (see(2.22)).

The following bilinear identities are particularly useful when
exploring supersymmetry. For any two Majorana\textit{\ }spinors
$\psi _{1},\psi _{2}$ we have

\begin{equation}
\overline{\psi }_{1}\mathit{\Gamma }\psi _{2}=\eta \overline{\psi }_{2}%
\mathit{\Gamma }\psi _{1}
\end{equation}

where $\eta =(1,1,-1,1,-1)$ for $\Gamma =(1,\gamma _{5},\gamma
_{\mu },\gamma _{\mu }\gamma _{5},\sigma _{\mu \nu }).$

To discover the Fermi-Bose symmetries of $\mathfrak{L}$, we make
the following infinitesimal transformations: $A\rightarrow
A^{^{\prime }}=A+\delta A,$etc., where

\begin{subequations}
\begin{eqnarray}
\delta A &=&\overline{\varepsilon }\psi , \\
\delta B &=&i\overline{\varepsilon }\gamma _{5}\psi , \\
\delta \psi &=&-i\gamma ^{\mu }\partial _{\mu }(A+i\gamma
_{5}B)\varepsilon .
\end{eqnarray}

$\varepsilon $ being a constant infinitesimal Majorana spinor that
anticommutes with $\psi $ and commutes with $A$ and $B.$ These
transformations are clearly Lorentz-covariant but otherwise $\delta A$ and $%
\delta B$ are just fairly obvious first guesses. The possibility
of just
constructing two independent invariant quantities $\overline{\varepsilon }%
\psi $ and $\overline{\varepsilon }\gamma _{5}\psi $ is the reason
for introducing both scalar and pseudoscalar fields. Since $A$ and
$\psi $ have mass dimensions $1$ and $\frac{3}{2},$ respectively,
$\varepsilon $ must have dimension $-\frac{1}{2}.$ The derivative
in $\delta \psi $ is therefore required to match these dimensions.
We have assumed that the transformations have to be linear in the
fields.

Under (2.35) the change in $\mathfrak{L}$ can be written in the
form

\end{subequations}
\begin{eqnarray}
\delta \mathfrak{L} &\mathfrak{=}&\partial ^{\mu }A\partial _{\mu
}(\delta
A)+\partial ^{\mu }B\partial _{\mu }(\delta B)+\frac{1}{2}i(\delta \overline{%
\psi })\gamma ^{\mu }\partial _{\mu }\psi +\frac{1}{2}i\overline{\psi }%
\gamma ^{\mu }\partial _{\mu }(\delta \psi )  \notag \\
&=&\partial ^{\mu }A\overline{\varepsilon }\partial _{\mu }\psi
+i\partial
^{\mu }B\overline{\varepsilon }\gamma _{5}\partial _{\mu }\psi -\frac{1}{2}%
\overline{\varepsilon }\gamma ^{\nu }\gamma ^{\mu }\partial _{\mu
}(A+i\gamma _{5}B)\partial _{\mu }\psi +\frac{1}{2}\overline{\psi
}\gamma ^{\mu }\partial _{\mu }[\gamma ^{\nu }\partial _{\nu
}(A+i\gamma
_{5}B)\varepsilon ]  \notag \\
&=&\partial _{\mu }\left[ \overline{\varepsilon }\{\partial ^{\mu
}(A+i\gamma _{5}B)-\frac{1}{2}\gamma ^{\nu }\gamma ^{\mu }\partial
_{\nu
}(A+i\gamma _{5}B)\}\psi \right]  \notag \\
&=&\partial _{\mu }[\frac{1}{2}\overline{\varepsilon }\gamma ^{\mu
}\{\partial (A+i\gamma _{5}B)\}\psi ],
\end{eqnarray}

where we have used the identities

\begin{equation}
\overline{\varepsilon }\psi =\psi \overline{\varepsilon },\text{ and }%
\overline{\varepsilon }\gamma ^{5}\psi =\overline{\psi }\gamma
^{5}\varepsilon ,
\end{equation}

of (2.34). Since $\delta \mathfrak{L}$ is a total derivative, it
integrates to zero when we form the action. hence, the action is
invariant under the combined global supersymmetric transformations
(2.35) that mix the fermion
and the boson fields. As usual, "global" is used ti indicate that $%
\varepsilon $ is independent of space-time.

We have remarked that the $\delta \psi $ transformations (2.35c)
contains a derivative. It thus relates the Fermi-Bose symmetry to
the Poincare$^{\prime }$ group. In particular, the appearance of
the time derivative gives an \textit{absolute }significance to the
total energy, which is normally absent in theories that do not
involve gravity.

Returning attention again to the global transformations (2.35), we
recall that the commutator of two successive transformations of a
symmetry group must itself be a symmetry transformation. In this
way we identify the algebra of the generators of the group. To
obtain the corresponding result for supersymmetry, we must
therefore consider two successive supersymmetric transformations
like (2.35). For example, if for the scalar field \textit{A} we
may take a transformation (2.35a) associated with parameter
$\varepsilon _{1},$ followed another with parameter $\varepsilon
_{2},$ then we obtain from (2.35c)

\begin{equation}
\delta _{2}(\delta _{1}A)=\delta _{2}(\overline{\varepsilon _{1}}\psi )=-i%
\overline{\varepsilon _{1}}\gamma ^{\mu }\partial _{\mu
}(A+i\gamma _{5}B)\varepsilon _{2}.
\end{equation}

Hence, the commutator

\begin{eqnarray}
(\partial _{2}\partial _{1}-\partial _{1}\partial _{2})A &=&-i\overline{%
\varepsilon _{1}}\gamma ^{\mu }\partial _{\mu }(A+i\gamma
_{5}B)\varepsilon _{2}+i\overline{\varepsilon _{2}}\gamma ^{\mu
}\partial _{\mu }\left(
A+i\gamma _{5}B\right) \varepsilon _{1}  \notag \\
&=&-2i\overline{\varepsilon _{1}}\gamma ^{\mu }\varepsilon
_{2}\partial
_{\mu }A  \notag \\
&=&-2\overline{\varepsilon _{1}}\gamma ^{\mu }\varepsilon
_{2}P_{\mu }A
\end{eqnarray}

as the terms involving \textit{B} cancel when we use the
identities for Majorana spinors (2.34) and $i\partial _{\mu
}=P_{\mu }.$

Now the generator of supersymmetric transformations $Q_{\alpha }$
is a four-component Majorana spinor, which we define by the
requirement that

\begin{equation}
\delta A=\overline{\varepsilon }QA,
\end{equation}

To make this consistent with (2.39), we form the commutator

\begin{eqnarray}
\left[ \partial _{2},\partial _{1}\right] A &=&\left[
\overline{\varepsilon _{2}}Q,\overline{\varepsilon _{1}}Q\right]
A=\left[ \overline{Q}\varepsilon
_{2},\overline{\varepsilon _{1}}Q\right] A  \notag \\
&=&(\overline{Q_{\beta }}\varepsilon _{2\beta }\overline{\varepsilon }%
_{1\alpha }Q_{\alpha }-\overline{\varepsilon }_{1\alpha }Q_{\alpha }%
\overline{Q}_{\beta }\varepsilon _{2\beta })A  \notag \\
&=&-\overline{\varepsilon }_{1\alpha }\varepsilon _{2\beta }\{Q_{\alpha },%
\overline{Q}_{\beta }\}A,
\end{eqnarray}

using (2.37). Writing (2.39) in component form and equating it
with (2.41) reveals the requirement

\begin{equation}
\{Q_{\alpha },\overline{Q}_{\beta }\}=2(\gamma ^{\mu })_{\alpha
\beta }P_{\mu }
\end{equation}

which is indeed part of the supersymmetry algebra (2.27). The same
commutator is found on applying successive supersymmetry
transformations to field $B$.

Finally, we must check that the algebra closes when acting on the
spinor field $\psi .$We find from (2.35c)

\begin{eqnarray}
\left[ \delta _{2},\delta _{1}\right] \psi &=&-i\gamma ^{\mu
}\partial _{\mu
}\delta _{2}(A+i\gamma _{5}B)\varepsilon _{1}-(1\leftrightarrow 2)  \notag \\
&=&-i\eth (\overline{\varepsilon }_{2}\psi \varepsilon _{1}+i\gamma _{5}%
\overline{\varepsilon }_{2}i\gamma _{5}\psi \varepsilon
_{1})\varepsilon
_{1}-(1\leftrightarrow 2)  \notag \\
&=&-2i\overline{\varepsilon }_{1}\gamma ^{\mu }\varepsilon
_{2}\partial _{\mu }\psi +i\overline{\varepsilon }_{1}\gamma ^{\nu
}\varepsilon _{2}\gamma _{\nu }\eth \psi ,
\end{eqnarray}

where the last equality uses a Fierz rearrangements to bring the two $%
\varepsilon $ together, as well as the Majorana identities (2.37).
If we use the field equation $\eth \psi =0$ for a free massless
fermion, the last term vanishes identically and (2.43) has exactly
the same form as (2.39) for the field $A$, and hence we obtain
(2.42) as before.

However, there is a problem with (2.43) because it gives the
required closure only when $\psi $ satisfies the Dirac equation,
but not for interacting fermions that are \textquotedblleft\
off-the mass-shell\textquotedblright . The reason is that for
off-mass-shell particles, the numbers of fermion and boson degrees
of freedom no longer match up. $A$ and $B$ still have two bosonic
degrees of freedom, whereas the Majorana spinor $\psi $ has four.
We can restore the symmetry by adding two extra bosonic fields,
$\mathfrak{F}$ and $\mathfrak{G}$ (called auxiliary fields), whose
free Lagrangian takes the form

\begin{equation}
\mathfrak{L}=\frac{1}{2}\mathfrak{F}^{2}+\frac{1}{2}\mathfrak{G}^{2}.
\end{equation}

This gives the field equations $\mathfrak{F=G=}0\mathfrak{,}$ so
these new fields have no on-mass-shell states. From (2.44) they
clearly must have mass dimension 2, so on dimensional grounds
their supersymmetry transformations can only take the forms

\begin{subequations}
\begin{equation}
\mathfrak{\delta F=}-i\overline{\varepsilon }\gamma ^{\mu
}\partial _{\mu
}\psi ,\text{ }\ \ \ \ \ \ \ \ \ \ \ \ \mathfrak{\delta G=}\overline{%
\varepsilon }\gamma ^{5}\gamma ^{\mu }\partial _{\mu }\psi
\end{equation}

and (2.35c) becomes,\qquad \qquad \qquad \qquad \qquad \qquad

\begin{equation}
\delta \psi =-i\gamma ^{\mu }\partial _{\mu }\left( A+i\gamma
_{5}B\right) \varepsilon +\left( \mathfrak{F}+i\gamma
_{5}\mathfrak{G}\right) \varepsilon .
\end{equation}

The mass dimensions prevent $\mathfrak{F}$ and $\mathfrak{G}$ from
occurring in

\begin{equation}
\delta A=\overline{\varepsilon }\psi ,\ \ \ \ \ \ \ \ \ \ \delta B=i%
\overline{\varepsilon }\gamma _{5}\psi .
\end{equation}

Under these modified supersymmetry transformations (2.45), we can
show that the unwanted term in (2.43) cancels and, moreover, that

\end{subequations}
\begin{equation}
\left[ \delta _{1},\delta _{2}\right] \mathfrak{F}=-2i\overline{\varepsilon }%
_{1}\gamma ^{\mu }\overline{\varepsilon }_{2}\partial _{\mu
}\mathfrak{F,}
\end{equation}

and similarly for $\mathfrak{G}$, as required by (2.42).

In this way, we have obtained the spin 0, spin-$\frac{1}{2}$
realization of supersymmetry originally found by Wess and Zumino
(1974).

\section{Mass $and$ Interaction Terms $in$ $the$ Lagrangian}

We have found that the free Lagrangian

\begin{equation}
\mathfrak{L}=\frac{1}{2}\partial _{\mu }A\partial _{\mu }A+\frac{1}{2}%
\partial ^{\mu }B\partial _{\mu }B+\frac{1}{2}i\overline{\psi }\eth \psi +%
\frac{1}{2}\mathfrak{F}^{2}+\frac{1}{2}\mathfrak{G}^{2},
\end{equation}

that describes the multiplet ($A,B,\psi
,\mathfrak{F},\mathfrak{G}$), is invariant (up to a total
derivative) under the supersymmetry transformations (2.45).
However, it is easy to check that supersymmetry invariance is
still preserved if the Lagrangian is extended to include a
quadratic mass term of the form

\begin{equation}
\mathfrak{L}_{m}=m(\mathfrak{F}A+\mathfrak{G}B-\frac{1}{2}\psi \overline{%
\psi })
\end{equation}

and a cubic interaction term

\begin{equation}
\mathfrak{L}_{i}=\frac{g}{\sqrt{2}}[\mathfrak{F}A^{2}-\mathfrak{F}B^{2}+2%
\mathfrak{G}AB-\overline{\psi }(A-i\gamma _{5}B)\psi ]
\end{equation}

Higher-order terms must be excluded because they are
nonrenormalizable. When we use the classical equations of motion,

\begin{equation}
\frac{\partial \mathfrak{L}}{\partial \mathfrak{F}}\mathfrak{=}\frac{%
\partial \mathfrak{L}}{\partial \mathfrak{G}}=0,
\end{equation}

for the complete Lagrangian, $\mathfrak{L=L}_{0}+\mathfrak{L}_{m}+\mathfrak{L%
}_{i},$ we find

\begin{subequations}
\begin{eqnarray}
\mathfrak{F}+mA+\frac{g}{\sqrt{2}}(A^{2}-B^{2}) &=&0, \\
\mathfrak{G}+mB+\sqrt{2}gAB &=&0.
\end{eqnarray}

These equations of motion are purely algebraic and so the dynamics
is
unchanged if we use them to eliminate the auxiliary fields $\mathfrak{F\ }$%
and $\mathfrak{G}$ from the Lagrangian. We obtain

\end{subequations}
\begin{eqnarray}
\mathfrak{L}\mathfrak{=\ } &&\frac{1}{2}\partial _{\mu }A\partial ^{\mu }A+%
\frac{1}{2}\partial _{\mu }B\partial ^{\mu }B+\frac{1}{2}i\overline{\psi }%
\eth \psi -\frac{1}{2}m\overline{\psi }\psi -\frac{1}{2}m^{2}(A^{2}+B^{2})-%
\frac{1}{\sqrt{2}}mgA(A^{2}+B^{2})  \notag \\
&&-\frac{1}{2}g^{2}(A^{2}+B^{2})^{2}-\frac{1}{\sqrt{2}}g\overline{\psi }%
(A-i\gamma ^{5}B)\psi
\end{eqnarray}

Several features of this Lagrangian, which are characteristics of
supersymmetric theories, are worth nothing. The masses of the
scalars and the fermions are all equal. There are cubic and
quartic couplings between
the scalar fields, and also a Yukawa-type interaction between the fermion $%
\psi $ and the scalars $A,$ and $B$ yet in total there are only
two free parameters: $m$ and $g.$ These interrelation between
fermion and boson masses and couplings is the essence of the
supersymmetry.

The model can also be shown to have some remarkable
renormalization properties in that, despite the presence of the
scalar fields, there is no renormalization of the mass and
coupling constant (although wave function renormailzation is still
necessary). the divergences arising from boson loops are cancelled
by those from fermion loops which have the opposite sign. This is
just the type of cancellation we need to stabilize the gauge
hierarchy. These powerful nonrenormailzation theorems make
supersymmetry particularly compelling. However when we break
supersymmetry, as well give the absence of fermion-boson mass
degeneracy in nature, we have to be careful to preserve the
relations between the couplings of particles of different spin
implied in (2.52).

\section{The Superpotential}

To see how this results generalize with higher symmetries, it is
convenient to work entirely with left handed fermion fields.
Recall from (2.29) that Majorana spinor $\psi $ can be formed
entirely from a left-handed Weyl spinor

\begin{equation}
\psi =\psi _{L}+C\overline{\psi }_{L}^{T}
\end{equation}

and that the mass term is

\begin{eqnarray}
\mathfrak{L} &=&m\overline{\psi }\psi =m\overline{\psi
}_{R}^{c}\psi
_{L}+herm.conj.  \notag \\
&=&m\psi _{L}^{T}C\psi _{L}+herm.conj.
\end{eqnarray}

using the relation

\begin{equation}
\overline{\psi }_{R}^{c}=\psi _{R}^{c+}\gamma ^{0}=\psi _{L}^{\ast
+}\gamma _{0}^{T+}C^{+}\gamma ^{0}=-\psi _{L}^{T}C^{-1}=\psi
_{L}^{T}C.
\end{equation}

For simplicity we have set $-C^{-1}=C,$ which is valid in all the
familiar representations of the Dirac matrices. We can rewrite the
supersymmetry Lagrangian of section (2.6) using just a left handed
field $\psi _{L},$ and complex fields $\phi $ and $F$ for its
scalar partner, viz.,

\begin{equation}
\phi \equiv \frac{1}{\sqrt{2}}(A+iB),\ \ \ \ \ \ \ \ \ \ \ \ \ \ \
and\ \ \
\ \ \ \ \ \ \ \ \ \ \ \ \ \ F\equiv \frac{1}{\sqrt{2}}(\mathfrak{F}-i%
\mathfrak{G})
\end{equation}

From $\mathfrak{L}$ $=\mathfrak{L}_{0}+\mathfrak{L}_{m}\ +\mathfrak{L}$%
\bigskip $_{i}$ we obtain

\begin{equation}
\mathfrak{L}=\partial _{\mu }\phi ^{\ast }\partial ^{\mu }\phi +i\overline{%
\psi }_{L}\eth \psi _{L}+FF^{\ast }+m(\phi F-\frac{1}{2}\psi
_{L}^{T}C\psi _{L})+herm.conj.+g(\phi ^{2}F-\phi \psi
_{L}^{T}C\psi _{L})+hem.conj.
\end{equation}

Then using the equation of motion $\frac{\partial
\mathfrak{L}}{\partial \mathfrak{F}}=0,$ which gives

\begin{equation}
F^{\ast }=-m\phi -g\phi ^{2},
\end{equation}

we can eliminate the auxiliary field $F^{\ast }$ and so the
Lagrangian becomes

\begin{equation}
\mathfrak{L}=\partial _{\mu }\phi ^{\ast }\partial ^{\mu }\phi +i\overline{%
\psi }_{L}\eth \psi _{L}-\left\vert m\phi +g\phi ^{2}\right\vert -(\frac{1}{2%
}m\psi _{L}^{T}C\psi _{L}+g\phi \psi _{L}^{T}C\psi
_{L}+herm.conj.)
\end{equation}

It is useful to re-examine the Lagrangian (2.57) in terms of an
analytic function W($\phi $), known as the \textquotedblleft
superpotential\textquotedblright , viz.,

\begin{equation}
\mathfrak{L}=\mathfrak{L}_{K.E.}+FF^{\ast }+F^{\ast }\frac{\partial W}{%
\partial \phi }+F^{\ast }\frac{\partial W^{\ast }}{\partial \phi ^{\ast }}-%
\frac{1}{2}(\frac{\partial ^{2}W}{\partial \phi ^{2}}\psi
_{L}^{T}C\psi _{L}+h.c.),
\end{equation}

where $\mathfrak{L}_{K.E.}$ denotes the sum of the kinetic energy
terms of the $\phi $ and $\psi _{L}$ fields. Note that $W$, which
is of dimension 3,
depends only on $\phi $and not on $\phi ^{\ast }.$ Upon using $\frac{%
\partial \mathfrak{L}}{\partial \mathfrak{F}}=\frac{\partial \mathfrak{L}}{%
\partial \mathfrak{F}^{\ast }}=0$ to eliminate the auxiliary fields, we find

\begin{equation}
\mathfrak{L}=\mathfrak{L}_{K.E.}-\left\vert \frac{\partial W}{\partial \phi }%
\right\vert ^{2}-\frac{1}{2}(\frac{\partial ^{2}W}{\partial \phi
^{2}}\psi _{L}^{T}C\psi _{L}+h.c.).
\end{equation}

For a renormalizable theory $W$ can be, at most, a cubic function
of $\phi ,$ since otherwise the Lagrangian would contain couplings
with dimension less than 0. Substituting

\begin{equation}
W=\frac{1}{2}m\phi ^{2}+\frac{1}{3}g\phi ^{3}
\end{equation}

into (2.61) immediately reproduces the Lagrangian of (2.59). The
superpotential is the only free function in the supersymmetry
Lagrangian and determines both the potential of the scalar fields,
and the masses and couplings of fermions and bosons.

In general there may be several chiral multiplets to consider, For
example, if $\psi ^{i}$ belongs to \ a representation of an
$SU(N)$ symmetry group, we will have the supermultiplets

\begin{equation}
(\phi ^{i},\psi _{L}^{i})
\end{equation}

where in the fundamental representation $i=1,2,...,N.$ From (2.61)
we readily obtain a Lagrangian that is invariant under the
additional symmetry and incorporates the new supermultiplets. It
is

\begin{equation}
\mathfrak{L}_{chiral}=\sum_{i}\left\vert \partial _{\mu }\phi
^{i}\right\vert ^{2}+i\sum_{i}\overline{\psi _{L}^{i}}\eth \psi
_{L}^{i}-\sum_{i}\left\vert \frac{\partial W}{\partial \phi
^{i}}\right\vert ^{2}-\frac{1}{2}\left( \sum_{i,j}\frac{\partial
^{2}W}{\partial \phi ^{i}\partial \phi ^{j}}\psi _{L}^{iT}C\psi
_{L}^{j}+herm.conj.\right) ,
\end{equation}

and the most general form of the superpotential $W$ is

\begin{equation}
W=\lambda _{i}\phi ^{i}+\frac{1}{2}m_{ij}\phi ^{i}\phi ^{j}+\frac{1}{3}%
g_{ijk}\phi ^{i}\phi ^{j}\phi ^{k},
\end{equation}

where the coefficients $m$ and \thinspace $g$ are completely
symmetric under interchange of indices. The relevance of the term
that is linear in the fields will be discussed below. Since $W$
must be invariant under $SU(N)$ symmetry transformations this term
can only occur if a gauge-singlet field exists.

\section{Supersymmetric Gauge Theory}

A combination of supersymmetry with gauge theory is clearly
necessary if these ideas are to make any contact with the real
world. In \ addition to the chiral multiplets of (2.63), we must
include the \textquotedblleft gauge\textquotedblright\
supermultiplets

\begin{equation}
\left( A_{\mu }^{a},\chi ^{a}\right) ,\text{ \ \ \ \ \ \ \ }%
a=1,2,3,...,N^{2}-1,
\end{equation}

where $A_{\mu }^{a}$ are the spin 1 gauge bosons of the gauge
group G (taken to be SU(N)) and $\chi ^{a}$ are their Majorana
fermion superpartners ( the so called \textquotedblleft
gauginos\textquotedblright ). These boson-fermion pairs, which in
the absence of symmetry breaking are assumed to be massless,
belong to the adjoint representation of the gauge group. Our task
is to find a supersymmetry, ad a gauge-invariant Lagrangian
containing all these chiral and gauge supermultiplets.

The gauge multiplets are described by the Lagrangian

\begin{equation}
\mathfrak{L}_{G}=\frac{-1}{4}F_{\mu \nu }^{a}F_{a}^{\mu \nu }+\frac{1}{2}i%
\overline{\chi ^{a}}(D\chi )_{a}+\frac{1}{2}(D^{a})^{2},
\end{equation}

where the gauge field-strength tensor is

\begin{equation}
F_{a}^{\mu \nu }=\partial ^{\mu }\chi _{a}-g_{G}f_{abc}A_{b}^{\mu
}A_{c}^{\nu },
\end{equation}

$D^{\mu }$ is the covariant derivative satisfying

\begin{equation}
(D^{\mu }\chi )_{a}=\partial ^{\mu }\chi
_{a}-g_{G}f_{abc}A_{b}^{\mu }\chi _{c}
\end{equation}

and $D^{\mu }$ are auxiliary scalar fields ( similar to $F_{i}$ of
the chiral multiplet).

Actually, for this pure gauge Lagrangian the equation of motion
$\partial \mathfrak{L}_{G}/\partial D^{a}=0,$ implies $D^{a}=0;$
however, it will become nonzero when the chiral fields are coupled
in the notation will be familiar: $g_{G}$ and $f_{abc}$ are the
coupling and the structure constants of the gauge group, and in
(2.96) the matrices $T^{b}$ representing the
generators in the adjoint representation have been replaced by (\thinspace $%
T^{b})=if_{abc.}$ It is straight forward to show that the lagrangian $%
\mathfrak{L}_{G}$ is invariant, and that the algebra closes, under
the supersymmetry transformation

\begin{eqnarray}
\delta A_{a}^{\mu } &=&-\overline{\varepsilon }\gamma ^{\mu
}\gamma ^{5}\chi
_{a},  \notag \\
\delta \chi ^{a} &=&-\frac{1}{2}\sigma ^{\mu \nu }F_{\mu \nu
}^{a}\gamma
^{5}\varepsilon +D^{a}\varepsilon , \\
\delta D^{a} &=&-i\overline{\varepsilon }(D\chi )^{a},  \notag
\end{eqnarray}

where $\varepsilon $ is a constant infinitesimal Majorana spinor.
This transformation is analogous to (2.35) for the chiral
multiplet.

To include the chiral fields ($\phi ^{i},\psi _{L}^{i}),$ we add $\mathfrak{L%
}_{chiral}$ of (2.64) but substituting the covariant derivative
$D_{\mu }$ for $\partial _{\mu }$ in the kinetic energy terms,
viz.,

\begin{equation}
\partial _{\mu }\rightarrow D_{\mu }=\partial _{\mu }+ig_{G}T^{a}A_{\mu
}^{a},
\end{equation}

where $T^{a}$ are the matrices representing the generators of the
gauge group in the representation to which $(\phi _{i},\psi _{i})$
belong. To ensure the supersymmetry of the combined
\textquotedblleft chiral + gauge\textquotedblright\ Lagrangian, \
we must include two further terms, and write

\begin{equation}
\mathfrak{L=L}_{chiral}+\mathfrak{L}_{G}-g_{G}\phi _{i}^{\ast
}(T^{a})_{ij}\phi _{j}D^{a}+[\sqrt{2}g_{G}\phi _{i}^{\ast
}\overline{\chi ^{a}}(T^{a})_{ij}P_{L}\psi _{j}+herm.conj.]
\end{equation}

where $P_{L}\equiv \frac{1}{2}(1-\gamma ^{5}),$ and replace
$\partial _{\mu } $ in (2.45) by $D_{\mu }.$Using $\partial
\mathfrak{L}_{G}/\partial D^{a}=0 $ to eliminate the auxiliary
field gives

\begin{equation}
D^{a}=g_{G}\phi _{i}^{\ast }(T^{a})_{ij}\phi _{j}.
\end{equation}

The term in the Lagrangian that contribute to the potential for
the scalar fields are evident by inspection of (2.61) and (2.67).
They are

\begin{eqnarray}
V(\phi ,\phi ^{\ast }) &=&\left\vert F_{i}\right\vert ^{2}+\frac{1}{2}%
D_{a}^{2}  \notag \\
&=&\sum_{i}\left\vert \frac{\partial W}{\partial \phi _{i}}\right\vert ^{2}+%
\frac{1}{2}\sum_{a}[g_{G}\sum_{i,j}\phi _{i}^{\ast
}(T^{a})_{ij}\phi _{j}]^{2},
\end{eqnarray}

which are known as $F$ \ and $D$ terms, respectively. This
potential will play a central role in the spontaneous breaking of
supersymmetry and the gauge symmetry.

Model building begins with the supersymmetry Lagrangian (2.72).
Apart from the choice of gauge group and the representations
formed by the chiral
multiplets, the only freedom lies in the choice of the superpotential $%
W(\phi _{i})$ which must, of course, be a single of the gauge
group.

\section{Spontaneous Breaking $of$ Supersymmetry}

The particles observed in nature show no sign whatsoever of
degeneracy between fermions and bosons. Even the photon and
neutrino, which appear to be degenerate in mass, can not be
supersymmetric partners. Hence, supersymmetry, is it to be
relevant in nature, must be broken.

The breaking could be either explicit or spontaneous. Explicit
breaking would be quite \textsl{ad hoc. }The supersymmetry
generators would no longer commute with the Hamiltonian,

\begin{equation}
\left[ Q_{\alpha },H\right] \neq 0
\end{equation}

and so the violation would have to be small enough to preserve the
good features of supersymmetry and yet large enough to push the
supersymmetric partners out of reach of current experiments.
However, we would inevitably lose the nice nonrenormalization
theorems and, even worse, any attempt to embrace gravity viz local
supersymmetry would be prohibited. So, instead, we prefer to
consider the spontaneous breaking of supersymmetry, not least
because this has proved so successful previously for breaking
gauge symmetries. Hence, we assume that the Lagrangian is
supersymmetric but that the vacuum state is not, that is

\begin{equation}
\left[ Q_{\alpha },H\right] =0\text{, \ \ \ \ but \ \ \ \ \ \ \ \
}Q_{\alpha }\left\vert 0\right\rangle \neq 0
\end{equation}

A new feature arises here, however. The Higgs mechanism of
spontaneous symmetric breaking is not available in supersymmetry
because, if we were to introduce a spin 0 field with negative
mass-squared, its fermionic superpartner would have an imaginary
mass. Also, using the anticommutator (2.27),

\begin{equation}
\left\{ Q_{\alpha },Q_{\delta }^{+}\right\} \gamma _{\delta \beta
}^{0}=2\gamma _{\alpha \beta }^{0}P_{\mu },
\end{equation}

we can directly establish a general and important theorem. If we
multiply (2.77) by $\gamma _{\beta \alpha }^{0}$ and sum over
$\beta $ and $\alpha $, we obtain

\begin{equation}
\sum_{\alpha }\{Q_{\alpha },Q_{\alpha }^{+}\}=8P_{0}=8H
\end{equation}

and hence

\begin{equation}
8\left\langle 0\right\vert H\left\vert 0\right\rangle
=\sum_{\alpha }\left\vert Q_{\alpha }\left\vert 0\right\rangle
\right\vert ^{2}+\sum_{\alpha }\left\vert Q_{\beta }^{+}\left\vert
0\right\rangle \right\vert ^{2}
\end{equation}

It follows immediately that

1- The vacuum energy must be greater than or equal to zero;

2- If the vacuum is supersymmetric, that is, if $Q_{\alpha
}\left\vert 0\right\rangle =Q_{\beta }^{+}\left\vert
0\right\rangle =0$ for all $\alpha , $ the vacuum energy is zero;
and

3- Conversely, if supersymmetry is spontaneously broken,
$Q_{\alpha }\left\vert 0\right\rangle \neq 0,$then the vacuum
energy is positive.

\bigskip

These results have a disappointing consequences. Conclusion (1)
gives an absolute meaning to the zero of energy, a fact that it
was hoped to use to explain why the vacuum energy of the universe
is zero or very close to zero. But now from (3) we see that broken
supersymmetry implies a positive vacuum energy.

Leaving this aside we can see from (3) that supersymmetry breaking
is rather special because it requires the ground-state energy to
be positive.

In The classical approximation, the energy of the ground state is
given by the minimum of the potential (2.74):

\begin{eqnarray}
V(\phi ,\phi ^{\ast }) &=&\left\vert F_{i}\right\vert ^{2}+\frac{1}{2}%
D_{a}^{2}  \notag \\
&=&\sum_{i}\left\vert \frac{\partial W}{\partial \phi _{i}}\right\vert ^{2}+%
\frac{1}{2}\sum_{\beta ,a}\left[ g_{\beta }\phi _{i}^{\ast
}(T_{\beta }^{\ast })_{ij}\phi _{j}+\eta \delta _{\beta 1}\right]
^{2},
\end{eqnarray}

with

\begin{equation}
W=\lambda _{i}\phi _{i}+\frac{1}{2}m_{ij}\phi _{i}\phi
_{j}+g_{ijk}\phi _{i}\phi _{j}\phi _{k}
\end{equation}

The sum over $\beta $ has been included to allow for the
possibility of different gauge groups with different couplings,
and the constant term $\eta $ can only occur if $\beta $ labels a
$U(1)$ factor.

It is evidently hard to break supersymmetry. The minimum $V=0$
will occur when $\phi _{i}=0$ for all $i$ ( and so supersymmetry
will be broken) unless one of the following conditions applies.

1- $\lambda _{i}\neq 0,$ that is, there exists a gauge singlet
field $\phi
_{i},$ so $W$ can contain a linear term yet still be gauge invariant ($%
F-type $ $breaking$);

2- $\eta \neq 0,$ so the gauge group contain an abelian $U(1)$ factor ($%
D-type$ $breaking$). This is a necessary but not a sufficient
requirement. This mechanism can not occur in $GUTs$ because they
are based on simple gauge groups that do not have $U(1)$ factor.

There is an alternative way of seeing that the spontaneous
symmetry breaking of supersymmetry can only be accomplished by
$\left\langle F\right\rangle \neq 0$ and/or $\left\langle
D\right\rangle \neq 0.$ If we look back at the general structure
of supersymmetric transformations (2.45) for the chiral multiplet
($\phi ,\psi ,F$), which takes the form

\begin{equation}
\delta \phi \sim \psi ,\text{ \ \ \ \ \ \ }\delta \psi \sim \eth \phi +F,%
\text{ \ \ \ \ \ \ }\delta F\sim \eth \phi
\end{equation}

and at (2.70) for the gauge multiplet ($A_{\mu },\chi ,D$), in
which

\begin{equation}
\delta A_{\mu }\sim \gamma _{\mu }\chi ,\text{ \ \ \ \ \ \ }\delta
\chi \sim \sigma ^{\mu \nu }F_{\mu \nu }+D,\text{ \ \ \ \ \ \
}\delta D\sim \eth \chi ,
\end{equation}

and note that the vacuum expectation values of the spinor and
tensor fields and $\partial _{\mu }\phi $ must be zero to preserve
the Lorentz invariance of the vacuum, then it is only to break the
symmetry through the non-zero VEVs of the auxiliary fields $F$ and
$D.$

The spontaneous breaking of supersymmetry requires

\begin{equation}
Q_{\alpha }\left\vert 0\right\rangle \neq 0
\end{equation}

and $Q_{\alpha }\left\vert 0\right\rangle $is necessarily a
fermionic state, which we denote by $\left\vert \psi
_{G}\right\rangle .$

Since the $Q_{\alpha }$ commute with $H$, the state $\left\vert
\psi _{G}\right\rangle $ must be degenerate with the vacuum. It
must therefore describe a massless fermion (with zero momentum).
The situation is thus exactly analogous to the spontaneous
breaking of ordinary global symmetry in which massless Goldstone
bosons are created out of the vacuum. Here the spontaneous
breaking of global supersymmetry implies the existence of a
massless fermion, which is called \textquotedblleft
Goldstino.\textquotedblright

We next consider examples of these two types of symmetry breaking
$F-type$ \ and $D-type$ \ introduced in (1) and (2) above.

\section{F-type Breaking (O'raifeartaigh Model)}

A simple example of supersymmetry breaking arising from the
presence of a linear term in the superpotential $W$ is provided by

\begin{equation}
W=-\lambda A+mBC+gAB^{2},
\end{equation}

which contain three complex scalar fields $A,B,$ and $C.$ In this
example the scalar potential (2.80) becomes

\begin{eqnarray}
V &=&\left\vert \frac{\partial W}{\partial A}\right\vert
^{2}+\left\vert
\frac{\partial W}{\partial B}\right\vert ^{2}+\left\vert \frac{\partial W}{%
\partial C}\right\vert ^{2}  \notag \\
&=&\left\vert -\lambda +gB^{2}\right\vert ^{2}+\left\vert
mC+2gAB^{2}\right\vert ^{2}+\left\vert mB\right\vert ^{2}
\end{eqnarray}

and we see that $V=0$ is excluded because the last term is only zero if $%
B=0, $but then the first term is positive-definite. We conclude
that $V>0$ and that supersymmetry is broken. Provided that
$m^{2}>2\lambda g$ \ the potential$V$ has a minimum when $B=C=0,$
independently of the value of $A.$ For simplicity, we set $A=0$ at
the minimum.

As usual, the scalar masses are determined by evaluating

\begin{equation}
V_{AB}\equiv \frac{\partial ^{2}V}{\partial A\partial B},\text{ \ \ \ \ \ }%
etc.,
\end{equation}

at the minimum. The only non-zero elements are

\begin{equation}
\left\langle V_{BB^{\prime }}\right\rangle B^{2}+2\left\langle
V_{BB^{\ast }}\right\rangle BB^{\ast }+\left\langle V_{B^{\ast
}B^{\ast }}\right\rangle B^{\ast 2}=(m^{2}-2g\lambda
)B_{1}^{2}+(m^{2}+2g\lambda )B_{2}^{2},
\end{equation}

where $B=(B_{1}+iB_{2})/\sqrt{2},$ and so the real scalar fields $B_{1}$and $%
B_{2}$ have $(mass)^{2}=m^{2}\mp 2g\lambda ,$respectively.

The fermion masses are obtained by evaluating $\partial
^{2}W/\partial A\partial B,$ etc., at the minimum ( see (2.64)).
From (2.85) we find that the only non-zero term is

\begin{equation}
\frac{\partial ^{2}W}{\partial B\partial C}=m
\end{equation}

and so the fermion mass matrix takes the form

\begin{equation}
M_{F}=%
\begin{pmatrix}
0 & 0 & 0 \\
0 & 0 & m \\
0 & m & 0%
\end{pmatrix}%
\end{equation}

in the bases of Majorana spinors $\psi _{A},\psi _{B},\psi _{C}.$
The massless Goldstino state $\psi _{A}$ is evident, and the
off-diagonal structure signals that the two Majorana spinors $\psi
_{B},$ $\psi _{C}$ \ will combine to give a single Dirac fermion
of mass $m$. \ Despite the supersymmetry breaking, there is still
\ an equality between the sum of the (mass)$^{2}$ of the bosons
and that of the fermions. Explicitly, for the each degree of
freedom we have the masses given in table (2.1)

\begin{equation*}
\underset{\text{\textbf{Table (2.1): Masses of bosons and their
fermionic
partners}}}{%
\begin{tabular}{|c|c|c|c|c|}
\hline \multicolumn{3}{|c|}{$Bosons$} &
\multicolumn{2}{|c|}{$Fernions$} \\ \hline $A$ & $B$ & $C$ & $\psi
_{A}$ & $\psi _{B},$ $\ \psi _{C}$ \\ \hline
$0,0$ & $m^{2}\pm 2\lambda g$ & $m^{2},m^{2}$ & $0,0$ & $%
m^{2},m^{2},m^{2},m^{2}$ \\ \hline
\end{tabular}%
}
\end{equation*}

\bigskip

Only $B$ suffers supersymmetry breaking. The reason is that it is
the only
field that couples to the Goldstino; its coupling $gB\overline{\psi }%
_{B}\psi _{A}$ appears when (2.85) is inserted into (2.64). The
value of the potential at the minimum can be written

\begin{equation}
\left\langle V\right\rangle =\lambda ^{2}\equiv M_{S}^{4}
\end{equation}

where the mass splitting with in the supermultiplet of table (2.1)
are therefore

\begin{equation}
\vartriangle m^{2}\approx gM_{S}^{2},
\end{equation}

where $g$ is the coupling to the Goldstino.

This simple model illustrates several more general results. The
mass relation is a particular example of the \textquotedblleft
super-trace relation\textquotedblright ,

\begin{equation}
STr(M^{2})\equiv \sum_{J}(2J+1)(-1)^{2J}Tr(M_{J}^{2})=0,
\end{equation}

which holds whether supersymmetry is spontaneously broken or not. Here $%
M_{J} $is the mass matrix for the fields of spin-$J$, and the sum
is over all the physical particles of the theory. Relation (2.93)
holds in
lowest-order perturbation theory. We say that it is a \textquotedblleft $%
tree-level$\textquotedblright\ results because it neglects
corrections due to the diagrams containing loops. This super-trace
mass relation is important because it ensures that the scalars are
not subject to quadratically divergent renormalization.

We may readily verify that (2.93) holds for an arbitrary multiplet
structure. If there are several chiral multiplets $(\phi _{i},\psi
_{i}),$ then it is convenient to arrange the scalar fields and
their complex conjugates as a column vector so that the boson mass
terms have the matrix structure

\begin{equation}
\begin{pmatrix}
\phi ^{\ast } & \phi%
\end{pmatrix}%
\begin{pmatrix}
X & Y \\
Y^{+} & X%
\end{pmatrix}%
\begin{pmatrix}
\phi \\
\phi ^{\ast }%
\end{pmatrix}%
.
\end{equation}

The block diagonal parts of the boson $(mass)^{2}$ matrix,
$M_{B}^{2},$have elements

\begin{align}
X_{ij}& =\frac{\partial ^{2}V}{\partial \phi _{i}\partial \phi _{j}^{\ast }}%
=\sum_{k}\left( \frac{\partial ^{2}W}{\partial \phi _{i}\partial \phi _{k}}%
\right) \left( \frac{\partial ^{2}W}{\partial \phi _{j}^{\ast
}\partial \phi
_{k}^{\ast }}\right)  \notag \\
& =\sum_{k}\left( M_{F}\right) _{ik}\left( M_{F}^{\ast }\right)
_{kj}=(M_{f}M_{F}^{\ast })_{ij}
\end{align}

where $M_{F}$ is the fermion mass matrix and so it follows that

\begin{equation}
Tr(M_{B}^{2})=2Tr(M_{F}^{2})
\end{equation}

at tree level.

We can also show that the fermion mass matrix has a zero
eigenvalue and hence identify the Goldstone. At the minimum of the
potential,

\begin{equation}
0=\frac{\partial V}{\partial \phi _{i}}=\frac{\partial }{\partial \phi _{i}}%
\left( \sum_{j}\left\vert \frac{\partial W}{\partial \phi
_{i}}\right\vert ^{2}\right) =\sum_{j}\left( \frac{\partial
^{2}W}{\partial \phi _{i}\partial \phi _{j}}\right) \left(
\frac{\partial W}{\partial \phi _{j}}\right) ^{\ast
}=\sum_{j}\left( M_{F}\right) _{ij}\left\langle F_{j}\right\rangle
^{\ast }.
\end{equation}

Thus, the mass matrix $M_{F}$ annihilates the fermion state

\begin{equation}
\psi _{G}=\sum_{j}\left\langle F_{j}\right\rangle ^{\ast }\psi
_{j}
\end{equation}

which is thus identified as the massless Goldstino. In our
example, $\psi _{G}=\psi _{A}$ since $\left\langle
F_{B}\right\rangle =\left\langle F_{c}\right\rangle =0.$

However, the equality (2.96), which is so desirable to ensure the
boson-fermion loop cancellations, is not supported experimentally.
The difficulty is that in these simple models, the relation
applies to each supermultiplet separately. Hence, for the
electron, for example, we require

\begin{equation}
2m_{e}^{2}=m_{A}^{2}+m_{B}^{2},
\end{equation}

which implies that one of the two scalar electrons ($A,B$) must
have a mass less than or equal to that of the electron. Such a
particle would have been detected long time ago if it existed.

\section{D-type Breaking (The Fayet-Iliopoulos Model)}

As a simple example of supersymmetry breaking caused by the presence of a $%
U(1)$ factor in the gauge group, we take a supersymmetric version
of $QED$ with two chiral multiplets $(\phi _{+},\psi _{+})$ and
$(\phi _{-},\psi _{-}),$ where the subscripts give the sign of the
charge. The $U(1)$ gauge-invariant superpotential is

\begin{equation}
W=m\phi _{+}\phi _{-}
\end{equation}

and so the scalar potential (2.80) becomes

\begin{eqnarray}
V &=&m^{2}\left\vert \phi _{+}\right\vert ^{2}+m^{2}\left\vert
\phi _{-}\right\vert ^{2}+\frac{1}{2}\left[ e\left( \left\vert
\phi
_{+}\right\vert ^{2}-\left\vert \phi _{-}\right\vert ^{2}\right) +\eta %
\right] ^{2} \\
&=&\frac{1}{2}e^{2}\left( \left\vert \phi _{+}\right\vert
^{2}-\left\vert \phi _{-}\right\vert ^{2}\right) ^{2}+\left(
m^{2}+e\eta \right) \left\vert \phi _{+}\right\vert ^{2}+\left(
m^{2}-e\eta \right) \left\vert \phi _{-}\right\vert
^{2}+\frac{1}{2}\eta ^{2}  \notag
\end{eqnarray}

Various possible forms for $V$ \ are shown in Figure (2.1)

\begin{equation*}
The\text{ \ }Pic
\end{equation*}

Provided $m^{2}>2\eta $ (where $e\eta >0$), the minimum occurs at

\begin{equation}
\phi _{+}=\phi _{-}=0,
\end{equation}

so $U(1)$ gauge invariance is not spontaneously broken, but
supersymmetry is broken since $V\neq 0.$ The boson masses are
split, $\ m_{\pm }^{2}=m^{2}\pm e\eta ,$ whereas the fermion
masses are unaffected by the breakdown of supersymmetry. Like
(2.90) the (off-diagonal)\ form of the fermion mass matrix in the
$\psi _{+},$ $\ \psi _{-}$ Majorana basis implies that these two
states combine together to give a Dirac fermion of \ mass $m$. The
fermion-boson mass splitting signals that the breakdown of
supersymmetry but the $(mass)^{2}$ equality still holds, since

\begin{equation}
m_{+}^{2}+m_{-}^{2}=2m^{2}
\end{equation}

For $m^{2}>e\eta ,$the $U(1)$ symmetry is unbroken and the gauge multiplet $%
\left( A_{\mu },\chi \right) $ remains massless. The fermion $\chi
$ is the \textquotedblleft Goldstino\textquotedblright\ arising
from the spontaneous supersymmetry breaking.

The case $m^{2}<e\eta $ is more interesting. The minimum of the
potential now occurs at

\begin{equation}
\phi _{+}=0\text{ \ \ \ \ \ \ \ \ \ \ \ \ \ }\phi _{-}=\upsilon
\end{equation}

where $e^{2}\upsilon ^{2}=\left( e\eta -m^{2}\right) .$ Now, both the $U(1)$%
gauge symmetry and supersymmetry are spontaneously broken; see
Figure (2.1.c). We find that the complex field $\phi _{+}$ has
$\left( mass\right) ^{2}=2m^{2},$ while one component of $\phi
_{-}$ is \textquotedblleft eaten\textquotedblright\ \ by the usual
Higgs mechanism to give $\left( mass\right) ^{2}=2e^{2}\upsilon
^{2}$ \ to the vector gauge field $A_{\mu },$ and the remaining
component also acquires $\left( mass\right) ^{2}=2e^{2}\upsilon
^{2}$. A linear combination of the $\psi _{+}$ and$\ \chi $
Majorana fields forms the massless \textquotedblleft
Goldstino\textquotedblright , whereas the two remaining combinations of $%
\psi _{+}$,$\psi _{-},$ and $\chi $ both have $\left( mass\right)
^{2}=m^{2}+2e^{2}\upsilon ^{2}.$ Despite the symmetry breaking,
the super-trace mass relation (2.93) remains true, that is

\begin{equation}
2\left( 2m^{2}\right) +2e^{2}\upsilon ^{2}-\left( 2+2\right)
\left( m^{2}+2e^{2}\upsilon ^{2}\right) +3\left( 2e^{2}\upsilon
^{2}\right)
\end{equation}

It is straight forward to show that, as in sec. (2.9), the
super-trace relation $STr\left( M^{2}\right) =0$ holds in general,
with just one
exception, and that the Goldstino can be identified as the combination%
\begin{equation}
\psi _{G}=\left\langle F_{j}\right\rangle \psi _{j}-\frac{1}{\sqrt{2}}%
\left\langle D^{a}\right\rangle \chi _{a}.
\end{equation}

Since the super-trace relation leads to problems, as we found in
(2.99), it is desirable to explore the exception. In the presence
of a $U(1)$ factor of the gauge group, we find that (2.93) becomes

\begin{equation}
STr\left( M^{2}\right) =2\left\langle D\right\rangle Tr\text{ }Q,
\end{equation}

where $Q$ is the $U(1)$ charge matrix of the chiral multiplet and
$D$ is the auxiliary field. Perhaps this extra contribution will
permit the superpartners to be sufficiently massive to escape
detection. Unfortunately, the $U(1)$ of the standard model is not
suitable, as the weak hypercharge $Y$ must satisfy $Tr$ $Y=0.$
This extra contribution is also absent in $GUTs,$
which have no $U\left( 1\right) $ factor. To have an additional $U(1)$ with $%
Tr$ $Q\neq 0$ \ would create new problems with triangle anomalies,
which can only be avoided by introducing new chiral multiplets.
Thus far, no satisfactory model of $D-type$ breaking has been
found.

\section{The Supersymmetric Standard Model}

The standard model $\left( SM\right) $ has 28 bosonic degrees of
freedom (12 massless gauge bosons and 2 complex scalars) together
with 90 fermionic degrees of freedom (3 families each with 15
two-component Weyl fermions). To make the model supersymmetric we
must clearly introduce additional particles. In fact, since none
of the observed particles pare off,

we have to double the number. In section 2.3 we saw that the gauge
bosons are partnered by spin-$\frac{1}{2}$ gauginos, and these
cannot be identified with any of the quarks and leptons. So the
latter have to be partnered by new spin 0 squarks and sleptons.

We also have to complete the Higgs supermultiplet. now the $Y=-1$
Higgs doublet has the same quantum numbers as the $\left( \nu
,e^{-}\right) _{L}$ doublet, so one might try to identify the
Higgs with a spin 0 slepton. Unfortunately even this is not
possible, because any attempt to partner a lepton with a Higgs, by
giving the latter a non-zero lepton number $L,$ leads to
$L$-violating processes and large $\bigtriangleup L=2$ Majorana
mass terms. Even worse, in the standard Higgs $\left( \phi \right)
$ generates masses for the down-type quarks and the charged
leptons, while its charge conjugate $\left( \phi _{c}=i\tau
_{2}\phi ^{\ast }\right) $ gives masses to the up- type quarks.
Now the superpotential $W$ is a function only of $\phi $ and not
$\phi ^{\ast },$ and so in supersymmetry we need to introduce a
second unrelated Higgs doublet. There is an alternative way to see
this. Under charge conjugation, the helicity of the
spin-$\frac{1}{2}$ partner of the Higgs (\textquotedblleft the
Higgsino\textquotedblright ) is reversed, and so it proves
impossible to use a single Higgs to give masses
to both up-type and down-type quarks. the second $\left( \func{complex}%
\right) $ doublet is also needed to cancel the anomalies that
would arise if there were only on Higgsino. As in the standard
model, three of the Higgs fields are absorbed to make the $W^{\pm
}$ and $Z$ bosons massive, and we are therefore left with two
charged and three neutral massive Higgs particles.

The particle content of the supersymmetric standard model is shown in table $%
\left( 2.2\right) .$

\begin{eqnarray*}
&&%
\begin{tabular}{|c|c|c|c|}
\hline \multicolumn{2}{|c|}{$Chiral$\textbf{\ \ }$Multiplets$} &
\multicolumn{2}{|c|}{$Gauge$\textbf{\ }$\ Multiplets$} \\ \hline
\textbf{Spin-}$\frac{1}{2}$ & \textbf{Spin 0} & \textbf{Spin 1} & \textbf{%
Spin-}$\frac{1}{2}$ \\ \hline
Quark $q_{L},q_{R}$ & Squark $\widetilde{q_{L}},\widetilde{q_{R}}$ & photon $%
\gamma $ & photino $\widetilde{\gamma }$ \\ \hline Lepton
$l_{L},l_{R}$ & Slepton $\widetilde{l_{L}},\widetilde{l_{R}}$ &
$W,Z$ \ bosons & Wino $\widetilde{W},$ Zino $\widetilde{Z}$ \\
\hline Higgsino $\widetilde{\phi }$ , $\widetilde{\phi ^{\prime
}}$ & Higgs $\phi ,\phi ^{\prime }$ & Gluon $g$ & Gluino
$\widetilde{g}$ \\ \hline
\end{tabular}
\\
&&\text{Table (2.2) Particle Multiplets in the Supersymmetric
Standard Model}
\end{eqnarray*}

There is no doubt that this table is a setback for supersymmetry.
To be economical, supersymmetry ought to unite the known fermionic
\textquotedblleft matter\textquotedblright\ $quarks$ and $\
leptons$ with the vector \textquotedblleft
forces\textquotedblright\ $\gamma ,g,W,Z,$ but we have been
compelled to keep them separate and to introduce a new
superpartner for each particle. A great deal of effort has gone
into the search for these superpartners but so far non has been
found.

\chapter{Minimal Supersymmetric Standard Model}

\section{Introduction}

The minimal supersymmetric extension of the standard model (MSSM)
[11] was generated by taking the standard model (SM) and adding
the corresponding
supersymmetric partners. In addition, the MSSM contains two hypercharge $%
Y=\pm 1$ Higgs doublets, which is the minimal structure for the
Higgs sector of an anomaly-free supersymmetric extension of the
standard model. The supersymmetric structure of the theory also
requires (at least) two Higgs doublets to generate mass for both
up-type and down-type quarks (and charged leptons). All
renormalizable supersymmetric interactions \ consistent with
(global) $B-L$ conservation ($B=baryon$ $number$ and $L=lepton$
$number$) are included. Finally, the most general
soft-supersymmetric-breaking terms are added.

If supersymmetry is relevant for explaining the scale of
electroweak interactions, then the mass parameters exist due to
the absence of supersymmetric-particle at current accelerators.
Additional constraints arise from limits on the contributions of
virtual supersymmetric particle exchange to a variety of SM
processes.

As a consequence of $B-L$ \ invariance, the MSSM processes a discreet $%
R-parity$ invariance, where $R=(-1)^{3(B-L)-2S}$ for a particle of
spin $S.$
This formula implies that all the ordinary SM particles have $even-R$ $%
parity,$ whereas the corresponding supersymmetric partners have odd $%
R-parity.$ The conservation of $R-parity$ in scattering and decay
processes has a crucial impact on supersymmetric phenomenology.
For example starting from an initial state involving ordinary
($R-even$) particles, it follows that supersymmetric particles
must be produced in pairs. In general, these particles are highly
unstable and decay quickly into lighter states. However R-parity
invariance also implies that the lightest supersymmetric particle
(LSP) is absolutely stable, and must eventually be produced at the
end of a heavy unstable supersymmetric particle.

In order to be consistent with cosmological constraints, the LSP
is almost certainly electrically and color neutral. Consequently,
the LSP is weakly-interacting in ordinary matter, \textsl{i.e. }it
behaves like a stable heavy neutrino and will escape detectors
without being directly observed. Thus, the canonical signature for
($R-parity$ conserving) supersymmetric theories is a missing
(transverse) energy, due to the escape of the LSP.

Some model builders attempt to relax the assumption of $R-parity$
conservation. Models of this type must break $B-L$ conservation
and are therefore constrained by experiment. Nevertheless, it is
still important to allow the possibility of $R-parity$ violation
processes in the search for supersymmetry. In such models the LSP
is unstable and supersymmetric
particles can be singly produced and destroyed in association with $B$ and $%
L $ violation. These features lead to a phenomenology of broken
$R-parity$ models that is very different from that of the MSSM

In the MSSM, supersymmetry breaking is accompanied by including
the soft-supersymmetry breaking terms. These terms parameterize
our ignorance of the fundamental mechanism of supersymmetry
breaking. If this breaking occurs spontaneously, then (in the
absence) of supergravity a massless goldstone fermion is called
the $\mathit{goldstino}$ ($\widetilde{G}$) must exist. The
goldstino would be the LSP and could play an important role in
supersymmetric phenomenology. In models that incorporates
supergravity (SUGRA), this picture changes. If supergravity is
spontaneously broken, the
goldstino is absorbed (eaten) by the \textsl{gravitino}, the spin-$\frac{3}{2%
}$ partner of the graviton. By this super\_Higgs mechanism, the
gravitino acquires a mass. In many models, the gravitino mass is
of order as the order of the electroweak-breaking scale, while its
coupling are gravitational in strength. Such a gravitino would
play no role in supersymmetric phenomenology at colliders. The
parameters of the MSSM are conveniently described by considering
separately the supersymmetric conserving sector and the
supersymmetry breaking sector. Among the parameters of the
supersymmetry conserving sector are:

1- gauge couplings: $g^{\prime },$ $g,$ and $g_{s},$ corresponding
to U(1), SU(2), and SU(3) subgroups of the SM respectively;

2- Higgs-Yukawa couplings: $\lambda _{e},$ $\lambda _{u},$ and
$\lambda _{d}$ (which are 3$\times $3 matrices in flavor space);
and

3- a supersymmetry-conserving Higgs mass parameter $\mu .$

The supersymmetric-breaking sector contains the following set of
parameters:

i- gauging Majorana masses $M_{1},$ $M_{2}$ and $M_{3}$ associated
with the U(1), SU(2), and SU(3) subgroups of the SM;

ii- scalar mass matrices for the squarks and sleptons.;

iii- Higgs-squark trilinear interaction terms (the so-called
$A-parameters$) and corresponding terms involving the sleptons;
and

iv- three scalar Higgs mass parameters- two-diagonal and one
off-diagonal mass terms for two Higgs doublets. These three mass
parameters can be
re-expressed in terms of the two Higgs vacuum expectation values (VEV), $%
\upsilon _{1}$ and $\upsilon _{2},$and one physical Higgs mass (usually, $%
m_{H_{3}^{0}}$). Here, $\upsilon _{1}$ ($\upsilon _{2}$) is the
vacuum
expectation values of the Higgs field which couples exclusively to $%
down-type $ ($up-type$) quarks and leptons. The value $\upsilon
_{1}^{2}+\upsilon _{2}^{2}$ is fixed by the $W$ mass (or
equivalently by the Fermi constant $G_{F}$),

\begin{equation}
\upsilon _{1}^{2}+\upsilon _{2}^{2}\approx (246\text{ }GeV)^{2}
\end{equation}

while the ratio $\upsilon _{2}/\upsilon _{1}$ is a free parameter
of the model in terms of the angle $\beta ;$

\begin{equation}
\tan \beta =\upsilon _{2}/\upsilon _{1}
\end{equation}

The supersymmetric constraints imply that the MSSM Higgs sector is
automatically $CP-conserving$ (at tree level). Thus, $\tan \beta $
is a real parameter (conventionally taken to be positive), and the
physical neutral Higgs scalars are $CP-eigenstates.$ Nevertheless,
the MSSM does contain a number of possible new sources of
$CP-violation.$ For example, gaugino-mass parameters, the
$A-parameters,$ and $\mu $ may be complex. Some combination of
these complex phases must be less than an order of
$10^{-2}-10^{-13}$ (for a supersymmetric-breaking scale of $100$
$GeV$) to avoid generating electric dipole moments for the
neutron, electron and atoms in conflict with observed data.
However, these complex phases have little impact on the direct
searches for supersymmetric particles, and are usually ignored in
experimental analysis.

\section{Extended Higgs Sectors}

The Higgs mechanism has solved the problem of having massive gauge
bosons in an invariant local gauge theory without spoiling
renormalizability and unitarity. This is achieved by means of a
spontaneous breaking of the gauge symmetry in which the ground
state (vacuum) loses part of the symmetry whereas the Lagrangian
itself remains fully symmetric.

In the SU(2)$\times $U(1) standard Glashow-Weinberg-Salam model ($GWS$ or $%
SM $) the spontaneous symmetry breaking is induced by the presence
of a doublet (under SU(2)) of complex scalar fields [12]

\begin{equation}
\phi =\binom{\phi ^{+}}{\phi ^{0}}.
\end{equation}

The new fields have Yukawa type interactions with matter fermion
fields and
also have self-interactions of the form%
\begin{equation}
V(\phi )\equiv -\mu ^{2}|\phi |^{2}+\lambda |\phi |^{2},
\end{equation}

where $\mu ^{2}$ and $\lambda $ are positive constants. After the
Higgs mechanism, the theory contains - apart from the fields- 3
massive gauge
bosons $(W^{+},$ $W^{-},$ $Z),$ 1 massless photon and 1 physical scalar $%
(H), $ the \textquotedblleft\ Higgs boson\textquotedblright . The
other three real scalars of the doublet (the \textquotedblleft\
Goldstino bosons\textquotedblright ) have become the longitudinal
components of the three massive gauge bosons.

Although the minimal Higgs sector of the SM is sufficient to
explain the generation of the fermion and gauge boson masses, more
complicated structures in the scalar sector can not be excluded
and are even unavoidable in many unifying extensions of the SM.
These extended Higgs sectors have potentially richer phenomenology
but are also subjected to phenomenological constraints, for
example, the electroweak $\rho -parameter,$ and the presence of
tree level couplings of the type $W^{-}Z^{0}H^{-}.$

\subsection{The $\protect\rho -$Parameter constraint}

The most important phenomenological constraint on the Higgs sector
is the value of the electroweak $\rho -$parameter which,
experimentally, is very close to 1

\begin{equation}
\rho \equiv \frac{m_{W}^{2}}{m_{Z}^{2}\cos ^{2}\theta _{W}}\approx
1
\end{equation}

With an arbitrary Higgs sector consisting of several scalar
multiplets $\phi
_{i}$ of weak isospin $T_{(i)}$ and weak hypercharge $Y_{(i)},$ the $\rho -$%
parameter is given by

\begin{equation}
\rho =\frac{\dsum\limits_{i}\left[ T_{(i)}\left( T_{(i)}+1\right)
-\left( Y_{(i)}/2\right) ^{2}\right] \upsilon
_{i}^{2}c_{i}}{2\dsum\limits_{i}\left( \left( Y_{(i)}/2\right)
^{2}\right) \upsilon _{i}^{2}c_{i}}
\end{equation}

where $\upsilon _{i}$ is the VEV of the multiplet $\phi _{i}$ and $c_{i}=1(%
\frac{1}{2})$ when $Y_{(i)}\neq 0$ $(=0).$

It is easy to check that below $T=10$ only the representations $(T,Y)=(0,0),(%
\frac{1}{2},\pm 1)$ and $(3,\pm 4)$ lead \textquotedblleft\
naturally\textquotedblright\ $(i.e.$ independently of the values of $%
\upsilon _{i})$ to $\rho =1.$Even if we allow $\rho $ to deviate
from 1 by 1\%, no new representations appear. Leaving aside the
case of $(T,Y)=(3,\pm 4)$ involving scalars of electric charge
$Q=5,$only doublets and singlets are acceptable.

Of course other representations are allowed if we only require
$\rho \approx 1$ for \textsl{some} values of the $\upsilon _{i}$.
The simplest cases are the models with a doublets and a (real or
complex) triplet but with $\rho $ differing slightly from 1 unless
$\upsilon _{i}=0$. The simplest \textquotedblleft\
unnatural\textquotedblright\ case with $\rho =1$ is a model with
one doublet and two triplets (one real and one complex) with equal
VEV's.

\subsection{The $W^{-}Z^{0}H^{-}$ Couplings}

a general feature of the extended Higgs sectors is the presence of $\mathit{%
physicsal}$ charged scalar fields ($H^{\pm }$). This fact implies
a potentially rich phenomenology. In particular one expects a tree
level
couplings of the type $W^{-}Z^{0}H^{-}$ in analogy to $W^{-}W^{+}H$ and $%
Z^{0}Z^{0}H$ \ in the SM. However, it turns out that this coupling
is absent (at the tree level) in the simplest\ \ \ \ \ \ \
\textquotedblleft\ natural\textquotedblright\ extensions of the
Higgs sector ($i.e.$ with doublets and singlets) and it is small
(proportional to $\sqrt{\left\vert 1-\rho \right\vert }$) in the
simplest \textquotedblleft unnatural \textquotedblright\
extensions. It is also easy to prove (using the electromagnetic
gauge invariance) that the $W^{-}\gamma H^{+}$ couplings vanish at
the tree level in all models.

\section{The Two Higgs Doublet Model}

A Higgs sector consisting of two scalar doublets

\begin{equation}
\phi _{1}=\binom{\phi _{1}^{+}}{\phi _{1}^{0}},\text{ \ \ \ \ \ \
\ \ \ and \ \ \ \ \ \ \ \ \ \ \ }\phi _{2}=\binom{\phi
_{2}^{+}}{\phi _{2}^{0}}
\end{equation}

is the simplest \textquotedblleft\ natural\textquotedblright\
$extension$ of the SM. As we have seen, in such case the
$W^{-}Z^{0}H^{+}$ coupling is automatically absent at the tree
level. The most general $CP-$conserving potential involving two
doublets is [13]

\begin{eqnarray}
V\left( \phi _{1},\phi _{2}\right) &=&\lambda _{1}(\phi
_{1}^{+}\phi _{1}-\upsilon _{1}^{2})^{2}+\lambda _{2}(\phi
_{2}^{+}\phi _{2}-\upsilon
_{2}^{2})^{2}  \notag \\
&&+\lambda _{3}\left[ (\phi _{1}^{+}\phi _{1}-\upsilon
_{1}^{2})^{2}+(\phi
_{2}^{+}\phi _{2}-\upsilon _{2}^{2})^{2}\right] ^{2}  \notag \\
&&+\lambda _{4}\left[ (\phi _{1}^{+}\phi _{1})(\phi _{2}^{+}\phi
_{2})-(\phi
_{1}^{+}\phi _{2})(\phi _{2}^{+}\phi _{1})\right] \\
&&+\lambda _{5}\left[ \func{Re}(\phi _{1}^{+}\phi _{2})-\upsilon
_{1}\upsilon _{2}\right] ^{2}+\lambda _{6}\left[ \func{Im}(\phi
_{1}^{+}\phi _{2})\right] ^{2},  \notag
\end{eqnarray}

where $\lambda _{i}$ are 6 arbitrary real parameters. If $\lambda _{i}>0,$%
the minimum of the potential corresponds to

\begin{equation}
\left\langle \phi _{i}\right\rangle =\binom{0}{\upsilon
_{1}},\text{ \ \ \ \
\ \ and \ \ \ \ \ \ \ }\left\langle \phi _{2}\right\rangle =\binom{0}{%
\upsilon _{2}}.
\end{equation}

In general the neutral components of the doublets can have flavor
changing couplings to the fermions. These flavor changing neutral
currents (FCNC)interactions can be suppressed ( as required
phenomenologically) either by giving large masses to these scalars
or by arranging their Yukawa couplings to the fermions.
\textit{Glashow} and \textit{Weinberg }proved a theorem stating
the sufficient condition for avoiding these FCNC effects;
\textquotedblleft\ (FCNC) interactions induced by neutral Higgs
scalars are absent if all fermions of a given charge receive their
masses from a single doublet\textquotedblright . This theorem is
trivially satisfied in the SM since there is only one doublet
available. It is also satisfied in the \textquotedblleft Minimal
supersymmetric model\textquotedblright .

Of the 8 available degrees of freedom (4 complex scalars with two
real components), 3 are the Goldstone bosons to become the
longitudinal components of the $W^{\pm }$ and $Z^{0}$ bosons that
become massive by the \textquotedblleft Higgs
mechanism\textquotedblright . The remaining five scalars are
\textit{physical} states (two are charged and three neutral). They
are

\begin{subequations}
\begin{eqnarray}
H^{\pm } &=&-\sin \beta \phi _{1}^{\pm }+\cos \beta \phi _{2}^{\pm }, \\
H_{1}^{0} &=&\sqrt{2}\left[ \left( \func{Re}\phi _{1}^{0}-\upsilon
_{1}\right) \cos \alpha +\left( \func{Re}\phi _{2}^{0}-\upsilon
_{2}\right)
-\sin \alpha \right] , \\
H_{2}^{0} &=&\sqrt{2}\left[ -\left( \func{Re}\phi
_{1}^{0}-\upsilon _{1}\right) \sin \alpha +\left( \func{Re}\phi
_{2}^{\pm }-\upsilon
_{2}\right) -\cos \alpha \right] , \\
H_{3}^{0} &=&\sqrt{2}\left[ \sin \beta \func{Im}\phi _{1}^{0}+\cos
\beta \func{Im}\phi _{2}^{\pm }\right] ,
\end{eqnarray}

with masses $m_{H^{\pm }},$ $m_{H_{1}^{0}},$ $m_{H_{2}^{0}},$ and $%
m_{H_{3}^{0}}$, respectively. The angle $\alpha $, and the masses
are functions of the parameters $\lambda _{i}$ and the VEV's
$\upsilon _{1,2}$

\end{subequations}
\begin{subequations}
\begin{eqnarray}
m_{H^{\pm }}^{2} &=&\lambda _{4}(\upsilon _{1}^{2}+\upsilon _{2}^{2}), \\
m_{H_{3}^{0}} &=&\lambda _{6}(\upsilon _{1}^{2}+\upsilon _{2}^{2}), \\
m_{H_{1}^{0},H_{2}^{0}}^{2} &=&\frac{1}{2}\left[ A+C\pm D\right] , \\
\sin 2\alpha &=&\frac{2B}{D}, \\
\cos 2\alpha &=&\frac{A-C}{D},
\end{eqnarray}

where

\end{subequations}
\begin{subequations}
\begin{eqnarray}
A &=&4\upsilon _{1}^{2}(\lambda _{2}+\lambda _{3})+\upsilon
_{2}^{2}\lambda
_{5}, \\
B &=&(4\lambda _{3}+\lambda _{5})\upsilon _{1}\upsilon _{2}, \\
C &=&4\upsilon _{2}^{2}(\lambda _{2}+\lambda _{3})+\upsilon
_{1}^{2}\lambda
_{5}, \\
D &=&\sqrt{(A-C)^{2}+4B^{2}}.
\end{eqnarray}

This model is completely specified by six parameters: $m_{H^{\pm }},$ $%
m_{H_{1}^{0}},$ $m_{H_{2}^{0}},$ $m_{H_{3}^{0}},$ $\alpha $, and
$\tan \beta \left( \equiv \upsilon _{2}/\upsilon _{1}\right) .$

In the absence of fermions, the Lagrangian (involving only gauge
bosons and scalars) is $C-$ and $P-$conserving and the gauge
bosons have the following quantum numbers:

\end{subequations}
\begin{equation}
J^{PC}=1^{-\text{ }-}(\gamma ),\text{ }1^{-\text{ }-}(Z),\text{ and }%
J^{P}=1^{-}(W).
\end{equation}

Similarly, the physical scalars are fixed to be

\begin{equation}
J^{PC}=0^{+\text{ }+}(H_{1}^{0}),\text{ }0^{+\text{ }+}(H_{2}^{0}),\text{ }%
0^{+\text{ }+}(H_{3}^{0}),\text{ and }J^{P}=0^{+}(H^{\pm }).
\end{equation}

As a consequence, the couplings $ZH_{1}^{0}H_{1}^{0}$ and $%
ZH_{2}^{0}H_{2}^{0}$ are zero since they would violate Bose
symmetry. The coupling $ZH_{1}^{0}H_{2}^{0}$ also vanishes due to
$CP-$conservation and
couplings $ZZH_{3}^{0}$ and $W^{-}W^{+}H_{3}^{0}$ are forbidden by $C-$%
conservation. These last tree results hold to all orders of
perturbation theory before fermions are introduced.

Other couplings are absent only at the tree level:

$\gamma H_{i}^{0}H_{j}^{0},$ $\gamma \gamma H_{i}^{0},$ $ggH_{i}^{0},$ $%
W^{\pm }\gamma H^{\mp },$ and $W^{\pm }ZH^{\mp },$ but they can be
generated in higher orders of perturbation theory and can lead to
interesting rare decays. All other couplings are in principle
allowed. In particular, the couplings $W^{-}W^{+}H_{1,2}^{0}$ exit
and satisfy the sum rule

\begin{equation}
g_{VVH_{1}^{0}}^{2}+g_{VVH_{2}^{0}}^{2}=g_{VVH}^{2}(SM),\text{ \ \
}V=(W,Z),
\end{equation}

$i.e.$ they are somewhat suppressed compared to the analogous SM
couplings.

When fermions are introduced, since the couplings to fermions are
not $C-$ and $P-$conserving (although $CP$ still approximately
conserved), scalar and
gauge bosons are regarded by fermions as mixtures of $J^{PC}$and $%
J^{(-P)(-C)}$ states. In particular, since a fermion-antifermion
pair with zero total angular momentum always has $C=+,$ in the
$H_{i}^{0}f\overline{f}$ couplings, the Higgs fields $H_{1}^{0},$
$H_{2}^{0},$ and $H_{3}^{0}$ acts respectively as $0^{+\text{
}+},$ $0^{+\text{ }+},$ and $0^{-\text{ }+}$ states.

\section{The Higgs Sector $of$ $the$ MSSM}

One peculiar fact of all supersymmetric gauge theories is that at
least two doublets are required. The simplest supersymmetric
extension of the SM is the MSSM where the Higgs sector consists of
just two doublets and it is a particular example of the
two-doublet models. In this case, SUSY imposes constrains and the
number of the independent parameters is reduced from six to two:
$m_{H_{3}^{0}}$ and $\tan \beta .$ The Higgs sector is then
completely specified by the values of these two parameters.

The Higgs potential of the MSSM can be written as [14], [15]:

\begin{eqnarray}
V\left( \phi _{1},\phi _{2}\right) &=&m_{1}^{2}\phi _{1}^{+}\phi
_{1}+m_{2}^{2}\phi _{2}^{+}\phi _{2}-m_{1,2}^{2}(\phi _{1}^{+}\phi
_{2}+\phi
_{2}^{+}\phi _{1})  \notag \\
&&+\frac{1}{8}(g^{\prime 2}+g^{2})\left[ \left( \phi _{1}^{+}\phi
_{1}\right) ^{2}+(\phi _{2}^{+}\phi _{2})^{2}\right] \\
&&+\frac{1}{4}(g^{\prime 2}-g^{2})(\phi _{1}^{+}\phi _{1})\left(
\phi _{2}^{+}\phi _{2}\right) -\frac{1}{2}g^{2}(\phi _{1}^{+}\phi
_{2})(\phi _{2}^{+}\phi _{1}),  \notag
\end{eqnarray}

where $g^{\prime }$ and $g$ are the U(1) and SU(2) gauge
couplings, respectively. Let $V$ be broken spontaneously, then by
comparing eq.(3.8) with eq.(3.16) we obtain the following results

\begin{subequations}
\begin{eqnarray}
\lambda _{2} &=&\lambda _{1} \\
\lambda _{3} &=&\frac{1}{8}(g^{2}+g^{\prime 2})-\lambda _{1} \\
\lambda _{4} &=&2\lambda _{1}-\frac{1}{2}g^{\prime 2} \\
\lambda _{5} &=&\lambda _{6}=2\lambda _{1}-\frac{1}{2}(g^{2}+g^{\prime 2}) \\
\lambda _{7} &=&-\frac{1}{8}(\upsilon _{1}^{2}-\upsilon
_{2}^{2})^{2}(g^{2}+g^{\prime 2}) \\
m_{1}^{2} &=&2\lambda _{1}\upsilon
_{2}^{2}-\frac{1}{4}(g^{2}+g^{\prime
2})(\upsilon _{1}^{2}+\upsilon _{2}^{2}) \\
m_{2}^{2} &=&2\lambda _{1}\upsilon
_{1}^{2}-\frac{1}{4}(g^{2}+g^{\prime
2})(\upsilon _{1}^{2}+\upsilon _{2}^{2}) \\
m_{1,2}^{2} &=&\frac{1}{2}\upsilon _{1}\upsilon _{2}(4\lambda
_{1}-g^{2}-g^{\prime 2})
\end{eqnarray}

Using eq. (3.17h) to eliminate $\lambda _{1}$ in eqs. (3.17f) and
(3.17g) we get

\end{subequations}
\begin{equation}
m_{1}^{2}=m_{1,2}^{2}\frac{\upsilon _{2}}{\upsilon _{1}}-\frac{1}{4}%
(g^{2}+g^{\prime 2})(\upsilon _{1}^{2}+\upsilon _{2}^{2})
\end{equation}

\begin{equation}
m_{2}^{2}=m_{1,2}^{2}\frac{\upsilon _{1}}{\upsilon _{2}}-\frac{1}{4}%
(g^{2}+g^{\prime 2})(\upsilon _{2}^{2}+\upsilon _{1}^{2})
\end{equation}

Hence

\begin{equation}
m_{1}^{2}+m_{2}^{2}=m_{1,2}^{2}(\tan \beta +\cot \beta )
\end{equation}

\begin{equation}
\upsilon _{1}^{2}+\upsilon _{2}^{2}=\frac{-4m_{1}^{2}\cos
^{2}\beta +4m_{2}^{2}\sin ^{2}\beta }{(g^{2}+g^{\prime 2})(\cos
^{2}\beta -\sin ^{2}\beta )}
\end{equation}

Using eqs. $\left( 3.12\right) $, and (3.17)-(3.21), we
immediately get the spectrum of physical Higgs particles. The
results are

\begin{eqnarray}
m_{H_{1}^{0}}^{2} &=&m_{1}^{2}+m_{2}^{2},  \notag \\
m_{H_{2}^{0}}^{2} &=&m_{W}^{2}+m_{H_{3}^{0}}^{2} \\
m_{_{H_{1}^{0},H_{2}^{0}}}^{2} &=&\frac{1}{2}\left[ \sqrt{%
m_{Z}^{2}+m_{_{H_{3}^{0}}}^{2}\pm
(m_{Z}^{2}+m_{_{H_{3}^{0}}}^{2})^{2}-4m_{Z}^{2}m_{_{H_{3}^{0}}}^{2}\cos
^{2}2\beta }\right] ,  \notag
\end{eqnarray}

where

\begin{eqnarray}
m_{W}^{2} &=&\frac{1}{2}g^{2}(\upsilon _{1}^{2}+\upsilon
_{2}^{2}),  \notag
\\
\text{ \ }m_{Z}^{2} &=&\frac{1}{2}(g^{2}+g^{\prime 2})(\upsilon
_{1}^{2}+\upsilon _{2}^{2})
\end{eqnarray}

are the squares of the masses of both $W$ and $Z$ bosons.

In this case the relations between $\alpha $ and $\beta $ are

\begin{eqnarray}
\sin 2\alpha &=&-\sin 2\beta \left( \frac{m_{H_{1}^{0}}^{2}+m_{H_{2}^{0}}^{2}%
}{m_{H_{1}^{0}}^{2}-m_{H_{2}^{0}}^{2}}\right) ,  \notag \\
\cos 2\alpha &=&-\cos 2\beta \left( \frac{m_{H_{3}^{0}}^{2}-m_{Z}^{2}}{%
m_{H_{1}^{0}}^{2}-m_{H_{2}^{0}}^{2}}\right) .
\end{eqnarray}

From these expressions, the following inequalities follow,

\begin{eqnarray}
m_{W} &<&m_{H^{\pm }},  \notag \\
m_{H_{2}^{0}} &<&m_{H_{3}^{0}}, \\
m_{H_{2}^{0}} &<&m_{Z}\,<m_{H_{1}^{0}}  \notag
\end{eqnarray}

The fact that one of the neutral scalars, $H_{2}^{0},$ is lighter than the $%
Z $ boson, is an interesting results of the MSSM. The heavy scalar $%
H_{1}^{0},$ on the other hand, has a $W^{-}W^{+}H_{1}^{0}$
coupling which is suppresses with respect to corresponding one in
the SM by a factor

\begin{equation}
\left[ \frac{m_{H_{2}^{0}}^{2}\left( m_{Z}^{2}-m_{H_{2}^{0}}^{2}\right) }{%
\left( m_{H_{1}^{0}}^{2}+m_{H_{2}^{0}}^{2}\right) \left(
m_{H_{1}^{0}}^{2}+m_{H_{2}^{0}}^{2}-m_{Z}^{2}\right) }\right]
^{2}.
\end{equation}

This factor decreases as $1/m_{H_{1}^{0}}^{2}$ where
$m_{H_{1}^{0}}^{2}$ increases, and it is lower than $\sim 0.15$
for $m_{H_{1}^{0}}^{2}>2m_{W}$. The heavy scalar then behaves
differently from the SM Higgs since the latter interacts more and
more strongly with the $W$ boson when $m_{H}$ increases. This
different behavior of SUSY theories is consistent with the fact
that the ultraviolet cut-off of the latter is far above $\sim
1TeV.$

These two results, namely, the existence of a light Higgs boson
and the decoupling of the heavy one from the gauge bosons are
general results which survive in the more general supersymmetric
models, including the \textquotedblleft superstring
inspired\textquotedblright\ ones.

Recent results from LEP experiments have restricted the allowed
region in the $m_{H_{1}^{0}},$ $\tan \beta $ plane and on
$m_{H^{\pm }}.$

To summarize, there are five physical Higgs particles in the MSSM,
a charged
Higgs pair ($H^{\pm }$), two $CP-$even neutral Higgs bosons (denoted by $%
H_{1}^{0}$ and $H_{2}^{0}$ where $m_{H_{1}^{0}}>m_{H_{2}^{0}}$) and one $CP-$%
odd neutral Higgs boson\footnote{%
In recent reviews of particle properties, the symbol $A^{0}$ replaces $%
H_{3}^{0}$, to denote the $CP-$odd neutral Higgs boson.
\par
{}} $H_{3}^{0}$. The properties of the Higgs sector of the MSSM
are determined by the Higgs potential which is made of quadratic
terms and quartic interaction terms. The strength of the
interaction terms are directly related to the gauge couplings by
supersymmetry (and are not affected at tree-level by supersymmetry
breaking). As a result, $\tan \beta $ and one Higgs mass
($m_{H_{3}^{0}}$) determine: The Higgs spectrum, an angle $\alpha
$ (which indicates the amount of mixing of the original $Y=\pm 1$
Higgs doublet states in the physical $CP-$even scalars), and the
Higgs boson couplings.

\section{The Supersymmetric Particle Sector $of$ $the$ MSSM}

The supersymmetric partner of the gauge and Higgs bosons are
fermions, whose names are obtained by appending \textquotedblleft
$ino$\textquotedblright\ at the end of the SM particle name. The
\textit{gluino }is the color octet
Majorana fermion partner of the gluon with mass $m_{\widetilde{g}%
}=\left\vert M_{3}\right\vert $. The supersymmetric partner of the
electroweak gauge and Higgs bosons (the \textit{gauginos} and \textit{%
Higgsinos}) can mix. As a result, the physical mass eigenstates
are
model-dependent linear combinations of these states, called \emph{charginos }%
and\emph{\ neutralinos}, which are obtained by diagonalizing the
corresponding mass matrix [16].

The chargino-mass matrix depends on $M_{2}$, $\mu ,$ $\tan \beta ,$ and $%
m_{W}$. The chargino mass eigenstates are denoted by $\widetilde{\chi }%
_{1}^{+},$ $\widetilde{\chi }_{2}^{+}$ according to the convention that $%
\widetilde{\chi }_{1}^{+}\leq \widetilde{\chi }_{2}^{+}$.

The neutralino mass matrix depends on $M_{1},$ $M_{2}$, $\mu $, $\tan \beta $%
, $m_{Z}$, and the weak mixing angle $\theta _{W}$. The
corresponding neutralino eigenstates are denoted by
$\widetilde{\chi }_{i}^{0}$
(\thinspace $i=1,...,4$), according to the convention that $\widetilde{\chi }%
_{1}^{0}\leq \widetilde{\chi }_{2}^{0}\,\leq \widetilde{\chi
}_{3}^{0}\leq \widetilde{\chi }_{4}^{0}$.

If a chargino or a neutralino eigenstate approximates a particular
gaugino or Higgsino state, it may be convenient to use the
corresponding
nomenclature. For example, if $M_{1}$ and $M_{2}$ are small compared to $%
m_{Z}$ (and $\mu $), then the lightest neutralino $\widetilde{\chi
}_{1}^{0}$ will be nearly a pure photino, $\widetilde{\gamma }$ (
the supersymmetric partner of the photon).

It is common to reduce the supersymmetric parameter freedom by
requiring that all three gaugino-mass parameters are equal at some
grand unification scale. Then, at the electroweak scale the
gaugino-mass parameter can be expressed in terms of on of them
which we choose to be $M_{2}\equiv M$. The other two gaugino-mass
parameters are given by

\begin{eqnarray}
M_{1} &=&\frac{3}{5}M^{\prime }=\left( \frac{g^{\prime
2}}{g^{2}}\right) M,
\notag \\
M_{3} &\equiv &m_{\widetilde{g}}=\left(
\frac{g_{s}^{2}}{g^{2}}\right) M,
\end{eqnarray}

where $M^{\prime }$, $M$ and $m_{\widetilde{g}}$ are the bino
masses respectively. Having made this assumption, the chargino and
neutralino masses and mixing angles depend only on three unknown
parameters: the \textit{wino} mass $M$, the Higgs mass parameter
$\mu $, and $\tan \beta $.

The supersymmetric partners of the squarks and leptons are spin
zero bosons: the \textit{squarks}, charged\textit{\ sleptons},
and\textit{\ sneutrinos.}

\subsection{The charginos}

The charginos, $\widetilde{\chi }_{i}^{\pm }$ (\thinspace
$i=1,2$), are four
component Dirac fermions which arise due to mixing of winos, $\widetilde{W}%
^{-}$ , $\widetilde{W}^{+}$ and the charged Higgsinos,
$\widetilde{H}^{-},$ and $\widetilde{H}^{+}$ [17], [18]. Because
there actually two independent mixings, $\left(
\widetilde{W}^{-},\widetilde{H}^{-}\right) $ and $\left(
\widetilde{W}^{+},\widetilde{H}^{+}\right) ,$ we shall need to
define two unitary mixing matrices. We define in two components
spinor notation:

\begin{equation}
\left( \psi _{j}^{\pm }\right) ^{T}=\left( -i\widetilde{W}^{\pm },\widetilde{%
H}^{\pm }\right) ,\text{ \ \ \ \ \ \ \ \ where }j=1,2
\end{equation}

The mass term in the Lagrangian is:

\begin{equation}
\mathfrak{L}_{m}=\left( \psi ^{-}\right) ^{T}\mathbf{X}\psi
^{+}+h.c.,
\end{equation}

where

\begin{equation}
\mathbf{X=}%
\begin{pmatrix}
M & \sqrt{2}m_{W}\sin \beta \\
\sqrt{2}m_{W}\cos \beta & \mu%
\end{pmatrix}%
.
\end{equation}

The mass matrix \textbf{X }is diagonalized by the unitary $2\times
2$ matrices \textbf{U }and \textbf{V}:

\begin{equation}
\mathbf{U}^{\ast }\mathbf{XV}^{-1}=\mathbf{M}_{D},
\end{equation}

where \textbf{M}$_{D}$ is the diagonal chargino mass matrix. In
particular, \textbf{U }and \textbf{V} can be cj=hosen so that the
elements of the diagonal matrix \textbf{M}$_{D}$ are real and
\textit{non-negative}. We define two component mass eigenstates
via:

\begin{eqnarray}
\chi _{1}^{+} &=&V_{ij}\psi _{j}^{+},  \notag \\
\chi _{1}^{-} &=&V_{ij}\psi _{j}^{-},\text{ \ \ \ \ \ \ \ \ \ \ \
\ \ \ \ \ \ \ \ \ \ where }i,j=1,2,
\end{eqnarray}

The proper four component mass eigenstates are the charginos which
are defined in terms of two component mass eigenstates as:

\begin{equation}
\widetilde{\chi }_{1}^{+}=\binom{\chi _{1}^{+}}{\chi
_{1}^{-}},\text{ \ \ \ \ \ \ \ }\widetilde{\chi
}_{2}^{+}=\binom{\chi _{2}^{+}}{\chi _{2}^{-}},
\end{equation}

The mass eigenvalues \textbf{M}$_{D}$ $_{i}$ (the two components
of the diagonal) are given by

\begin{equation}
\mathbf{M}_{D\text{ }\ 1,2}^{2}=\frac{1}{2}\left\{
\begin{array}{c}
\left\vert \mu ^{2}\right\vert +\left\vert M^{2}\right\vert +2m_{W}^{2} \\
\mp \sqrt{\left( \left\vert \mu ^{2}\right\vert +\left\vert
M^{2}\right\vert +2m_{W}^{2}-4\left\vert \mu ^{2}\right\vert
\left\vert M^{2}\right\vert
\right) ^{2}-4m_{W}^{4}\sin ^{2}2\beta +8m_{W}^{2}\sin ^{2}2\beta \func{Re}%
\left( \mu M\right) }%
\end{array}%
\right\}
\end{equation}

If $CP-$violation effects are ignored (in such case, $M$ and $\mu
$ are real parameters), then one can choose a convention where
$\tan \beta $ and $M$ are positive- note that the relative sign of
$M$ and $\mu $ are meaningful.
The sign of $\mu $ is convention dependent\footnote{%
Notice that both sign conventions appear in literature.} Now eq.
(3.34) becomes

\begin{equation}
\mathbf{M}_{D\text{ }\ 1,2}^{2}=\frac{1}{2}\left\{
\begin{array}{c}
\mu ^{2}+M^{2}+2m_{W}^{2} \\
\pm \sqrt{\left( M^{2}-\mu ^{2}\right) ^{2}+4m_{W}^{4}\cos
^{2}2\beta
+4m_{W}^{2}\left( M^{2}+\mu ^{2}+2M\mu \sin 2\beta \right) }%
\end{array}%
\right\}
\end{equation}

and it has the roots

\begin{equation}
M_{D\text{ }1,2}=\frac{1}{2}\left( \sqrt{\left( M-\mu \right)
^{2}+2m_{W}^{2}\left( 1+\sin 2\beta \right) }\mp \sqrt{\left(
M+\mu \right) ^{2}+2m_{W}^{2}\left( 1-\sin 2\beta \right) }\right)
\end{equation}

We write the chargino mass eigenvalues in the form $M_{D\text{
}i}=\eta
_{i}m_{\widetilde{\chi }_{i}^{\pm }},$ $i,j=1,2,$ with $m_{\widetilde{\chi }%
_{i}^{\pm }}=\left\vert M_{D\text{ }i}\right\vert ,$ and $\eta
_{i}=sign\left( M_{D\text{ }i}\right) =\pm 1.$

Assuming $CP-$conservation we choose the matrices \textbf{U} and \textbf{V }%
real. The matrix elements $U_{ij}$ and $V_{ij}$ are given by

\begin{subequations}
\begin{eqnarray}
U_{1,2} &=&U_{2,1}=\frac{\theta
_{1}}{\sqrt{2}}\sqrt{1+\frac{M^{2}-\mu
^{2}-2m_{W}^{2}\cos 2\beta }{W}} \\
U_{2,2} &=&-U_{1,1}=\frac{\theta
_{2}}{\sqrt{2}}\sqrt{1-\frac{M^{2}-\mu
^{2}-2m_{W}^{2}\cos 2\beta }{W}} \\
V_{2,1} &=&-V_{1,2}=\frac{\theta
_{3}}{\sqrt{2}}\sqrt{1+\frac{M^{2}-\mu
^{2}+2m_{W}^{2}\cos 2\beta }{W}} \\
V_{2,2} &=&V_{1,1}=\frac{\theta
_{4}}{\sqrt{2}}\sqrt{1-\frac{M^{2}-\mu ^{2}+2m_{W}^{2}\cos 2\beta
}{W}}
\end{eqnarray}

Where the sign factors $\theta _{i}$, $i=1,...,4$, are given in
Table 3.1, and

\end{subequations}
\begin{equation}
W=\sqrt{\left( M^{2}+\mu ^{2}+2m_{W}^{2}\right) ^{2}-4\left( M\mu
-m_{W}^{2}\sin 2\beta \right) ^{2}}
\end{equation}

\begin{center}
\bigskip

\begin{tabular}[t]{|c|c|c|}
\hline\hline \multicolumn{1}{||c|}{$\theta _{i}$ \ \ \ \ \ \ \ \ \
\ } & \multicolumn{1}{||c|}{$\tan \beta >1$} &
\multicolumn{1}{||c||}{$\tan \beta <1$} \\ \hline\hline $\theta
_{1}$ & 1 & $\varepsilon _{B}$ \\ \hline $\theta _{2}$ &
$\varepsilon _{B}$ & 1 \\ \hline $\theta _{3}$ & $\varepsilon
_{A}$ & 1 \\ \hline $\theta _{4}$ & 1 & $\varepsilon _{A}$ \\
\hline
\end{tabular}

Table 3.1. Sign factors $\theta _{i},i=1,...,$ where $\varepsilon
_{A}=sign\left( M\sin \beta +\mu \cos \beta \right) $

and $\varepsilon _{B}=sign\left( M\cos \beta +\mu \sin \beta
\right) .$
\end{center}

\subsection{The Neutralinos}

The neutralinos, $\widetilde{\chi }_{i}^{0}\left( i=1,...,4\right)
,$ are four-component Majorana fermions which arise due to mixing
of the two neutral gauginos $\widetilde{B}$ $\left(
\text{Bino}\right) ,$ $\widetilde{W} $ $^{3}\left( \text{neutral
}W-ino\right) ,$ and the two neutral Higgsinos, \
$\widetilde{H}_{1}^{0},$ and
$\widetilde{H}_{2}^{0}$[19],[20],[21]. As basis of the neutral
gaugino-Higgsino system we conveniently take

\begin{equation}
\left( \psi ^{0}\right) ^{T}=\left( -i\widetilde{B},-i\widetilde{W}^{3},%
\widetilde{H}_{1}^{0},\widetilde{H}_{2}^{0}\right) ,
\end{equation}

The mass term in the Lagrangian is:

\begin{equation}
\mathfrak{L}_{m}=\frac{1}{2}\left( \psi ^{0}\right)
^{T}\mathbf{Y}\psi ^{0}+h.c.,
\end{equation}

where

\begin{equation}
\mathbf{Y}=%
\begin{pmatrix}
M^{\prime } & 0 & -m_{Z}\sin \theta _{W}\cos \beta & -m_{Z}\sin
\theta
_{W}\cos \beta \\
0 & M & m_{Z}\cos \theta _{W}\sin \beta & -m_{Z}\cos \theta
_{W}\sin \beta
\\
-m_{Z}\sin \theta _{W}\cos \beta & m_{Z}\cos \theta _{W}\cos \beta
& 0 & -\mu
\\
m_{Z}\sin \theta _{W}\sin \beta & -m_{Z}\cos \theta _{W}\sin \beta
& -\mu
^{2} & 0%
\end{pmatrix}%
\end{equation}

on the $\left( -i\widetilde{B},-i\widetilde{W}^{3},\widetilde{H}_{1}^{0},%
\widetilde{H}_{2}^{0}\right) $ basis. The two-component mass
eigenstates can be obtained by diagonalizing the mass matrix
$\mathbf{Y}$

\begin{equation}
\widetilde{\chi }_{i}^{0}=N_{ij}\psi _{j}^{0},
\end{equation}

where $N$ is a complex unitary matrix
$(\mathbf{N}^{+}\mathbf{N=1)}$ satisfying

\begin{equation}
\mathbf{N}^{\ast }\mathbf{YN}^{-1}=\mathbf{N}_{D},
\end{equation}

and $\mathbf{N}_{D}$ is the diagonal neutralino mass matrix. the
four-component mass eigenstates are the neutralinos which are
defined in terms of two-component mass eigenstates:

\begin{equation}
\widetilde{\chi }_{i}^{0}=\binom{\chi _{i}^{0}}{\overline{\chi }_{i}^{0}},%
\text{ \ \ \ \ \ \ }i=1,...,4.
\end{equation}

Assuming $CP-$invariance $\mathbf{N}$ is replaced by another matrix $\mathbf{%
Z.}$ This implies the changing of eq.$\left( 3.43\right) $ to

\begin{equation}
\mathbf{Z}^{\ast }\mathbf{YZ}^{-1}=\mathbf{M}_{D},
\end{equation}

and we write the neutralino mass eigenvalues in the form $M_{D\text{ }%
i}=\varepsilon _{i}m_{\widetilde{\chi }_{i}^{0}},i=1,...,4,$ with $m_{%
\widetilde{\chi }_{i}^{0}}=\left\vert M_{D\text{ }i}\right\vert ,$ and $%
\varepsilon _{i}=sign\left( M_{D\text{ }i}\right) =\pm 1.$ The
relation between the $\mathbf{N}$ and $\mathbf{Z}$ matrices is

\begin{equation}
N_{ij}=\sqrt{\varepsilon _{i}}Z_{ij},\text{ \ \ \ \ \ }\left( no\text{ }sum%
\text{ }over\text{ }i\right) .
\end{equation}

Using the theory of equations, the expressions for the $M_{D\text{ }i}$ $%
\left( \text{the four components of the diagonal}\right) $ are
given by

\begin{subequations}
\begin{eqnarray}
M_{D\text{ }1} &=&-A+B_{+}+\frac{1}{4}C_{+}, \\
M_{D\text{ }2} &=&+A-B_{\_}+\frac{1}{4}C_{+}, \\
M_{D\text{ }3} &=&-A-B_{+}+\frac{1}{4}C_{+}, \\
M_{D\text{ }4} &=&+A+B_{\_}+\frac{1}{4}C_{+},
\end{eqnarray}

where

\end{subequations}
\begin{subequations}
\begin{eqnarray}
A &=&\sqrt{\frac{1}{2}a-\frac{1}{6}b}, \\
B_{\pm } &=&\sqrt{-\frac{1}{2}a-\frac{1}{3}b\pm \frac{c}{\sqrt{8a-\frac{8}{3}%
b}}}, \\
C_{\pm } &=&M^{\prime }\pm M,
\end{eqnarray}

and

\end{subequations}
\begin{subequations}
\begin{align}
a& =\frac{1}{\sqrt[3]{2}}\func{Re}\left( c^{2}+\frac{2}{27}b^{3}-\frac{8}{3}%
bd+i\sqrt{\frac{D}{27}}\right) ^{\frac{1}{3}}, \\
b& =E-\frac{3}{8}C_{+}^{2}, \\
c& =\frac{1}{8}C_{+}^{3}+\frac{1}{2}C_{+}E+C_{+}\mu
^{2}+Fm_{Z}^{2}-\mu
m_{Z}^{2}\sin 2\beta , \\
d& =F\mu m_{Z}^{2}\sin 2\beta -M^{\prime }M\mu ^{2}+\frac{1}{16}EC_{+}^{2}-%
\frac{3}{256}C_{+}^{4}+\frac{1}{4}C_{+}(C_{+}\mu
^{2}+Fm_{Z}^{2}-\mu m_{Z}^{2}\sin 2\beta ),
\end{align}

where

\end{subequations}
\begin{subequations}
\begin{eqnarray}
D &=&-4\left( -\frac{1}{3}b^{3}-4d\right) ^{3}-27\left( -c^{2}-\frac{2}{27}%
b^{3}+\frac{8}{3}bd\right) ^{2}, \\
E &=&M^{\prime }M-m_{Z}^{2}-\mu ^{2}, \\
F &=&M^{\prime }\cos ^{2}\theta _{W}+M\sin ^{2}\theta _{W}.
\end{eqnarray}

The elements of the mixing matrix $\mathbf{Z}$ are given by

\end{subequations}
\begin{subequations}
\begin{eqnarray}
Z_{i1} &=&\frac{1}{\sqrt{1+G_{i}^{2}+H_{i}^{2}+I_{i}^{2}}} \\
Z_{i2} &=&G_{i}Z_{i1}, \\
Z_{i3} &=&H_{i}Z_{i1}, \\
Z_{i4} &=&I_{i}Z_{i1},\text{ \ \ \ \ \ \ \ }i=1,...,4,
\end{eqnarray}

where

\end{subequations}
\begin{subequations}
\begin{eqnarray}
G_{i} &=&-\frac{J_{i}^{\prime }}{J_{i}\tan \theta _{W}}, \\
H_{i} &=&\frac{\mu J_{i}J_{i}^{\prime }-\frac{1}{2}K_{i}}{L_{i}}, \\
I_{i} &=&\frac{-M_{D\text{ }i}J_{i}J_{i}^{\prime }-K_{i}}{L_{i}}, \\
J_{i} &=&M-M_{D\text{ }i},\text{ \ }J_{1}^{\prime }=M^{\prime }-M_{D\text{ }%
i}, \\
K_{i} &=&m_{Z}^{2}\sin 2\beta \left( C_{-}\cos ^{2}\theta
_{W}+J_{i}\right) ,
\\
L_{i} &=&m_{Z}J_{i}\sin \theta _{i}\left( \mu \cos \beta
+M_{D\text{ }i}\sin \beta \right) .
\end{eqnarray}

\subsection{The Sferminos}

For a given fermion $f$, there are two supersymmetric partners $\widetilde{f}%
_{L}$ and $\widetilde{f}_{R}$ (sfermions) which are scalar
parameters of the
corresponding left and right-handed fermion. There are no $\widetilde{\nu }%
_{R}$. However, in general, $\widetilde{f}_{L}$ and
$\widetilde{f}_{R}$ are not mass-eigenstates since there is
$\widetilde{f}_{L}-\widetilde{f}_{R}$ mixing which is proportional
in strength to the corresponding element of the scalar
mass-squared matrix [22]:

\end{subequations}
\begin{equation}
M_{LR}^{2}=\QATOPD\{ \} {m_{d}\left( A_{d}-\mu \tan \beta \right)
,\text{ \ for "down"-type }f}{m_{u}\left( A_{u}-\mu \cot \beta
\right) ,\text{ \ \ for "up"type \ \ \ \ \ \ }f},
\end{equation}

where $m_{d}\left( m_{u}\right) $ is the mass of the appropriate "down" $%
\left( \text{"up"}\right) $ type quark or lepton. Here, $A_{d}$
and $A_{u}$
are (unknown) soft-supersymmetric-breaking $A$-parameters and $\mu $ and $%
\tan \beta $ have been defined earlier. The signs of the
$A$-parameters are also convenient-dependent due to the appearance
of the \textsl{fermion }mass in eq.(3.53), one expects $M_{LR}$ to
be small compared to the diagonal squark and slepton masses, with
the possible exception of the top-squark, since $m_{t}$ is large,
and the bottom-squark and tau-slepton if $\tan \beta
>>1.$

The (diagonal) \ $L-and$ $R-type$ squark and slepton masses are
given by

\begin{subequations}
\begin{eqnarray}
m_{\widetilde{u}_{L}}^{2}
&=&M_{\widetilde{Q}}^{2}+m_{u}^{2}+m_{Z}^{2}\cos
2\beta \left( \frac{1}{2}-\frac{2}{3}\sin ^{2}\theta _{W}\right) , \\
m_{\widetilde{u}_{R}}^{2} &=&M_{\widetilde{U}}^{2}+m_{u}^{2}+\frac{2}{3}%
m_{Z}^{2}\cos 2\beta \sin ^{2}\theta _{W}, \\
m_{\widetilde{d}_{L}}^{2}
&=&M_{\widetilde{Q}}^{2}+m_{d}^{2}-m_{Z}^{2}\cos
2\beta \left( \frac{1}{2}-\frac{1}{3}\sin ^{2}\theta _{W}\right) , \\
m_{\widetilde{d}_{R}}^{2} &=&M_{\widetilde{D}}^{2}+m_{d}^{2}-\frac{1}{3}%
m_{Z}^{2}\cos 2\beta \sin ^{2}\theta _{W}, \\
m_{\widetilde{\nu }}^{2}
&=&M_{\widetilde{L}}^{2}+\frac{1}{3}m_{Z}^{2}\cos
2\beta , \\
m_{\widetilde{e}_{L}}^{2}
&=&M_{\widetilde{L}}^{2}+m_{e}^{2}-m_{Z}^{2}\cos
2\beta \left( \frac{1}{2}-\sin ^{2}\theta _{W}\right) , \\
m_{\widetilde{e}_{R}}^{2}
&=&M_{\widetilde{E}}^{2}+m_{e}^{2}-m_{Z}^{2}\cos 2\beta \sin
^{2}\theta _{W},
\end{eqnarray}

The soft-supersymmetry-breaking parameters: $M_{\widetilde{Q}}^{2}$ ,$M_{%
\widetilde{U}}^{2}$ ,$M_{\widetilde{D}}^{2}$ ,$M_{\widetilde{L}}^{2}$ ,$M_{%
\widetilde{E}}^{2}$ are unknown parameters. In the equations
above, the notation of the first generation fermions has been used
and generational indices have been suppressed. further
complications such as integrational mixing are possible, although
there are some constraints from the non-observation of
flavor-changing current (FCNC).

\section{Reducing \textsl{the }MSSM Parameter Freedom}

One way to guarantee the absence of FCNC's mediated by virtual
supersymmetric-particle exchange is to posit that the diagonal
soft-supersymmetry-breaking scalar squared-masses are universal in
flavor space at some energy scale (normally taken to be at or near
the Plank scale).\ Renormalization group evolution is used to
determine the low energy values for the scalar mass parameters
listed above. This assumption reduces the MSSM parameter freedom.
For example, the supersymmetric grand unified
models with universal scalar masses at Plank scale typically give $M_{%
\widetilde{L}}\approx M_{\widetilde{E}}<M_{\widetilde{Q}}\approx M_{%
\widetilde{U}}\approx M_{\widetilde{D}}$ with the squark masses
somewhere between a factor of 1-3 larger than the slepton masses
(neglecting generational distinction). More specifically, the
first two generations are
thought to be degenerate in mass, while $M_{\widetilde{Q}_{3}}$ and $M_{%
\widetilde{U}_{3}}$are typically reduced by a factor of 1-3 from
the other soft supersymmetric breaking masses because of
renormalization effects due to the heavy top quark masses.

As a result, four flavors of the squarks (with two squarks
eigenstates per flavor) and $\widetilde{b}_{R}$ will be nearly
mass-degenerate and somewhat heavier than six flavors of nearly
mass-degenerate sleptons (with two per flavor for the charged
sleptons and one for the sneutrinos). On the other
hand, $\widetilde{b}_{L}$ mass and the diagonal $\widetilde{t}_{L}$ and $%
\widetilde{t}_{R}$ masses are reduced compared to the common
squark mass of the first two generations. In addition, third
generation squark masses and
tau-slepton masses are sensitive to the respective \thinspace $\widetilde{f}%
_{L}-\widetilde{f}_{R}$ mixing as discussed before.

Two additional theoretical frameworks are often introduced to
reduce further the MSSM parameter freedom. the first involves
grand unified theories (GUTs) and the desert hypothesis ($i.e.$ \
no new physics between the TeV-scale and the GUT-scalae). Perhaps
one of the most compelling hints for low energy supersymmetry is
the unification of $SU(3)\times SU(2)\times U(1)$ gauge coupling
predicted by supersymmetric GUT models (with the supersymmetry
breaking scale of order 1 TeV or below).

The unification, which takes place at an energy scale of order
$10^{16}GeV$ is quite robust (and depends weakly on the details of
the GUT-scale theory). For example, a recent analysis finds that
supersymmetric GUT unification implies that $\alpha
_{s}(m_{Z})=0.129\pm 0.010,$not including threshold
corrections due to GUT-scale particles (which could diminish the value of $%
\alpha _{s}(m_{Z})$). This result is compatible with the world average of $%
\alpha _{s}(m_{Z})=0.118\pm 0.003$. In contrast, gauge coupling
unification in the simplest non-supersymmetric GUT models fails by
many standard deviations.

\subsection{Minimization \textsl{of the }Higgs potential}

In the MSSM, the Higgs sector has two unknown parameters, usually
taken to be $\tan \beta \equiv \upsilon _{2}/\upsilon _{1}$ and
$m_{H_{3}^{0}}$, the mass of its one physical pseudoscalar
particle.\ numerous phenomenological studies have been made using
these parameters as continuous variables. However, there is an
argument for $m_{H_{3}^{0}}=m_{Z}$ at the tree level and perhaps
also $\tan \beta >\sqrt{3}$, by minimizing the minimum of the
Higgs potential along a certain direction in parameter space [23].

The part of $V$ (see eq.(316)) involving only neutral fields
depends on four
parameters: $m_{1}^{2}$, $m_{2}^{2}$, $m_{1,2}^{2}$, and $%
g_{1}^{2}+g_{2}^{2} $. At its minimum $V_{0}$, we can choose to keep $%
m_{1}^{2}$ and $m_{2}^{2}$, but replace $m_{1,2}^{2}$ with $\tan
\beta $ through eq. (3.20) and $g_{1}^{2}+g_{2}^{2}$ with
$\upsilon _{1}^{2}+\upsilon _{2}^{2}$ through eq.(3.21). In that
case,

\end{subequations}
\begin{equation}
V_{0}=\frac{1}{2}(\upsilon _{1}^{2}+\upsilon _{2}^{2})\left( \cos
^{2}\beta -\sin ^{2}\beta )(m_{1}^{2}\cos ^{2}\beta -m_{2}^{2}\sin
^{2}\beta \right)
\end{equation}

We now seek to minimize $V_{0}$ in parameter space. This is based
on the assumption that the dynamic mechanism responsible for the
soft breaking of supersymmetry may be such that the lowest
possible value of $V_{0}$ is automatically chosen. It is also
clear that $\upsilon _{1}^{2}+\upsilon _{2}^{2}$, $m_{1}^{2}$,
$m_{2}^{2}$ set the energy scale of the symmetry breaking and
$V_{0}$ has no lower bound as a function of these parameters. We
should therefore consider them as fixed and vary $\sin ^{2}\beta $
to minimize $V_{0}.$ Let $x\equiv \sin ^{2}\beta $, then

\begin{equation}
\frac{\partial V_{0}}{\partial x}=\frac{1}{2}\left( \upsilon
_{1}^{2}+\upsilon _{2}^{2}\right) [-\left(
3m_{1}^{2}+m_{2}^{2}\right) +4\left( m_{1}^{2}+m_{2}^{2}\right)
x],
\end{equation}

and

\begin{equation}
\frac{\partial ^{2}V_{0}}{\partial x^{2}}=2\left( \upsilon
_{1}^{2}+\upsilon _{2}^{2}\right) \left(
m_{1}^{2}+m_{2}^{2}\right) .
\end{equation}

Hence, the minimization of $V_{0}$ is achieved if

\begin{equation}
x=\frac{3m_{1}^{2}+m_{2}^{2}}{4\left( m_{1}^{2}+m_{2}^{2}\right) }
\end{equation}

and $m_{1}^{2}+m_{2}^{2}>0$, which is consistent with eq.$\left(
3.22\right) .$ Using eq. $\left( 3.21\right) $ we then find

\begin{equation}
m_{1}^{2}+m_{2}^{2}=\frac{1}{2}\left( g_{1}^{2}+g_{2}^{2}\right)
\left( \upsilon _{1}^{2}+\upsilon _{2}^{2}\right) ,
\end{equation}

or equivalently,%
\begin{equation}
m_{H_{3}^{0}}=m_{Z}\text{.}
\end{equation}

This implies

\begin{equation}
m_{H^{\pm }}=\sqrt{m_{W}^{2}+m_{Z}^{2}}\approx 121GeV
\end{equation}

and

\begin{equation}
m_{H_{1}^{0},H_{2}^{0}}=m_{Z}^{2}\sqrt{1\pm \sin 2\beta }
\end{equation}

From eq.$\left( 3.58\right) ,$we find

\begin{equation}
\tan ^{2}\beta =\frac{3m_{1}^{2}+m_{2}^{2}}{m_{1}^{2}+3m_{2}^{2}},
\end{equation}

which shows that if $m_{1}^{2}>0$ and $m_{2}^{2}>0$, then $\frac{1}{\sqrt{3}}%
<\tan \beta <\sqrt{3}$. However, because $\phi _{2}$ couples to the \textit{%
top} quark with a large Yukawa coupling, $m_{2}^{2}$ is expected
to differ from $m_{1}^{2}$ by a large negative contribution from
the
renormalization-group equations, hence the case $m_{1}^{2}>0$ and $%
m_{2}^{2}<0$ should be considered. We then obtain

\begin{equation}
\tan \beta >\sqrt{3}\approx 1.731,
\end{equation}

where $m_{1}^{2}>3\left\vert m_{2}^{2}\right\vert $ has also been
assumed or else $V_{0}$ would have been minimized at $\sin
^{2}\beta >1$, which is impossible. Using eq.$\left( 3.62\right) $
we find $m_{H_{1}^{0}}>33GeV$ and
$m_{H_{2}^{0}}<125GeV$, with the constraints that $%
m_{H_{1}^{0}}^{2}+m_{H_{2}^{0}}^{2}=2m_{Z}^{2}$. Experimentally,
there is no evidence for the existence of any of the five scalar
particles of the MSSM from the $Z$ decay or in any other process.

\subsection{Diophantine Analysis $of$\textsl{\ }$the$ Higgs Boson}

Diophantine quantization involves treating mass relations as
Diophantine equations and seeking solutions in integers, analogous
to the sets $\left(
3,4,5\right) ,\left( 5,12,13\right) ,etc.,$ for the Pythagorean equation $%
x^{2}+y^{2}=z^{2}.$ It was first applied to the Gell-Mann-Okubo
meson-mass relation [24]

\begin{equation}
m_{\pi }^{2}+3m_{\eta }^{2}=4m_{K}^{2},
\end{equation}

for which the simplest nontrivial solution was the set $\left( 2,8,7\right) $%
. The procedure not only gave integers proportional to the
experimental masses $m_{\pi }=135-140MeV,$ $m_{\eta }=547MeV,$
$m_{K}=494-498MeV$ but
also set a unit mass- the GCF of the three mesons masses- of the $70MeV$ $%
\left( =\left( \hbar c/e^{2}\right) m_{e}c^{2}\right) ,$ as
originally proposed by Nambu.

There is an analogy to this in the standard model. If one looks
carefully at the values of the $W$ and $Z$ masses

\begin{equation}
m_{W}=80.22\pm 0.26GeV,
\end{equation}

\begin{equation}
m_{Z}=91.173\pm 0.020GeV,
\end{equation}

one finds that within experimental limits,

\begin{equation}
\cos \theta _{W}=m_{W}/m_{Z}=15/17.
\end{equation}

not only is this the ratio of the two integers, but these two
particular two integers $\left( 15,17\right) $ happen to be two
sides of a right-angled triangle, then $\theta _{W}$ is, quite
interestingly, a rational angle. A simple consequence of this,
which we shall use below, is that the sine and cosine of such an
angle must be rational.

A lengthy Diophantine analysis [25] applied to the scalar Higgs
gauge boson mass relations of eq.$\left( 3.22\right) $ of minimal
supersymmetry results in the following relations

\begin{equation}
\sin \beta =\QATOPD\{ . {\frac{8}{17},\text{ \ \ \ \ \ \ \ \ \
}\beta <\pi /4}{\frac{15}{17},\text{ \ \ \ \ \ \ \ \ \ }\beta >\pi
/4},
\end{equation}

\begin{subequations}
\begin{eqnarray}
m_{H_{3}^{0}} &=&m_{Z}=91.2GeV, \\
m_{H_{1}^{0}} &=&37.5GeV, \\
m_{H_{2}^{0}} &=&123GeV, \\
m_{H^{\pm }} &=&121.5GeV.
\end{eqnarray}

Furthermore, most surprising of all, if we recall that the Weinberg angle $%
\theta _{W}$ is characterized by eq.$\left( 3.68\right) ,$ we have
a relation between Higgs and Weinberg angles

\end{subequations}
\begin{equation}
\QATOPD\{ . {\theta _{W},\text{ \ \ \ \ \ \ \ \ \ \ \ \ }\beta <\pi /4}{%
\frac{\pi }{4}-\theta _{W},\text{ \ \ \ \ \ }\beta >\pi /4.}
\end{equation}

Finally, if $\beta >\pi /4,$ $\sin \beta =\frac{15}{17},$ where
the infrared quasi-fixed-point solution yields a
\textit{top}-quark mass

\begin{equation}
m_{t}\cong \left( 190-210\right) \sin \beta GeV\approx 168-175GeV
\end{equation}

within the range of recent data [26].

\section{Conclusion}

The minimal supersymmetric standard model $\left( MSSM\right) $ is
the simplest extension of the standard model $\left( SM\right) $
that includes soft broken supersymmetry $\left( SUSY\right) $. The
word \textquotedblleft minimal\textquotedblright\ refers to
minimal content of particles and gauge groups.

The $MSSM$ contains a great number of free parameters which
considerably limit the predictive power of the model. These are
some commonly used ways to reduce the number of free constants and
parameters in this theory. The most often employed method is to
obtain values of parameters at the scale of the order of $m_{W}$
by renormalization-group equations from the coupling constants of
the supergravity theories investigated at the Plank mass. usually
such theories are much more unified and contain typically only few
free numbers. Of course, there are also many constraints
originating from the experimental data-first of all the masses of
the superpartners are bounded from below by their absence in the
present experiments, but one can find also many more subtle
limits.

The usual $MSSM$ framework assumptions are:

\begin{enumerate}
\item \textit{R}-parity is conserved. this assumption is commonly
made in supersymmetry studies, as it prevents protons from
decaying too quickly.

\item The lightest supersymmetric particle (LSP) is the lightest
neutralino.

\item The intergenerational mixing in the squark, slepton, and
quark sectors is small and can be neglected.

\item The four \textit{left-handed} squarks of the first two
generations are nearly degenerate at low energy with mass
$m_{\widetilde{q}}$ as are all six left-handed\textit{\ }sleptons
with mass $m_{\widetilde{l}}:$
\end{enumerate}

\begin{eqnarray}
m_{\widetilde{u}_{L}} &\approx &m_{\widetilde{d}_{L}}\approx m_{\widetilde{c}%
_{L}}\approx m_{\widetilde{s}_{L}}\approx m_{\widetilde{q}_{L}},  \notag \\
m_{\widetilde{\nu }_{e_{L}}} &\approx &m_{\widetilde{e}_{L}}\approx m_{%
\widetilde{\nu }_{\mu _{L}}}\approx m_{\widetilde{\mu }_{L}}\approx m_{%
\widetilde{\nu }_{\tau _{L}}}\approx m_{\widetilde{\tau }_{L}}\approx m_{%
\widetilde{l}}.
\end{eqnarray}

\begin{enumerate}
\item[5.] The gaugino mass $M_{i},$the parameter $\mu $ and $\tan
\beta $ may be taken to be real, so that $CP-violation$ plays no
role.
\end{enumerate}

Unification of gaugino mass, $i.e.$ $M_{1}$and $M_{2}$ are not
independent.

\part{Phenomenological Calculations}

\chapter{Introduction}

\bigskip The phenomenological predictions of supersymmetry (SUSY) may be
divided into three categories [27], [28], [29]:

\begin{enumerate}
\item Reflections of supersymmetric lagrangian in Standard Model
(SM) phenomenology, including relations among the gauge coupling
constants from supersymmetric grand unification and the presence
of a heavy top quark and light Higgs scalar;

\item The prediction of new particles with the correct spin and
quantum numbers assignments to be superpartners of the standard
model particles; and

\item Well-defined quantitative relations among the couplings and
masses of these new particles.
\end{enumerate}

While the predictions of (1) are of great interest, their
verification is clearly no substitute for direct evidence. The
discovery of a large number of particles in category (2) would be
strong support for supersymmetry. On the other hand, the most
compelling confirmation of supersymmetry would likely be the
precise verification of the relations of category (3). This would
be specially true if, initially, only a small set of candidate
supersymmetric partners are observed.

Most discussions of supersymmetry at future high energy colliders
have concentrated on the question of particle searches. From one
point of view, this is reasonable, because the existence of
supersymmetric partners is unproven and this is a prerequisite for
any further analysis. On the other hand, the discovery of the
first evidence for supersymmetry, or for any other theoretical
extension of the standard model, will begin a program of detailed
experimental investigation of the new sector of particles required
by this extension.

Supersymmetry provides a particularly interesting subject for
studies of the detailed analysis of physics beyond the standard
model. Supersymmetry are weakly coupled, and so their consequences
can be worked out straightforwardly using perturbative
computations. At the same time, supersymmetric models depend on a
large number of unknown parameters, and different choices for
these parameters yield qualitatively different realizations of
possible new physics. Thus the phenomenology of supersymmetry is
quite complex. Eventually, if supersymmetry does give a correct
model of nature, the colliders of the next generation will be
expected to determine the supersymmetric parameters, and their
values will become clues that take us a step closer to a
fundamental theory.

In the minimal supersymmetric extension of the standard model,
MSSM, among the lightest supersymmetric particles there are four
neutralinos (the supersymmetric partners of the neural electroweak
gauge and Higgs bosons). In most scenarios, apart from the
lightest supersymmetric particle (LSP),
which is in general assumed to be the lightest neutralino ($\widetilde{\chi }%
_{1}^{o}$) (stable and invisible), the particles that could be
first
observed at future colliders are the next-to-lightest neutralino ($%
\widetilde{\chi }_{2}^{o}$) and the light chargino ($\widetilde{\chi }%
_{1}^{\pm }$) [30]. Therefore, any reasonable large supersymmetric
signal
must involve either the second lightest neutralino $\widetilde{\chi }%
_{2}^{o} $ or the lighter charginos $\widetilde{\chi }_{1}^{\pm
}$. In general, we can not assume that the second lightest
neutralino is heavier than the lighter chargino, since,
$m_{\widetilde{\chi }_{2}^{o}}$ is not
independent of $m_{\widetilde{\chi }_{1}^{o}}$ and $m_{\widetilde{\chi }%
_{1}^{\pm }}$. In fact, in the region of parameter space in which
charginos
production is accessible to he future $e^{-}e+$ colliders, $m_{\widetilde{%
\chi }_{2}^{o}}$ and $m_{\widetilde{\chi }_{1}^{\pm }}$ are very
roughly degenerate, with the mass difference typically in the
range

\begin{equation*}
-10GeV\leq m_{\widetilde{\chi }_{2}^{o}}-m_{\widetilde{\chi
}_{1}^{\pm }}\leq 20GeV.
\end{equation*}

When $m_{\widetilde{\chi }_{2}^{o}}<m_{\widetilde{\chi }_{1}^{\pm
}}$, it is
possible for the lighter chargino to decay through a cascade decays to a $%
\widetilde{\chi }_{2}^{o}$, which in urn decays to an LSP.

The $e^{-}e+$ colliders have been playing complementary roles to
the hadron colliders in supersymmetry searches. In general,
$e^{-}e+$ colliders have reasonable signal rates in a clean
environment with a definite center-of-mass energy, enabling us to
perform precision measurements of particles' masses, lifetimes,
and various different cross-sections, while hadron colliders
provide opportunities to quickly survey high energy
frontier. In particular, the production of $\widetilde{\chi }_{1}^{o}$ $%
\widetilde{\chi }_{2}^{o}$ pairs at $e^{-}e+$ colliders could
allow the study of a wide region of the supersymmetry parameter
space.

Owing to the relatively large cross-section of two body final
state reactions, they can be used to search for supersymmetric
particles with masses up to the beam energy. In this study, the
production of certain three body final state reactions are
calculated to improve the sensitivity in searching for
supersymmetric particles.
\newpage
\begin{center}
{\huge Outline of Part Two}
\end{center}

In this part "the calculation part", reactions' cross sections are
calculated and the corresponding curves are plotted with the
results tabulated for a convenient reference. Work went as follows:

\begin{enumerate}
\item In chapter five, the reaction $e^- e^+ \rightarrow
H^- \widetilde{\chi }_{1}^+ \widetilde{\chi }_{1}^o$
is considered.

\item In chapter six, the reaction $e^{-}e^{+}\rightarrow h\widetilde{\chi }%
_{1}^{+}\widetilde{\chi }_{1}^{-}$ is considered.

\item In chapter seven, the reaction $e^{-}e^{+}\rightarrow
h\widetilde{\chi }_{1}^{o}\widetilde{\chi }_{1}^{o}$ is
considered$.$

\item In chapter eight, the reaction $e^{-}e^{+}\rightarrow
hH^{+}H^{-}$ is considered.

\end{enumerate}

\chapter{Production of a charged Higgs boson with a chargino and a neutralino%
}

\section{Introduction}

In this chapter, the total cross section for the process $e^-(p1)
e^+(p2) \rightarrow  H^-(p3) \widetilde{\chi }_{1}^+(p4)
\widetilde{\chi }_{1}^o(p5)$ for different topologies and
propagators (see Appendix A) are calculated and represented
graphically. There are 28 different Feynman diagrams (tree level
approximation) for which we gave the matrix element corresponding
to each diagram. Diagrams with the same topology that can be
obtained by changing the indices were represented once.

Work will go as follows:

\begin{enumerate}
\item Feynman diagrams are given,

\item Diagrams with the same topology are represented once, but
has been taken into considerations  when calculating the cross
section.

\item  Matrix elements are written, all the four momenta squares
are defined to be mass squared $(>0)$,

\item Matrix elements are squared,

\item An average over the initial spin polarizations of the
electron and positron pair and a sum over the final spin states of
the outgoing particles arising from each initial spin state is
carried out.

\item Results are represented graphically, and summarized in
subsequent tables.

\end{enumerate}

\section{Feynman Diagrams}

The follwoing is the set of Feynman diagrams which were used to
calculate the cross section of the associated production of a
charged Higgs boson with
a chargino and a neutralino. Our momentum notation is: $%
e^-(p_1) $, $e^+(p_2)$, $H^-(p_3)$, $\widetilde{\chi}^+_1(p_4)$
and $\widetilde{\chi}^o_1(p_3)$.

\begin{figure}[tph]
\begin{center}
\vskip-5.5cm
\mbox{\hskip-3.5cm\centerline{\epsfig{file=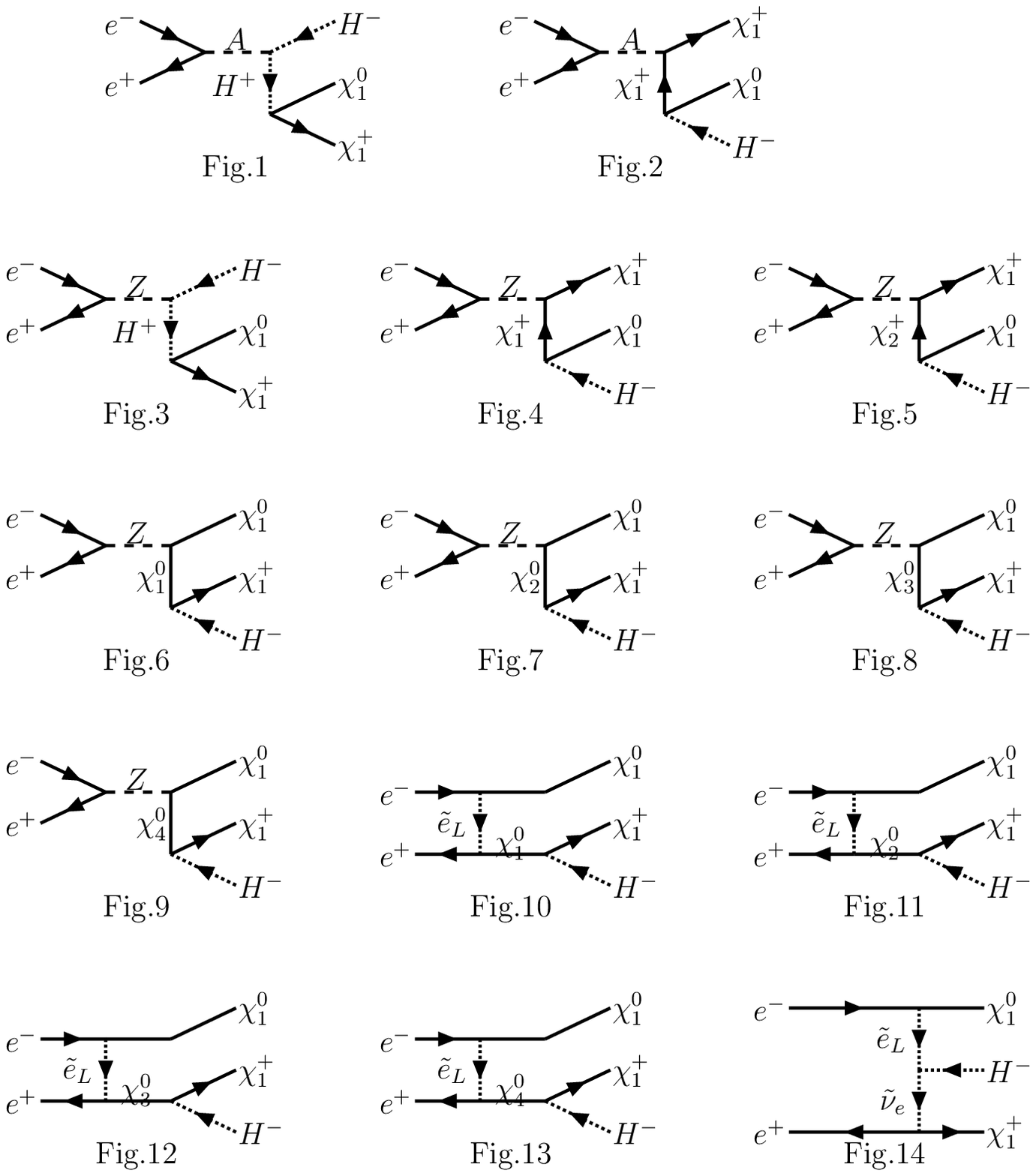,width=17cm}}}
\end{center}
\caption{Feynman diagrams for the reaction: $e^-(p1) e^+(p2)
\rightarrow  H^-(p3) \widetilde{\chi }_{1}^+(p4) \widetilde{\chi
}_{1}^o(p5)$} \label{feyn1}
\end{figure}

\begin{figure}[tph]
\begin{center}
\vskip-5.5cm
\mbox{\hskip-3.5cm\centerline{\epsfig{file=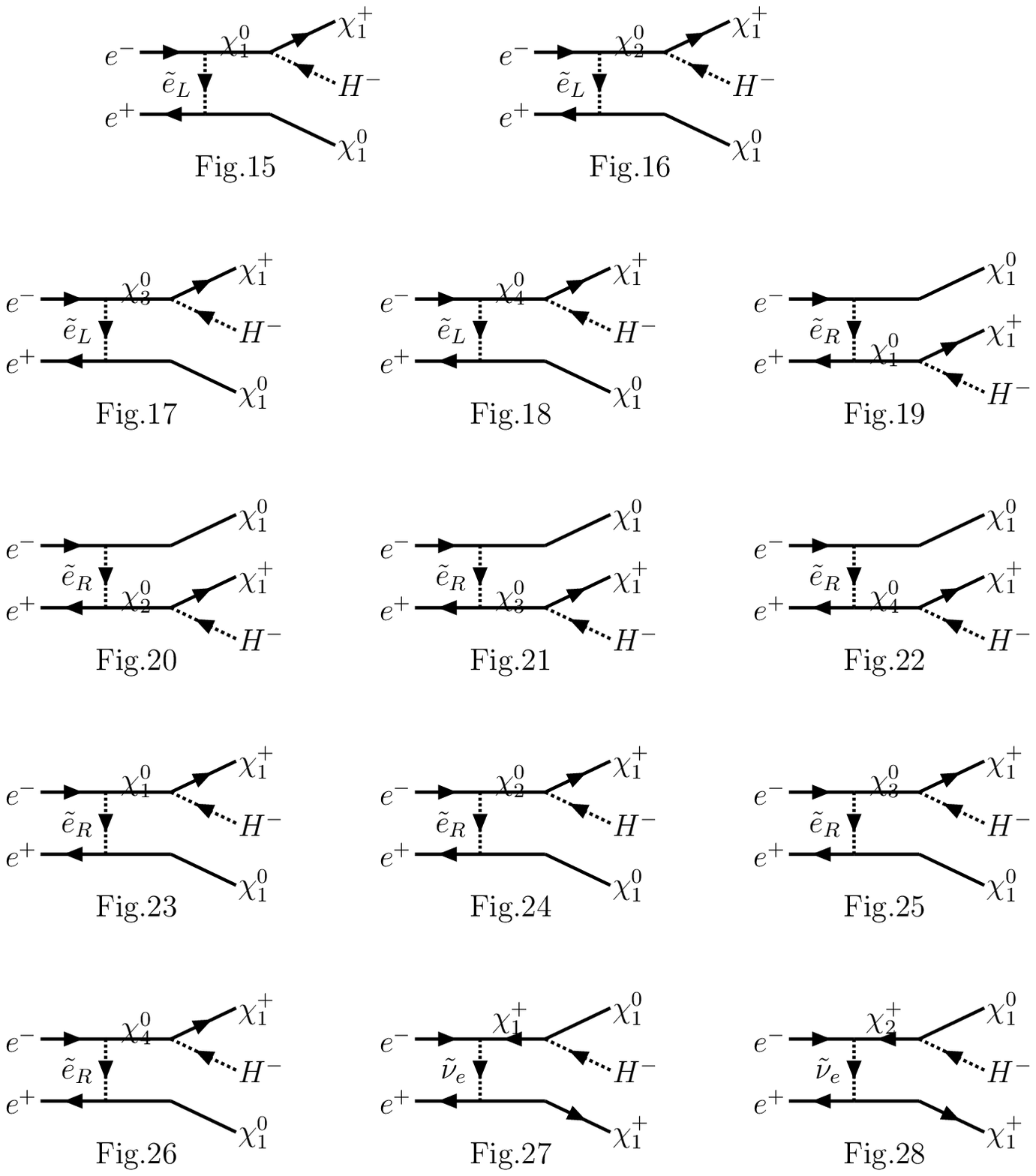,width=17cm}}}
\caption{Cont. Feynman diagrams for the reaction: $e^-(p1) e^+(p2)
\rightarrow  H^-(p3) \widetilde{\chi }_{1}^+(p4) \widetilde{\chi
}_{1}^o(p5)$}
\end{center}
\label{feyn2}
\end{figure}

\newpage

\section{Matrix Elements}

The following is the set of matrix elements corresponding to
feynman diagrams in figures \ref{feyn1} and \ref{feyn2} used in
our calculations:
\begin{eqnarray*}
\mathcal{M}_{1} &=&\overline{v}(p_{2})A\gamma _{\mu
}u(p_{1})P_{\gamma
}^{\mu \nu }(p_{1}+p_{2})J(p_{3}+p_{4}-p_{5})_{\nu }D_{H^{+}}(p_{3}+p_{4})%
\overline{u}(p_{4})  \notag \\
&&(Q_{ij}^{L}P_{L}+Q_{ij}^{R}P_{R})v(p_{3})
\end{eqnarray*}

\begin{center}
\begin{eqnarray*}
\mathcal{M}_{2} &\mathbf{=}&\text{ }\overline{v}(p_{2})A\gamma
_{\mu }u(p_{1})P_{\gamma }^{\mu \nu
}(p_{1}+p_{2})\overline{u}(p_{4})V\gamma _{\nu
}D_{\overline{\chi }_{1}^{+}}(p_{3}+p_{4})(\NEG{p}_{3}+\NEG{p}_{4}+m_{%
\overline{\chi }_{1}^{+}})  \notag \\
&&(Q_{ij}^{L}P_{L}+Q_{ij}^{R}P_{R})v(p_{3})
\end{eqnarray*}
\end{center}

\begin{eqnarray*}
\mathcal{M}_{3} &=&\overline{v}(p_{2})\gamma _{\mu
}(B^{L}P_{L}+B^{R}P_{R})u(p_{1})P_{Z}^{\mu \nu
}(p_{1}+p_{2})H(p_{1}+p_{2})_{\nu }D_{H^{+}}(p_{3}+p_{4})  \notag \\
&&\overline{u}(p_{4})(Q_{ij}^{L}P_{L}+Q_{ij}^{R}P_{R})v(p_{3})
\end{eqnarray*}

\begin{eqnarray*}
\mathcal{M}_{4,5} &=&\text{ }\overline{v}(p_{2})\gamma _{\mu
}(B^{L}P_{L}+B^{R}P_{R})u(p_{1})P_{Z}^{\mu \nu }(p_{1}+p_{2})\overline{u}%
(p_{4})\gamma _{\nu }(W_{ij}^{L}P_{L}+W_{ij}^{R}P_{R})  \notag \\
&&D_{_{\overline{\chi }_{k}^{+}}}(p_{3}+p_{4})(\NEG{p}_{3}+\NEG%
{p}_{4}+m_{_{_{\overline{\chi }%
_{k}^{+}}}})(Q_{ij}^{L}P_{L}+Q_{ij}^{R}P_{R})v(p_{3})
\end{eqnarray*}

Where $k=1,2.$

\begin{eqnarray*}
\mathcal{M}_{6,7,8,9} &=&\overline{v}(p_{2})\gamma _{\mu
}(B^{L}P_{L}+B^{R}P_{R})u(p_{1})P_{Z}^{\mu \nu }(p_{1}+p_{2})\overline{u}%
(p_{3})\gamma _{\nu }(S_{ij}+S_{ij}^{^{\prime }}\gamma _{5})  \notag \\
&&D_{_{\overline{\chi }_{k}^{o}}}(p_{4}+p_{5})(\NEG{p}_{4}+\NEG%
{p}_{5}+m_{_{_{\overline{\chi }_{k}^{o}}}})(Q_{ij}^{L}P_{L}+Q_{ij}^{R}P_{R})%
\overline{u}(p_{4})
\end{eqnarray*}

Where $k=1,2,3,4.$

\begin{eqnarray*}
\mathcal{M}_{10,16,17,18,19,24,25,26} &=&\overline{v}%
(p_{2})(N_{hi}+N_{hi}^{^{\prime }}\gamma _{5})v(p_{3})D_{\widetilde{e}%
_{h}}(p_{1}-p_{3})\overline{u}(p_{3})(Q_{ij}^{L}P_{L}+Q_{ij}^{R}P_{R})
\notag \\
&&D_{_{\overline{\chi }_{k}^{o}}}(p_{3}+p_{5})(\NEG{p}_{3}+\NEG%
{p}_{5}+m_{_{_{\overline{\chi
}_{k}^{o}}}})(N_{hi}-N_{hi}^{^{\prime }}\gamma _{5})u(p_{1})
\end{eqnarray*}

Where $k=1,2,3,4.$

and $h=L,R.$

\begin{eqnarray*}
\mathcal{M}_{11,12,13,14,20,21,22,23} &=&\overline{v}%
(p_{2})(N_{hi}+N_{hi}^{^{\prime }}\gamma _{5})(\NEG{p}_{4}+\NEG%
{p}_{5}+m_{_{_{\overline{\chi }_{k}^{o}}}})D_{_{\overline{\chi }%
_{k}^{o}}}(p_{4}+p_{5})\overline{u}(p_{4})(Q_{ij}^{L}P_{L}+Q_{ij}^{R}P_{R})
\notag \\
&&D_{\widetilde{e}_{h}}(p_{1}-p_{3})\overline{u}(p_{3})(N_{hi}-N_{hi}^{^{%
\prime }}\gamma _{5})u(p_{1})
\end{eqnarray*}

again, $k=1,2,3,4.$

and $h=L,R.$

\begin{eqnarray*}
\mathcal{M}_{27,28} &=&\overline{v}%
(p_{2})(T_{i}^{L}P_{L}+T_{i}^{R}P_{R})v(p_{4})D_{\widetilde{\nu }%
_{e}}(p_{2}-p_{4})\overline{u}(p_{3})(Q_{ij}^{L}P_{L}+Q_{ij}^{R}P_{R})(\NEG%
{p}_{3}+\NEG{p}_{5}+m_{_{_{\overline{\chi }_{k}^{+}}}})  \notag \\
&&D_{\widetilde{\chi }%
_{k}^{+}}(p_{3}+p_{5})(T_{i}^{L}P_{L}+T_{i}^{R}P_{R})u(p_{1})
\end{eqnarray*}

here $k=1,2.$

\begin{eqnarray*}
\mathcal{M}_{8} &=&\overline{v}(p_{2})T_{i}^{R}P_{R}v(p_{4})D_{\widetilde{e}%
_{L}}(p_{2}-p_{4})L(p_{1}+p_{2}-p_{3}-p_{4})D_{\widetilde{\nu }%
_{e}}(p_{1}-p_{3})  \notag \\
&&\overline{v}(p_{3})(T_{i}^{L}P_{L}+T_{i}^{R}P_{R})u(p_{1})
\end{eqnarray*}

Where:

$D_{X}(q)=\frac{1}{(q^{2}-m_{X}^{2})},$

$P_{Z,\gamma }^{\mu \nu }=\frac{-g^{\mu \nu }+\frac{q^{\mu }q^{\nu }}{%
m_{Z,\gamma }^{2}}}{q^{2}-m_{Z,\gamma }^{2}+im_{Z,\gamma }\Gamma
_{Z}}.$

\noindent For the definitions of the constants used here, the
reader is referred to Appendix A.

\section{Cross Sections}

To be able to calculate the differential cross sections, and
hence, the total cross section, we need first to obtain the
squared matrix element for each Feynman diagram, where use of the
trace theorems was made. Later an average over the initial spin
polarizations of the electron and the positron pair and the sum
over the final spin states of the outgoing particles arising from
each initial spin state is carried out. The total cross section as
a function of the center of mass energy (see Appendix
B) is then calculated. The calculations were done for the following cases:\\
$tan\beta = 05$, $tan\beta = 15$ and $tan\beta = 35$ where $M_2 =
150$ or $M_2 = 300$ for each case of $tan\beta$. All results are
given in the following figures.
\begin{figure}[th]
\vspace{-4.5cm}
\centerline{\epsfxsize=4.3truein\epsfbox{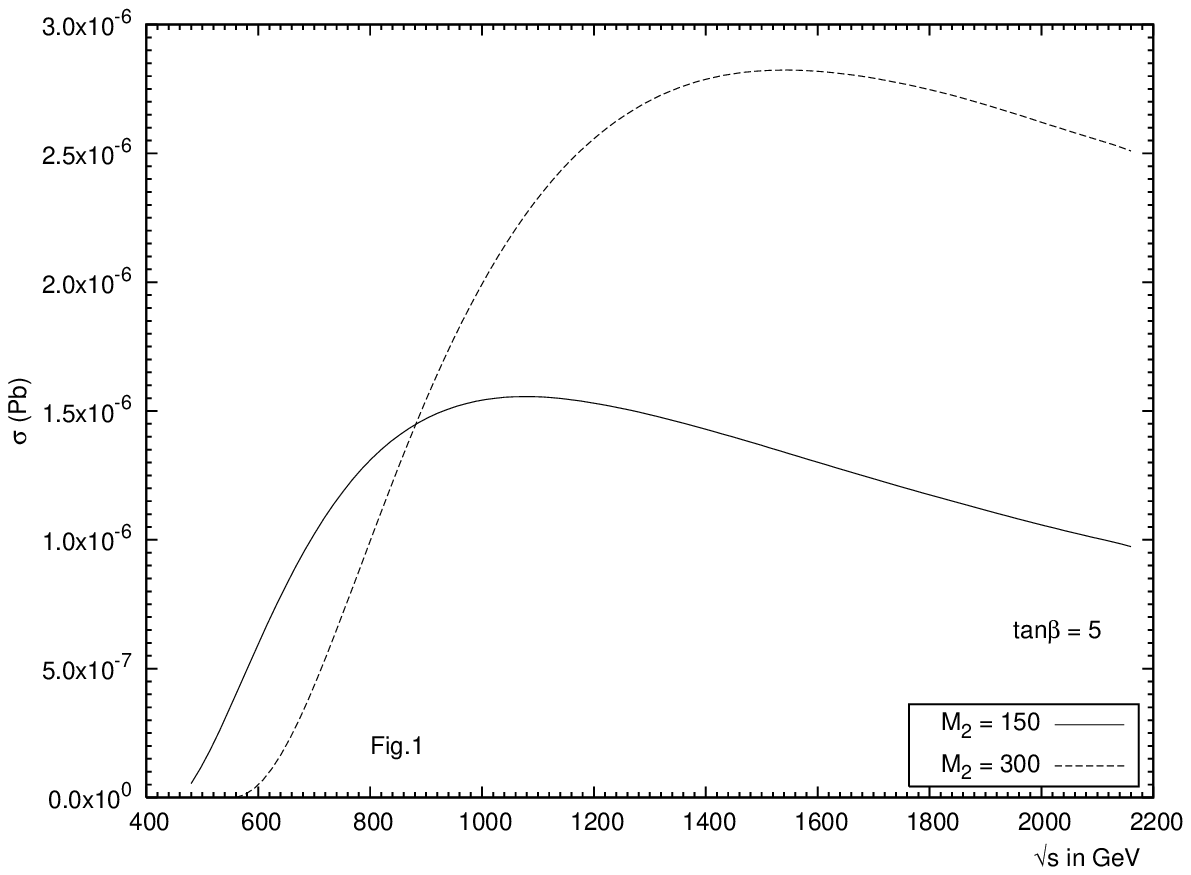}}
\vspace{-0.1cm}
\centerline{\epsfxsize=4.3truein\epsfbox{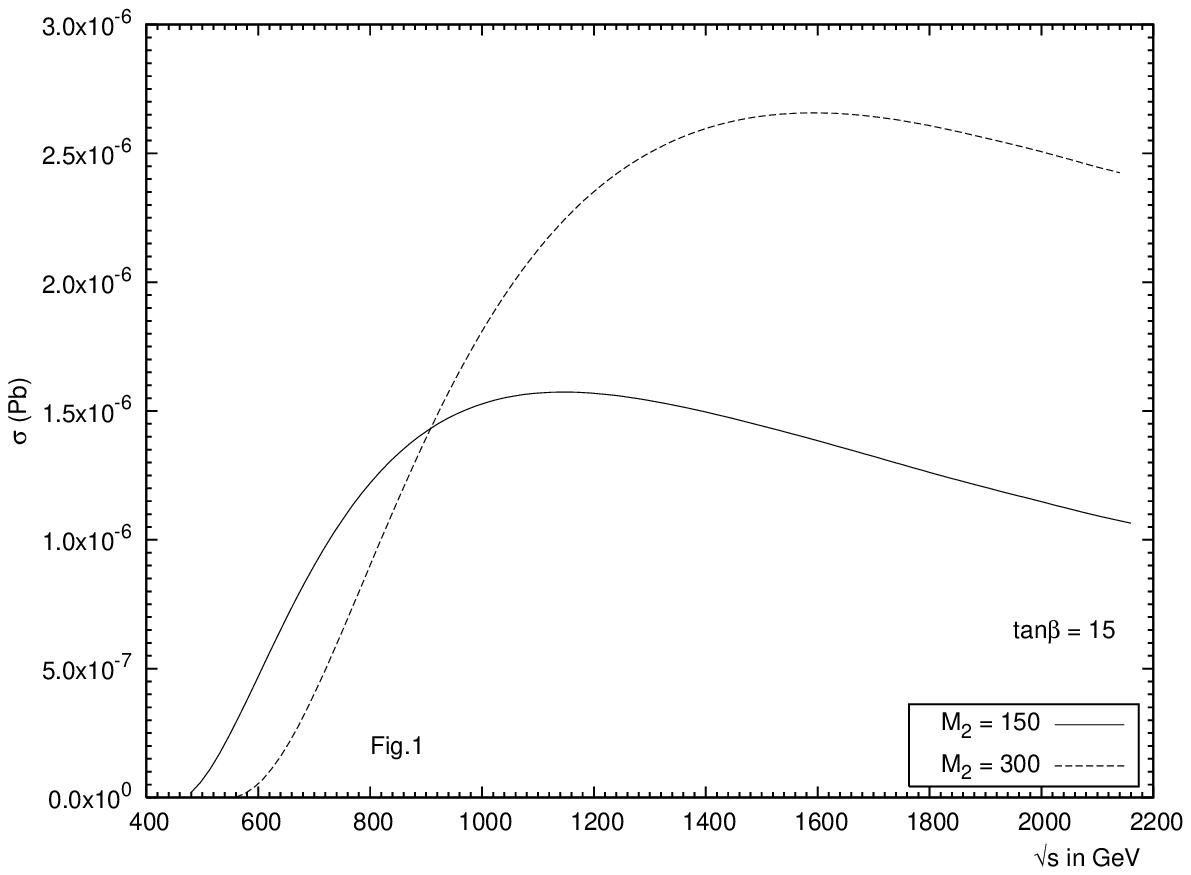}}
\vspace{-0.1cm}
\centerline{\epsfxsize=4.3truein\epsfbox{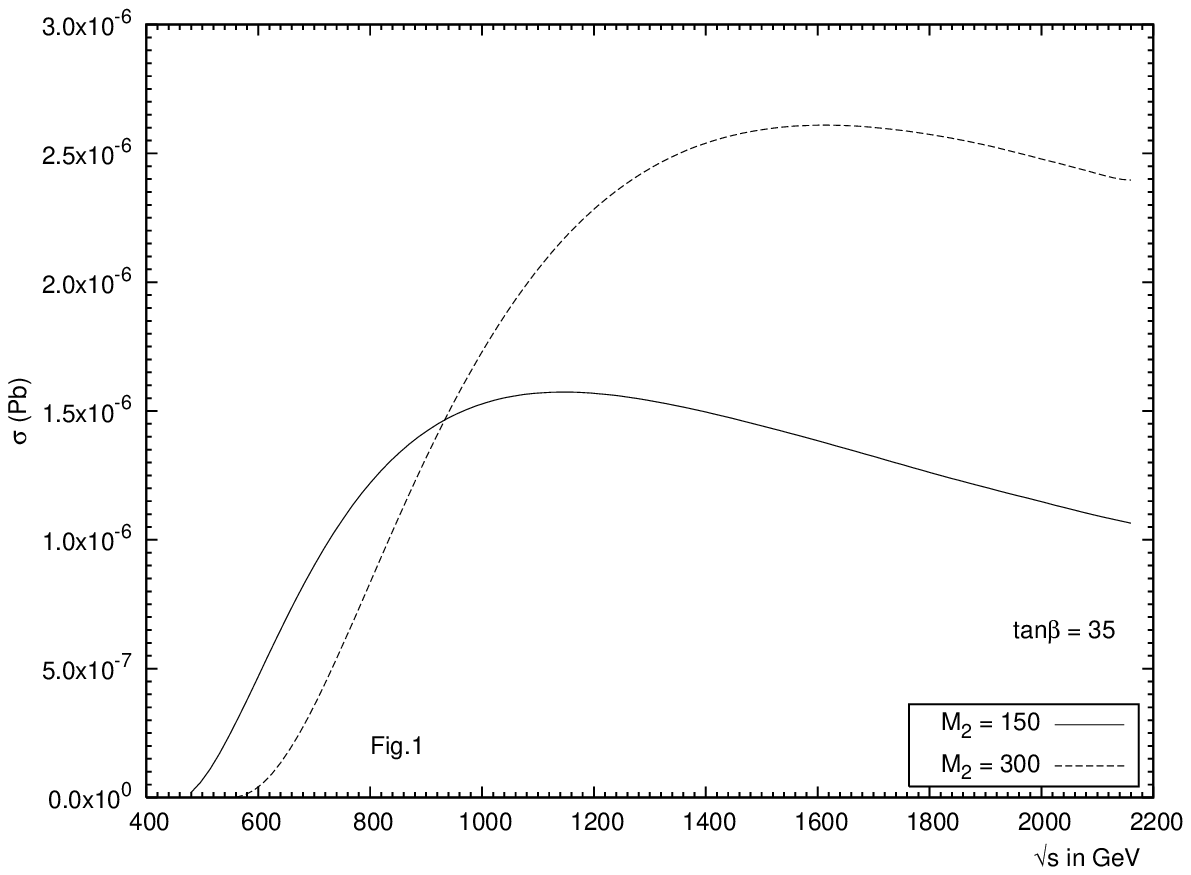}}
\vspace{0.5cm} \caption{ \small Cross sections for
diagram no. 1 in figure \ref{feyn1}} \label{fig.1}
\end{figure}

\begin{figure}[th]
\vspace{-4.5cm}
\centerline{\epsfxsize=4.3truein\epsfbox{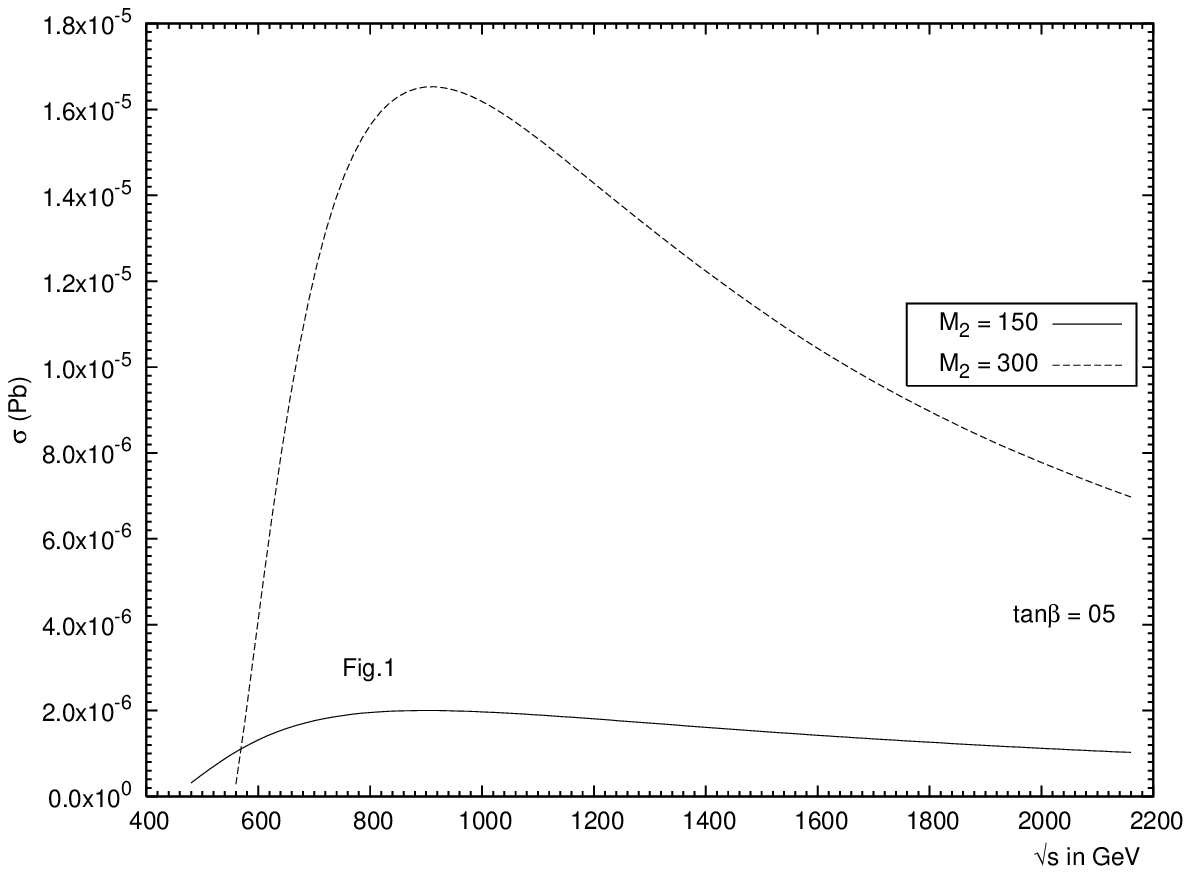}}
\vspace{-0.1cm}
\centerline{\epsfxsize=4.3truein\epsfbox{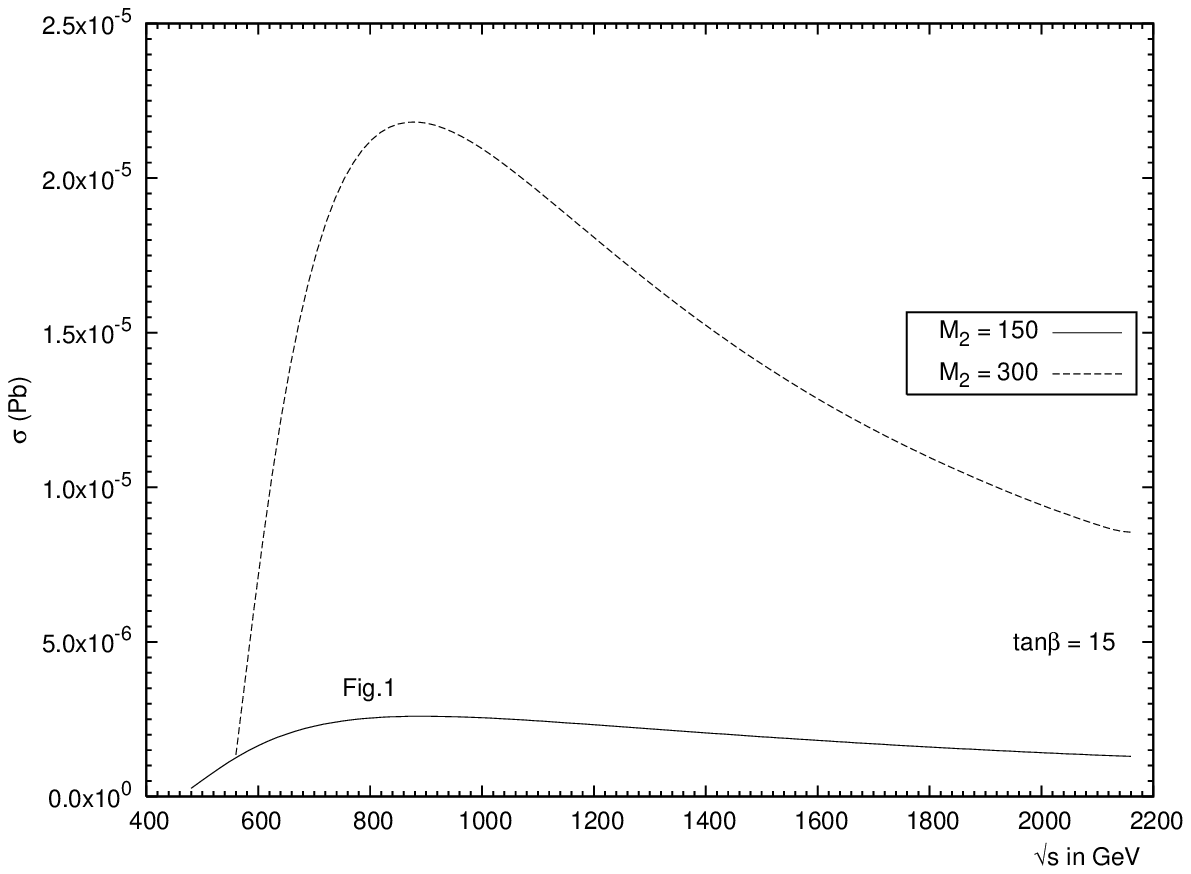}}
\vspace{-0.1cm}
\centerline{\epsfxsize=4.3truein\epsfbox{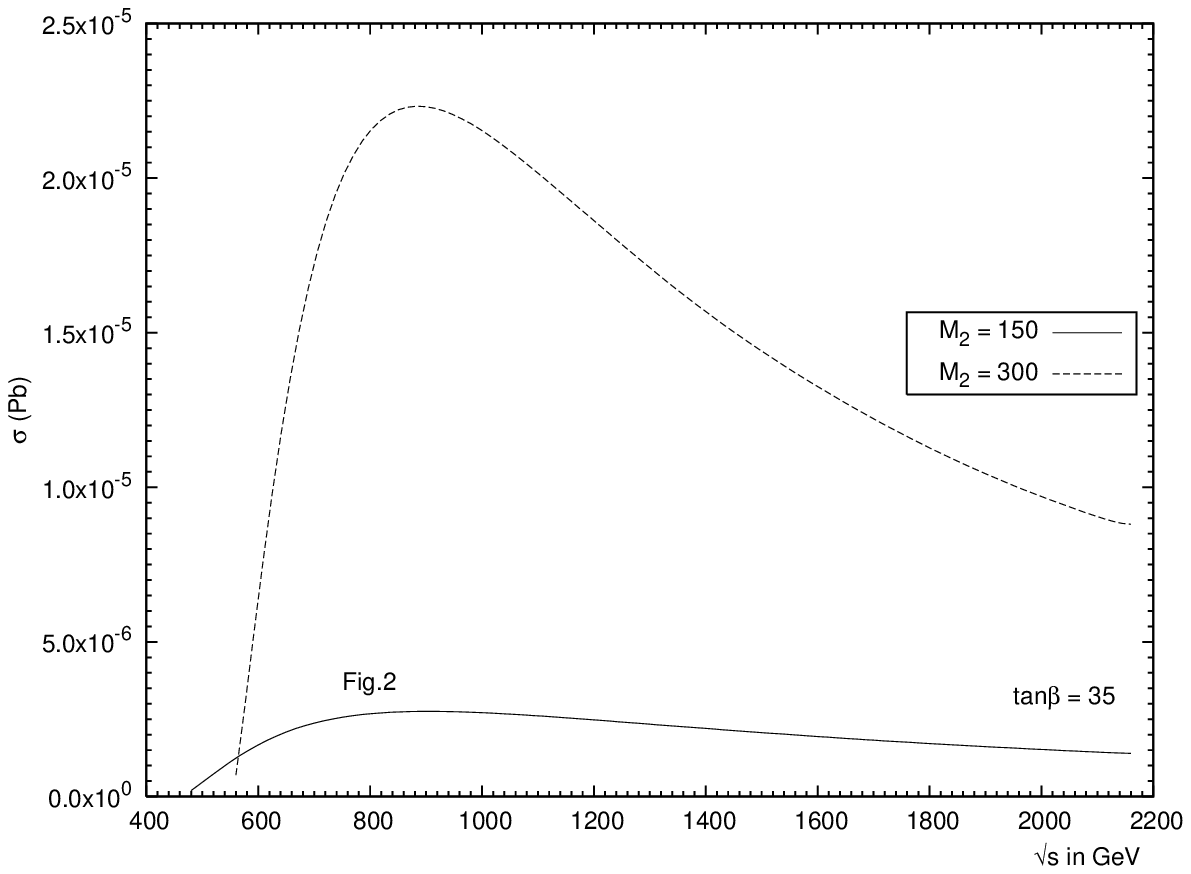}}
\vspace{0.5cm} \caption{ \small Cross sections for
diagram no. 2 in figure \ref{feyn1}} \label{fig.2}
\end{figure}

\begin{figure}[th]
\vspace{-4.5cm}
\centerline{\epsfxsize=4.3truein\epsfbox{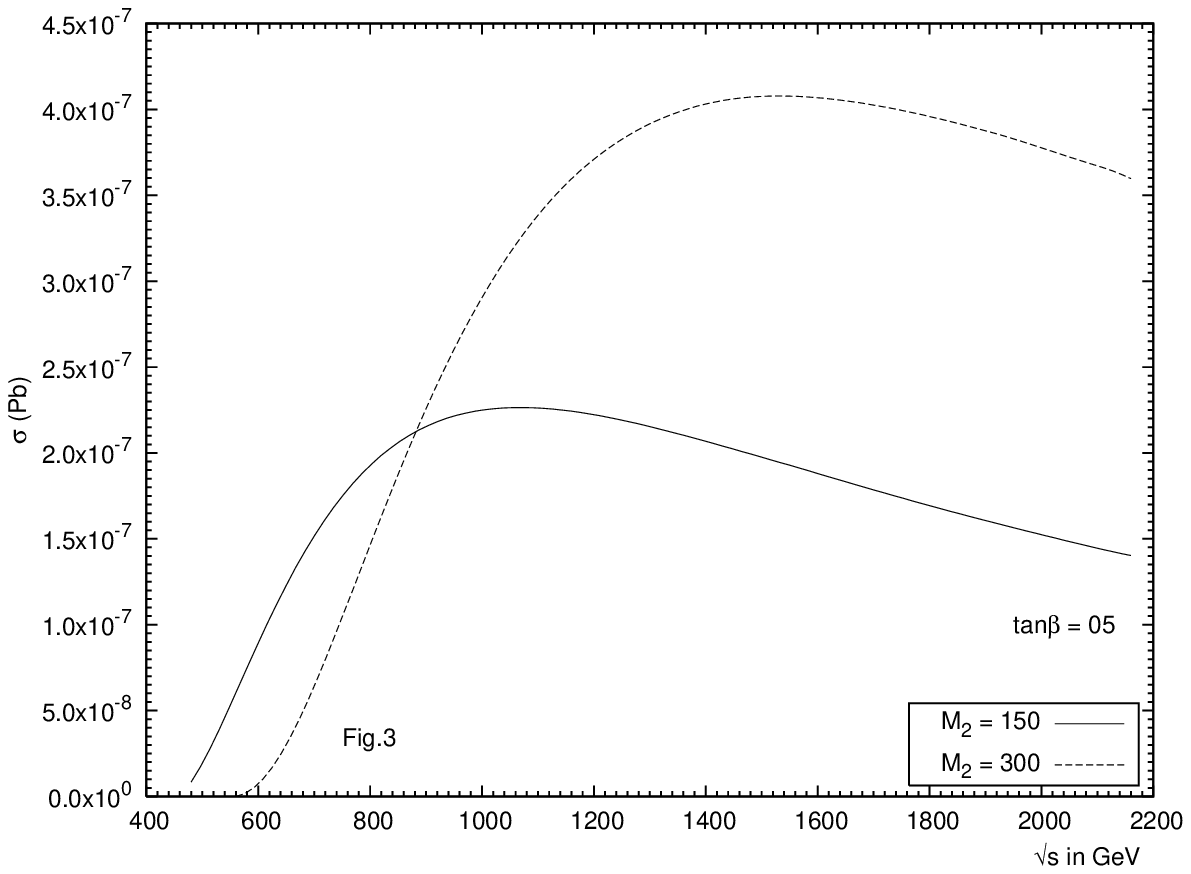}}
\vspace{-0.1cm}
\centerline{\epsfxsize=4.3truein\epsfbox{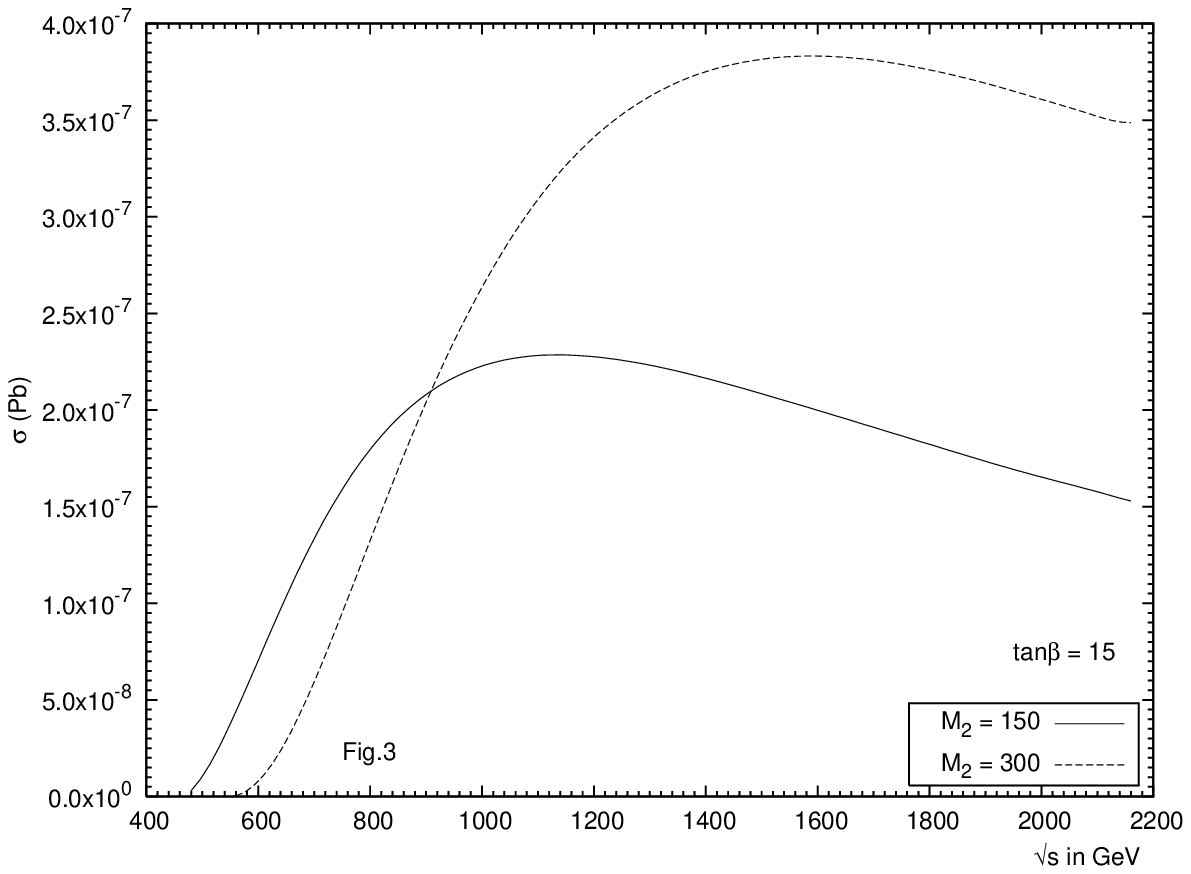}}
\vspace{-0.1cm}
\centerline{\epsfxsize=4.3truein\epsfbox{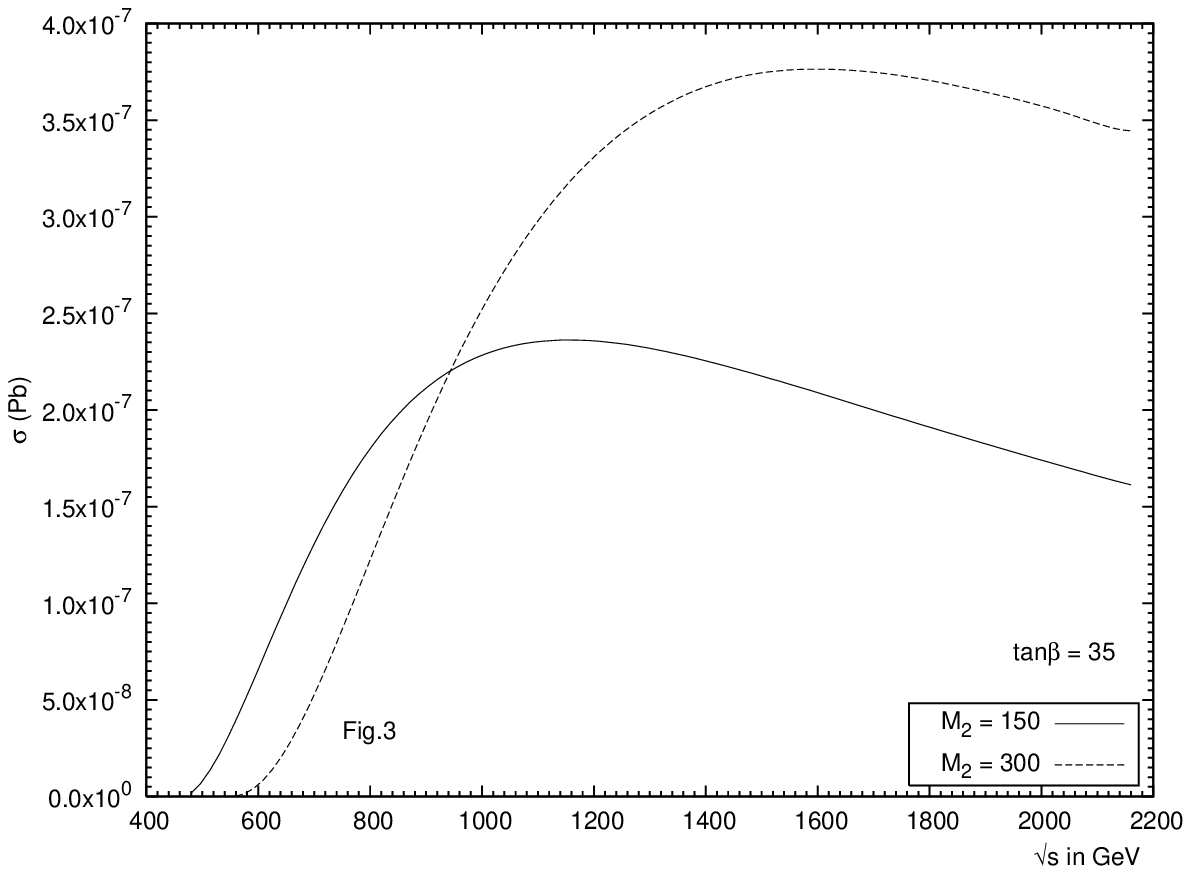}}
\vspace{0.5cm} \caption{ \small Cross sections for
diagram no. 3 in figure \ref{feyn1}} \label{fig.3}
\end{figure}

\begin{figure}[th]
\vspace{-4.5cm}
\centerline{\epsfxsize=4.3truein\epsfbox{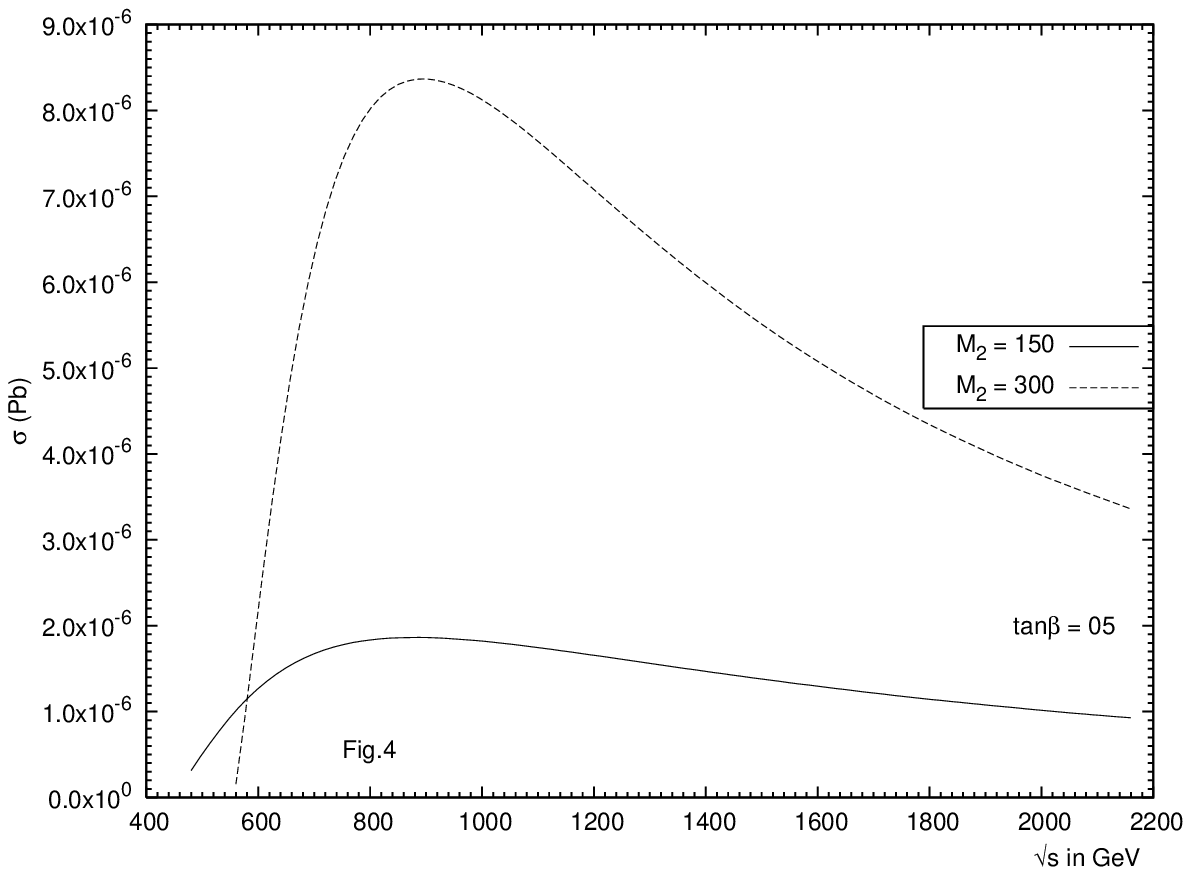}}
\vspace{-0.1cm}
\centerline{\epsfxsize=4.3truein\epsfbox{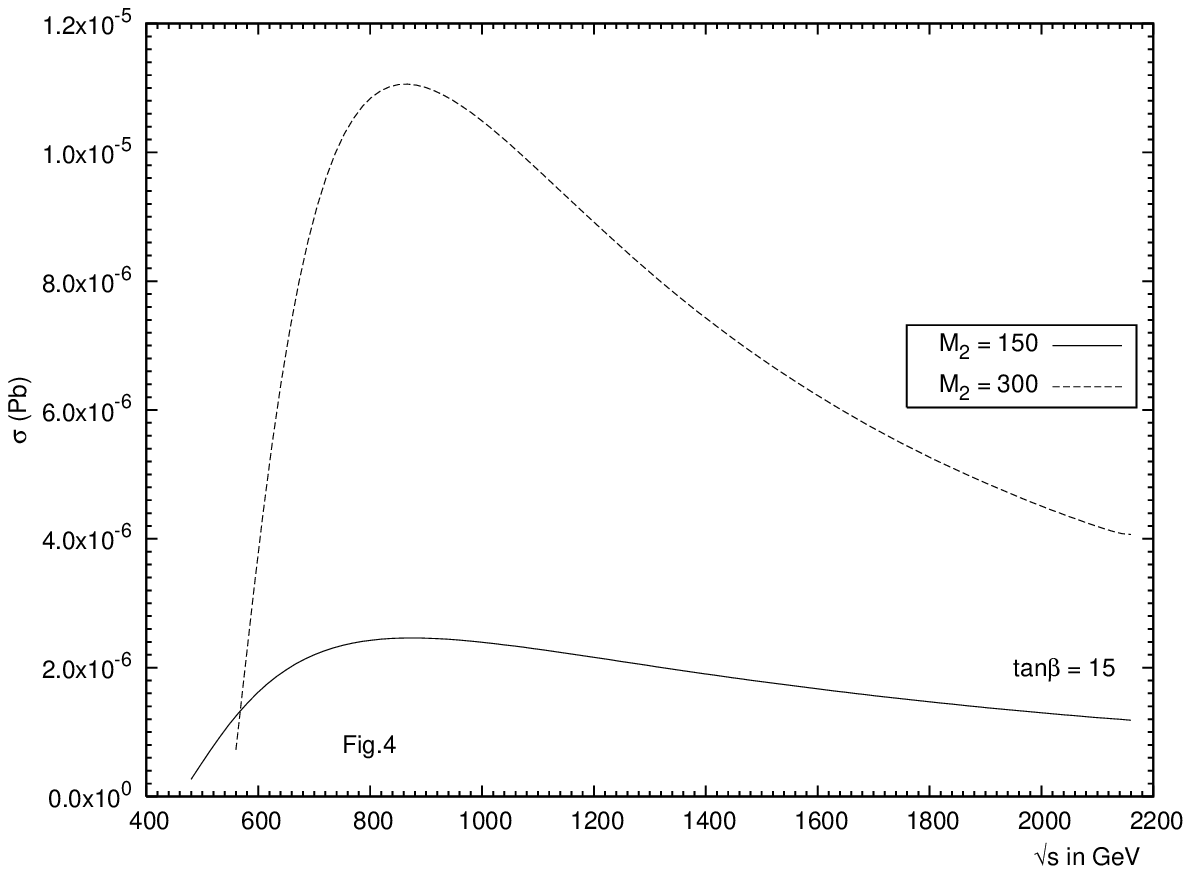}}
\vspace{-0.1cm}
\centerline{\epsfxsize=4.3truein\epsfbox{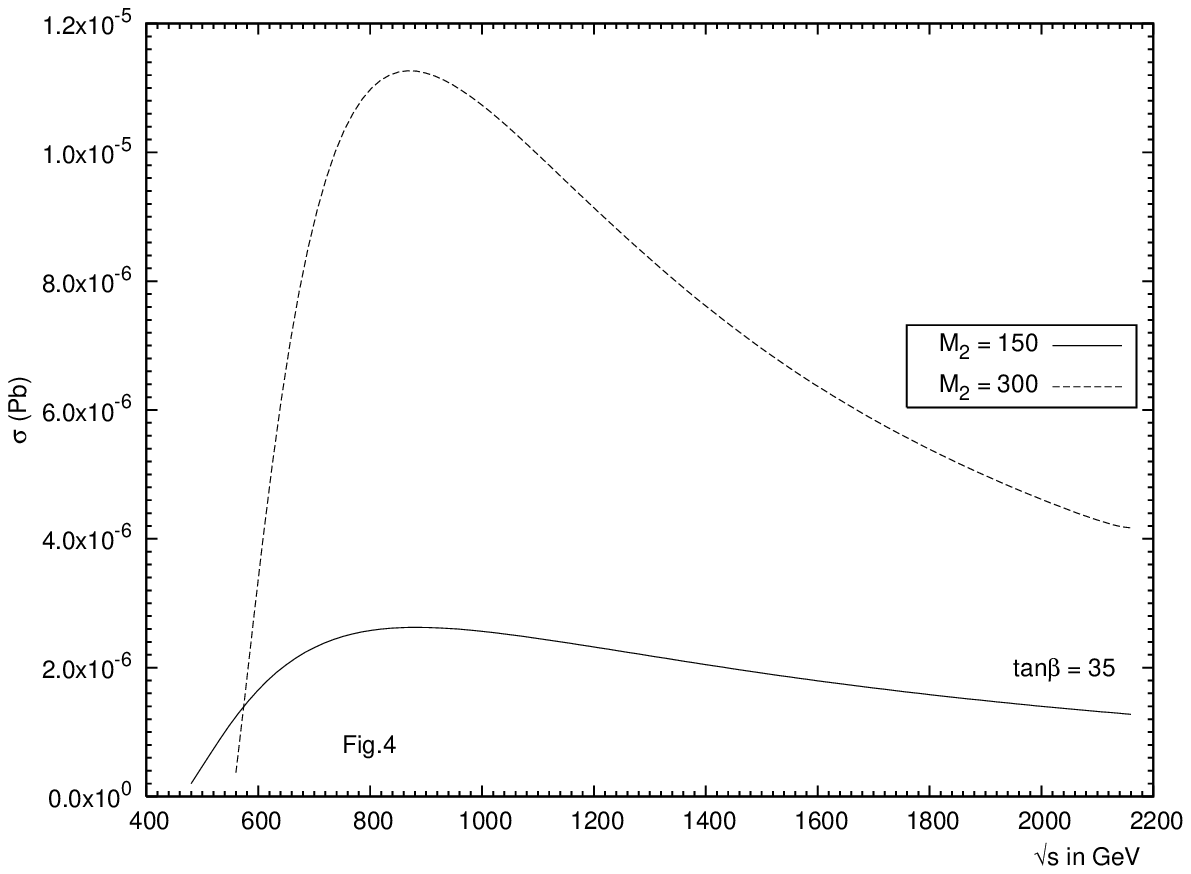}}
\vspace{0.5cm} \caption{ \small Cross sections for
diagram no. 4 in figure \ref{feyn1}} \label{fig.4}
\end{figure}

\begin{figure}[th]
\vspace{-4.5cm}
\centerline{\epsfxsize=4.3truein\epsfbox{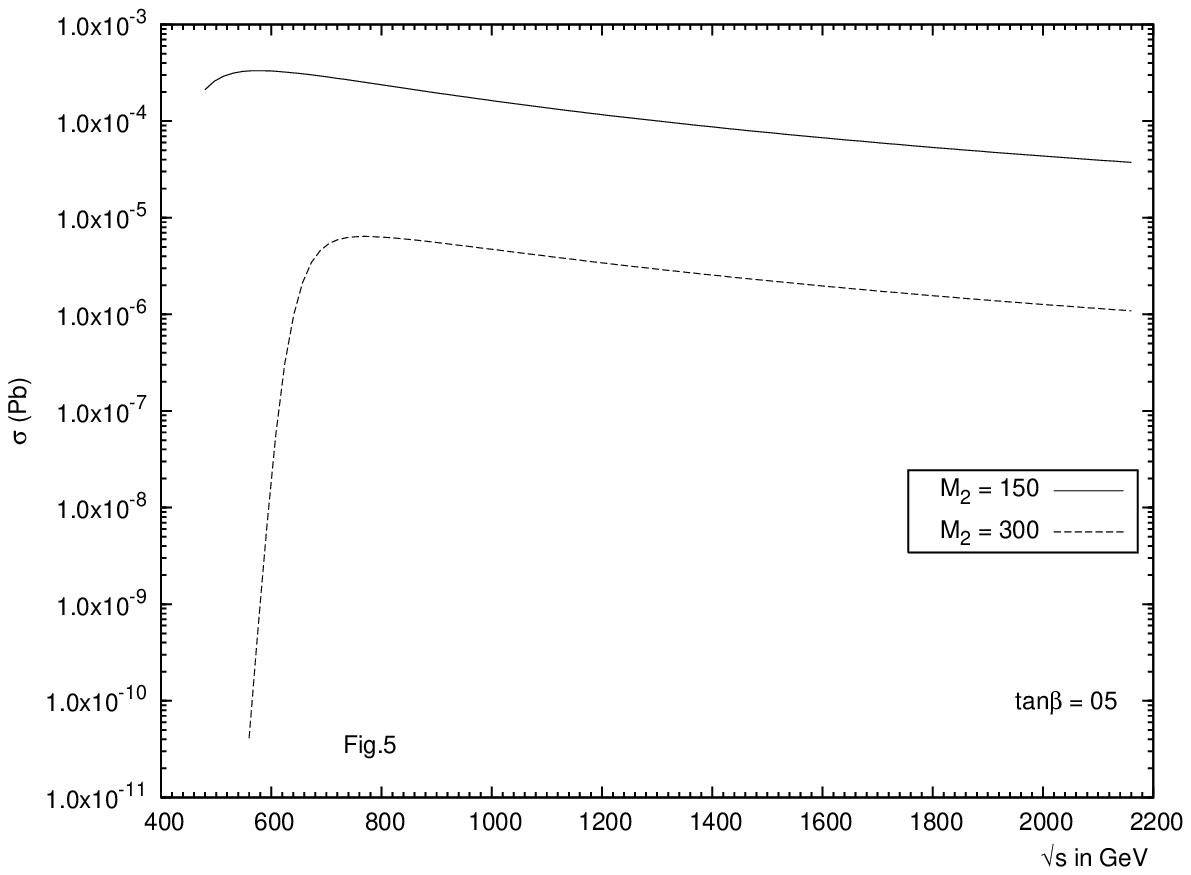}}
\vspace{-0.1cm}
\centerline{\epsfxsize=4.3truein\epsfbox{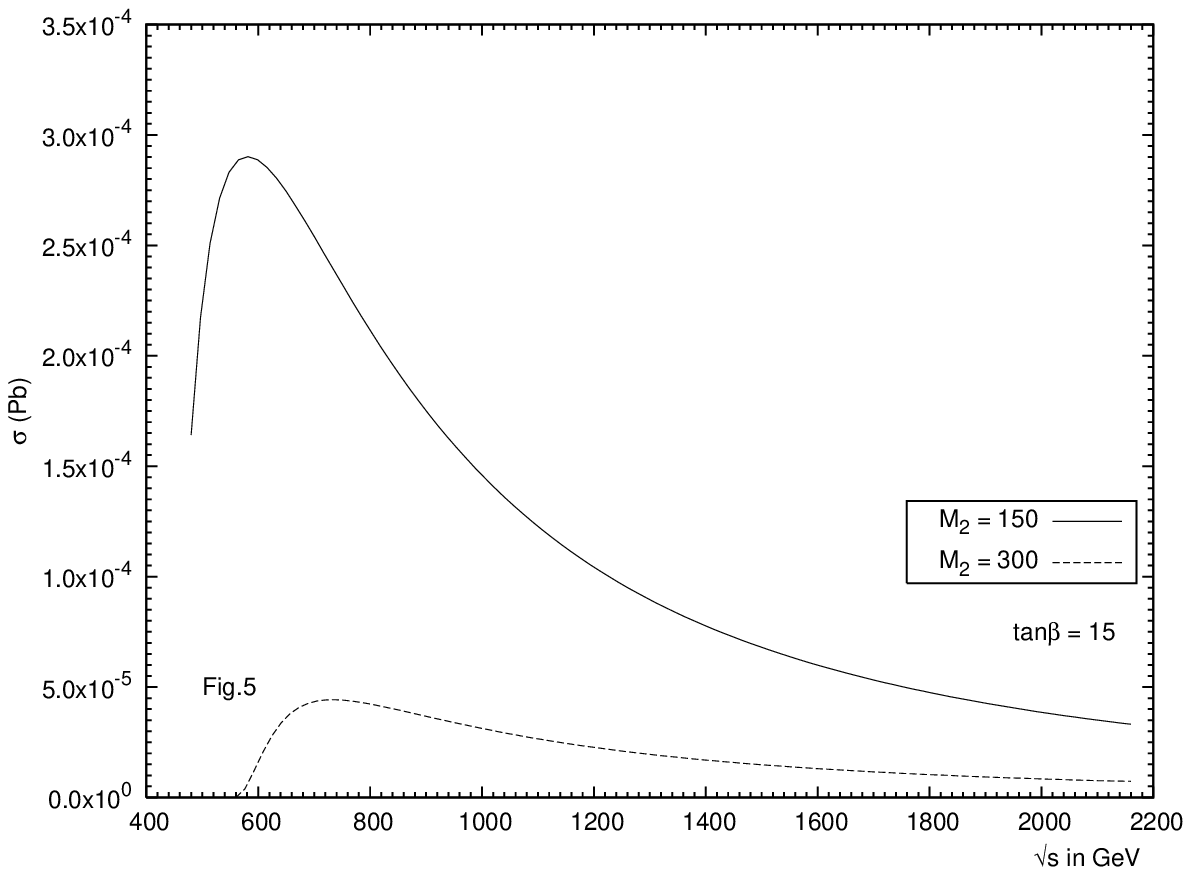}}
\vspace{-0.1cm}
\centerline{\epsfxsize=4.3truein\epsfbox{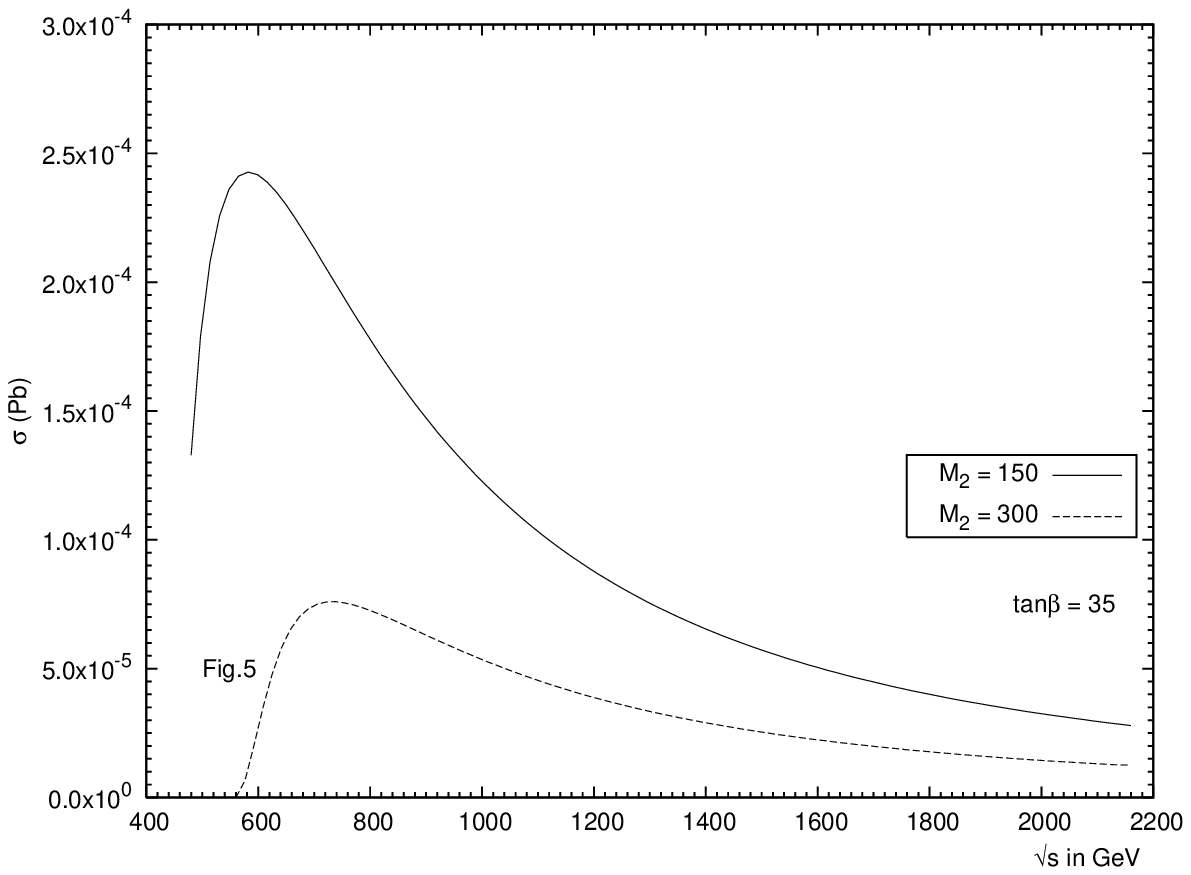}}
\vspace{0.5cm} \caption{ \small Cross sections for
diagram no. 5 in figure \ref{feyn1}} \label{fig.5}
\end{figure}

\begin{figure}[th]
\vspace{-4.5cm}
\centerline{\epsfxsize=4.3truein\epsfbox{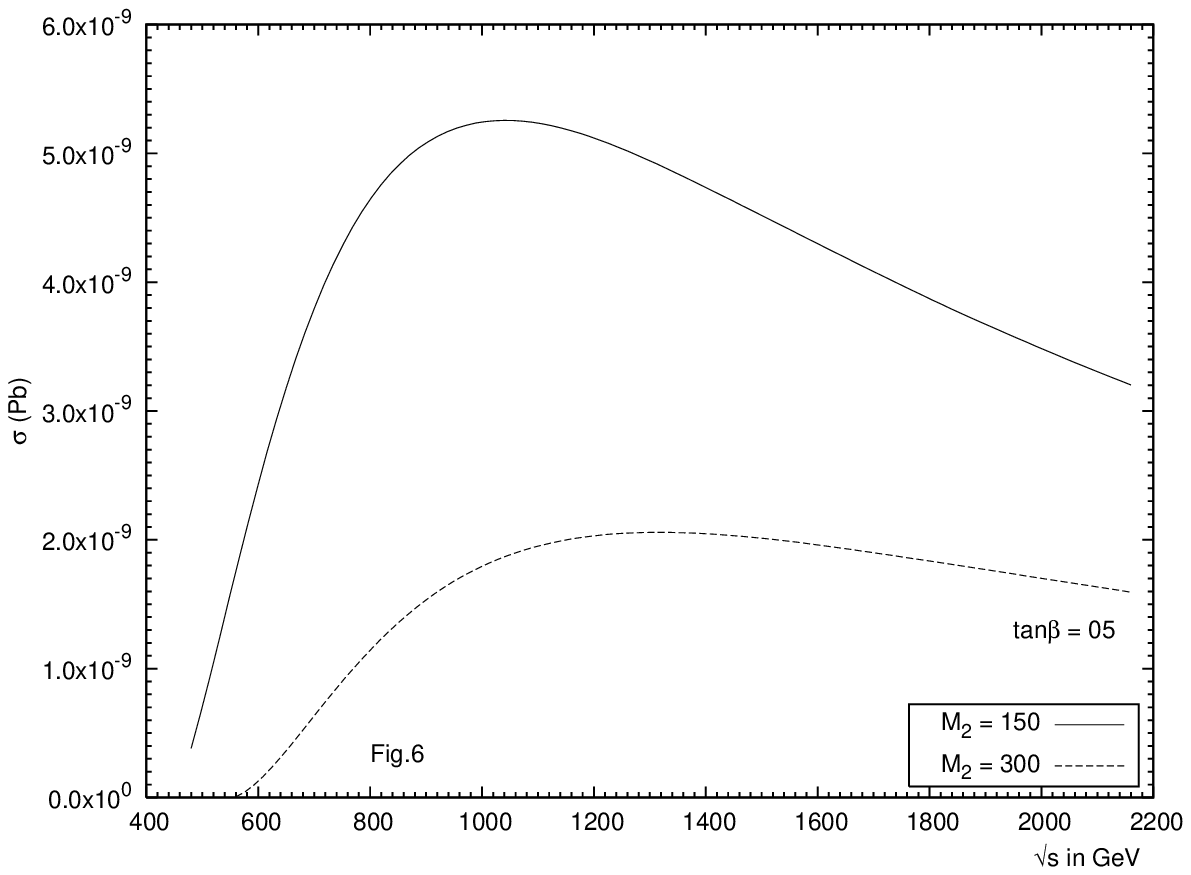}}
\vspace{-0.1cm}
\centerline{\epsfxsize=4.3truein\epsfbox{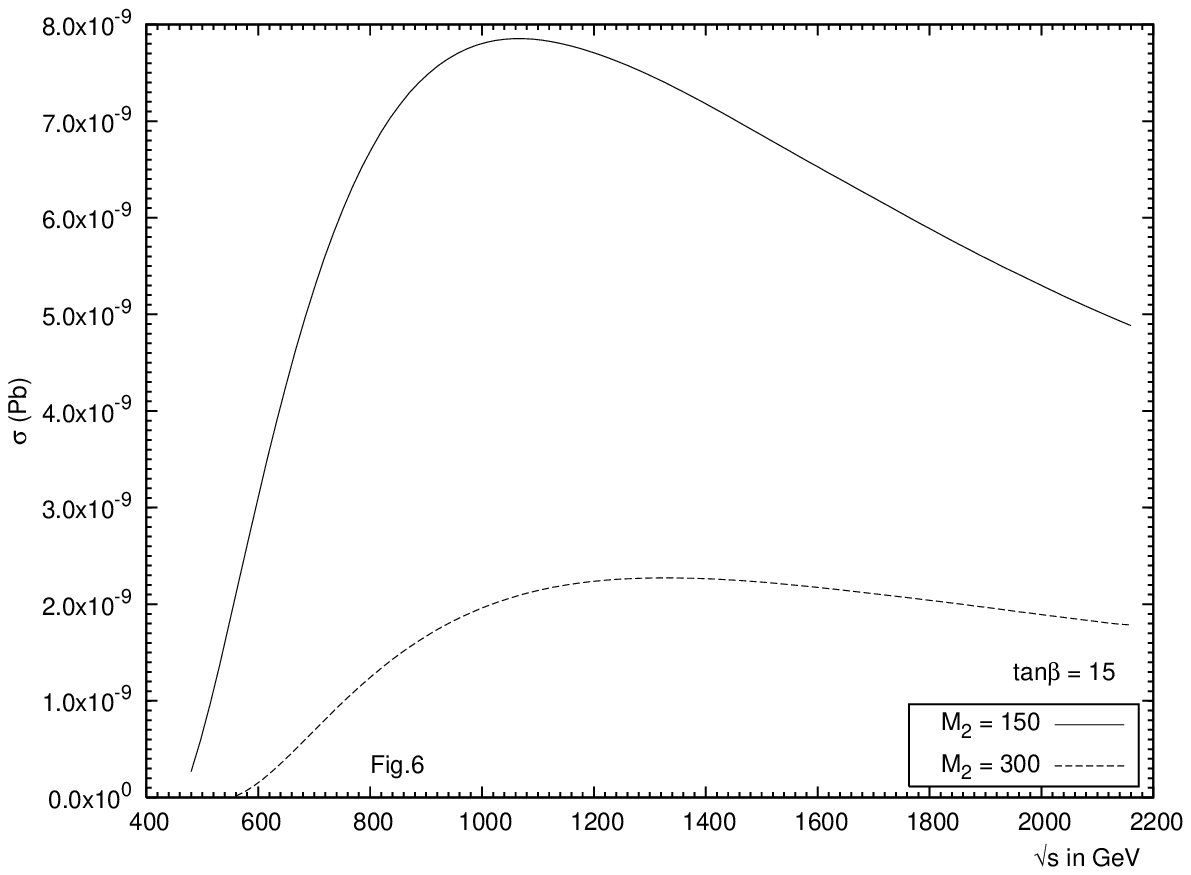}}
\vspace{-0.1cm}
\centerline{\epsfxsize=4.3truein\epsfbox{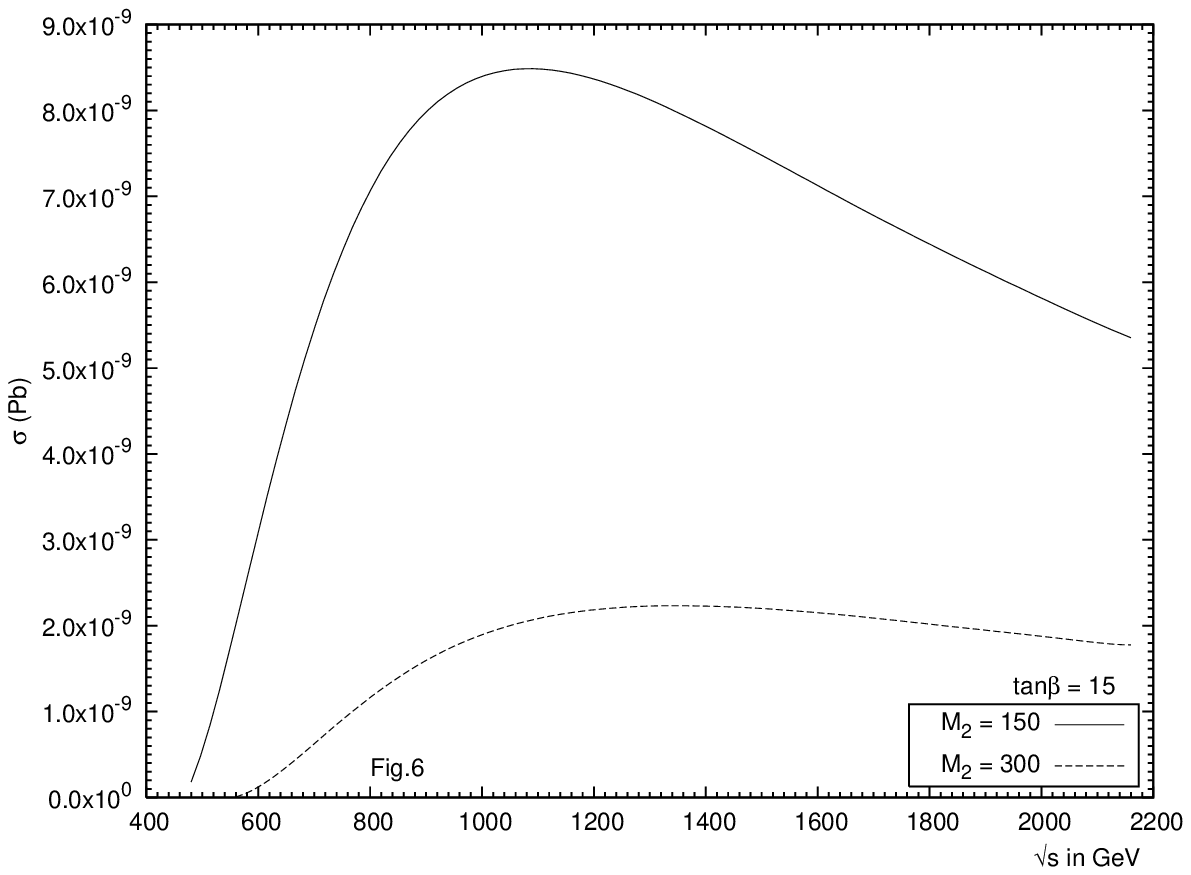}}
\vspace{0.5cm} \caption{ \small Cross sections for
diagram no. 6 in figure \ref{feyn1}} \label{fig.6}
\end{figure}

\begin{figure}[th]
\vspace{-4.5cm}
\centerline{\epsfxsize=4.3truein\epsfbox{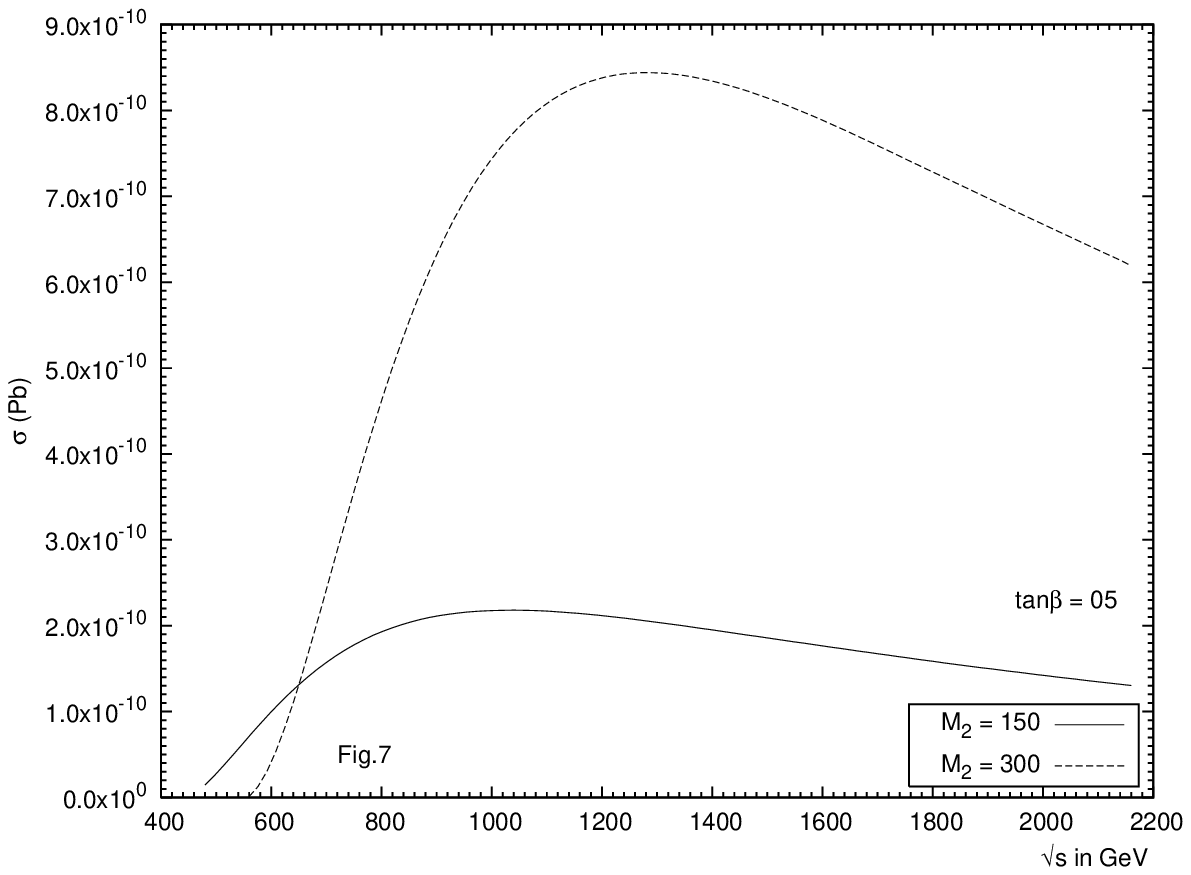}}
\vspace{-0.1cm}
\centerline{\epsfxsize=4.3truein\epsfbox{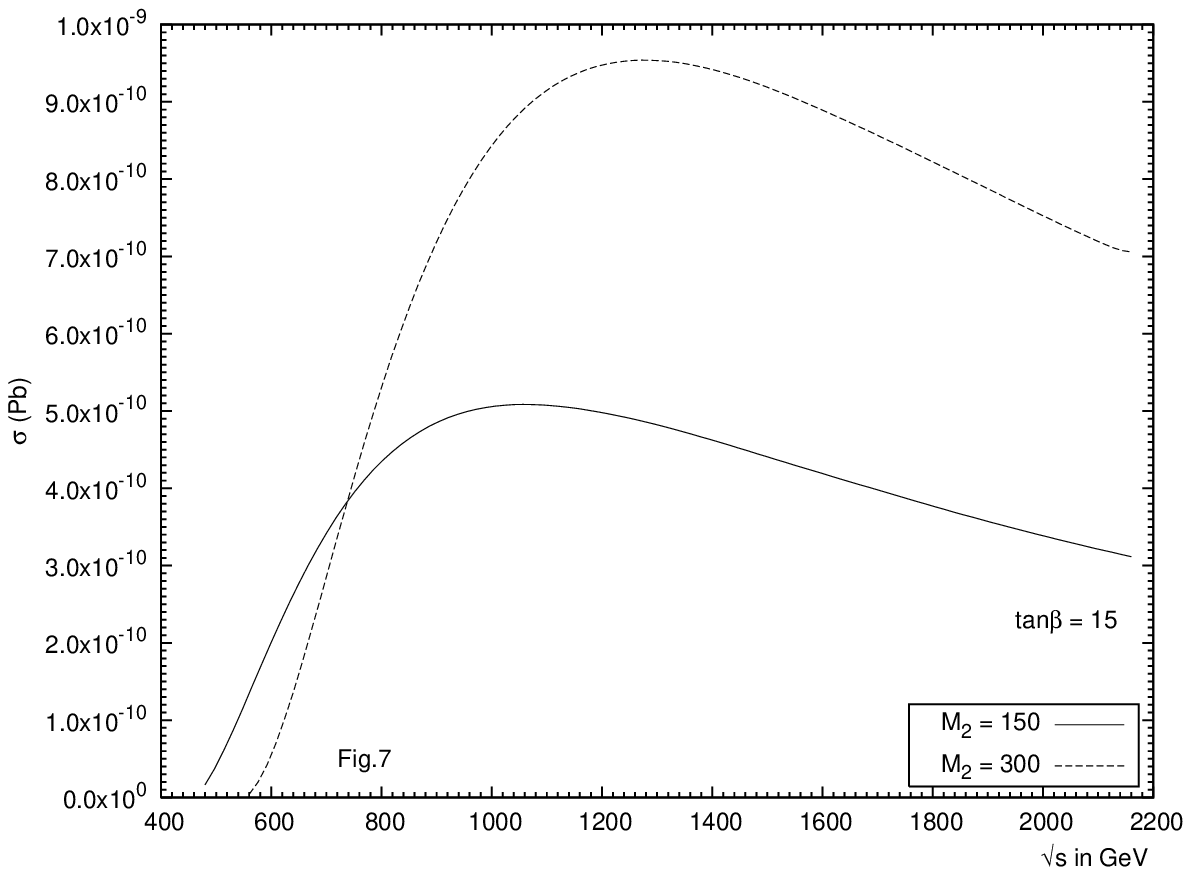}}
\vspace{-0.1cm}
\centerline{\epsfxsize=4.3truein\epsfbox{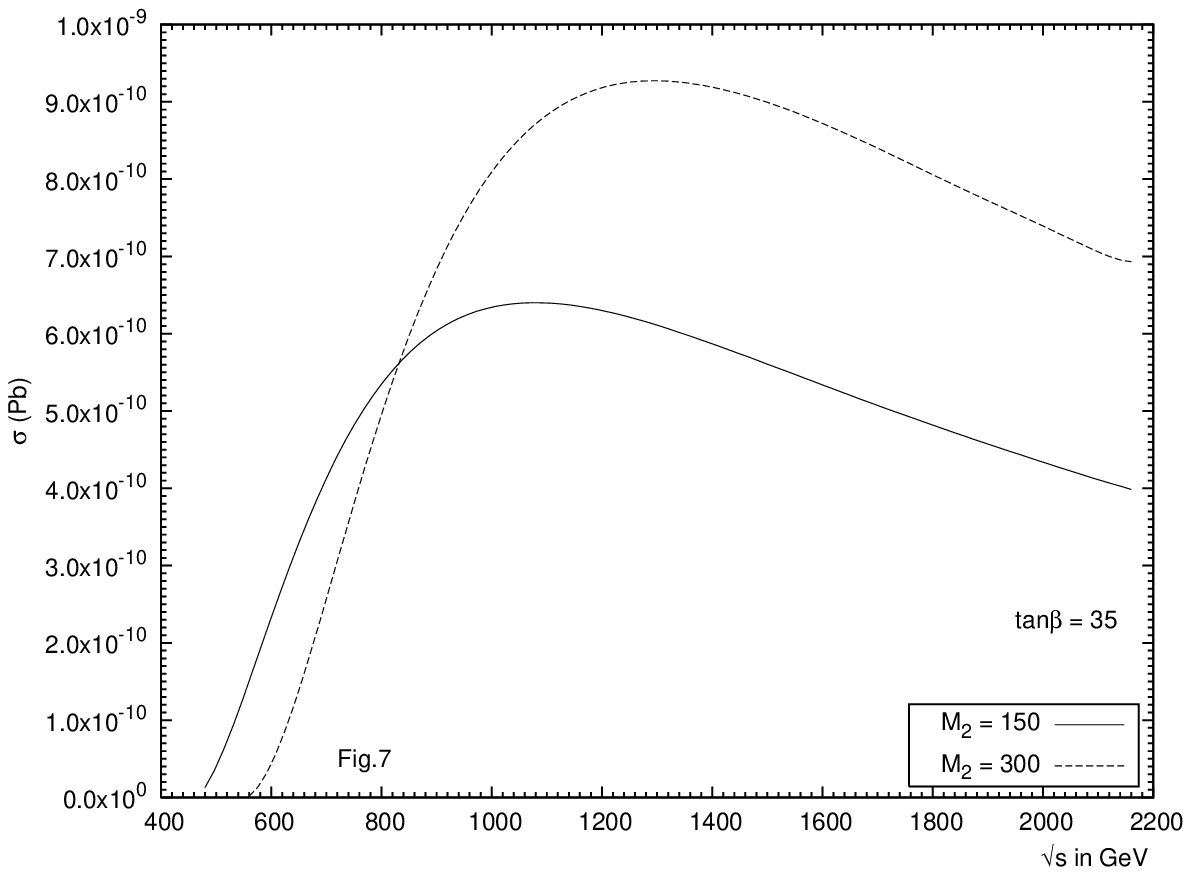}}
\vspace{0.5cm} \caption{ \small Cross sections for
diagram no. 7 in figure \ref{feyn1}} \label{fig.7}
\end{figure}

\begin{figure}[th]
\vspace{-4.5cm}
\centerline{\epsfxsize=4.3truein\epsfbox{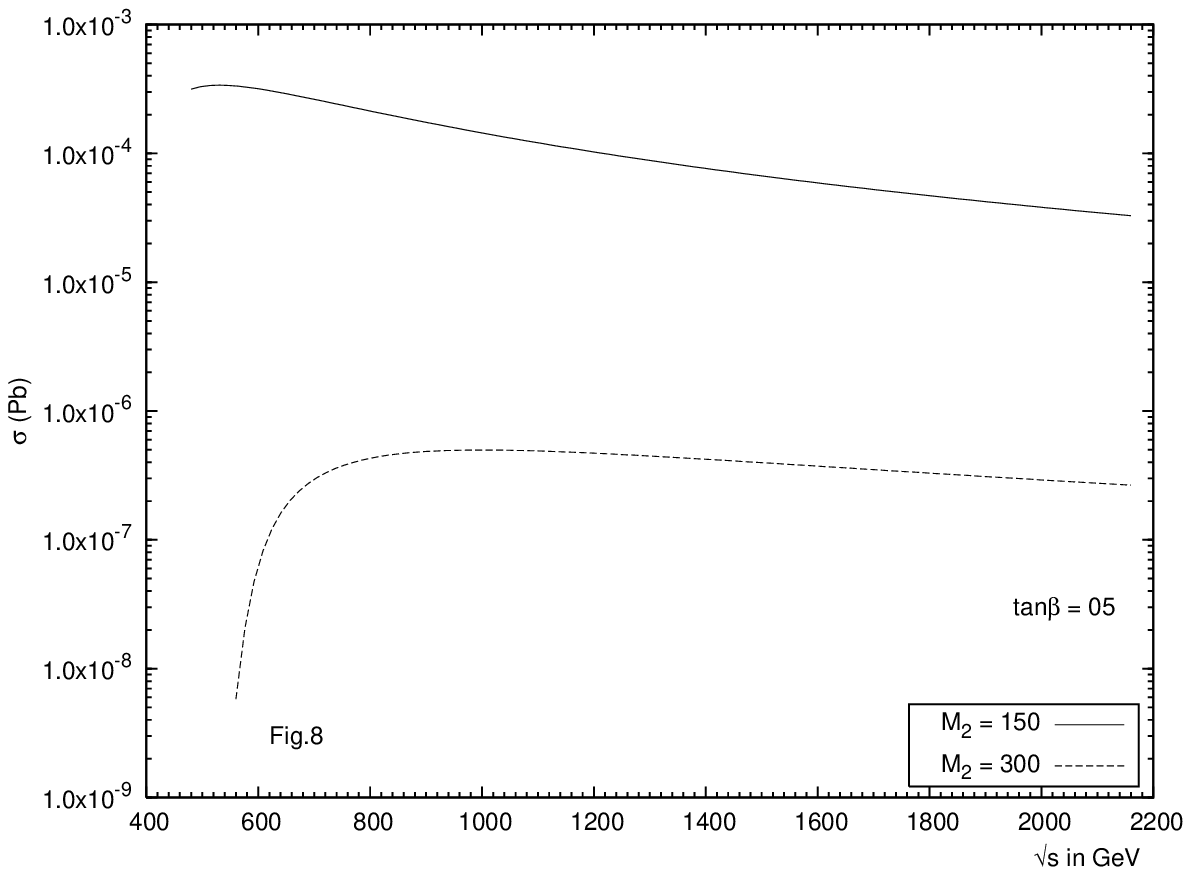}}
\vspace{-0.1cm}
\centerline{\epsfxsize=4.3truein\epsfbox{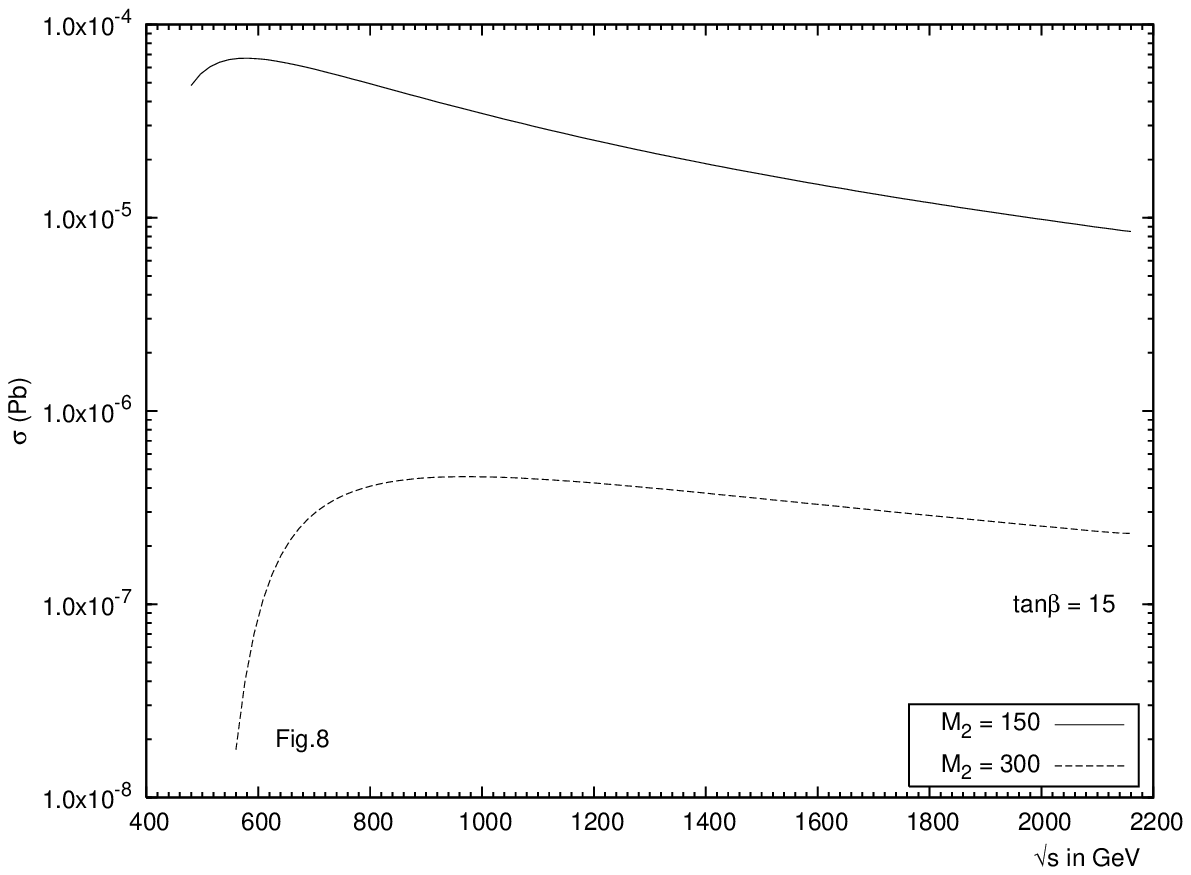}}
\vspace{-0.1cm}
\centerline{\epsfxsize=4.3truein\epsfbox{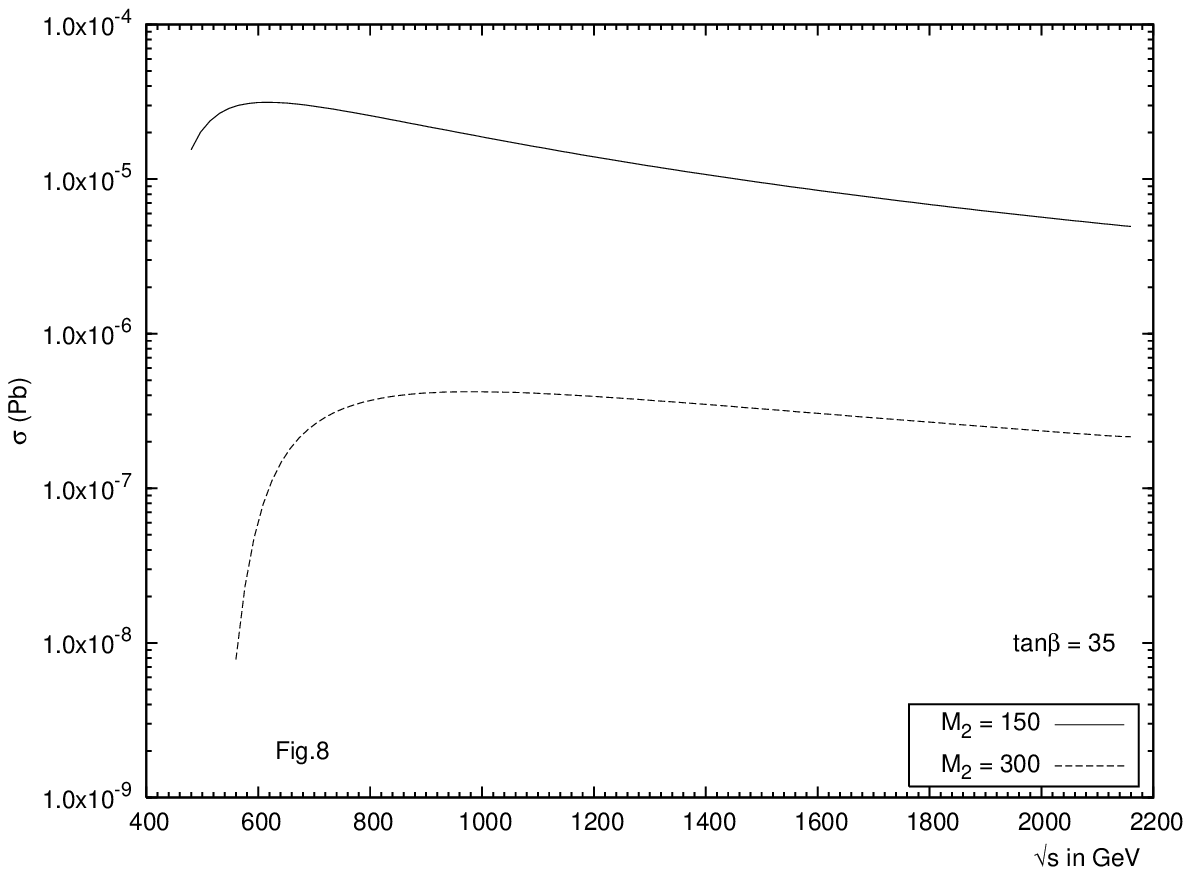}}
\vspace{0.5cm} \caption{ \small Cross sections for
diagram no. 8 in figure \ref{feyn1}} \label{fig.8}
\end{figure}

\begin{figure}[th]
\vspace{-4.5cm}
\centerline{\epsfxsize=4.3truein\epsfbox{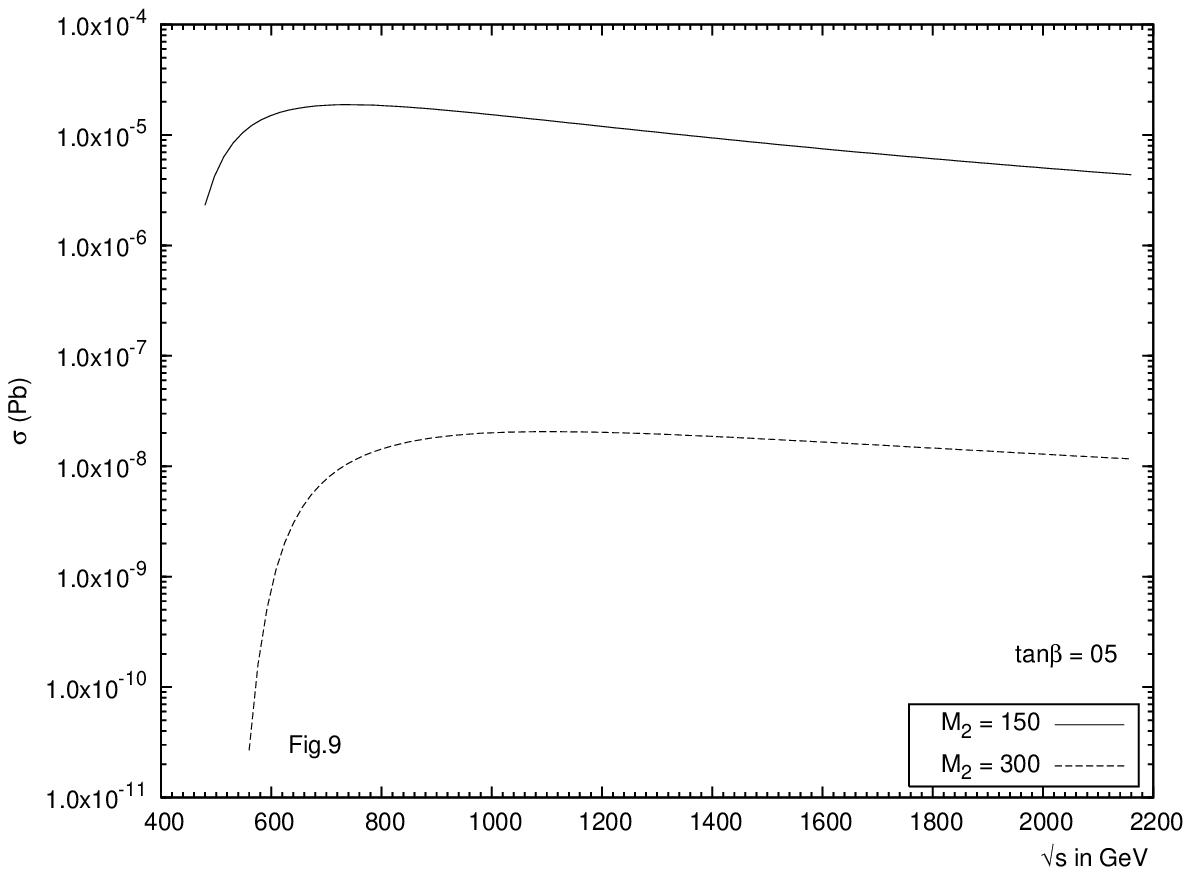}}
\vspace{-0.1cm}
\centerline{\epsfxsize=4.3truein\epsfbox{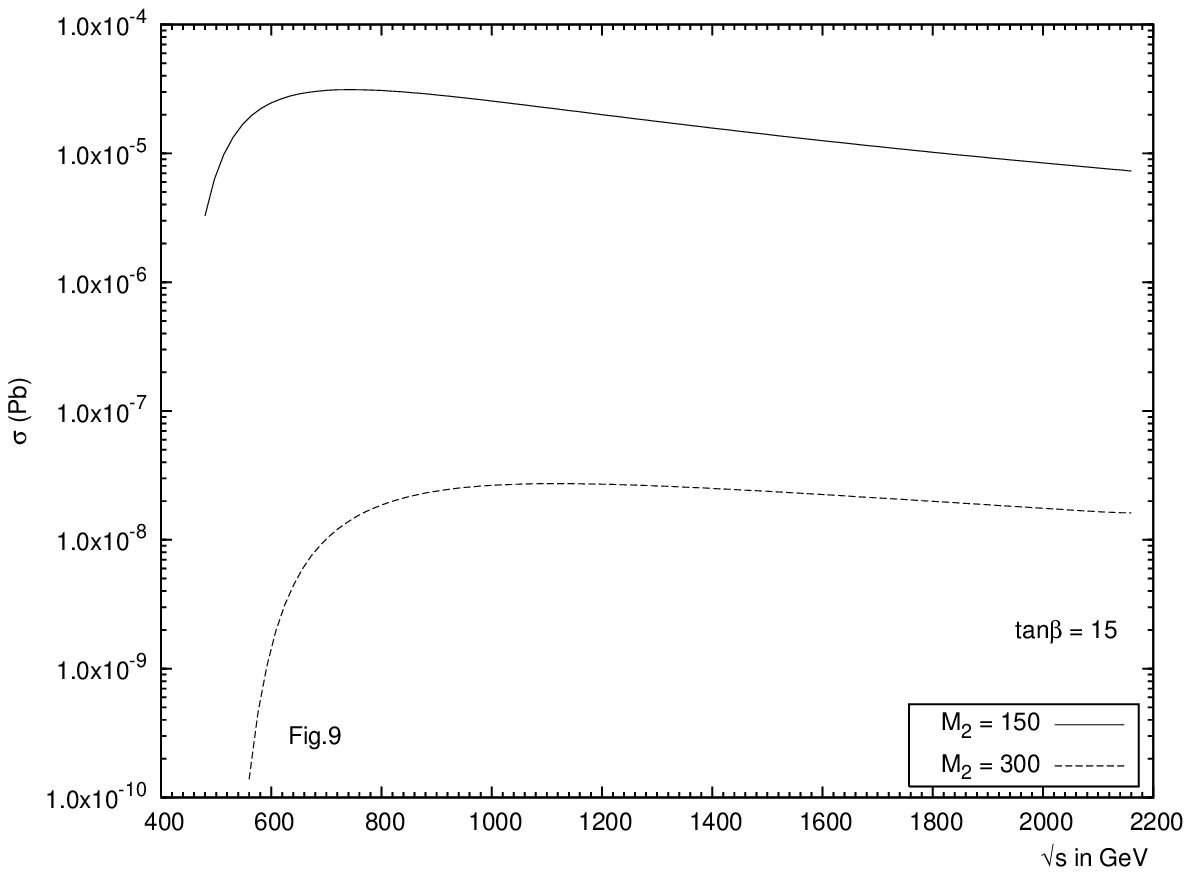}}
\vspace{-0.1cm}
\centerline{\epsfxsize=4.3truein\epsfbox{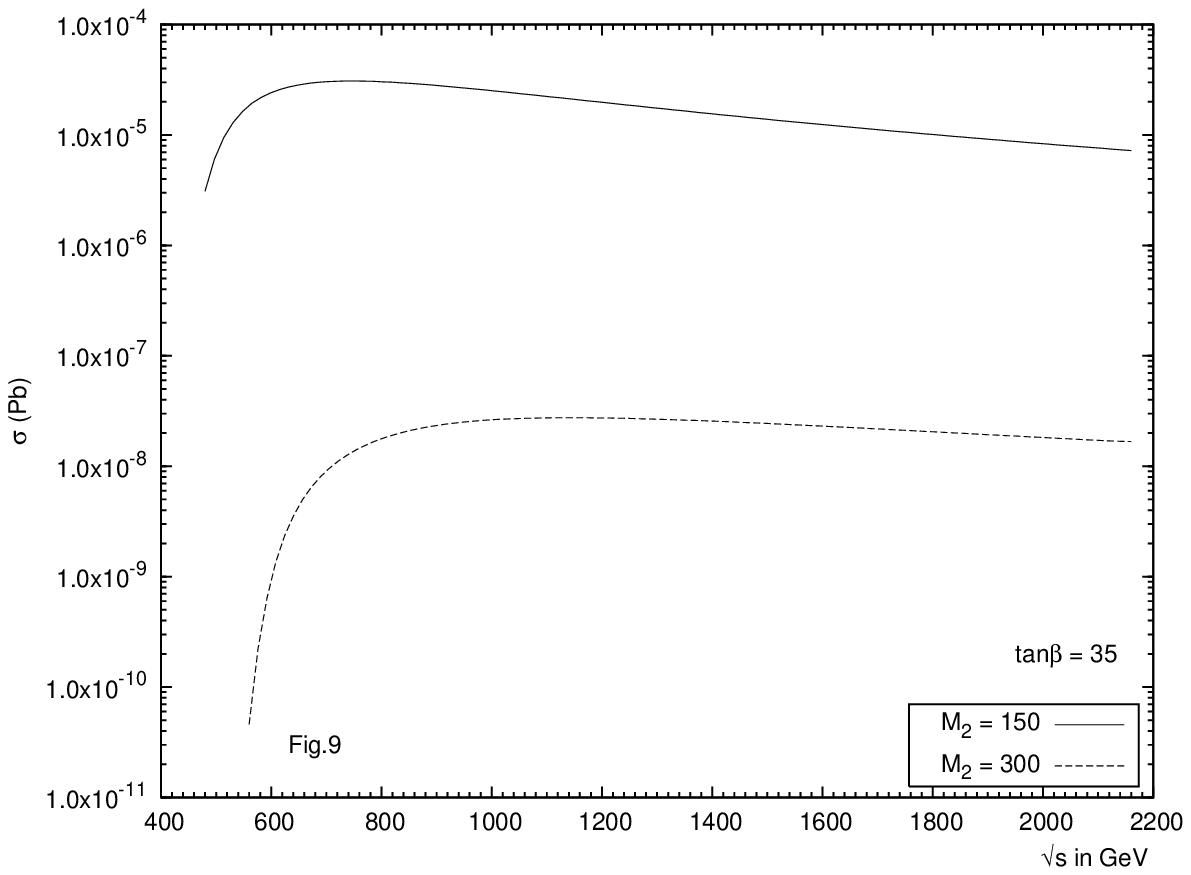}}
\vspace{0.5cm} \caption{ \small Cross sections for
diagram no. 9 in figure \ref{feyn1}} \label{fig.9}
\end{figure}

\begin{figure}[th]
\vspace{-4.5cm}
\centerline{\epsfxsize=4.3truein\epsfbox{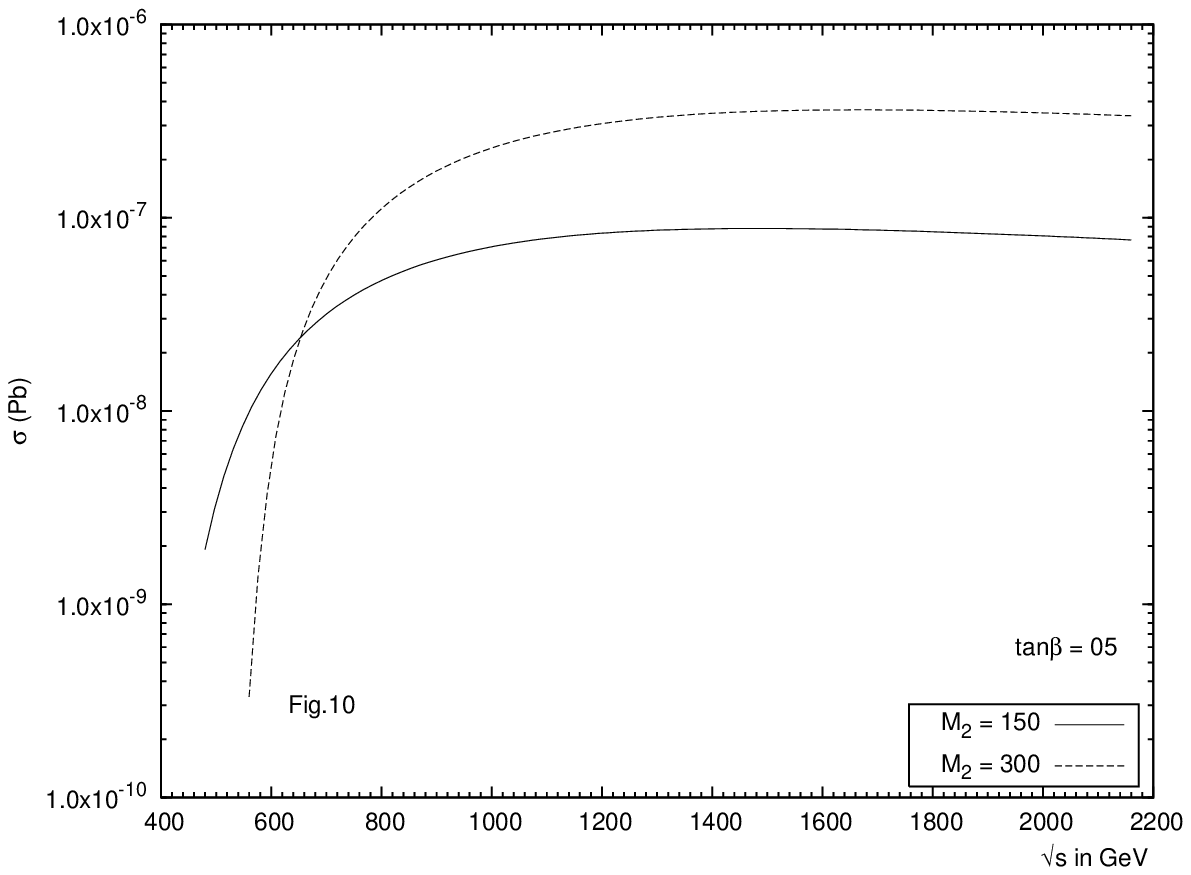}}
\vspace{-0.1cm}
\centerline{\epsfxsize=4.3truein\epsfbox{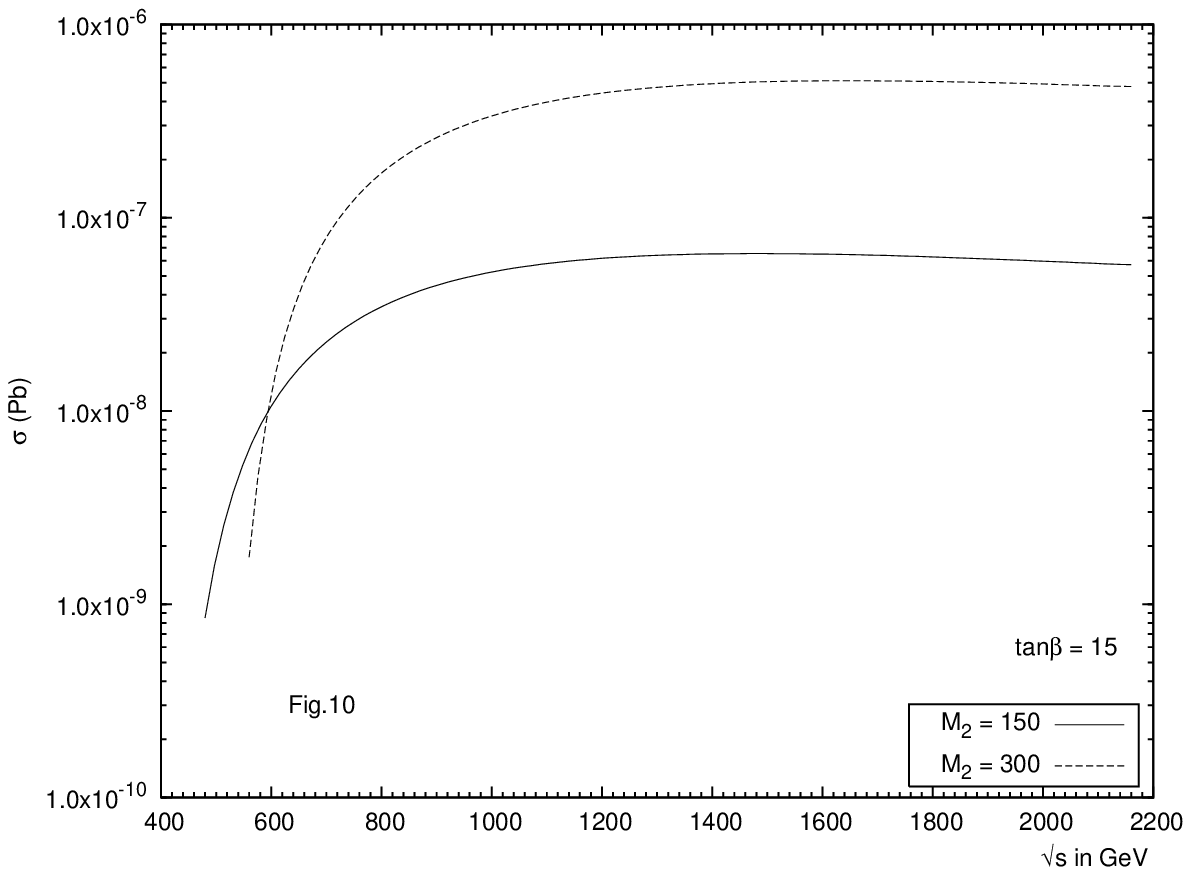}}
\vspace{-0.1cm}
\centerline{\epsfxsize=4.3truein\epsfbox{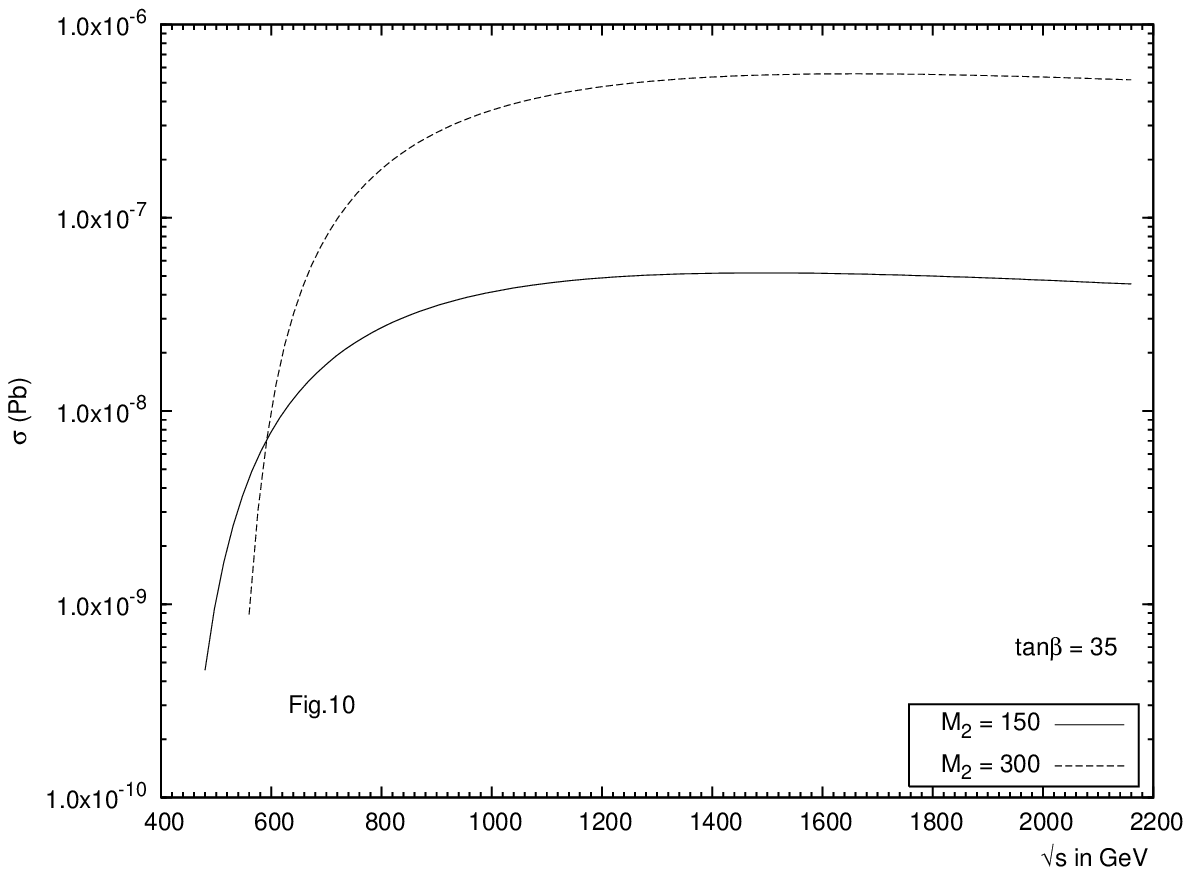}}
\vspace{0.5cm} \caption{ \small Cross sections for
diagram no. 10 in figure \ref{feyn1}} \label{fig.10}
\end{figure}

\begin{figure}[th]
\vspace{-4.5cm}
\centerline{\epsfxsize=4.3truein\epsfbox{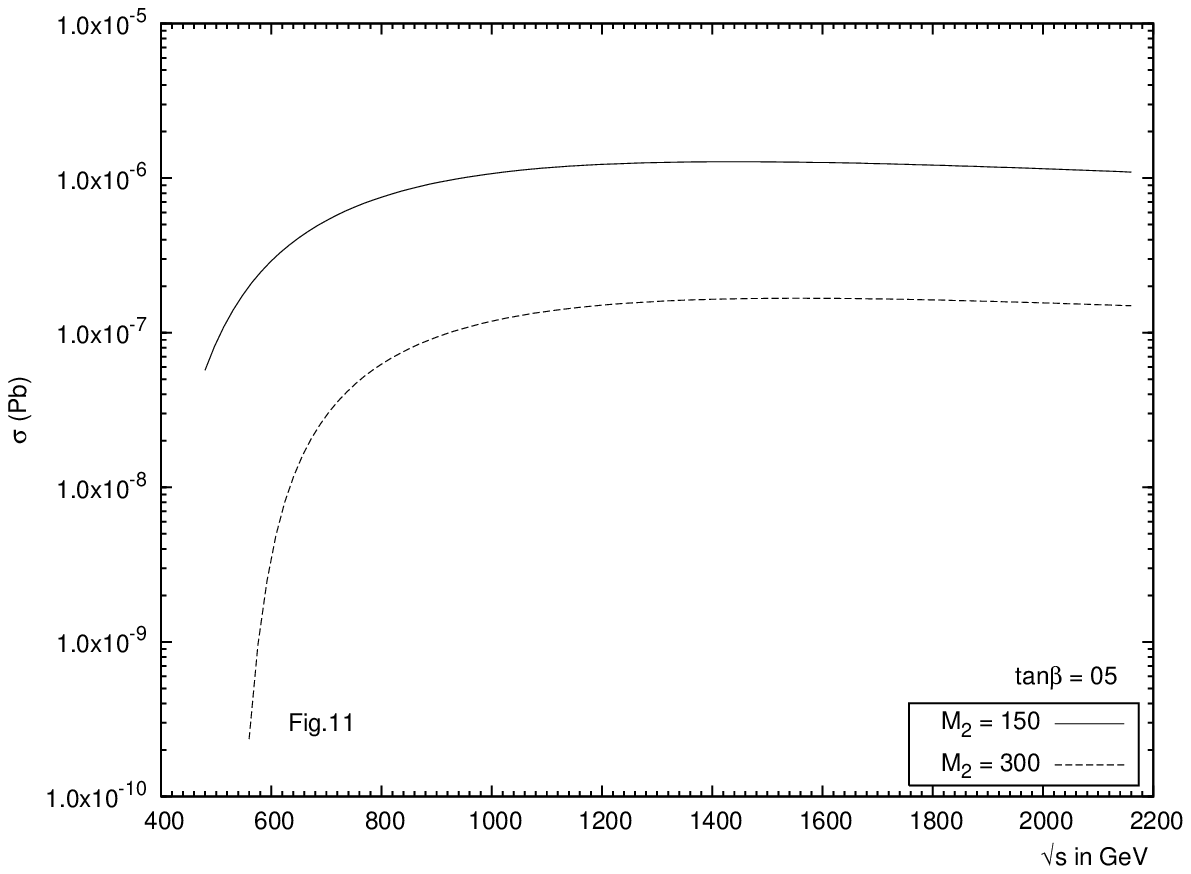}}
\vspace{-0.1cm}
\centerline{\epsfxsize=4.3truein\epsfbox{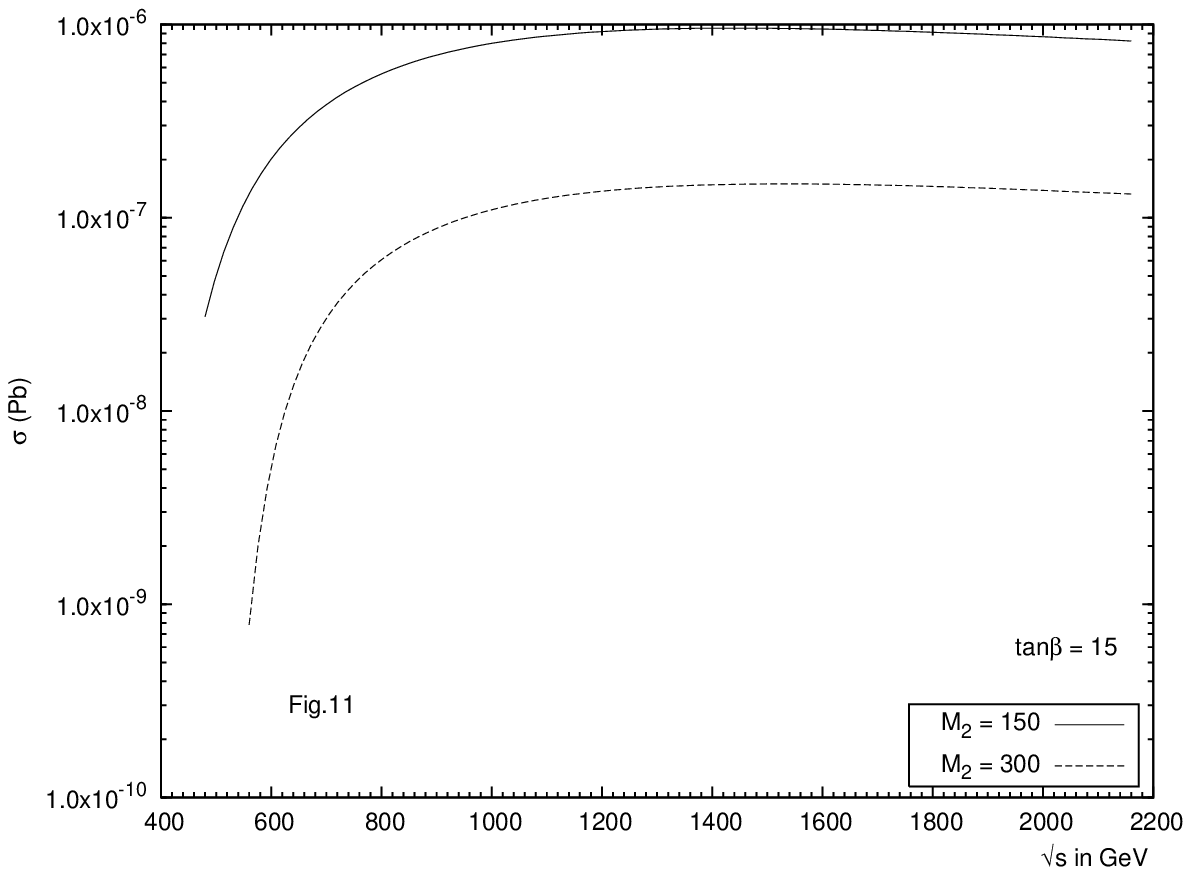}}
\vspace{-0.1cm}
\centerline{\epsfxsize=4.3truein\epsfbox{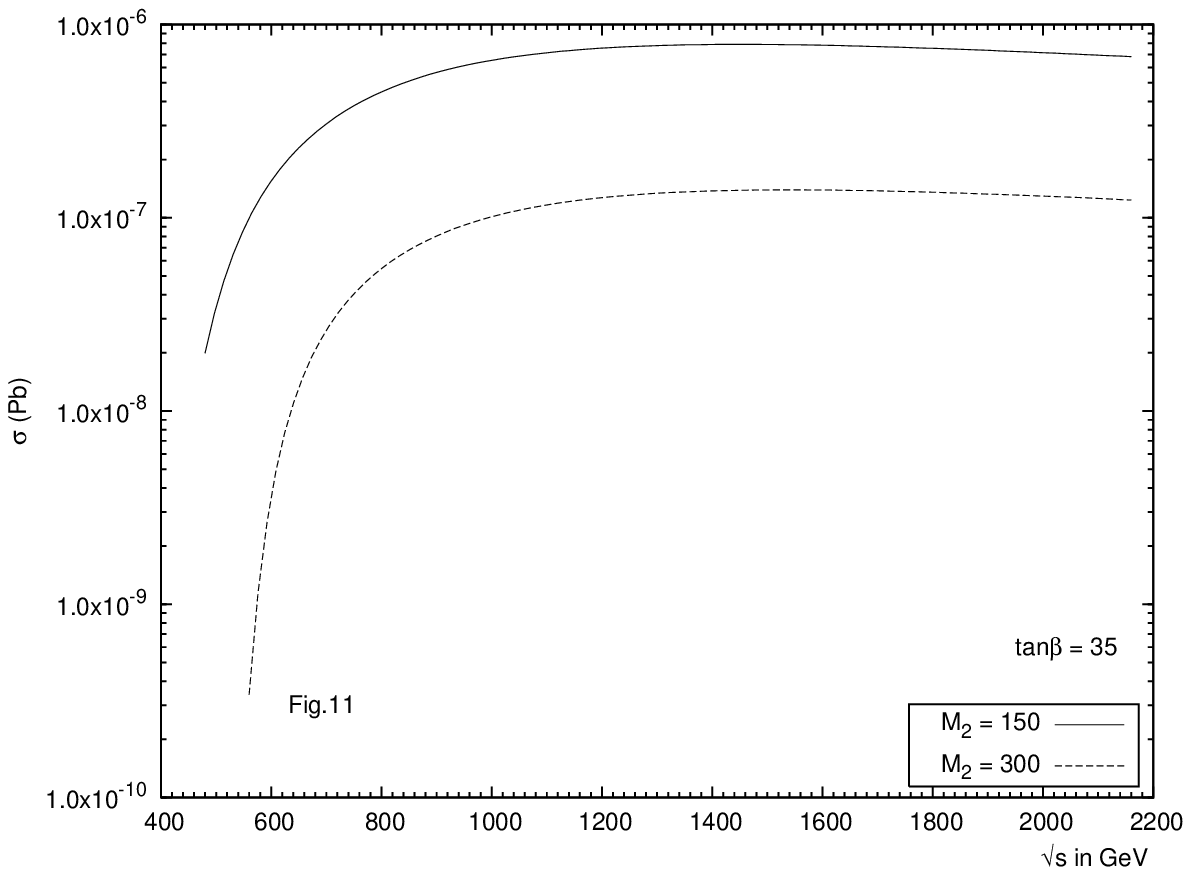}}
\vspace{0.5cm} \caption{ \small Cross sections for
diagram no. 11 in figure \ref{feyn1}} \label{fig.11}
\end{figure}

\begin{figure}[th]
\vspace{-4.5cm}
\centerline{\epsfxsize=4.3truein\epsfbox{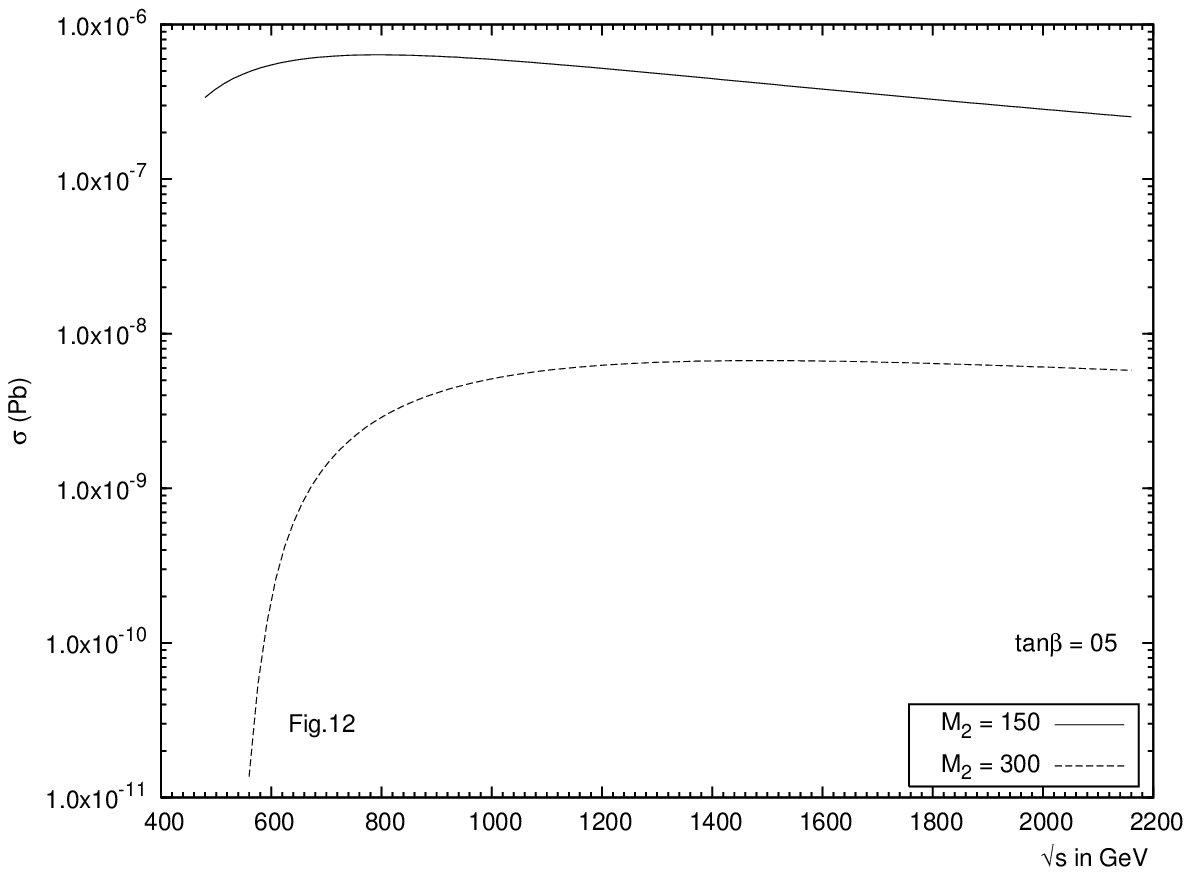}}
\vspace{-0.1cm}
\centerline{\epsfxsize=4.3truein\epsfbox{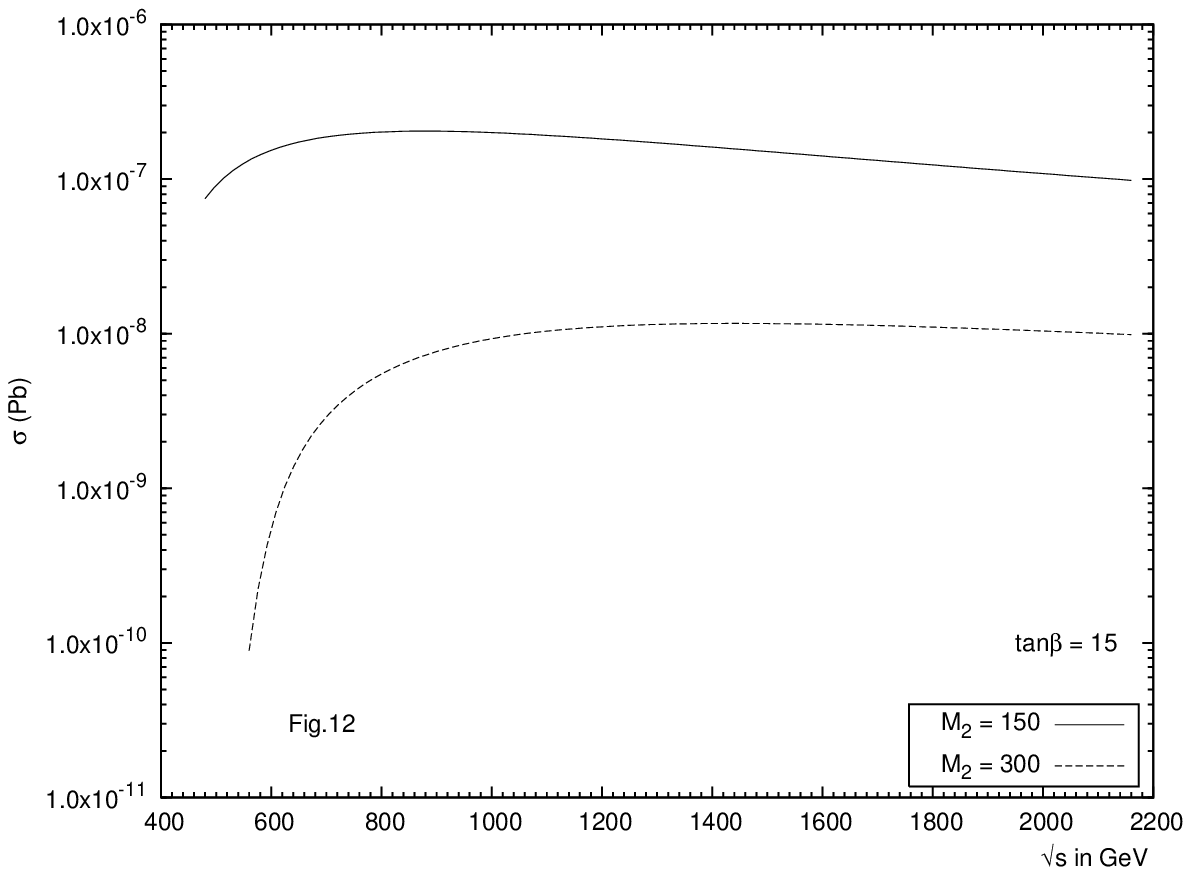}}
\vspace{-0.1cm}
\centerline{\epsfxsize=4.3truein\epsfbox{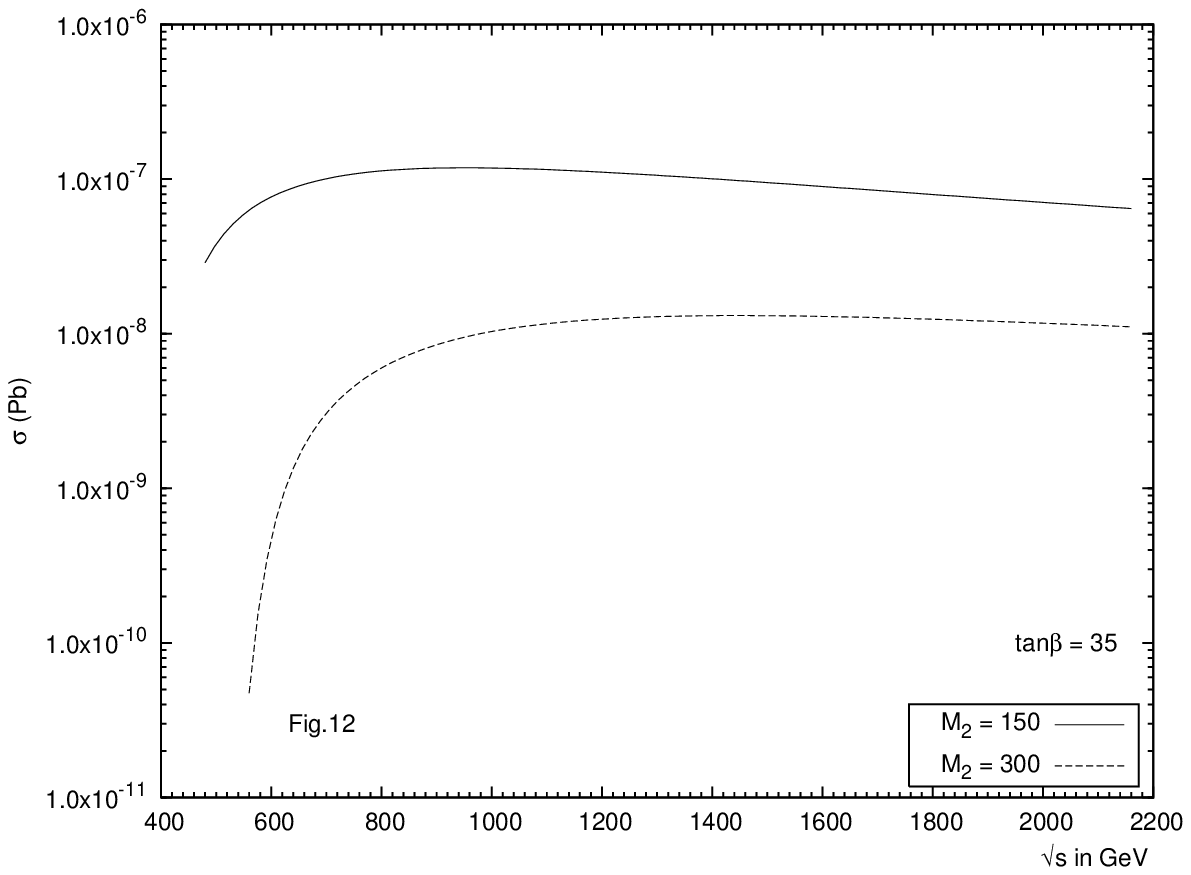}}
\vspace{0.5cm} \caption{ \small Cross sections for
diagram no. 12 in figure \ref{feyn1}} \label{fig.12}
\end{figure}

\begin{figure}[th]
\vspace{-4.5cm}
\centerline{\epsfxsize=4.3truein\epsfbox{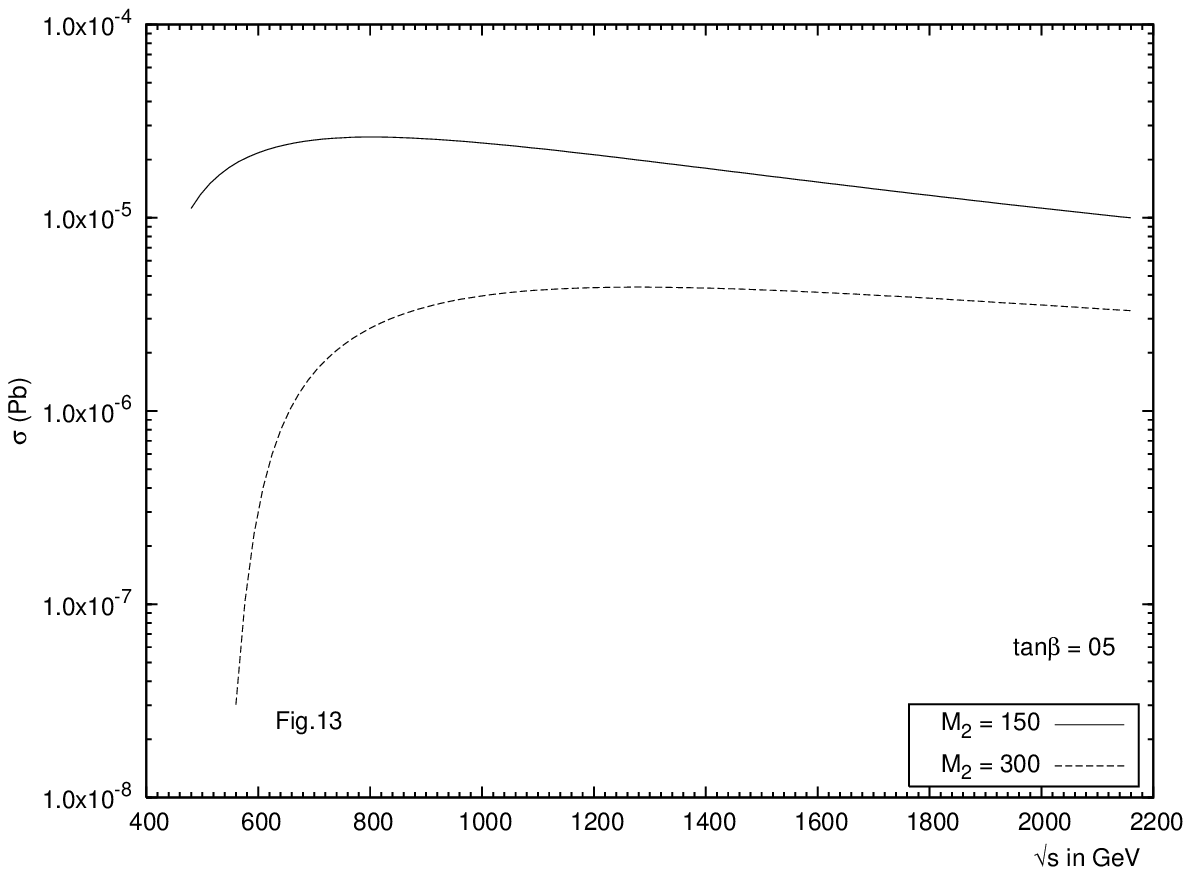}}
\vspace{-0.1cm}
\centerline{\epsfxsize=4.3truein\epsfbox{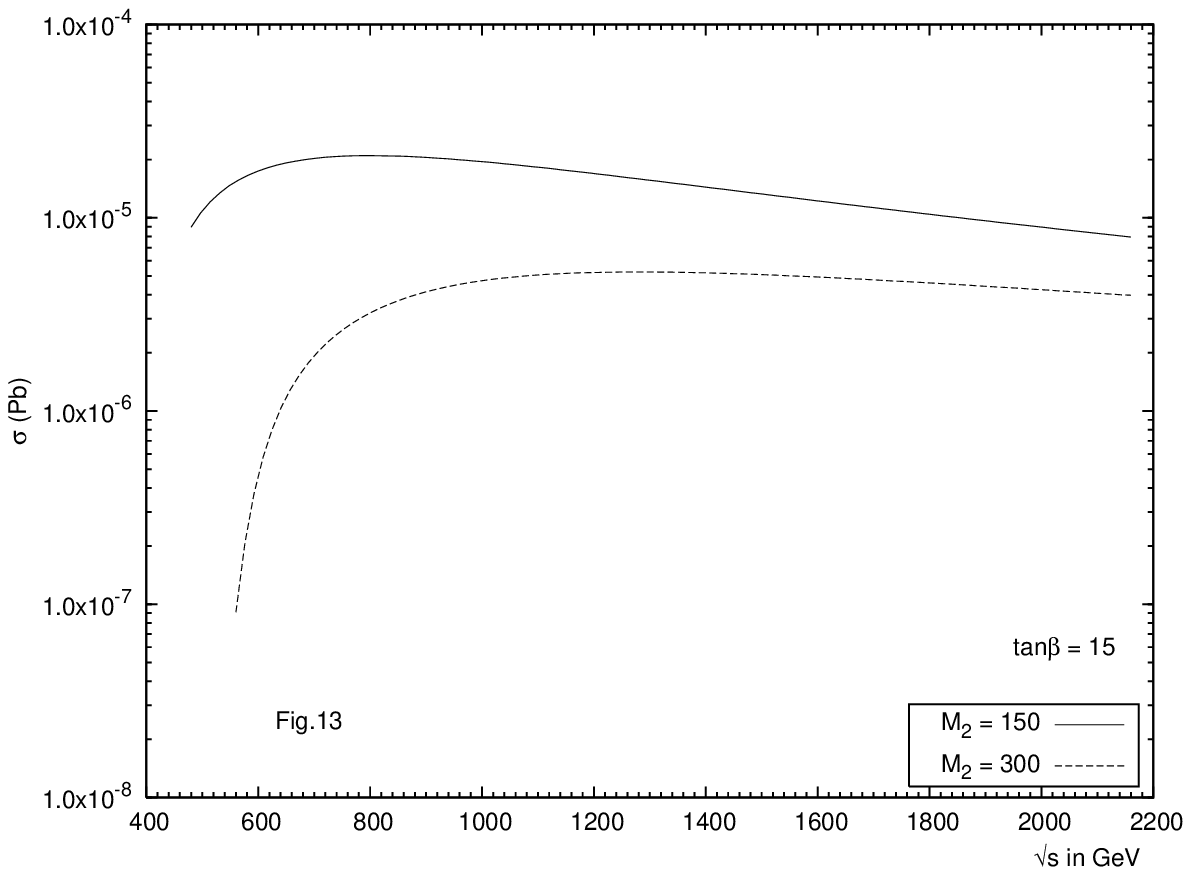}}
\vspace{-0.1cm}
\centerline{\epsfxsize=4.3truein\epsfbox{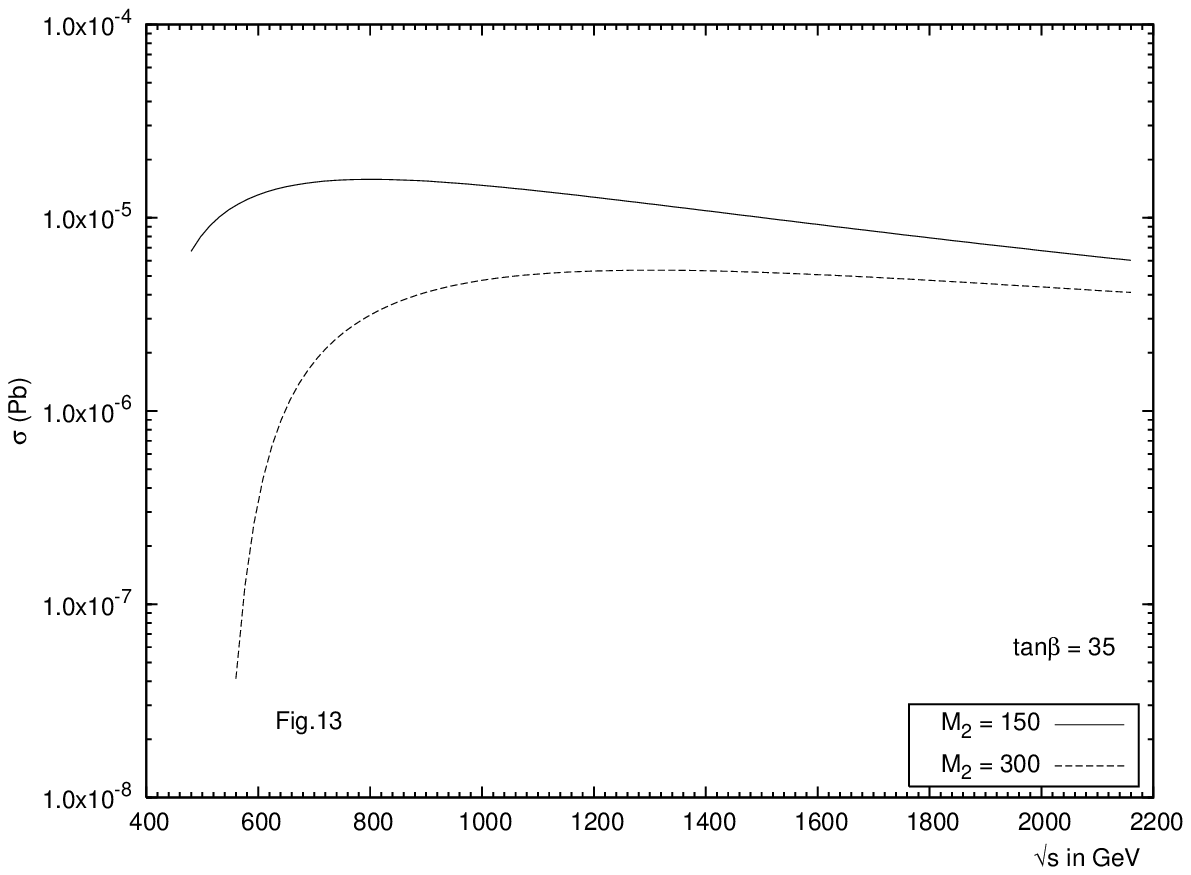}}
\vspace{0.5cm} \caption{ \small Cross sections for
diagram no. 13 in figure \ref{feyn1}} \label{fig.13}
\end{figure}

\begin{figure}[th]
\vspace{-4.5cm}
\centerline{\epsfxsize=4.3truein\epsfbox{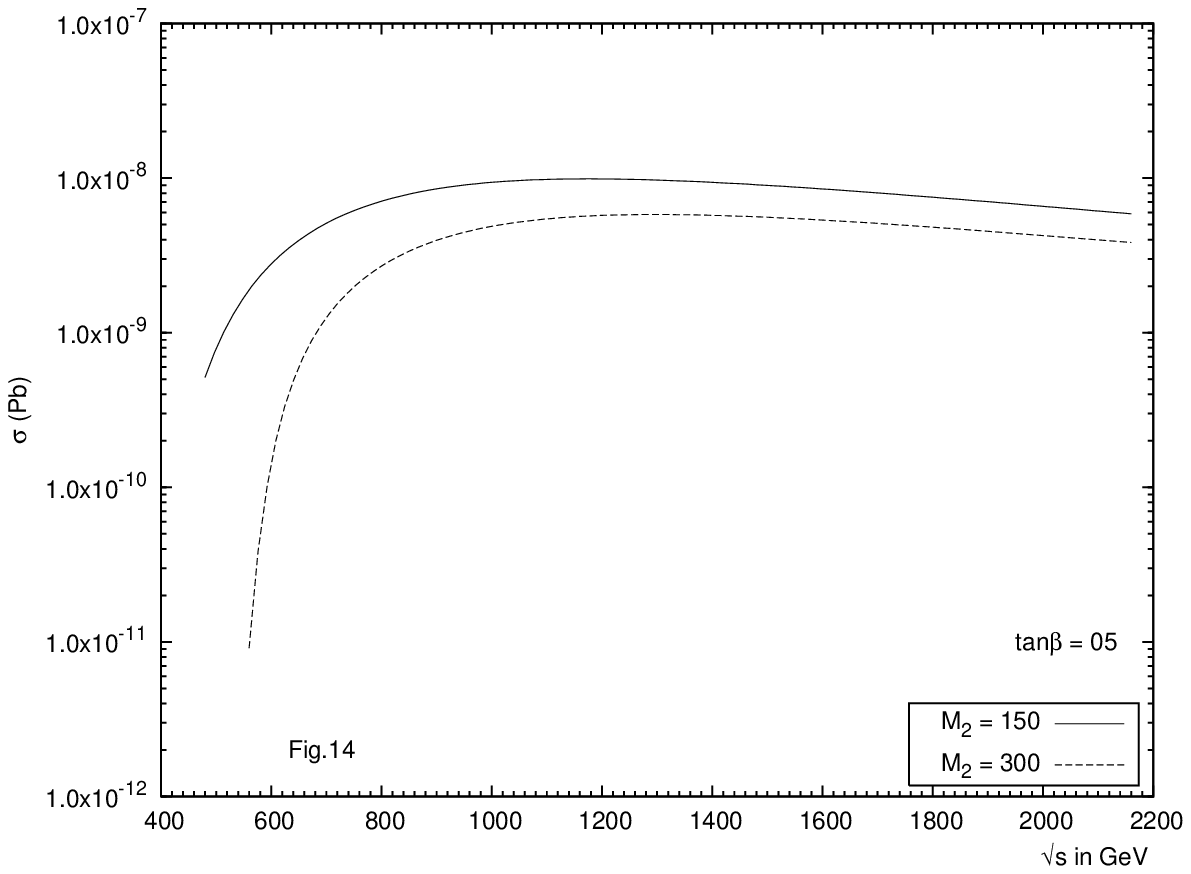}}
\vspace{-0.1cm}
\centerline{\epsfxsize=4.3truein\epsfbox{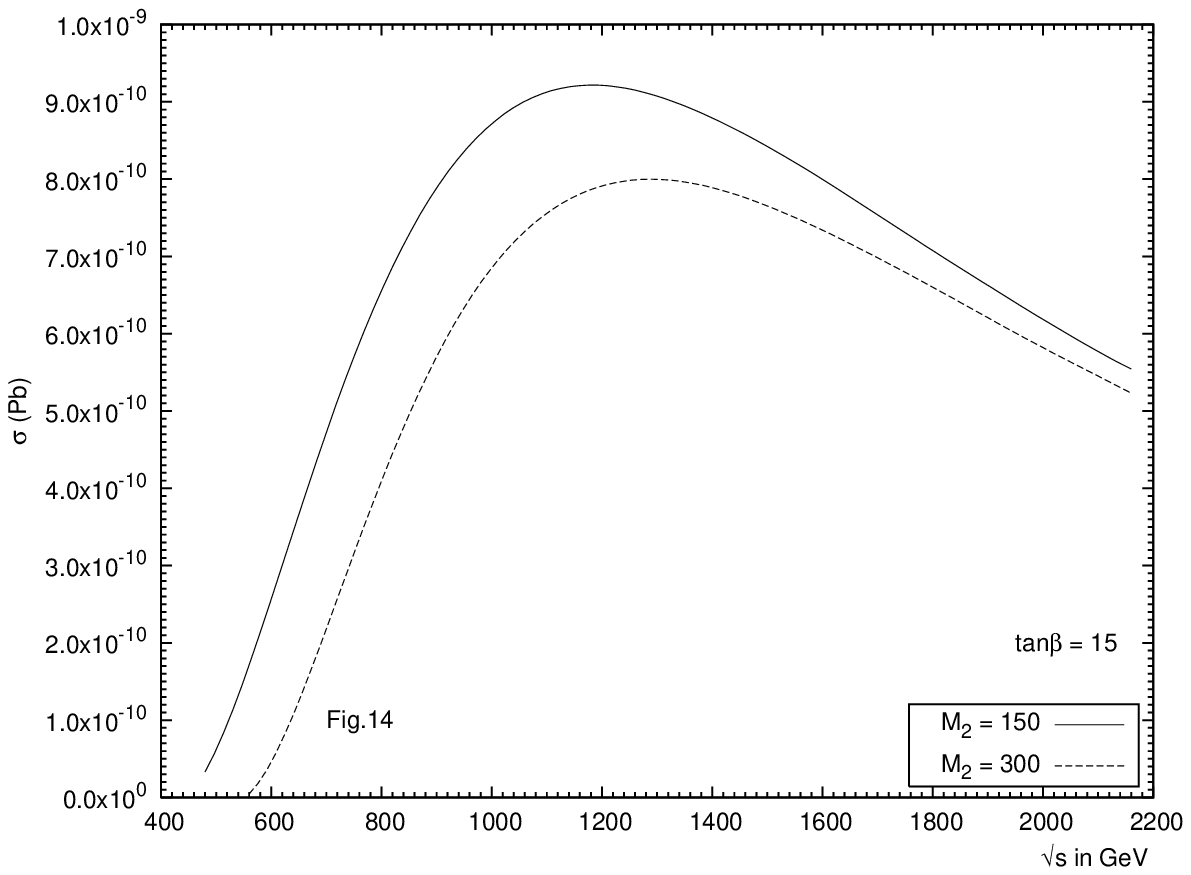}}
\vspace{-0.1cm}
\centerline{\epsfxsize=4.3truein\epsfbox{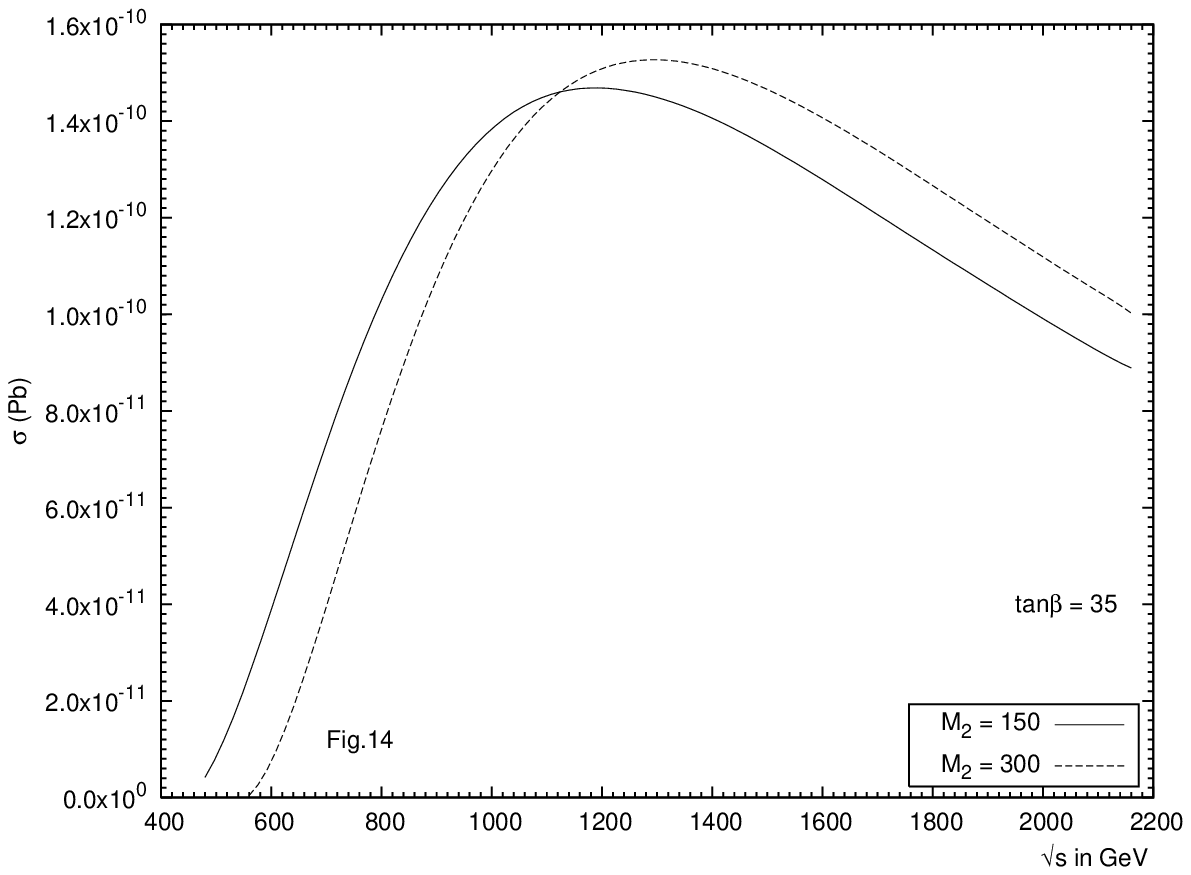}}
\vspace{0.5cm} \caption{ \small Cross sections for
diagram no. 14 in figure \ref{feyn1}} \label{fig.14}
\end{figure}

\begin{figure}[th]
\vspace{-4.5cm}
\centerline{\epsfxsize=4.3truein\epsfbox{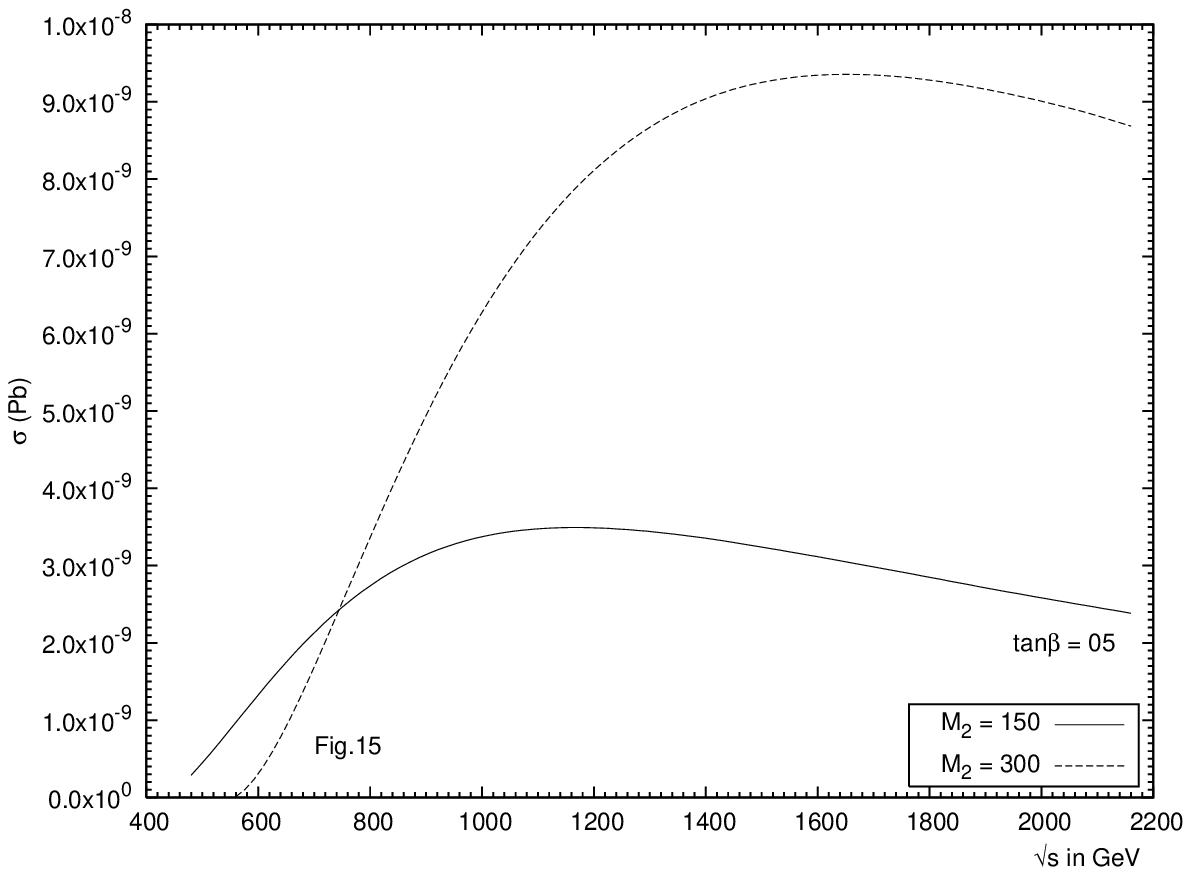}}
\vspace{-0.1cm}
\centerline{\epsfxsize=4.3truein\epsfbox{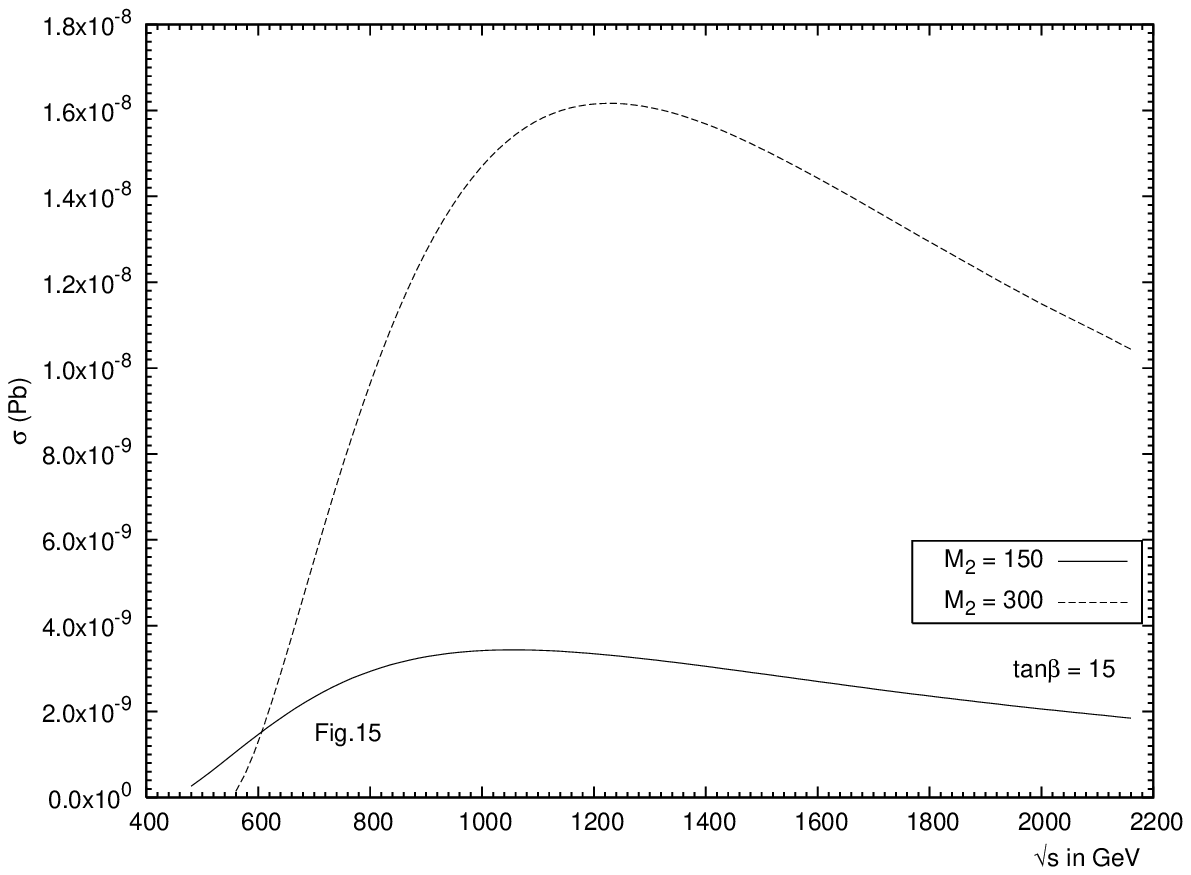}}
\vspace{-0.1cm}
\centerline{\epsfxsize=4.3truein\epsfbox{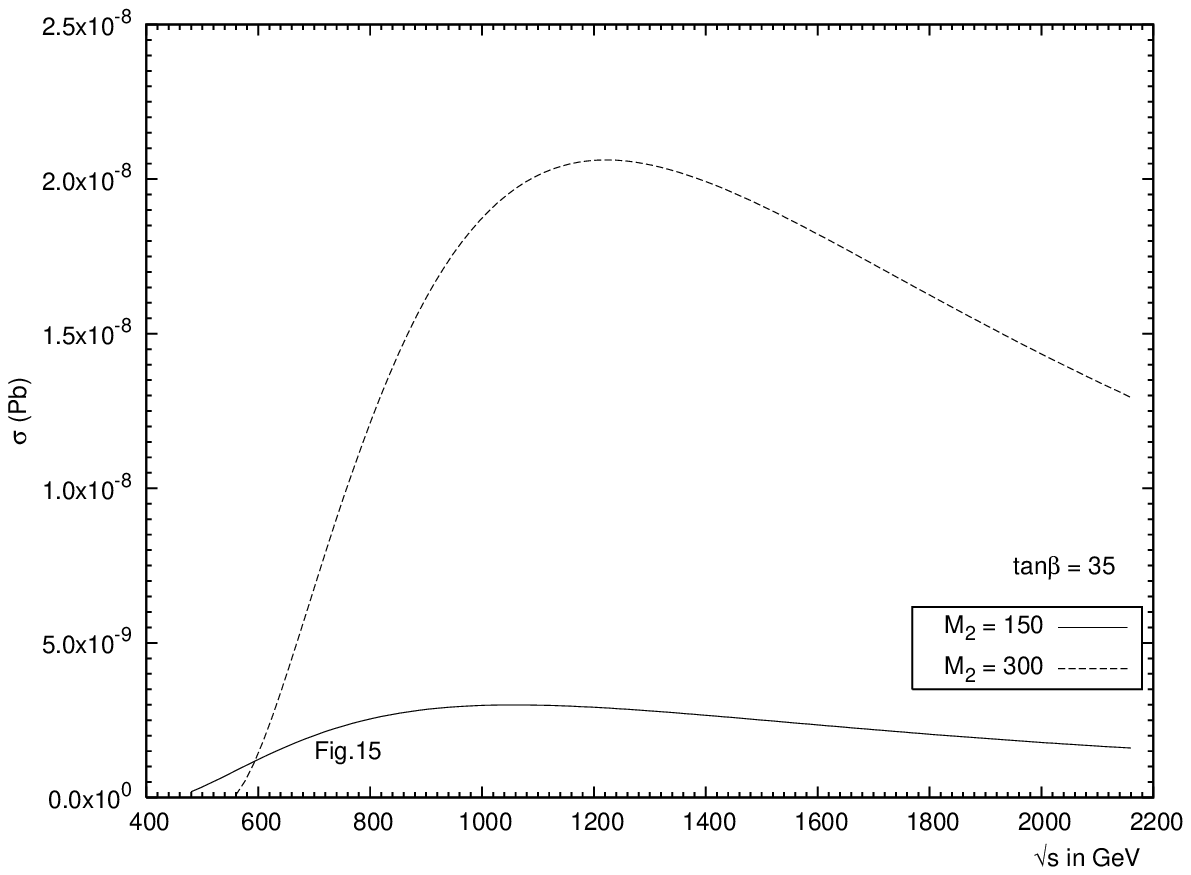}}
\vspace{0.5cm} \caption{ \small Cross sections for
diagram no. 15 in figure \ref{feyn2}} \label{fig.15}
\end{figure}

\begin{figure}[th]
\vspace{-4.5cm}
\centerline{\epsfxsize=4.3truein\epsfbox{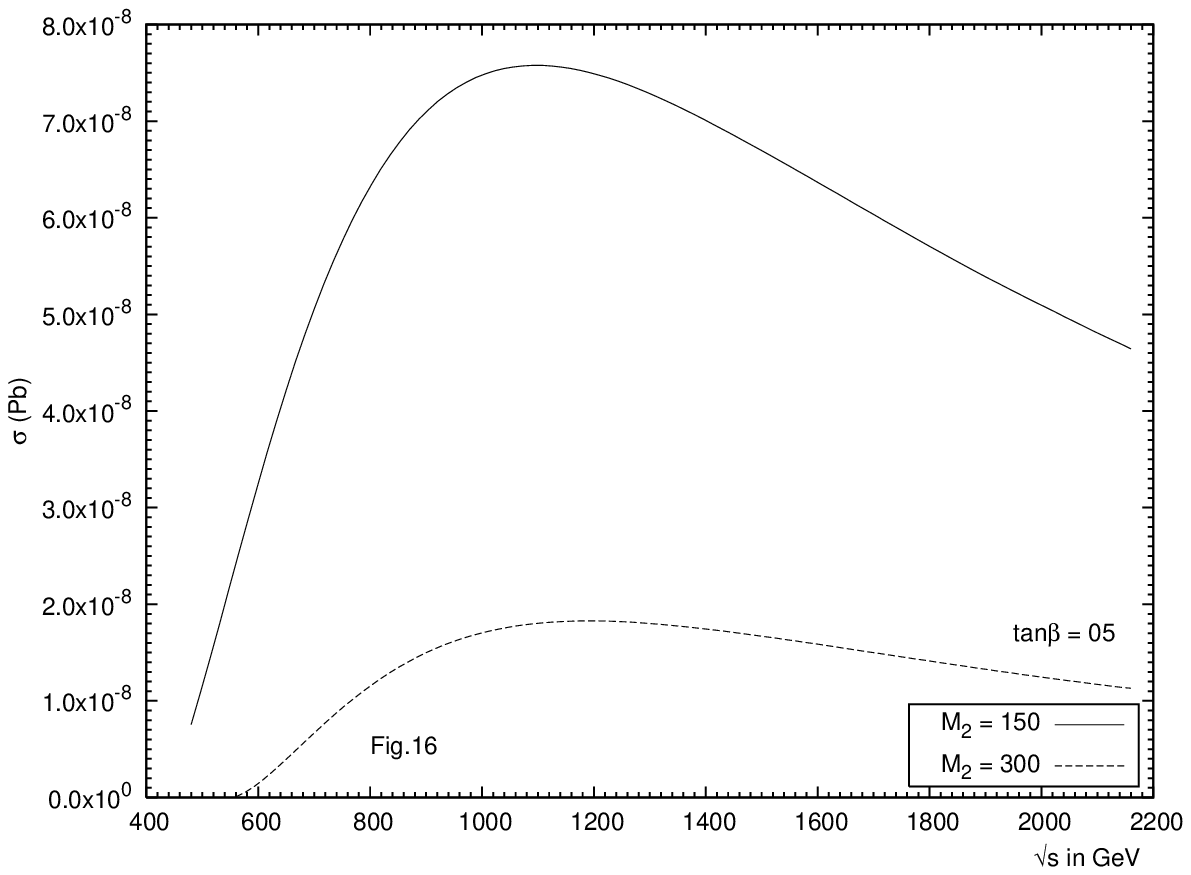}}
\vspace{-0.1cm}
\centerline{\epsfxsize=4.3truein\epsfbox{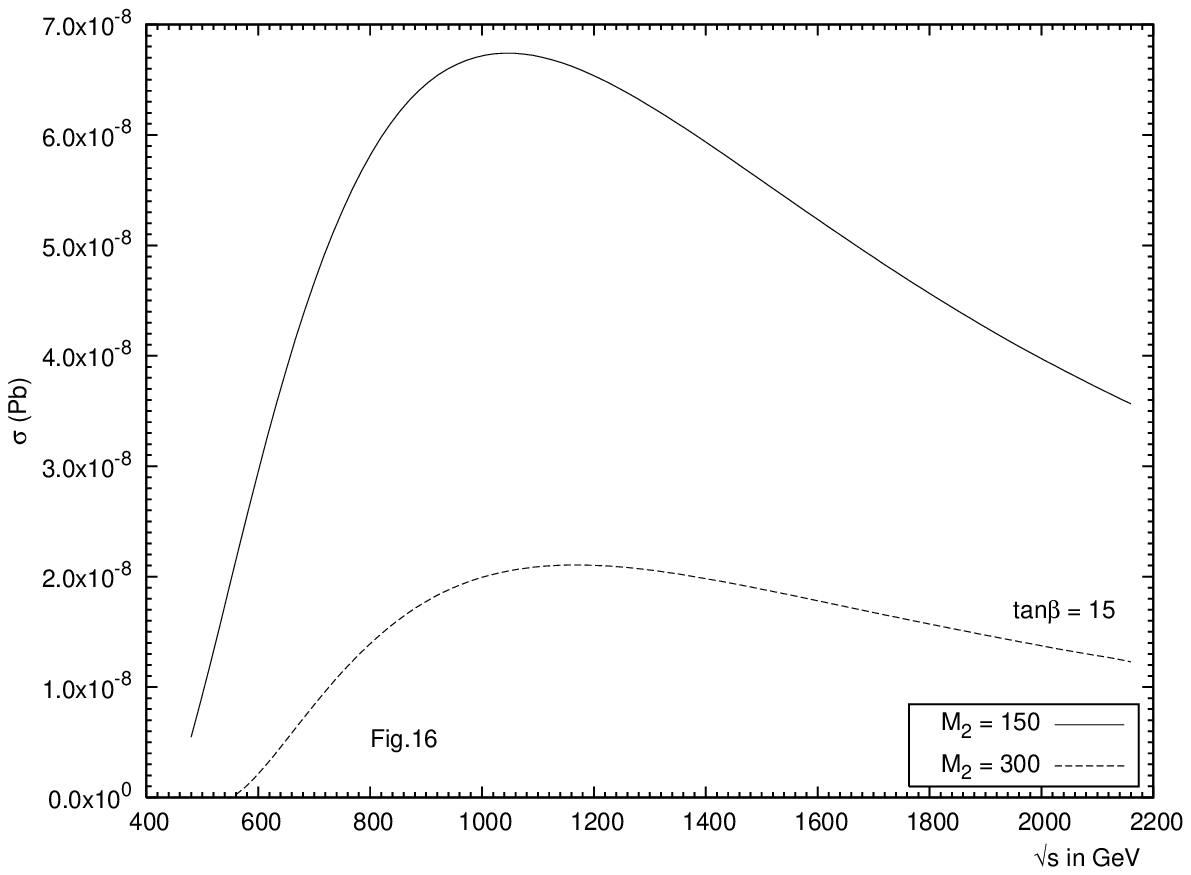}}
\vspace{-0.1cm}
\centerline{\epsfxsize=4.3truein\epsfbox{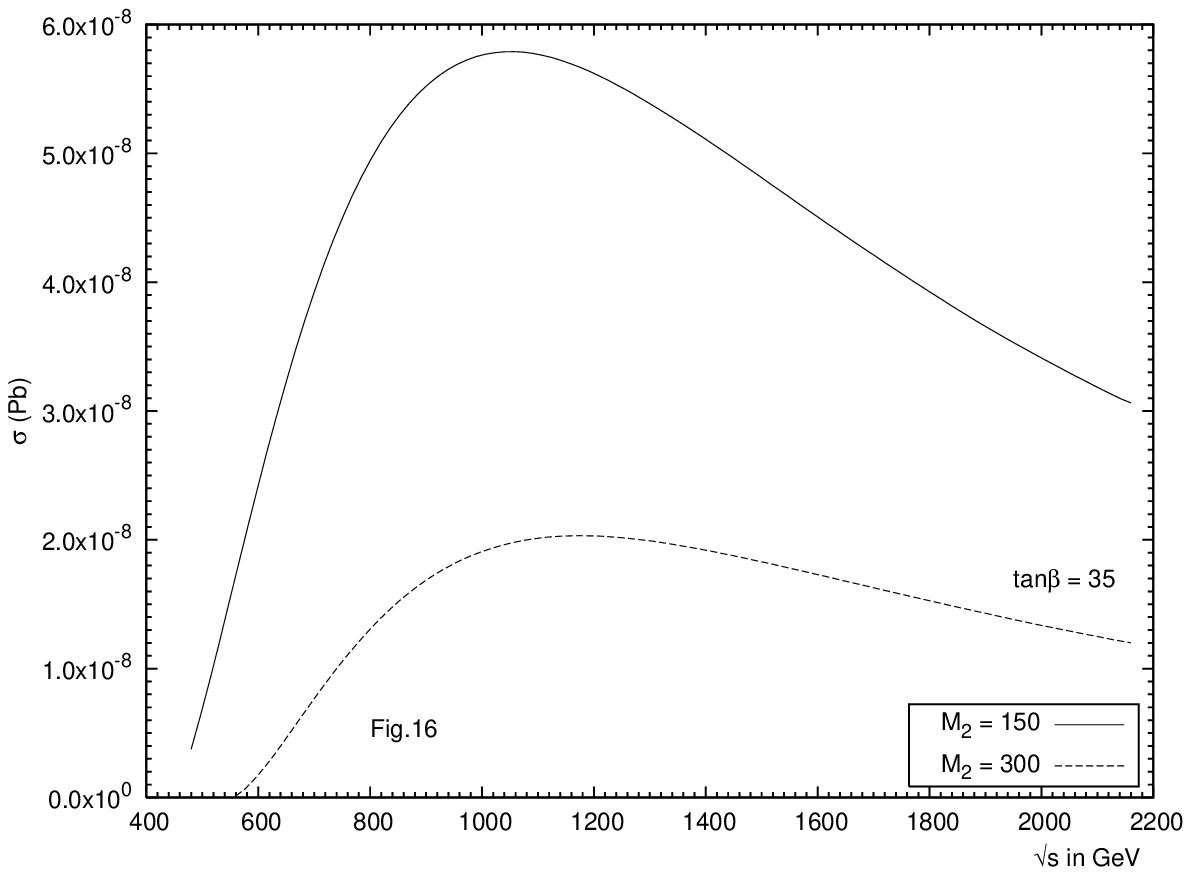}}
\vspace{0.5cm} \caption{ \small Cross sections for
diagram no. 16 in figure \ref{feyn2}} \label{fig.16}
\end{figure}

\begin{figure}[th]
\vspace{-4.5cm}
\centerline{\epsfxsize=4.3truein\epsfbox{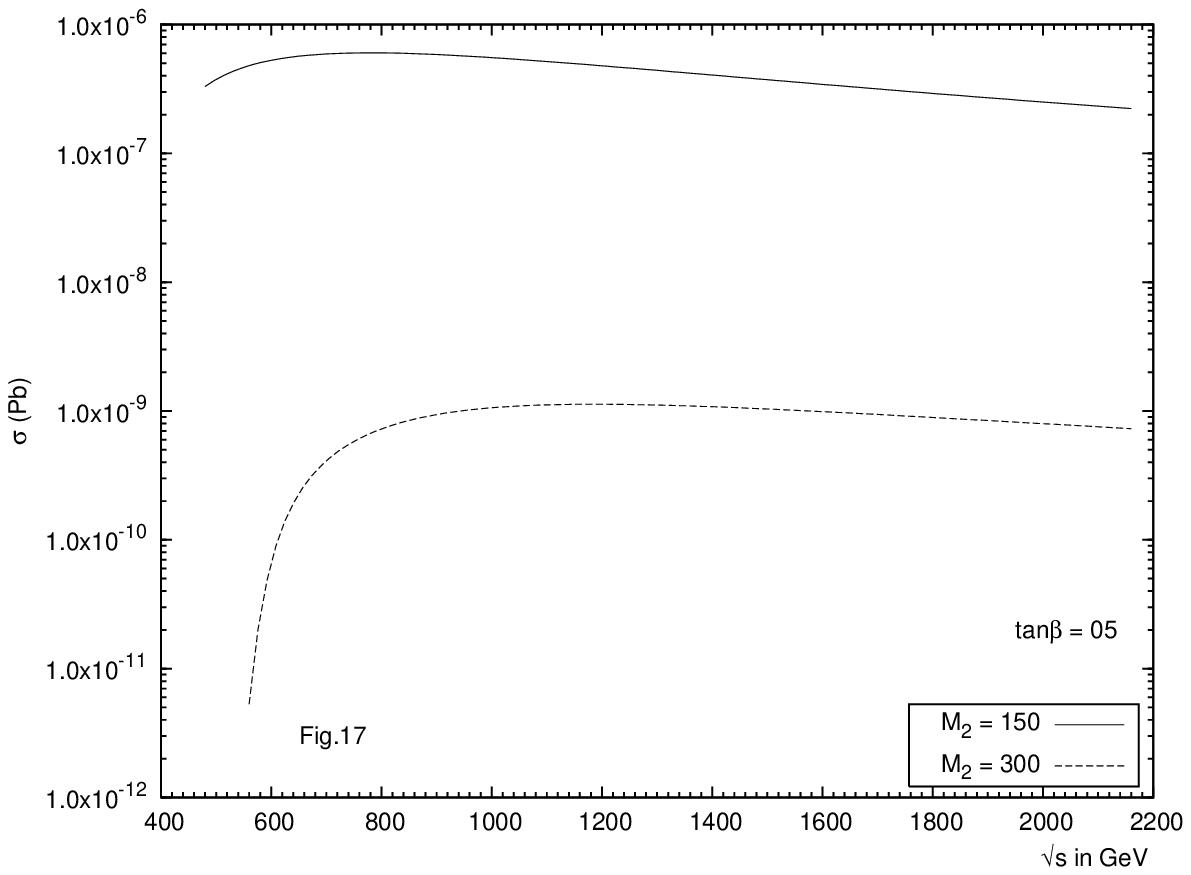}}
\vspace{-0.1cm}
\centerline{\epsfxsize=4.3truein\epsfbox{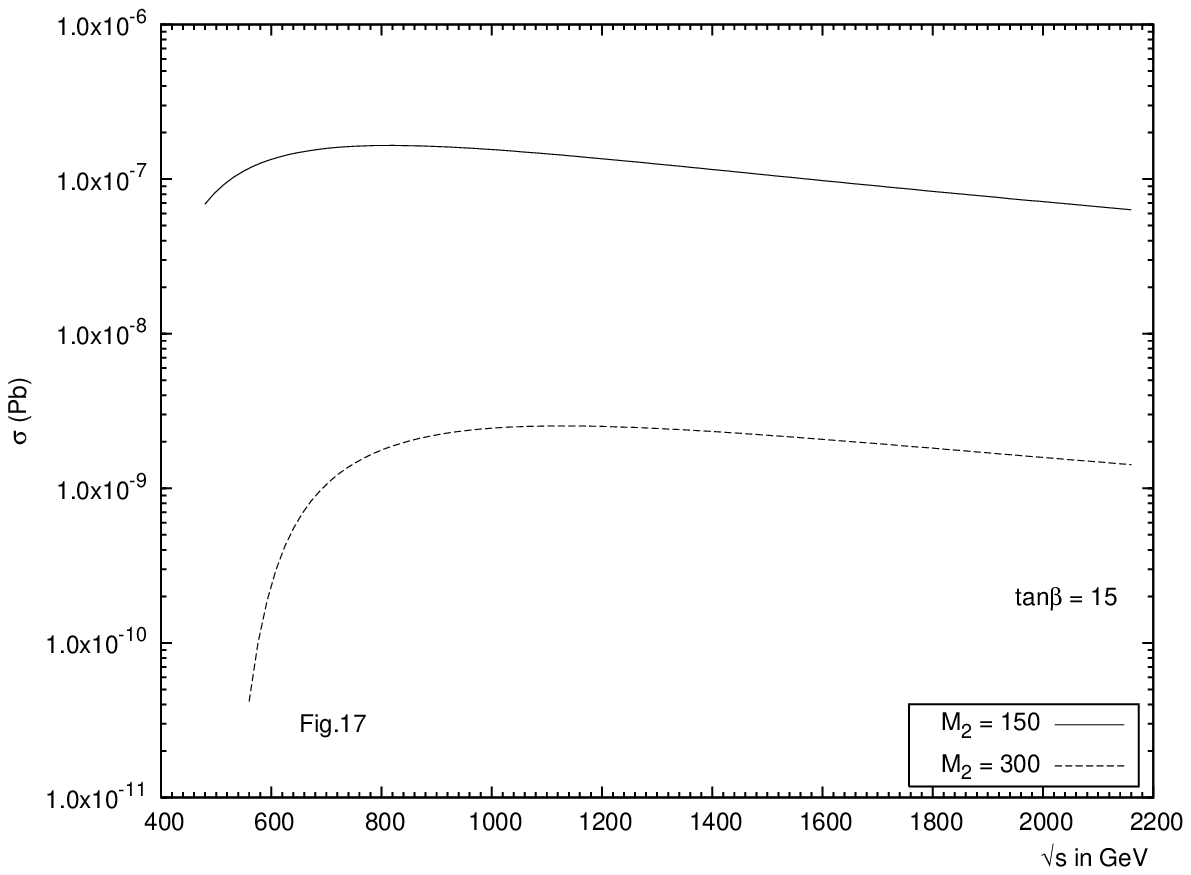}}
\vspace{-0.1cm}
\centerline{\epsfxsize=4.3truein\epsfbox{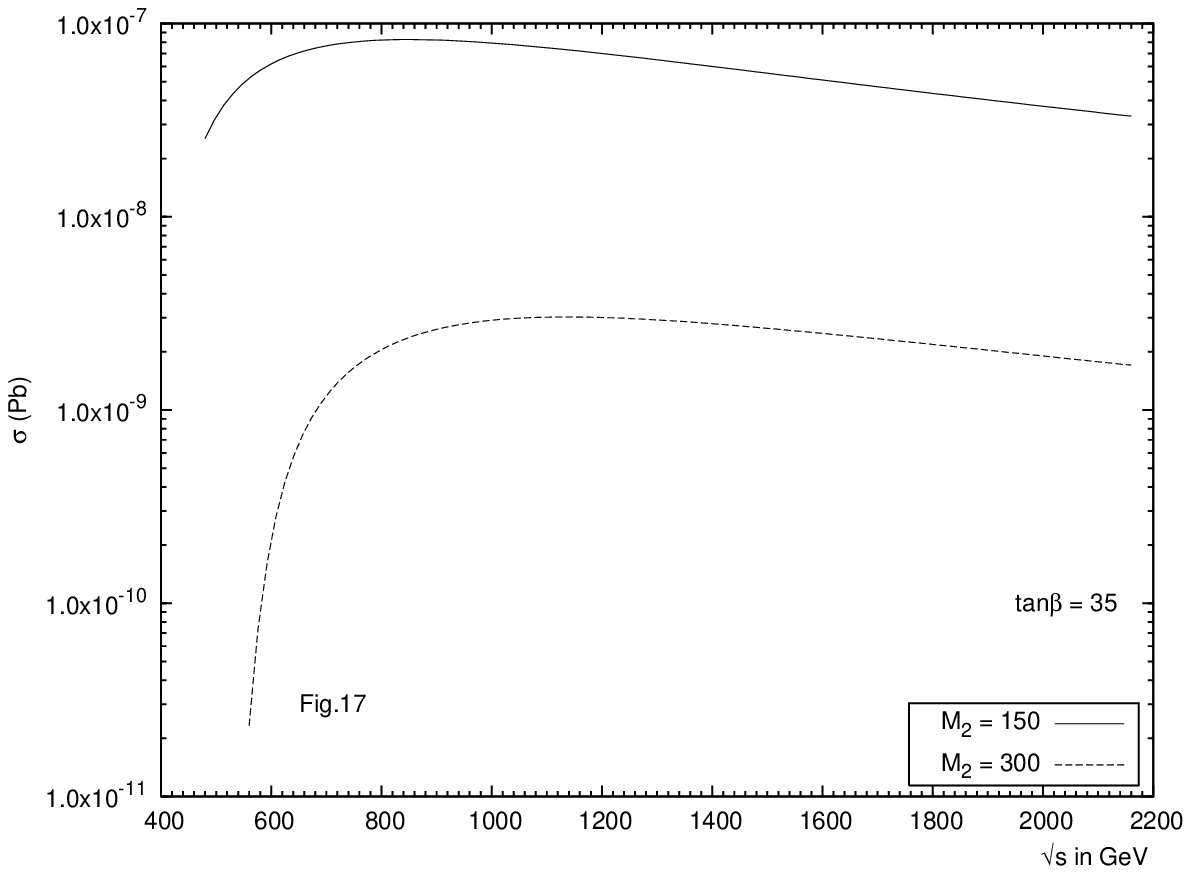}}
\vspace{0.5cm} \caption{ \small Cross sections for
diagram no. 17 in figure \ref{feyn2}} \label{fig.17}
\end{figure}

\begin{figure}[th]
\vspace{-4.5cm}
\centerline{\epsfxsize=4.3truein\epsfbox{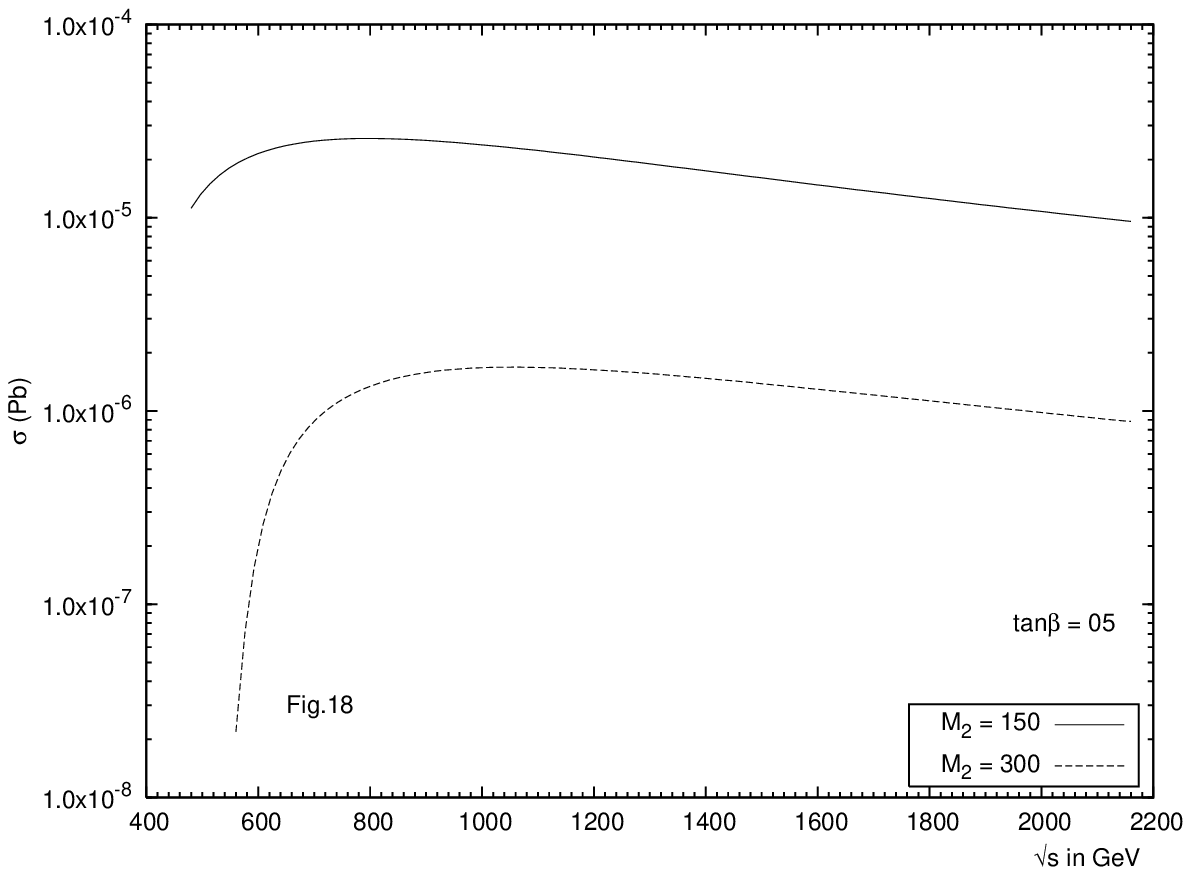}}
\vspace{-0.1cm}
\centerline{\epsfxsize=4.3truein\epsfbox{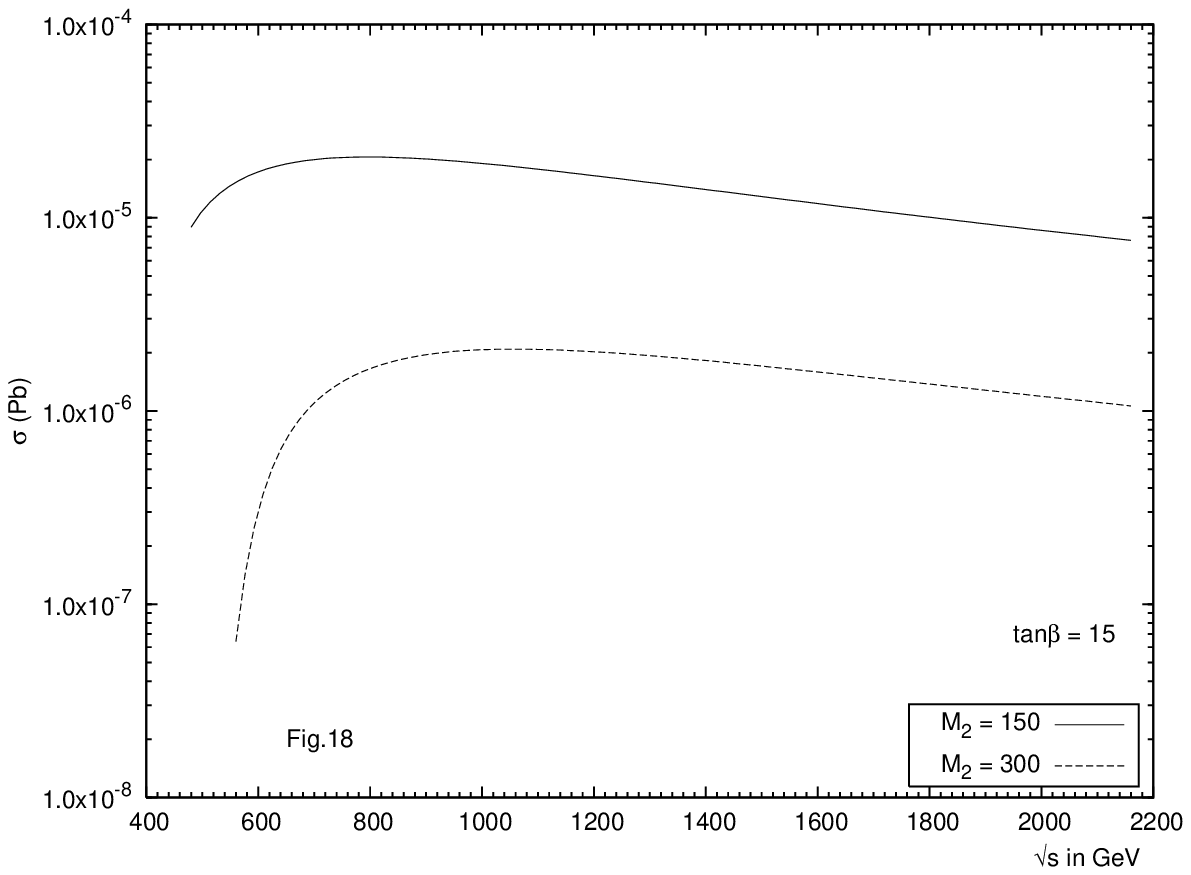}}
\vspace{-0.1cm}
\centerline{\epsfxsize=4.3truein\epsfbox{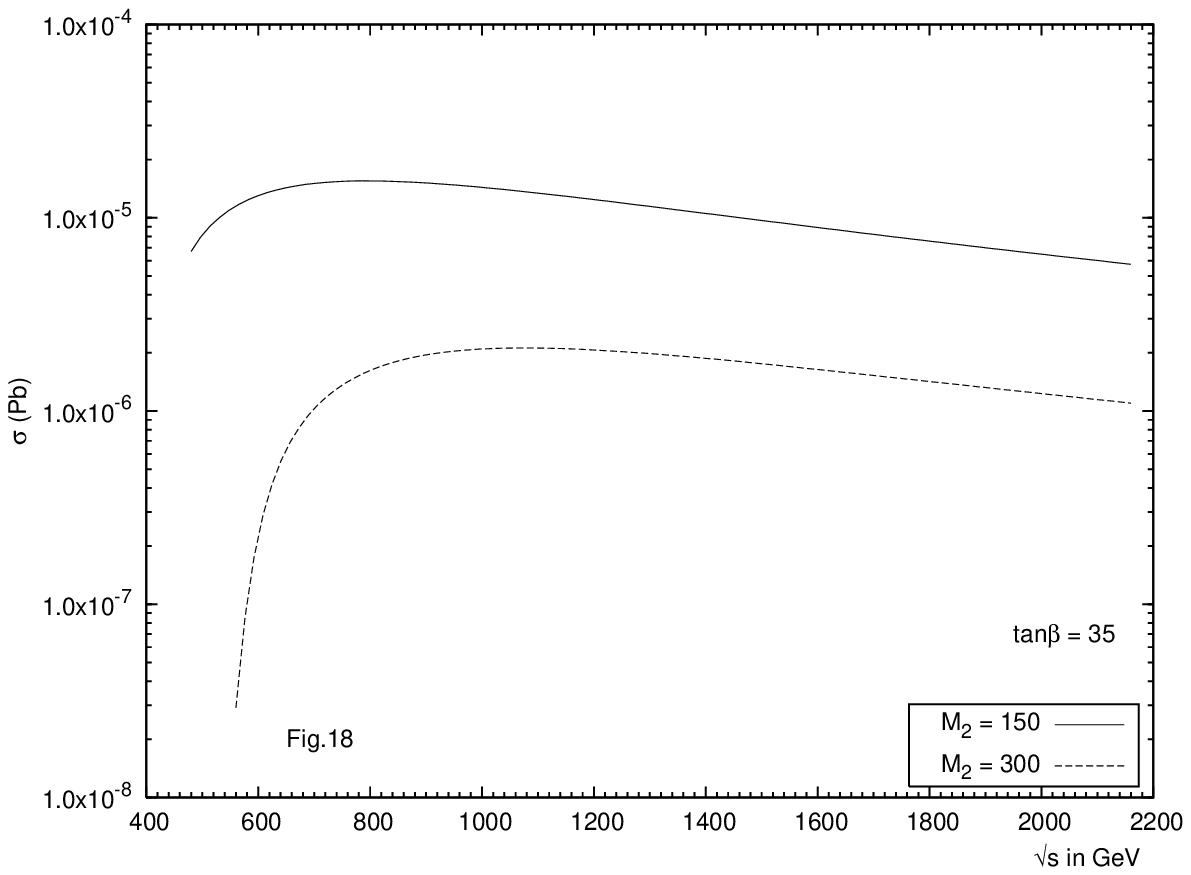}}
\vspace{0.5cm} \caption{ \small Cross sections for
diagram no. 18 in figure \ref{feyn2}} \label{fig.18}
\end{figure}

\clearpage

\begin{figure}[th]
\vspace{-4.5cm}
\centerline{\epsfxsize=4.3truein\epsfbox{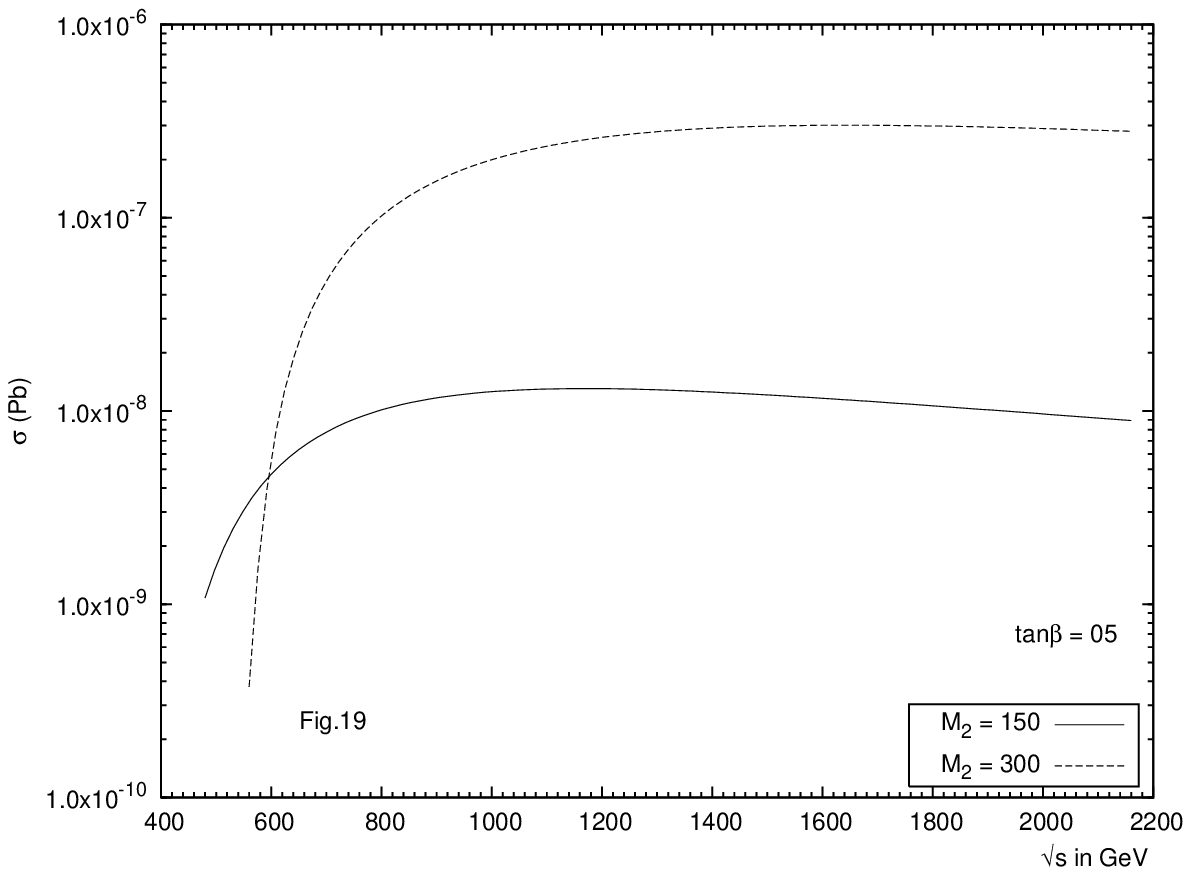}}
\vspace{-0.1cm}
\centerline{\epsfxsize=4.3truein\epsfbox{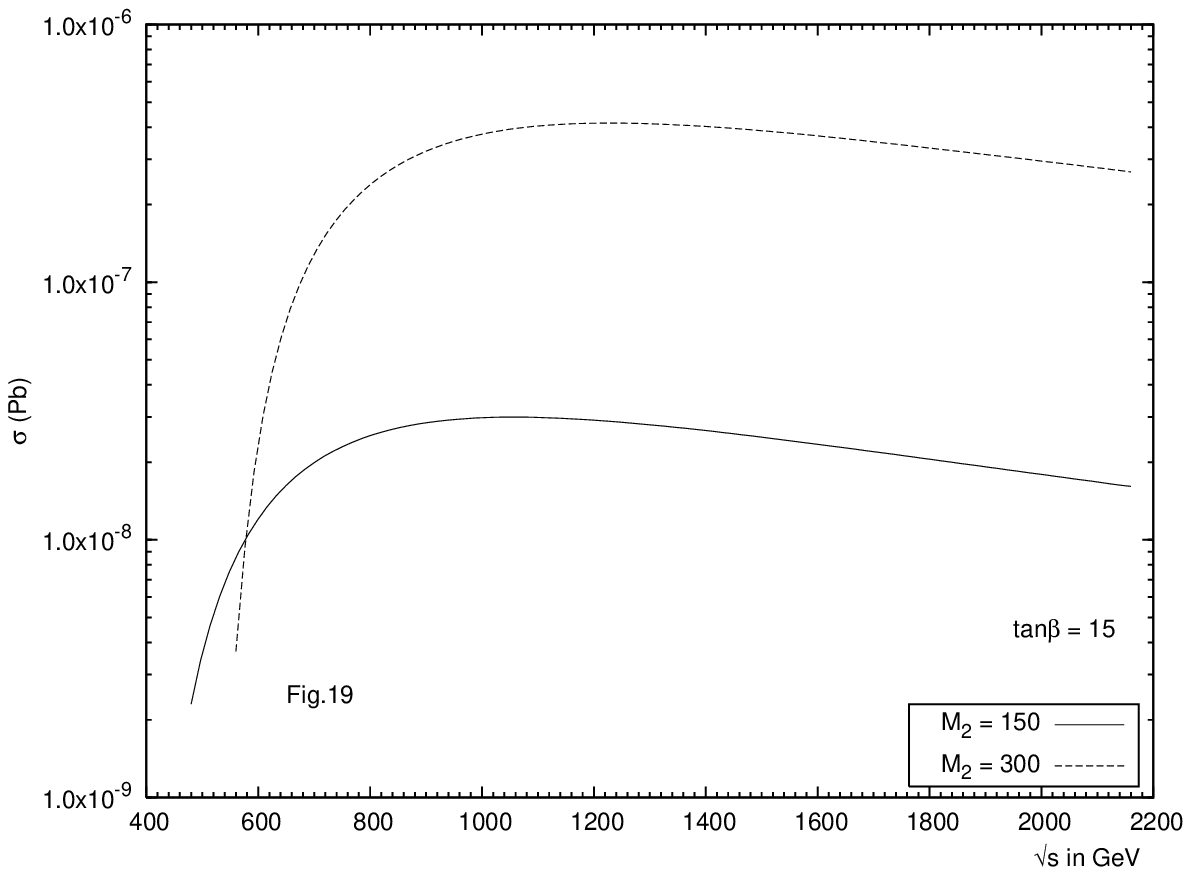}}
\vspace{-0.1cm}
\centerline{\epsfxsize=4.3truein\epsfbox{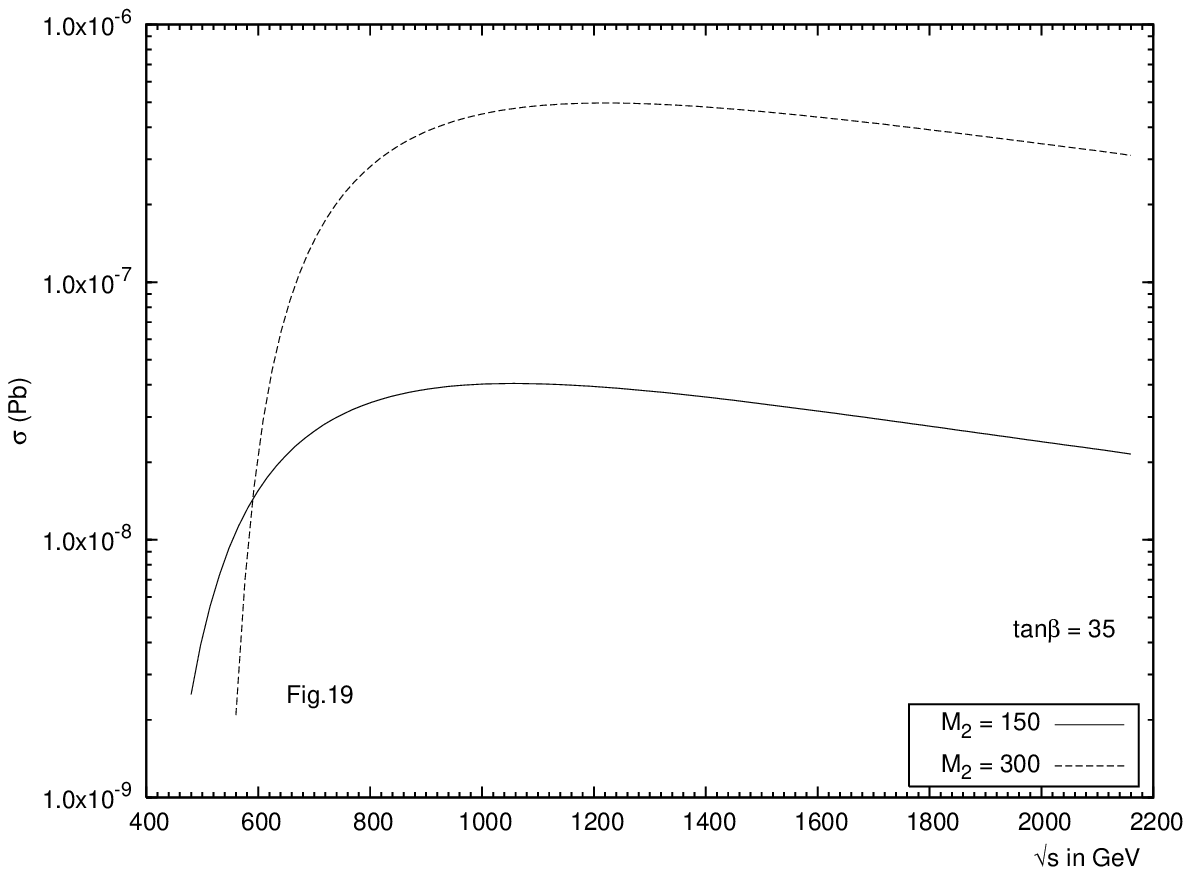}}
\vspace{0.5cm} \caption{ \small Cross sections for
diagram no. 19 in figure \ref{feyn2}} \label{fig.19}
\end{figure}

\begin{figure}[th]
\vspace{-4.5cm}
\centerline{\epsfxsize=4.3truein\epsfbox{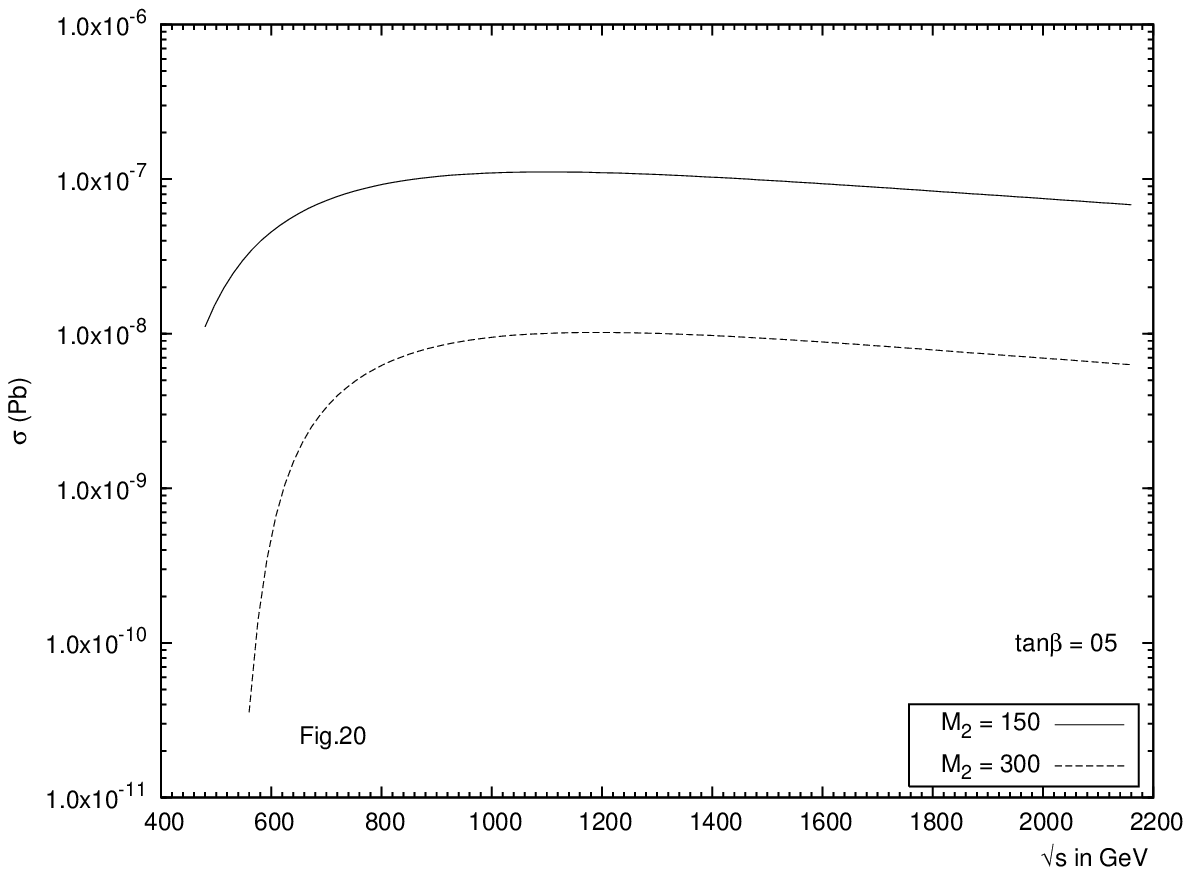}}
\vspace{-0.1cm}
\centerline{\epsfxsize=4.3truein\epsfbox{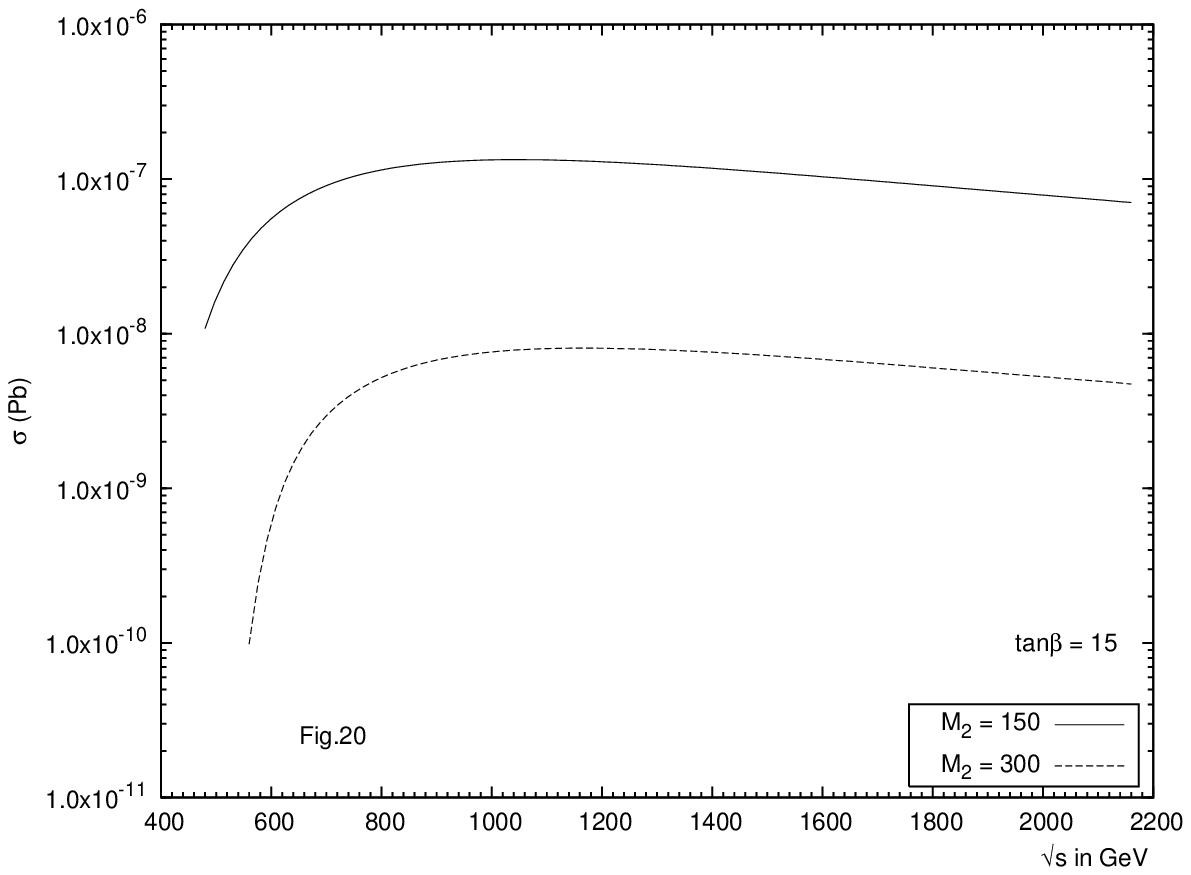}}
\vspace{-0.1cm}
\centerline{\epsfxsize=4.3truein\epsfbox{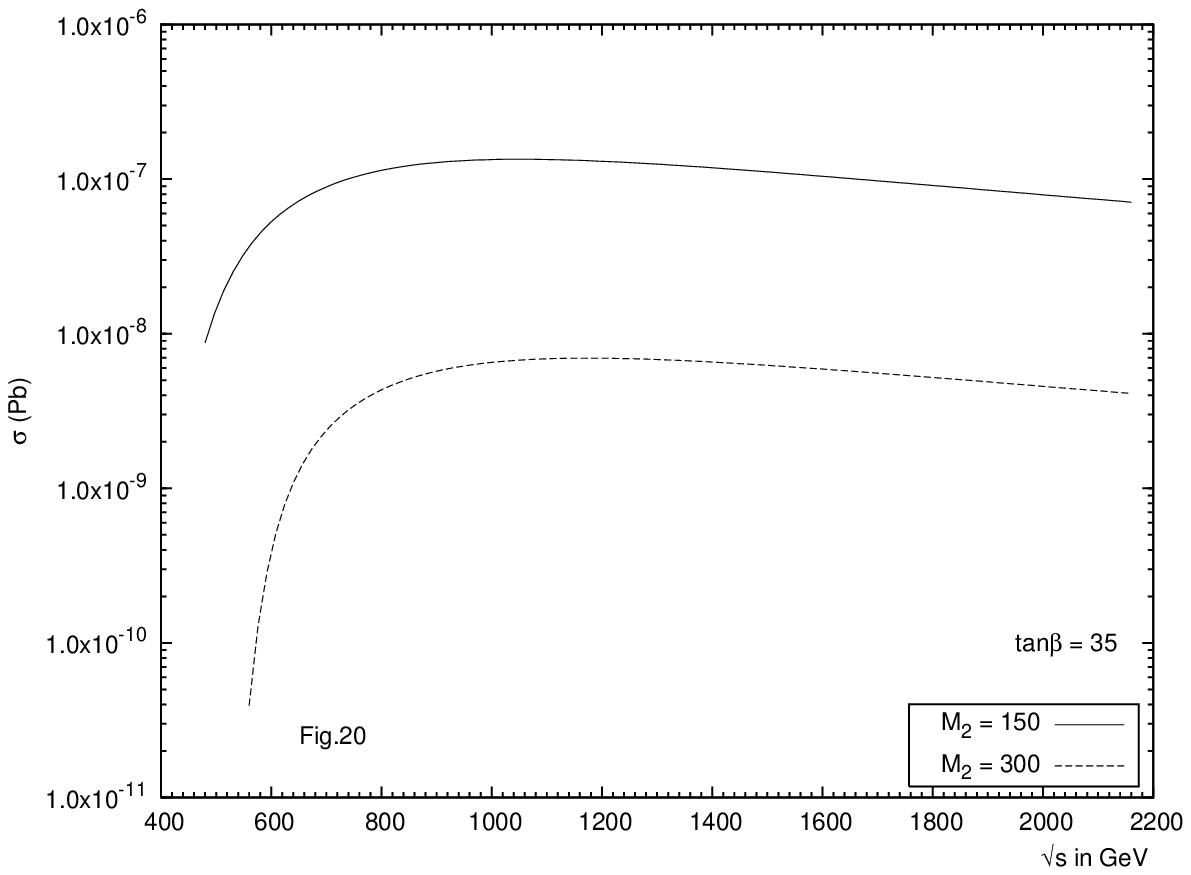}}
\vspace{0.5cm} \caption{ \small Cross sections for
diagram no. 20 in figure \ref{feyn2}} \label{fig.20}
\end{figure}

\begin{figure}[th]
\vspace{-4.5cm}
\centerline{\epsfxsize=4.3truein\epsfbox{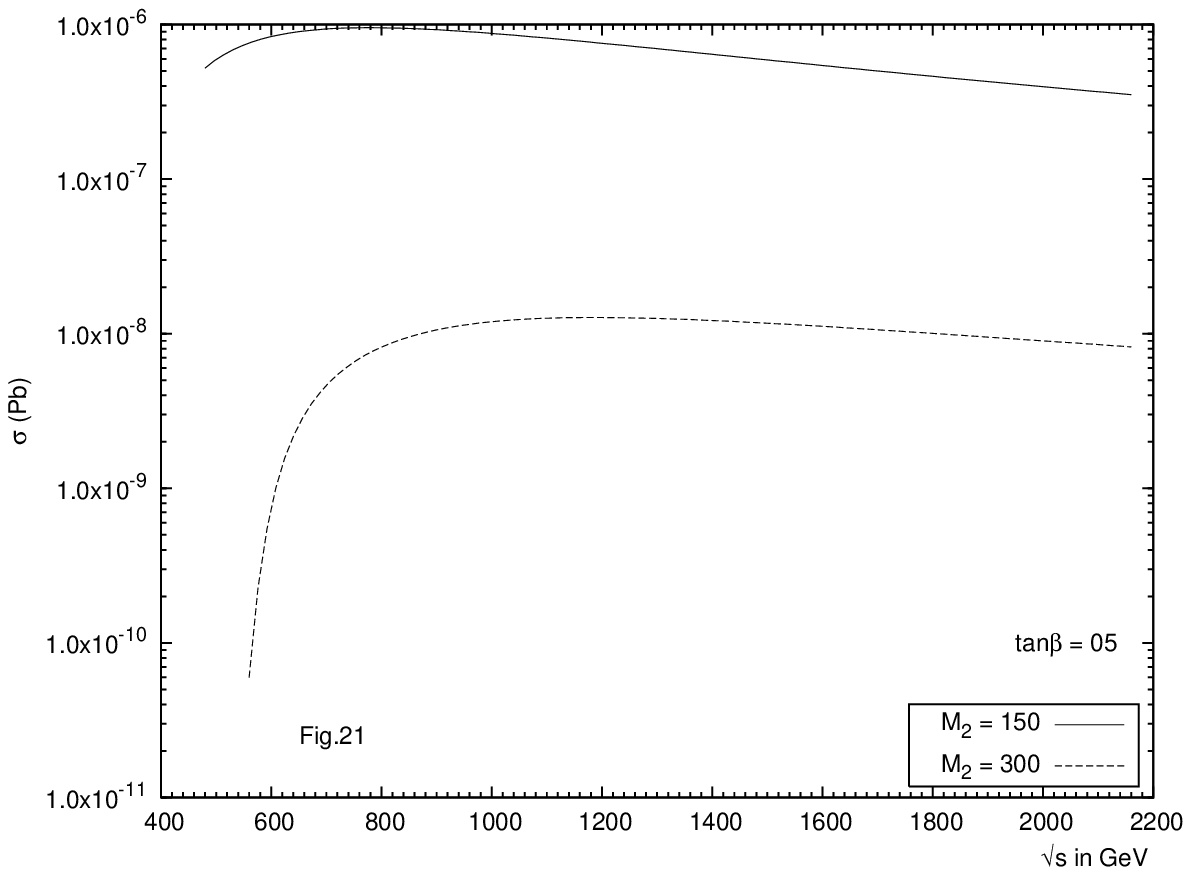}}
\vspace{-0.1cm}
\centerline{\epsfxsize=4.3truein\epsfbox{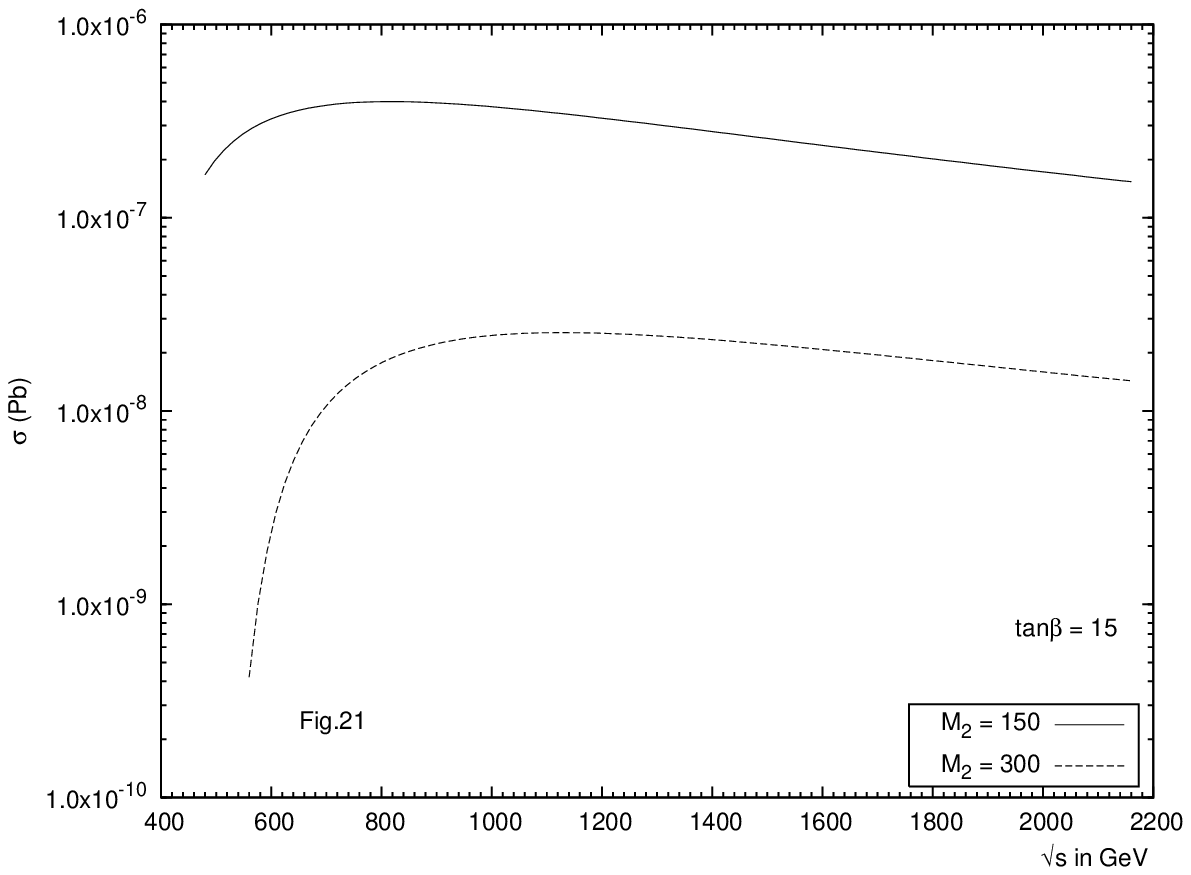}}
\vspace{-0.1cm}
\centerline{\epsfxsize=4.3truein\epsfbox{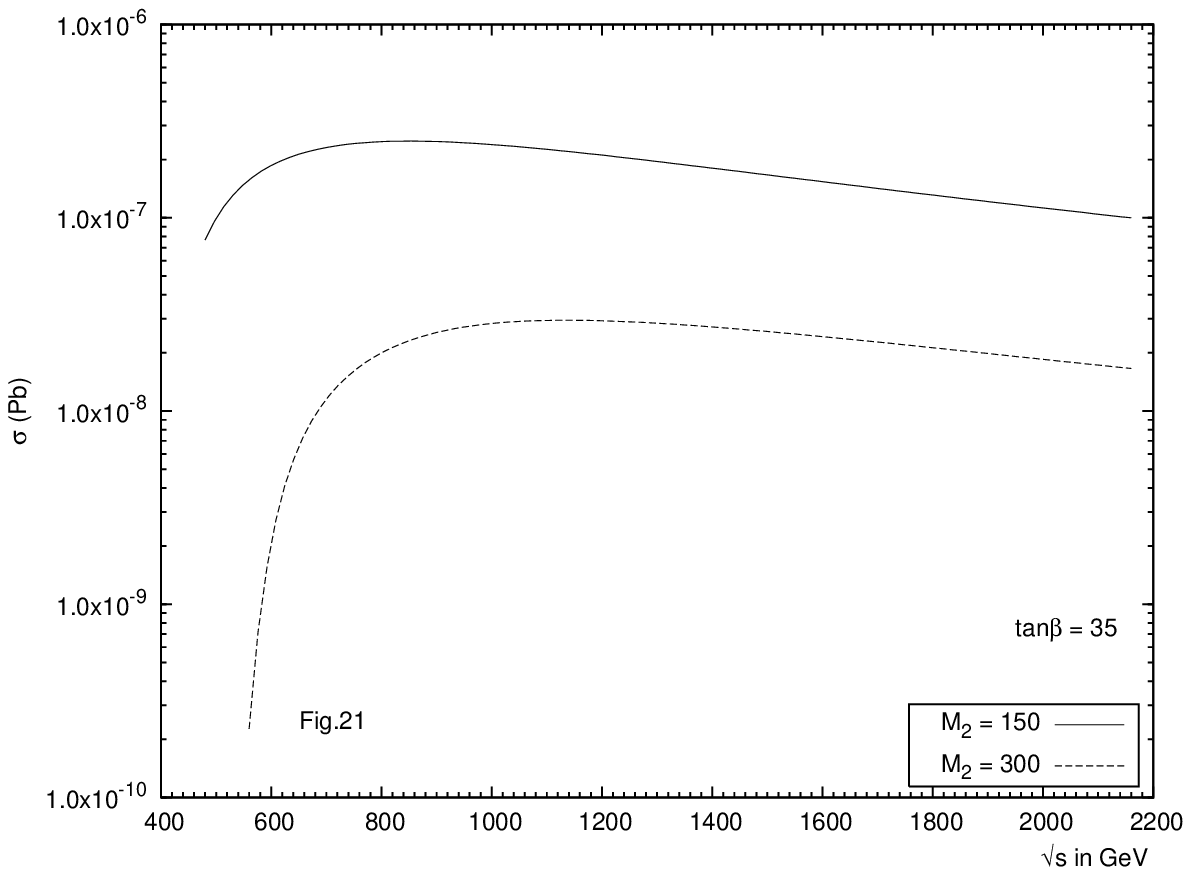}}
\vspace{0.5cm} \caption{ \small Cross sections for
diagram no. 21 in figure \ref{feyn2}} \label{fig.21}
\end{figure}

\begin{figure}[th]
\vspace{-4.5cm}
\centerline{\epsfxsize=4.3truein\epsfbox{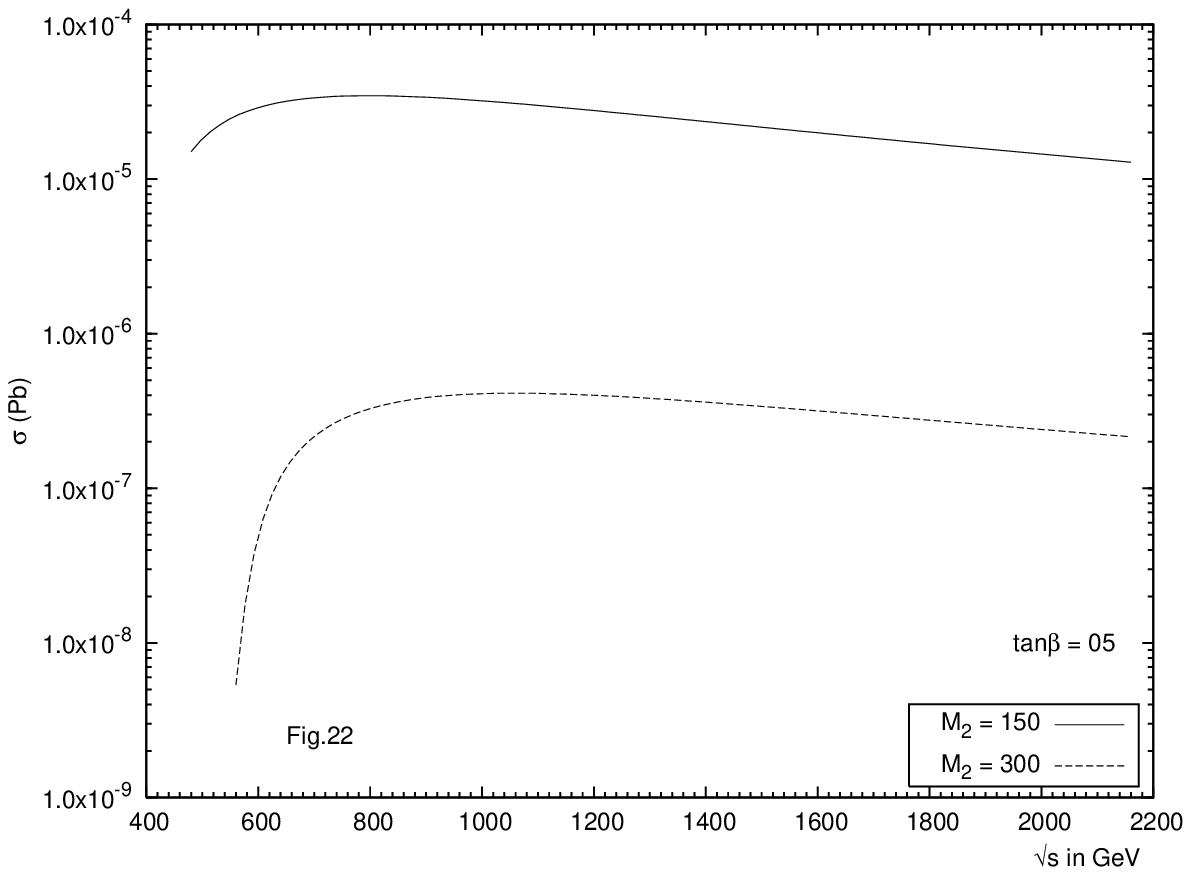}}
\vspace{-0.1cm}
\centerline{\epsfxsize=4.3truein\epsfbox{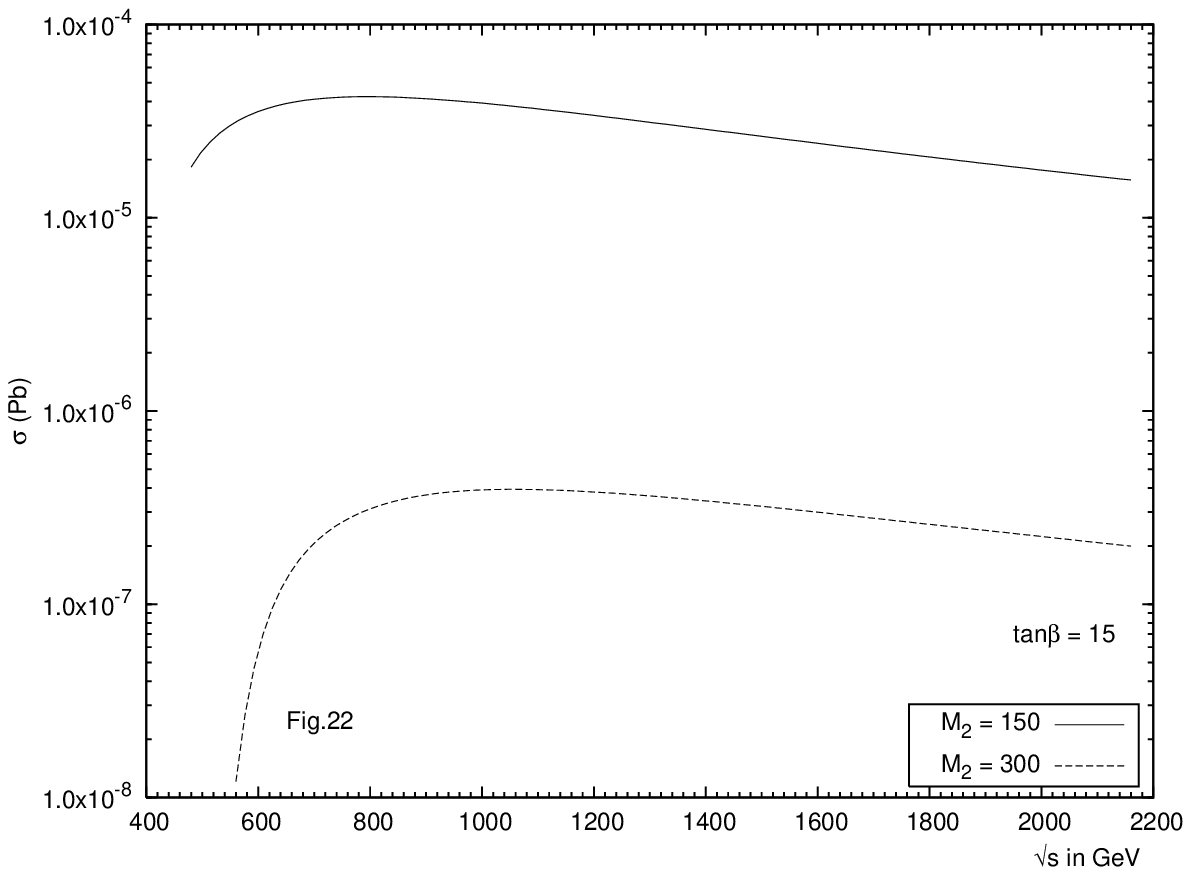}}
\vspace{-0.1cm}
\centerline{\epsfxsize=4.3truein\epsfbox{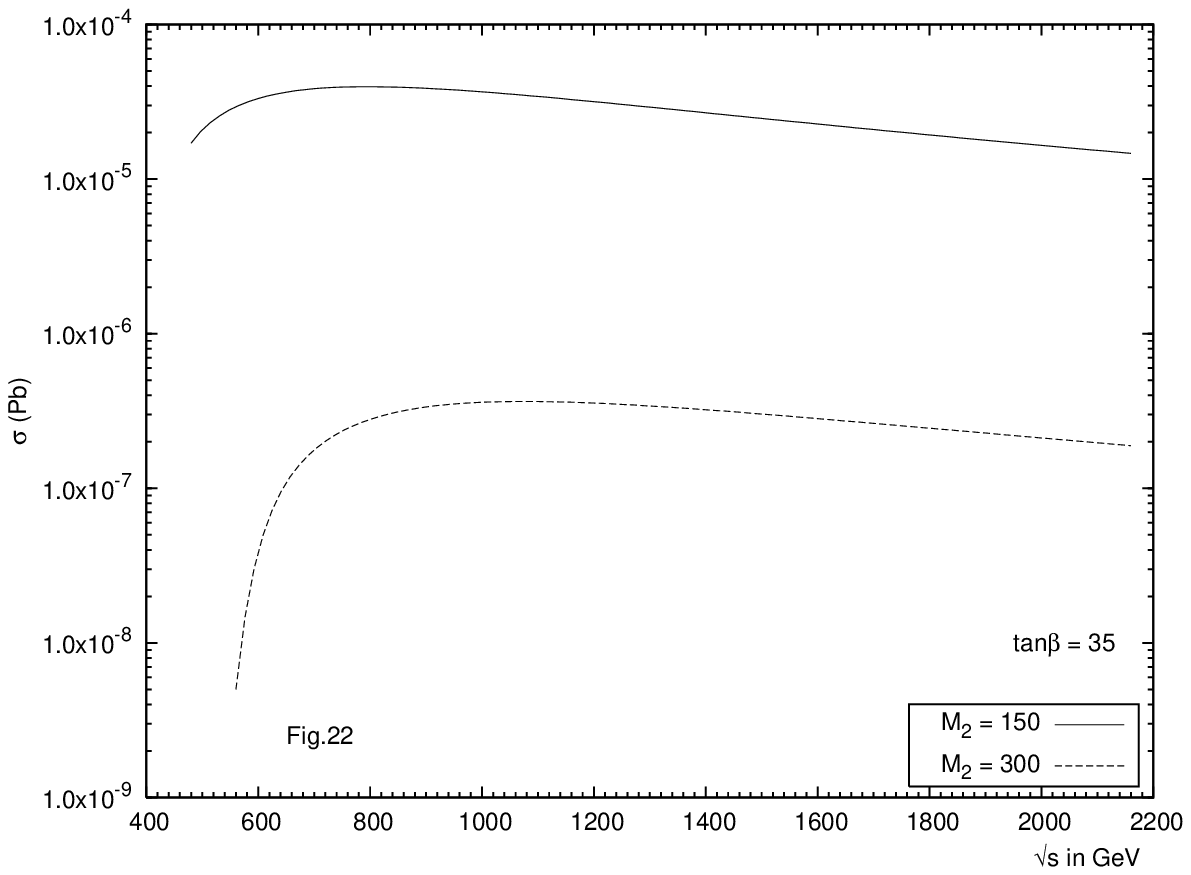}}
\vspace{0.5cm} \caption{ \small Cross sections for
diagram no. 22 in figure \ref{feyn2}} \label{fig.22}
\end{figure}

\begin{figure}[th]
\vspace{-4.5cm}
\centerline{\epsfxsize=4.3truein\epsfbox{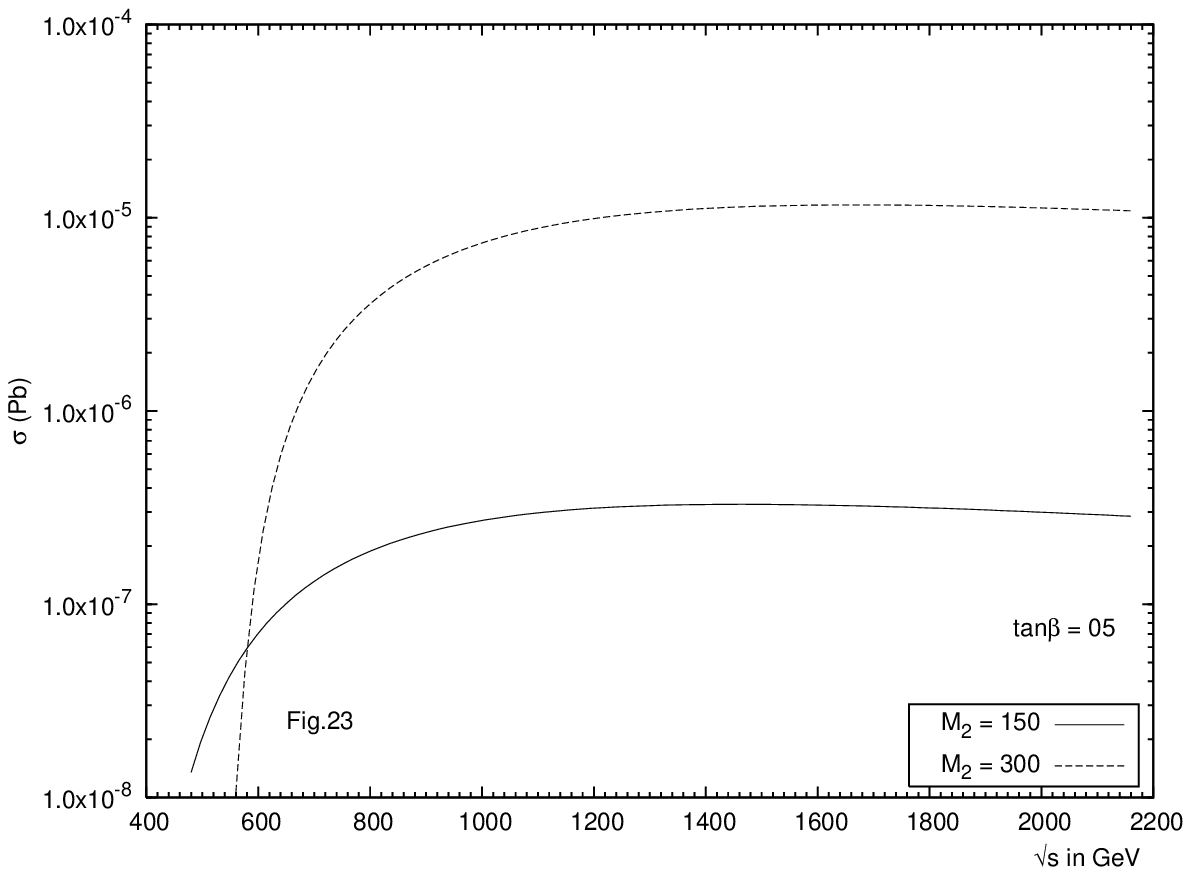}}
\vspace{-0.1cm}
\centerline{\epsfxsize=4.3truein\epsfbox{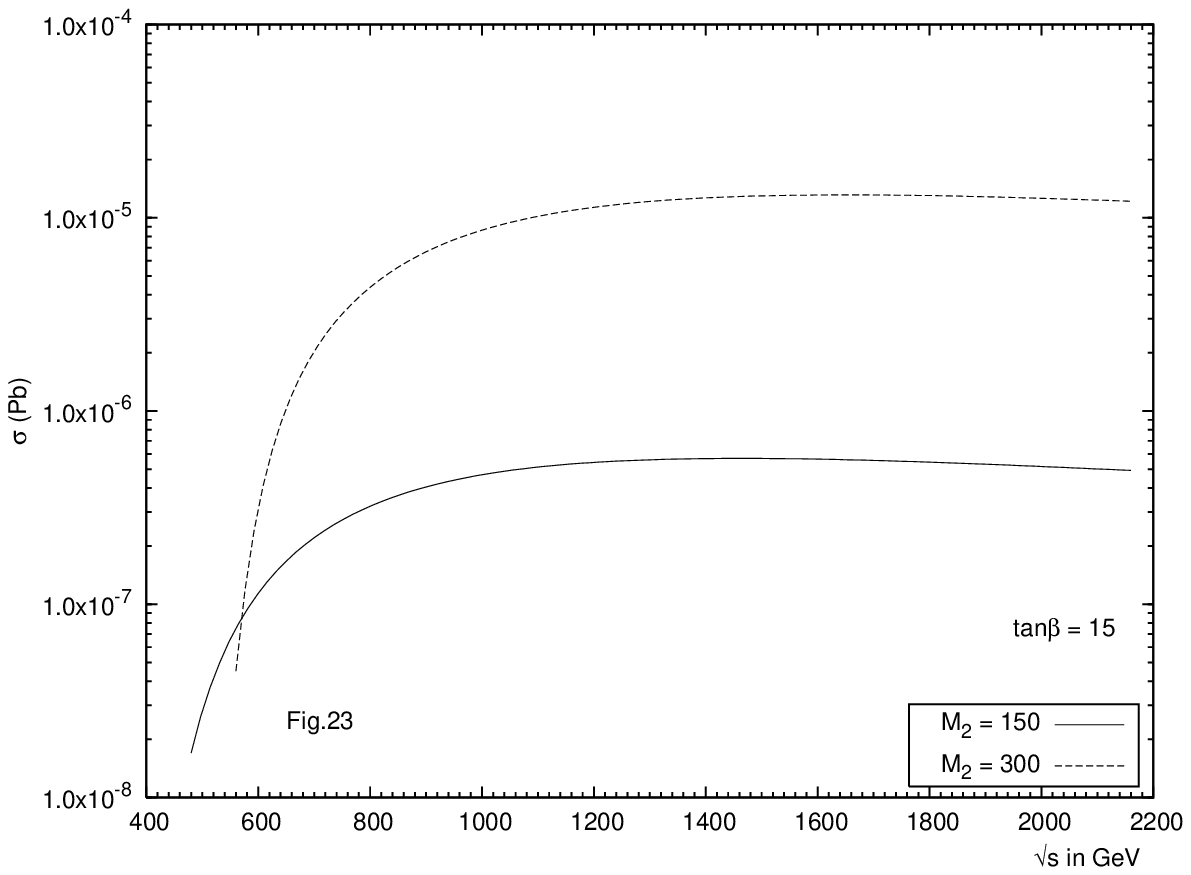}}
\vspace{-0.1cm}
\centerline{\epsfxsize=4.3truein\epsfbox{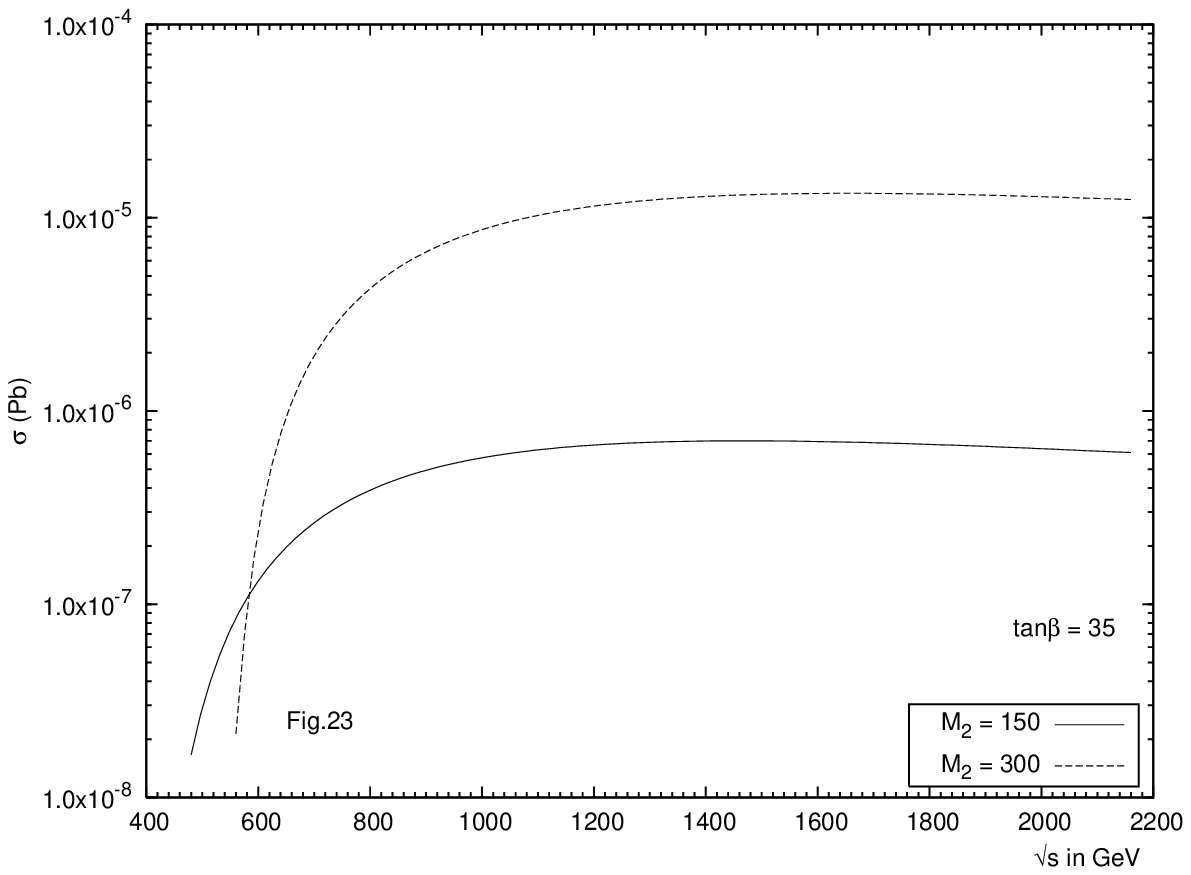}}
\vspace{0.5cm} \caption{ \small Cross sections for
diagram no. 23 in figure \ref{feyn2}} \label{fig.23}
\end{figure}

\begin{figure}[th]
\vspace{-4.5cm}
\centerline{\epsfxsize=4.3truein\epsfbox{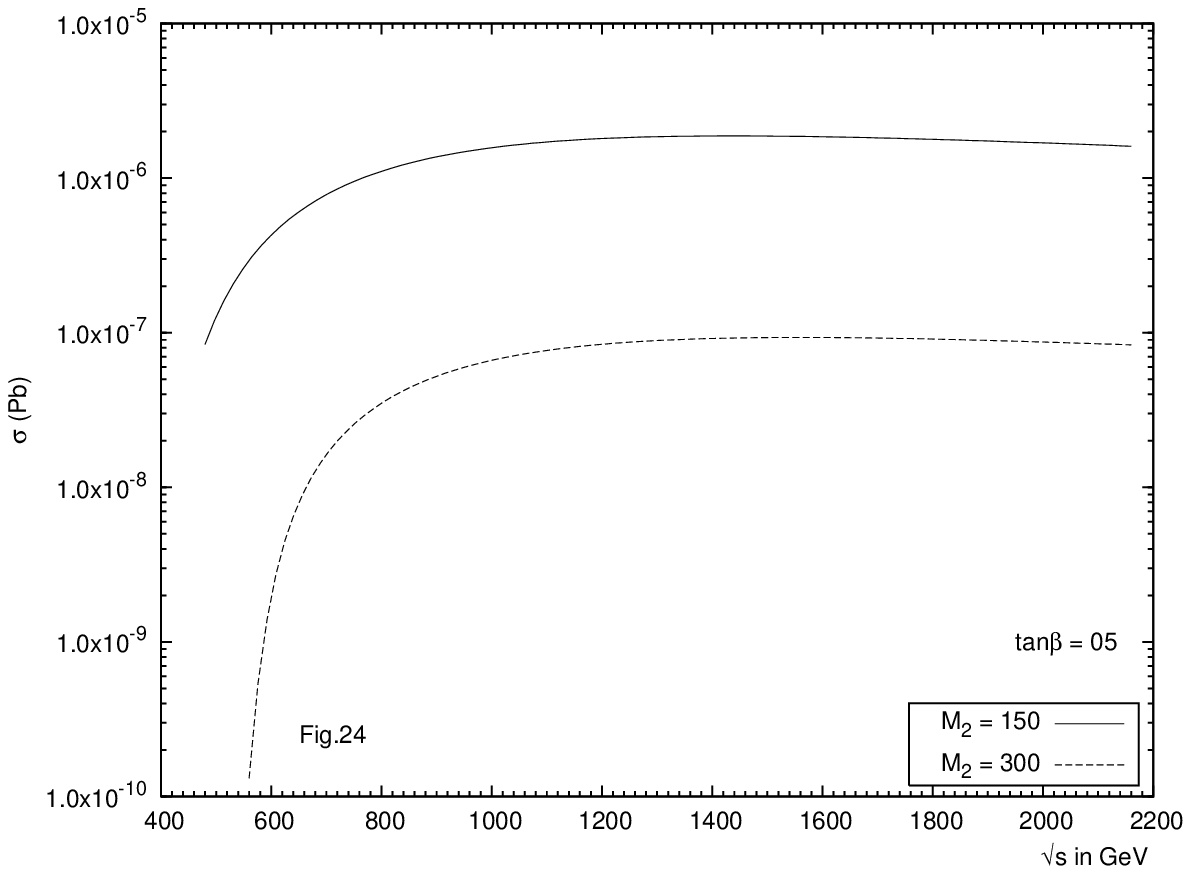}}
\vspace{-0.1cm}
\centerline{\epsfxsize=4.3truein\epsfbox{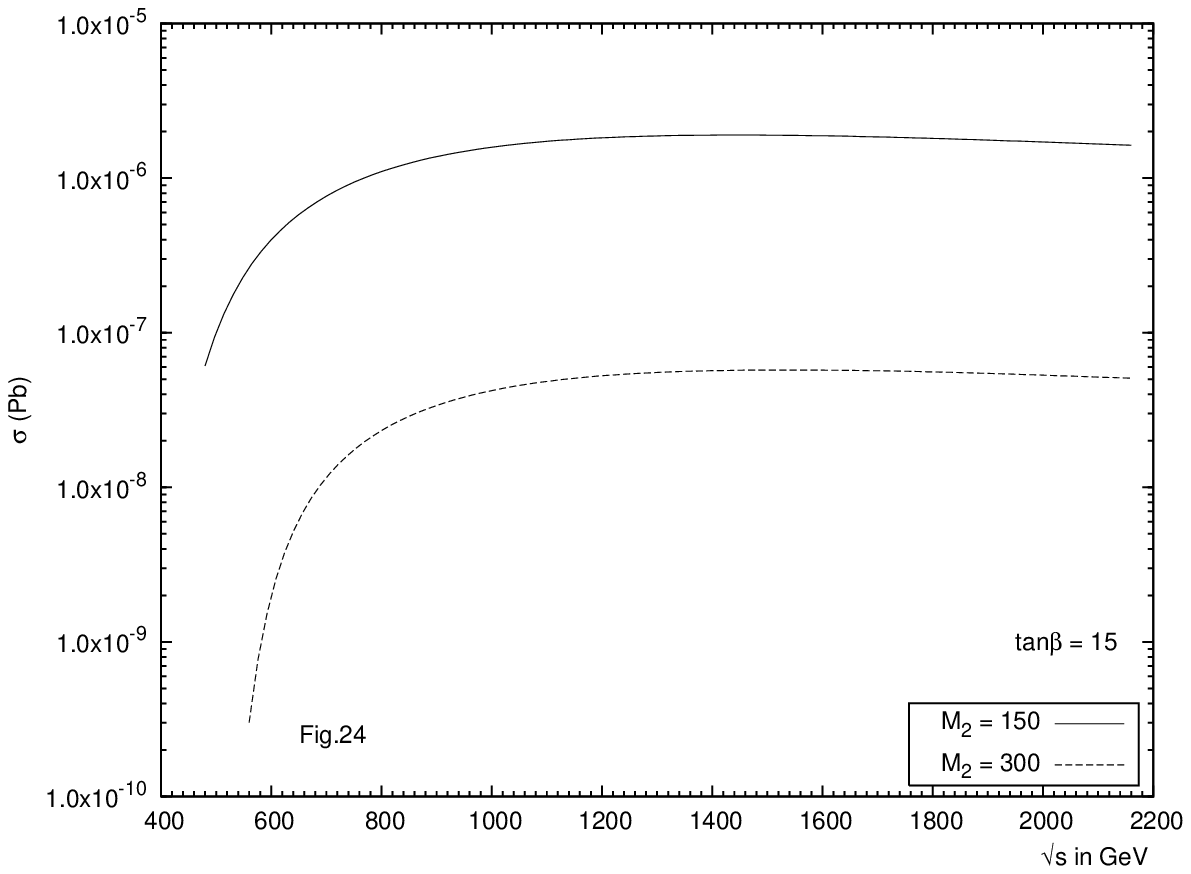}}
\vspace{-0.1cm}
\centerline{\epsfxsize=4.3truein\epsfbox{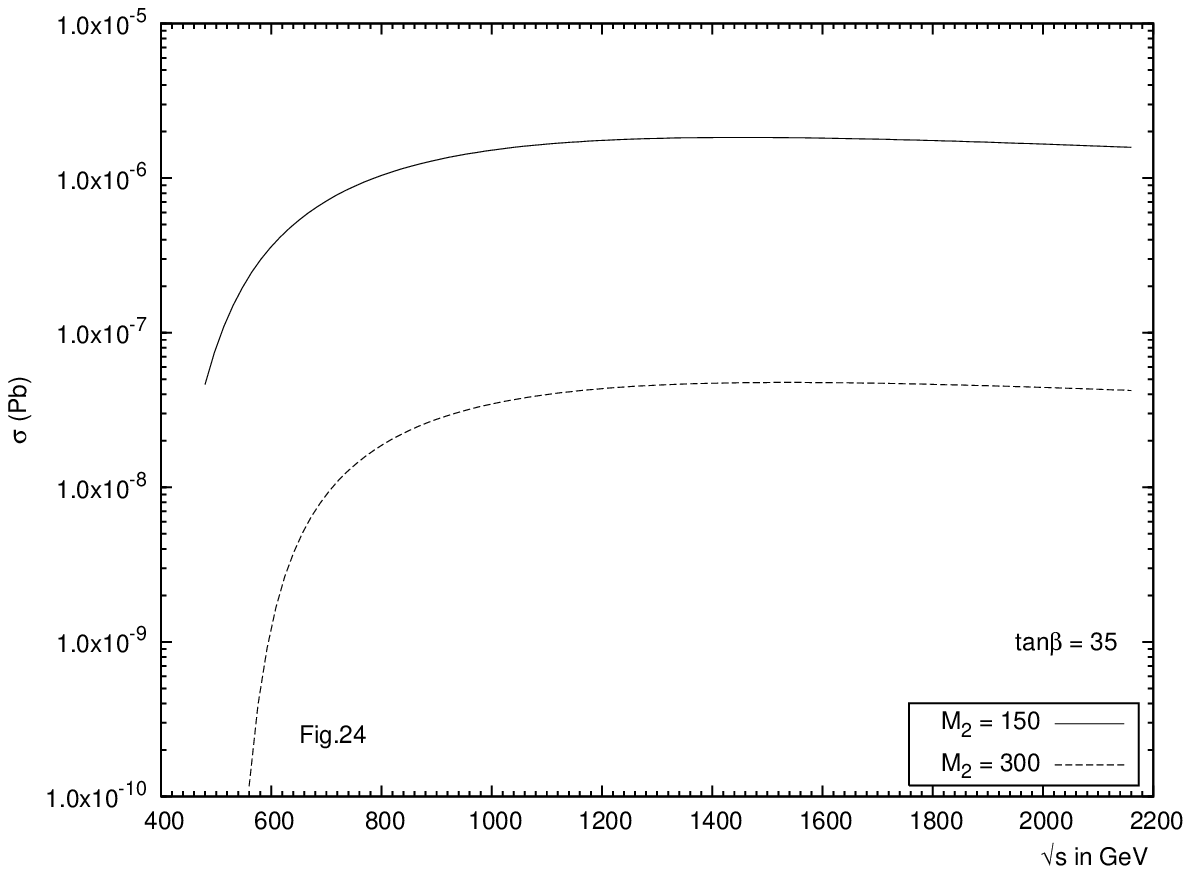}}
\vspace{0.5cm} \caption{ \small Cross sections for
diagram no. 24 in figure \ref{feyn2}} \label{fig.24}
\end{figure}

\begin{figure}[th]
\vspace{-4.5cm}
\centerline{\epsfxsize=4.3truein\epsfbox{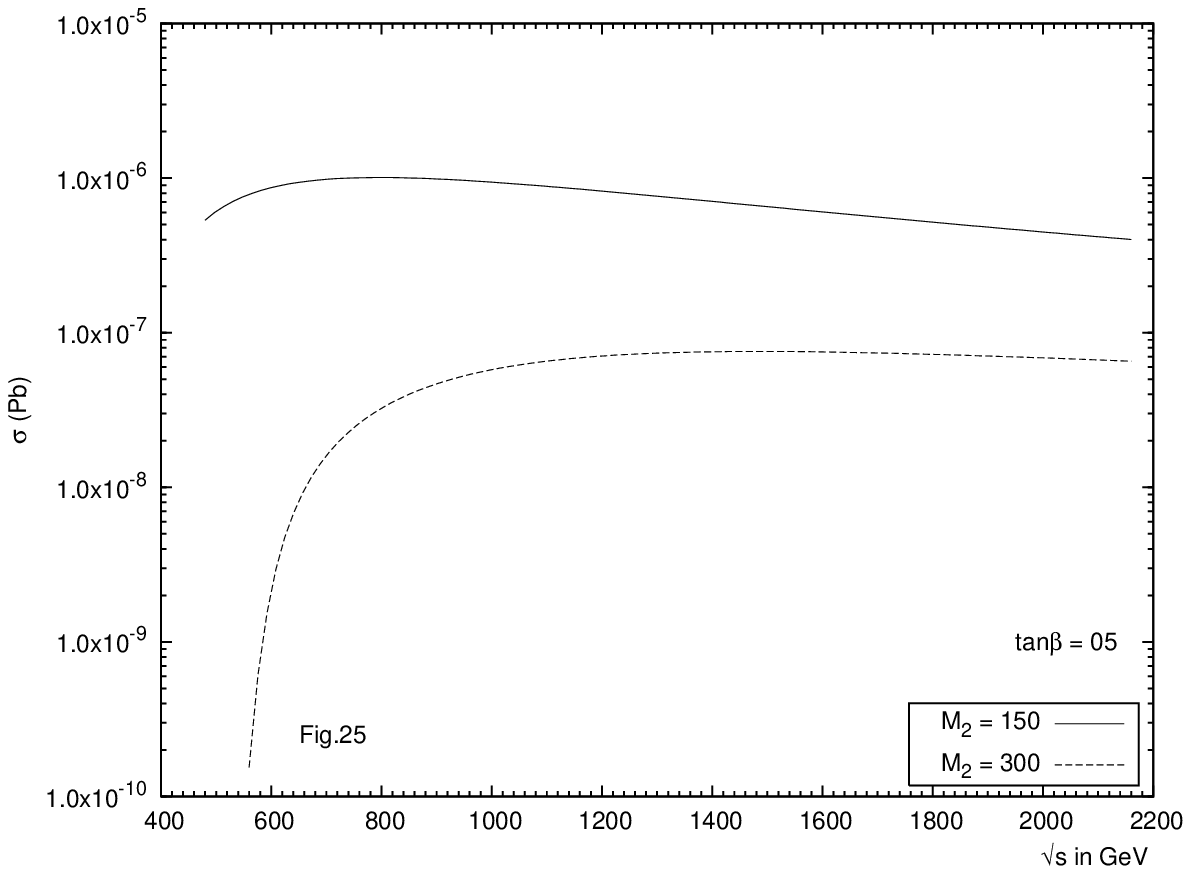}}
\vspace{-0.1cm}
\centerline{\epsfxsize=4.3truein\epsfbox{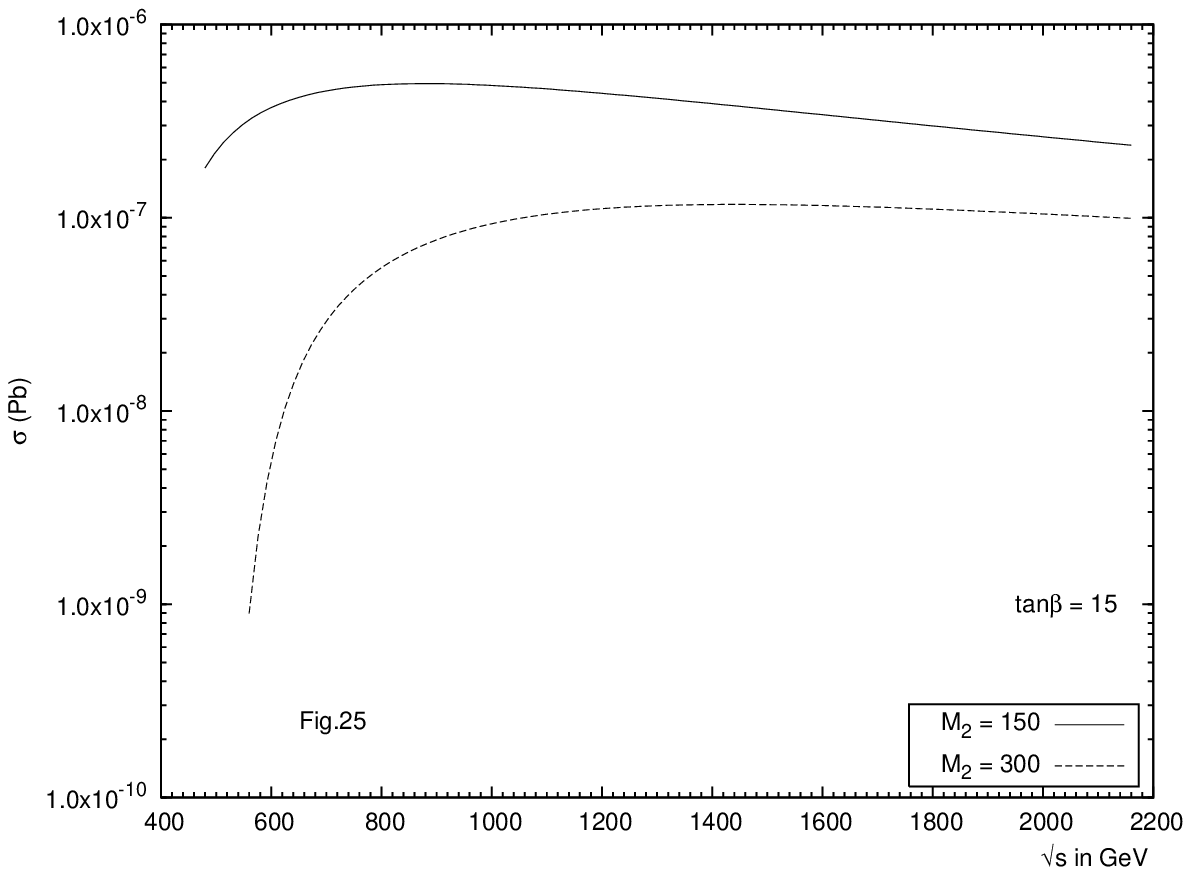}}
\vspace{-0.1cm}
\centerline{\epsfxsize=4.3truein\epsfbox{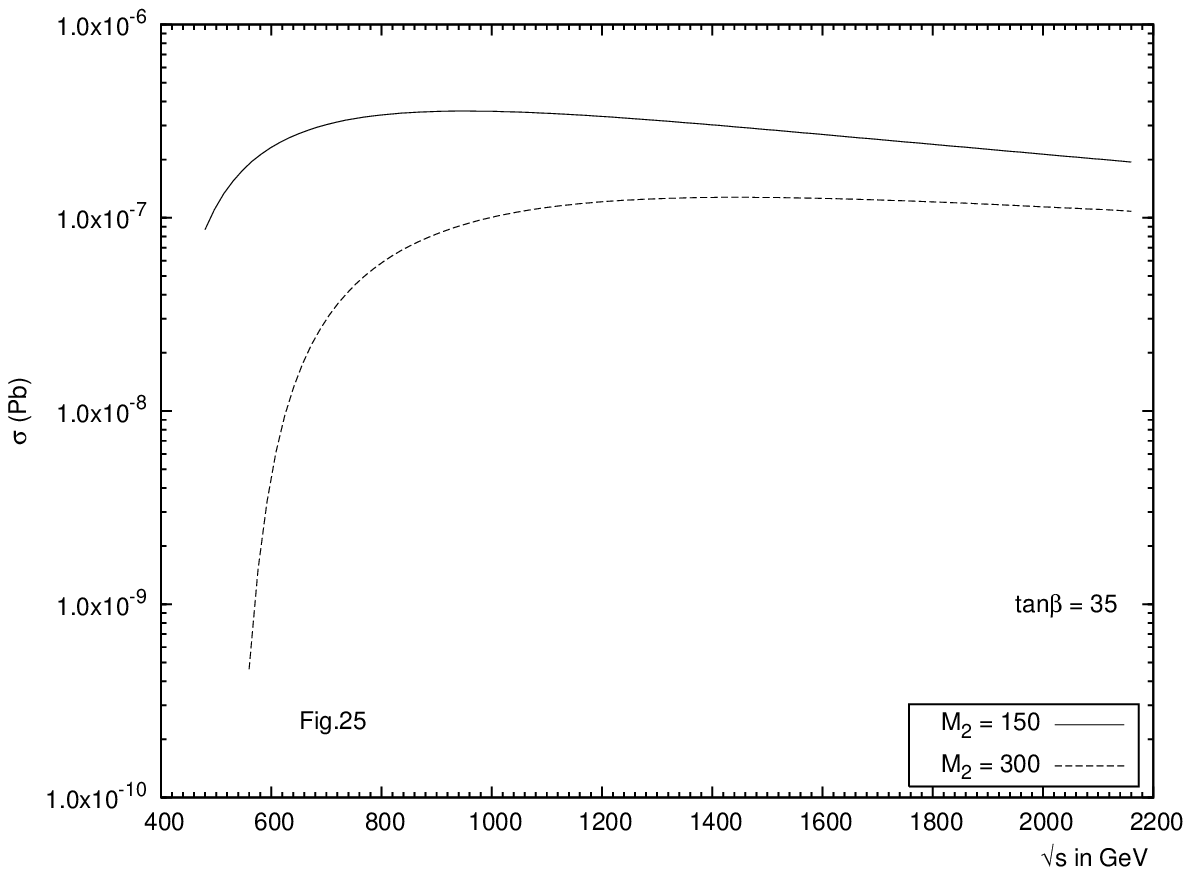}}
\vspace{0.5cm} \caption{ \small Cross sections for
diagram no. 25 in figure \ref{feyn2}} \label{fig.25}
\end{figure}

\begin{figure}[th]
\vspace{-4.5cm}
\centerline{\epsfxsize=4.3truein\epsfbox{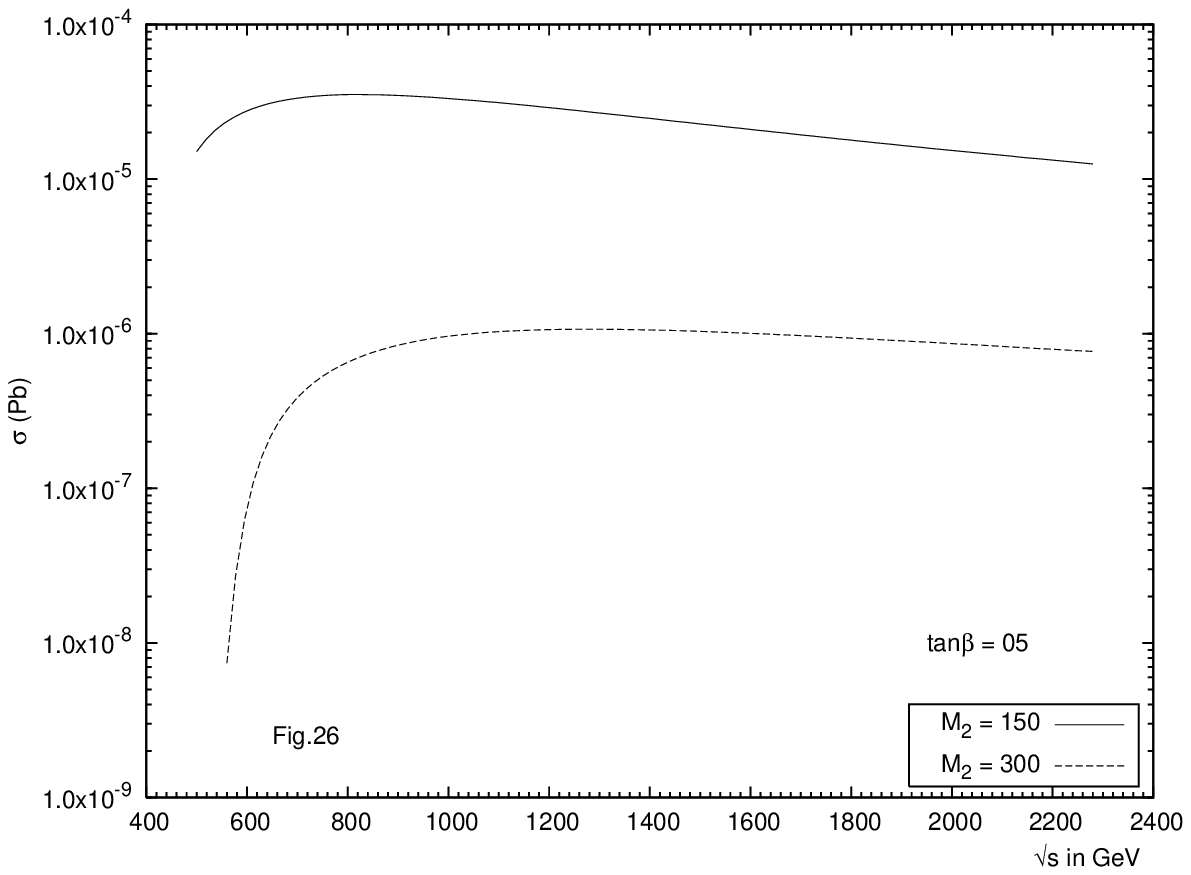}}
\vspace{-0.1cm}
\centerline{\epsfxsize=4.3truein\epsfbox{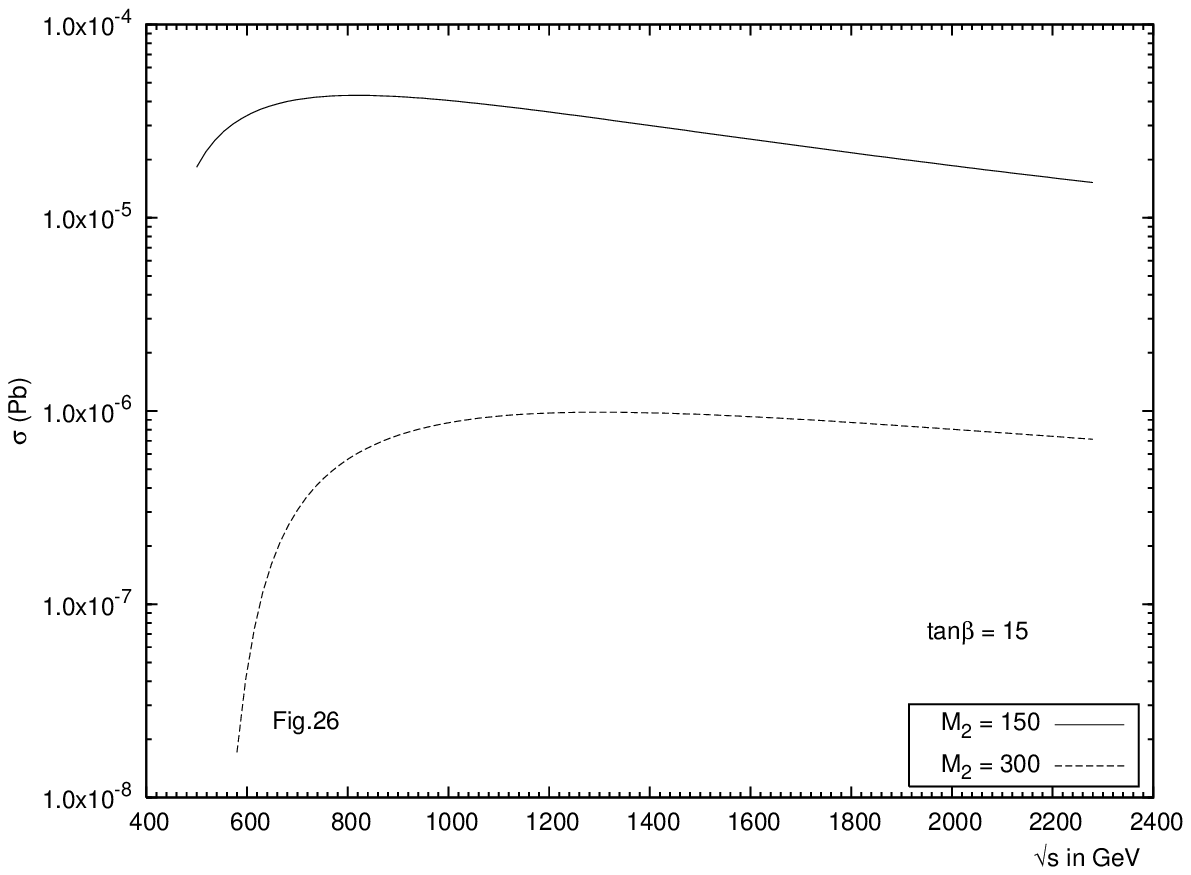}}
\vspace{-0.1cm}
\centerline{\epsfxsize=4.3truein\epsfbox{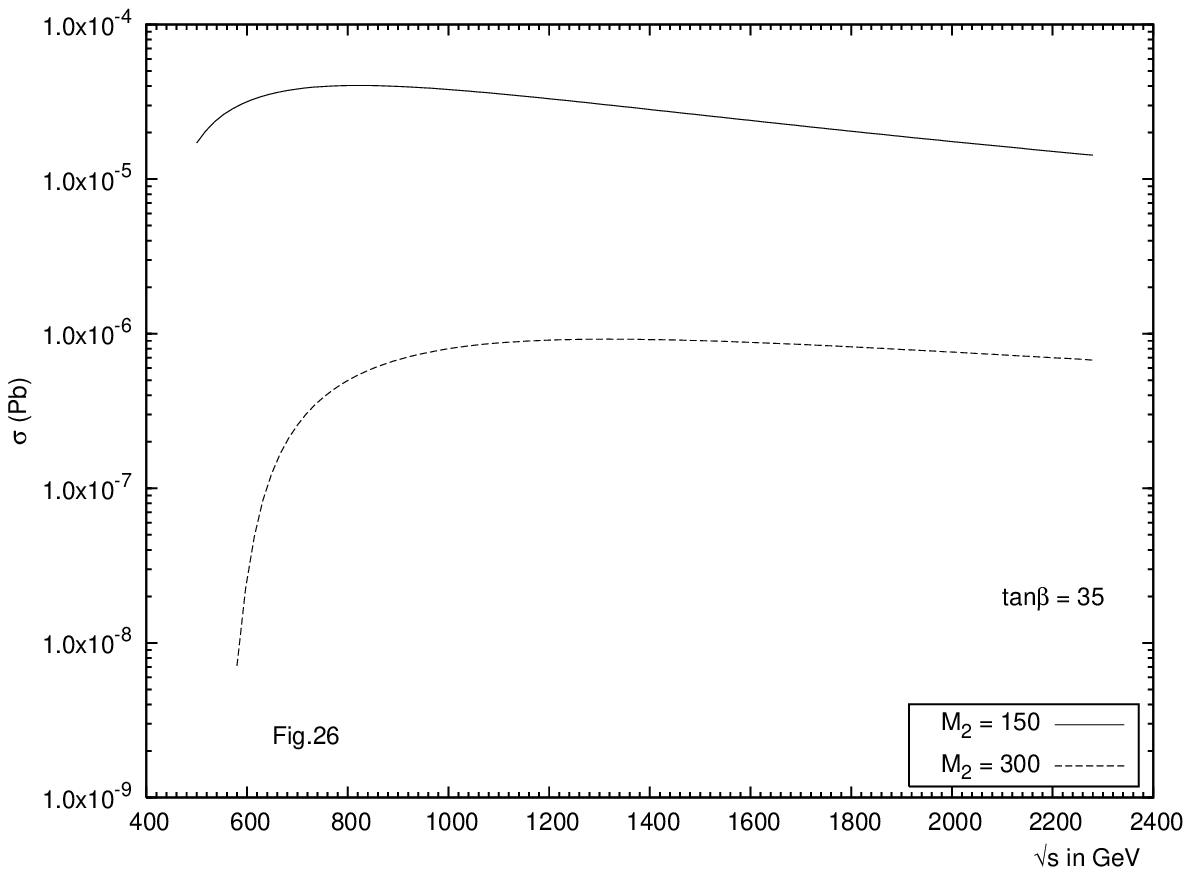}}
\vspace{0.5cm} \caption{ \small Cross sections for
diagram no. 26 in figure \ref{feyn2}} \label{fig.26}
\end{figure}

\begin{figure}[th]
\vspace{-4.5cm}
\centerline{\epsfxsize=4.3truein\epsfbox{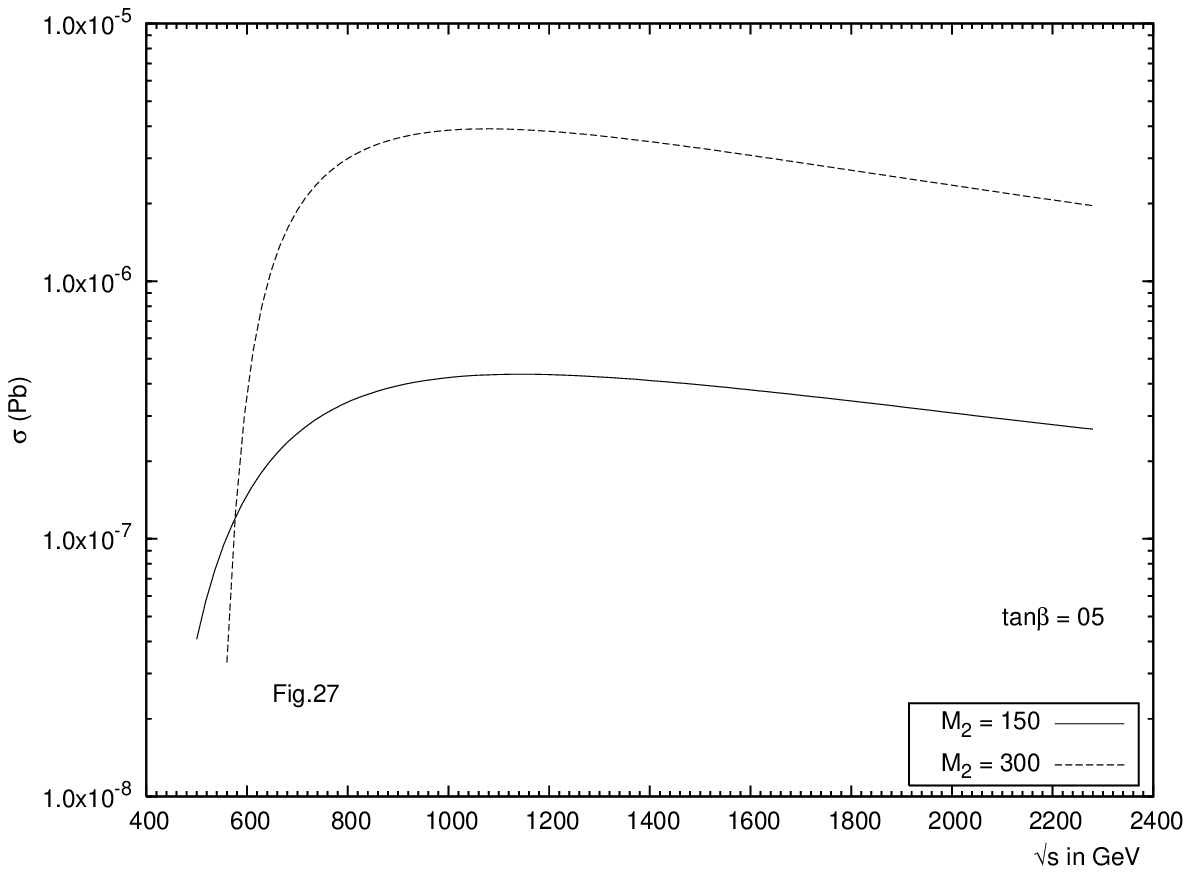}}
\vspace{-0.1cm}
\centerline{\epsfxsize=4.3truein\epsfbox{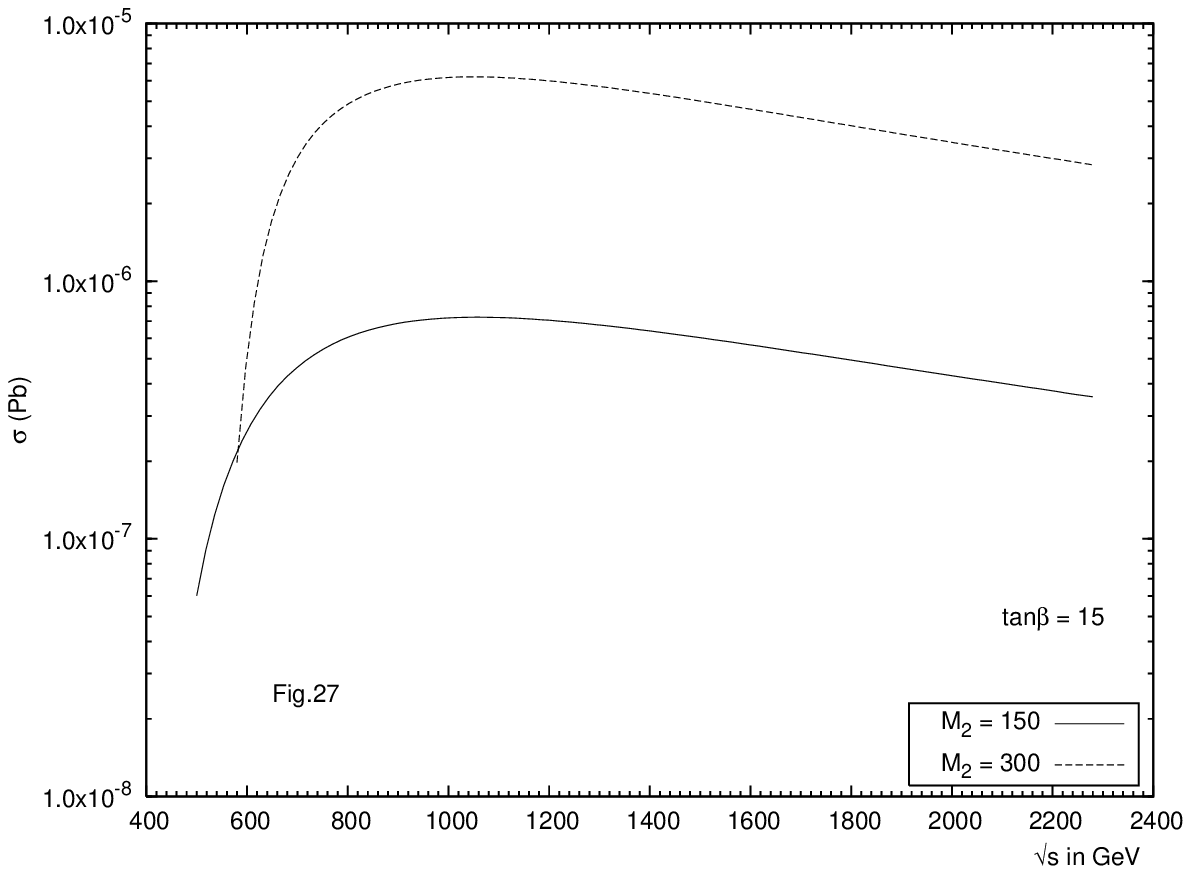}}
\vspace{-0.1cm}
\centerline{\epsfxsize=4.3truein\epsfbox{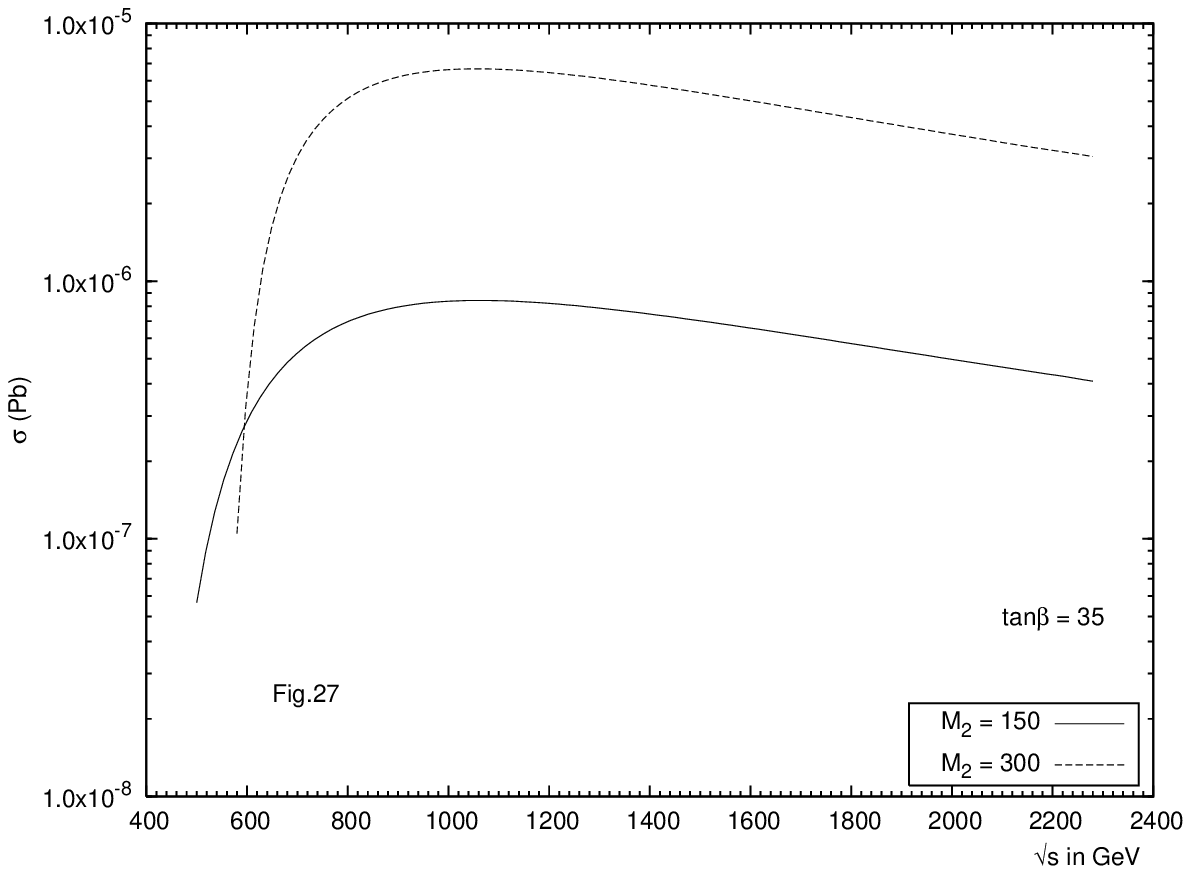}}
\vspace{0.5cm} \caption{ \small Cross sections for
diagram no. 27 in figure \ref{feyn2}} \label{fig.27}
\end{figure}

\begin{figure}[th]
\vspace{-4.5cm}
\centerline{\epsfxsize=4.3truein\epsfbox{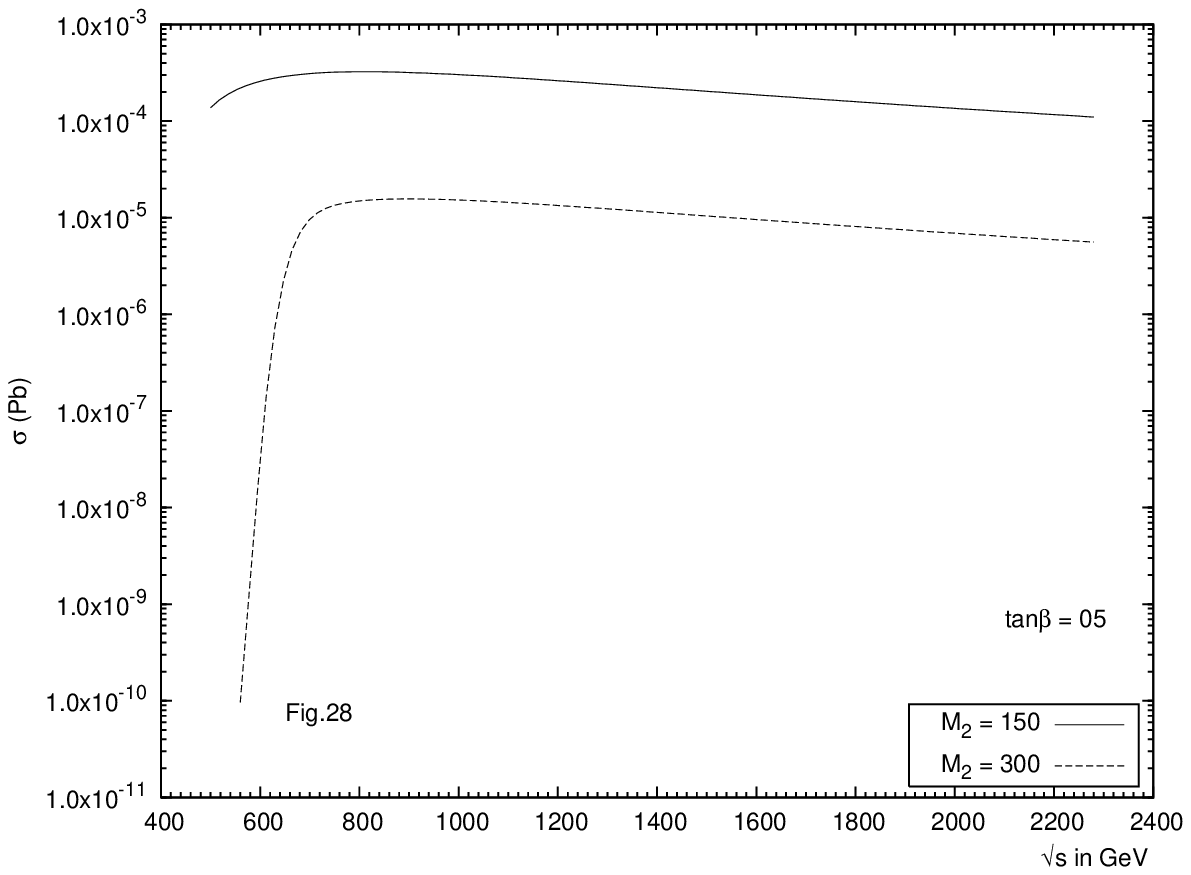}}
\vspace{-0.1cm}
\centerline{\epsfxsize=4.3truein\epsfbox{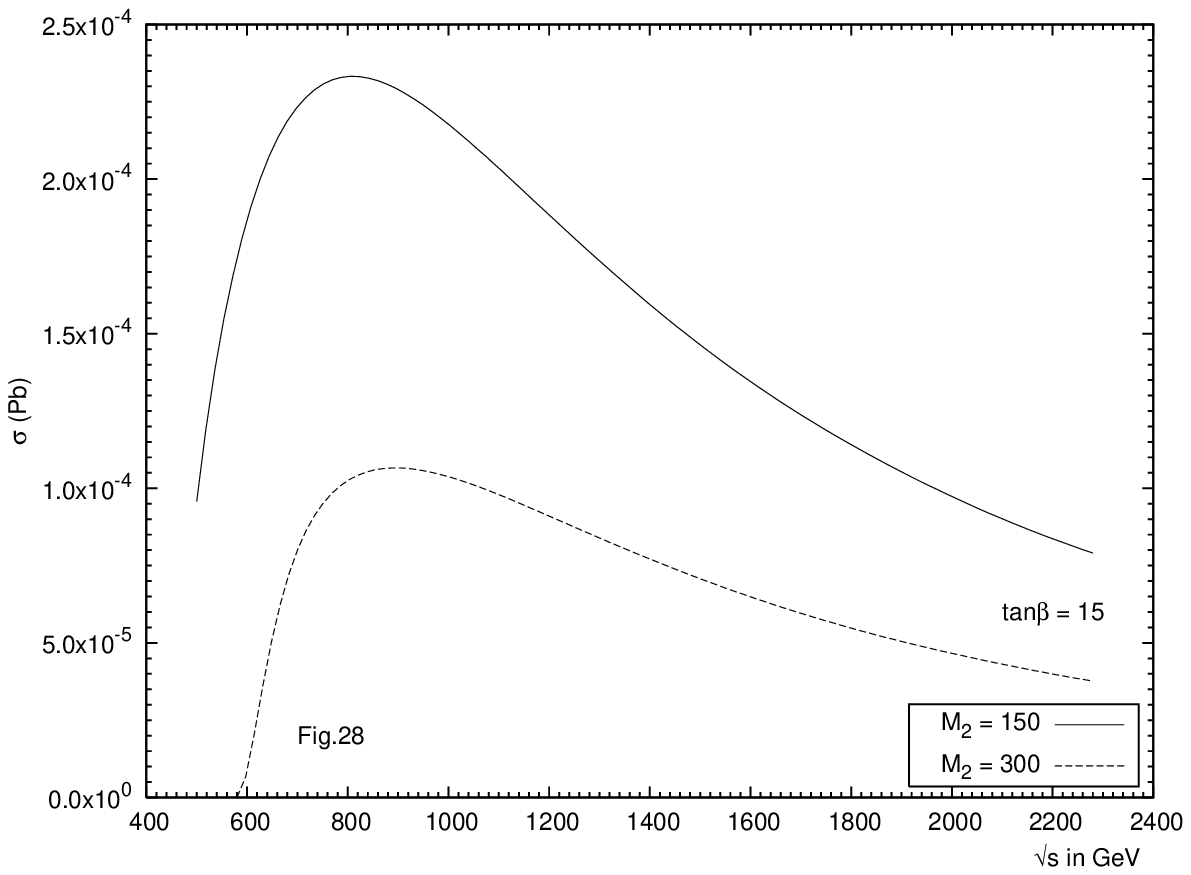}}
\vspace{-0.1cm}
\centerline{\epsfxsize=4.3truein\epsfbox{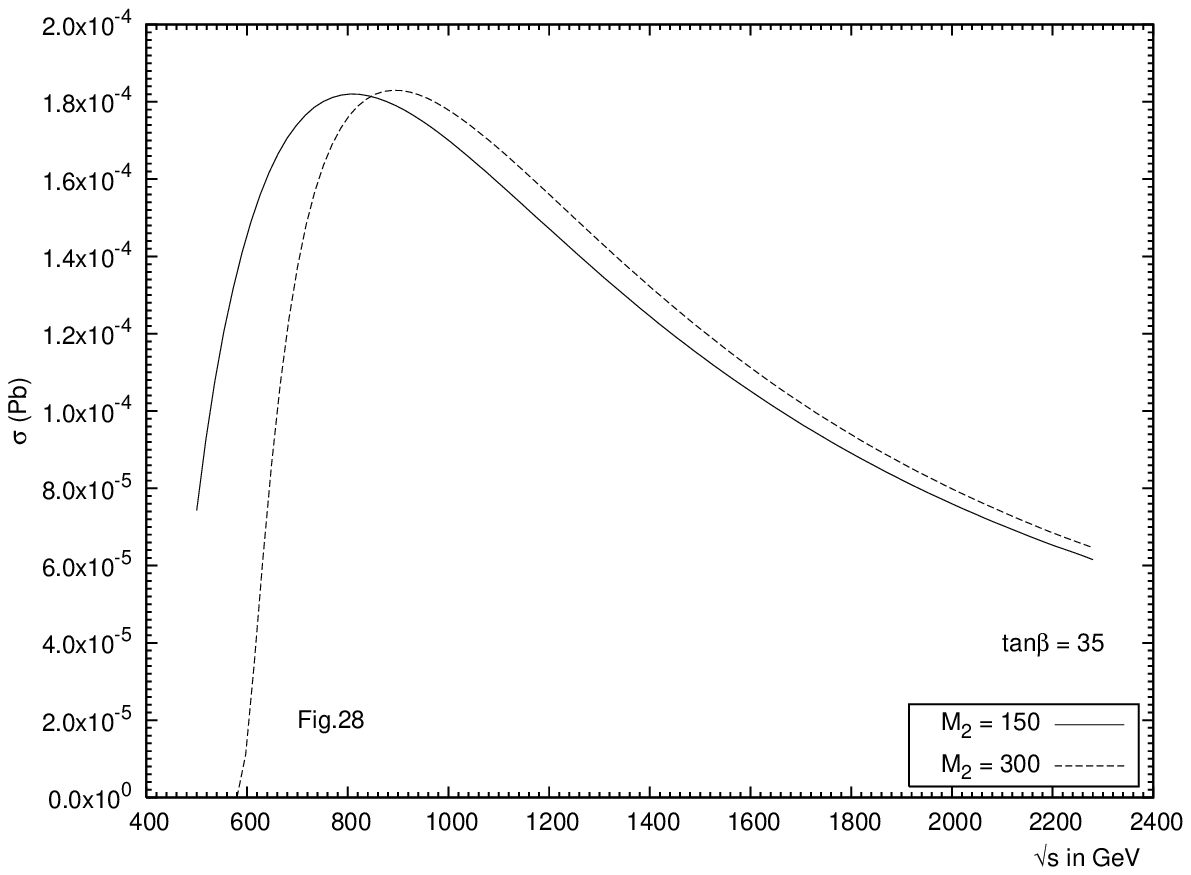}}
\vspace{0.5cm} \caption{ \small Cross sections for
diagram no. 28 in figure \ref{feyn2}} \label{fig.28}
\end{figure}

\begin{figure}[th]
\vspace{-4.5cm}
\centerline{\epsfxsize=4.3truein\epsfbox{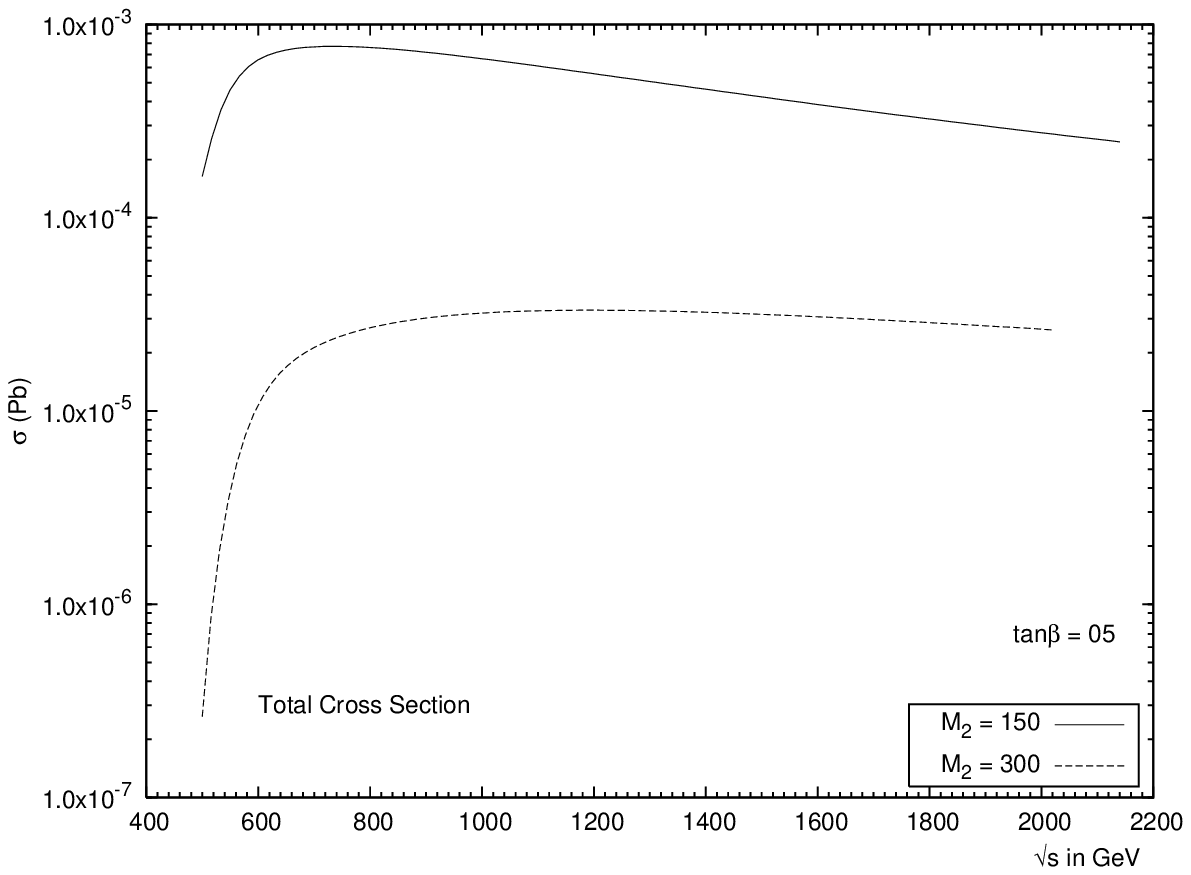}}
\vspace{-0.1cm}
\centerline{\epsfxsize=4.3truein\epsfbox{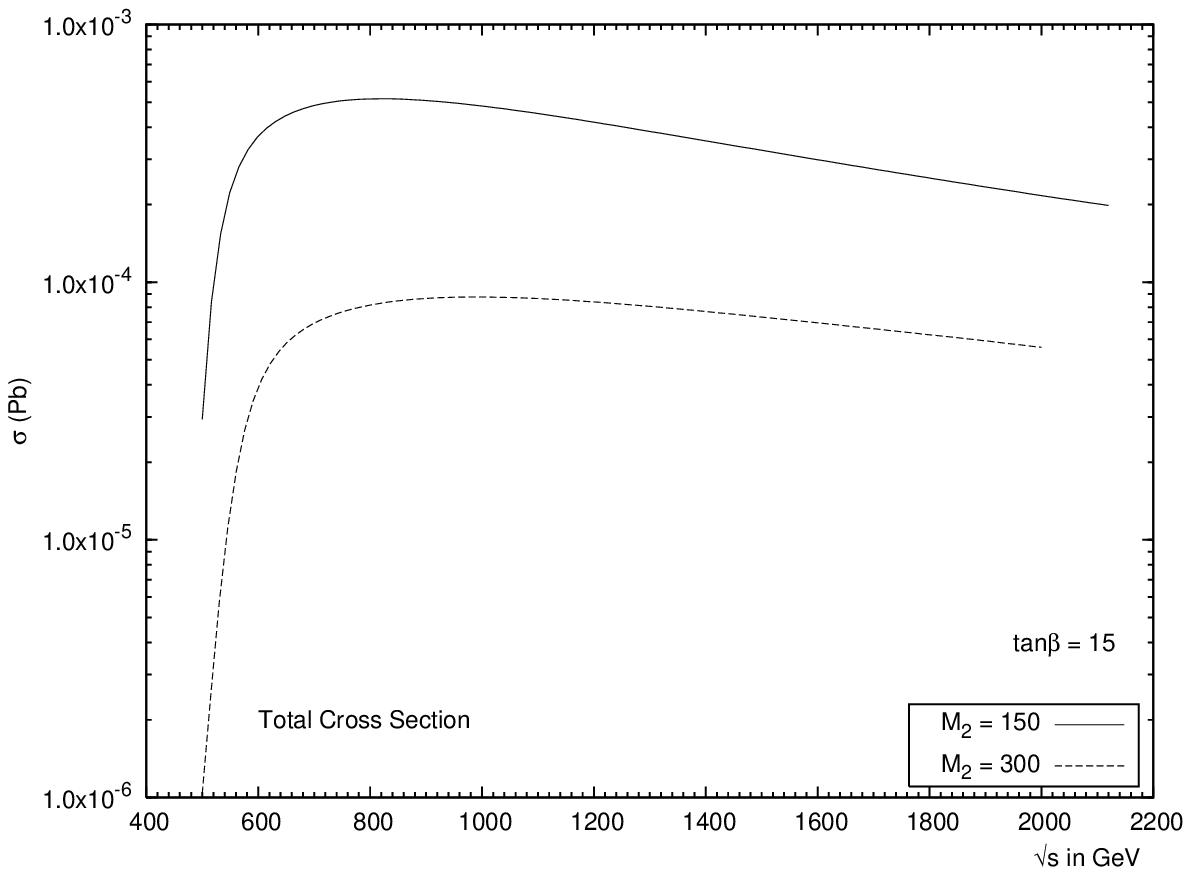}}
\vspace{-0.1cm}
\centerline{\epsfxsize=4.3truein\epsfbox{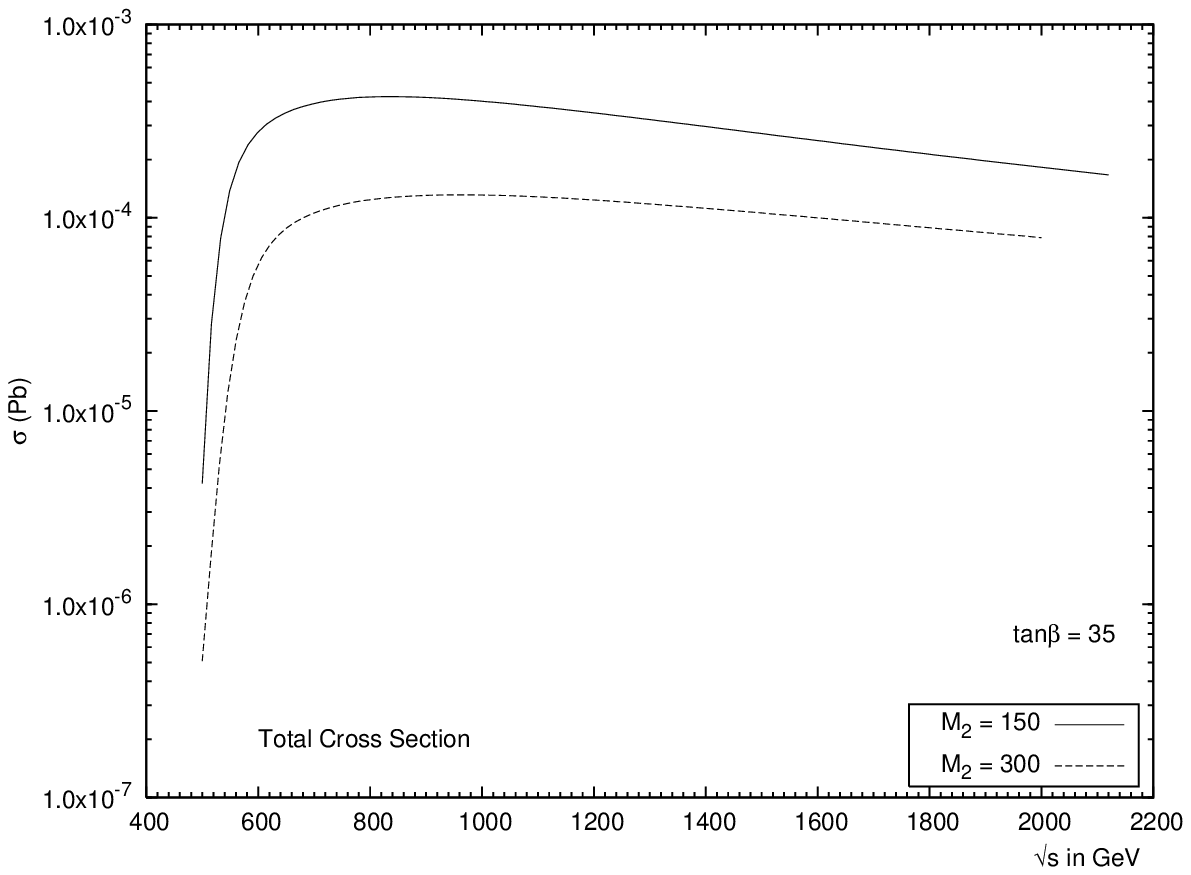}}
\vspace{0.5cm} \caption{ \small Total cross section for the
reaction $e^-(p1) e^+(p2) \rightarrow  H^-(p3) \widetilde{\chi
}_{1}^+(p4) \widetilde{\chi }_{1}^o(p5)$} \label{total}
\end{figure}
\clearpage
\section{Conclusion}
Results of the previous section are summarized in tables
\ref{table1} and \ref{table2} for $M_2 = 150$ GeV, and tables
\ref{table3} and \ref{table4} for $M_2 = 300$ GeV.\\ 
From these results, it is clear that the reaction is most probably to proceed
through diagram no. 5 through the exchange of a $Z$ boson. The maximum cross sextion achieved reached the value of $3.4611\times 10^{-4}$ [pb] for $tan\beta$ = 5 and $M_2$ = 150 GeV at a center of mass energy, $E_{CM}$ = 560 GeV. A maximum of $3.0094\times 10^{_4}$ [pb] was obtained for $tan\beta$ = 15 and $M_2$ = 150 GeV at a center of mass energy, $E_{CM}$ = 560 GeV and again for the diagram no. 5. For $tan\beta$ = 35 and $M_2$ = 150 GeV at a center of mass energy, $E_{CM}$ = 580 GeV, the cross section obtained, reached the value of $2.5145\times 10^{-4}$ [pb].\\
The maximum cross sextion achieved reached the value of $1.6853\times 10^{-5}$ [pb] for $tan\beta$ = 5 and $M_2$ = 300 GeV at a center of mass energy, $E_{CM}$ = 880 GeV for the diagram no. 2 which proceeds through the photon propagator, $\gamma$. A maximum of $1.0836\times 10^{_4}$ [pb] was obtained for $tan\beta$ = 15 and $M_2$ = 300 GeV at a center of mass energy, $E_{CM}$ = 860 GeV for the diagram no. 28 which proceeds through the exchange of the scalar neutrino propagator, $\widetilde{\nu}_e$.
For $tan\beta$ = 35 and $M_2$ = 300 GeV at a center of mass energy, $E_{CM}$ = 900 GeV, the cross section obtained, reached the value of $1.8623\times 10^{-4}$ [pb] for diagram no. 28.\\
total cross section achieved the following values,
\begin{enumerate}
\item For $tan\beta$ = 5 and $M_2$ = 150, $\sigma_{max}$ = $7.8383\times 10^{-4}$ [pb] at $E_{CM}$ = 720 GeV.
\item For $tan\beta$ = 15 and $M_2$ = 150, $\sigma_{max}$ = $5.2422\times 10^{-4}$ [pb] at $E_{CM}$ = 820 GeV.
\item For $tan\beta$ = 35 and $M_2$ = 150, $\sigma_{max}$ = $4.3095\times 10^{-4}$ [pb] at $E_{CM}$ = 800 GeV.
\item For $tan\beta$ = 5 and $M_2$ = 300, $\sigma_{max}$ = $3.3550\times 10^{-4}$ [pb] at $E_{CM}$ = 1080 GeV.
\item For $tan\beta$ = 15 and $M_2$ = 300, $\sigma_{max}$ = $8.8536\times 10^{-5}$ [pb] at $E_{CM}$ = 960 GeV.
\item For $tan\beta$ = 35 and $M_2$ = 300, $\sigma_{max}$ = $1.3291\times 10^{-4}$ [pb] at $E_{CM}$ = 940 GeV.
\end{enumerate}

\begin{table}[htbp]
\begin{center}
\begin{tabular}[htbp]{|c||c|c||c|c||c|c|}
  \hline
  \hline
  Figure No. & \multicolumn{2}{c|}{$\sigma_{tan\beta = 5}$} & \multicolumn{2}{c|}{$\sigma_{tan\beta = 15}$} & \multicolumn{2}{c|}{$\sigma_{tan\beta = 35}$}\\
  \cline{2-7}
  &$E_{CM}$& $\sigma$ (Pb)&$E_{CM}$& $\sigma$ (Pb)&$E_{CM}$& $\sigma$ (Pb)\\
  \hline
  1 & 1040 & 1.5779e-06 & 1140& 1.5906e-06& 1160& 1.6421e-06\\
  2 & 900 & 2.0289e-06 & 880& 2.6311e-06& 920& 2.7928e-06\\
  3 & 1080 & 2.2854e-07 & 1140& 2.3085e-07& 1120& 2.3836e-07\\
  4 & 860 & 1.8844e-06 & 840& 2.4976e-06& 860& 2.6618e-06\\
  5 & 560 & 0.00034611 & 560& 0.00030094& 580& 0.00025145\\
  6 & 1000 & 5.3209e-09 & 1060& 7.9594e-09& 1060& 8.5888e-09\\
  7 & 1000 & 2.2084e-10 & 1060& 5.1532e-10& 1060& 6.4742e-10\\
  8 & 520 & 0.000346 & 560& 6.9141e-05& 600& 3.2496e-05\\
  9 & 720 & 1.9348e-05 & 720& 3.2102e-05& 740& 3.167e-05\\
  10 & 1480 & 8.8544e-08 & 1500& 6.5752e-08& 1480& 5.2112e-08\\
  11 & 1420 & 1.2839e-06 & 1420& 9.6548e-07& 1460& 7.9427e-07\\
  12 & 780 & 6.4527e-07 & 880& 2.0701e-07& 940& 1.1992e-07\\
  13 & 800 & 2.6592e-05 & 800& 2.125e-05& 780& 1.6044e-05\\
  14 & 1160 & 1.0001e-08 & 1160& 9.3249e-10& 1160& 1.4909e-10\\
  15 & 1200 & 3.5258e-09 & 1020& 3.4863e-09& 1040& 3.04e-09\\
  16 & 1080 & 7.6575e-08 & 1040& 6.8391e-08& 1040& 5.8846e-08\\
  \hline
\end{tabular}
\caption{Summary of the results obtained for the reaction,
$e^-(p1) e^+(p2) \rightarrow  H^-(p3) \widetilde{\chi }_{1}^+(p4)
\widetilde{\chi }_{1}^o(p5)$ for $M_2 = 150$ GeV} \label{table1}
\end{center}
\end{table}

\begin{table}[thbp]
\begin{center}
\begin{tabular}[thbp]{|c||c|c||c|c||c|c|}
  \hline
  Figure No. & \multicolumn{2}{c|}{$\sigma_{tan\beta = 5}$} & \multicolumn{2}{c|}{$\sigma_{tan\beta = 15}$} & \multicolumn{2}{c|}{$\sigma_{tan\beta = 35}$}\\
  \cline{2-7}
  &$E_{CM}$& $\sigma$ (Pb)&$E_{CM}$& $\sigma$ (Pb)&$E_{CM}$& $\sigma$ (Pb)\\
  \hline
  \hline
  17 & 760 & 6.1116e-07 & 800& 1.6734e-07& 820& 8.3768e-08\\
  18 & 800 & 2.6135e-05 & 780& 2.0918e-05& 800& 1.5747e-05\\
  19 & 1180 & 1.3204e-08 & 1060& 3.0367e-08& 1040& 4.1023e-08\\
  20 & 1060 & 1.1262e-07 & 1060& 1.3572e-07& 1020& 1.3639e-07\\
  21 & 780 & 9.6615e-07 & 800& 4.0522e-07& 820& 2.5265e-07\\
  22 & 780 & 3.5148e-05 & 760& 4.295e-05& 780& 4.0162e-05\\
  23 & 1440 & 3.3033e-07 & 1420& 5.7192e-07& 1460& 7.0432e-07\\
  24 & 1440 & 1.8847e-06 & 1420& 1.9127e-06& 1460& 1.8439e-06\\
  25 & 780 & 1.023e-06 & 880& 5.0101e-07& 960& 3.6067e-07\\
  26 & 820 & 3.5797e-05 & 820& 4.3613e-05& 820& 4.0866e-05\\
  27 & 1120 & 4.4032e-07 & 1040& 7.3516e-07& 1080& 8.5312e-07\\
  28 & 800 & 0.00033039 & 780& 0.00023661& 800& 0.00018492\\
  \hline
\end{tabular}
\caption{Cont. Summary of the results obtained for the reaction,
$e^-(p1) e^+(p2) \rightarrow  H^-(p3) \widetilde{\chi }_{1}^+(p4)
\widetilde{\chi }_{1}^o(p5)$ for $M_2 = 150$ GeV} \label{table2}
\end{center}
\end{table}

\begin{table}[htbp]
\begin{center}
\begin{tabular}[htbp]{|c||c|c||c|c||c|c|}
  \hline
  \hline
  Figure No. & \multicolumn{2}{c|}{$\sigma_{tan\beta = 5}$} & \multicolumn{2}{c|}{$\sigma_{tan\beta = 15}$} & \multicolumn{2}{c|}{$\sigma_{tan\beta = 35}$}\\
  \cline{2-7}
  &$E_{CM}$& $\sigma$ (Pb)&$E_{CM}$& $\sigma$ (Pb)&$E_{CM}$& $\sigma$ (Pb)\\
  \hline
  1 & 1460 & 2.8416e-06 & 1580& 2.6736e-06& 1640& 2.6225e-06\\
  2 & 880 & 1.6853e-05 & 880& 2.2289e-05& 860& 2.2811e-05\\
  3 & 1560 & 4.1064e-07 & 1560& 3.8557e-07& 1560& 3.7849e-07\\
  4 & 860 & 8.5308e-06 & 860& 1.1303e-05& 840& 1.1522e-05\\
  5 & 720 & 6.843e-06 & 720& 4.6316e-05& 700& 7.9708e-05\\
  6 & 1280 & 2.0786e-09 & 1300& 2.293e-09& 1360& 2.25e-09\\
  7 & 1280 & 8.5288e-10 & 1300& 9.6192e-10& 1260& 9.378e-10\\
  8 & 980 & 5.045e-07 & 980& 4.6375e-07& 960& 4.2704e-07\\
  9 & 1060 & 2.0822e-08 & 1100& 2.7708e-08& 1120& 2.7887e-08\\
  10 & 1660 & 3.6361e-07 & 1560& 5.1425e-07& 1680& 5.5852e-07\\
  11 & 1520 & 1.6809e-07 & 1460& 1.5064e-07& 1540& 1.4015e-07\\
  12 & 1460 & 6.7452e-09 & 1400& 1.1721e-08& 1400& 1.3222e-08\\
  13 & 1220 & 4.4132e-06 & 1280& 5.285e-06& 1280& 5.3928e-06\\
  14 & 1260 & 5.8849e-09 & 1280& 8.0866e-10& 1280& 1.5408e-10\\
  15 & 1620 & 9.3888e-09 & 1220& 1.6336e-08& 1240& 2.0865e-08\\
  16 & 1200 & 1.8497e-08 & 1120& 2.1328e-08& 1140& 2.061e-08\\
  \hline
\end{tabular}
\caption{Summary of the results obtained for the reaction,
$e^-(p1) e^+(p2) \rightarrow  H^-(p3) \widetilde{\chi }_{1}^+(p4)
\widetilde{\chi }_{1}^o(p5)$ for $M_2 = 300$ GeV} \label{table3}
\end{center}
\end{table}

\begin{table}[thbp]
\begin{center}
\begin{tabular}[thbp]{|c||c|c||c|c||c|c|}
  \hline
  Figure No. & \multicolumn{2}{c|}{$\sigma_{tan\beta = 5}$} & \multicolumn{2}{c|}{$\sigma_{tan\beta = 15}$} & \multicolumn{2}{c|}{$\sigma_{tan\beta = 35}$}\\
  \cline{2-7}
  &$E_{CM}$& $\sigma$ (Pb)&$E_{CM}$& $\sigma$ (Pb)&$E_{CM}$& $\sigma$ (Pb)\\
  \hline
  \hline
  17 & 1180 & 1.1424e-09 & 1120& 2.5658e-09& 1100& 3.0713e-09\\
  18 & 1020 & 1.7106e-06 & 1060& 2.12e-06& 1080& 2.1515e-06\\
  19 & 1640 & 3.0247e-07 & 1240& 4.1924e-07& 1200& 5.0226e-07\\
  20 & 1160 & 1.0304e-08 & 1140& 8.1575e-09& 1140& 7.0367e-09\\
  21 & 1160 & 1.2888e-08 & 1120& 2.5779e-08& 1160& 2.9906e-08\\
  22 & 1020 & 4.1835e-07 & 1040& 3.9874e-07& 1060& 3.7029e-07\\
  23 & 1600 & 1.1700e-05 & 1660& 1.3191e-05& 1640& 1.3441e-05\\
  24 & 1540 & 9.3783e-08 & 1520& 5.7649e-08& 1580& 4.7951e-08\\
  25 & 1480 & 7.6122e-08 & 1400& 1.1802e-07& 1480& 1.2841e-07\\
  26 & 1300 & 1.0781e-06 & 1300& 9.9456e-07& 1320& 9.3037e-07\\
  27 & 1040 & 3.9654e-06 & 1040& 6.3122e-06& 1060& 6.788e-06\\
  28 & 880  & 1.5994e-05 & 860 & 0.00010836& 900 & 0.00018623\\
  \hline
\end{tabular}
\caption{Cont. Summary of the results obtained for the reaction,
$e^-(p1) e^+(p2) \rightarrow  H^-(p3) \widetilde{\chi }_{1}^+(p4)
\widetilde{\chi }_{1}^o(p5)$ for $M_2 = 300$ GeV} \label{table4}
\end{center}
\end{table}


\chapter{Production of a light neutral Higgs boson with a chargino and a neutralino}

\section{Introduction}
In this chapter, the production of a light neutral Higgs boson is
considered through the reaction, $e^{-}(p1)e^{+}(p2)\rightarrow
h(p3) \widetilde{\chi }_{1}^+(p4) \widetilde{\chi }_{1}^-(p5)$,
for different topologies and different propagators (see Appendix
A). There are a total of 13 feynman diagrams for this reaction
(tree level approximation) for which we gave the matrix element
corresponding to each diagram. Again, diagrams with the same
topology which can be obtained by interchanging the indices were
represented once. Our work will proceed as before,

\begin{enumerate}
\item Feynman diagrams are given,

\item Diagrams with the same topology are represented once, but
has been taken into considerations  when calculating the cross
section.

\item  Matrix elements are written, all the four momenta squares
are defined to be mass squared $(>0)$,

\item Matrix elements are squared,

\item An average over the initial spin polarizations of the
electron and positron pair and a sum over the final spin states of
the outgoing particles arising from each initial spin state is
carried out.

\item Results are represented graphically, and summarized in
subsequent tables.
\end{enumerate}

\section{Feynman Diagrams}

The follwoing is the set of Feynman diagrams which were used to
calculate the cross section of the associated production of a
charged Higgs boson with a chargino and a neutralino. Our momentum
notation is: $e^{-}(p1)$, $e^{+}(p2)$, $h(p3)$ $\widetilde{\chi
}_{1}^+(p4)$ and $\widetilde{\chi }_{1}^-(p5)$.

\begin{figure}[tph]
\begin{center}
\vskip-5.5cm
\mbox{\hskip-3.5cm\centerline{\epsfig{file=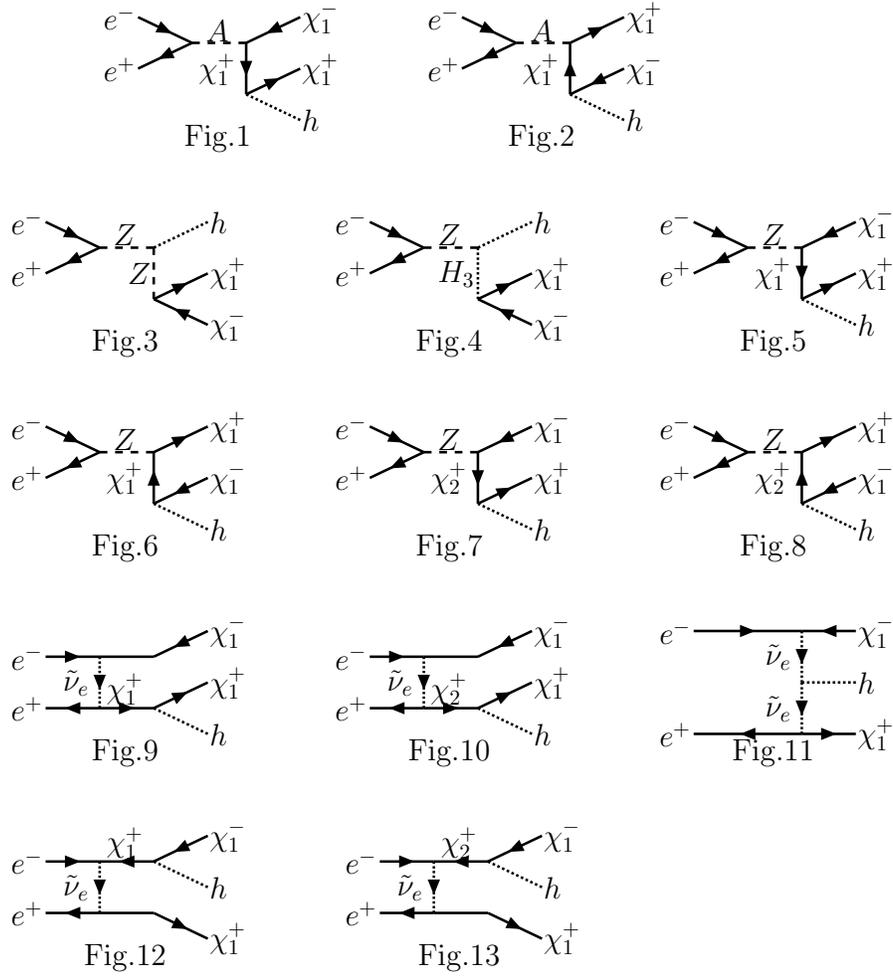,width=17cm}}}
\end{center}
\caption{Feynman diagrams for the reaction: $e^{-}(p1)$,
$e^{+}(p2)$, $h(p3)$ $\widetilde{\chi }_{1}^+(p4)$ and
$\widetilde{\chi }_{1}^-(p5)$} \label{feyn3}
\end{figure}

\newpage

\section{Matrix Elements}
The following is the set of matrix elements corresponding to
diagrams in figure \ref{feyn3} used in our calculations:

\begin{eqnarray*}
\mathcal{M}_{1}=\overline{v}(p_{2})\gamma _{\mu }u(p_{1})\frac{ige^{2}}{%
\left( p_{1}+p_{2}\right) ^{2}+i\epsilon }\overline{u}(p4)\left(
C_{11}^{L}P_{L}+C_{11}^{R}P_{R}\right) \frac{\left( \NEG{p}_{3}+\NEG%
{p}_{4}+m_{\widetilde{\chi }_{1}^{+}}\right) }{\left(
p_{3}+p_{4}\right) ^{2}+m_{\widetilde{\chi
}_{1}^{+}}^{2}+i\epsilon }\gamma ^{\mu }v(p_{5})
\end{eqnarray*}

\begin{eqnarray*}
\mathcal{M}_{2}=\overline{v}(p_{2})\gamma _{\mu }u(p_{1})\frac{ige^{2}}{%
\left( p_{1}+p_{2}\right) ^{2}+i\epsilon }\overline{u}(p5)\gamma ^{\mu }%
\frac{\left( \NEG{p}_{3}+\NEG{p}_{5}+m_{\widetilde{\chi }_{1}^{-}}\right) }{%
\left( p_{3}+p_{5}\right) ^{2}+m_{\widetilde{\chi }_{1}^{-}}^{2}+i\epsilon }%
\left( C_{11}^{L}P_{L}+C_{11}^{R}P_{R}\right) v(p_{4})
\end{eqnarray*}

\begin{eqnarray*}
\mathcal{M}_{3} &=&\overline{v}(p_{2})\gamma _{\sigma
}(g_{V}-\gamma
_{5})u(p_{1})\frac{ig^{3}M_{Z}\sin (\beta -\alpha )}{4\cos ^{3}\theta _{w}}%
\overline{u}(p_{5})\gamma ^{\mu }\left(
O_{11}^{L}P_{L}+O_{11}^{R}P_{R}\right) v(p_{4}) \\
&&\times \frac{(g_{\mu \nu }-q_{\mu }q_{\nu }/m_{Z}^{2})(g^{\nu
\sigma
}-k^{\nu }k^{\sigma }/M_{Z}^{2})}{((p_{1}+p_{2}-p_{5})^{2}-M_{Z}^{2}+i%
\epsilon )}
\end{eqnarray*}

\begin{eqnarray*}
\mathcal{M}_{4} &=&\overline{v}(p_{2})\gamma ^{\nu }(g_{V}-\gamma
_{5})u(p_{1})\frac{ig^{3}\cos (\alpha -\beta )}{8\cos ^{2}\theta _{w}}%
\overline{u}(p_{5})\left(
C_{11}^{A,L}P_{L}+C_{11}^{A,R}P_{R}\right) v(p_{4})
\\
&&\times \frac{(q_{\mu }-h_{\mu })}{q^{2}-M_{H_{3}}^{2}+i\epsilon }\frac{%
(g^{\mu \nu }-k^{\mu }k^{\nu }/M_{Z}^{2})}{((p_{1+}p_{2})^{2}-M_{Z}^{2}+i%
\epsilon )}
\end{eqnarray*}

\begin{eqnarray*}
\mathcal{M}_{5} &=&\overline{v}(p_{2})\gamma ^{\nu }(g_{V}-\gamma
_{5})u(p_{1})\frac{-ig^{3}}{4\cos ^{2}\theta
_{w}((p_{1+}p_{2})^{2}-M_{Z}^{2}+i\epsilon
)}\overline{u}(p_{4})\left(
C_{11}^{L}P_{L}+C_{11}^{R}P_{R}\right)  \\
&&\times \frac{\left( \NEG{p}_{3}+\NEG{p}_{4}+m_{\widetilde{\chi }%
_{1}^{+}}\right) }{\left( p_{3}+p_{4}\right) ^{2}+m_{\widetilde{\chi }%
_{1}^{+}}^{2}+i\epsilon }\gamma ^{\mu }\left(
O_{11}^{L}P_{L}+O_{11}^{R}P_{R}\right) v(p_{5})\left( g_{\mu \nu }-\frac{%
k_{\mu }k_{\nu }}{M_{Z}^{2}}\right)
\end{eqnarray*}

\bigskip
\begin{eqnarray*}
\mathcal{M}_{6} &=&\overline{v}(p_{2})\gamma ^{\nu }(g_{V}-\gamma
_{5})u(p_{1})\frac{-ig^{3}}{4\cos ^{2}\theta
_{w}((p_{1+}p_{2})^{2}-M_{Z}^{2}+i\epsilon
)}\overline{u}(p_{5})\gamma ^{\mu
}\left( O_{11}^{L}P_{L}+O_{11}^{R}P_{R}\right)  \\
&&\times \frac{-(\NEG{p}_{3}+\NEG{p}_{5}+m_{\chi
_{1}^{+}})}{\left( p_{3}+p_{5}\right) ^{2}+m_{\widetilde{\chi
}_{1}^{+}}^{2}+i\epsilon }\left(
C_{11}^{L}P_{L}+C_{11}^{R}P_{R}\right) v(p_{4})\left( g_{\mu \nu }-\frac{%
k_{\mu }k_{\nu }}{M_{Z}^{2}}\right)
\end{eqnarray*}

\begin{eqnarray*}
\mathcal{M}_{7} &=&\overline{v}(p_{2})\gamma ^{\nu }(g_{V}-\gamma
_{5})u(p_{1})\frac{-ig^{3}}{4\cos ^{2}\theta
_{w}((p_{1+}p_{2})^{2}-M_{Z}^{2}+i\epsilon
)}\overline{u}(p_{4})\left(
C_{12}^{L}P_{L}+C_{12}^{R}P_{R}\right)  \\
&&\times \frac{\NEG{p}_{3}+\NEG{p}_{4}+m_{\chi _{2}^{+}}}{\left(
p_{3}+p_{4}\right) ^{2}+m_{\widetilde{\chi
}_{2}^{+}}^{2}+i\epsilon }\gamma
^{\mu }\left( O_{21}^{L}P_{L}+O_{21}^{R}P_{R}\right) \left( g_{\mu \nu }-%
\frac{k_{\mu }k_{\nu }}{M_{Z}^{2}}\right)
\end{eqnarray*}

\begin{eqnarray*}
\mathcal{M}_{8} &=&\overline{v}(p_{2})\gamma ^{\nu }(g_{V}-\gamma
_{5})u(p_{1})\frac{-ig^{3}}{4\cos ^{2}\theta
_{w}((p_{1+}p_{2})^{2}-M_{Z}^{2}+i\epsilon
)}\overline{u}(p_{5})\gamma ^{\mu
}\left( O_{12}^{L}P_{L}+O_{12}^{R}P_{R}\right)  \\
&&\times \frac{\NEG{p}_{3}+\NEG{p}_{5}+m_{\chi _{2}^{+}}}{\left(
p_{3}+p_{5}\right) ^{2}+m_{\widetilde{\chi
}_{2}^{+}}^{2}+i\epsilon }\left(
C_{21}^{L}P_{L}+C_{21}^{R}P_{R}\right) v(p_{4})\left( g_{\mu \nu }-\frac{%
k_{\mu }k_{\nu }}{M_{Z}^{2}}\right)
\end{eqnarray*}

\begin{eqnarray*}
\mathcal{M}_{9}=\frac{ig^{3}\left\vert V_{11}\right\vert ^{2}}{%
(p_{1}-p_{5})^{2}-m_{\widetilde{\nu }}^{2}}\overline{v}(p_{2})P_{L}\frac{\NEG%
{p}_{3}+\NEG{p}_{4}+m_{\widetilde{\chi }_{1}^{+}}}{((p_{3}+p_{4})^{2}-m_{%
\widetilde{\chi }_{1}^{+}}^{2}+i\epsilon }\left(
C_{11}^{L}P_{L}+C_{11}^{R}P_{R}\right) u(p_{5})\overline{v}(p_{4})P_{R}%
\overline{u}(p_{1})
\end{eqnarray*}

\begin{eqnarray*}
\mathcal{M}_{10}=\frac{ig^{3}\left\vert V_{11}\right\vert
\left\vert
V_{21}\right\vert }{(p_{1}-p_{5})^{2}-m_{\widetilde{\nu }}^{2}}\overline{v}%
(p_{2})P_{L}\frac{\NEG{p}_{3}+\NEG{p}_{4}+m_{\widetilde{\chi }_{1}^{+}}}{%
((p_{3}+p_{4})^{2}-m_{\widetilde{\chi }_{1}^{+}}^{2}+i\epsilon
}\left(
C_{21}^{L}P_{L}+C_{21}^{R}P_{R}\right) u(p_{5})\overline{v}(p_{4})P_{R}%
\overline{u}(p_{1})
\end{eqnarray*}

\begin{eqnarray*}
\mathcal{M}_{11}=\frac{ig^{3}M_{W}\sin (\alpha +\beta )\left\vert
V_{11}\right\vert ^{2}}{2\cos ^{2}\theta _{w}((p_{1}-p_{5})^{2}-m_{%
\widetilde{\nu }}^{2})((p_{1}-p_{4})^{2}-m_{\widetilde{\nu }}^{2})}\overline{%
v}(p_{2})P_{L}u(p_{5})\overline{v}(p_{4})P_{R}\overline{u}(p_{1})
\end{eqnarray*}

\begin{eqnarray*}
\mathcal{M}_{12}=\frac{ig^{3}\left\vert V_{11}\right\vert
\left\vert
V_{21}\right\vert }{((p_{1}-p_{4})^{2}-m_{\widetilde{\nu }}^{2})}\overline{v}%
(p_{2})P_{L}u(p_{5})\overline{v}(p_{4})\left(
C_{21}^{L}P_{L}+C_{21}^{R}P_{R}\right) \frac{-(\NEG{p}_{3}+\NEG{p}_{5})+m_{%
\widetilde{\chi }_{2}^{+}}}{((p_{3}+p_{5})^{2}-m_{\widetilde{\chi }%
_{2}^{+}}^{2}+i\epsilon }P_{R}\overline{u}(p_{1})
\end{eqnarray*}

\begin{eqnarray*}
\mathcal{M}_{13}=\frac{ig^{3}\left\vert V_{11}\right\vert ^{2}}{%
((p_{1}-p_{4})^{2}-m_{\widetilde{\nu }}^{2})}\overline{v}(p_{2})P_{L}u(p_{5})%
\overline{v}(p_{4})\left( C_{11}^{L}P_{L}+C_{11}^{R}P_{R}\right) \frac{-(\NEG%
{p}_{3}+\NEG{p}_{5})+m_{\widetilde{\chi }_{1}^{+}}}{((p_{3}+p_{5})^{2}-m_{%
\widetilde{\chi }_{1}^{+}}^{2}+i\epsilon }P_{R}\overline{u}(p_{1})
\end{eqnarray*}

\noindent For the definitions of the constants used here, the
reader is referred to Appendix A.

\section{Cross Sections}
As before, to calculate the differential cross sections, and
hence, the total cross section, we need first to obtain the
squared matrix element for each Feynman diagram, where use of the
trace theorems was made. Later an average over the initial spin
polarizations of the electron and the positron pair and the sum
over the final spin states of the outgoing particles arising from
each initial spin state is carried out. The total cross section as
a function of the center of the mass energy (see Appendix
B) is then calculated.\\
Calculations were done with the following set of parameters:\\
$tan\beta = 10$, and $tan\beta = 15$ where $M_2 = 150$ or $M_2 = 300$.\\
All results are given in the following figures.
\begin{figure}[th]
\vspace{-4.5cm}
\centerline{\epsfxsize=5.5truein\epsfbox{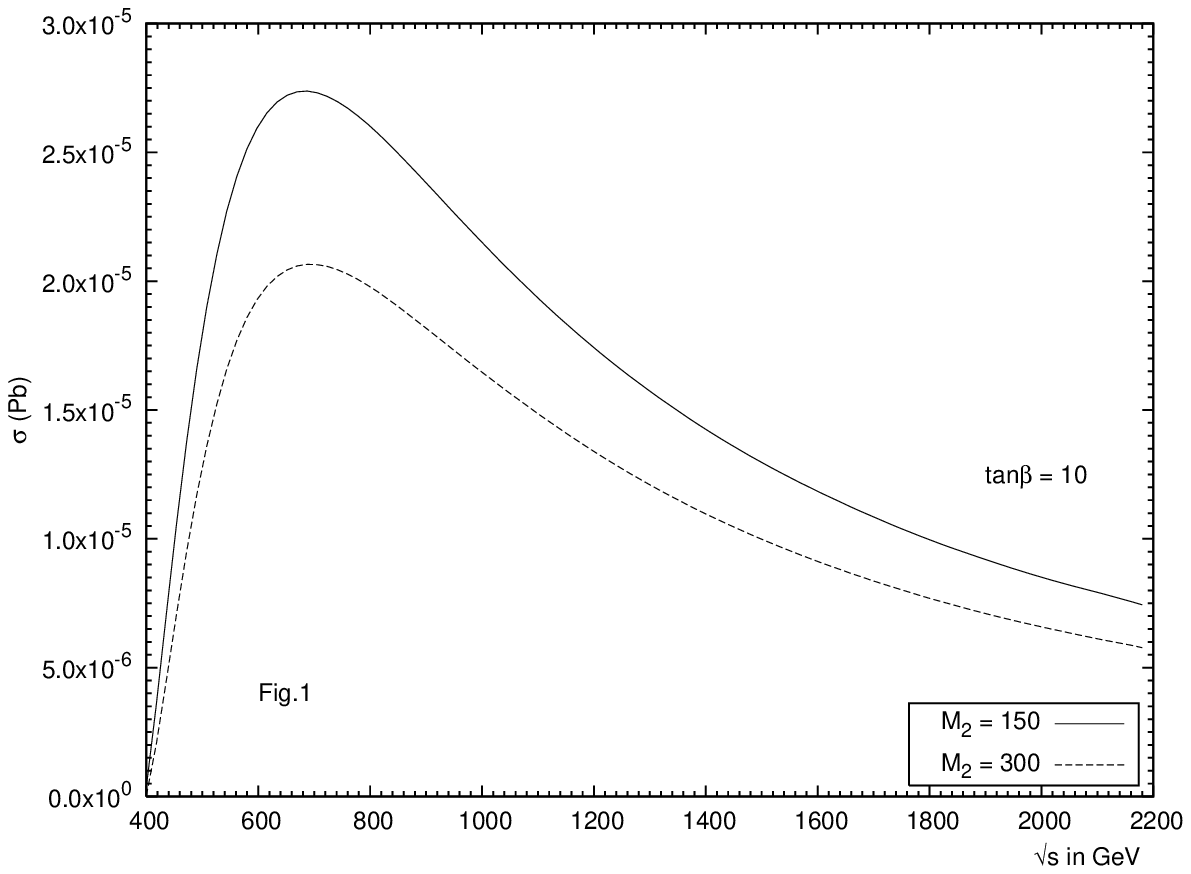}}
\vspace{-0.1cm}
\centerline{\epsfxsize=5.5truein\epsfbox{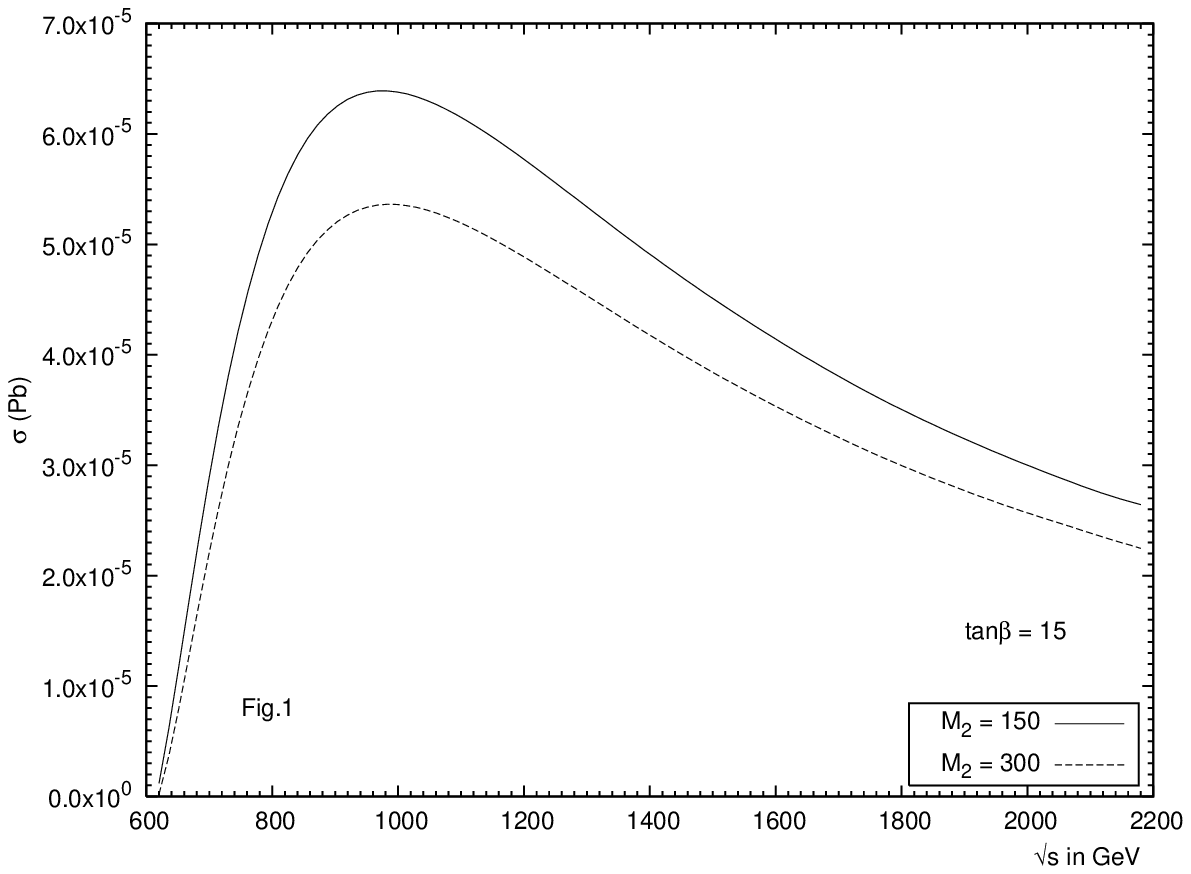}}
\vspace{0.5cm} \caption{\small Cross sections for
diagram no. 1 in figure \ref{feyn3}} \label{hXX1}
\end{figure}

\begin{figure}[th]
\vspace{-4.5cm}
\centerline{\epsfxsize=5.5truein\epsfbox{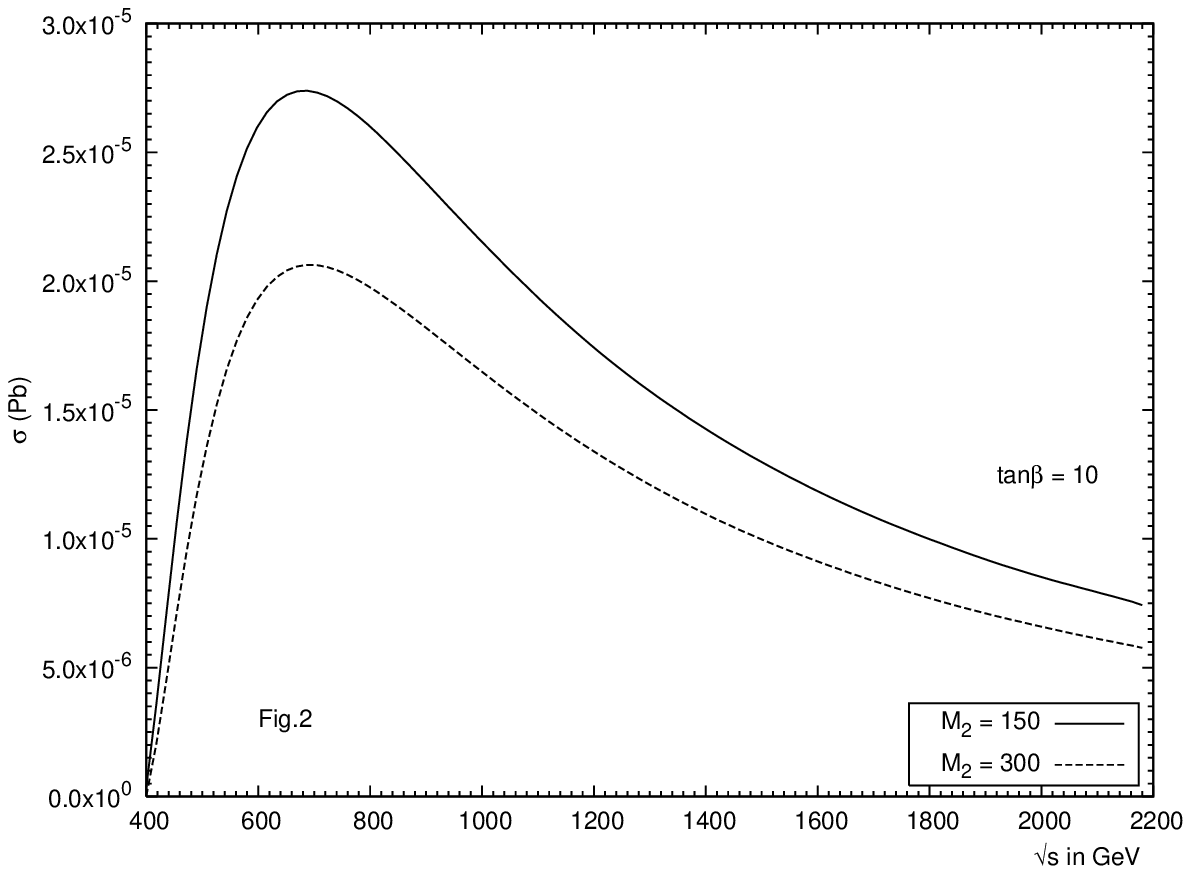}}
\vspace{-0.1cm}
\centerline{\epsfxsize=5.5truein\epsfbox{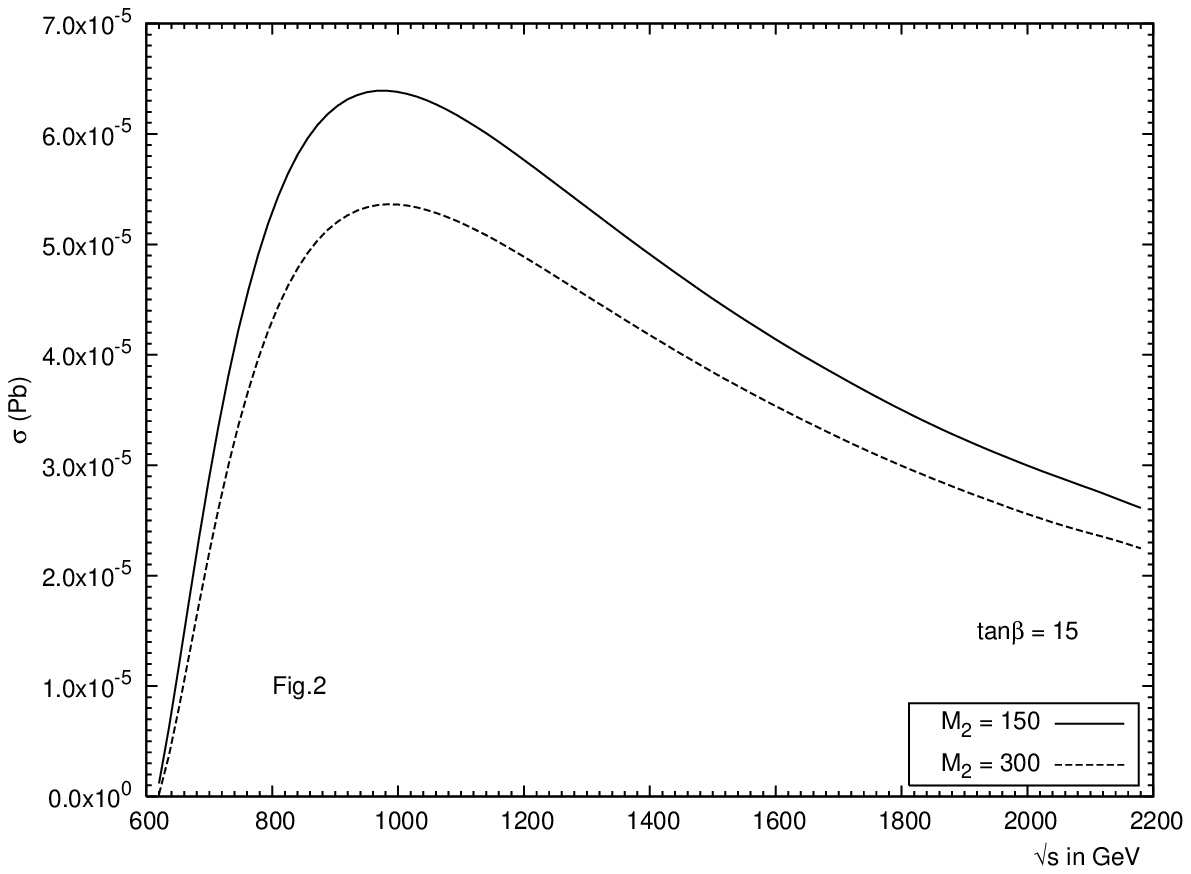}}
\vspace{0.5cm} \caption{\small Cross sections for
diagram no. 2 in figure \ref{feyn3}} \label{hXX2}
\end{figure}

\begin{figure}[th]
\vspace{-4.5cm}
\centerline{\epsfxsize=5.5truein\epsfbox{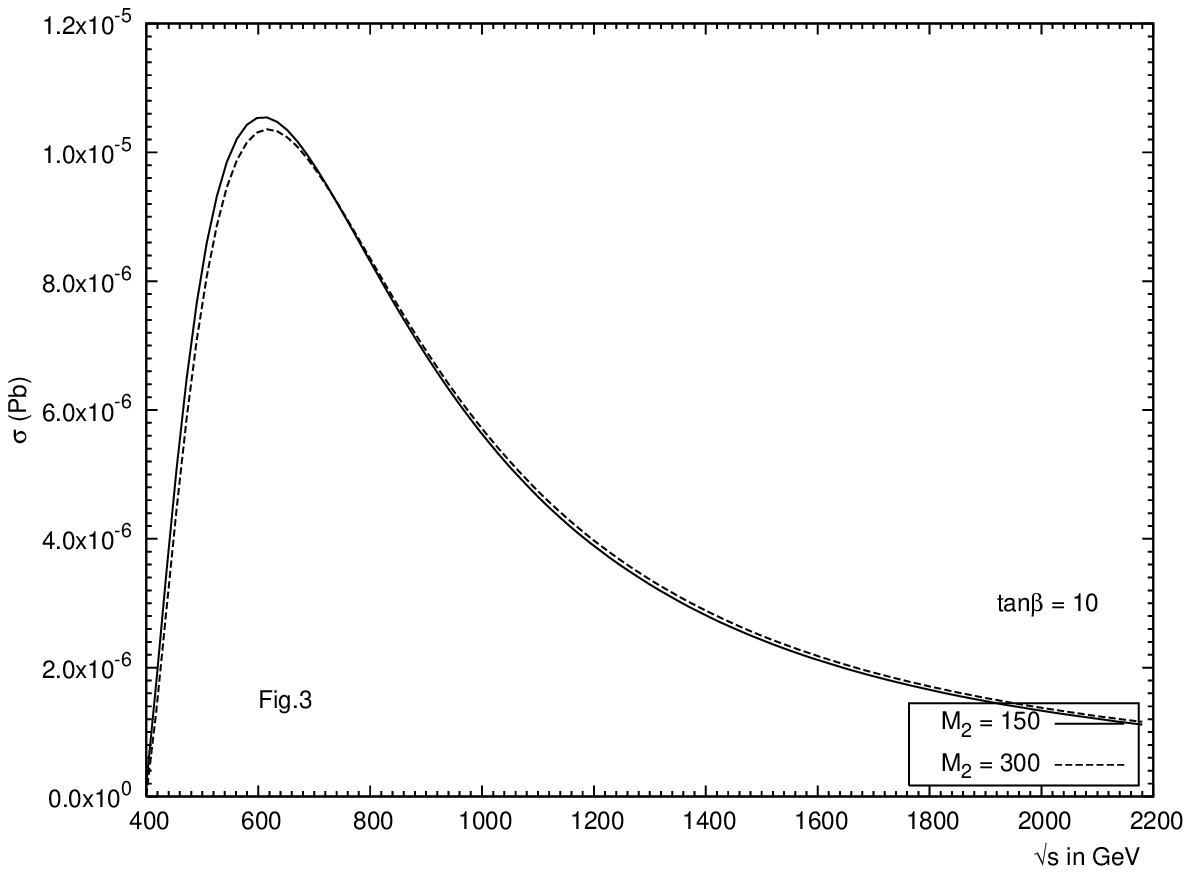}}
\vspace{-0.1cm}
\centerline{\epsfxsize=5.5truein\epsfbox{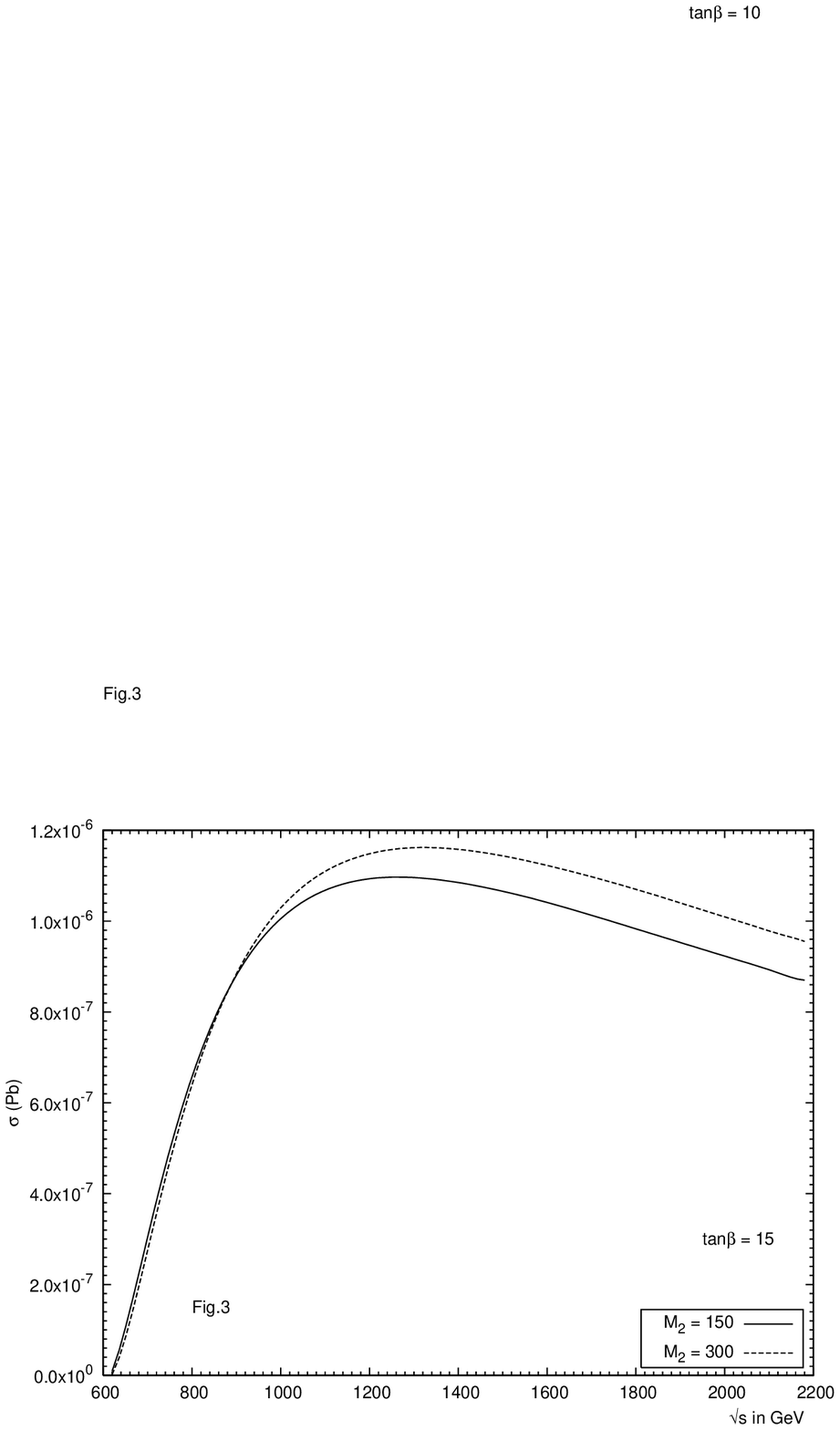}}
\vspace{0.5cm} \caption{\small Cross sections for
diagram no. 3 in figure \ref{feyn3}} \label{hXX3}
\end{figure}

\begin{figure}[th]
\vspace{-4.5cm}
\centerline{\epsfxsize=5.5truein\epsfbox{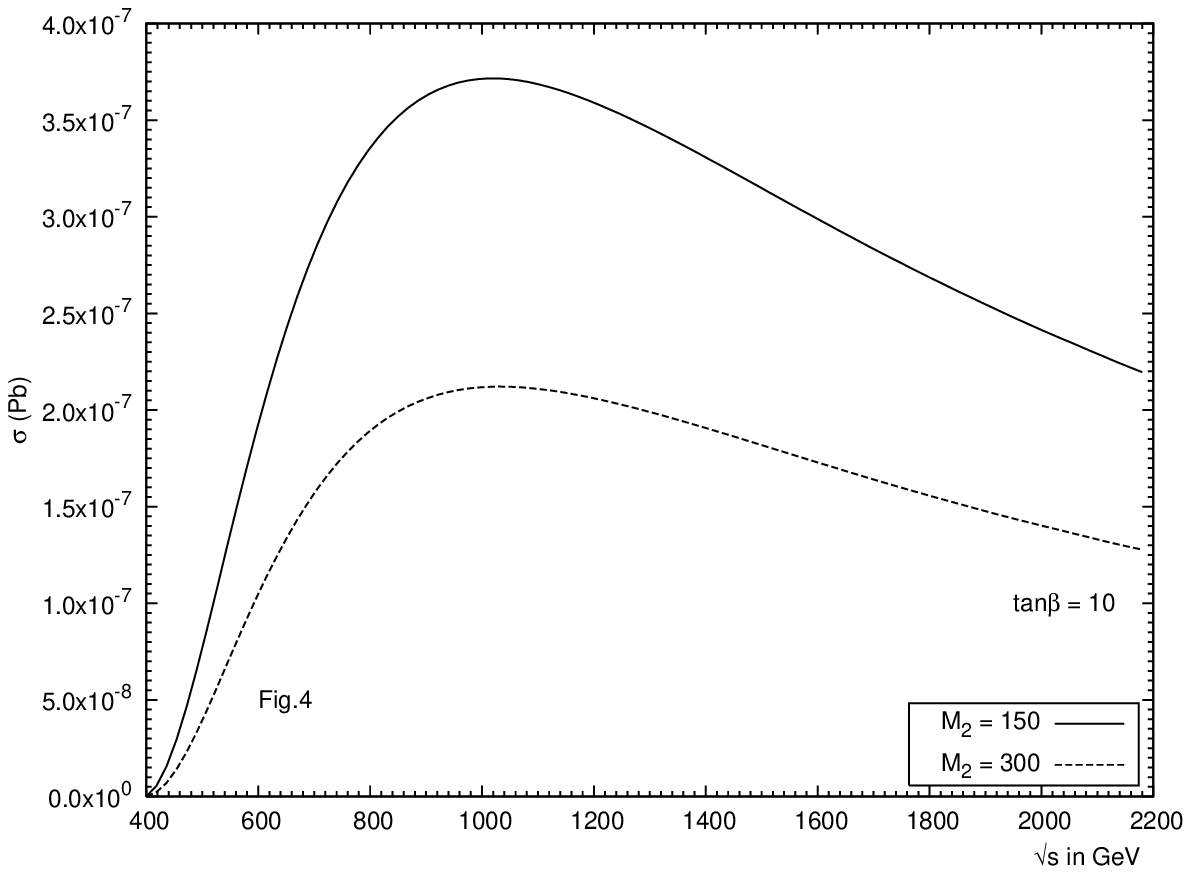}}
\vspace{-0.1cm}
\centerline{\epsfxsize=5.5truein\epsfbox{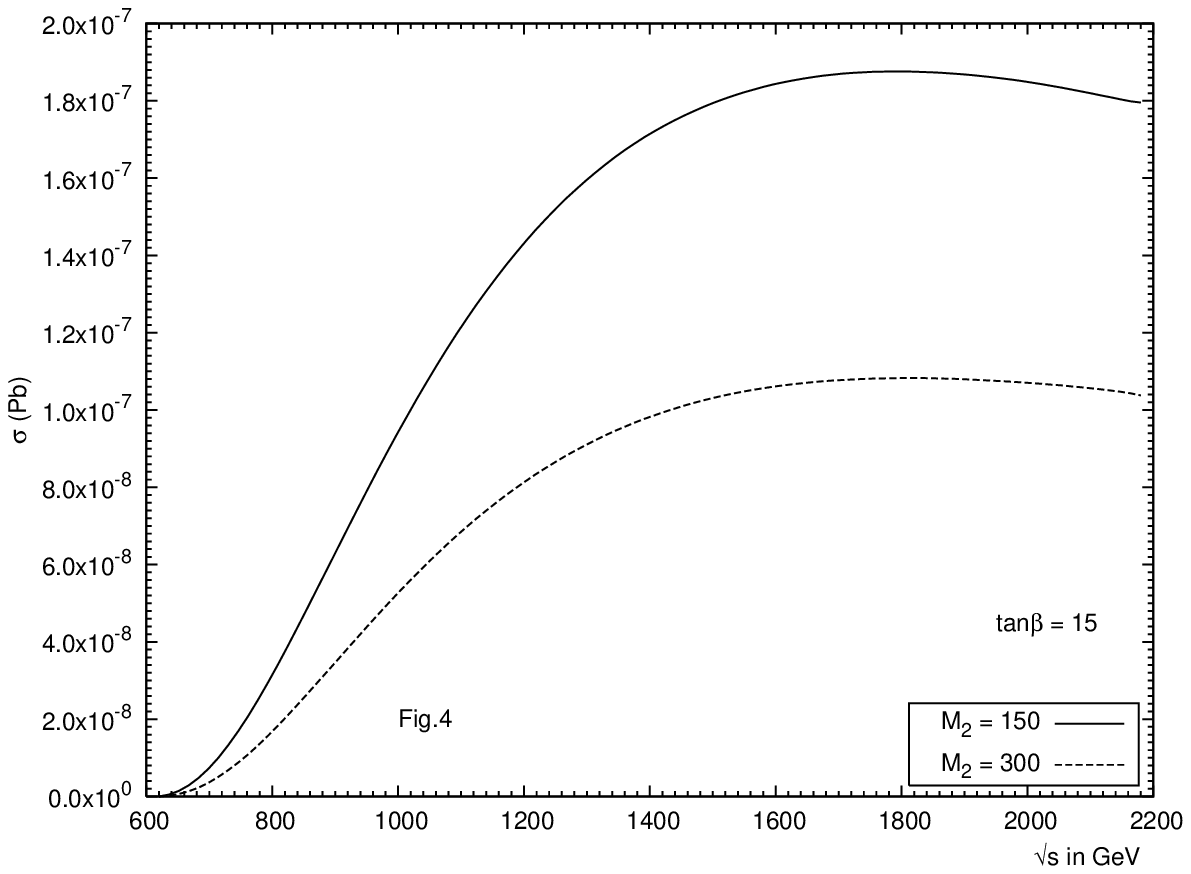}}
\vspace{0.5cm} \caption{\small Cross sections for
diagram no. 4 in figure \ref{feyn3}} \label{hXX4}
\end{figure}

\begin{figure}[th]
\vspace{-4.5cm}
\centerline{\epsfxsize=5.5truein\epsfbox{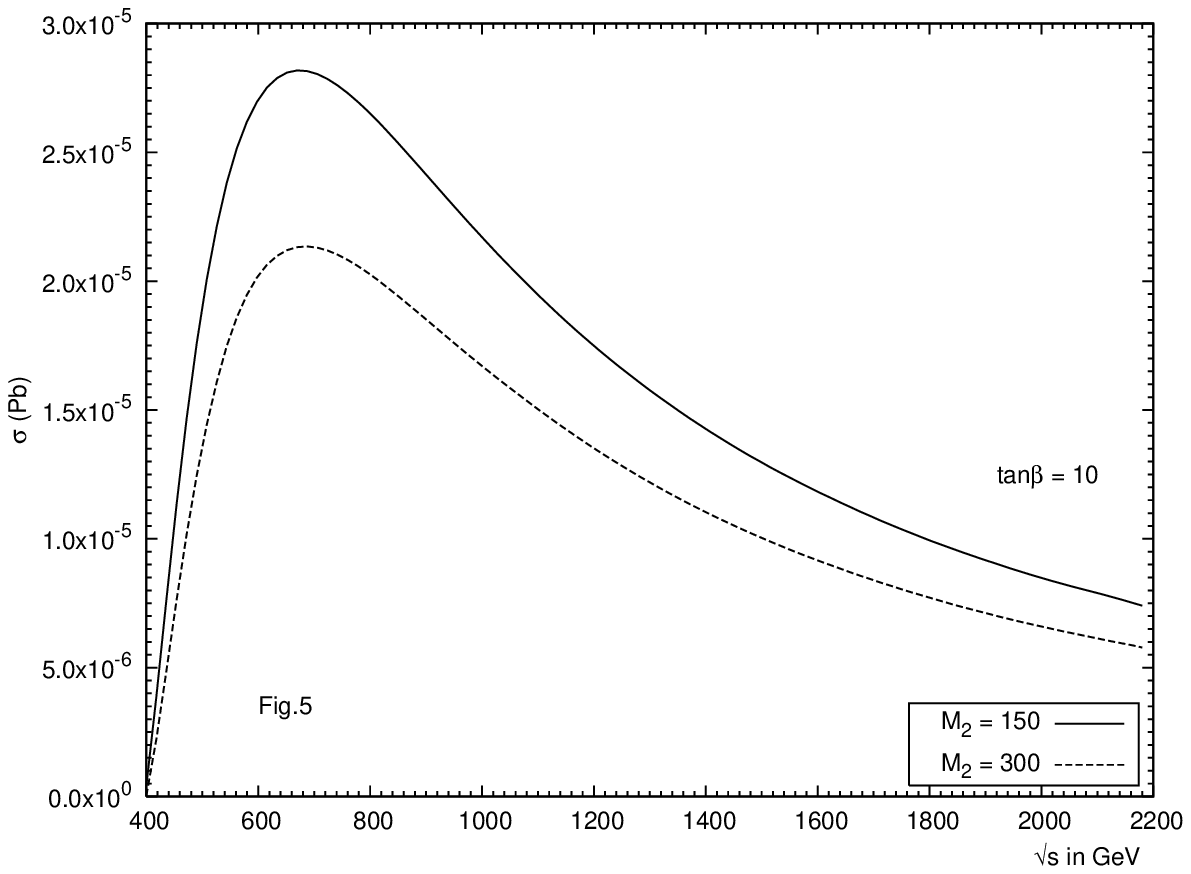}}
\vspace{-0.1cm}
\centerline{\epsfxsize=5.5truein\epsfbox{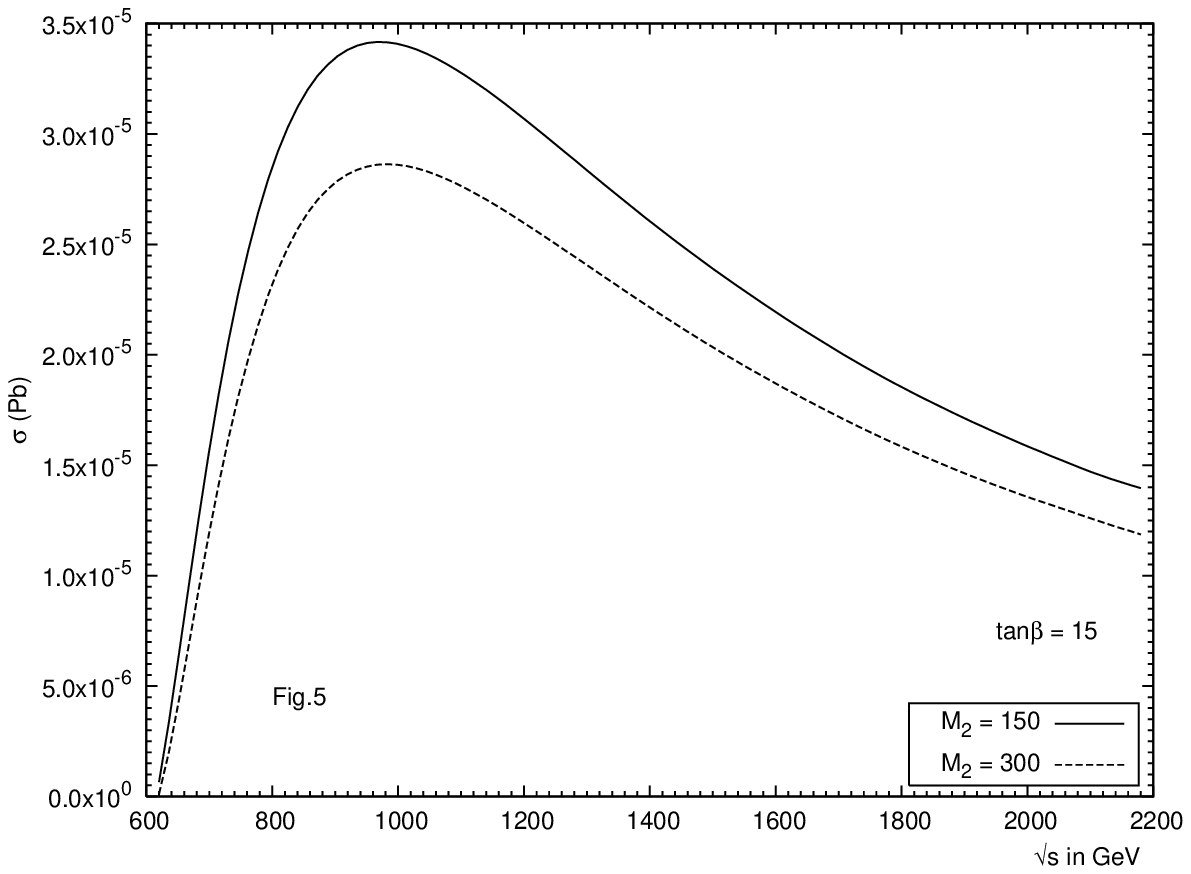}}
\vspace{0.5cm} \caption{\small Cross sections for
diagram no. 5 in figure \ref{feyn3}} \label{hXX5}
\end{figure}

\begin{figure}[th]
\vspace{-4.5cm}
\centerline{\epsfxsize=5.5truein\epsfbox{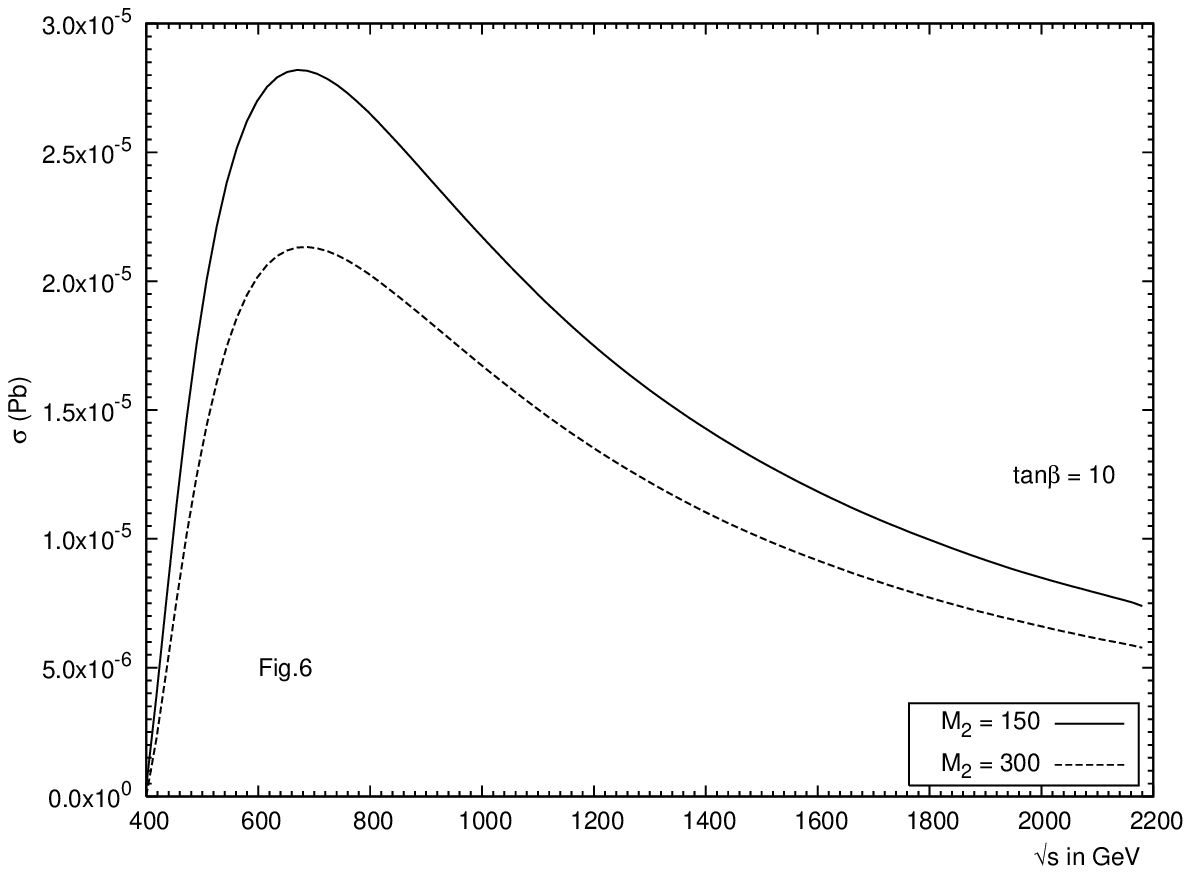}}
\vspace{-0.1cm}
\centerline{\epsfxsize=5.5truein\epsfbox{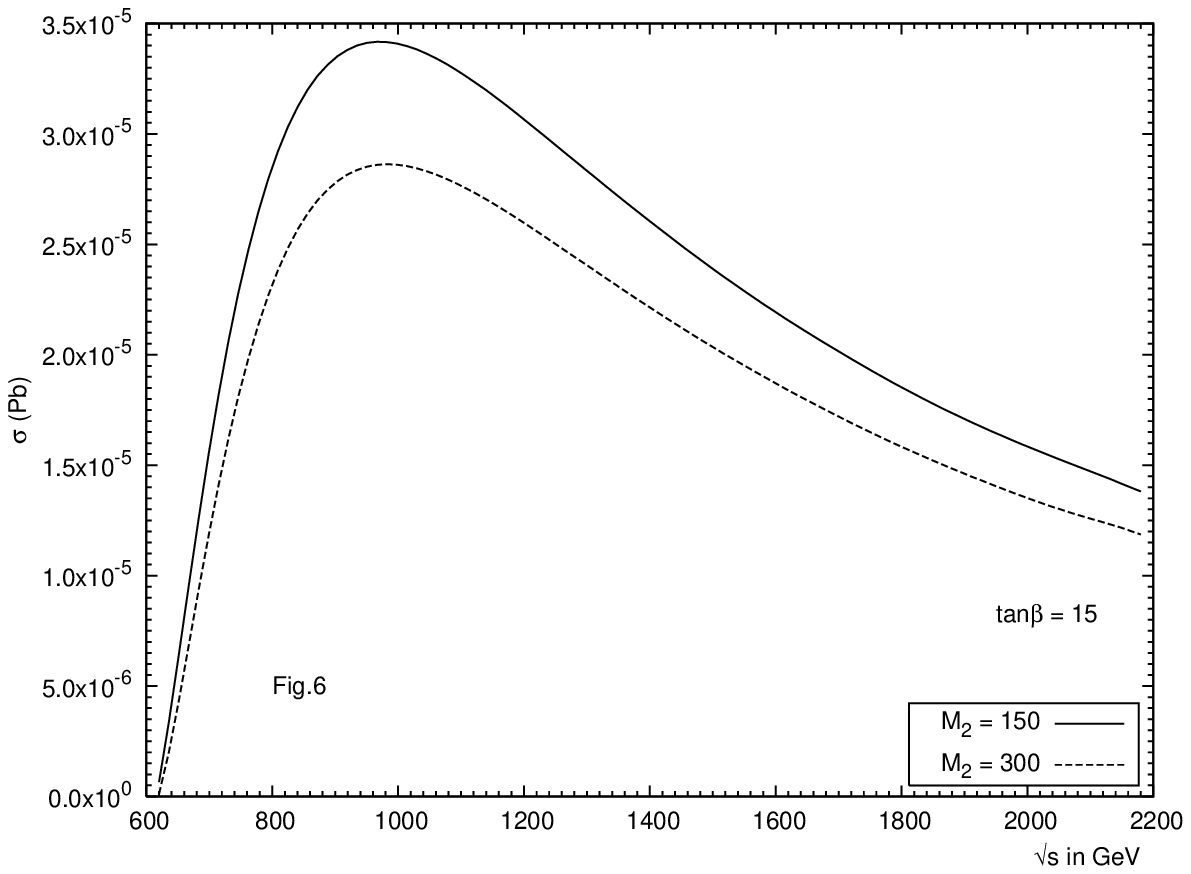}}
\vspace{0.5cm} \caption{\small Cross sections for
diagram no. 6 in figure \ref{feyn3}} \label{hXX6}
\end{figure}

\begin{figure}[th]
\vspace{-4.5cm}
\centerline{\epsfxsize=5.5truein\epsfbox{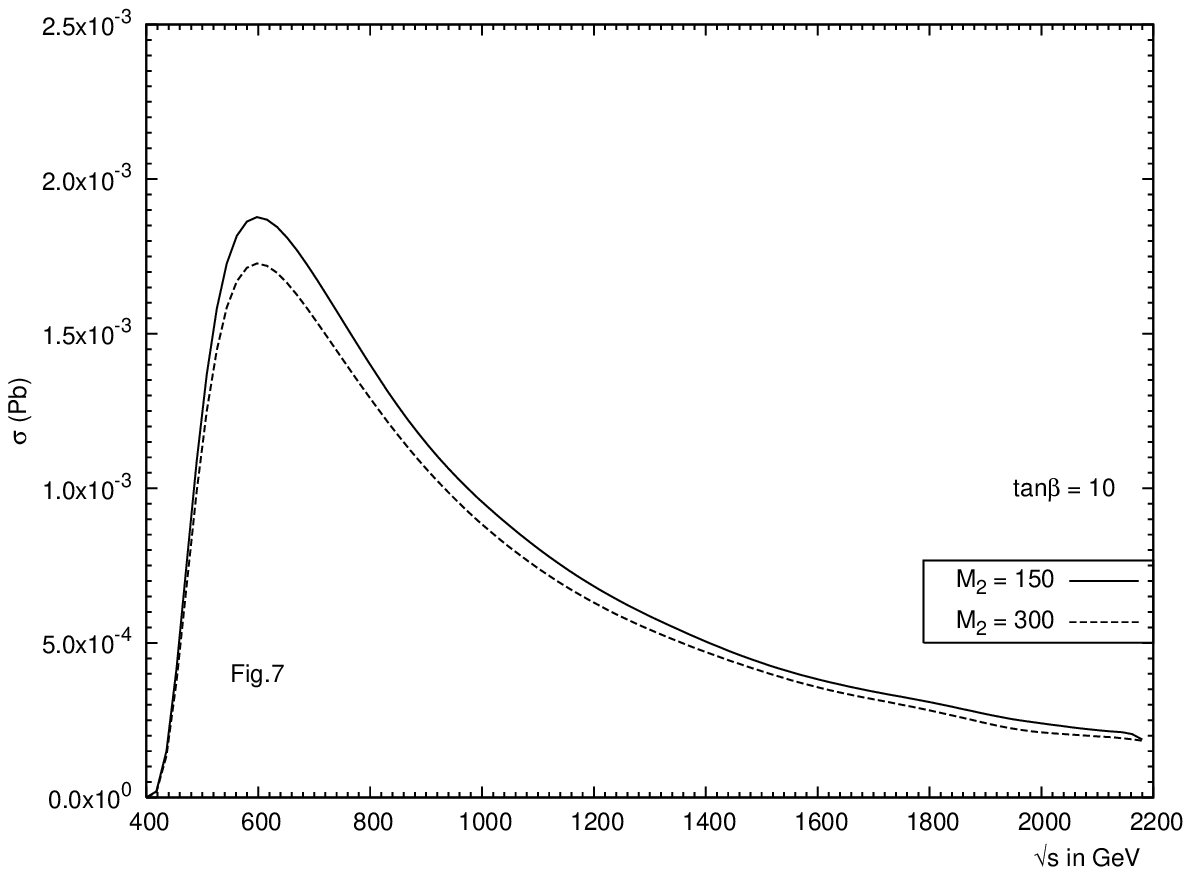}}
\vspace{-0.1cm}
\centerline{\epsfxsize=5.5truein\epsfbox{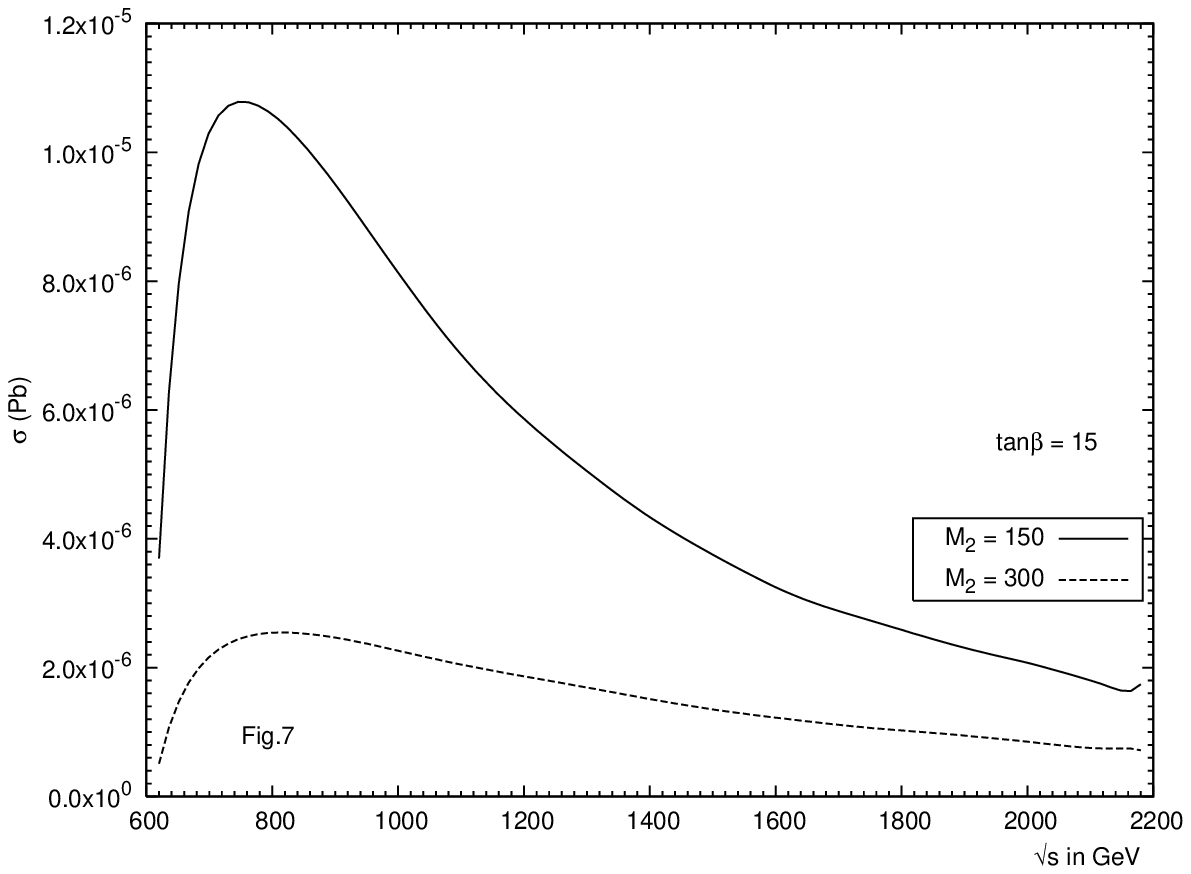}}
\vspace{0.5cm} \caption{\small Cross sections for
diagram no. 7 in figure \ref{feyn3}} \label{hXX7}
\end{figure}

\begin{figure}[th]
\vspace{-4.5cm}
\centerline{\epsfxsize=5.5truein\epsfbox{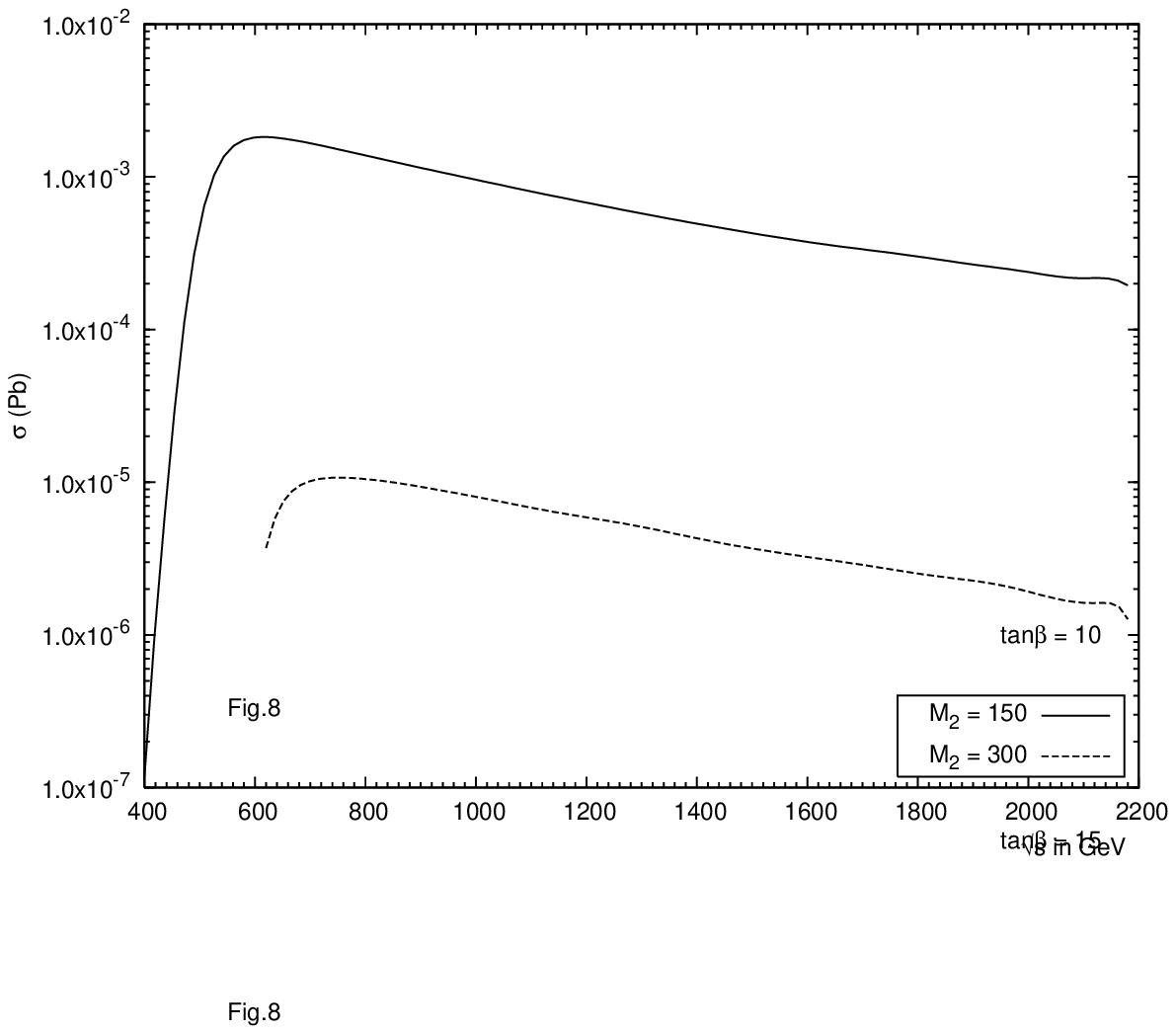}}
\vspace{-0.1cm}
\centerline{\epsfxsize=5.5truein\epsfbox{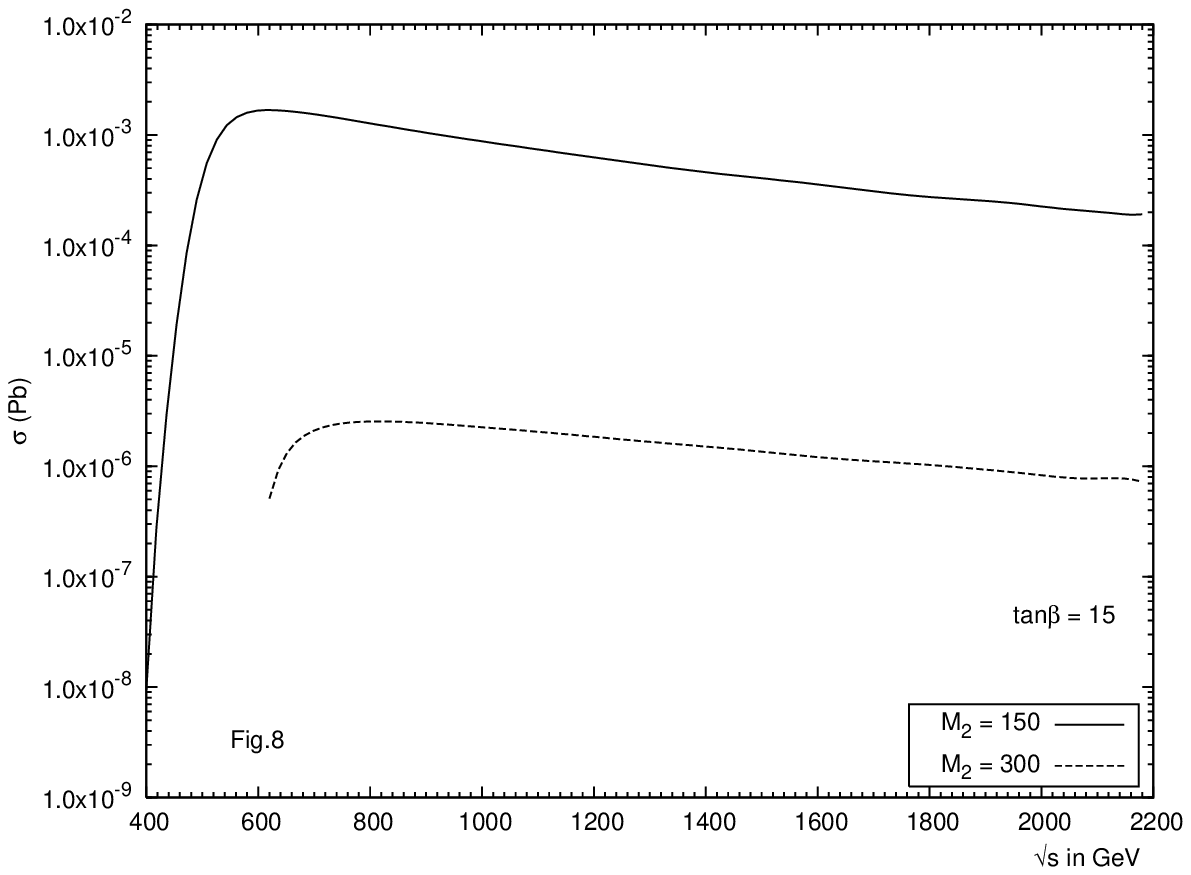}}
\vspace{0.5cm} \caption{\small Cross sections for
diagram no. 8 in figure \ref{feyn3}} \label{hXX8}
\end{figure}

\begin{figure}[th]
\vspace{-4.5cm}
\centerline{\epsfxsize=5.5truein\epsfbox{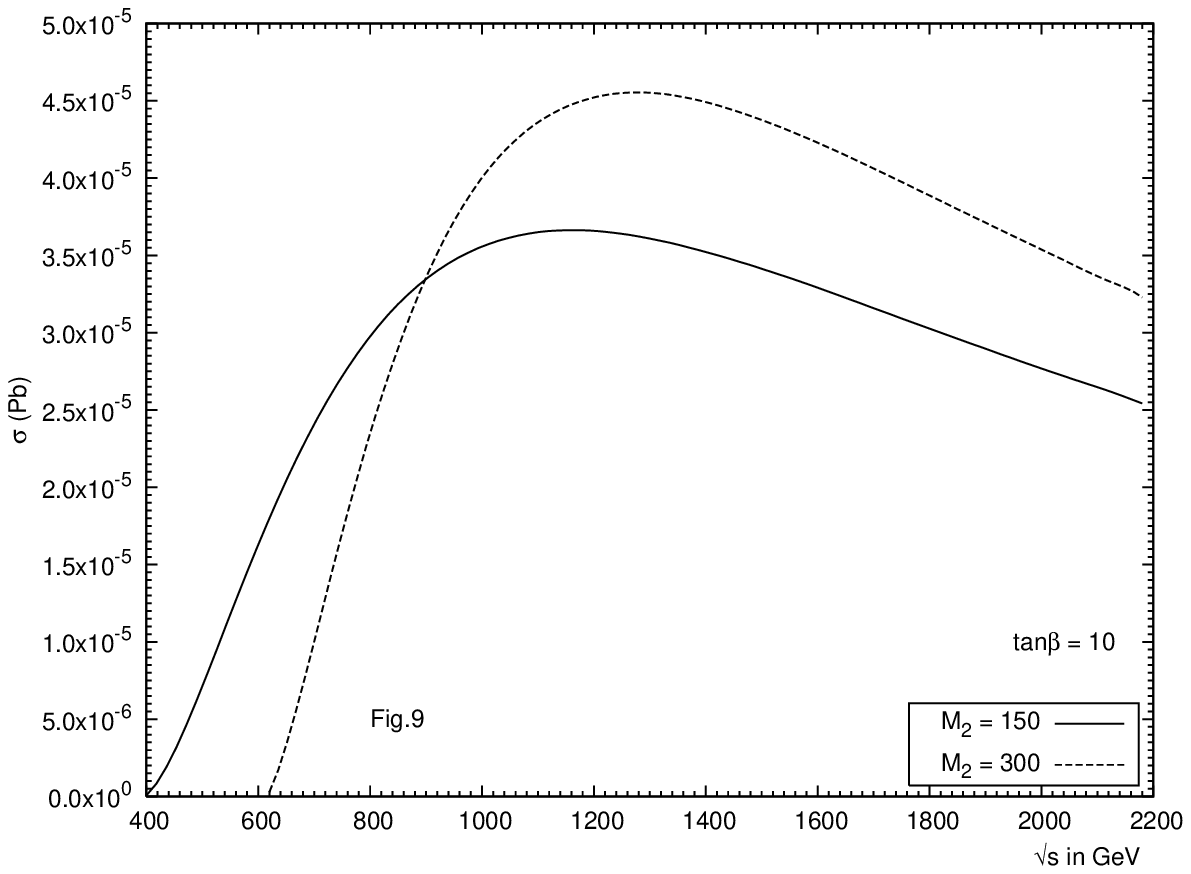}}
\vspace{-0.1cm}
\centerline{\epsfxsize=5.5truein\epsfbox{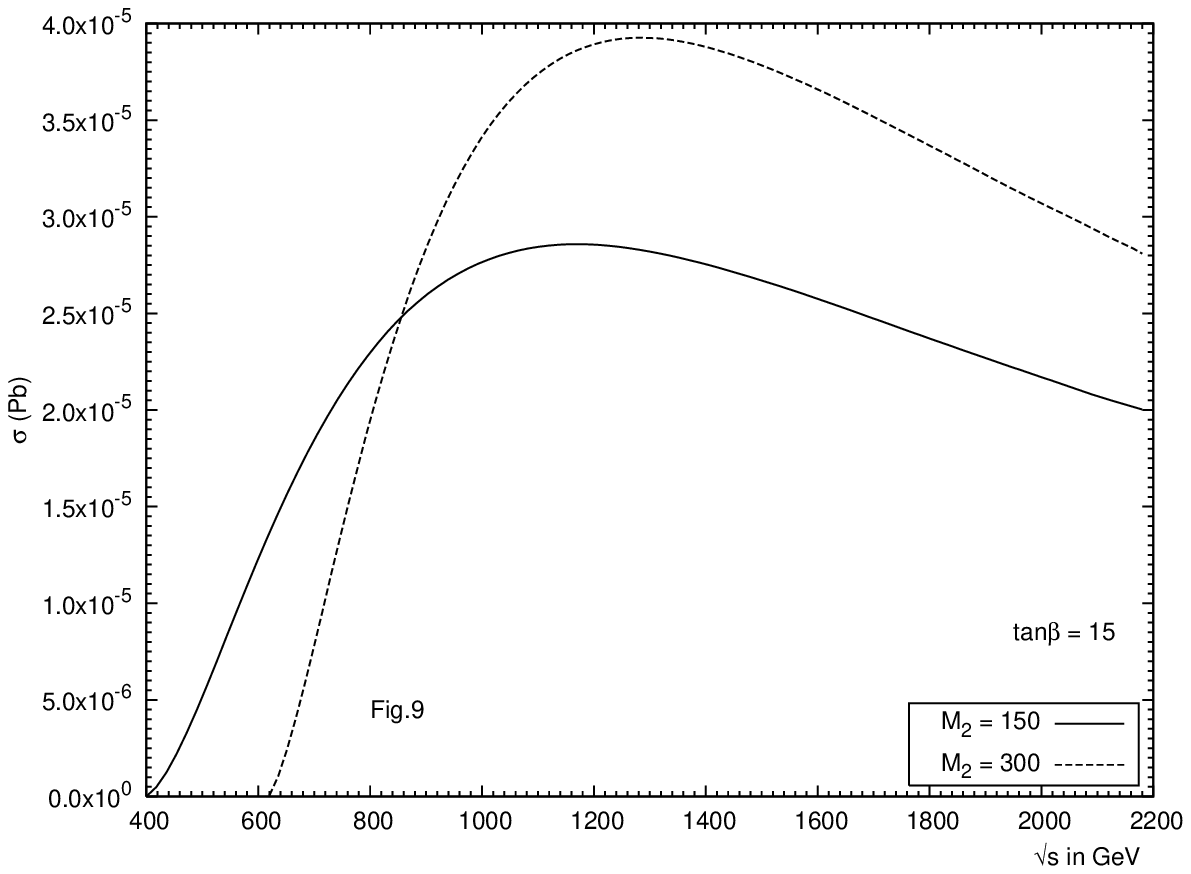}}
\vspace{0.5cm} \caption{\small Cross sections for
diagram no. 09 in figure \ref{feyn3}} \label{hXX09}
\end{figure}

\begin{figure}[th]
\vspace{-4.5cm}
\centerline{\epsfxsize=5.5truein\epsfbox{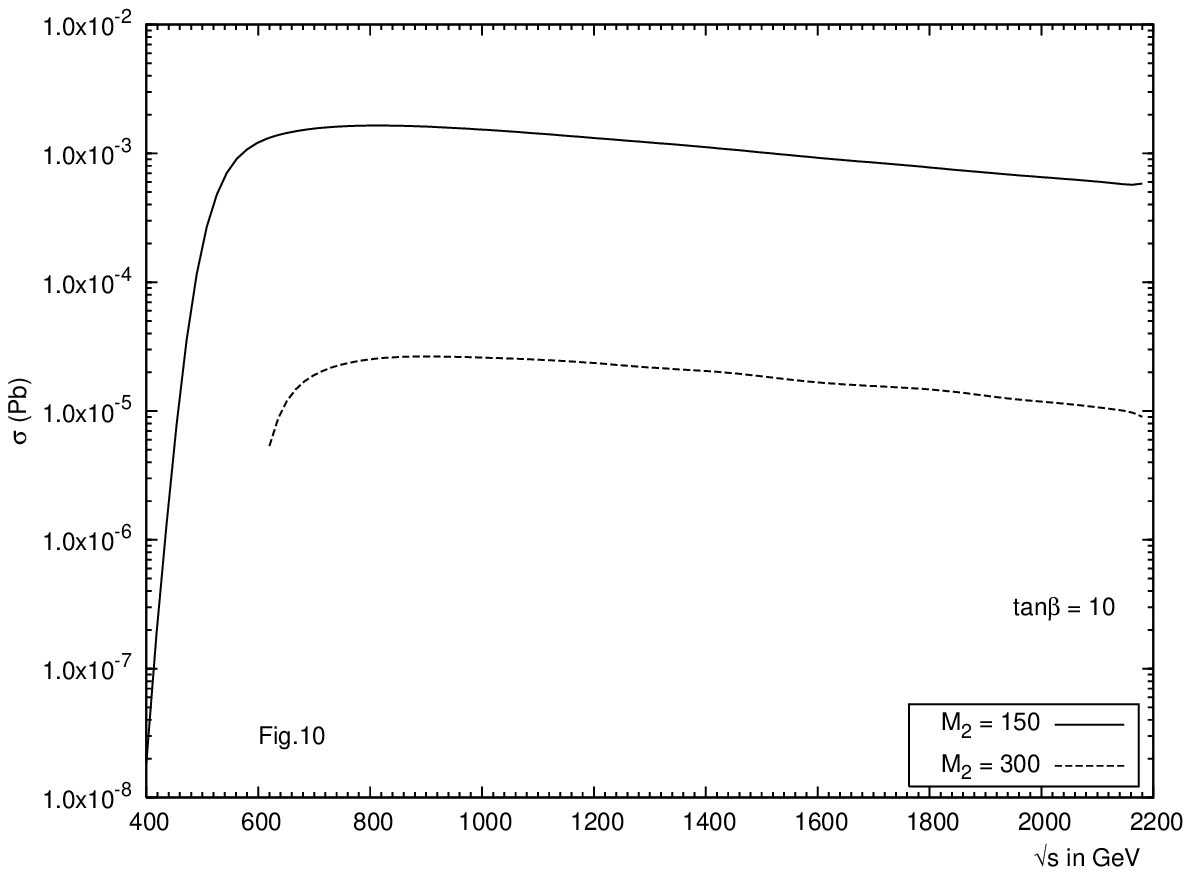}}
\vspace{-0.1cm}
\centerline{\epsfxsize=5.5truein\epsfbox{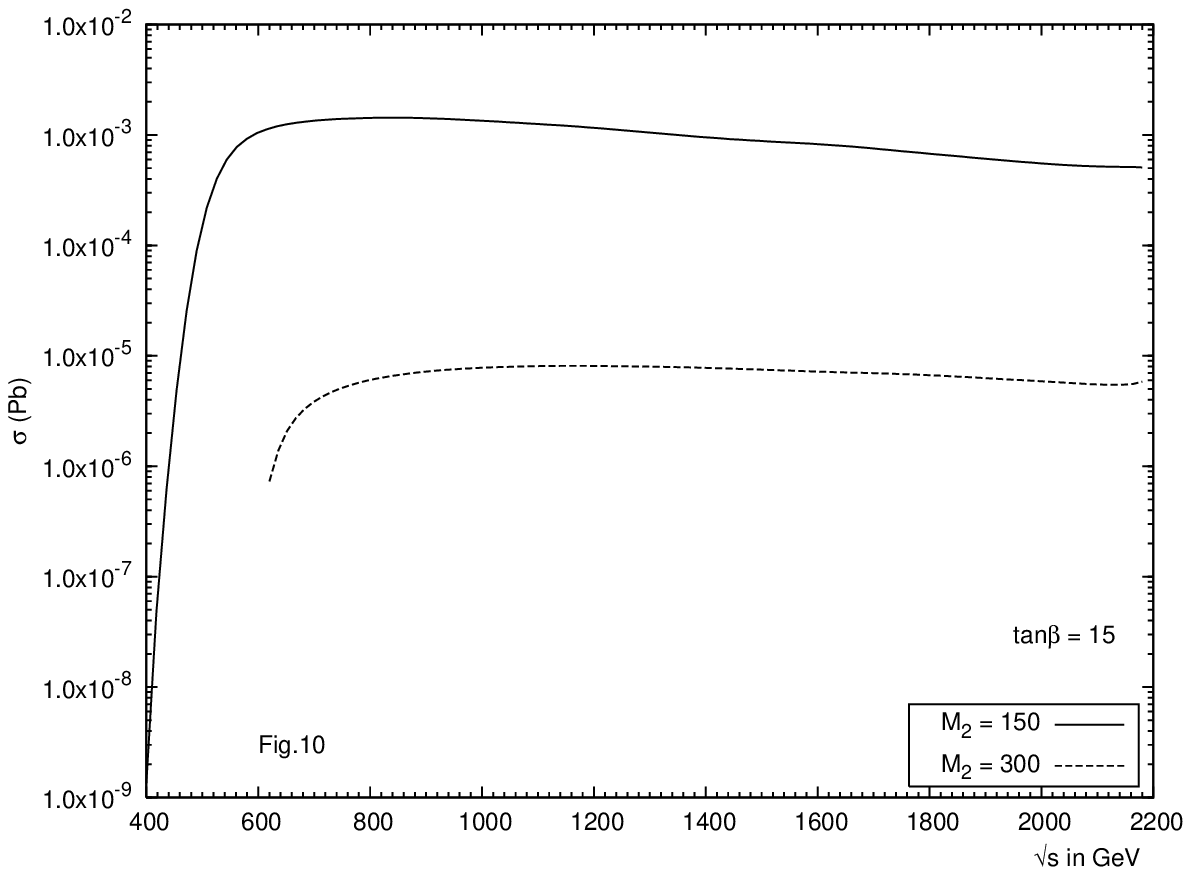}}
\vspace{0.5cm} \caption{\small Cross sections for
diagram no. 10 in figure \ref{feyn3}} \label{hXX10}
\end{figure}

\begin{figure}[th]
\vspace{-4.5cm}
\centerline{\epsfxsize=5.5truein\epsfbox{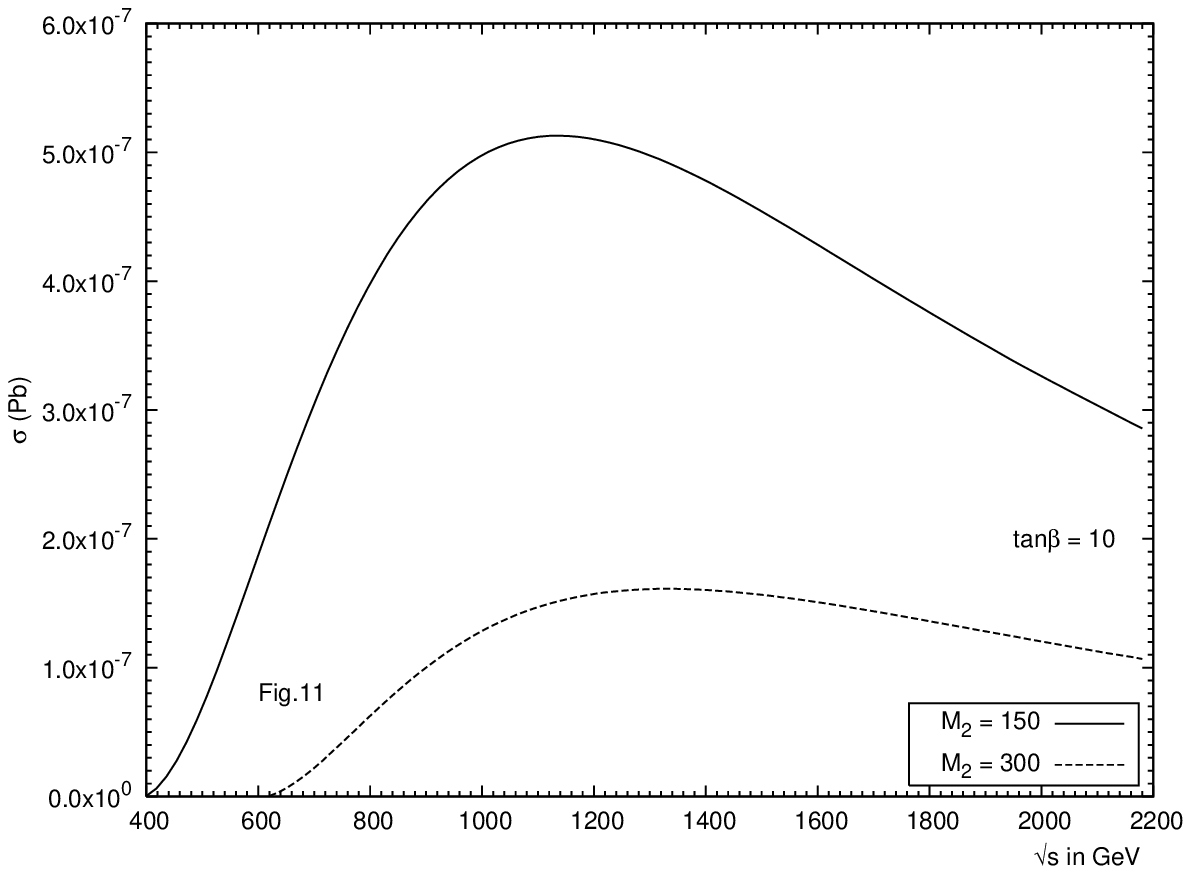}}
\vspace{-0.1cm}
\centerline{\epsfxsize=5.5truein\epsfbox{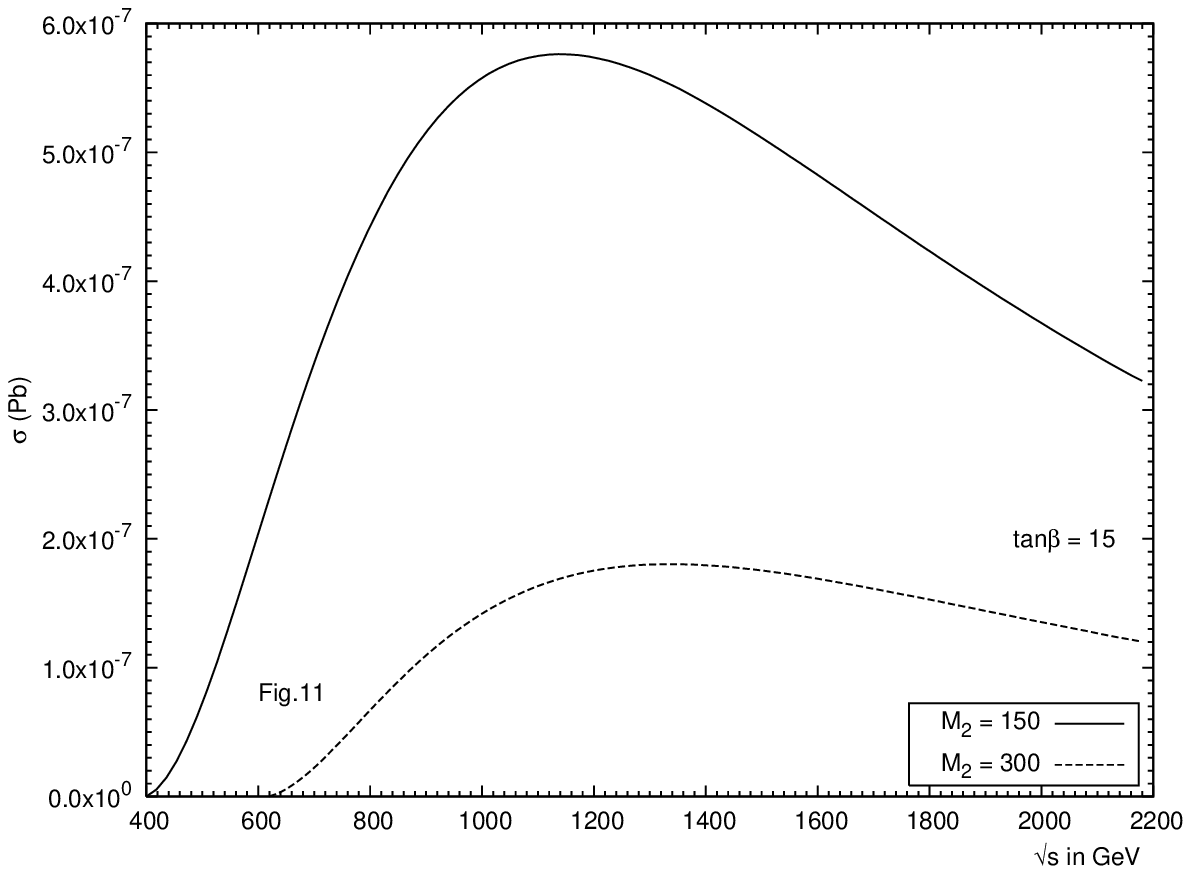}}
\vspace{0.5cm} \caption{\small Cross sections for
diagram no. 11 in figure \ref{feyn3}} \label{hXX11}
\end{figure}

\begin{figure}[th]
\vspace{-4.5cm}
\centerline{\epsfxsize=5.5truein\epsfbox{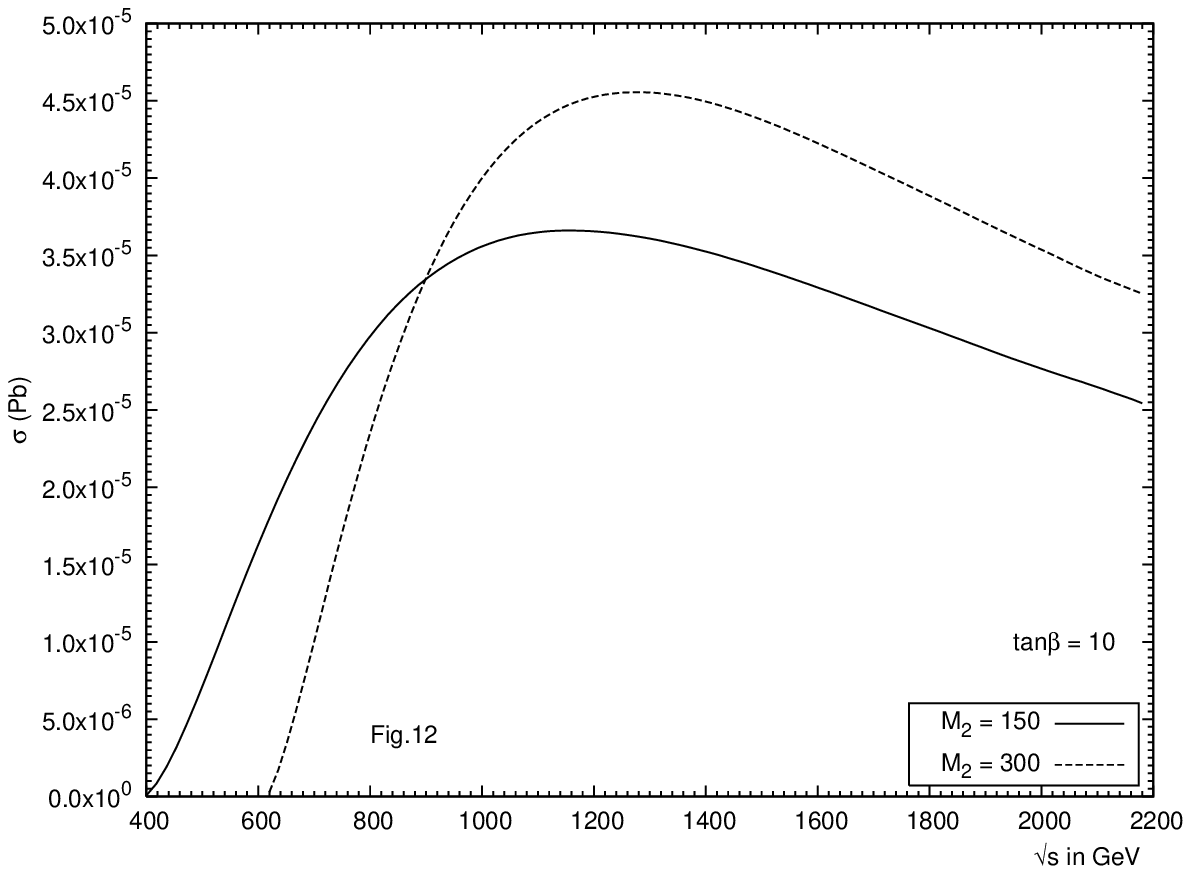}}
\vspace{-0.1cm}
\centerline{\epsfxsize=5.5truein\epsfbox{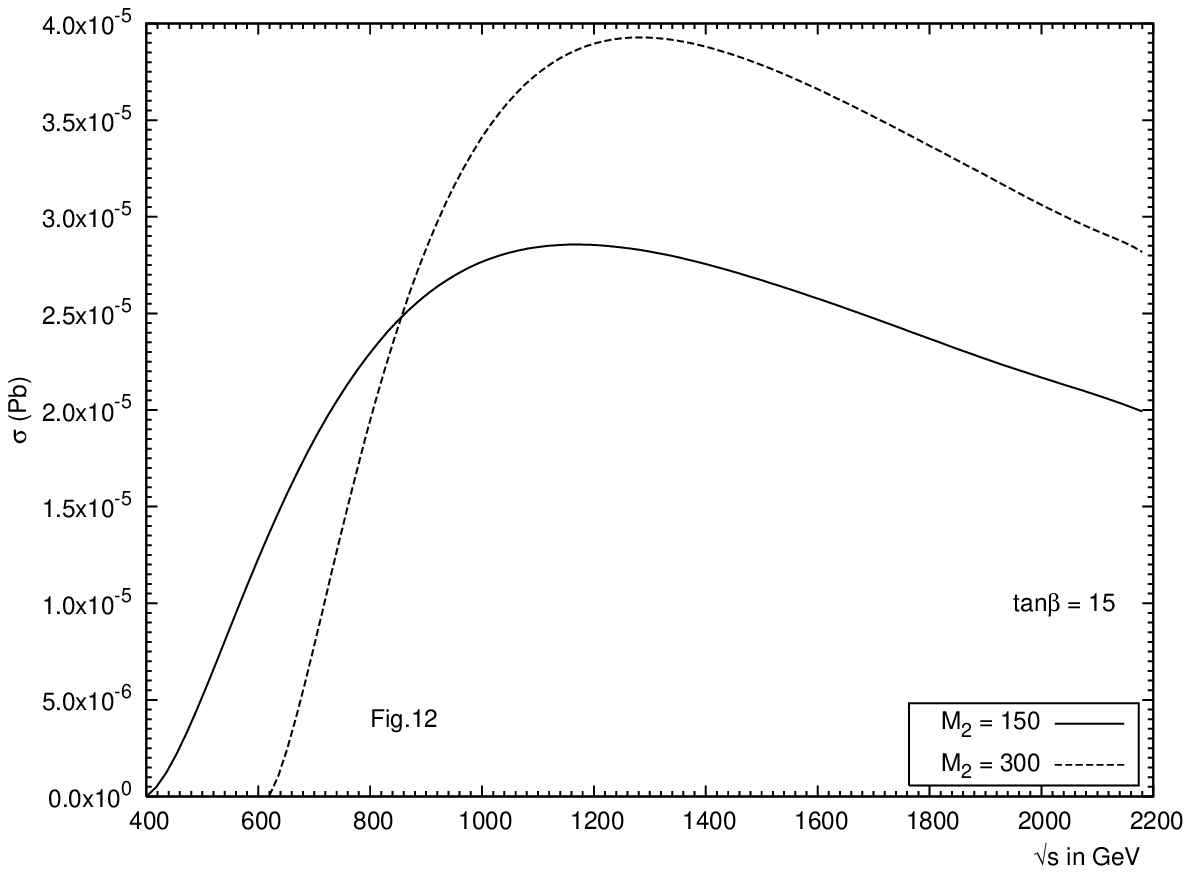}}
\vspace{0.5cm} \caption{\small Cross sections for
diagram no. 12 in figure \ref{feyn3}} \label{hXX12}
\end{figure}

\begin{figure}[th]
\vspace{-4.5cm}
\centerline{\epsfxsize=5.5truein\epsfbox{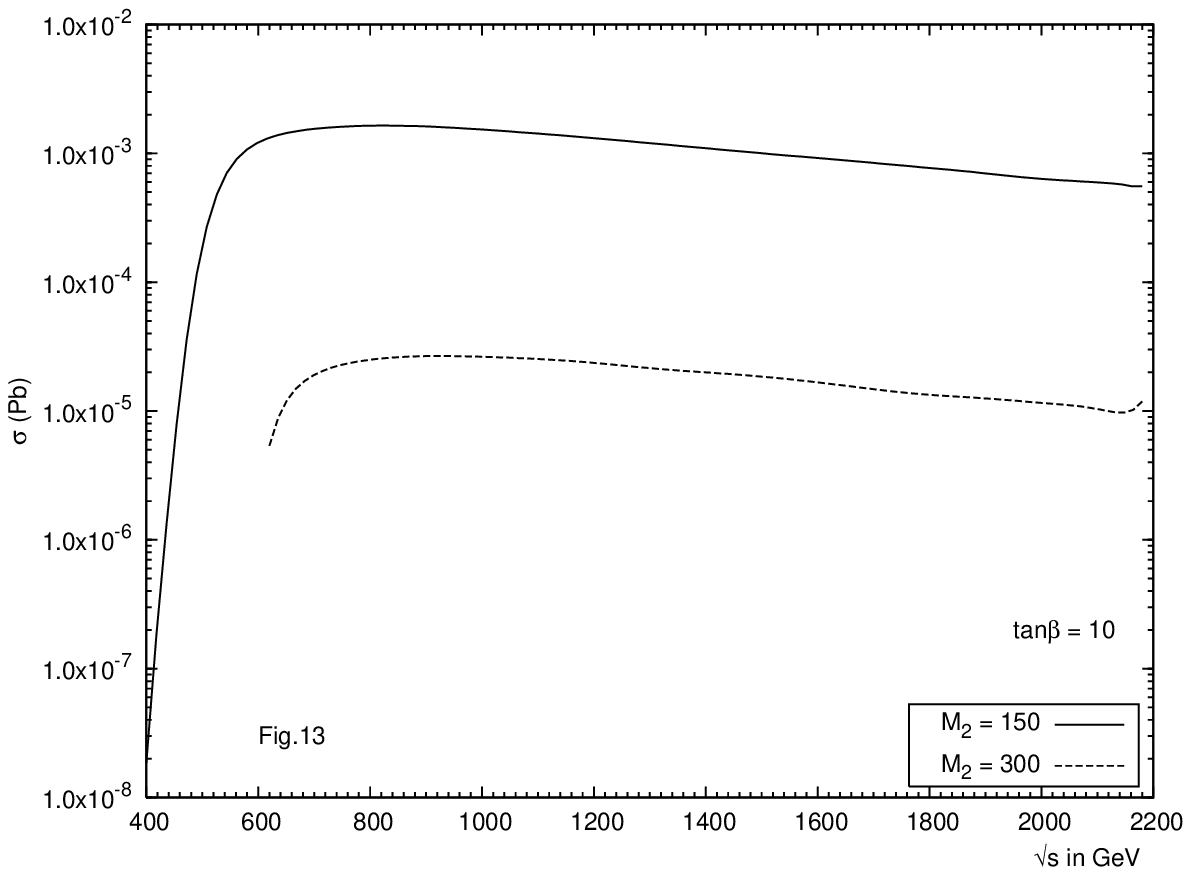}}
\vspace{-0.1cm}
\centerline{\epsfxsize=5.5truein\epsfbox{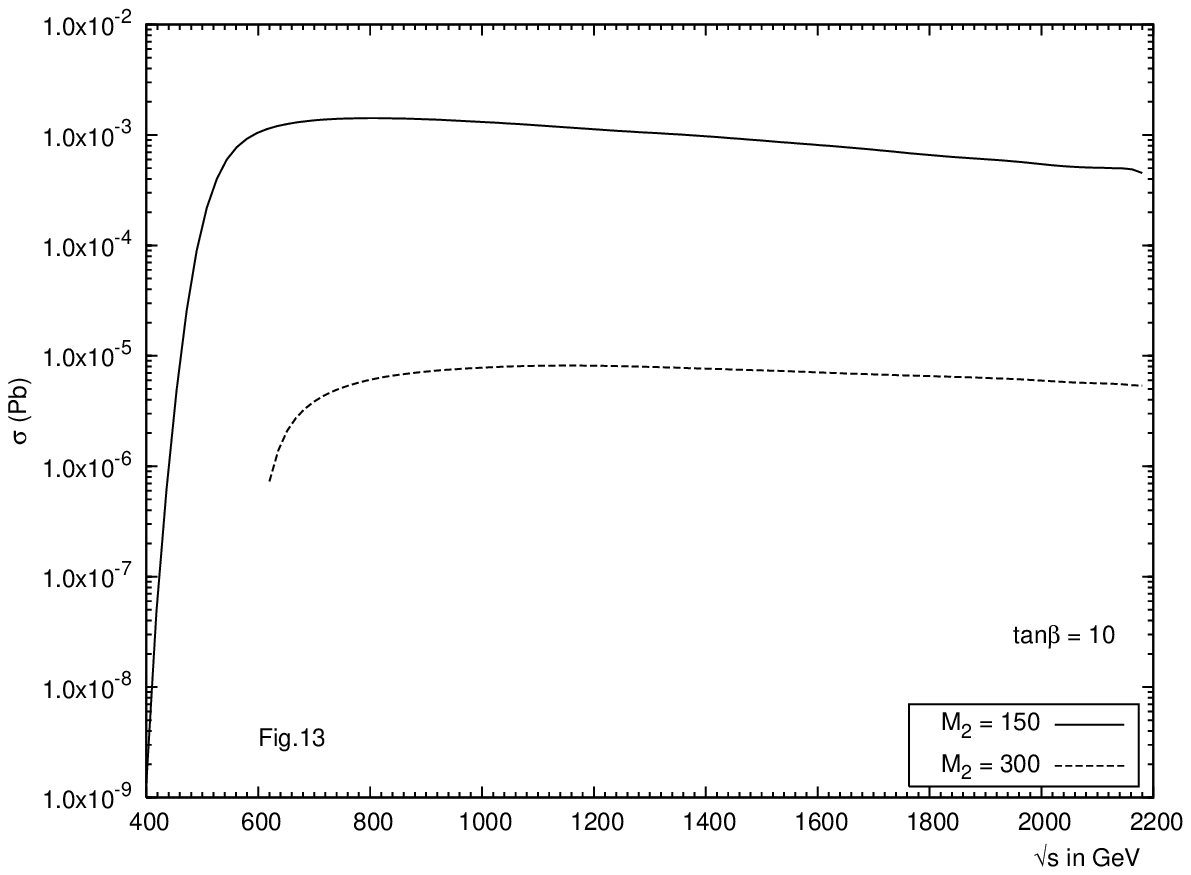}}
\vspace{0.5cm} \caption{\small Cross sections for
diagram no. 13 in figure \ref{feyn3}} \label{hXX13}
\end{figure}

\begin{figure}[th]
\vspace{-4.5cm}
\centerline{\epsfxsize=5.5truein\epsfbox{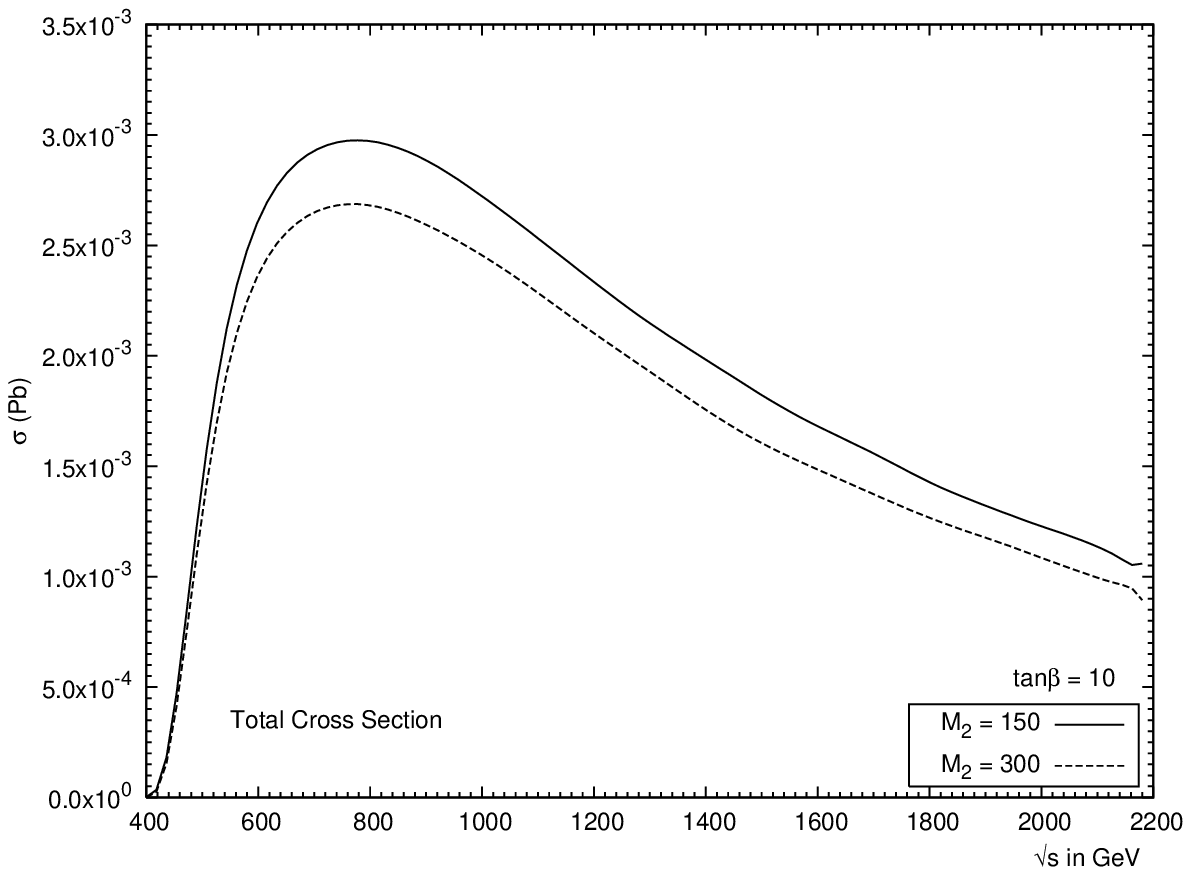}}
\vspace{-0.1cm}
\centerline{\epsfxsize=5.5truein\epsfbox{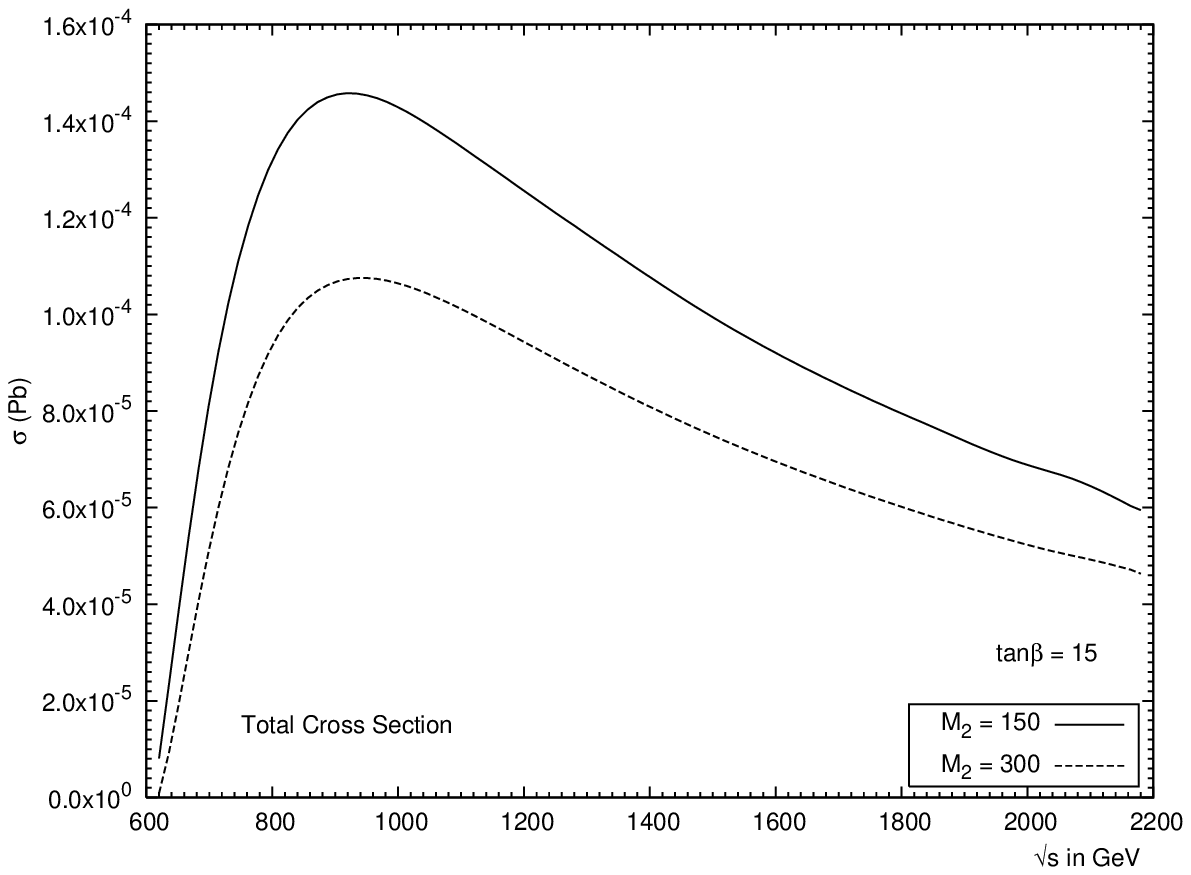}}
\vspace{0.5cm} \caption{\small Total cross sections for the
reaction $e^{-}e^{+}\rightarrow h \widetilde{\chi }_{1}^+
\widetilde{\chi }_{1}^-$} \label{hXXtot}
\end{figure}

\clearpage
\section{Conclusion}
Results of the previous section are summarized in tables
\ref{table5} and \ref{table6} for $M_2 = 150$ GeV and $M_2 = 300$
GeV respectively.\\ 
From these results, it was found that the maximum cross section obtained is $2.0371\times 10^{-3}$ [pb] at $tan\beta$ = 10 and $M_2$ = 150 GeV for diagram number 7, in which, the reaction procees through the $Z$ boson propagator. It takes place at a center of mass energy, $E_{CM}$ = 560 GeV. For $tan\beta$ = 15, the maximum cross section is $1.8340\times 10^{-3}$ [pb] where $M_2$ = 150 GeV for the diagram no. 8, in which, the reaction proceeds, again, through the $Z$ boson propagator. It occurs at a center of mass energy, $E_{CM}$ = 560 GeV.\\
The maximum cross section obtained by this reaction for $tan\beta$ = 10 and $M_2$ = 300 GeV is $1.8605\times 10^{-3}$ [pb] at a center of mass energy, $E_{CM}$ = 580 GeV for the feynman diagram no. 7 which proceeds through the $Z$ boson propagator. While the maximum cross section takes the value $5.4781\times 10^{-5}$ [pb] for $tan\beta$ = 15 and $M_2$ = 300 GeV and at a center of mass energy, $E_{CM}$ = 960 GeV. It occurs at diagram no. 1, which proceeds through the photon propagator.\\
Total cross section of this reaction is $3.0430\times 10^{-3}$ [pb] for $tan\beta$ = 10 and $M_2$ = 150 GeV at a center of mass energy, $E_{CM}$ = 780 GeV, and is $1.4490\times 10^{-4}$ [pb] for $tan\beta$ = 15 and $M_2$ = 150 GeV at a center of mass energy, $E_{CM}$ = 920 GeV.\\
Total cross section also assumes the value $2.7681\times 10^{-3}$ [pb] for $tan\beta$ = 15 and $M_2$ = 300 GeV at a center of mass energy, $E_{CM}$ = 840 GeV, and $1.0993\times 10^{-4}$ [pb] for $tan\beta$ = 15 and $M_2$ = 300 GeV at a center of mass energy, $E_{CM}$ = 920 GeV.
\begin{table}[htbp]
\begin{center}
\begin{tabular}[htbp]{|c||c|c||c|c|}
  \hline
  \hline
  Figure No.&\multicolumn{2}{c|}{$\sigma_{tan\beta = 10}$}&\multicolumn{2}{c|}{$\sigma_{tan\beta =15}$}\\
  \cline{2-5}
  &$E_{CM}$& $\sigma$ (Pb)&$E_{CM}$& $\sigma$ (Pb)\\
  \hline
  1 & 660 & 2.8134e-05 & 960& 6.5301e-05\\
  2 & 660 & 2.819e-05 & 960 & 6.5243e-05\\
  3 & 600 & 1.1079e-05 & 1240& 1.1034e-06\\
  4 & 1000 & 3.7658e-07 & 1720& 1.8818e-07\\
  5 & 660 & 2.8988e-05 & 960& 3.4915e-05\\
  6 & 660 & 2.9043e-05 & 960& 3.4883e-05\\
  7 & 560 & 0.0020371 & 720& 1.1311e-05\\
  8 & 560 & 0.0020166 & 560& 0.001834\\
  9 & 1180 & 3.7073e-05 & 1160& 2.8931e-05\\
  10 & 860 & 0.0017111 & 800& 0.0014741\\
  11 & 1100 & 5.2035e-07 & 1120& 5.8349e-07\\
  12 & 1140 & 3.6974e-05 & 1100& 2.885e-05\\
  13 & 780 & 0.0016977 & 860& 0.0014679\\
  total& 780& 0.003043& 920& 0.0001493\\
  \hline
\end{tabular}
\caption{Summary of the results obtained for the reaction,
$e^-(p1) e^+(p2) \rightarrow  h(p3) \widetilde{\chi }_{1}^+(p4)
\widetilde{\chi }_{1}^-(p5)$ for $M_2 = 150$ GeV} \label{table5}
\end{center}
\end{table}

\begin{table}[htbp]
\begin{center}
\begin{tabular}[htbp]{|c||c|c||c|c|}
  \hline
  \hline
  Figure No.&\multicolumn{2}{c|}{$\sigma_{tan\beta = 10}$}&\multicolumn{2}{c|}{$\sigma_{tan\beta =15}$}\\
  \cline{2-5}
  &$E_{CM}$& $\sigma$ (Pb)&$E_{CM}$& $\sigma$ (Pb)\\
  \hline
  1 & 680 & 2.1275e-05 & 960& 5.4781e-05\\
  2 & 660 & 2.1232e-05 & 960& 5.4754e-05\\
  3 & 600 & 1.0909e-05 & 1340& 1.1712e-06\\
  4 & 1060 & 2.1442e-07 & 1820& 1.0861e-07\\
  5 & 680 & 2.1987e-05 & 960& 2.9262e-05\\
  6 & 660 & 2.1988e-05 & 960& 2.9257e-05\\
  7 & 580 & 0.0018605 & 780& 2.6224e-06\\
  8 & 720 & 1.118e-05 & 800& 2.6398e-06\\
  9 & 1280 & 4.6013e-05 & 1260& 3.9595e-05\\
  10 & 840 & 2.8019e-05 & 1100& 8.4166e-06\\
  11 & 1280 & 1.6293e-07 & 1340& 1.8258e-07\\
  12 & 1280 & 4.5969e-05 & 1260& 3.962e-05\\
  13 & 920 & 2.8152e-05 & 1160& 8.646e-06\\
  total & 840 & 0.0027681 & 920 & 0.00010993\\
  \hline
\end{tabular}
\caption{Summary of the results obtained for the reaction,
$e^-(p1) e^+(p2) \rightarrow  h(p3) \widetilde{\chi }_{1}^+(p4)
\widetilde{\chi }_{1}^-(p5)$ for $M_2 = 300$ GeV} \label{table6}
\end{center}
\end{table}

\chapter{Production of a light neutral Higgs boson with a pair of neutralinos}

\section{Introduction}
In this chapter, the production of a light neutral Higgs boson is considered through the reaction, $e^{-}(p1)e^{+}(p2)\rightarrow h(p3) \widetilde{\chi }_{1}^o(p4) \widetilde{\chi }_{1}^o(p5)$, for different topologies and different propagators (see Appendix A). There are a total of 24 Feynman diagrams for this reaction (tree level approximation) for which we gave the matrix element corresponding to each diagram. Again, diagrams with the same topology which can be obtained by interchanging the indices were represented once.
Our work will proceed as before,

\begin{enumerate}
\item Feynman diagrams are given,

\item Diagrams with the same topology are represented once, but has been taken into considerations  when calculating the cross section.

\item  Matrix elements are written, all the four momenta squares are
defined to be mass squared $(>0)$,

\item Matrix elements are squared,

\item An average over the initial spin polarizations of the electron and
positron pair and a sum over the final spin states of the outgoing particles
arising from each initial spin state is carried out.

\item Results are represented graphically, and summarized in subsequent tables.
\end{enumerate}

\section{Feynman Diagrams}

The following is the set of Feynman diagrams which were used to calculate
the cross section of the associated production of a charged Higgs boson with
a pair of neutralinos. Our momentum notation is:
$e^{-}(p1)$, $e^{+}(p2)$, $h(p3)$ $\widetilde{\chi }_{1}^o(p4)$ and $\widetilde{\chi }_{1}^o(p5)$.

\begin{figure}[tph]
\begin{center}
\vskip-5.5cm \mbox{\hskip-3.5cm\centerline{\epsfig{file=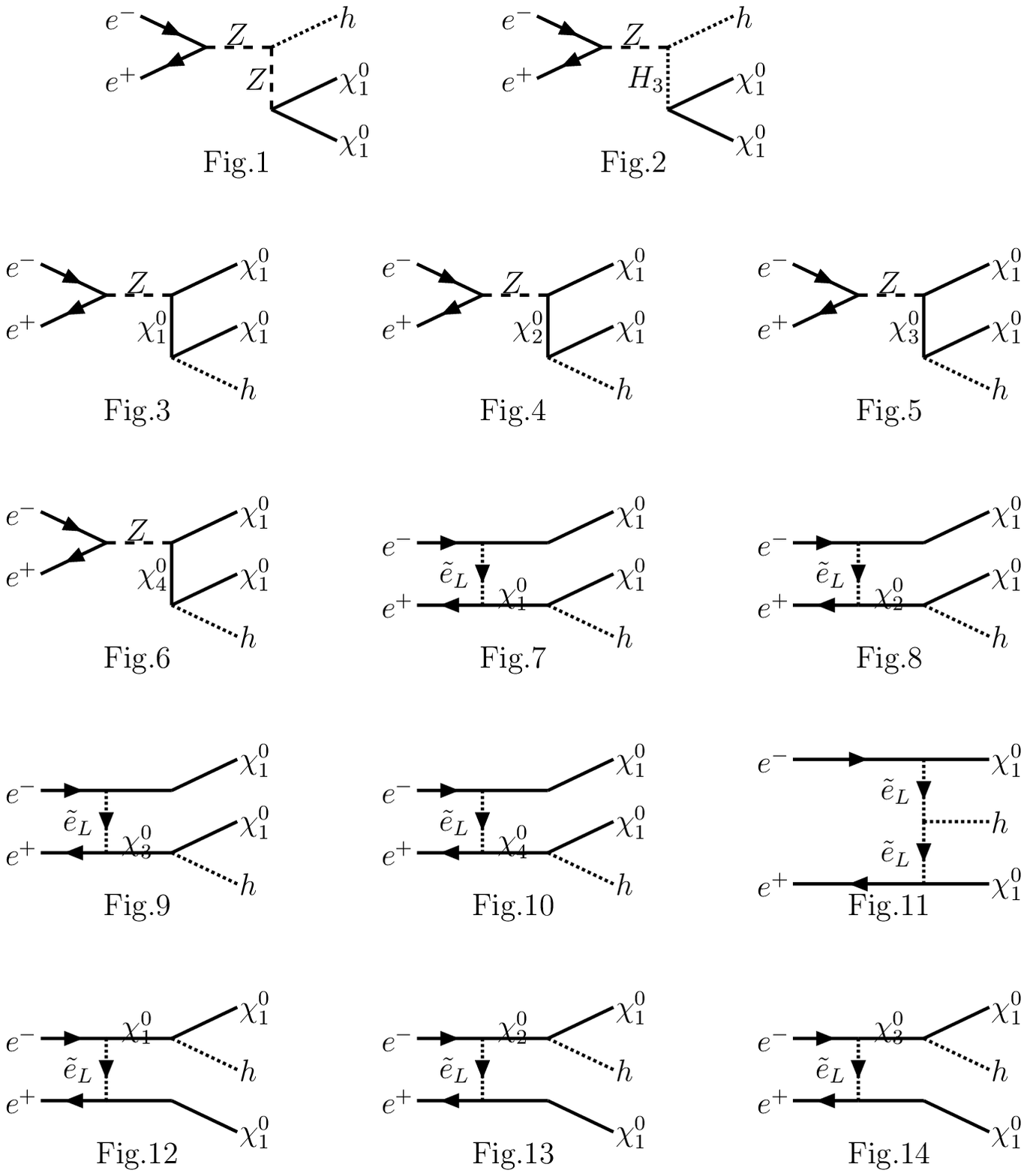,width=17cm}}}
\end{center}
\caption{Feynman diagrams for the reaction: $e^{-}(p1)e^{+}(p2)\rightarrow h(p3) \widetilde{\chi }_{1}^o(p4) \widetilde{\chi }_{1}^o(p5)$}
\label{feyn4}
\end{figure}

\begin{figure}[tph]
\begin{center}
\vskip-5.5cm \mbox{\hskip-3.5cm\centerline{\epsfig{file=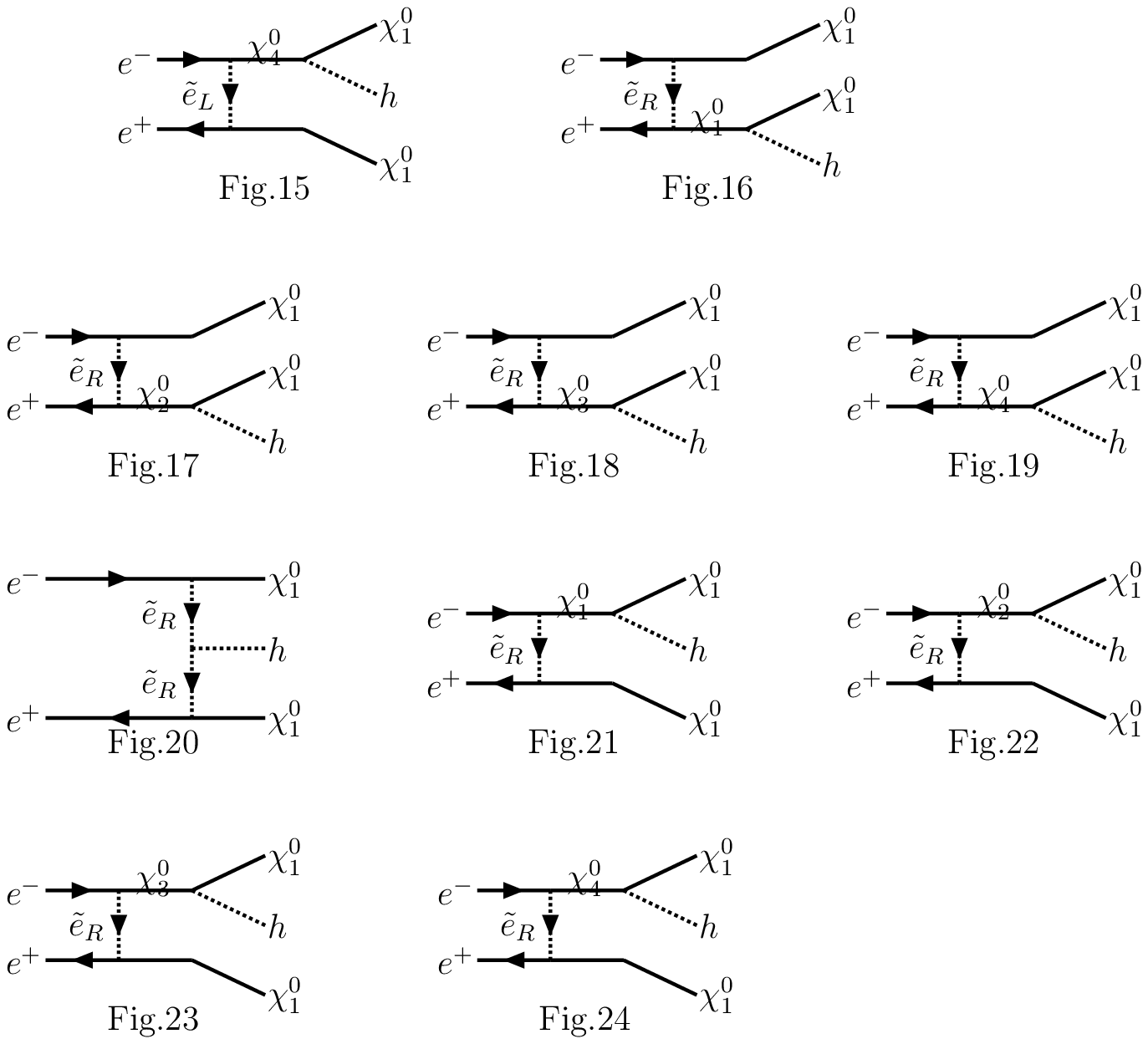,width=17cm}}}
\caption{Cont. Feynman diagrams for the reaction: $e^{-}(p1)e^{+}(p2)\rightarrow h(p3) \widetilde{\chi }_{1}^o(p4) \widetilde{\chi }_{1}^o(p5)$}
\end{center}
\label{feyn5}
\end{figure}

\newpage

\section{Matrix Elements}
The following is the set of matrix elements corresponding to diagrams in figures \ref{feyn4} and \ref{feyn5} used in our calculations:

\begin{eqnarray*}
\mathcal{M}_{1} &=&-i\overline{v}(p_{2})\gamma ^{\mu
}(B^{L}P_{L}+B^{R}P_{L})u(p_{1})\frac{g^{\mu \nu }-k_{\mu }k_{\nu }/M_{Z}^{2}%
}{(p_{1}+p_{2})^{2}-M_{Z}^{2}+i\epsilon } \\&&J_{1}g^{\nu \rho }\frac{g^{\rho
\sigma }-k_{\rho }k_{\sigma }/M_{Z}^{2}}{(p_{4}+p_{5})^{2}-M_{Z}^{2}+i%
\epsilon }\times \overline{u}(p_{5})\gamma _{\sigma }(S_{11}+S_{11}^{`}\gamma
_{5})v(p_{4})
\end{eqnarray*}

\begin{eqnarray*}
\mathcal{M}_{2} &=&i\overline{v}(p_{2})\gamma ^{\mu
}(B^{L}P_{L}+B^{R}P_{L})u(p_{1})\frac{g^{\mu \nu }-k_{\mu }k_{\nu }/M_{Z}^{2}%
}{(p_{1}+p_{2})^{2}-M_{Z}^{2}+i\epsilon } \times \\ && G_{1}(p_{4}+p_{5}-p_{3})^{\nu}
\frac{1}{(p_{4}+p_{5})^{2}-m_{H_{3}}^{2}}\overline{u}(p_{4})(R_{311}+R_{311}^{`}\gamma _{5})v(p_{5})
\end{eqnarray*}

\begin{eqnarray*}
\mathcal{M}_{3,4,5,6} &=&\overline{v}(p_{2})\gamma ^{\mu
}(B^{L}P_{L}+B^{R}P_{L})u(p_{1})\frac{g^{\mu \nu }-p_{1}^{\mu }p_{2}^{\nu
}/M_{Z}^{2}}{(p_{1}+p_{2})^{2}-M_{Z}^{2}+i\epsilon }\overline{u}%
(p_{5})(R_{111}+R_{111}^{`}\gamma _{5}) \\
&&\times \frac{\NEG{p}_{3}+\NEG{p}_{5}+m_{\widetilde{\chi }_{1}^{o}}}{%
(p_{3}+p_{5})^{2}-m_{\widetilde{\chi }_{1}^{o}}^{2}+i\epsilon }\gamma _{\nu
}(S_{11}+S_{11}^{`}\gamma _{5})v(p_{4})
\end{eqnarray*}

where $k$ assumes the values, $1,2,3,4$.

\begin{eqnarray*}
\mathcal{M}_{7,8,9,10} &=&-i\overline{v}(p_{2})(N_{L1}+N_{L1}^{`}\gamma _{5})%
\frac{\NEG{p}_{3}+\NEG{p}_{4}+m_{\widetilde{\chi }_{k}^{o}}}{%
(p_{3}+p_{4})^{2}-m_{\widetilde{\chi }_{k}^{o}}^{2}+i\epsilon }%
(R_{111}+R_{111}^{`}\gamma _{5})u(p_{4}) \\ && \times \frac{\NEG{p}_{1}-\NEG{p}_{5}+m_{%
\widetilde{e}}}{(p_{1}-p_{5})^{2}-m_{\widetilde{e}}^{2}+i\epsilon } \overline{u}(p_{5})(N_{L1}+N_{L1}^{`}\gamma _{5})u(p_{1})
\end{eqnarray*}

again, here $k$ assumes the values, $1,2,3,4$.

\begin{eqnarray*}
\mathcal{M}_{11} &=&-\overline{v}(p_{2})(N_{L1}+N_{L1}^{`}\gamma
_{5})u(p_{5})\frac{\NEG{p}_{2}-\NEG{p}_{5}+m_{\widetilde{e}}}{%
(p_{2}-p_{5})^{2}-m_{\widetilde{e}}^{2}+i\epsilon }\psi _{k}\frac{\NEG%
{p}_{1}-\NEG{p}_{4}+m_{\widetilde{e}}}{(p_{1}-p_{4})^{2}-m_{\widetilde{e}%
}^{2}+i\epsilon } \\
&&\times \overline{u}(p_{4})(N_{L1}+N_{L1}^{`}\gamma _{5})u(p_{1})
\end{eqnarray*}

Here, $\psi =\frac{igM_{Z}\cos (\alpha +\beta )}{\cos \theta _{w}}(\frac{1}{2%
}+\sin ^{2}\theta _{w})$.

\begin{eqnarray*}
\mathcal{M}_{12,13,14,15} &=&-i\overline{v}(p_{2})(N_{L1}+N_{L1}^{`}\gamma
_{5})u(p_{4})\frac{\NEG{p}_{2}-\NEG{p}_{4}+m_{\widetilde{e}}}{%
(p_{2}-p_{4})^{2}-m_{\widetilde{e}}^{2}+i\epsilon }\overline{u}%
(p_{5})(R_{111}+R_{111}^{`}\gamma _{5}) \\
&&\times \frac{\NEG{p}_{3}+\NEG{p}_{5}+m_{\widetilde{\chi }_{k}^{o}}}{%
(p_{3}+p_{5})^{2}-m_{\widetilde{\chi }_{k}^{o}}^{2}+i\epsilon }%
(N_{L1}+N_{L1}^{`}\gamma _{5})u(p_{1})
\end{eqnarray*}

\begin{eqnarray*}
\mathcal{M}_{16,17,18,19} &=&-i\overline{v}(p_{2})(N_{R1}+N_{R1}^{`}\gamma
_{5})\frac{\NEG{p}_{1}-\NEG{p}_{5}+m_{\widetilde{\chi }_{k}^{o}}}{%
(p_{1}-p_{5})^{2}-m_{\widetilde{\chi }_{k}^{o}}^{2}+i\epsilon }%
(R_{111}+R_{111}^{`}\gamma _{5})u(p_{4}) \\
&& \times \frac{\NEG{p}_{1}-\NEG{p}_{5}+m_{\widetilde{e}}}{(p_{1}-p_{5})^{2}-m_{%
\widetilde{e}}^{2}+i\epsilon }\overline{u}(p_{5})(N_{R1}+N_{R1}^{`}%
\gamma _{5})u(p_{1})
\end{eqnarray*}

As above, $k$ takes the vlaues, $1,2,3,4$.

\begin{eqnarray*}
\mathcal{M}_{20} &=&-\overline{v}(p_{2})(N_{R1}+N_{R1}^{`}\gamma
_{5})u(p_{4})\frac{\NEG{p}_{2}-\NEG{p}_{4}+m_{\widetilde{e}}}{%
(p_{2}-p_{4})^{2}-m_{\widetilde{e}}^{2}+i\epsilon }\psi _{k}\frac{\NEG%
{p}_{1}-\NEG{p}_{5}+m_{\widetilde{e}}}{(p_{1}-p_{5})^{2}-m_{\widetilde{e}%
}^{2}+i\epsilon } \\
&&\times \overline{u}(p_{5})(N_{R1}+N_{R1}^{`}\gamma _{5})u(p_{1})
\end{eqnarray*}

and also, $\psi =\frac{igM_{Z}\cos (\alpha +\beta )}{\cos \theta _{w}}(\frac{%
1}{2}+\sin ^{2}\theta _{w})$.

\begin{eqnarray*}
\mathcal{M}_{21,22,23,24} &=&\overline{v}(p_{2})(N_{R1}+N_{R1}^{`}\gamma
_{5})u(p_{5})\frac{\NEG{p}_{2}-\NEG{p}_{5}+m_{\widetilde{e}}}{%
(p_{2}-p_{5})^{2}-m_{\widetilde{e}}^{2}+i\epsilon }\overline{u}%
(p_{4})(R_{111}+R_{111}^{`}\gamma _{5}) \\
&&\times (R_{111}+R_{111}^{`}\gamma _{5})\frac{\NEG{p}_{3}+\NEG{p}_{4}+m_{%
\widetilde{\chi }_{k}^{o}}}{(p_{3}+p_{4})^{2}-m_{\widetilde{\chi }%
_{k}^{o}}^{2}+i\epsilon }(N_{R1}+N_{R1}^{`}\gamma _{5})u(p_{1})
\end{eqnarray*}
and $k=1,2,3,4$.

\noindent For the definitions of the constants used here, the
reader is referred to Appendix A.

\section{Cross Sections}
As before, to calculate the differential cross sections, and
hence, the total cross section, we need first to obtain the
squared matrix element for each Feynman diagram, where use of the
trace theorems was made. Later an average over the initial
spin polarizations of the electron and the positron pair and the
sum over the final spin states of the outgoing particles arising
from each initial spin state is carried out. The total cross
section as a function of the center of the mass energy (see Appendix
B) is then calculated.\\ 
Calculations were done with the following set of parameters:\\
$tan\beta = 10$, and $tan\beta = 15$ where $M_2 = 150$ or $M_2 = 300$.\\
All results are given in the following figures.
\begin{figure}[th]
\vspace{-4.5cm}
\centerline{\epsfxsize=5.5truein\epsfbox{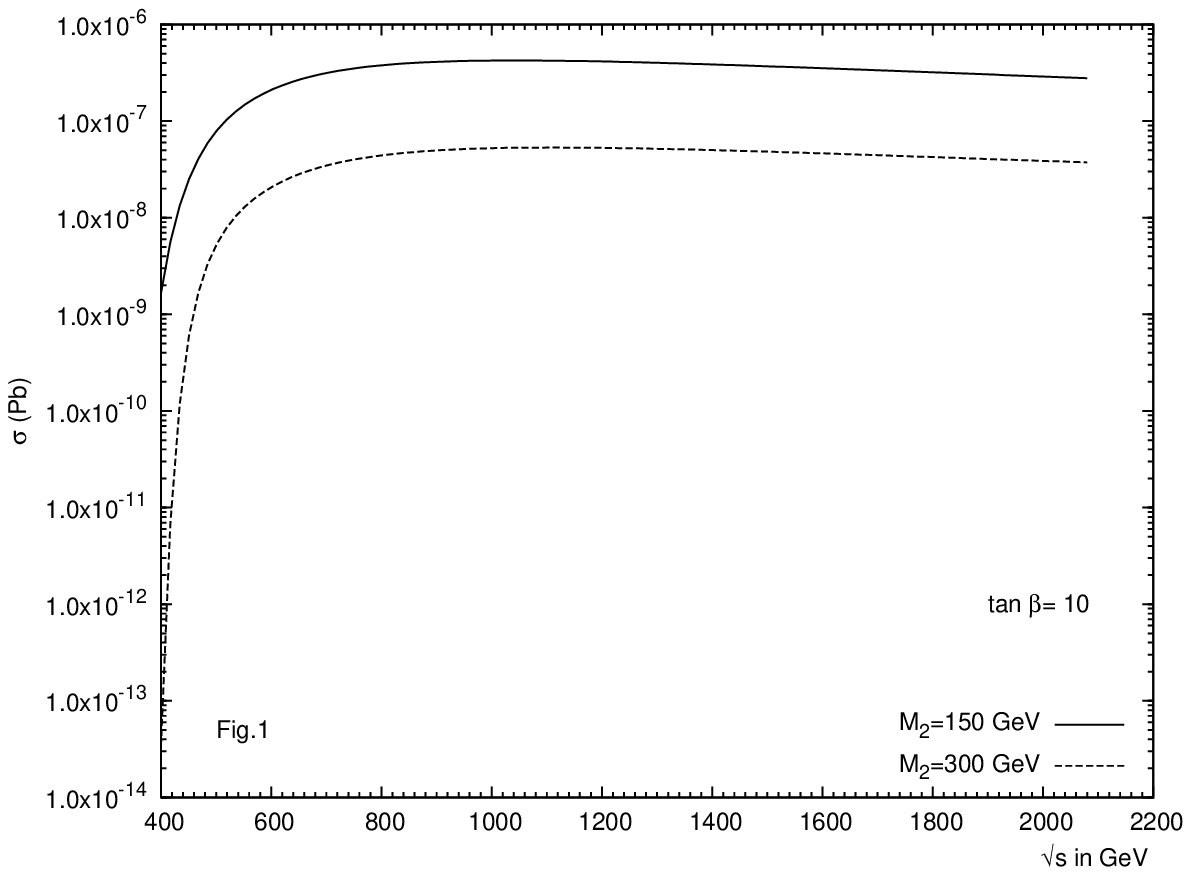}}
\vspace{-0.1cm}
\centerline{\epsfxsize=5.5truein\epsfbox{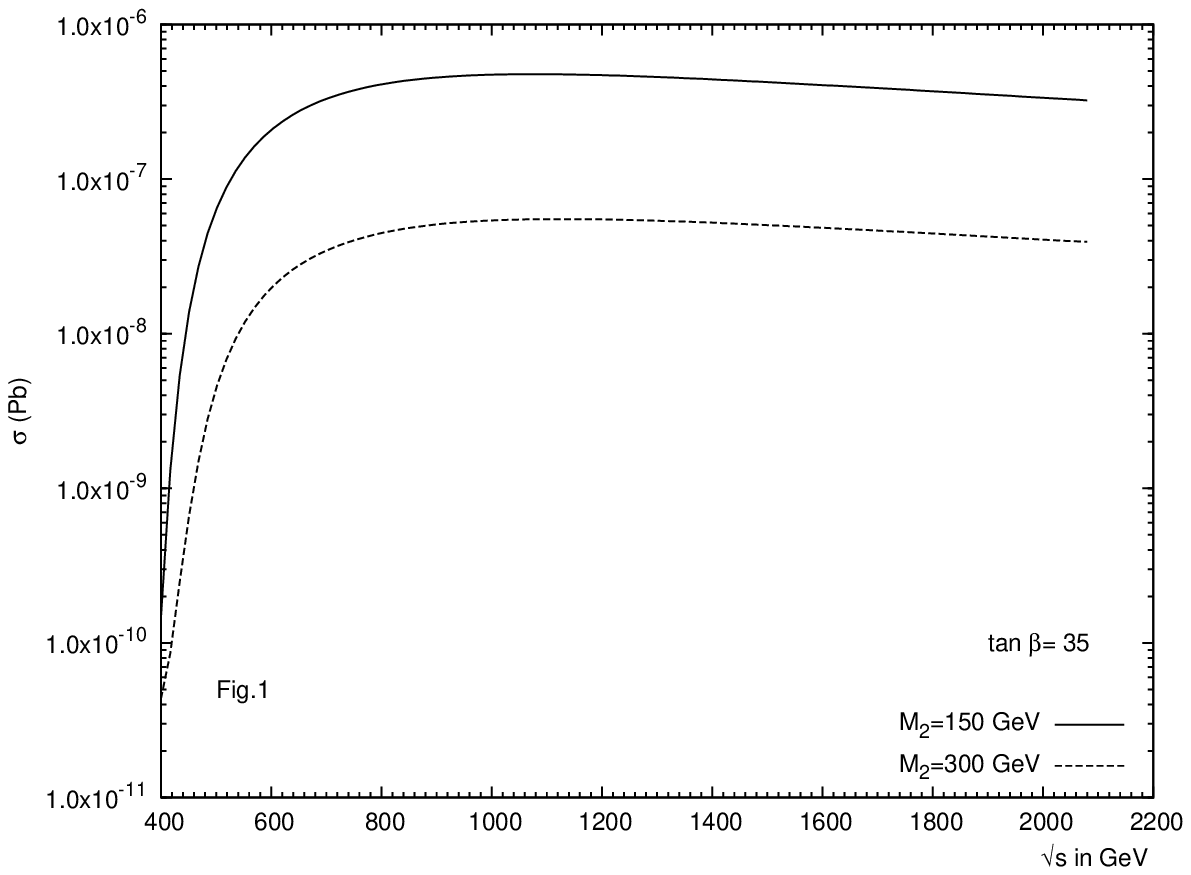}}
\vspace{0.5cm} 
\caption{\small Cross sections for
diagram no. 1 in figure \ref{feyn4}} 
\label{hxx1}
\end{figure}

\begin{figure}[th]
\vspace{-4.5cm}
\centerline{\epsfxsize=5.5truein\epsfbox{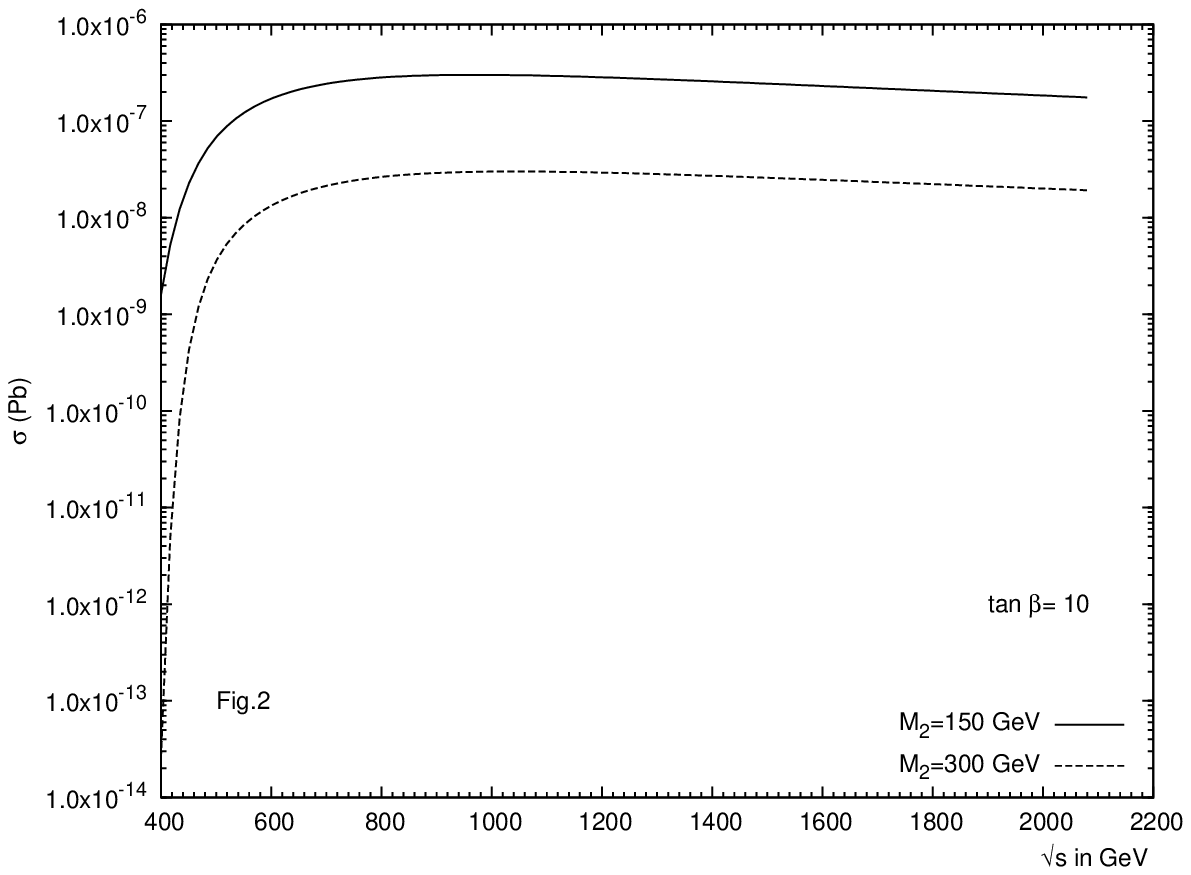}}
\vspace{-0.1cm}
\centerline{\epsfxsize=5.5truein\epsfbox{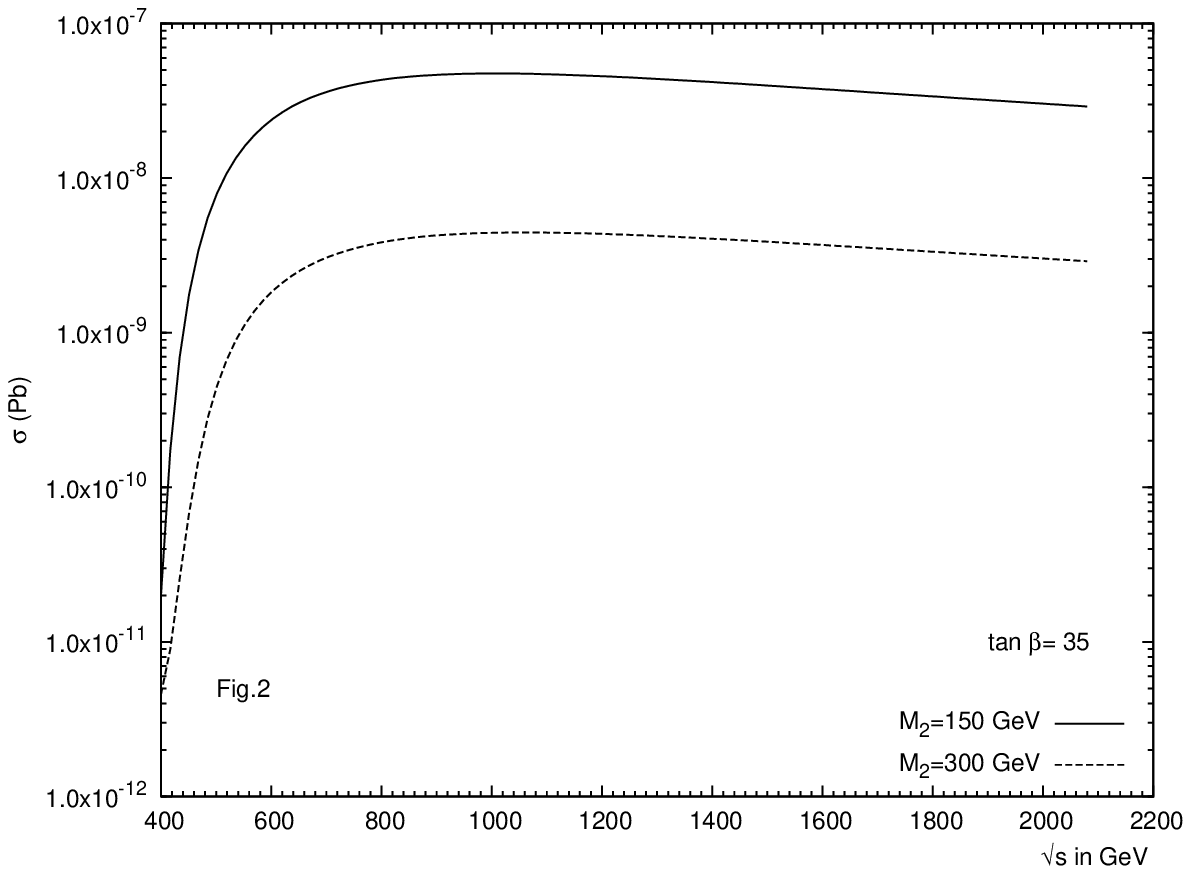}}
\vspace{0.5cm} 
\caption{\small Cross sections for
diagram no. 2 in figure \ref{feyn4}} 
\label{hxx2}
\end{figure}

\begin{figure}[th]
\vspace{-4.5cm}
\centerline{\epsfxsize=5.5truein\epsfbox{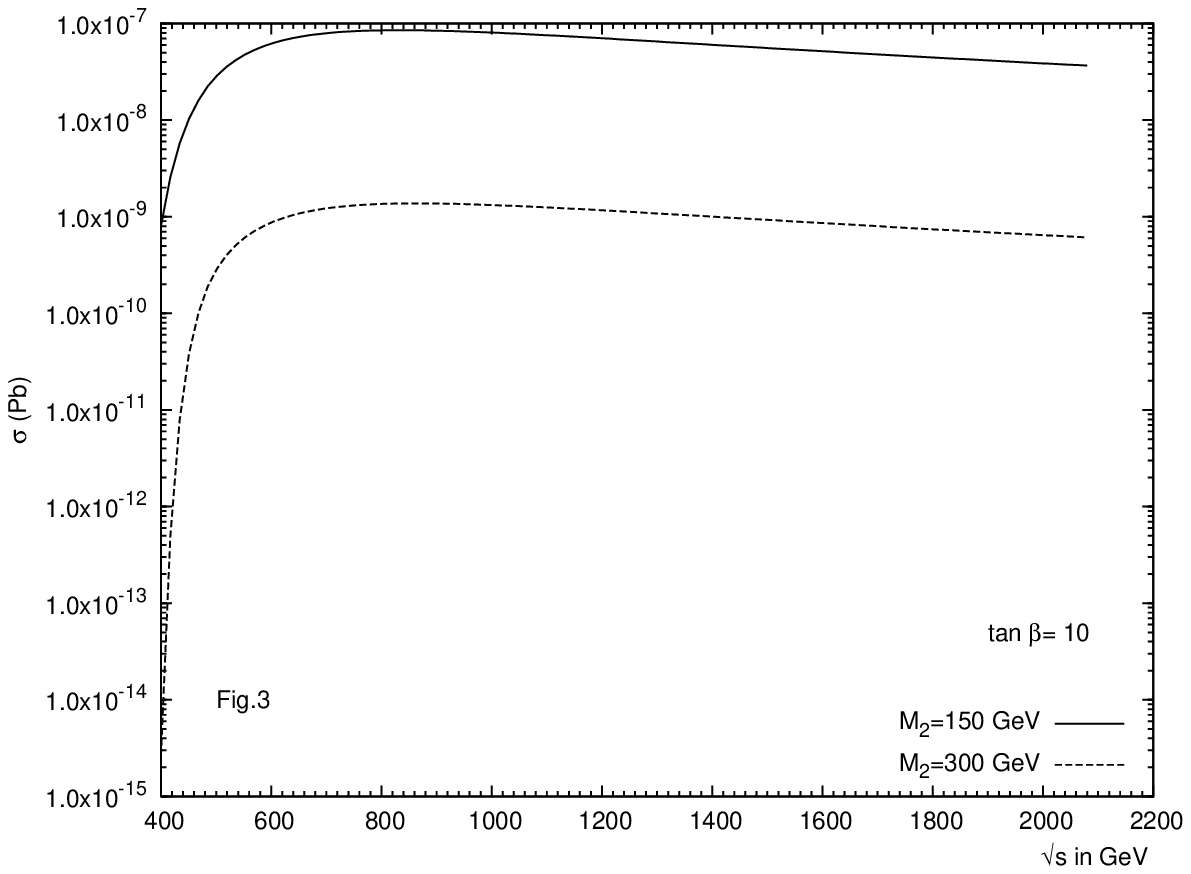}}
\vspace{-0.1cm}
\centerline{\epsfxsize=5.5truein\epsfbox{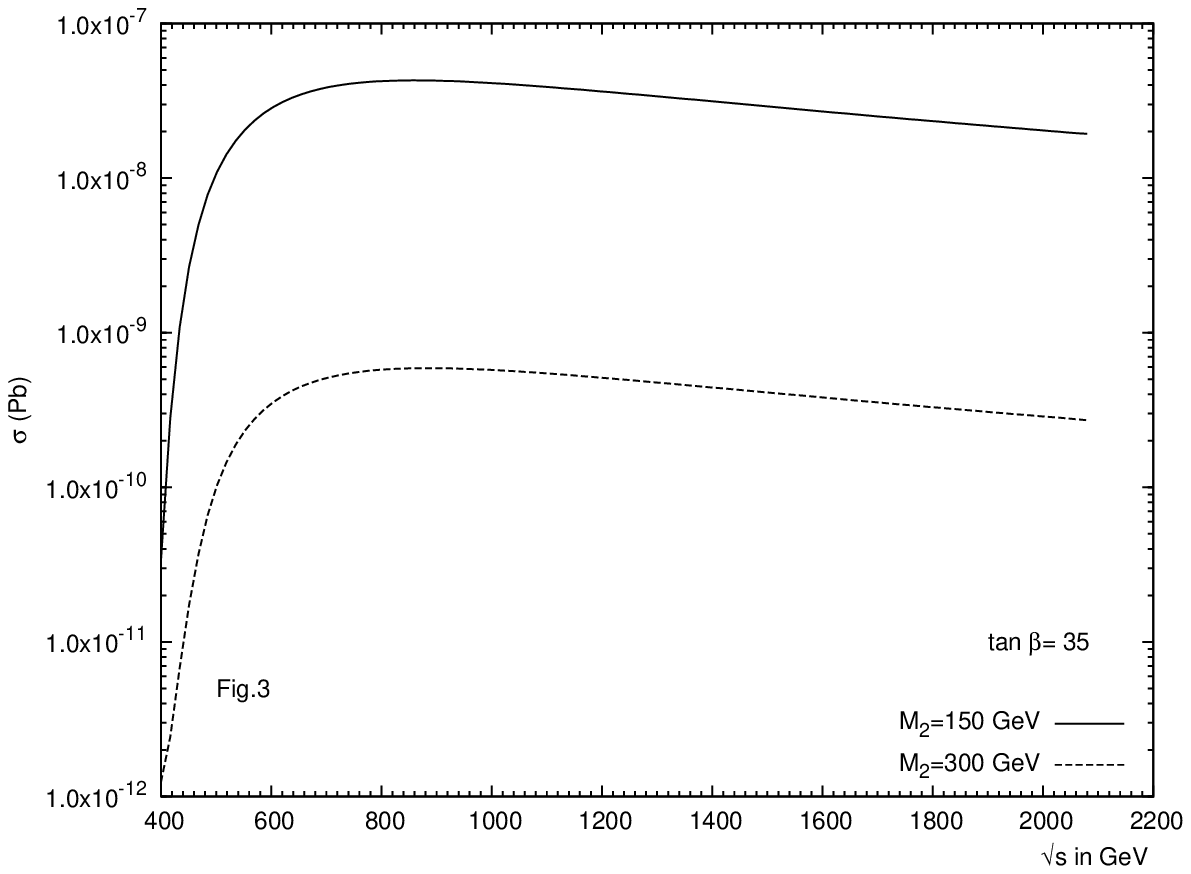}}
\vspace{0.5cm} 
\caption{\small Cross sections for
diagram no. 3 in figure \ref{feyn4}} 
\label{hxx3}
\end{figure}

\begin{figure}[th]
\vspace{-4.5cm}
\centerline{\epsfxsize=5.5truein\epsfbox{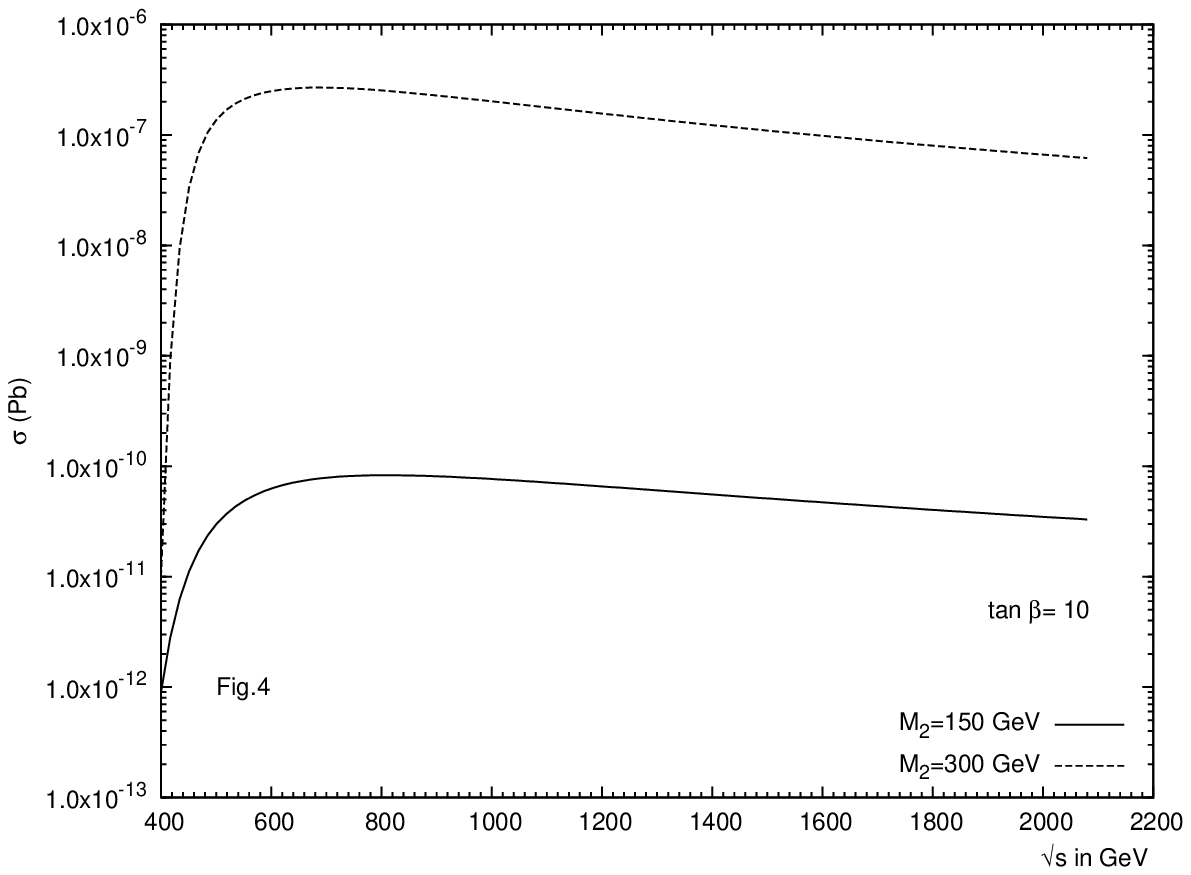}}
\vspace{-0.1cm}
\centerline{\epsfxsize=5.5truein\epsfbox{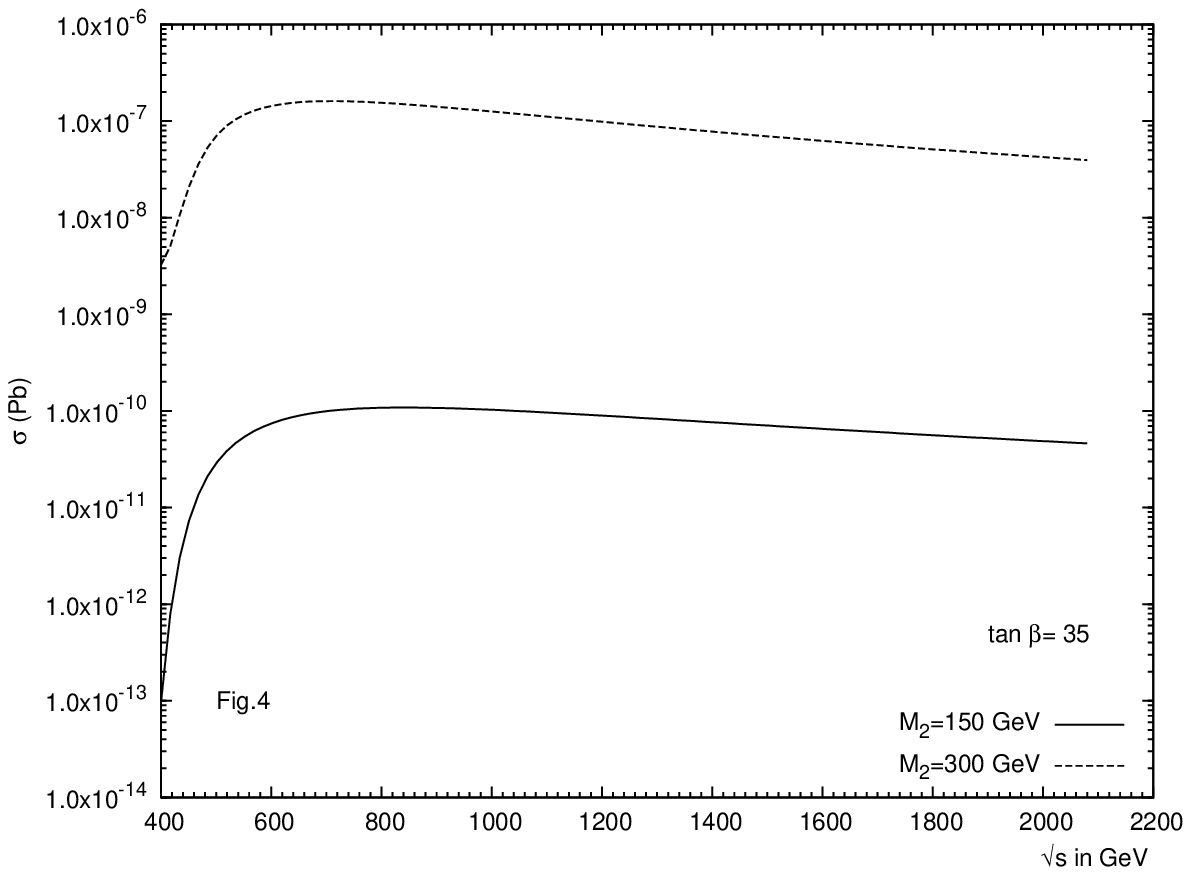}}
\vspace{0.5cm} 
\caption{\small Cross sections for
diagram no. 4 in figure \ref{feyn4}} 
\label{hxx4}
\end{figure}

\begin{figure}[th]
\vspace{-4.5cm}
\centerline{\epsfxsize=5.5truein\epsfbox{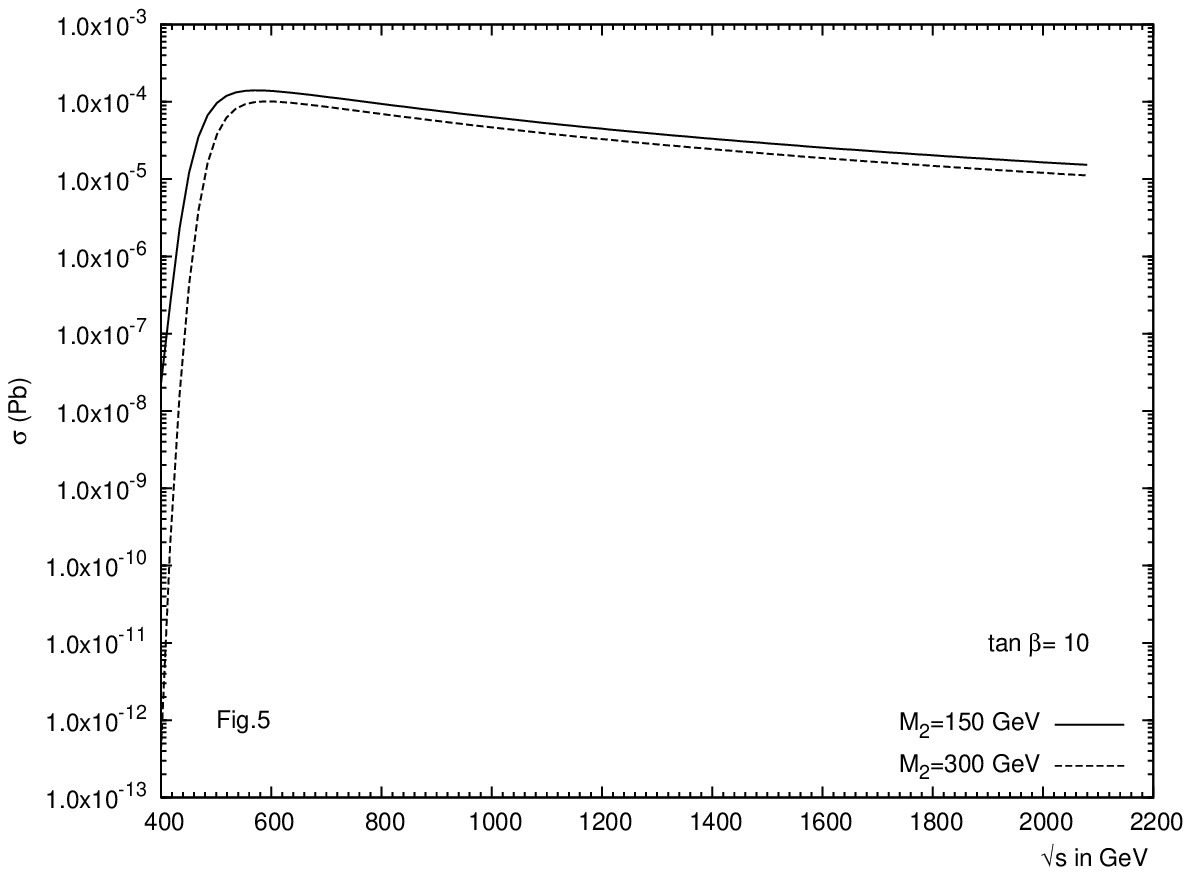}}
\vspace{-0.1cm}
\centerline{\epsfxsize=5.5truein\epsfbox{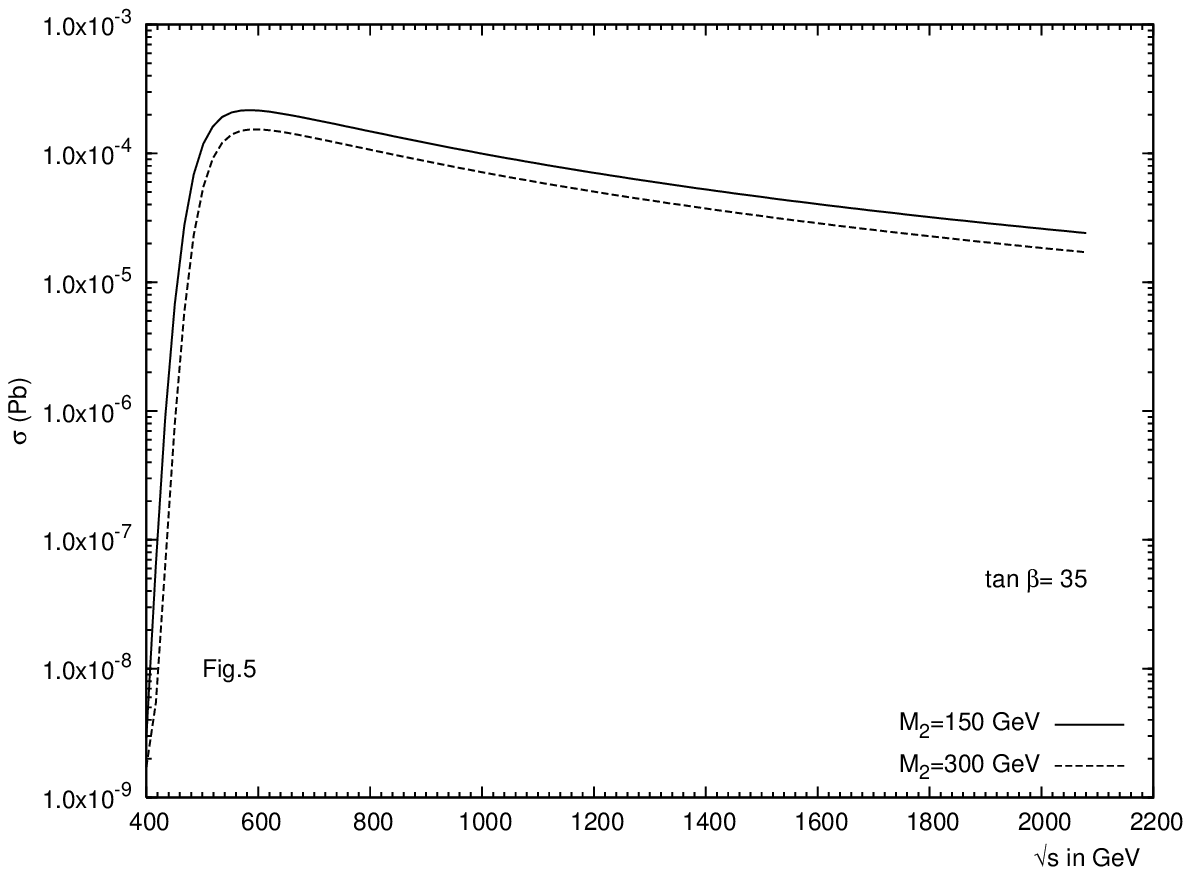}}
\vspace{0.5cm} 
\caption{\small Cross sections for
diagram no. 5 in figure \ref{feyn4}} 
\label{hxx5}
\end{figure}

\begin{figure}[th]
\vspace{-4.5cm}
\centerline{\epsfxsize=5.5truein\epsfbox{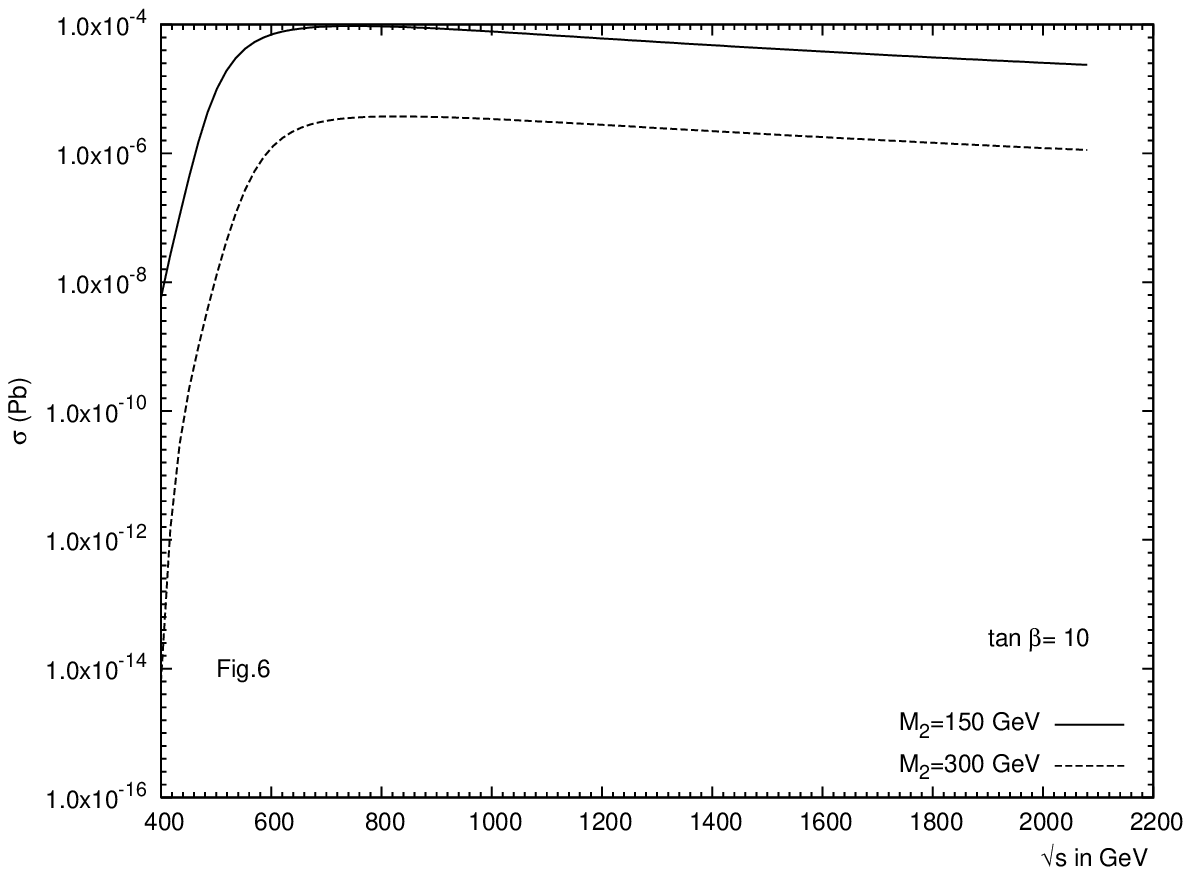}}
\vspace{-0.1cm}
\centerline{\epsfxsize=5.5truein\epsfbox{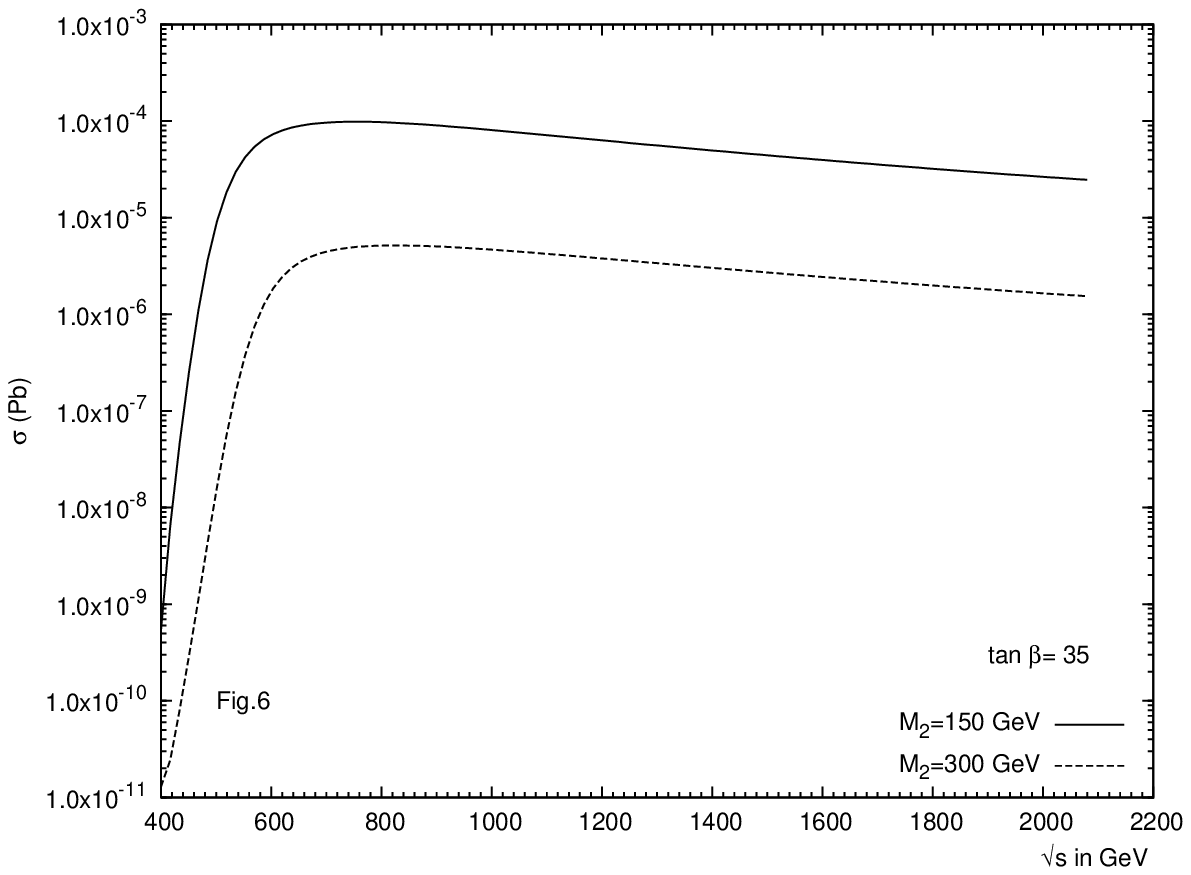}}
\vspace{0.5cm} 
\caption{\small Cross sections for
diagram no. 6 in figure \ref{feyn4}} 
\label{hxx6}
\end{figure}

\begin{figure}[th]
\vspace{-4.5cm}
\centerline{\epsfxsize=5.5truein\epsfbox{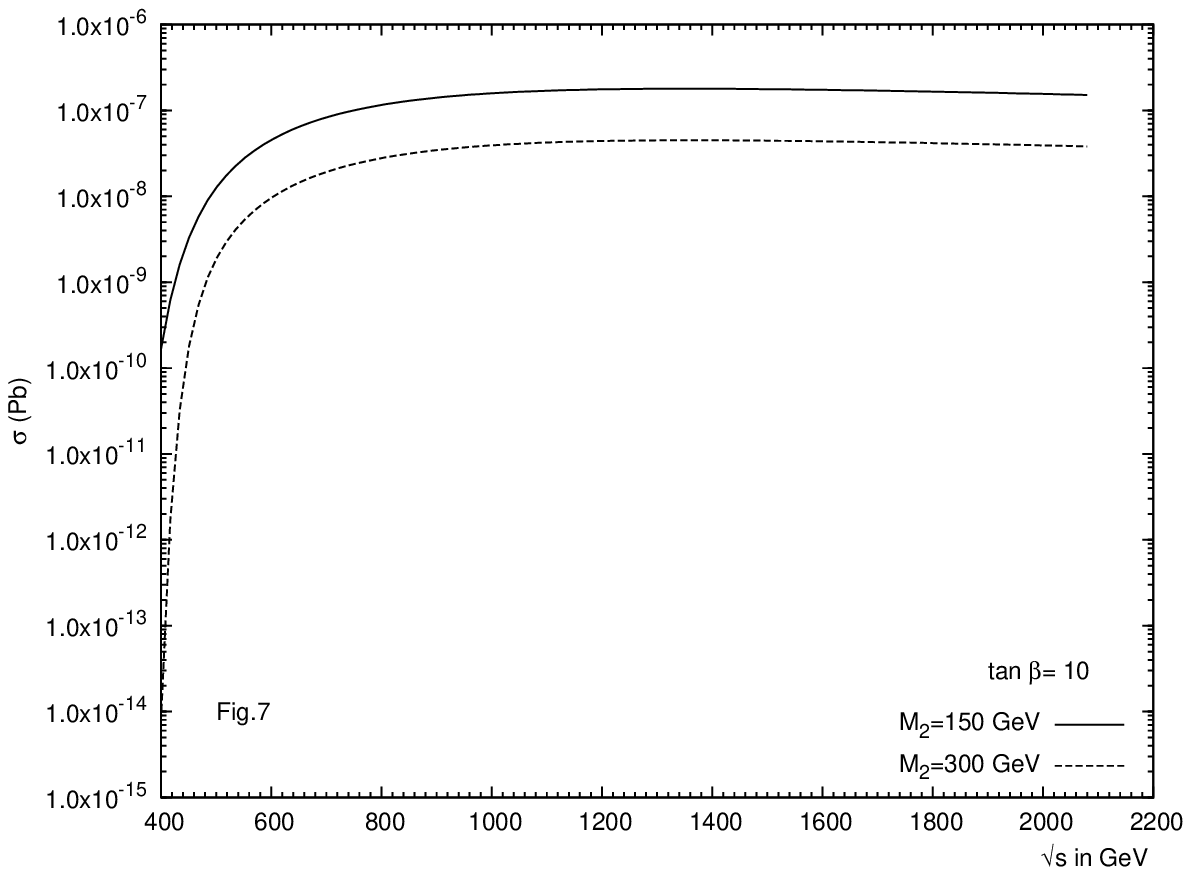}}
\vspace{-0.1cm}
\centerline{\epsfxsize=5.5truein\epsfbox{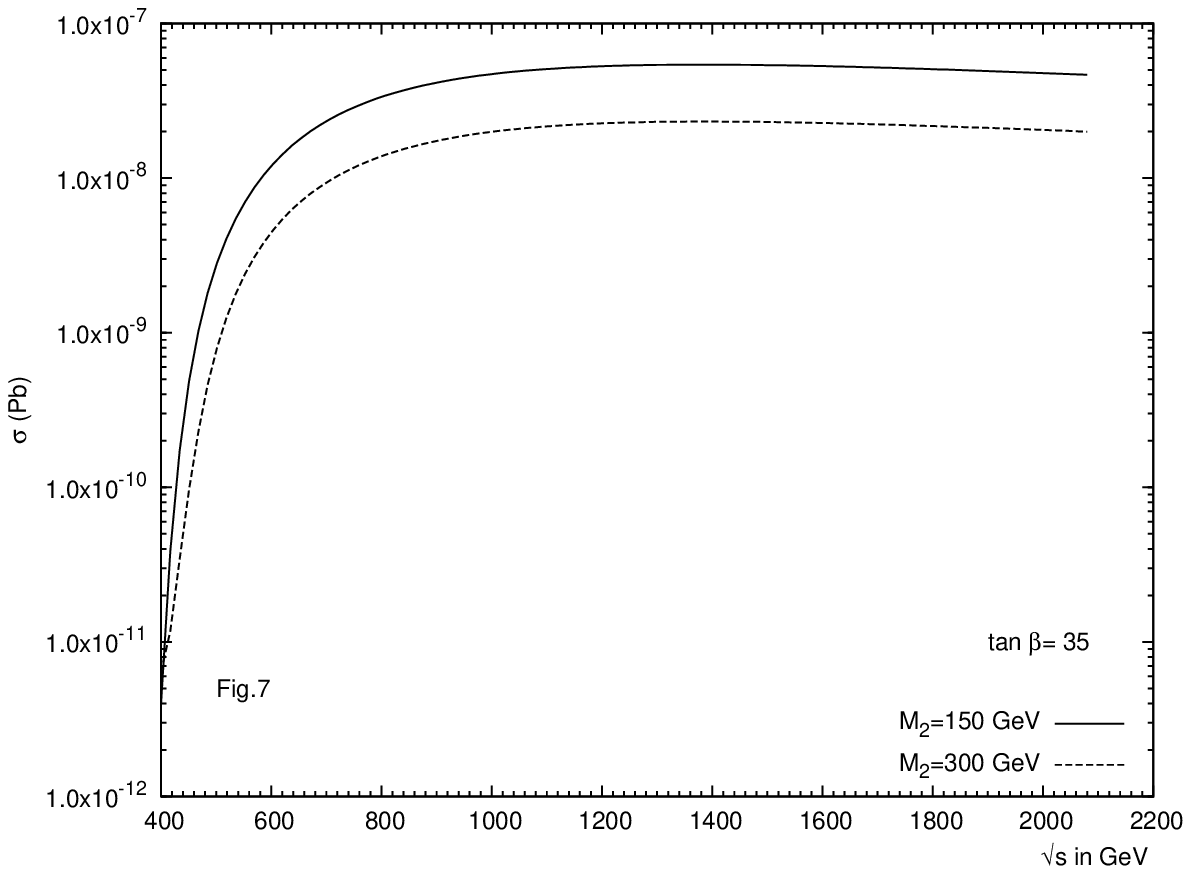}}
\vspace{0.5cm} 
\caption{\small Cross sections for
diagram no. 7 in figure \ref{feyn4}} 
\label{hxx7}
\end{figure}

\begin{figure}[th]
\vspace{-4.5cm}
\centerline{\epsfxsize=5.5truein\epsfbox{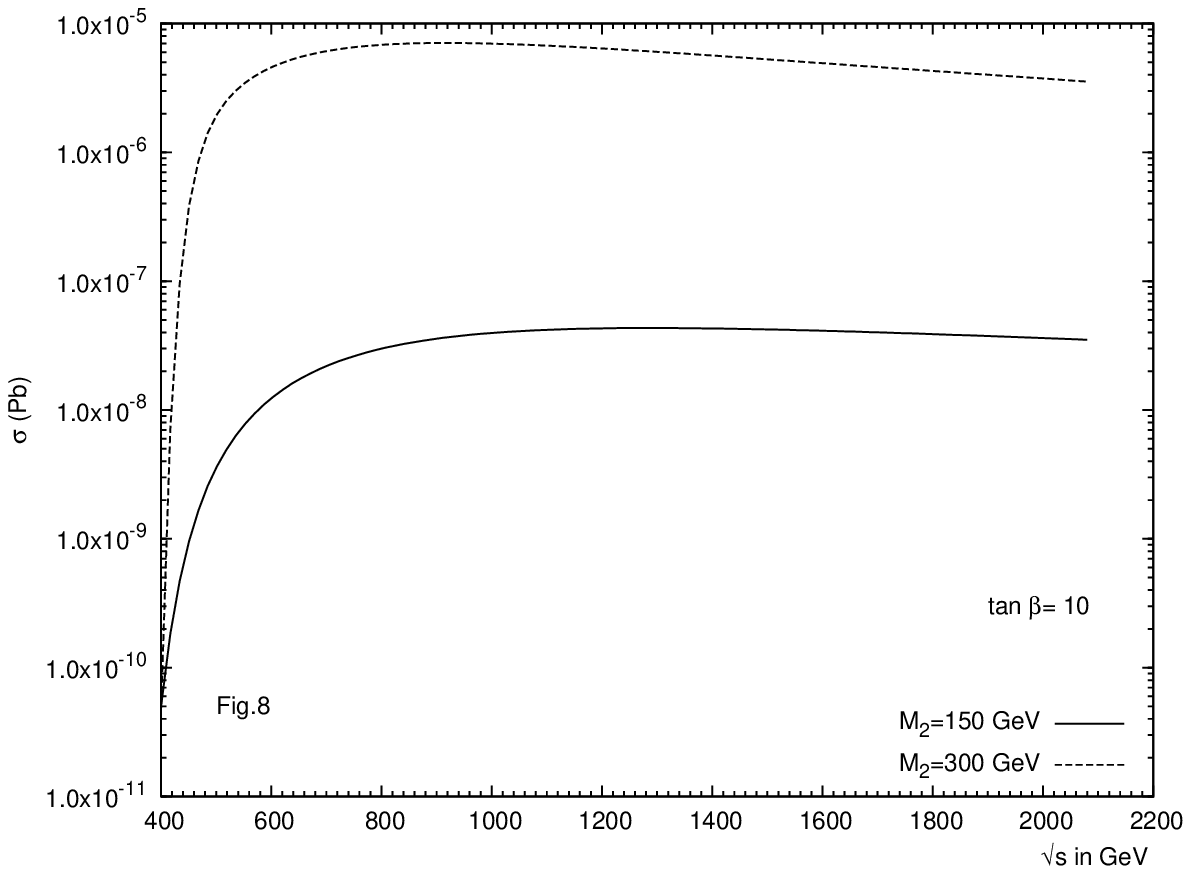}}
\vspace{-0.1cm}
\centerline{\epsfxsize=5.5truein\epsfbox{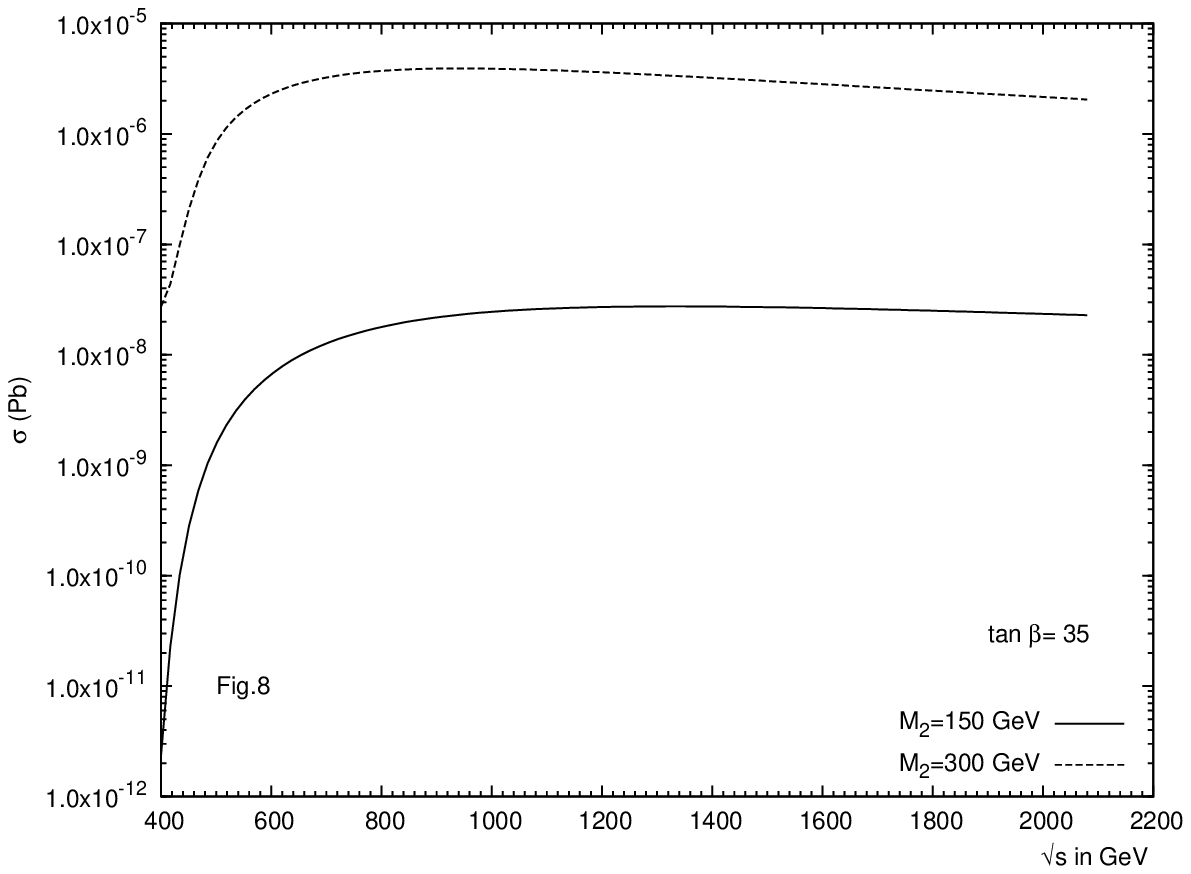}}
\vspace{0.5cm} 
\caption{\small Cross sections for
diagram no. 8 in figure \ref{feyn4}} 
\label{hxx8}
\end{figure}

\begin{figure}[th]
\vspace{-4.5cm}
\centerline{\epsfxsize=5.5truein\epsfbox{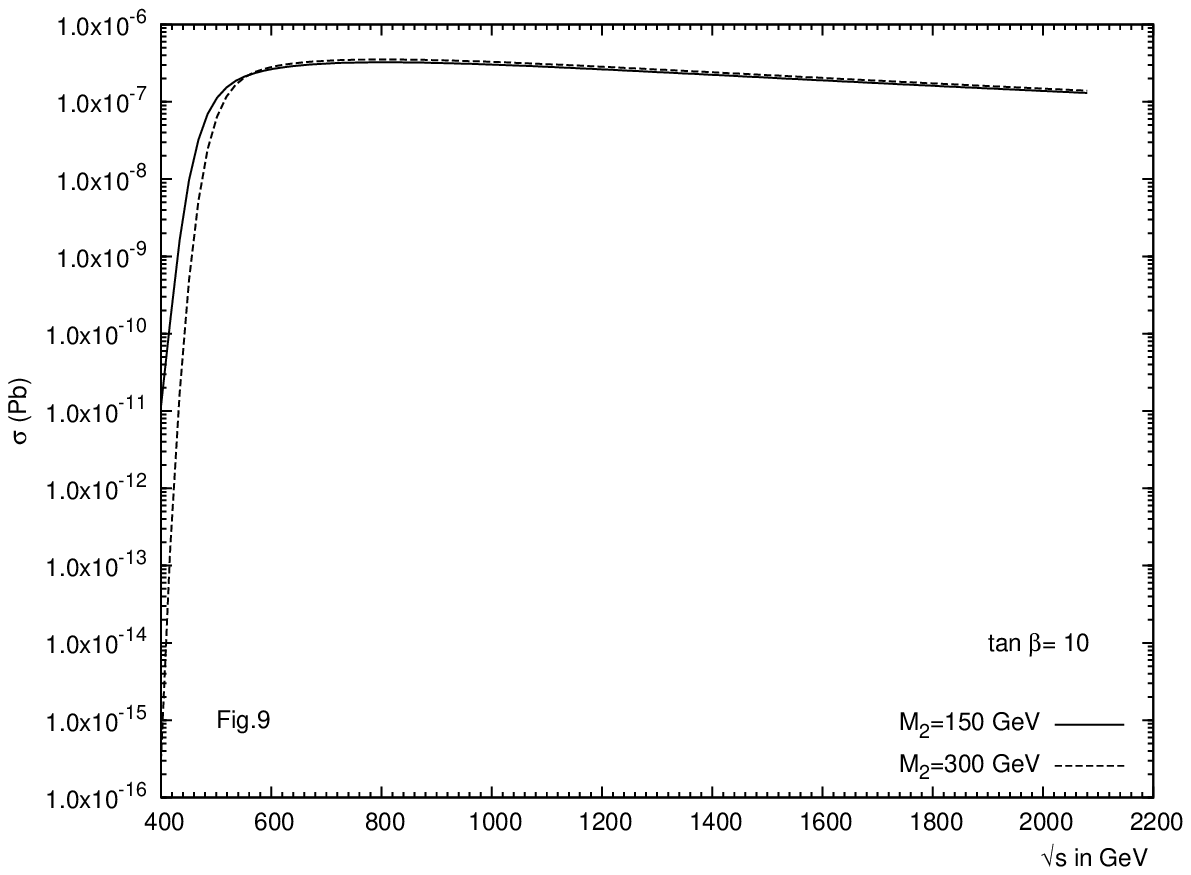}}
\vspace{-0.1cm}
\centerline{\epsfxsize=5.5truein\epsfbox{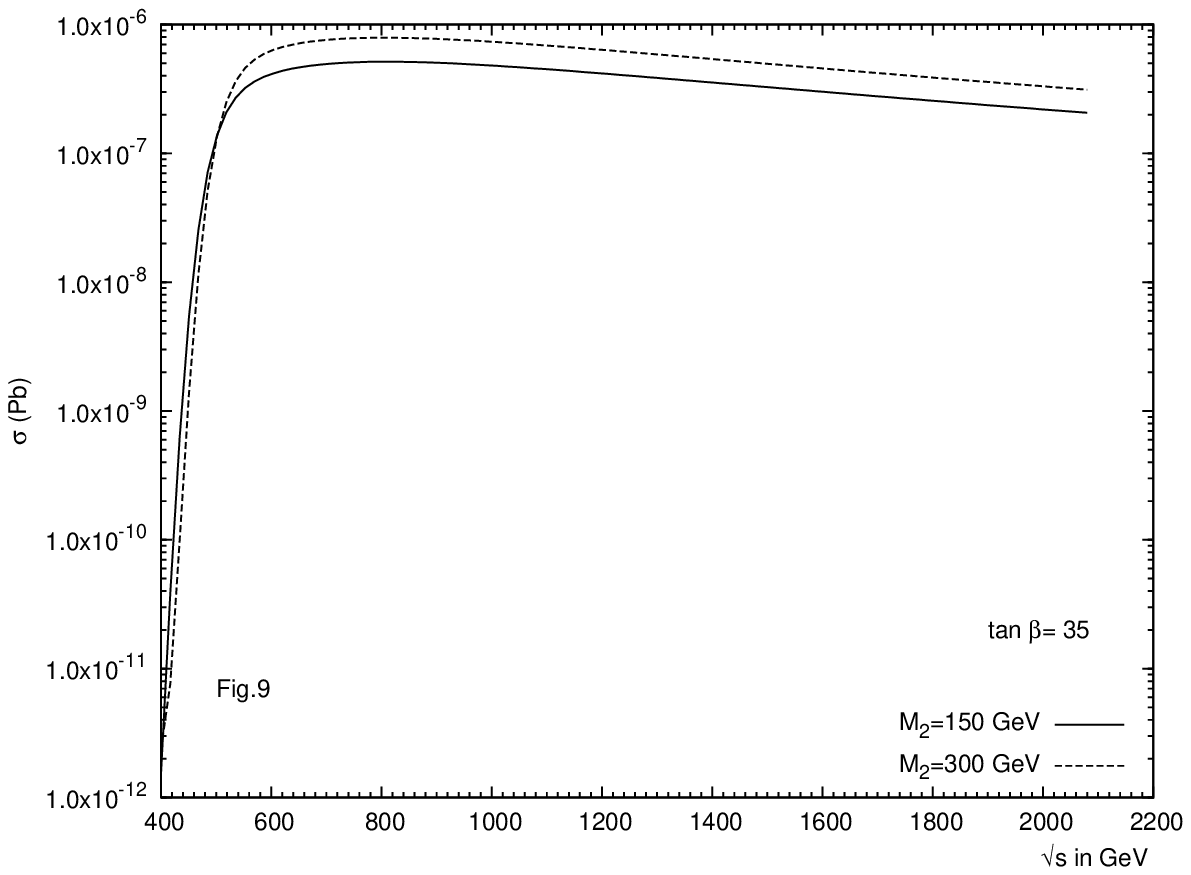}}
\vspace{0.5cm} 
\caption{\small Cross sections for
diagram no. 9 in figure \ref{feyn4}} 
\label{hxx9}
\end{figure}

\begin{figure}[th]
\vspace{-4.5cm}
\centerline{\epsfxsize=5.5truein\epsfbox{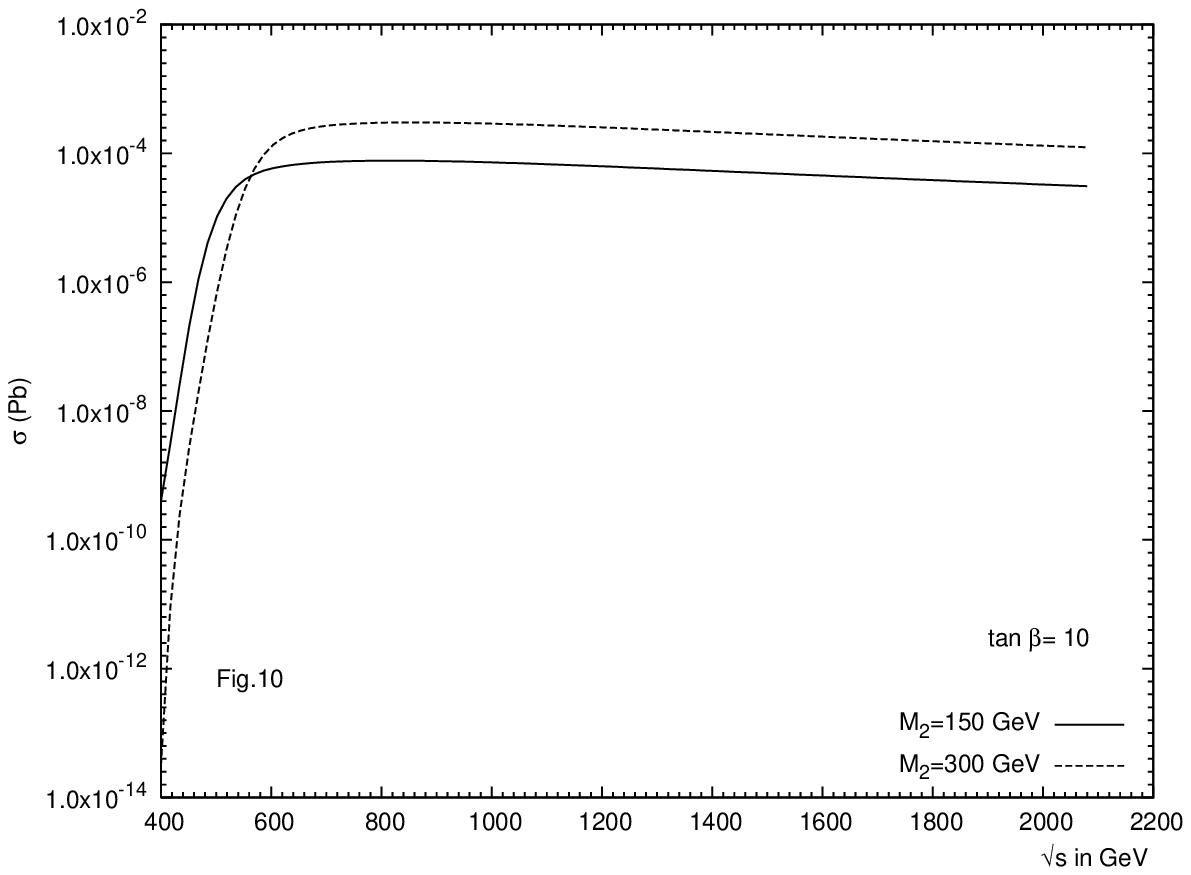}}
\vspace{-0.1cm}
\centerline{\epsfxsize=5.5truein\epsfbox{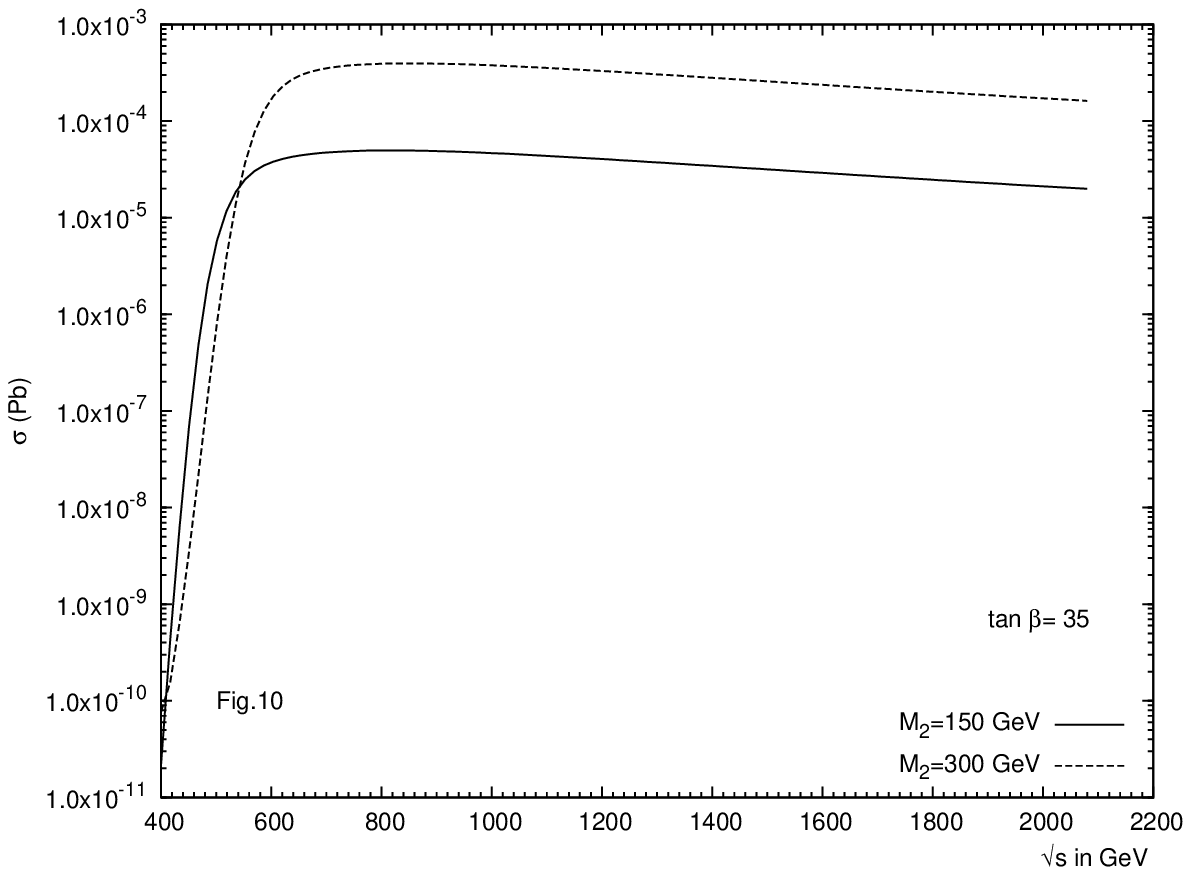}}
\vspace{0.5cm} 
\caption{\small Cross sections for
diagram no. 10 in figure \ref{feyn4}} 
\label{hxx10}
\end{figure}

\begin{figure}[th]
\vspace{-4.5cm}
\centerline{\epsfxsize=5.5truein\epsfbox{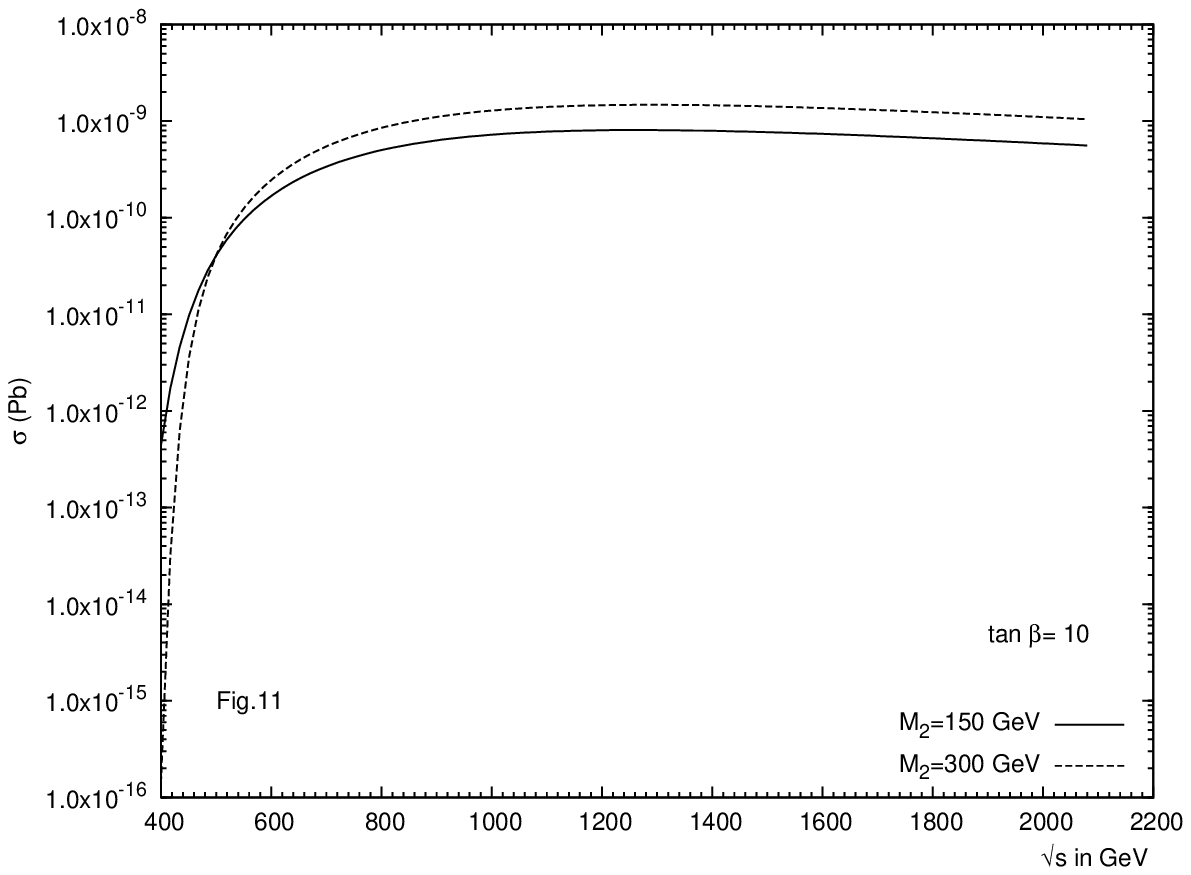}}
\vspace{-0.1cm}
\centerline{\epsfxsize=5.5truein\epsfbox{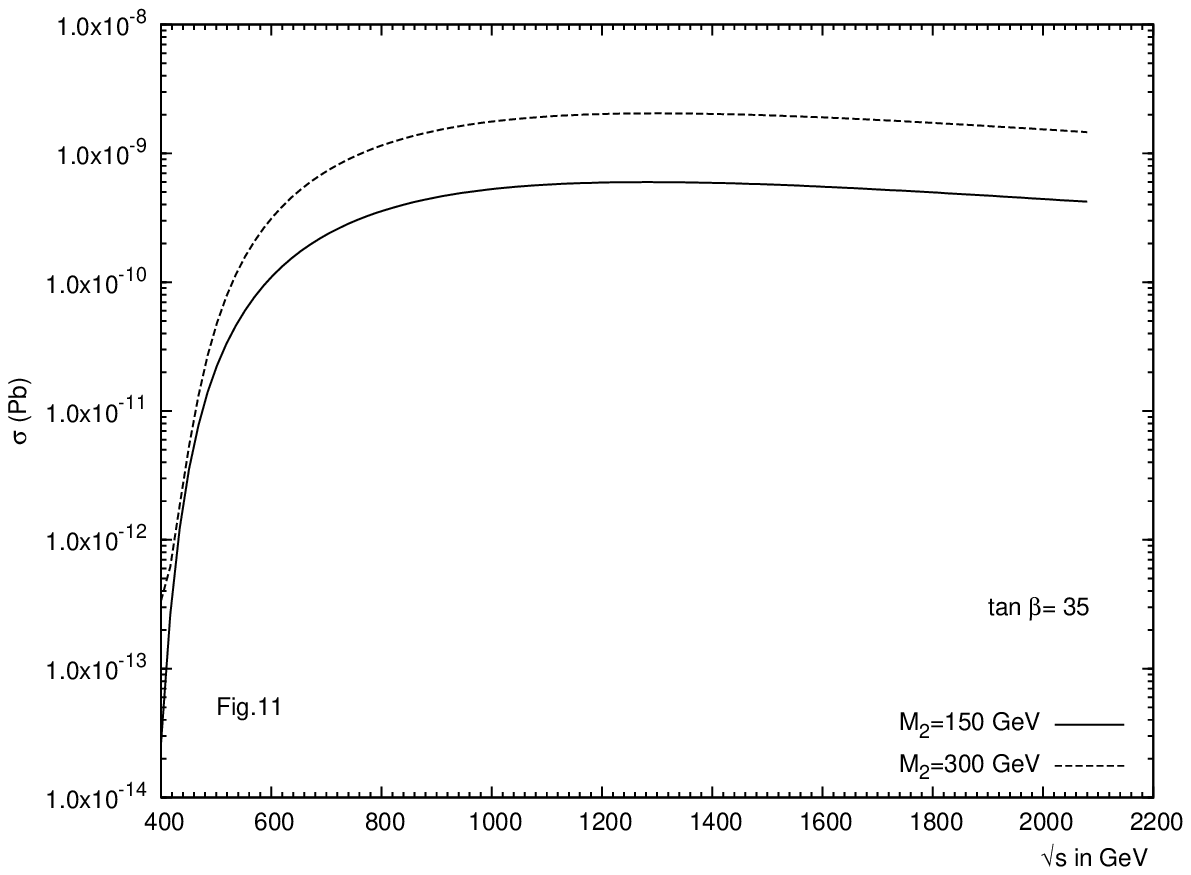}}
\vspace{0.5cm} 
\caption{\small Cross sections for
diagram no. 11 in figure \ref{feyn4}} 
\label{hxx11}
\end{figure}

\begin{figure}[th]
\vspace{-4.5cm}
\centerline{\epsfxsize=5.5truein\epsfbox{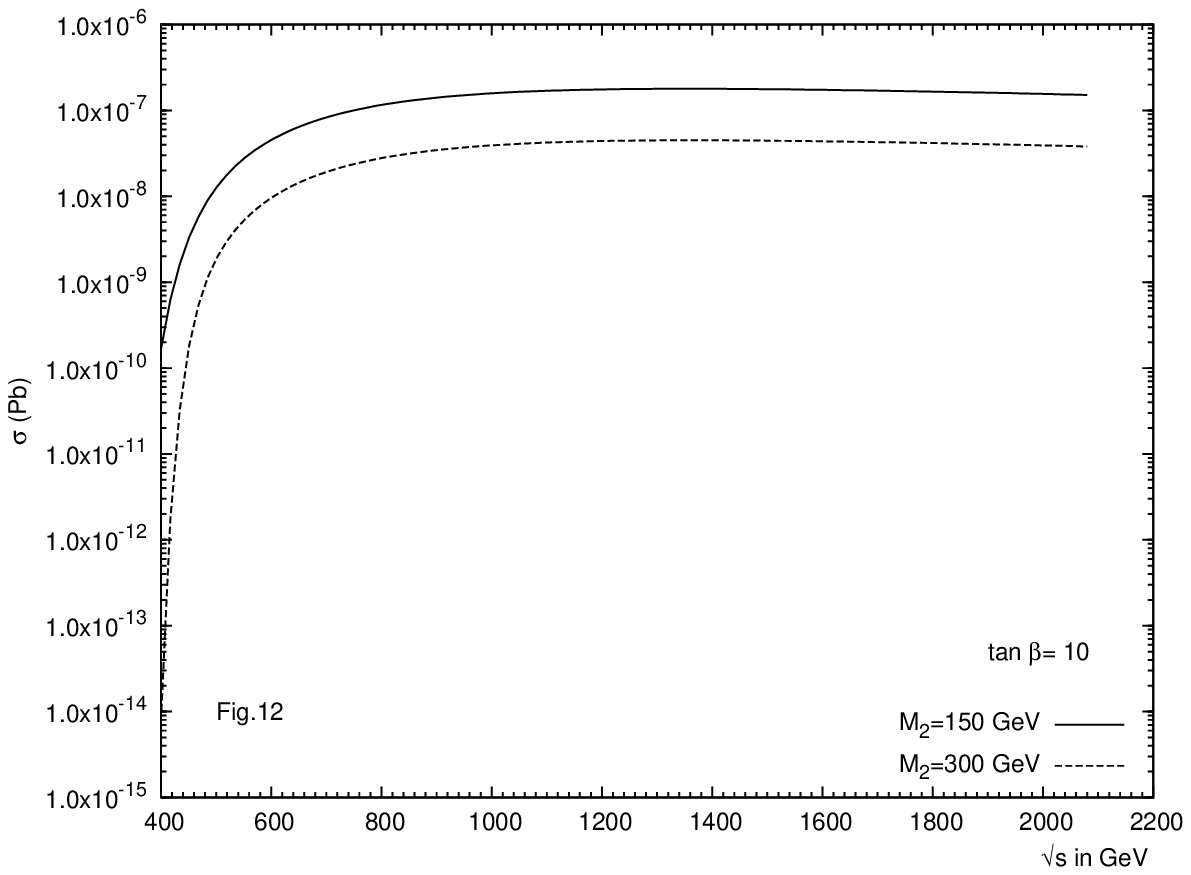}}
\vspace{-0.1cm}
\centerline{\epsfxsize=5.5truein\epsfbox{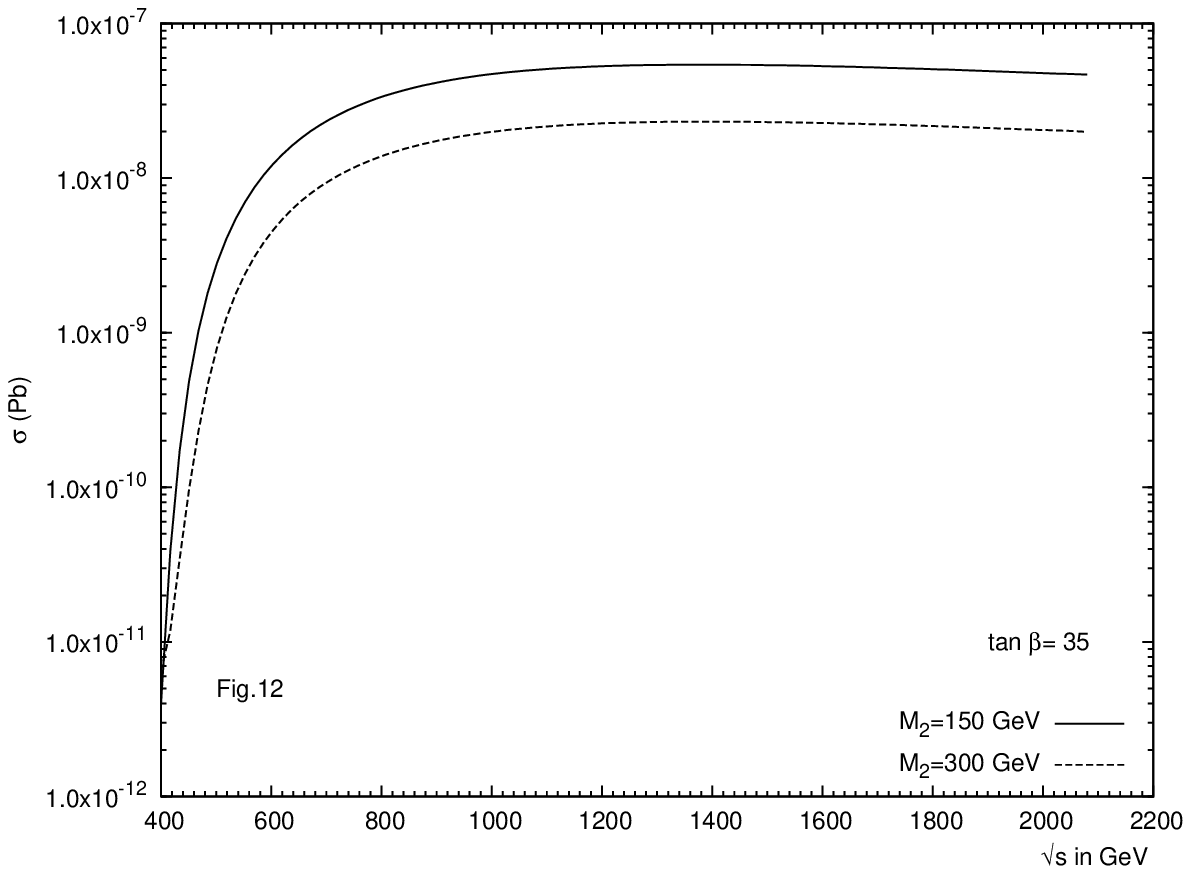}}
\vspace{0.5cm} 
\caption{\small Cross sections for
diagram no. 12 in figure \ref{feyn4}} 
\label{hxx12}
\end{figure}

\begin{figure}[th]
\vspace{-4.5cm}
\centerline{\epsfxsize=5.5truein\epsfbox{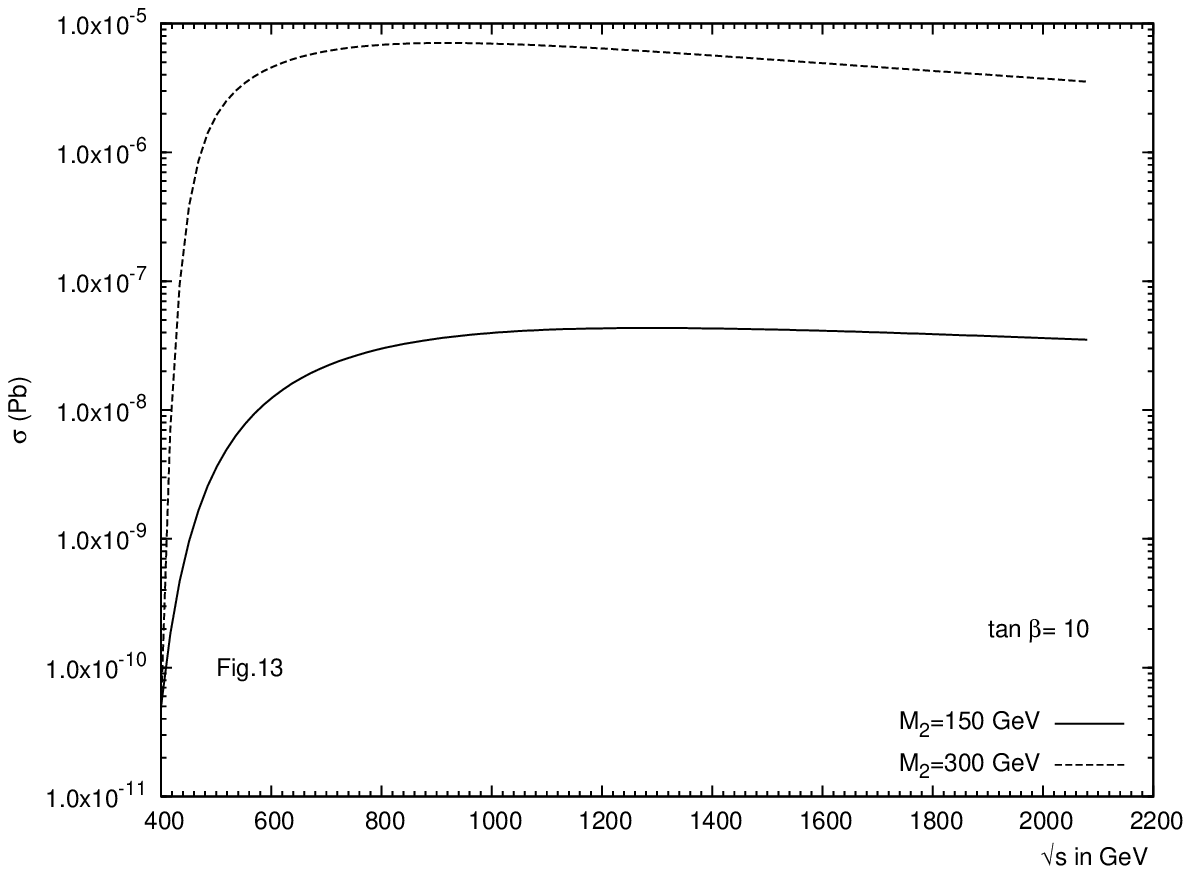}}
\vspace{-0.1cm}
\centerline{\epsfxsize=5.5truein\epsfbox{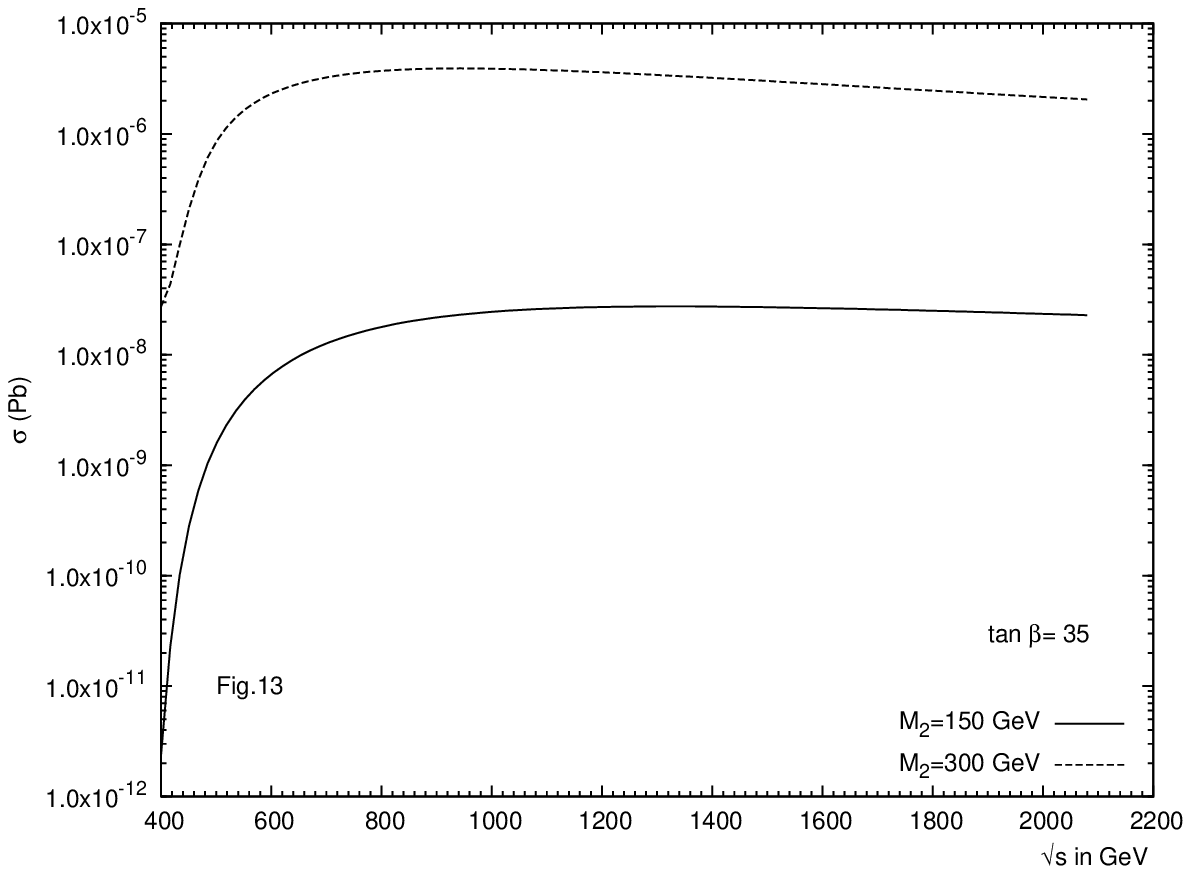}}
\vspace{0.5cm} 
\caption{\small Cross sections for
diagram no. 13 in figure \ref{feyn4}} 
\label{hxx13}
\end{figure}

\begin{figure}[th]
\vspace{-4.5cm}
\centerline{\epsfxsize=5.5truein\epsfbox{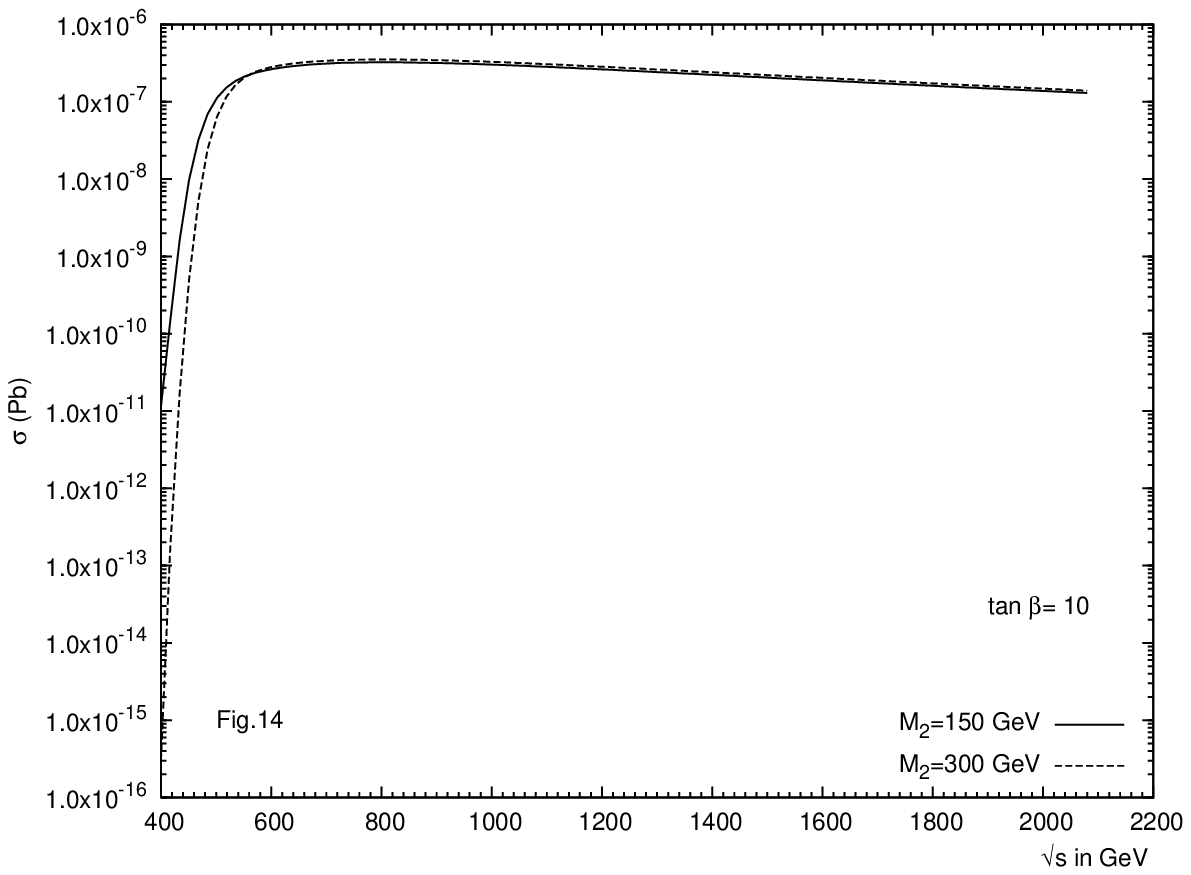}}
\vspace{-0.1cm}
\centerline{\epsfxsize=5.5truein\epsfbox{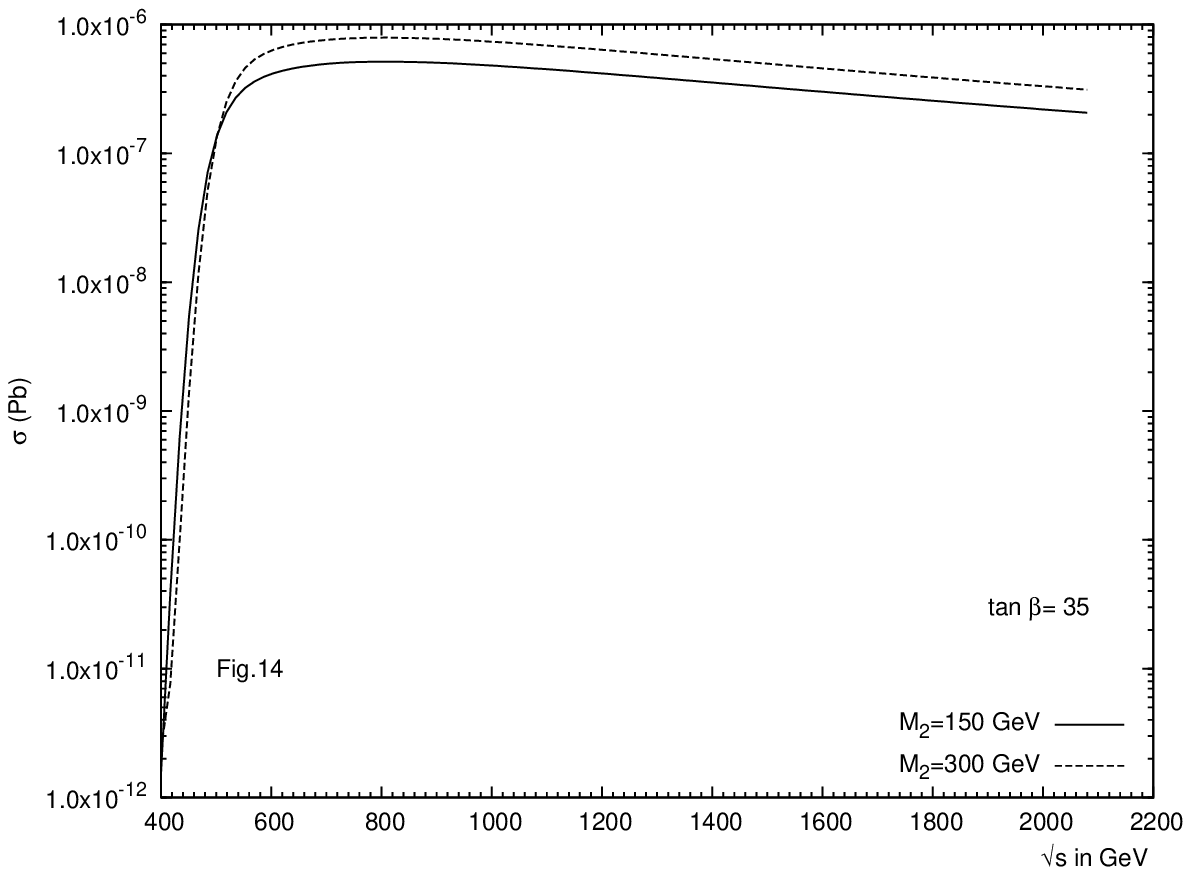}}
\vspace{0.5cm} 
\caption{\small Cross sections for
diagram no. 14 in figure \ref{feyn4}} 
\label{hxx14}
\end{figure}

\begin{figure}[th]
\vspace{-4.5cm}
\centerline{\epsfxsize=5.5truein\epsfbox{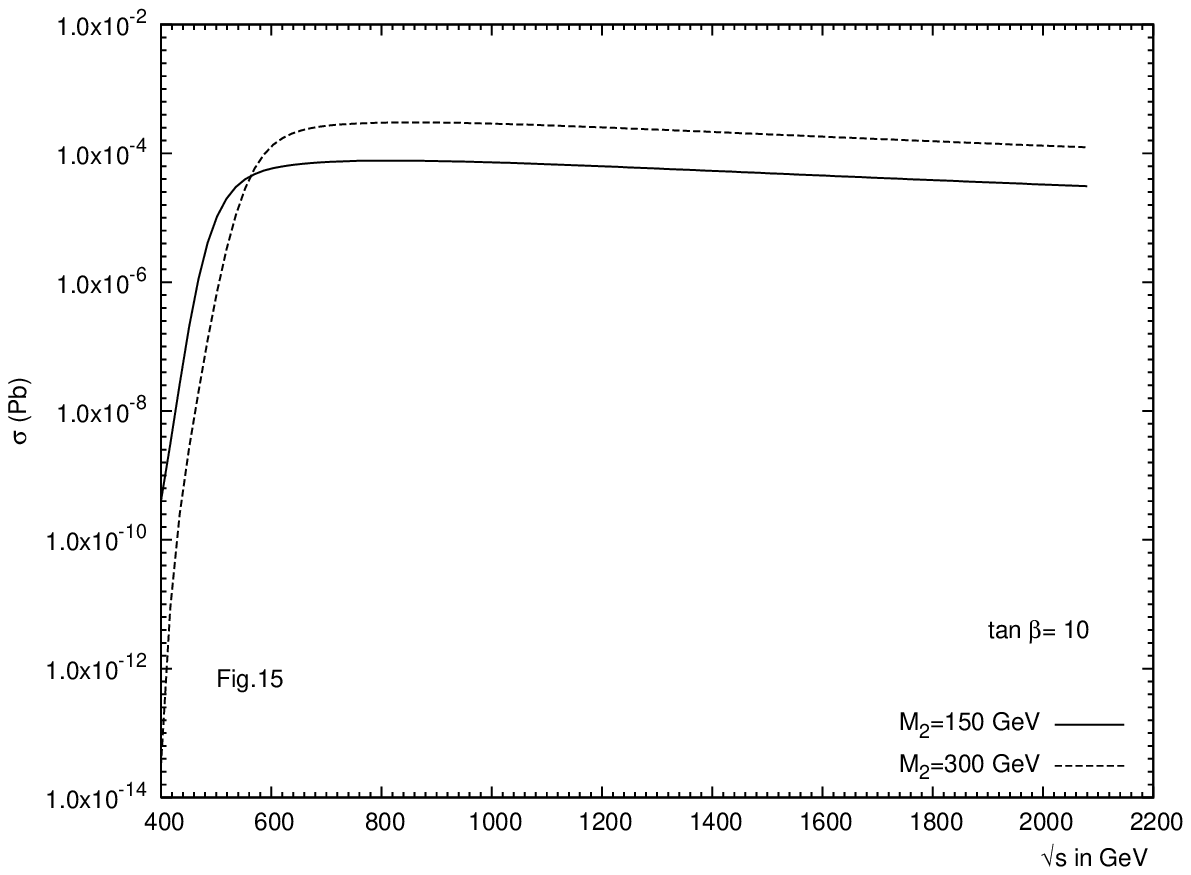}}
\vspace{-0.1cm}
\centerline{\epsfxsize=5.5truein\epsfbox{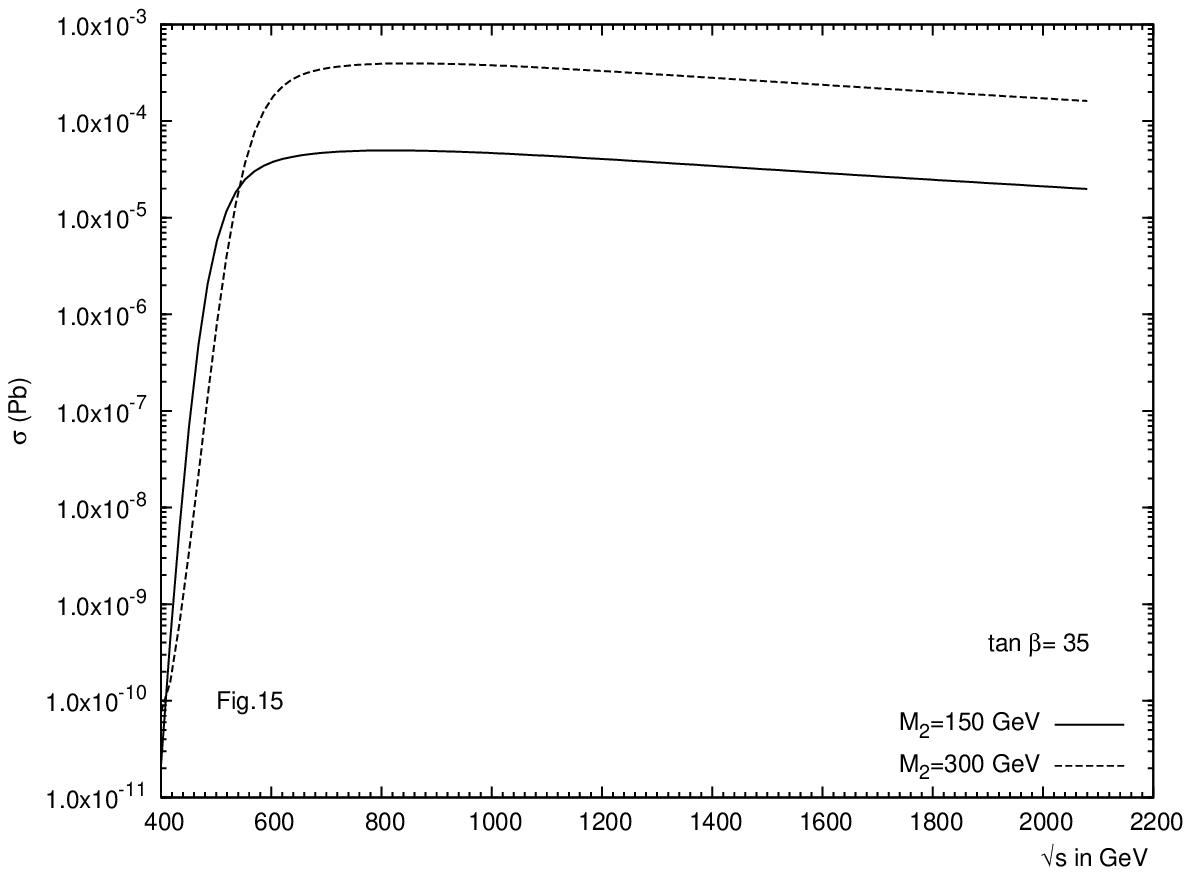}}
\vspace{0.5cm} 
\caption{\small Cross sections for
diagram no. 15 in figure \ref{feyn5}} 
\label{hxx15}
\end{figure}

\begin{figure}[th]
\vspace{-4.5cm}
\centerline{\epsfxsize=5.5truein\epsfbox{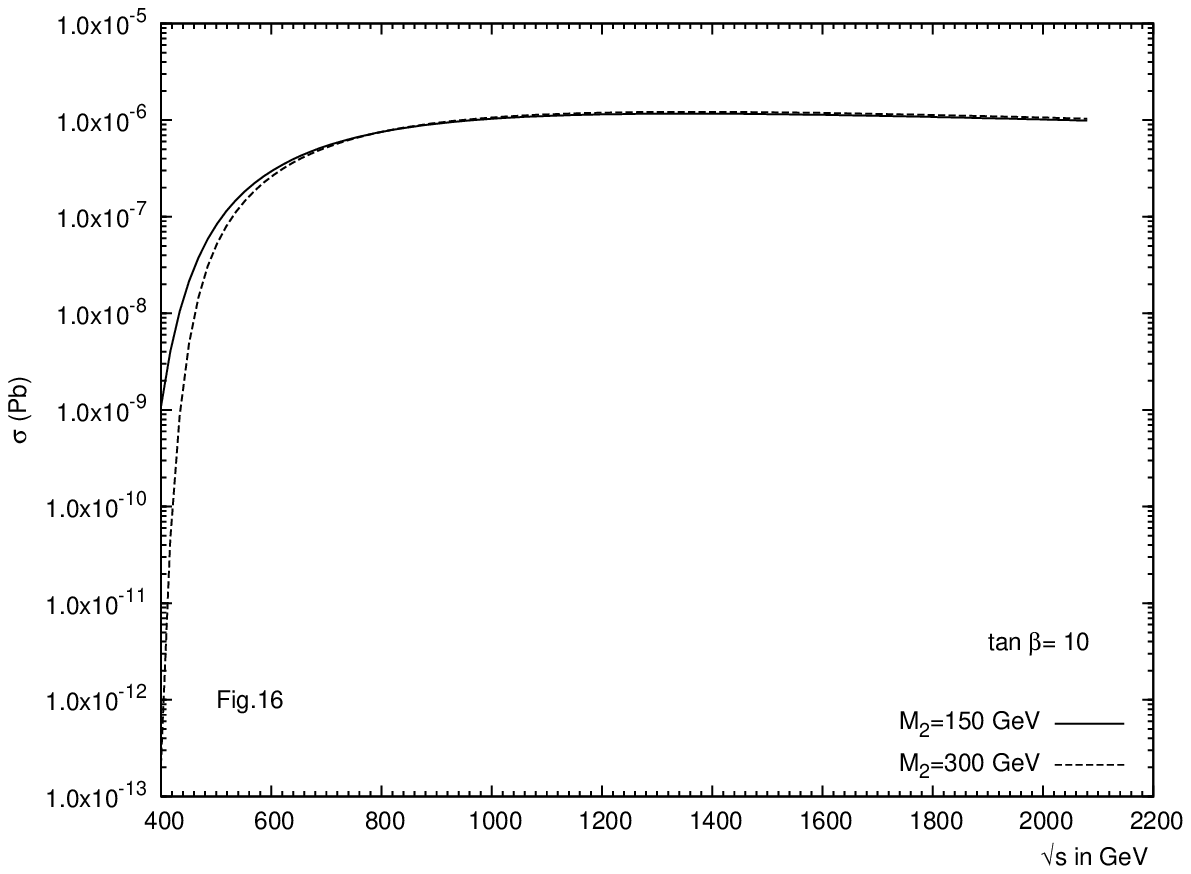}}
\vspace{-0.1cm}
\centerline{\epsfxsize=5.5truein\epsfbox{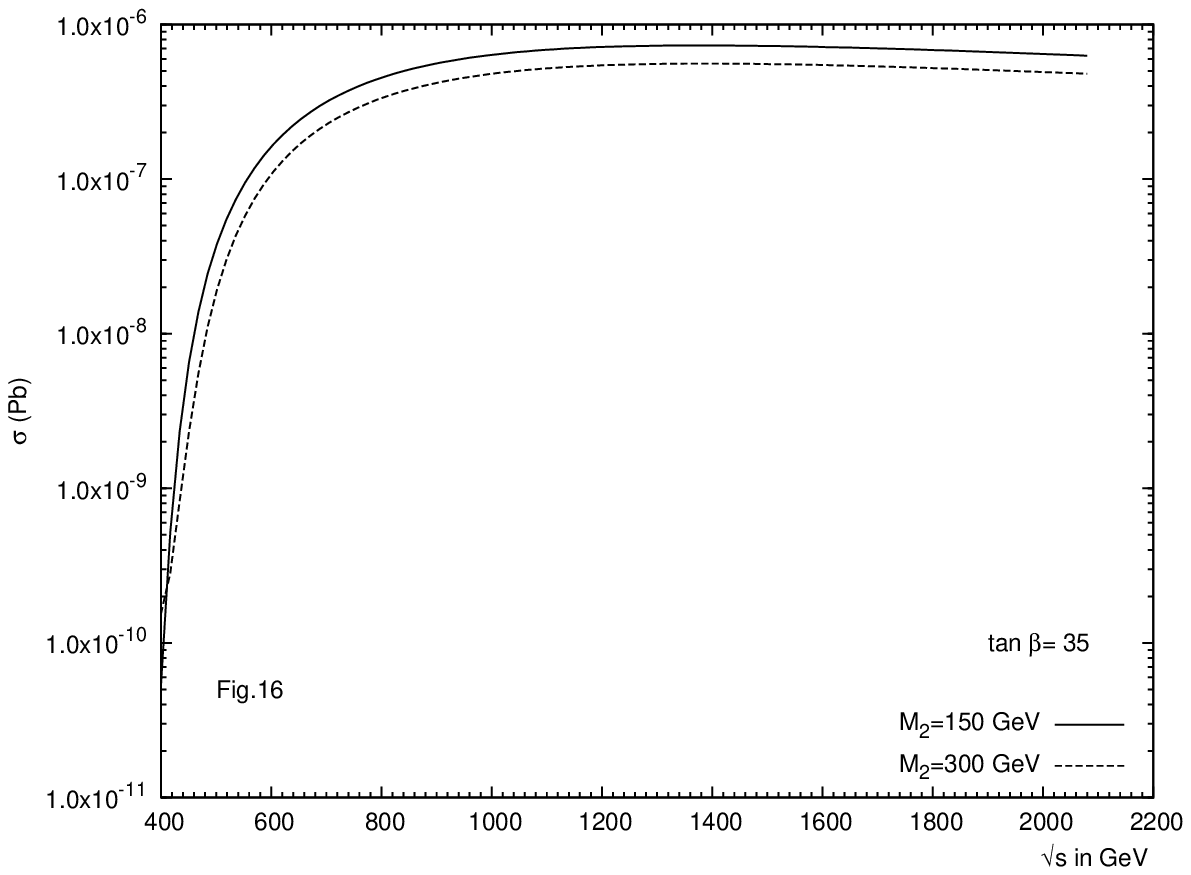}}
\vspace{0.5cm} 
\caption{\small Cross sections for
diagram no. 16 in figure \ref{feyn5}} 
\label{hxx16}
\end{figure}

\begin{figure}[th]
\vspace{-4.5cm}
\centerline{\epsfxsize=5.5truein\epsfbox{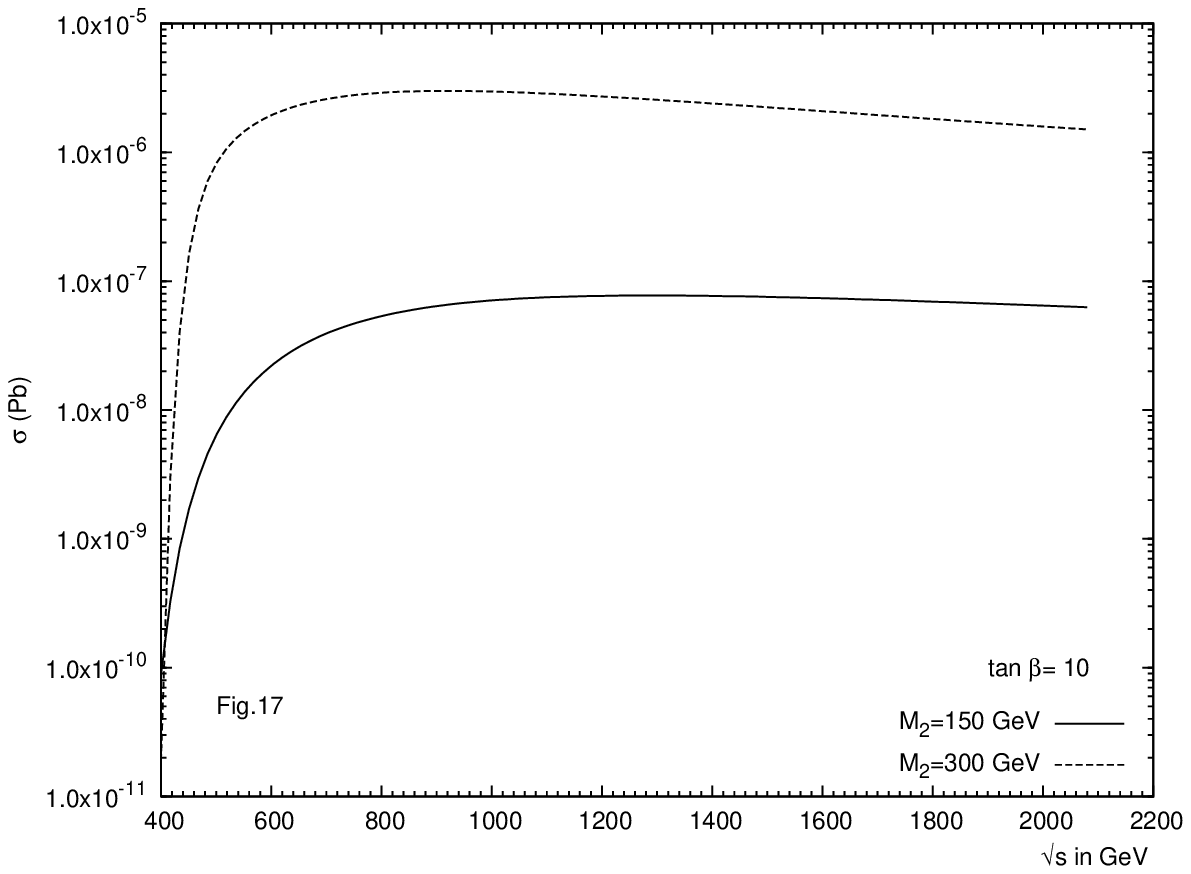}}
\vspace{-0.1cm}
\centerline{\epsfxsize=5.5truein\epsfbox{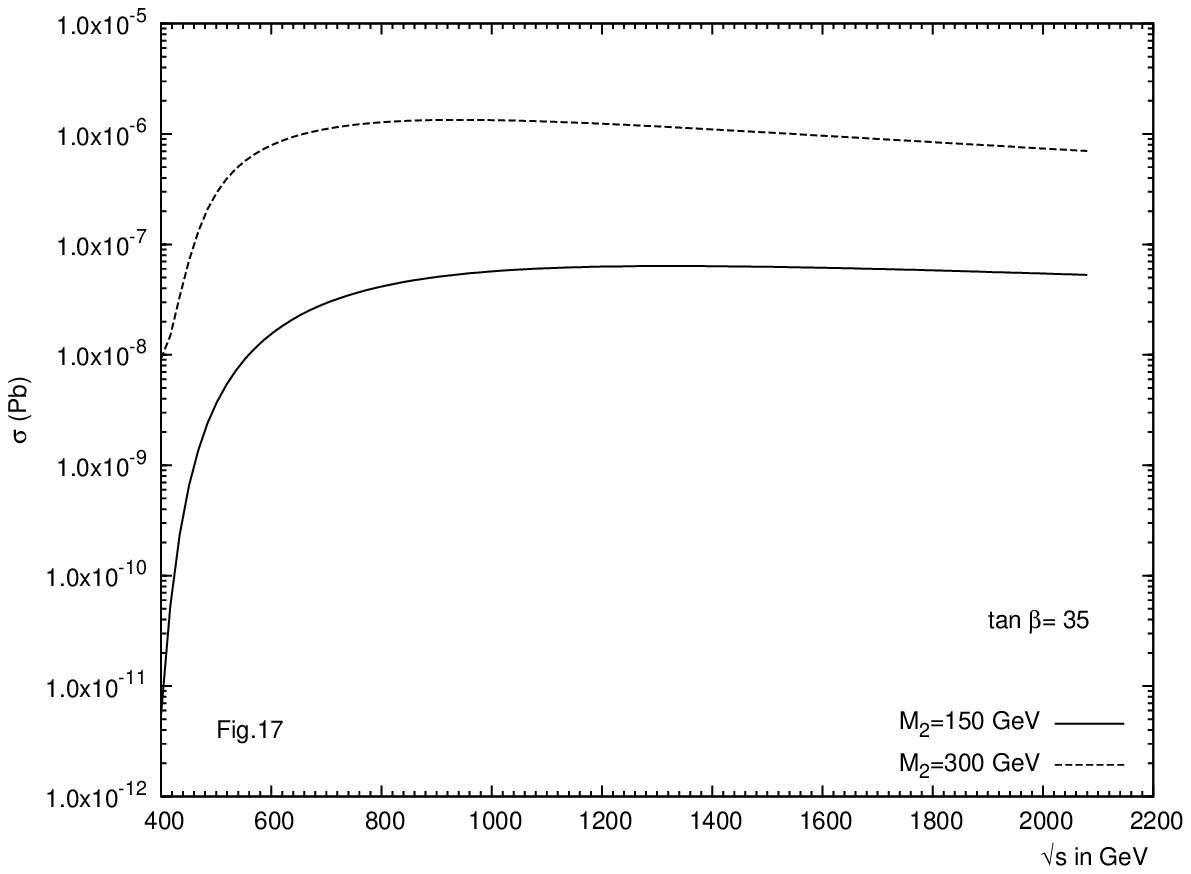}}
\vspace{0.5cm} 
\caption{\small Cross sections for
diagram no. 17 in figure \ref{feyn5}} 
\label{hxx17}
\end{figure}

\begin{figure}[th]
\vspace{-4.5cm}
\centerline{\epsfxsize=5.5truein\epsfbox{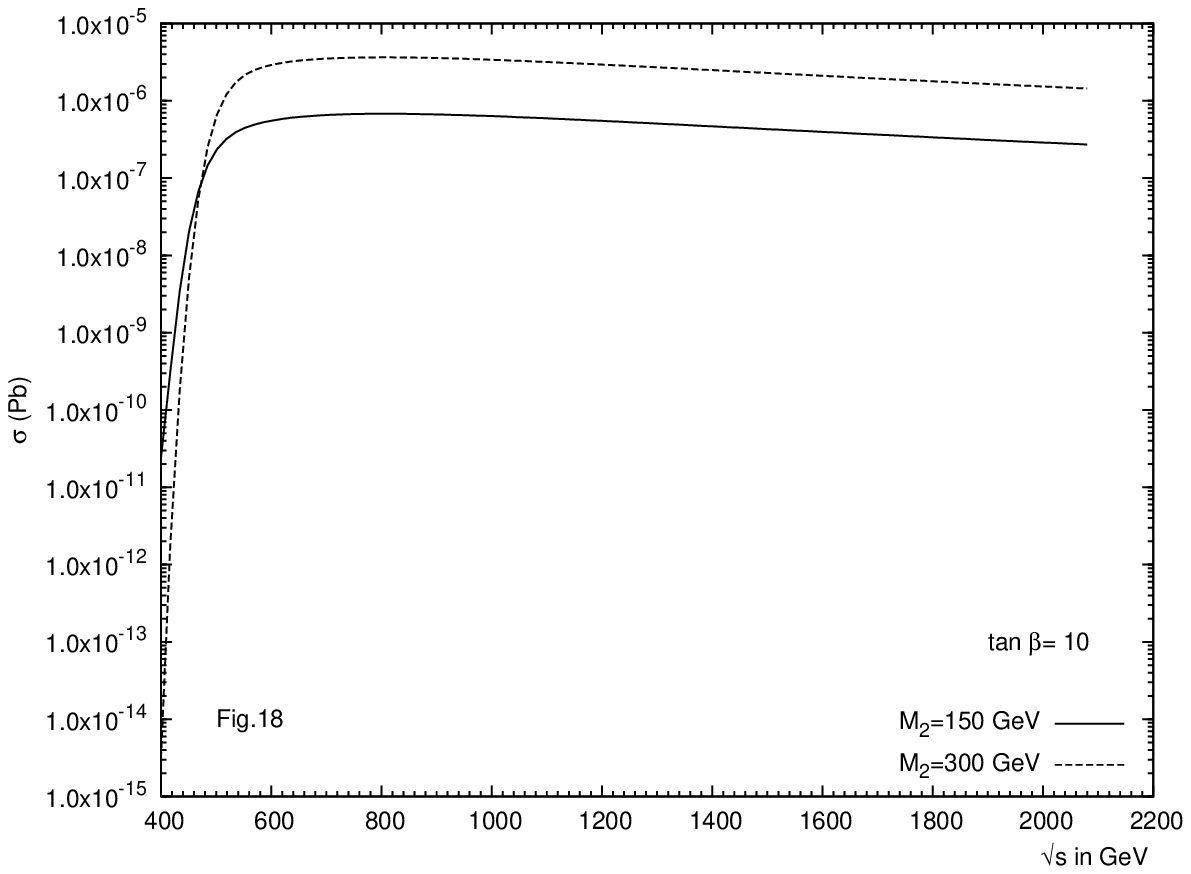}}
\vspace{-0.1cm}
\centerline{\epsfxsize=5.5truein\epsfbox{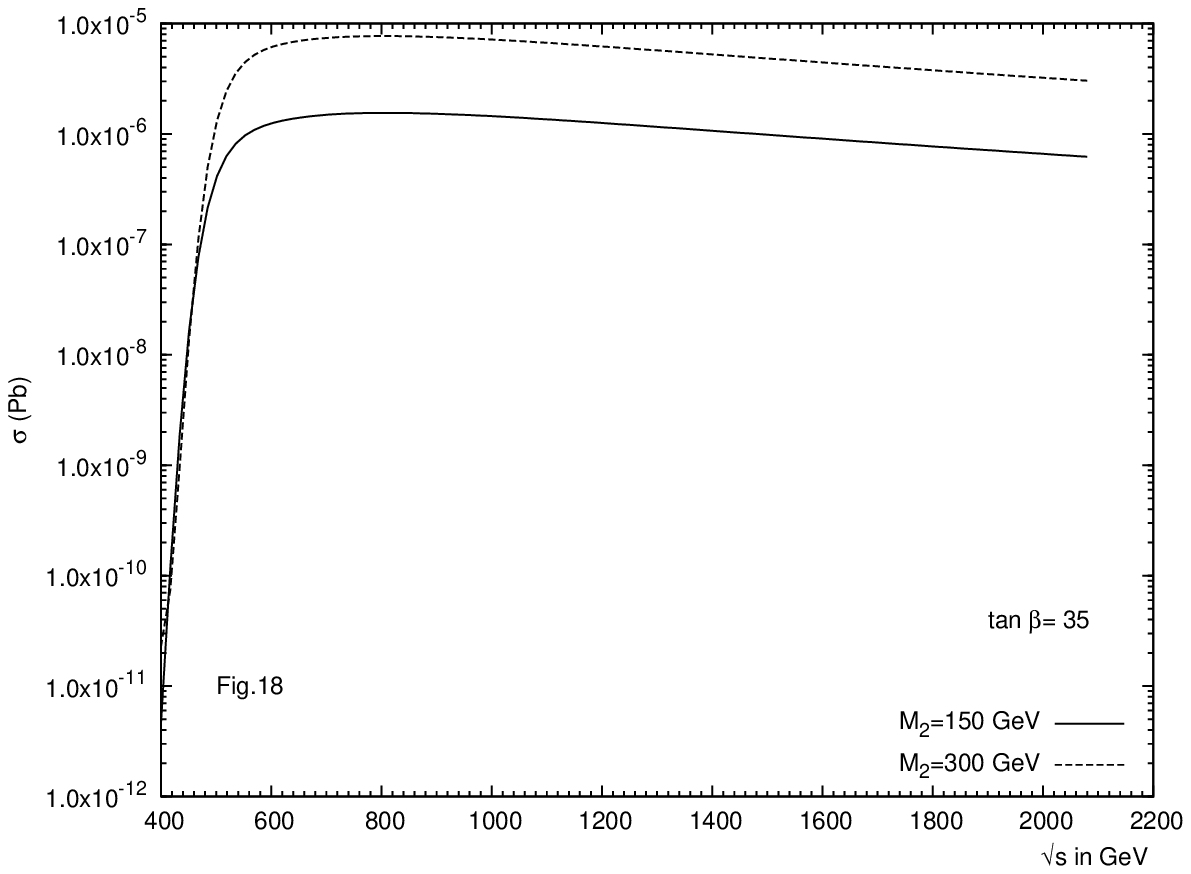}}
\vspace{0.5cm} 
\caption{\small Cross sections for
diagram no. 18 in figure \ref{feyn5}} 
\label{hxx18}
\end{figure}

\clearpage

\begin{figure}[th]
\vspace{-4.5cm}
\centerline{\epsfxsize=5.5truein\epsfbox{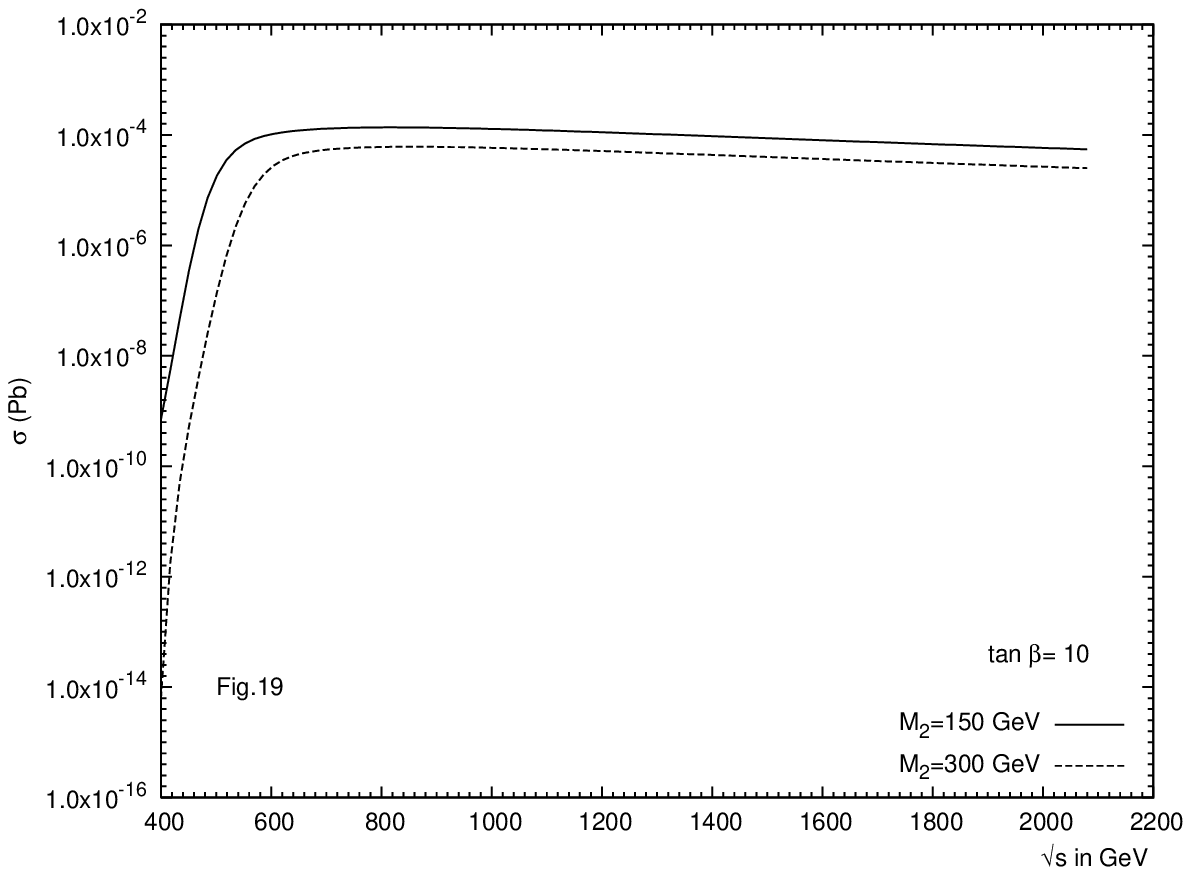}}
\vspace{-0.1cm}
\centerline{\epsfxsize=5.5truein\epsfbox{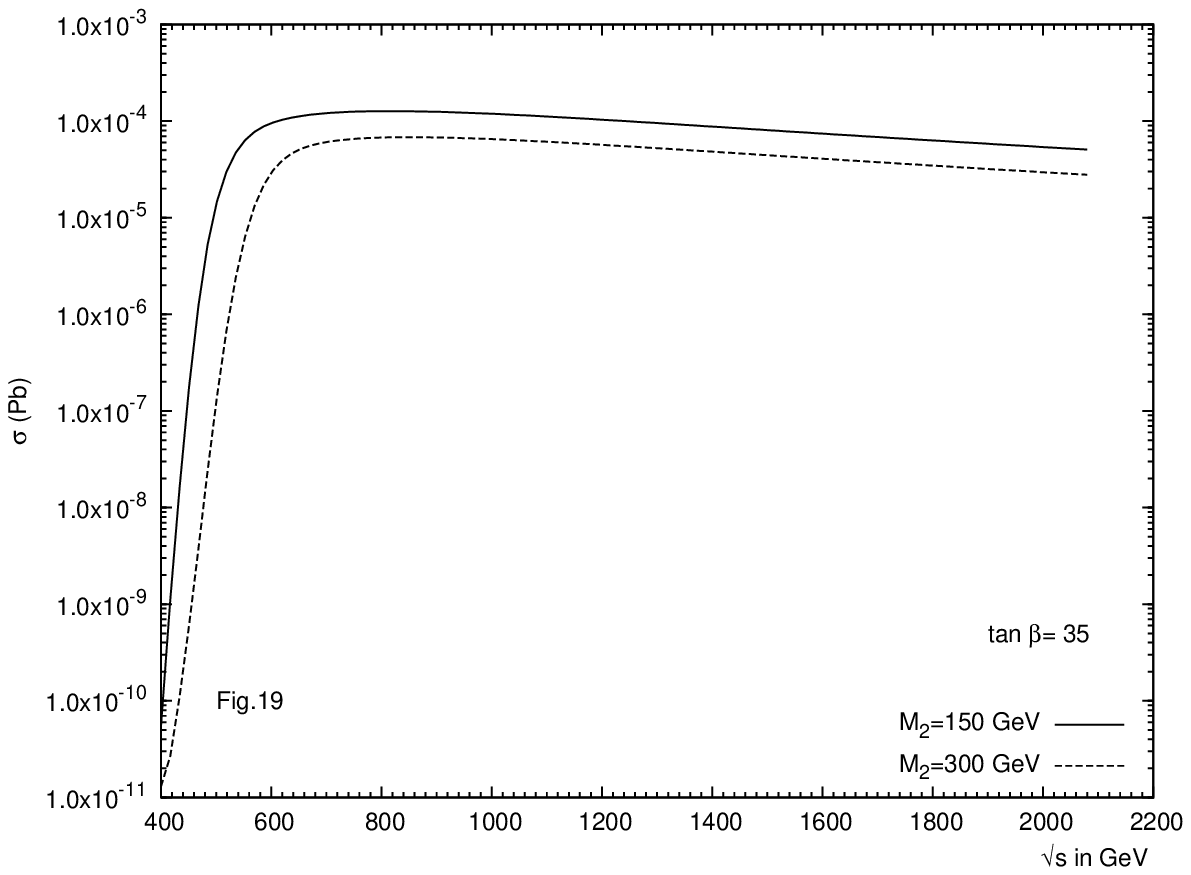}}
\vspace{0.5cm} 
\caption{\small Cross sections for
diagram no. 19 in figure \ref{feyn5}} 
\label{hxx19}
\end{figure}

\begin{figure}[th]
\vspace{-4.5cm}
\centerline{\epsfxsize=5.5truein\epsfbox{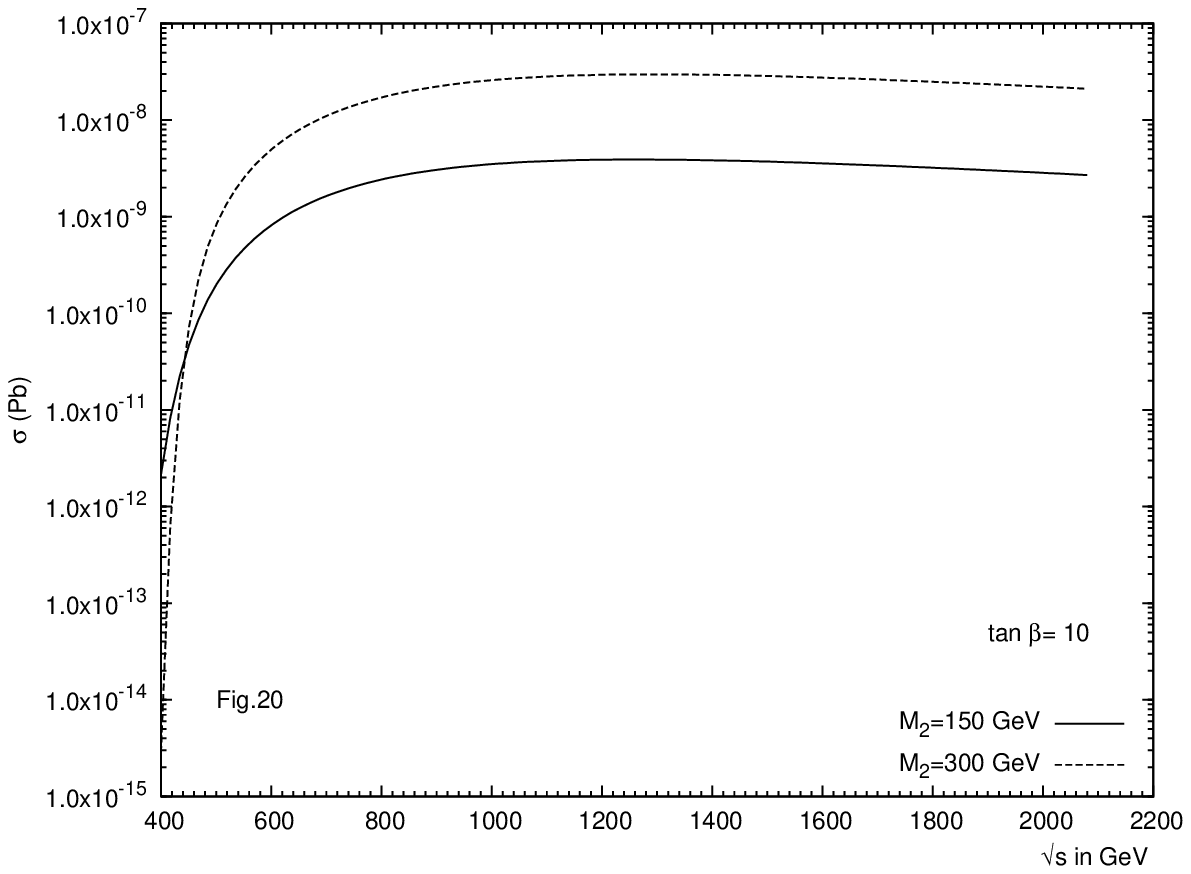}}
\vspace{-0.1cm}
\centerline{\epsfxsize=5.5truein\epsfbox{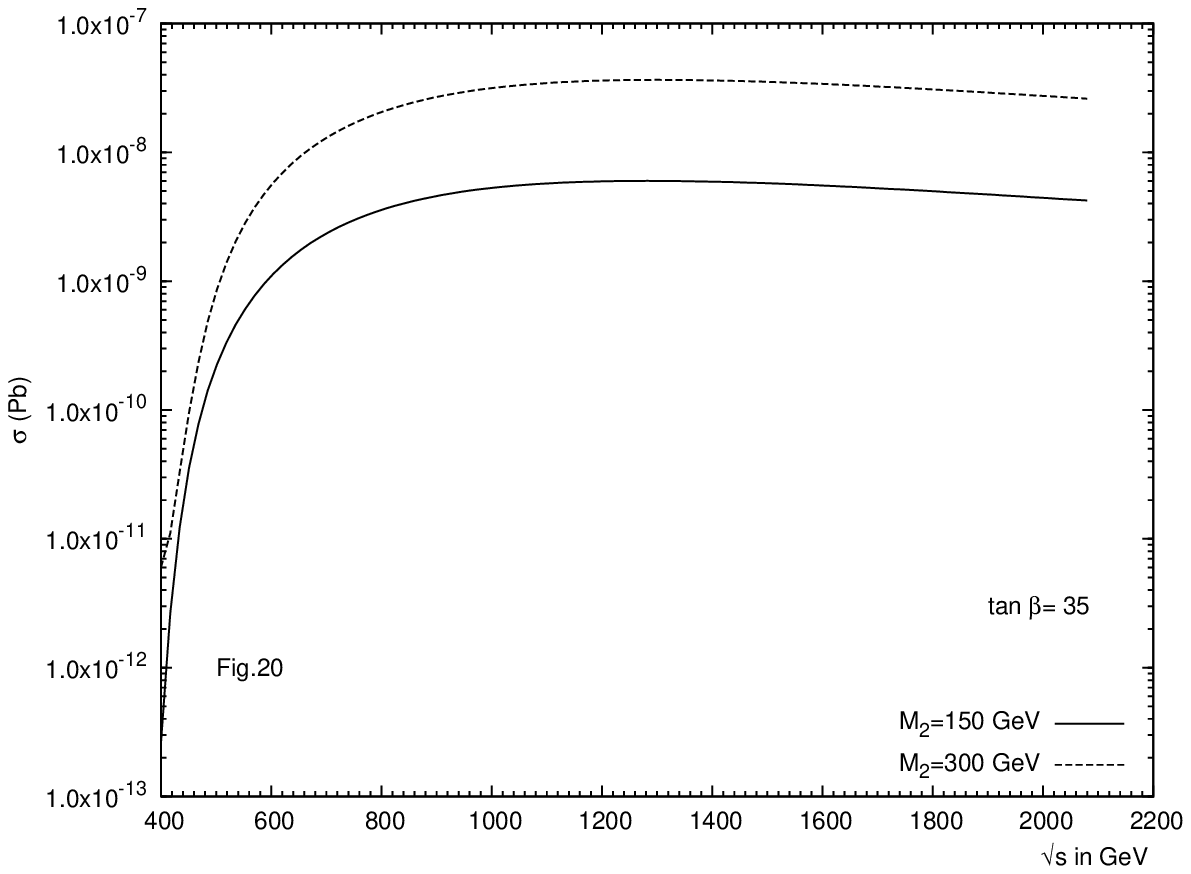}}
\vspace{0.5cm} 
\caption{\small Cross sections for
diagram no. 20 in figure \ref{feyn5}} 
\label{hxx20}
\end{figure}

\begin{figure}[th]
\vspace{-4.5cm}
\centerline{\epsfxsize=5.5truein\epsfbox{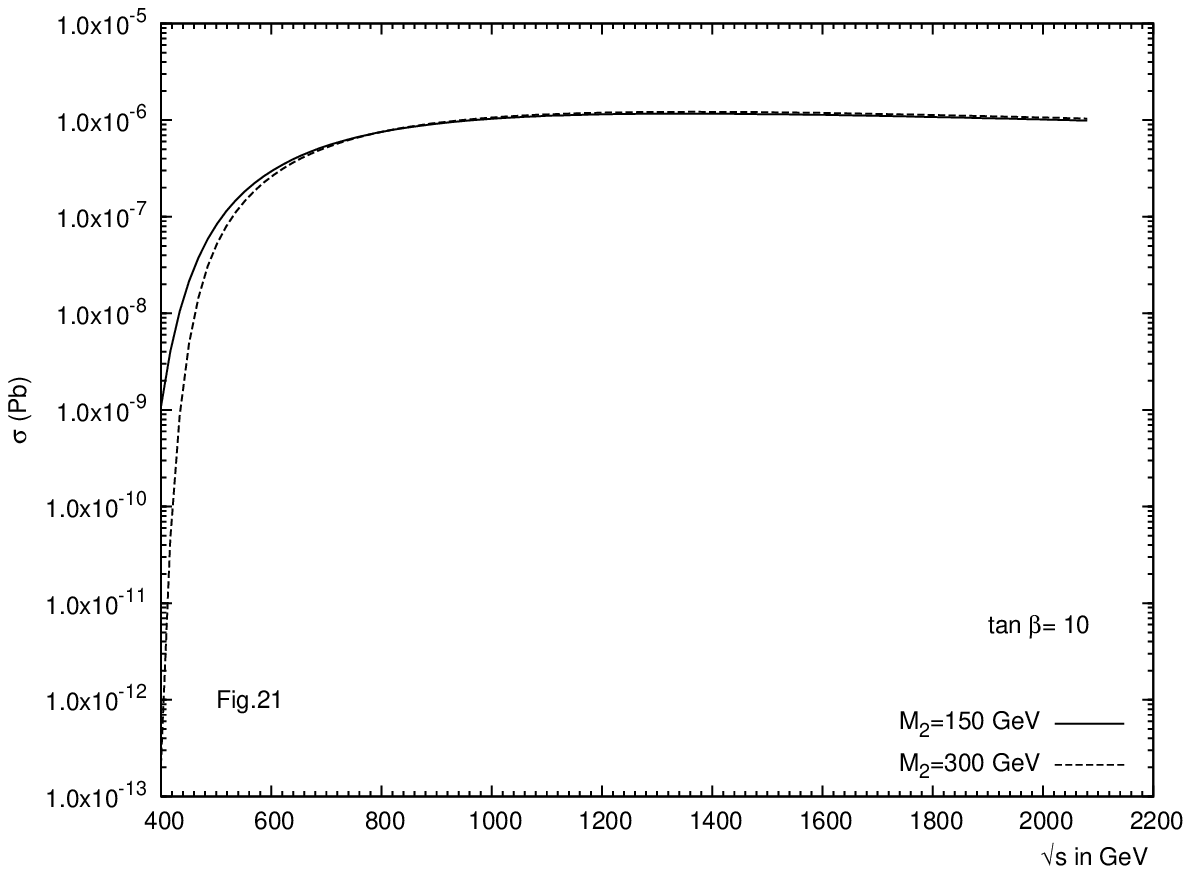}}
\vspace{-0.1cm}
\centerline{\epsfxsize=5.5truein\epsfbox{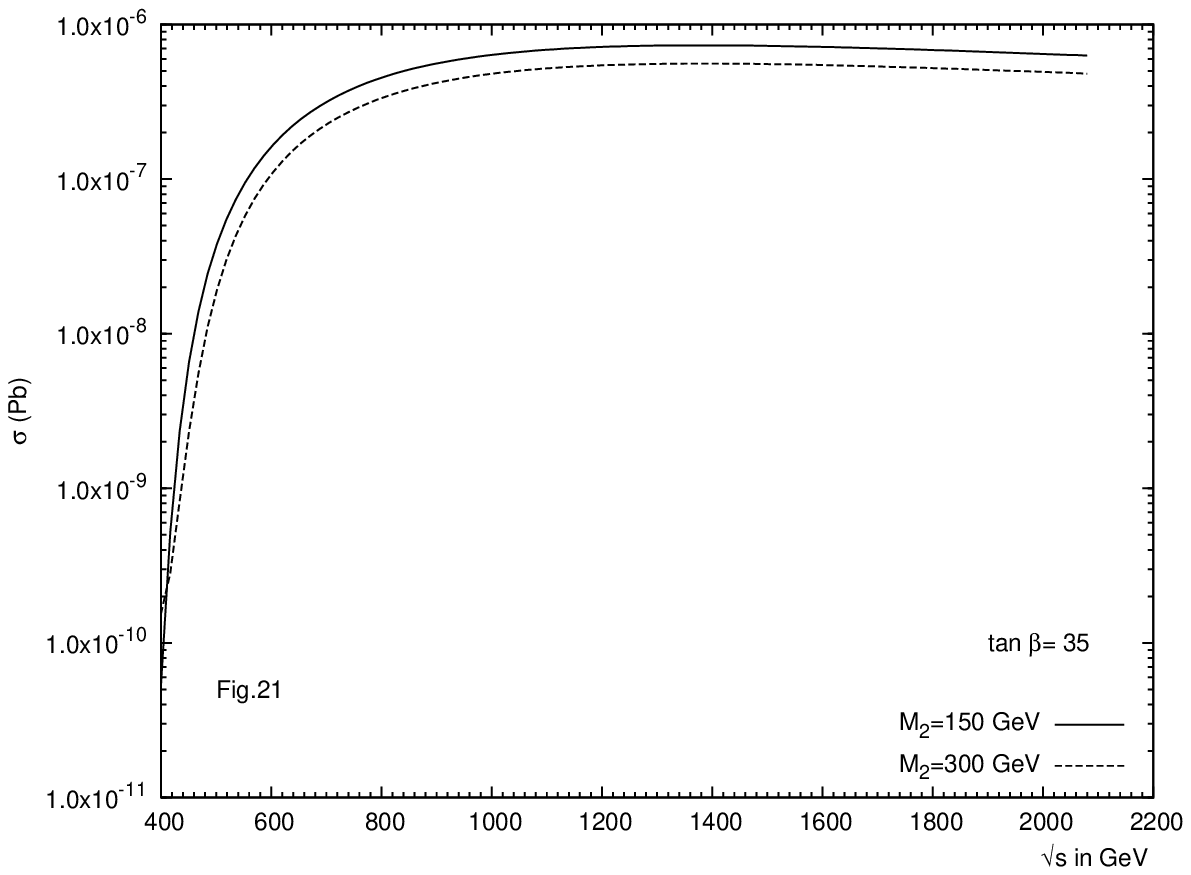}}
\vspace{0.5cm} 
\caption{\small Cross sections for
diagram no. 21 in figure \ref{feyn5}} 
\label{hxx21}
\end{figure}

\begin{figure}[th]
\vspace{-4.5cm}
\centerline{\epsfxsize=5.5truein\epsfbox{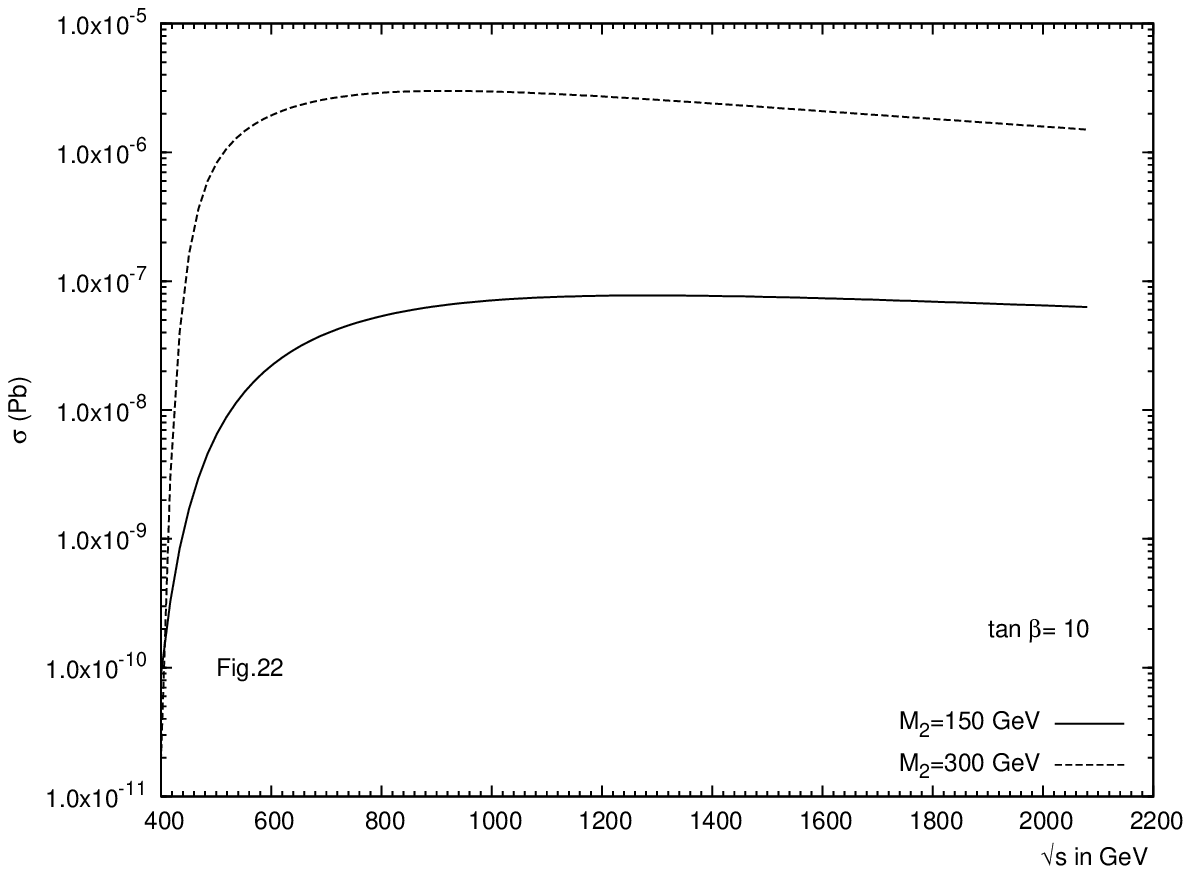}}
\vspace{-0.1cm}
\centerline{\epsfxsize=5.5truein\epsfbox{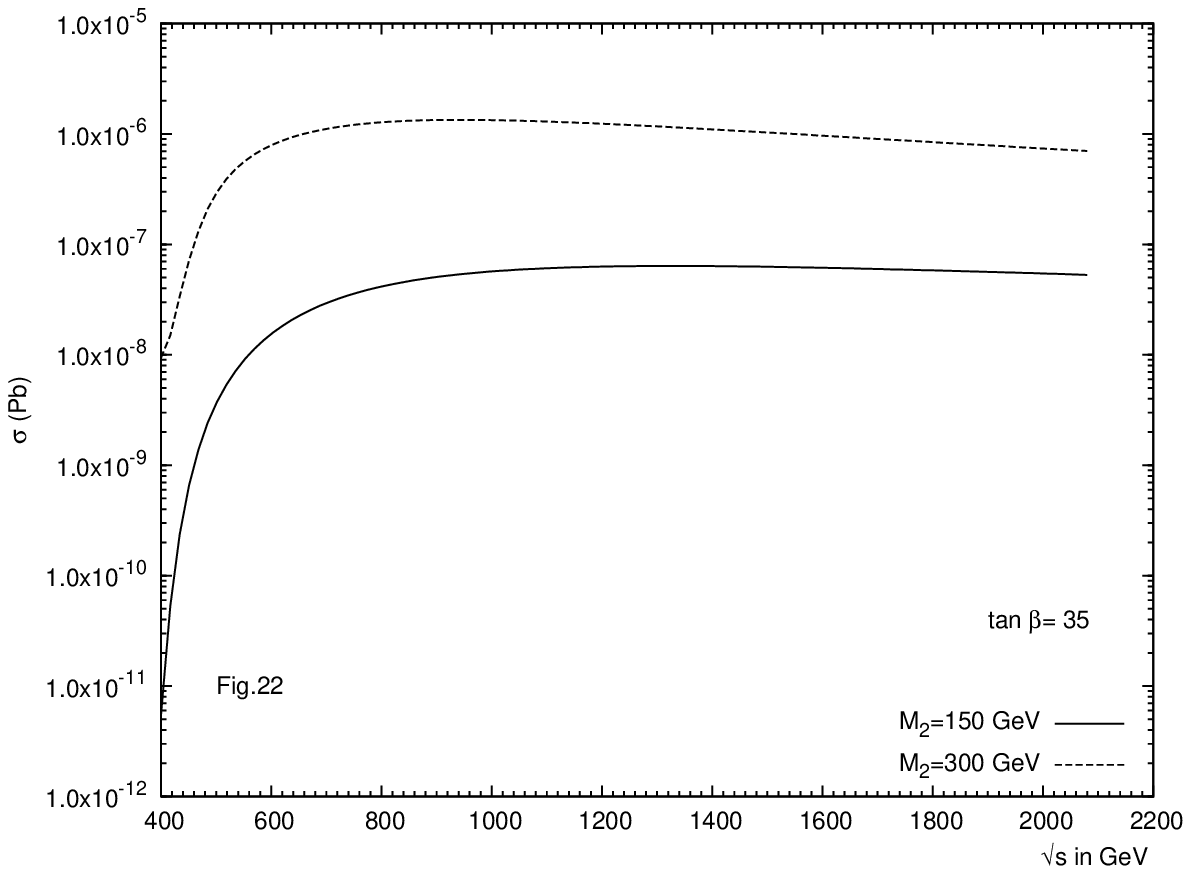}}
\vspace{0.5cm} 
\caption{\small Cross sections for
diagram no. 22 in figure \ref{feyn5}} 
\label{hxx22}
\end{figure}

\begin{figure}[th]
\vspace{-4.5cm}
\centerline{\epsfxsize=5.5truein\epsfbox{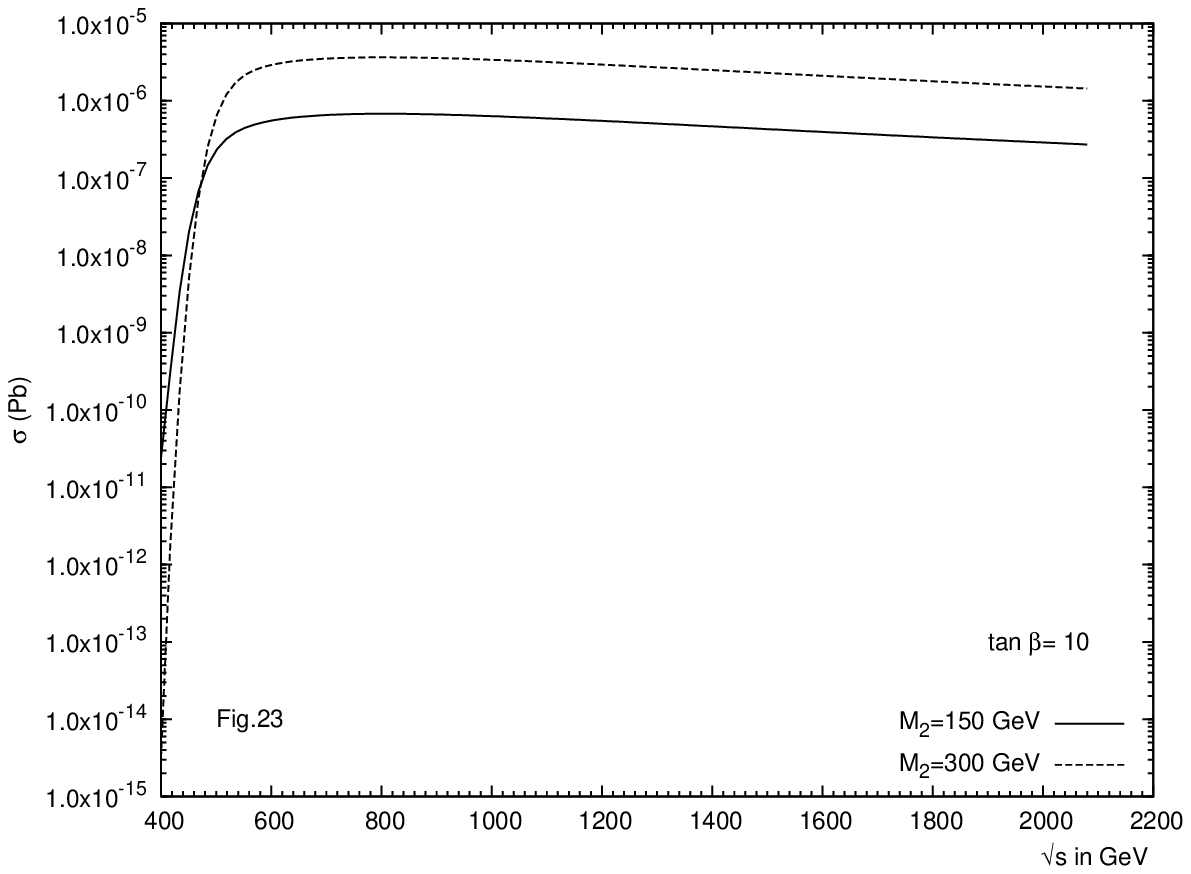}}
\vspace{-0.1cm}
\centerline{\epsfxsize=5.5truein\epsfbox{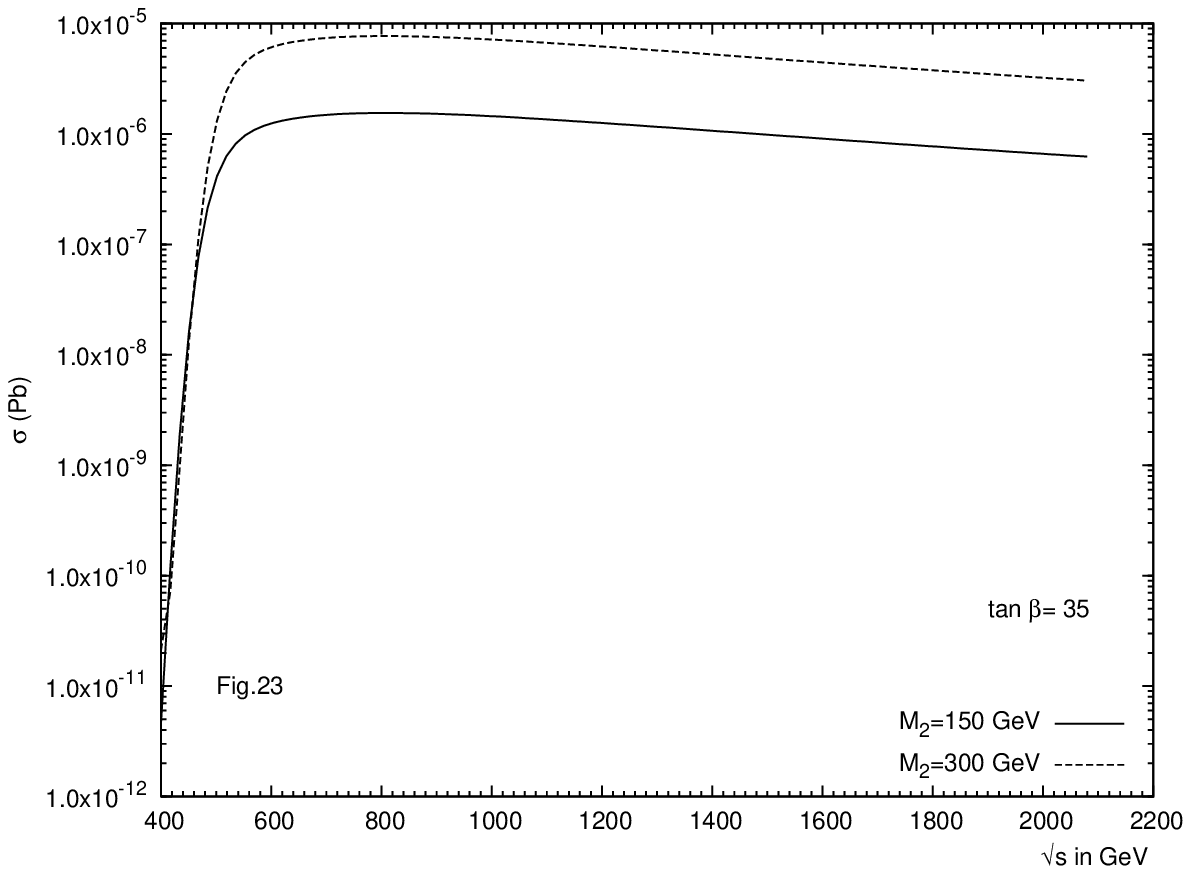}}
\vspace{0.5cm} 
\caption{\small Cross sections for
diagram no. 23 in figure \ref{feyn5}} 
\label{hxx23}
\end{figure}

\begin{figure}[th]
\vspace{-4.5cm}
\centerline{\epsfxsize=5.5truein\epsfbox{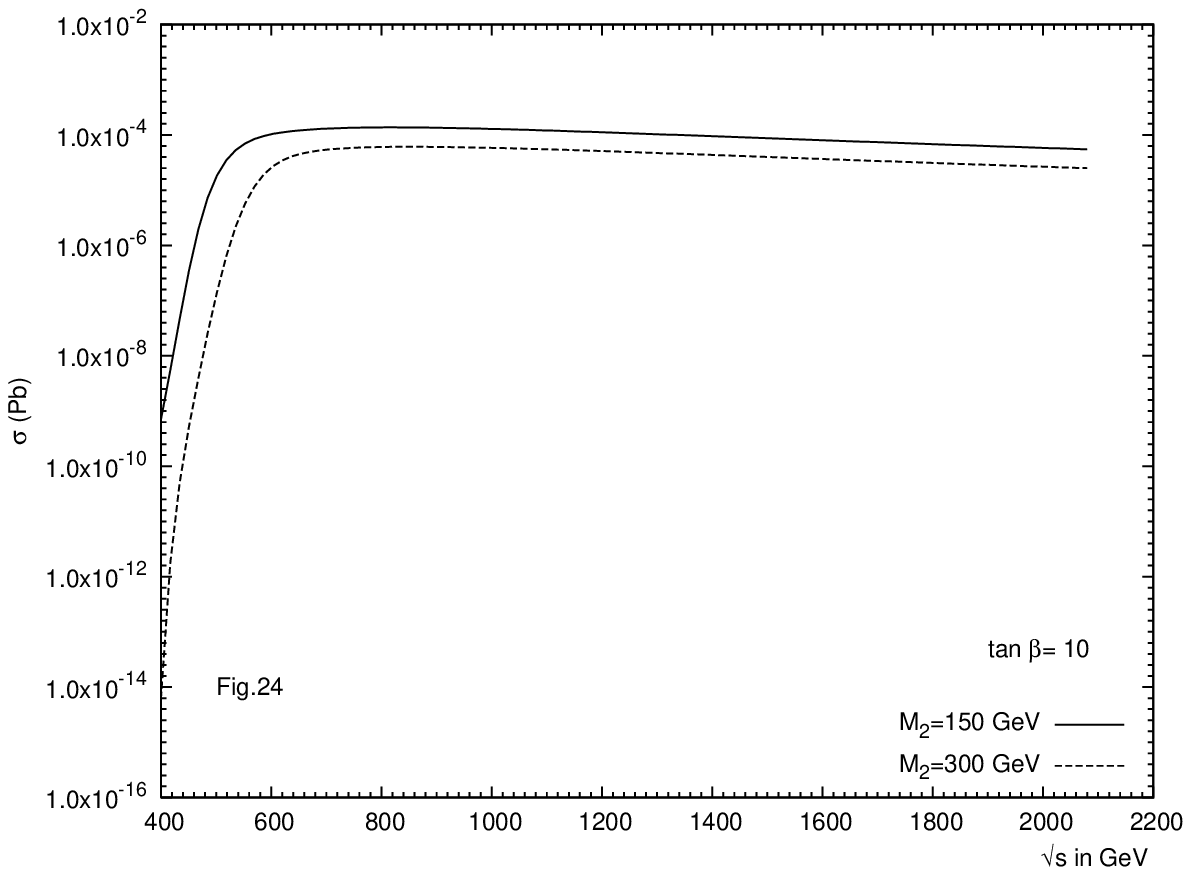}}
\vspace{-0.1cm}
\centerline{\epsfxsize=5.5truein\epsfbox{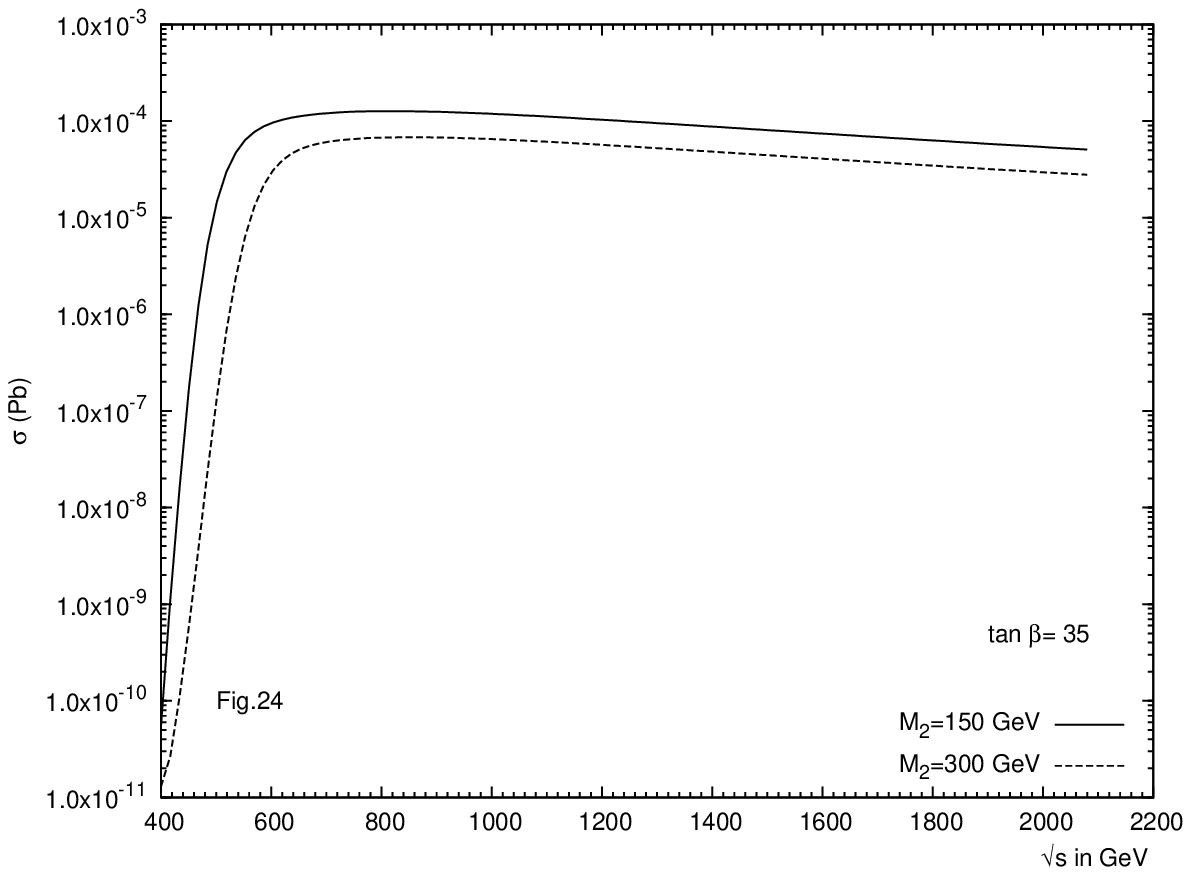}}
\vspace{0.5cm} 
\caption{\small Cross sections for
diagram no. 24 in figure \ref{feyn5}} 
\label{hxx24}
\end{figure}
\begin{figure}[th]
\vspace{-4.5cm}
\centerline{\epsfxsize=5.5truein\epsfbox{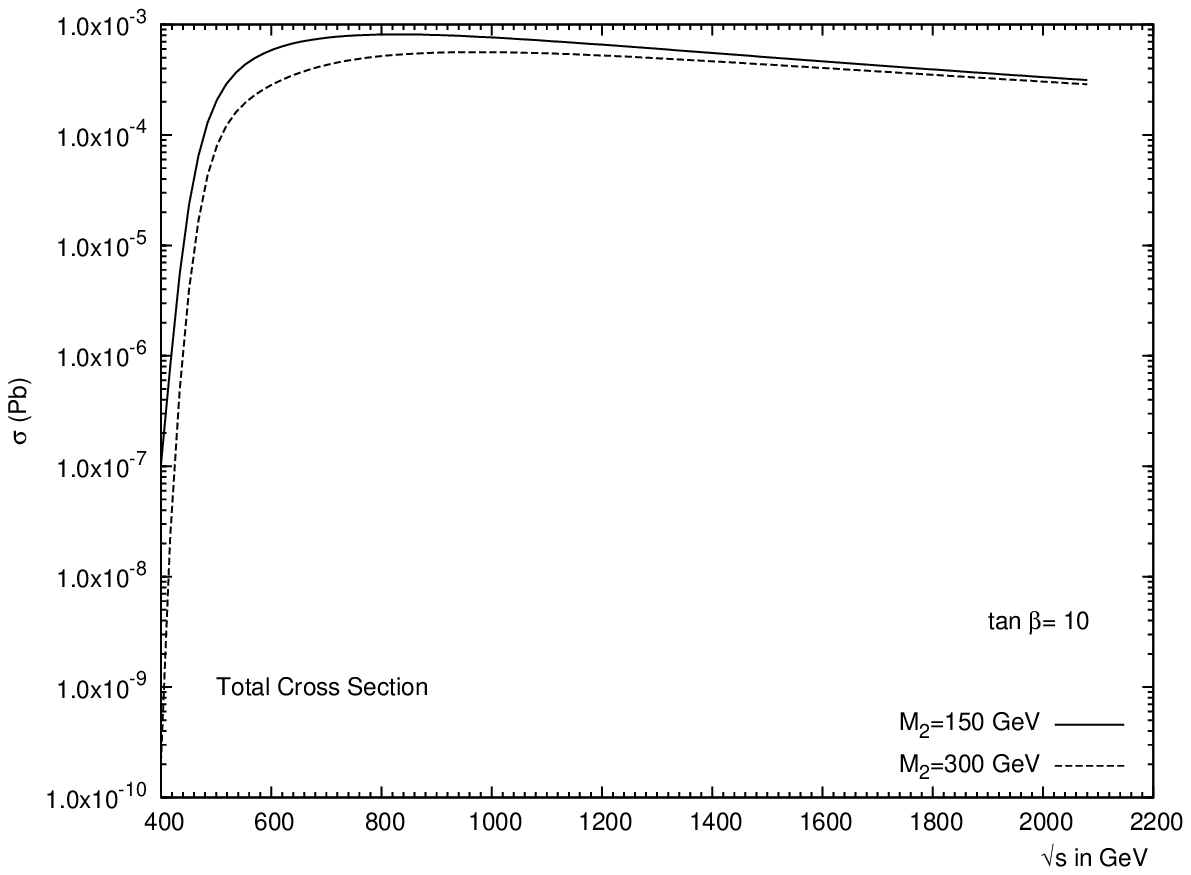}}
\vspace{-0.1cm}
\centerline{\epsfxsize=5.5truein\epsfbox{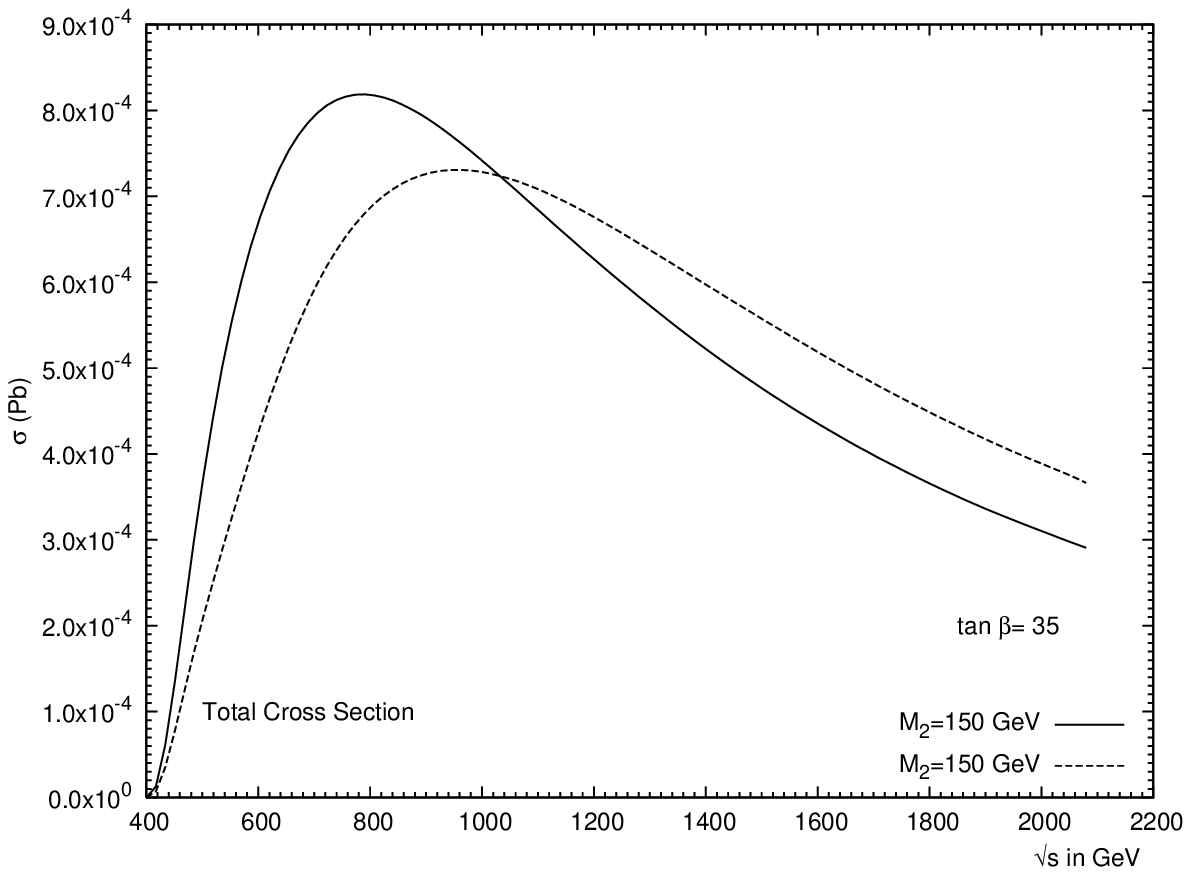}}
\vspace{0.5cm} 
\caption{\small Total cross sections for
the reaction $e^{-}e^{+}\rightarrow h \widetilde{\chi }_{1}^o \widetilde{\chi }_{1}^o$} 
\label{hxxtot}
\end{figure}

\clearpage
\section{Conclusion}
Results of the previous section are summarized in tables
\ref{table7} and \ref{table8} for $M_2 = 150$ GeV and in tables \ref{table9} and \ref{table10} for   $M_2 = 300$ GeV respectively. 
From these results, it is clear that the reaction is most probably to proceed through Feynman diagram no. 5. The cross section of this reaction achieved a value of $1.5336\times 10^{-4}$ [pb] at $tan\beta$ = 10 and for $M_2$ = 150 GeV. And the cross section takes the vlaue of $2.3852\times {10^-4}$ [pb] at $tan\beta$ = 35 and for $M_2$ = 150 GeV.
for $M_2$ = 300 GeV, the value of the cross section achieved $3.0865\times 10^{-4}$ [pb] at $tan\beta$ = 10, and the value $4.0505\times 10^{-4}$ [pb] at $tan\beta$ = 35.
The total cross section of this reaction reaches its maximum value, $8.3901\times {10^-4}$ [pb], at $tan\beta$ = 35 and for $M_2$ = 150 GeV.
Interference effects which are not studied here, should be taken into accounts when dealing with the total cross section. for example the interference terms emerging from diagrams 20 \& 5 ($M_2=150 GeV$) would decrease the value of the total cross section effectively.
\begin{table}[htbp]
\begin{center}
\begin{tabular}[htbp]{|c||c|c||c|c|}
  \hline
  \hline
  Figure No.&\multicolumn{2}{c|}{$\sigma_{tan\beta = 10}$}&\multicolumn{2}{c|}{$\sigma_{tan\beta =35}$}\\
  \cline{2-5}
  &$E_{CM}$& $\sigma$ (Pb)&$E_{CM}$& $\sigma$ (Pb)\\
  \hline
  1 & 1000 & 4.2976e-07 & 1060 & 4.8279e-07\\
  2 & 920 & 3.0571e-07 & 980 & 4.8221e-08\\
  3 & 820 & 8.7088e-08 & 840 & 4.3811e-08\\
  4 & 760 & 8.4636e-11 & 800 & 1.1105e-10\\
  5 & 520 & 0.00015336 & 540 & 0.00023852\\
  6 & 720 & 9.8329e-05 & 740 & 0.00010254\\
  7 & 1300 & 1.7996e-07 & 1340 & 5.4578e-08\\
  8 & 1260 & 4.3651e-08 & 1340 & 2.7638e-08\\
  9 & 800 & 3.3079e-07 & 800 & 5.2466e-07\\
  10 & 780 & 7.8608e-05 & 780 & 5.0584e-05\\
  11 & 1260 & 8.1842e-10 & 1280 & 6.0505e-10\\
  12 & 1300 & 1.7990e-07 & 1340 & 5.45e-08\\
  13 & 1240 & 4.3619e-08 & 1340& 2.7632e-08\\
  14 & 780 & 3.3081e-07 & 800 & 5.2398e-07\\
  15 & 800 & 7.8692e-05 & 800 & 5.0708e-05\\
  16 & 1280 & 1.176e-06 & 1360 & 7.3614e-07\\
  \hline
\end{tabular}
\caption{Summary of the results obtained for the reaction, $e^-(p1) e^+(p2) \rightarrow  h(p3) \widetilde{\chi }_{1}^o(p4) \widetilde{\chi }_{1}^o(p5)$ for $M_2 = 150$ GeV}
\label{table7}
\end{center}
\end{table}
\begin{table}[htbp]
\begin{center}
\begin{tabular}[htbp]{|c||c|c||c|c|}
  \hline
  \hline
  Figure No.&\multicolumn{2}{c|}{$\sigma_{tan\beta = 10}$}&\multicolumn{2}{c|}{$\sigma_{tan\beta =35}$}\\
  \cline{2-5}
  &$E_{CM}$& $\sigma$ (Pb)&$E_{CM}$& $\sigma$ (Pb)\\
  \hline
  17 & 1280 & 7.8242e-08 & 1320& 6.4104e-08\\
  18 & 780 & 6.9236e-07 & 800 & 1.581e-06\\
  19 & 780 & 0.00013973 & 800 & 0.0001293\\
  20 & 1260 & 3.9622e-09 & 1280 & 6.0674e-09\\
  21 & 1340 & 1.1762e-06 & 1340 & 7.3803e-07\\
  22 & 1240 & 7.8012e-08 & 1340 & 6.4162e-08\\
  23 & 780 & 6.9132e-07 & 780 & 1.576e-06\\
  24 & 780 & 0.00013979 & 800 & 0.00012911\\
  \hline
\end{tabular}
\caption{Summary of the results obtained for the reaction, $e^-(p1) e^+(p2) \rightarrow  h(p3) \widetilde{\chi }_{1}^o(p4) \widetilde{\chi }_{1}^o(p5)$ for $M_2 = 150$ GeV}
\label{table8}
\end{center}
\end{table}
\begin{table}[htbp]
\begin{center}
\begin{tabular}[htbp]{|c||c|c||c|c|}
  \hline
  \hline
  Figure No.&\multicolumn{2}{c|}{$\sigma_{tan\beta = 10}$}&\multicolumn{2}{c|}{$\sigma_{tan\beta =35}$}\\
  \cline{2-5}
  &$E_{CM}$& $\sigma$ (Pb)&$E_{CM}$& $\sigma$ (Pb)\\
  \hline
  1 & 1080 & 5.3869e-08 & 1120& 5.5643e-08\\
  2 & 1040 & 3.0443e-08 & 1020& 4.506e-09\\
  3 & 860 & 1.4074e-09 & 880& 6.0213e-10\\
  4 & 680 & 2.7783e-07 & 680& 1.6657e-07\\
  5 & 540 & 0.00011284 & 540& 0.00017088\\
  6 & 800& 3.8825e-06 & 800& 5.3689e-06\\
  7 & 1380 & 4.5216e-08 & 1380& 2.3421e-08\\
  8 & 880 & 7.155e-06 & 940& 3.9801e-06\\
  9 & 780 & 3.5995e-07 & 800& 8.0505e-07\\
  10 & 840 & 0.00030865 & 840& 0.00040393\\
  11 & 1260 & 1.4936e-09 & 1300& 2.0679e-09\\
  12 & 1340 & 4.5271e-08 & 1360& 2.3366e-08\\
  13 & 880 & 7.1686e-06 & 940& 3.9779e-06\\
  14 & 800 & 3.5916e-07 & 800& 8.0607e-07\\
  15 & 860 & 0.00030897 & 860& 0.00040505\\
  16 & 1300 & 1.2266e-06 & 1340& 5.621e-07\\
  \hline
\end{tabular}
\caption{Summary of the results obtained for the reaction, $e^-(p1) e^+(p2) \rightarrow  h(p3) \widetilde{\chi }_{1}^o(p4) \widetilde{\chi }_{1}^o(p5)$ for $M_2 = 300$ GeV}
\label{table9}
\end{center}
\end{table}
\begin{table}[htbp]
\begin{center}
\begin{tabular}[htbp]{|c||c|c||c|c|}
  \hline
  \hline
  Figure No.&\multicolumn{2}{c|}{$\sigma_{tan\beta = 10}$}&\multicolumn{2}{c|}{$\sigma_{tan\beta =35}$}\\
  \cline{2-5}
  &$E_{CM}$& $\sigma$ (Pb)&$E_{CM}$& $\sigma$ (Pb)\\
  \hline
  17 & 940 & 3.0376e-06 & 940& 1.3584e-06\\
  18 & 800 & 3.7196e-06 & 800& 7.8338e-06\\
  19 & 820 & 6.2326e-05 & 840& 6.9613e-05\\
  20 & 1260 & 3.0118e-08 & 1300& 3.6928e-08\\
  21 & 1360 & 1.2275e-06 & 1340& 5.627e-07\\
  22 & 880 & 3.0427e-06 & 940& 1.3604e-06\\
  23 & 780 & 3.7302e-06 & 800& 7.8293e-06\\
  24 & 840 & 6.2285e-05 & 860& 6.9524e-05\\
  \hline
\end{tabular}
\caption{Summary of the results obtained for the reaction, $e^-(p1) e^+(p2) \rightarrow  h(p3) \widetilde{\chi }_{1}^o(p4) \widetilde{\chi }_{1}^o(p5)$ for $M_2 = 300$ GeV}
\label{table10}
\end{center}
\end{table}
\chapter{Production of a light neutral Higgs boson with a pair of charged Higgs bosons}

\section{Introduction}
In this chapter, the production of a light neutral Higgs boson is considered through the reaction, $e^{-}(p1)e^{+}(p2)\rightarrow h(p3) H^+(p4) H^-(p5)$, for different topologies and different propagators (see Appendix A). There are a total of 5 Feynman diagrams for this reaction (tree level approximation) for which we gave the matrix element corresponding to each diagram. Again, diagrams with the same topology which can be obtained by interchanging the indices were represented once.
Our work will proceed as usual,

\begin{enumerate}
\item Feynman diagrams are given,

\item Diagrams with the same topology are represented once, but has been taken into considerations  when calculating the cross section.

\item  Matrix elements are written, all the four momenta squares are
defined to be mass squared $(>0)$,

\item Matrix elements are squared,

\item An average over the initial spin polarizations of the electron and
positron pair and a sum over the final spin states of the outgoing particles
arising from each initial spin state is carried out.

\item Results are represented graphically, and summarized in subsequent tables.
\end{enumerate}

\section{Feynman Diagrams}

The following is the set of Feynman diagrams which were used to calculate
the cross section of the associated production of a charged Higgs boson with
a pair of neutralinos. Our momentum notation is:
$e^{-}(p1)$, $e^{+}(p2)$, $h(p3)$, $H^+(p4)$, and $H^-(p5)$.

\begin{figure}[h]
\begin{center}
\vskip-2.5cm \mbox{\hskip-3.5cm\centerline{\epsfig{file=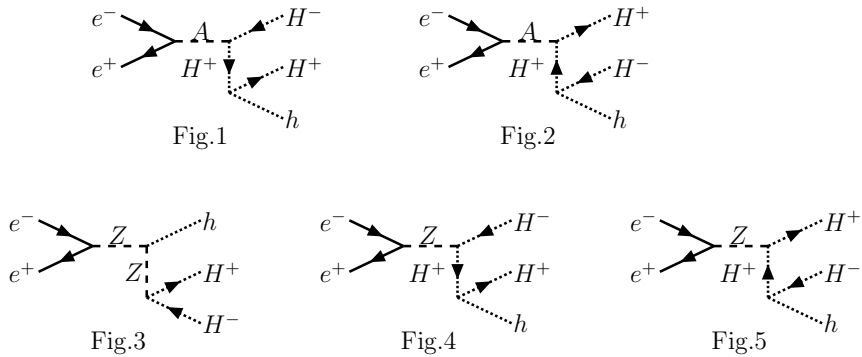,width=17cm}}}
\end{center}
\vskip-12cm
\caption{Feynman diagrams for the reaction: $e^{-}(p1)e^{+}(p2)\rightarrow h(p3) \widetilde{\chi }_{1}^o(p4) \widetilde{\chi }_{1}^o(p5)$}
\label{feyn6}
\end{figure}

\newpage

\section{Matrix Elements}
The following is the set of matrix elements corresponding to diagrams in figures \ref{feyn6} used in our calculations:

\begin{eqnarray*}
\mathcal{M}_{1} &=&\overline{v}(\vec{p}_{2})A^{\mu }\gamma _{\mu }\frac{%
g^{\mu \nu }}{(p_{1}+p_{2})^{2}}e(p_{1}+p_{2})^{\mu }\frac{1}{%
(p_{4}+p_{5})^{2}-m_{H_{3}}^{2}} \\
&&\times g\left[ M_{w}\cos (\beta -\alpha )-\frac{M_{Z}}{2\cos \theta _{w}}%
\cos 2\beta \cos (\beta +\alpha )\right] u(p_{1})
\end{eqnarray*}

\begin{eqnarray*}
\mathcal{M}_{2} &=&\overline{v}(\vec{p}_{2})A^{\mu }\gamma _{\mu }\frac{%
g^{\mu \nu }}{(p_{1}+p_{2})^{2}}e(p_{1}+p_{2})^{\mu }\frac{1}{%
(p_{3}+p_{5})^{2}-m_{H_{3}}^{2}} \\
&&\times g\left[ M_{w}\cos (\beta -\alpha )-\frac{M_{Z}}{2\cos \theta _{w}}%
\cos 2\beta \cos (\beta +\alpha )\right] u(p_{1})
\end{eqnarray*}

\begin{eqnarray*}
\mathcal{M}_{3} &=&\overline{v}(\vec{p}_{2})\gamma _{\mu
}(B^{L}P_{L}+B^{R}P_{R})^{\mu }\frac{g^{\mu \nu }-k_{\mu }k_{\nu }/M_{Z}}{%
(p_{1}+p_{2})^{2}-M_{Z}^{2}+i\epsilon }J_{1}g^{\nu \rho } \\
&&\times \frac{g^{\rho \sigma }-k_{\rho }k_{\sigma }/M_{Z}}{%
(p_{3}+p_{4})^{2}-M_{Z}^{2}+i\epsilon }\frac{g\cos 2\theta _{w}}{2\cos
\theta _{w}}(p_{3}+p_{4})^{\sigma }u(p_{1})
\end{eqnarray*}

\begin{eqnarray*}
\mathcal{M}_{4} &=&-\overline{v}(\vec{p}_{2})\gamma _{\mu
}(B^{L}P_{L}+B^{R}P_{R})^{\mu }\frac{g^{\mu \nu }-k_{\mu }k_{\nu }/M_{Z}}{%
(p_{1}+p_{2})^{2}-M_{Z}^{2}+i\epsilon }\frac{g\cos 2\theta _{w}}{2\cos
\theta _{w}}(p_{3}+p_{4}+p_{5})^{\nu } \\
&&\times \frac{1}{(p_{4}+p_{5})^{2}-m_{H_{3}}^{2}}g\left[ M_{w}\cos (\beta
-\alpha )-\frac{M_{Z}}{2\cos \theta _{w}}\cos 2\beta \cos (\beta +\alpha )%
\right] u(p_{1})
\end{eqnarray*}

\begin{eqnarray*}
\mathcal{M}_{5} &=&-\overline{v}(\vec{p}_{2})\gamma _{\mu
}(B^{L}P_{L}+B^{R}P_{R})^{\mu }\frac{g^{\mu \nu }-k_{\mu }k_{\nu }/M_{Z}}{%
(p_{1}+p_{2})^{2}-M_{Z}^{2}+i\epsilon }\frac{g\cos 2\theta _{w}}{2\cos
\theta _{w}}(p_{3}+p_{4}+p_{5})^{\nu } \\
&&\times \frac{1}{(p_{3}+p_{5})^{2}-m_{H_{3}}^{2}}g\left[ M_{w}\cos (\beta
-\alpha )-\frac{M_{Z}}{2\cos \theta _{w}}\cos 2\beta \cos (\beta +\alpha )%
\right] u(p_{1})
\end{eqnarray*}

\noindent For the definitions of the constants used here, the
reader is referred to Appendix A.

\section{Cross Sections}
As before, to calculate the differential cross sections, and
hence, the total cross section, we need first to obtain the
squared matrix element for each Feynman diagram, where use of the
trace theorems was made. Later an average over the initial
spin polarizations of the electron and the positron pair and the
sum over the final spin states of the outgoing particles arising
from each initial spin state is carried out. The total cross
section as a function of the center of the mass energy (see Appendix
B) is then calculated.\\ 
Calculations were done with the following set of parameters:\\
$tan\beta = 10$, and $tan\beta = 15$ where $M_2 = 150$ or $M_2 = 300$.\\
All results are given in the following figures.
\begin{figure}[th]
\vspace{-4.5cm}
\centerline{\epsfxsize=5.5truein\epsfbox{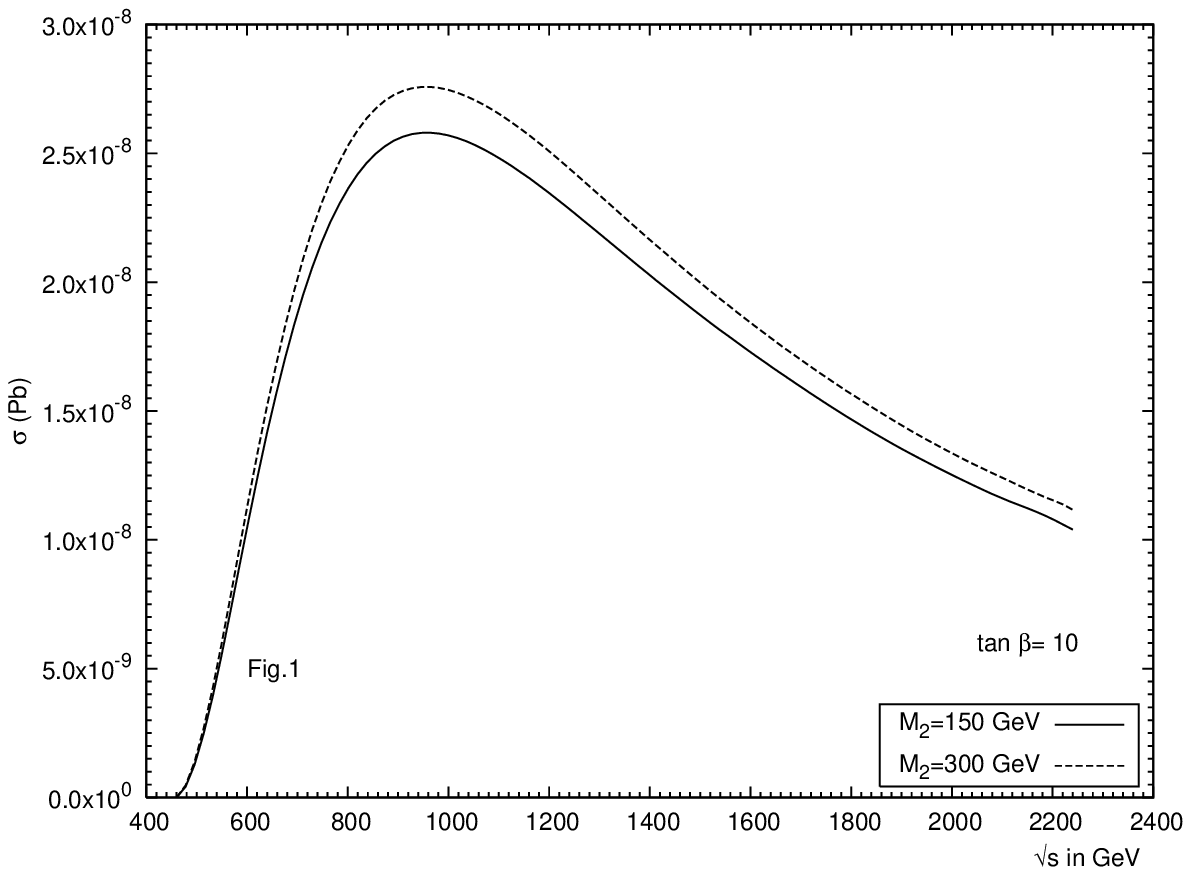}}
\vspace{-0.1cm}
\centerline{\epsfxsize=5.5truein\epsfbox{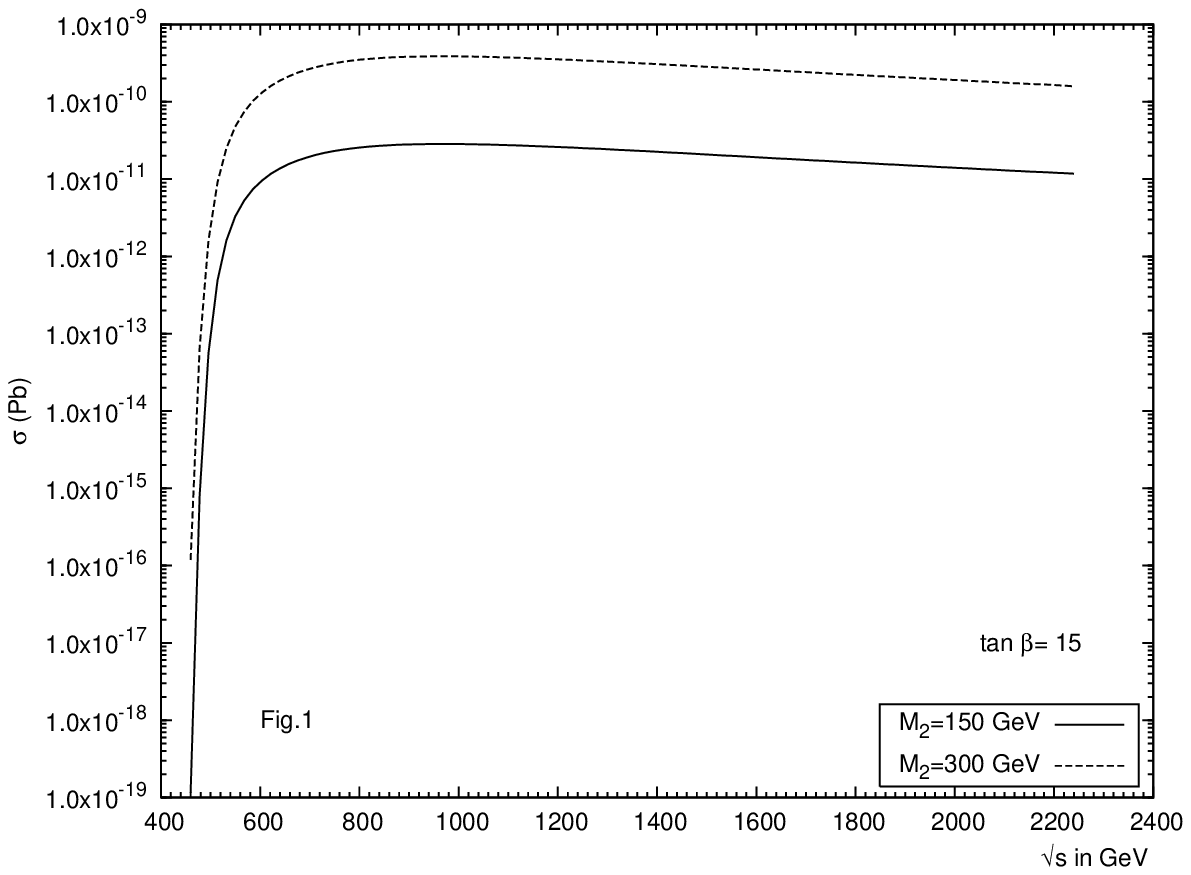}}
\vspace{0.5cm} 
\caption{\small Cross sections for
diagram no. 1 in figure \ref{feyn6}} 
\label{hHH1}
\end{figure}

\begin{figure}[th]
\vspace{-4.5cm}
\centerline{\epsfxsize=5.5truein\epsfbox{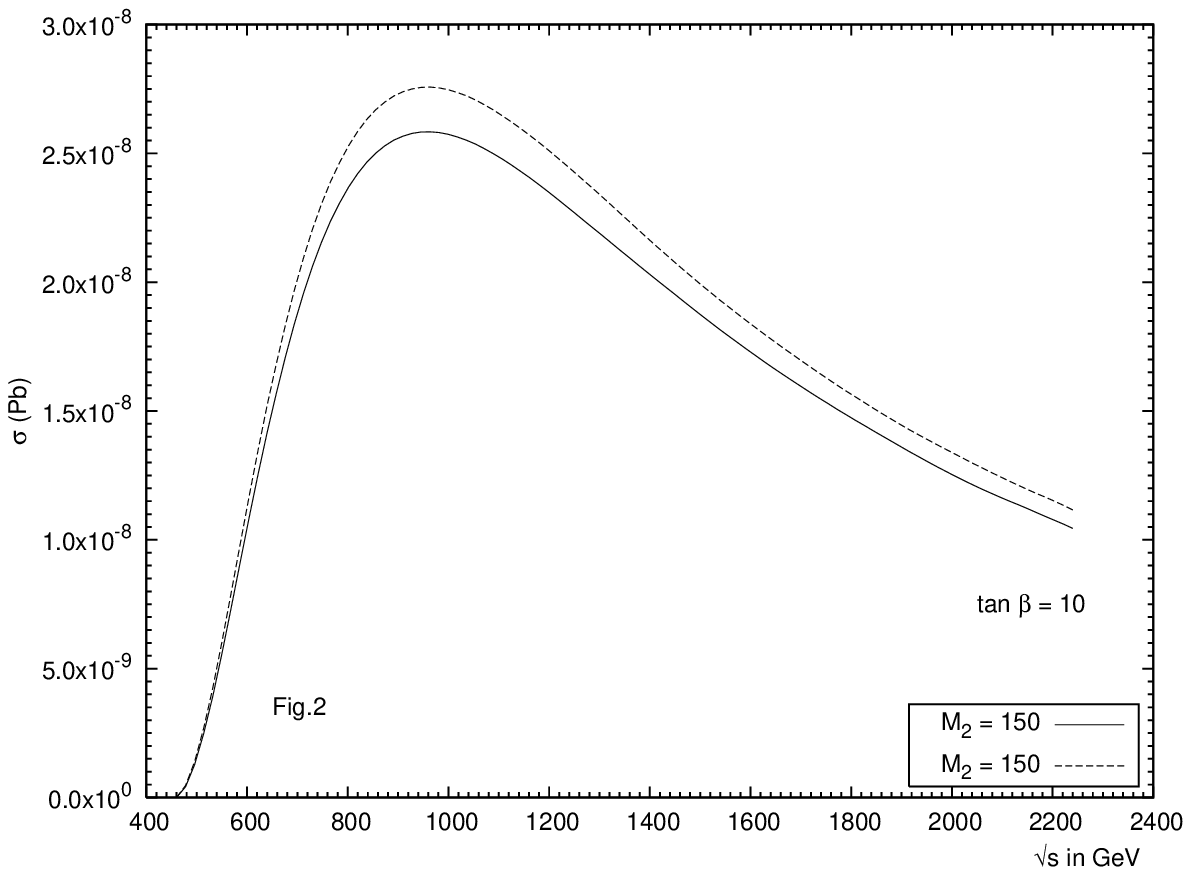}}
\vspace{-0.1cm}
\centerline{\epsfxsize=5.5truein\epsfbox{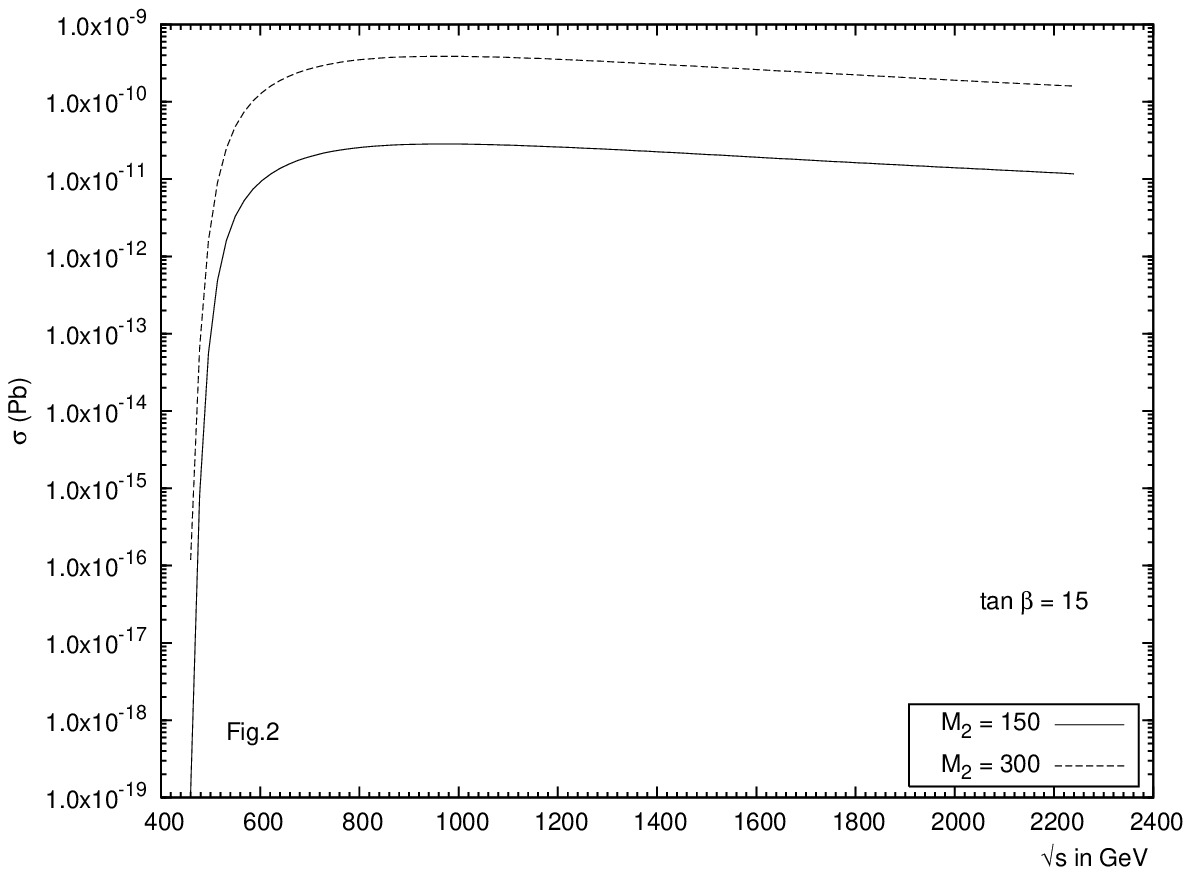}}
\vspace{0.5cm} 
\caption{\small Cross sections for
diagram no. 2 in figure \ref{feyn6}} 
\label{hHH2}
\end{figure}

\begin{figure}[th]
\vspace{-4.5cm}
\centerline{\epsfxsize=5.5truein\epsfbox{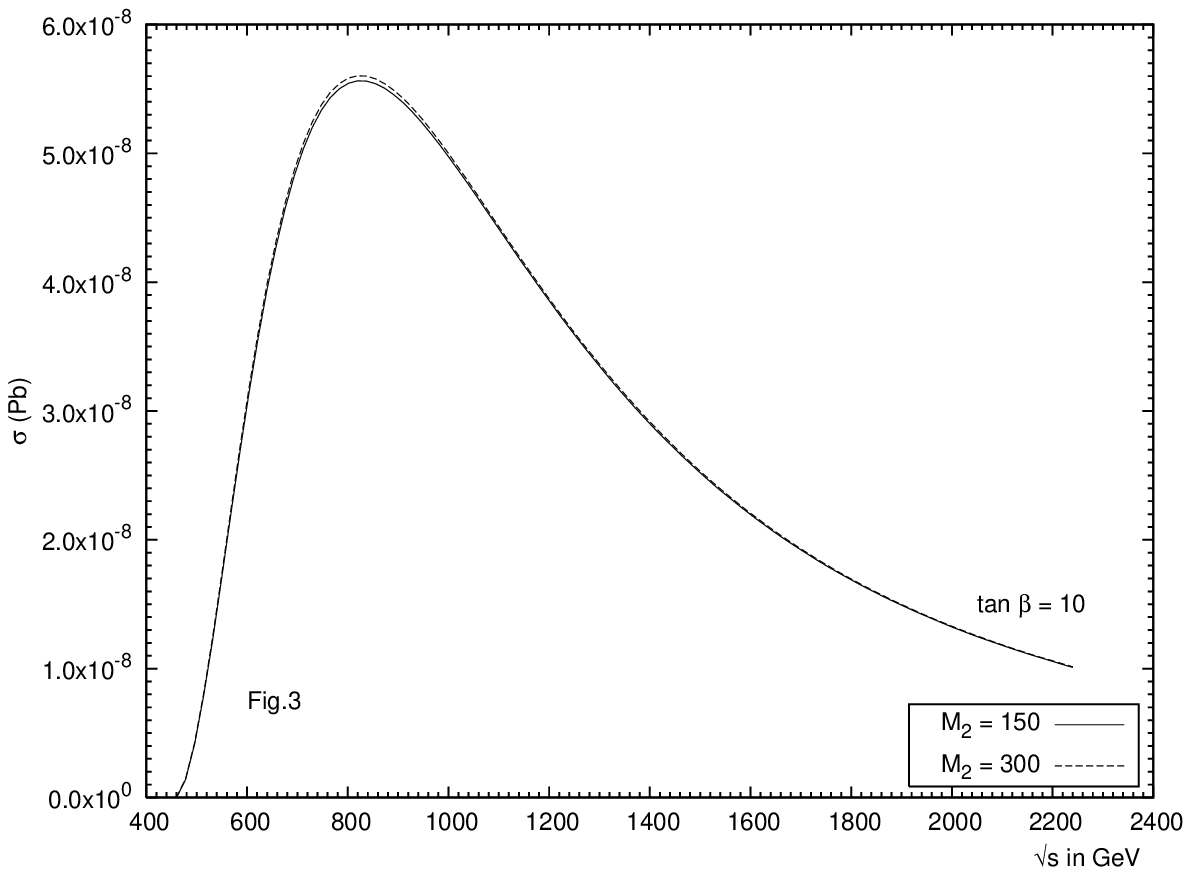}}
\vspace{-0.1cm}
\centerline{\epsfxsize=5.5truein\epsfbox{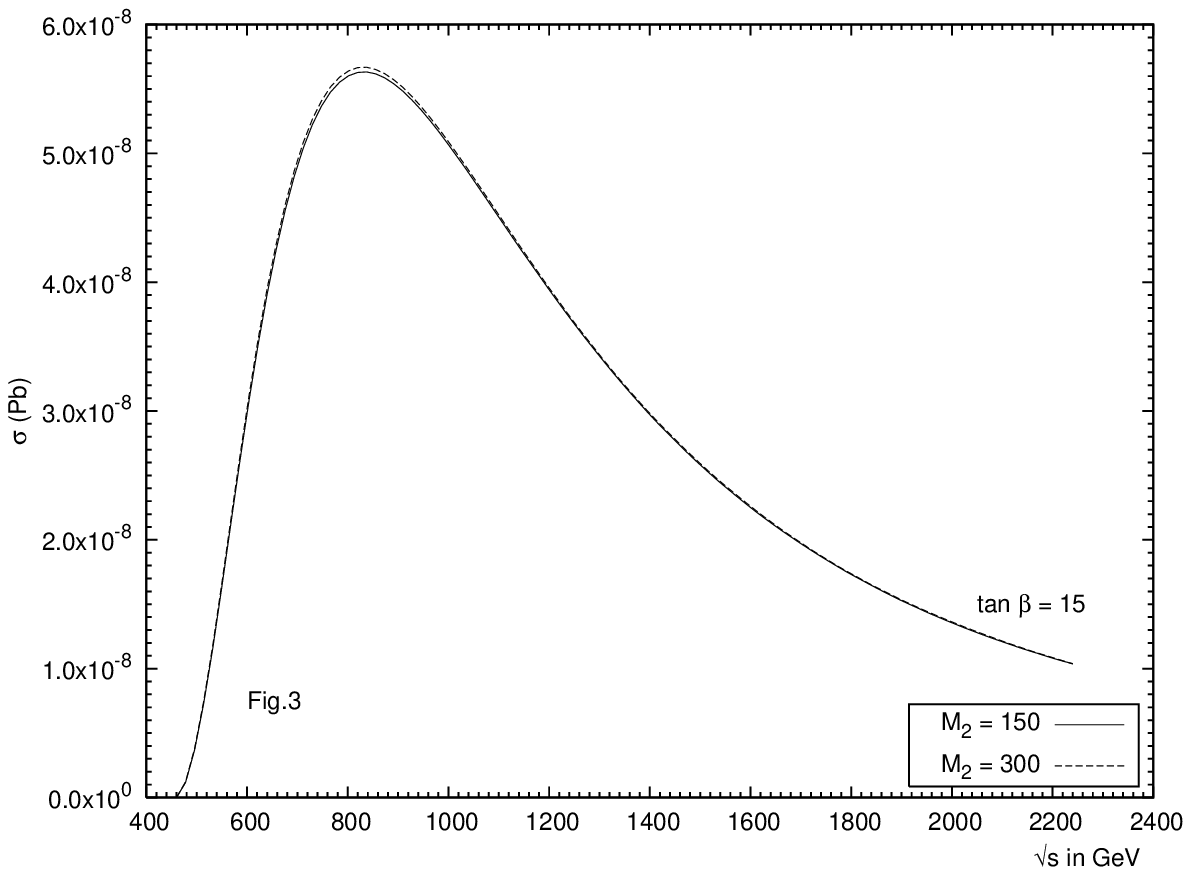}}
\vspace{0.5cm} 
\caption{\small Cross sections for
diagram no. 3 in figure \ref{feyn6}} 
\label{hHH3}
\end{figure}

\begin{figure}[th]
\vspace{-4.5cm}
\centerline{\epsfxsize=5.5truein\epsfbox{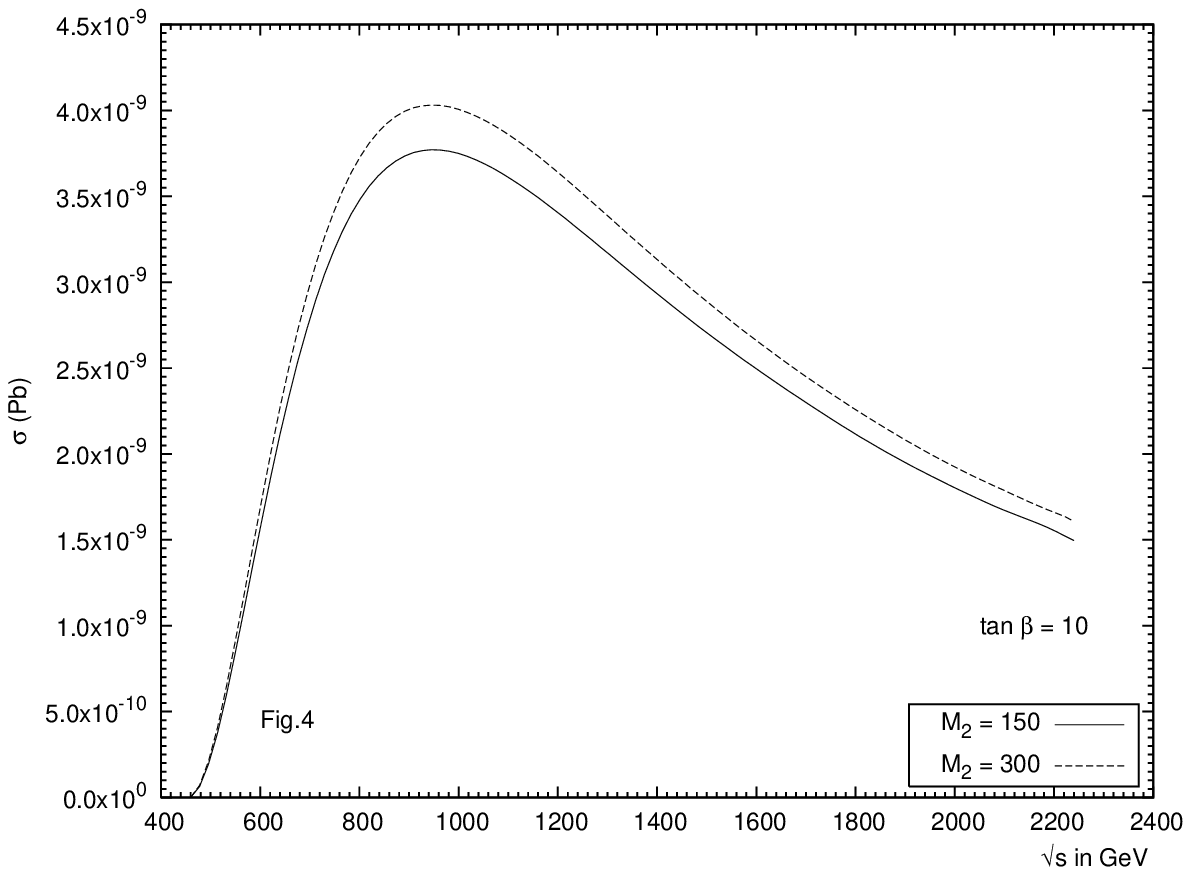}}
\vspace{-0.1cm}
\centerline{\epsfxsize=5.5truein\epsfbox{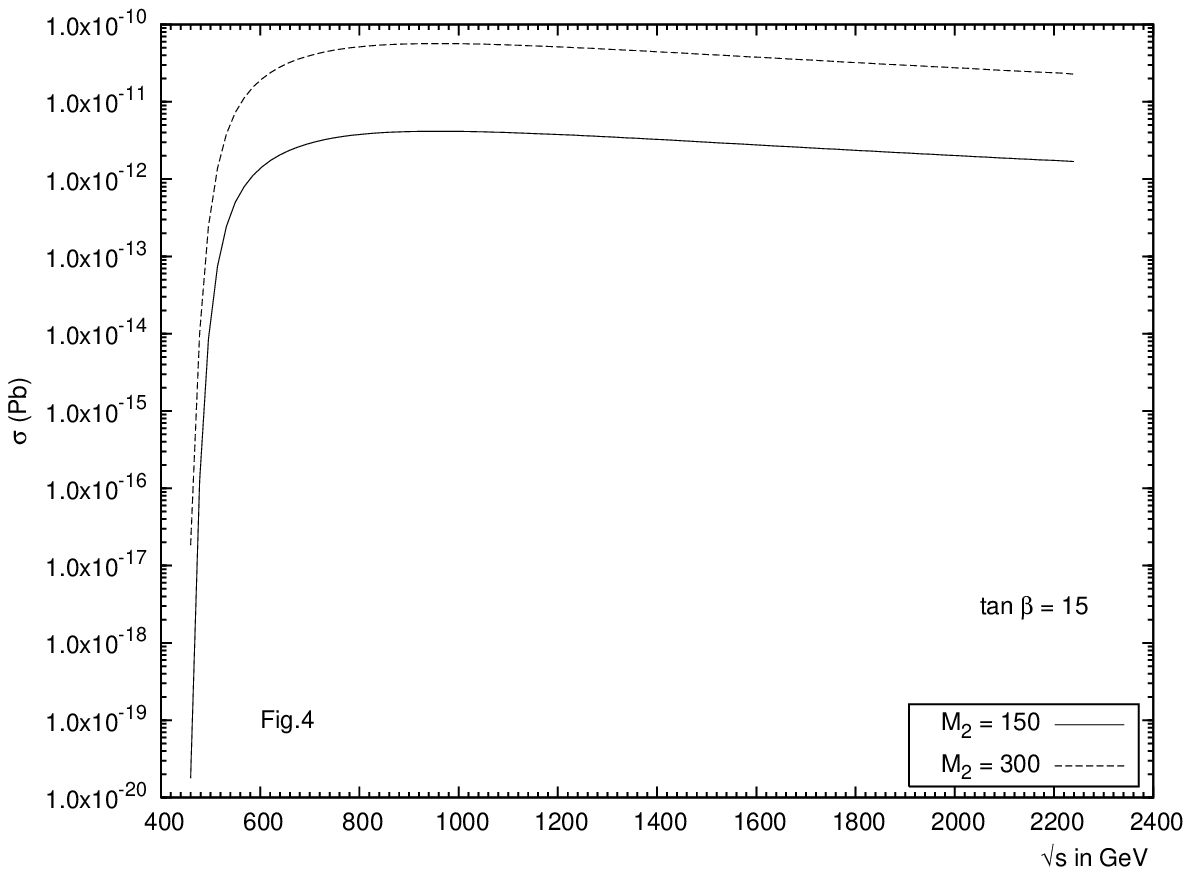}}
\vspace{0.5cm} 
\caption{\small Cross sections for
diagram no. 4 in figure \ref{feyn6}} 
\label{hHH4}
\end{figure}

\begin{figure}[th]
\vspace{-4.5cm}
\centerline{\epsfxsize=5.5truein\epsfbox{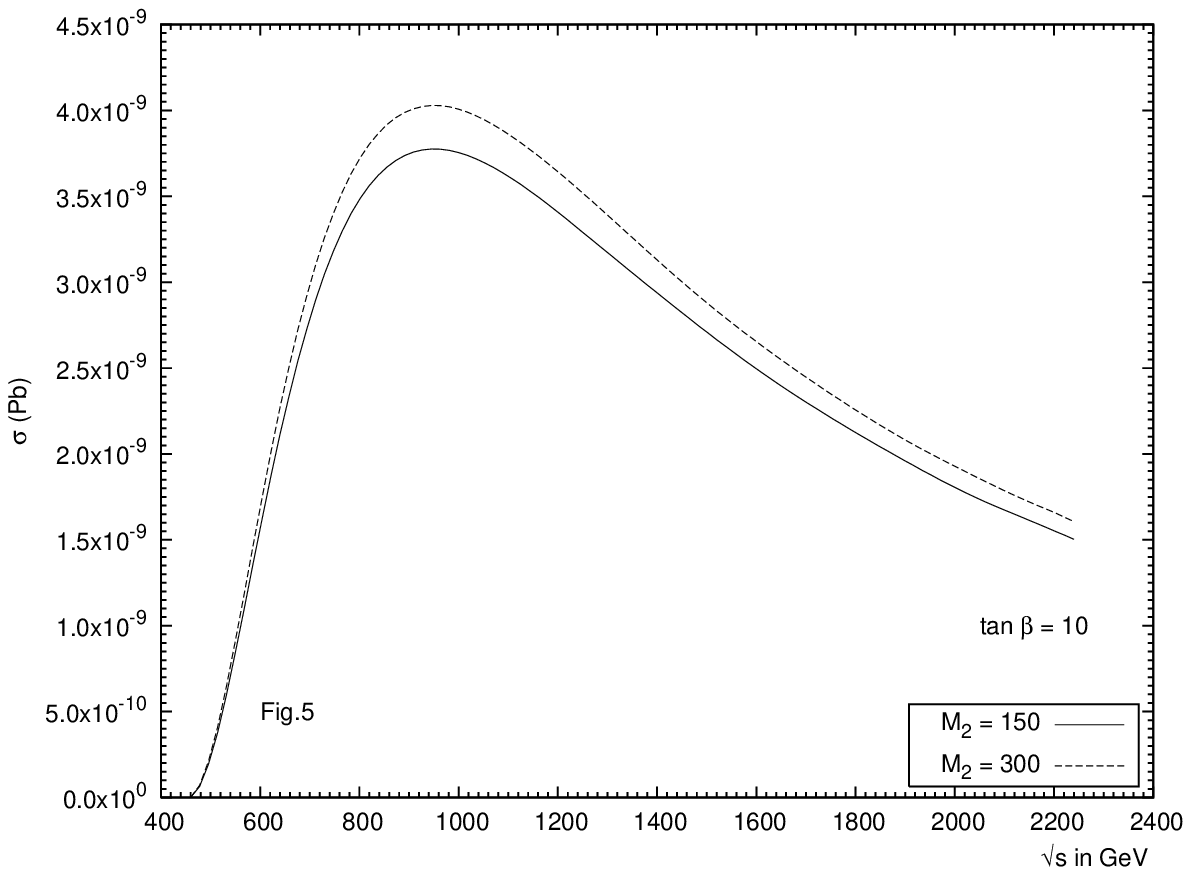}}
\vspace{-0.1cm}
\centerline{\epsfxsize=5.5truein\epsfbox{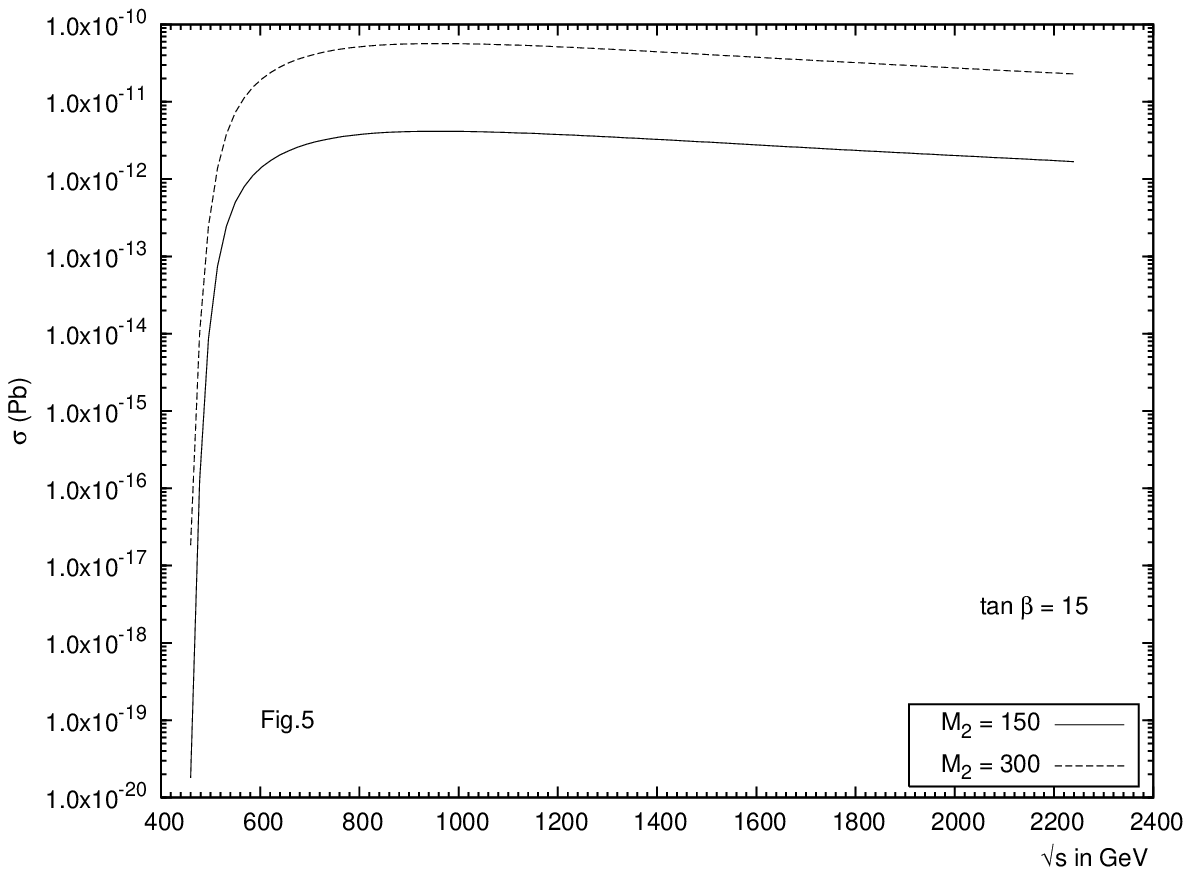}}
\vspace{0.5cm} 
\caption{\small Cross sections for
diagram no. 5 in figure \ref{feyn6}} 
\label{hHH5}
\end{figure}

\begin{figure}[th]
\vspace{-4.5cm}
\centerline{\epsfxsize=5.5truein\epsfbox{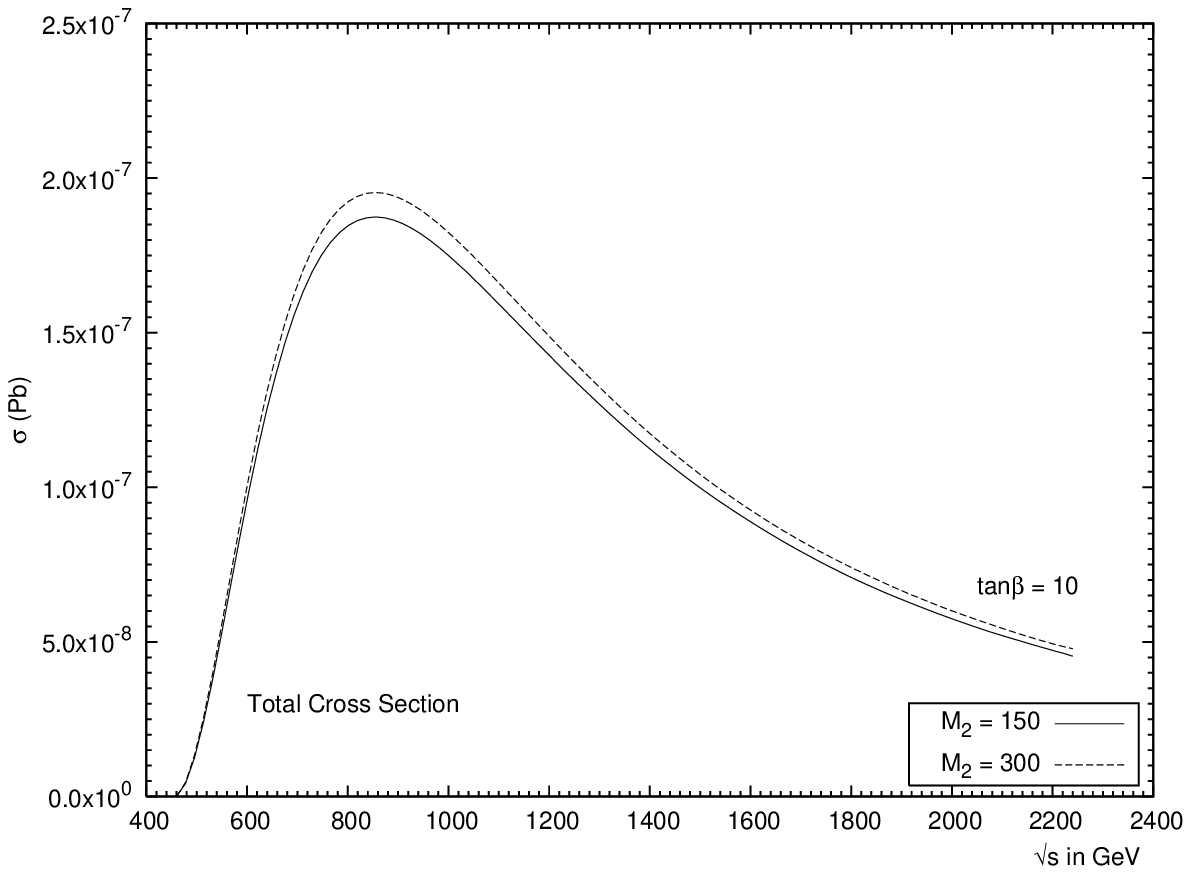}}
\vspace{-0.1cm}
\centerline{\epsfxsize=5.5truein\epsfbox{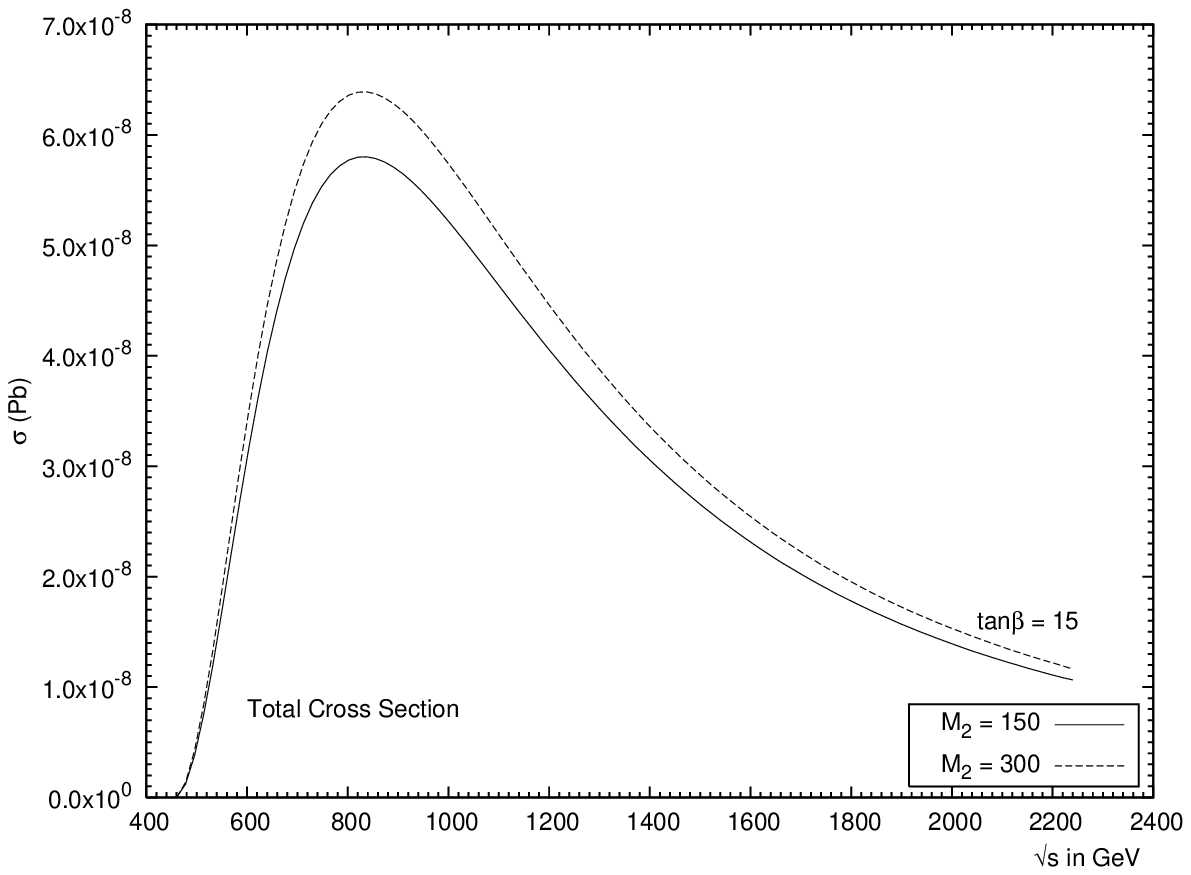}}
\vspace{0.5cm} 
\caption{\small Total cross section for the reaction $e^{-}(p1)e^{+}(p2)\rightarrow h(p3) H^+(p4) H^-(p5)$} 
\label{tothHH}
\end{figure}

\clearpage
\section{Conclusion}
Results of the previous section are summarized in tables
\ref{table11} for $M_2 = 150$ GeV and in tables \ref{table12} for $M_2 = 300$
GeV respectively. From these results, it is clear that the reaction
is most probably to proceed through diagram no. 3 at $tan\beta$ = 10 and $tan\beta$ = 15 where $M_2$ = 150 GeV. at these values the maximum cross section achieved is $5.7627\times 10^{-8}$ [pb] and $5.8264\times 10^{-8}$ [pb] respectively.\\ 
The total cross section achieved by this reaction is $1.9263\times 10^{-7}$ [pb] at $E_{CM}$ = 820 GeV for $tan\beta$ = 10 and $M_2$ = 150 GeV, and is $5.9956\times 10^{-8}$ [pb] at $E_{CM}$ = 840 GeV for $tan\beta$ = 15 and $M_2$ = 300 GeV
the maximum cross section at $M_2$ = 300 GeV for both values of $tan\beta$, that is 10 \& 15, takes the values $5.8647\times 10^{-8}$ at $E_{CM}$ = 800 GeV, and $5.8883\times 10^{-8}$ at $E_{CM}$ = 820 GeV.\\
The total cross section achieved by this reaction is $2.0128\times 10^{-7}$ [pb] at $E_{CM}$ = 860 GeV for $tan\beta$ = 10 and $M_2$ = 150 GeV, and is $6.6248\times 10^{-8}$ [pb] at $E_{CM}$ = 820 GeV for $tan\beta$ = 15 and $M_2$ = 300 GeV

\vskip2cm
\begin{table}[htbp]
\begin{center}
\begin{tabular}[htbp]{|c||c|c||c|c|}
  \hline
  \hline
  Figure No.&\multicolumn{2}{c|}{$\sigma_{tan\beta = 10}$}&\multicolumn{2}{c|}{$\sigma_{tan\beta =15}$}\\
  \cline{2-5}
  &$E_{CM}$& $\sigma$ (Pb)&$E_{CM}$& $\sigma$ (Pb)\\
  \hline
  1 & 940 & 2.6354e-08 & 960& 2.9167e-11\\
  2 & 940 & 2.6350e-08 & 960& 2.9131e-11\\
  3 & 800 & 5.7627e-08 & 800& 5.8246e-08\\
  4 & 940 & 3.8517e-09 & 960& 4.2596e-12\\
  5 & 940 & 3.8512e-09 & 940& 4.2561e-12\\
  total & 820 & 1.9263e-07& 840 & 5.9956e-08\\ 
  \hline
\end{tabular}
\caption{Summary of the results obtained for the reaction,
$e^{-}(p1)e^{+}(p2)\rightarrow h(p3) H^+(p4) H^-(p5)$ for $M_2 = 150$ GeV} \label{table11}
\end{center}
\end{table}
\begin{table}[htbp]
\begin{center}
\begin{tabular}[htbp]{|c||c|c||c|c|}
  \hline
  \hline
  Figure No.&\multicolumn{2}{c|}{$\sigma_{tan\beta = 10}$}&\multicolumn{2}{c|}{$\sigma_{tan\beta =15}$}\\
  \cline{2-5}
  &$E_{CM}$& $\sigma$ (Pb)&$E_{CM}$& $\sigma$ (Pb)\\
  \hline
  1 & 940 & 2.8116e-08 & 960 & 3.9597e-10\\
  2 & 940 & 2.8066e-08 & 940 & 3.9717e-10\\
  3 & 800 & 5.8047e-08 & 820& 5.8883e-08\\
  4 & 900 & 4.1112e-09 & 960& 5.7828e-11\\
  5 & 940 & 4.102e-09 & 940& 5.8048e-11\\
  total & 860 & 2.0128e-07& 820 & 6.6248e-08\\
  \hline
\end{tabular}
\caption{Summary of the results obtained for the reaction,
$e^{-}(p1)e^{+}(p2)\rightarrow h(p3) H^+(p4) H^-(p5)$ for $M_2 = 300$ GeV} \label{table12}
\end{center}
\end{table}
\appendix

\textbf{introduction:}
In this appendix, we summarize the Feynman rulesin a convenient
form, required to construct the covariant amplitudes in this study.
The given rules are computed by using the unitary (physical) gauge [33],
[34], [35]. This gauge is the prefered choice in the tree approximation
level. For Higher order corrections, we must include the ghost loops, and in
this case, it is much more effecient and easier to use the so called 't
Hooft-Feynman gauge [36], [37] than the unitary gauge.

\section{Appendix A: Feynman Rules}

The Feynman diagrams at the tree level for two particles in the initial
state and three particles at the final state are classified physically into
three different topologies according to the interaction of the two initial
particles (or annihilation of particle-antiparticle pair) through no
propagators, one propagator, and two propagators.

\textbf{External Lines (Incoming and Outgoing Particles)}

\subsection{Fermions}

\begin{center}
\begin{tabular}{ccc}
& Incoming & Outgoing \\ 
Particle & $u(p,s)$ & $\overline{u}(p,s)$ \\ 
AntiParticle & $\overline{v}(p,s)$ & $v(p,s)$%
\end{tabular}
\end{center}

\subsection{Vector Bosons}

\begin{center}
\begin{tabular}{cc}
Incoming & Outgoing \\ 
$\varepsilon _{\sigma }(p,s)$ & $\varepsilon _{\sigma }^{\ast }(p,s)$%
\end{tabular}
\end{center}

\subsection{Scalar (Higgs) Bosons}

Scalar bosons as external lines take the value \textbf{1 }in general.

\textbf{Internal Lines (Propagators)}

\subsection{fermions}

\[
i\frac{\NEG{p}+m}{p^{2}-m^{2}+i\epsilon }
\]

\subsection{Massive Vector Bosons}

\[
-i\frac{g^{\mu \nu }-k^{\mu }k^{\nu }/M^{2}}{p^{2}-m^{2}+i\epsilon }
\]

\subsection{Massless Vector Bosons (Photons)}

\[
-i\frac{g^{\mu \nu }}{p^{2}}
\]

\subsection{Scalar (Higgs) Bosons}

\[
i\frac{1}{p^{2}-m^{2}}
\]

\textbf{vertices}

\newpage
\section{Appendix B: Numerical Values of Constants}

The different factors of vertices are model dependent constants. The
numerical values of constants are given by

\[
A=e
\]

\[
B^{L}=g^{`}\left( \frac{\frac{1}{2}-\sin ^{2}\theta _{w}}{\sin \theta _{w}}%
\right) 
\]

\[
B^{R}=-g^{`}\sin \theta _{w}
\]%
\[
C=-\frac{g}{\sqrt{2}}
\]%
\[
D=e
\]%
\[
E=g\cos \theta _{w}
\]%
\[
F_{1}^{L}=-g\left( \frac{m_{e}\cos \alpha }{2M_{w}\cos \beta }\right) 
\]%
\[
F_{2}^{L}=g\left( \frac{m_{e}\cos \alpha }{2M_{w}\cos \beta }\right) 
\]%
\[
F_{3}^{L}=0
\]%
\[
F_{1}^{R}=0
\]%
\[
F_{2}^{R}=0
\]%
\[
F_{3}^{R}=ig\left( \frac{m_{e}\tan \beta }{2M_{w}}\right) 
\]%
\[
G_{1}=ig\left[ \frac{\sin (\beta -\alpha )}{2\cos \theta _{w}}\right] 
\]%
\[
G_{2}=-ig\left[ \frac{\cos (\beta -\alpha )}{2\cos \theta _{w}}\right] 
\]%
\[
H_{1}=gM_{w}\cos (\beta -\alpha )
\]%
\[
H_{2}=gM_{w}\sin (\beta -\alpha )
\]%
\[
I_{1}=\frac{g}{2}\sin (\alpha -\beta )
\]%
\[
I_{2}=\frac{g}{2}\cos (\alpha -\beta )
\]%
\[
I_{3}=-i\frac{g}{2}
\]%
\[
J_{1}=g\left[ \frac{M_{Z}\cos (\beta -\alpha )}{\cos \theta _{w}}\right] 
\]%
\[
J_{2}=g\left[ \frac{M_{Z}\sin (\beta -\alpha )}{\cos \theta _{w}}\right] 
\]%
\[
K_{L}=g\left( \frac{\frac{1}{2}-\sin ^{2}\theta _{w}}{\cos \theta _{w}}%
\right) 
\]%
\[
K_{R}=-g\left( \frac{\sin ^{2}\theta _{w}}{\cos \theta _{w}}\right) 
\]%
\[
L=\frac{g}{\sqrt{2}}
\]%
\[
M=\frac{g}{2\cos \theta _{w}}
\]%
\[
N_{Li}=\frac{-g}{2\sqrt{2}}\left( \varepsilon _{i}\frac{m_{e}Z_{i3}}{%
M_{w}\cos \beta }-Z_{i2}-Z_{i1}\tan \theta _{w}\right) 
\]%
\[
N_{Ri}=\frac{-g}{2\sqrt{2}}\left( \frac{m_{e}Z_{i3}}{M_{w}\cos \beta }%
+2\varepsilon _{i}Z_{i1}\tan \theta _{w}\right) 
\]%
\[
N_{Li}^{`}=\frac{-g}{2\sqrt{2}}\left( -\varepsilon _{i}\frac{m_{e}Z_{i3}}{%
M_{w}\cos \beta }-Z_{i2}-Z_{i1}\tan \theta _{w}\right) 
\]%
\[
N_{Ri}^{`}=\frac{-g}{2\sqrt{2}}\left( \frac{m_{e}Z_{i3}}{M_{w}\cos \beta }%
-2\varepsilon _{i}Z_{i1}\tan \theta _{w}\right) 
\]%
\[
O=\frac{g}{\sqrt{2}}\left( Z_{i1}\tan \theta _{w}-Z_{i2}\right) 
\]%
\[
P_{ij}^{L}=\varepsilon _{i}g\left( Z_{i2}U_{j1}+\frac{Z_{i4}V_{j2}}{\sqrt{2}}%
\right) 
\]%
\[
P_{ij}^{R}=g\left( Z_{i2}U_{j1}+\frac{Z_{i3}U_{j2}}{\sqrt{2}}\right) 
\]%
\[
Q_{ij}^{R}=\varepsilon _{i}g\sin \beta \left[ Z_{i3}U_{j1}-\frac{%
(Z_{i2}+Z_{i1}\tan \theta _{w})U_{i2}}{\sqrt{2}}\right] 
\]

\[
R_{lij}=-\frac{g}{2\sin \beta }\left\{ \frac{m_{\widetilde{\chi }%
_{i}^{o}}\delta _{ij}\sin \alpha }{M_{w}}+\left( \varepsilon
_{i}+\varepsilon _{j}\right) \left[ Q_{ij}^{``}\sin (\beta -\alpha
)-R_{ij}^{``}\sin \alpha \right] \right\} 
\]%
\[
R_{2ij}=-\frac{g}{2\sin \beta }\left\{ \frac{m_{\widetilde{\chi }%
_{i}^{o}}\delta _{ij}\cos \alpha }{M_{w}}-\left( \varepsilon
_{i}+\varepsilon _{j}\right) \left[ Q_{ij}^{``}\cos (\beta -\alpha
)+R_{ij}^{``}\cos \alpha \right] \right\} 
\]%
\[
R_{3ij}=\frac{ig}{2\sin \beta }\left( \varepsilon _{i}-\varepsilon
_{j}\right) \left[ Q_{ij}^{``}\cos 2\beta -R_{ij}^{``}\cos \beta \right] 
\]%
\[
R_{1ij}^{`}=-\frac{g}{2\sin \beta }\left[ \left( \varepsilon
_{i}-\varepsilon _{j}\right) Q_{ij}^{``}\sin (\beta -\alpha )-(\varepsilon
_{i}-\varepsilon _{j})R_{ij}^{``}\sin \alpha \right] 
\]%
\[
R_{2ij}^{`}=\frac{g}{2\sin \beta }\left( \varepsilon _{i}-\varepsilon
_{j}\right) \left[ Q_{ij}^{``}\cos (\beta -\alpha )+R_{ij}^{``}\cos \alpha %
\right] 
\]%
\[
R_{3ij}^{`}=-\frac{ig}{2\sin \beta }\left\{ \frac{m_{\widetilde{\chi }%
_{i}^{o}}\delta _{ij}\cos \alpha }{M_{w}}-\left( \varepsilon
_{i}+\varepsilon _{j}\right) \left[ Q_{ij}^{``}\cos 2\beta +R_{ij}^{``}\cos
\beta \right] \right\} 
\]

where

\[
Q_{ij}^{``}=\frac{1}{2}\left[ Z_{i3}\left( gZ_{j2}-g^{`}Z_{j1}\right)
+Z_{j3}(gZ_{i2}-g^{`}Z_{i1})\right] 
\]

and

\[
R_{ij}^{``}=\frac{1}{2M_{w}}\left[ MZ_{i2}Z_{j2}+M^{`}Z_{i1}Z_{j1}-\mu
\left( Z_{i3}Z_{j4}+Z_{i4}Z_{j3}\right) \right] 
\]

\[
S_{ij}=\frac{g}{4\cos \theta _{w}}\left( 1-\varepsilon _{i}\varepsilon
_{j}\right) \left( Z_{i3}Z_{j3}-Z_{i4}Z_{j4}\right) 
\]

\[
S_{ij}^{`}=\frac{g}{4\cos \theta _{w}}\left( 1+\varepsilon _{i}\varepsilon
_{j}\right) \left( Z_{i3}Z_{j3}-Z_{i4}Z_{j4}\right) 
\]

\[
T_{i}^{L}=g\left( \frac{m_{e}U_{i2}}{\sqrt{2}M_{w}\cos \beta }\right) 
\]%
\[
T_{i}^{R}=-gV_{i1}
\]%
\[
U_{i}=-gU_{i1}
\]

\[
V=e
\]

\[
W_{ij}^{L}=\frac{g}{\cos \theta _{w}}\left( \delta _{ij}\sin ^{2}\theta
_{w}-V_{i1}V_{j1}-\frac{1}{2}V_{i2}V_{j2}\right) 
\]

\[
W_{ij}^{R}=\frac{g}{\cos \theta _{w}}\left( \delta _{ij}\sin ^{2}\theta
_{w}-U_{i1}U_{j1}-\frac{1}{2}U_{i2}U_{j2}\right) 
\]%
\[
X_{1ij}=-\frac{g}{2\sin \beta }\left\{ \frac{m_{\widetilde{\chi }%
_{i}^{o}}\delta _{ij}\sin \alpha }{M_{w}}+2\left[ Q_{ij}\sin (\beta -\alpha
)-R_{ij}\sin \alpha \right] \right\} 
\]%
\[
X_{2ij}=-\frac{g}{2\sin \beta }\left\{ \frac{m_{\widetilde{\chi }%
_{i}^{o}}\delta _{ij}\cos \alpha }{M_{w}}-2\left[ Q_{ij}\cos (\beta -\alpha
)+R_{ij}\cos \alpha \right] \right\} 
\]

\[
X_{3ij}=0
\]

\[
X_{1ij}^{`}=0
\]

\[
X_{2ij}^{`}=0
\]

\[
X_{3ij}^{`}=-\frac{ig}{2\sin \beta }\left\{ \frac{m_{\widetilde{\chi }%
_{i}^{o}}\delta _{ij}\cos \beta }{M_{w}}-2\left[ Q_{ij}\cos 2\beta
+R_{ij}\cos \beta \right] \right\} 
\]

where

\[
Q_{ij}=\frac{U_{i2}V_{j1}}{\sqrt{2}}
\]

\[
R_{ij}=\frac{MU_{i1}V_{j1}+\mu U_{i2}V_{j2}}{2M_{w}}
\]

the values of the constants $e$, $\theta _{w}$, $g$, $g^{`}$, given in
natural uints ($\hbar =c=1$) are

\[
e=0302822
\]

\[
\theta _{w}=0.495293
\]

\[
g=\frac{e}{\sin \theta _{w}}=0.637132
\]

\[
g^{`}=\frac{e}{\cos \theta _{w}}=0.344183
\]

The masses of the electron, the $Z$, and $W$, bosons are

\[
m_{e}=0.51099906\text{ }MeV/c^{2}
\]

\[
M_{W}=80.33\text{ }GeV/c^{2}
\]

\[
M_{Z}=91.187\text{ }GeV/c^{2}
\]

The width of $Z$ and $W$ bosons are

\[
\Gamma _{W}=2.07\text{ }GeV/c^{2}
\]

\[
\Gamma _{Z}=2.490\text{ }GeV/c^{2}
\]

The mixing angles $\alpha $ and $\beta $ and the masses $M$ and $M^{`}$ and $%
\mu $ are model dependent free parameters and they are defined in chapter 3
together with the sign factors $\eta _{i}$ and $\varepsilon _{j}$, masses $%
m_{\widetilde{\chi }_{i}^{\pm }}$ and $m_{\widetilde{\chi }_{i}^{o}}$ and
the mixing matrices $U$, $V$ and $Z$ of charginos and neutralinos.
\textbf{Three body kinematics}

\textbf{Introduction}
The energy $E$ and the 3-momentum\textbf{\ }$\mathbf{p}$ of a particle of
mass $m$ form a 4-vector momentum $p$ $\equiv (E,\mathbf{p)}$ whose square $%
p^{2}\equiv E^{2}-\left\vert \mathbf{p}^{2}\right\vert =m^{2}$. The scalear
dot product of two 4-momenum $p_{1}.p_{2}=E_{1}E_{2}-\mathbf{p}_{1}.\mathbf{p%
}_{2}$ is invariant (frame independent).

In this appendix we describe the three body reaction kinematics and
cross-sections in terms of invariant 4-momenta dot products [39], [40].

\textbf{Dot Products of Momenta}

Three body reactions consists of two particles of 4-momenta $p_{1}$ and $%
p_{2}$ and masses $m_{1}$ and $m_{2}$ in the initial state which scatter to
three particles of 4-momenta $p_{3},p_{4},$ and $p_{5}$ and masses $m_{3}$, $%
m_{4}$ and $m_{5}$ in the final state. When the particles in the initial
state are light with respect to their energies (e.g. electrons and
positrons); the masses $m_{1}$ and $m_{2}$ are ignored and we take $m_{1}$=$%
m_{2}\simeq 0.$

The Lorentz covariant variables, or the dot products in the case of three
body reactions with light particles in the initial state are defined as

$p_{1}.p_{1}=m_{1}^{2}\approx 0,$

$p_{1}.p_{2}=\frac{1}{2}(s-m_{1}^{2}-m_{2}^{2})\simeq \frac{s}{2}$

$p_{1}.p_{3}\equiv \zeta $

$p_{1}.p_{4}=\frac{1}{2}(s+m_{1}^{2}-m_{2}^{2}-2x-2\zeta )\approx \frac{1}{2}%
(s-2x-2\zeta )$

$p_{1}.p_{5}\equiv x$

$p_{2}.p_{2}=m_{2}^{2}\approx 0$

$p_{2}.p_{3}=\frac{1}{2}(s^{\prime }+2y-2\zeta +m_{3}^{2}-m_{4}^{2})$

$p_{2}.p_{4}=\frac{1}{2}(2\zeta
+2x-2y-m_{1}^{2}+m_{2}^{2}-m_{3}^{2}+m_{4}^{2}-m_{5}^{2})$

\qquad \qquad $\approx \frac{1}{2}(2\zeta
+2x-2y-m_{3}^{2}+m_{4}^{2}-m_{5}^{2})$

$p_{2}.p_{5}=\frac{1}{2}(s-s^{\prime }-2x+m_{5}^{2})$

$p_{3}.p_{3}=m_{3}^{2}$

$p_{3}.p_{4}=\frac{1}{2}(s^{\prime }-m_{3}^{2}-m_{4}^{2})$

$p_{3}.p_{5}\equiv y$

$p_{4}.p_{4}=m_{4}^{2}$

$p_{4}.p_{5}=\frac{1}{2}(s-s^{\prime }-2y-m_{5}^{2})$

$p_{5}.p_{5}=m_{5}^{2}$

where the variables $s$ (square of the center of mass energy) and $s^{\prime
}$ are defined from

\[
\sqrt{s}=p_{1}+p_{2}
\]

\[
\sqrt{s^{^{\prime }}}=p_{3}+p_{4}
\]

The function $\zeta $ is defined as

\[
\zeta =\frac{1}{\Lambda \left( \sqrt{s},\sqrt{s^{^{\prime }}}%
,m_{5}^{2}\right) }\left[ 
\begin{array}{c}
-m_{5}^{2}(s+m_{1}^{2}-m_{2}^{2}) \\ 
+x(s^{^{\prime }}+m_{3}^{2}-m_{4}^{2})(s-s^{^{\prime }}+m_{5}^{2}) \\ 
+y(s+m_{1}^{2}-m_{2}^{2})(s-s^{^{\prime }}-m_{5}^{2}) \\ 
-2xy(s+s^{^{\prime }})%
\end{array}%
\right] 
\]

For massless or light particles in the initial state; $\zeta $ is
approximated as

\[
\zeta =\frac{1}{\Lambda \left( \sqrt{s},\sqrt{s^{^{\prime }}}%
,m_{5}^{2}\right) }%
\begin{array}{c}
-m_{5}^{2}s(s^{^{\prime }}+m_{3}^{2}-m_{4}^{2}) \\ 
+x(s^{^{\prime }}+m_{3}^{2}-m_{4}^{2})(s-s^{^{\prime }}+m_{5}^{2}) \\ 
+ys(s-s^{^{\prime }}-m_{5}^{2}) \\ 
-2xy(s+s^{^{\prime }})%
\end{array}%
\]

where

\[
\Lambda (a,b,c)=\sqrt{a^{4}+b^{4}+c^{4}-2a^{2}b^{2}-2a^{2}c^{2}-2b^{2}c^{2}}
\]

\textbf{Differential Cross Section}

The covariant phase space differential cross-section for three body
reactions can be written as

\[
d\sigma =\frac{(2\pi )^{4}\left\vert M\right\vert ^{2}}{4\sqrt{%
(p_{1}.p_{2})^{2}-m_{1}^{2}m_{2}^{2}}}d\Phi _{3}
\]

where $M$ is the matrix element (amplitude) for the reaction or scattering
process and $d\Phi _{3}$ is an element of the three body phase space given by

\[
d\Phi _{3}=\frac{1}{(2\pi )^{9}}d\rho _{3}
\]

where $d\rho _{3}$ is the Lorentz invariant phase space volume element

\[
d\rho _{3}=\delta ^{4}(p_{1}+p_{2}-p_{3}-p_{4}-p_{5})\frac{d^{3}\left\vert 
\mathbf{p}_{3}\right\vert }{2E_{3}}\frac{d^{3}\left\vert \mathbf{p}%
_{4}\right\vert }{2E_{4}}\frac{d^{3}\left\vert \mathbf{p}_{5}\right\vert }{%
2E_{5}}
\]

the energy and momenta are related by

\begin{eqnarray*}
p_{i} &=&(E_{i},\mathbf{p}_{i}) \\
p_{i}^{2} &=&E_{i}^{2}-\mathbf{p}_{i}^{2}=m_{i}^{2}
\end{eqnarray*}

In center-of-mass frame

\[
\sqrt{(p_{1}.p_{2})^{2}-m_{1}^{2}m_{2}^{2}}=\left\vert \mathbf{p}%
_{lcn}\right\vert \sqrt{s}
\]

for light particles in the initial state, this expression is approximated as

\[
\sqrt{(p_{1}.p_{2})^{2}-m_{1}^{2}m_{2}^{2}}\approx \sqrt{(p_{1}.p_{2})^{2}}%
=p_{1}.p_{2}=\frac{1}{2}(s-m_{1}^{2}-m_{2}^{2})\approx \frac{s}{2}
\]

\textbf{Total Cross Section}

The Lorentz invariant three body phase space volume element in terms of
covariant dot products is given by

\[
d\rho _{3}=\frac{\pi }{2}\frac{ds^{^{\prime }}dxdyd\phi }{\Lambda \left( 
\sqrt{s},m_{1},m_{2}\right) \Lambda \left( \sqrt{s},\sqrt{s^{^{\prime }}}%
,m_{5}\right) }
\]

for light articles in the initial state

\[
d\rho _{3}\approx \frac{\pi }{2}\frac{ds^{^{\prime }}dxdyd\phi }{\Lambda
\left( \sqrt{s},0,0\right) \Lambda \left( \sqrt{s},\sqrt{s^{^{\prime }}}%
,m_{5}\right) }=\frac{\pi }{2}\frac{ds^{^{\prime }}dxdyd\phi }{s\Lambda
\left( \sqrt{s},\sqrt{s^{^{\prime }}},m_{5}\right) }
\]

where $\phi $ is the polar angle of $\mathbf{p}_{1}$ with respect to $%
\mathbf{p}_{5}$.

The total cross-section as a function of the center-of-mass energy $\sqrt{s}$
given by

\[
\sigma (\sqrt{s})=\int\limits_{s_{-}^{^{\prime }}}^{s_{+}^{^{\prime
}}}ds^{^{\prime
}}\int\limits_{x_{-}}^{x_{+}}dx\int\limits_{y_{-}}^{y_{+}}dy\int%
\limits_{z_{-}}^{z_{+}}dz\int\limits_{0}^{2\pi }d\phi \left[ \frac{1}{128\pi
^{4}}\frac{\left\vert M\right\vert ^{2}}{\left\vert \mathbf{p}%
_{lcn}\right\vert \sqrt{s}\Lambda \left( \sqrt{s},m_{1},m_{2}\right) \Lambda
\left( \sqrt{s},\sqrt{s^{^{\prime }}},m_{5}\right) }\right] 
\]%
and for light particles in the initial state

\[
\sigma (\sqrt{s})=\int\limits_{s_{-}^{^{\prime }}}^{s_{+}^{^{\prime
}}}ds^{^{\prime
}}\int\limits_{x_{-}}^{x_{+}}dx\int\limits_{y_{-}}^{y_{+}}dy\int%
\limits_{z_{-}}^{z_{+}}dz\int\limits_{0}^{2\pi }d\phi \left[ \frac{1}{128\pi
^{4}}\frac{\left\vert M\right\vert ^{2}}{s^{2}\Lambda \left( \sqrt{s},\sqrt{%
s^{^{\prime }}},m_{5}\right) }\right] 
\]

If the variable angle $\phi $ is cyclic, the integration over $\phi $ just
contributes a factor of $2\pi $ to the phase space. \ The expression then
becomes

\[
\sigma (\sqrt{s})=\int\limits_{s_{-}^{^{\prime }}}^{s_{+}^{^{\prime
}}}ds^{^{\prime }}\int\limits_{x_{-}}^{x_{+}}dx\int\limits_{y_{-}}^{y_{+}}dy%
\left[ \frac{1}{64\pi ^{3}}\frac{\left\vert M\right\vert ^{2}}{s^{2}\Lambda
\left( \sqrt{s},\sqrt{s^{^{\prime }}},m_{5}\right) }\right] 
\]

The domain of integration for $s^{^{\prime }},x,$ and $y$ are given by

$s_{-}^{^{\prime }}=(m_{3}+m_{4})^{2}$

$s_{+}^{^{\prime }}=(s-m_{5})^{2}$

$x_{\pm }=\frac{1}{4s}\left[ \left( s+m_{1}^{2}+m_{2}^{2}\right) \left(
s-s^{^{\prime }}+m_{5}^{2}\right) \pm \Lambda \left( \sqrt{s}%
,m_{1},m_{2}\right) \Lambda \left( \sqrt{s},\sqrt{s^{^{\prime }}}%
,m_{5}\right) \right] $

\qquad $\approx \frac{1}{4s}\left[ \left( s+0+0\right) \left( s-s^{^{\prime
}}+m_{5}^{2}\right) \pm \Lambda \left( \sqrt{s},0,0\right) \Lambda \left( 
\sqrt{s},\sqrt{s^{^{\prime }}},m_{5}\right) \right] $

\qquad $\approx \frac{1}{4s}\left[ \left( s-s^{^{\prime }}+m_{5}^{2}\right)
\pm \Lambda \left( \sqrt{s},\sqrt{s^{^{\prime }}},m_{5}\right) \right] ,$

$y_{\pm }=\frac{1}{4s^{^{\prime }}}\left[ \left( s^{^{\prime
}}+m_{3}^{2}+m_{4}^{2}\right) \left( s-s^{^{\prime }}+m_{5}^{2}\right) \pm
\Lambda \left( \sqrt{s^{^{\prime }}},m_{3},m_{4}\right) \Lambda \left( \sqrt{%
s},\sqrt{s^{^{\prime }}},m_{5}\right) \right] .$

The resulting total cross section is in terms of $GeV^{-2}$. To convert it
into pbarn, we multiply it by the conversion constant

\[
(\hbar c)^{-2}=0.38937966\times 10^{9}GeV^{2}\text{ pbarn.}
\]
\newpage
\begin{center}
\Large{Refrences}
\end{center}
\begin{enumerate}
    \item I. J. Aitchison and A. J. Hey, \textbf{Gauge Theories in Particle
    Physics}, Adam Hilger Ltd., Bristol, 1983.
    \item S. M. Bilenky and J. Hosek, Phys. Reports 90 (1982) 73.
    \item S. Weinberg, Rev. of Mod. Phys. 52 (1980) 515.
    \item A. Salam, Rev. of Mod. Phys. 52 (1980) 525.
    \item S. L. Glashow, Rev. of Mod. Phys. 52 (1980) 539.
    \item M. A. B. Beg and Sirlin, Phys. Reports 88 (1980) 1.
    \item M. Chen et. al., Phys. Reports 159 (1988) 201.
    \item M. F. Sohnius, Phys. Reports 128 (1985) 39.
    \item H. Baer et. al. Phys. Rev. D 48 (1993) R2978.
    \item H. J. W. Mueller-Kirsten and A. Wiedemann,
    Supersymmetry, World Scientific Ltd., Singapore, 1987;\\
          P. D. B. Collins, Particle Physics and Cosmology, John
          Wiley \& Sons Inc., USA, 1989;\\
          P. Fayet and S. Ferrara, Phys. Reports 32 (1977) 249;\\
          B. Zumino, Phys. Reports 104 (1984) 113;\\
          H. P. Nilles, Phys. Reports 110 (1984) 1.
    \item H. E. Haber, et. al., Nucl. Phys. B 161 (1979) 493.
    \item A. Mendez, Mod. Phys. Lett. 5 (1990) 2087.
    \item P. Fayet, Phys. Reports 105 (1984) 21.
    \item H. Georgi, Hadr. Jour. of Phys. 1 (1978) 155.
    \item S. Bertolini, Nucl. Phys. B 272 (1986) 77.
    \item J. F. Gunion and H. E. Haber, Phys. Rev. D 37 (1988)
    2515.
    \item A. Bartl et. al., Zeit. f. Phys. C 30 (1986) 441.
    \item A. Bartl et. al., Zeit. f. Phys. C 35 (1992) 257.
    \item A. Bartl et. al., Nucl. Phys. B 278 (1986) 1.
    \item A. Bartl et. al., Phys. Rev.  D 40 (1989) 1594.
    \item M. M. El-Kheishin, Phys. Rev. D 45 (1992) 4345.
    \item H. E. Haber, Phys. Rev. D 54 (1996) 687.
    \item E. Ma, Phys. Rev. D 49 (1994) 1663.
    \item S. Okubo, Prog. of Theor. Phys. 27 (1962) 1949.
    \item R. L. Pease, Inter. Jour. of Theor. Phys. 34 (1995) 183
    \item F. Abe at. al., Fermilab Pub. 94/097-E CDF.
    \item T. Tsukamoto et. al., Phys. Rev. D 51 (1995) 3153.
    \item J. L. Feng and M. J. Strassler, Phys. Rev. D 51 (1995)
    4661.
    \item J. L. Feng at. al., Phys. Rev. 52 (1995) 1418.
    \item S. Ambrosanio and B. Mele, Phys. Rev. D 53 (1996) 2541.
    \item C. Itzykson and J. B. Zuber, Quantum Field Theory, Mc
    Graw-Hill Inc., USA, 1980.
    \item M. Veltman, Diagramatica, Cambridge University Press,
    1994.
    \item H.E. Haber and G. L. Kane, Phys. Rev. D 117 (1985) 75.
    \item J. F. Gunion and H. E. Haber, Nucl. Phys. B 272 (1986)
    1;\\
          J. F. Gunion and H. E. Haber, Nucl. Phys. B 278 (1986)
          449;\\
          J. F. Gunion and H. E. Haber, Nucl. Phys. B 307(1988)
          445.
    \item H.E. Haber and D. Wyler, Nucl. Phys. B 323 (1989) 267.
    \item J. Rosiek, Phys. Rev. D 41 (1990) 3464.
    \item J. W. Williamson, Am. Jour. of Phys. 33 (1965)987.
    \item Review of Particle Properties, Phys. Rev. D 54 Part II
    (1996) 175.
    \item Alexander Pukhov, CalcHEP 3.2: MSSM, structure functions, event generation, batchs, and generation of matrix elements for other packages, hep-ph/0412191
    \item Abdelhak Djouadi, Jean-Loic Kneur, Gilbert Moultaka, SuSpect: a Fortran Code for the Supersymmetric and Higgs Particle Spectrum in the MSSM, hep-ph/0211331
\end{enumerate}

\end{document}